\documentclass[%
reprint,
nofootinbib,
 amsmath,amssymb,
 aps,
]{revtex4-1}

\makeatletter
\renewcommand{\p@subsection}{}
\renewcommand{\p@subsubsection}{}
\makeatother

\usepackage{tabularx}                                 
\usepackage{graphicx}
\usepackage{xcolor}
\usepackage{eurosym}
\usepackage{subfigure}
\usepackage{wasysym}
\usepackage{amssymb,amsmath,array}
\usepackage{latexsym, amsfonts, wasysym}
\usepackage[colorlinks = true,
            linkcolor = blue,
            urlcolor  = blue,
            citecolor = blue,
            anchorcolor = blue]{hyperref}

\def\beq{\begin{equation}}
\def\eeq#1{\label{#1}\end{equation}}
\def\eeqn{\end{equation}}

\newenvironment{Eqnarray}%
   {\arraycolsep 0.14em\begin{eqnarray}}{\end{eqnarray}}
\def\beqa{\begin{Eqnarray}}
\def\eeqa#1{\label{#1}\end{Eqnarray}}
\def\eeqan{\end{Eqnarray}}
\def\CR{\nonumber \\ }

\def\leqn#1{(\ref{#1})}

\let\bar=\overbar

\def\ie{{\it i.e.}}
\def\eg{{\it e.g.}}
\def\etc{{\it etc.}}

\def\ifb{ fb$^{-1}$ }
\def\iab{ ab$^{-1}$ } 

\def\lsim{\mathrel{\raise.3ex\hbox{$<$\kern-.75em\lower1ex\hbox{$\sim$}}}}
\def\gsim{\mathrel{\raise.3ex\hbox{$>$\kern-.75em\lower1ex\hbox{$\sim$}}}}

\def\L{{\cal L}}

\def\O{{\cal O}}

\def\half{\frac{1}{2}}

\def\del{\partial}
\def\Dslash{\not{\hbox{\kern-4pt $D$}}}
\def\dslash{\not{\hbox{\kern-2pt $\del$}}}

\def\ee{e^+e^-}

\def\msb{{\bar{\scriptsize M \kern -1pt S}}}

\def\drb{{\bar{\scriptsize D \kern -1pt R}}}

\def\GeV{ \mbox{GeV} }
\def\TeV{  \mbox{TeV} } 
\def\MeV{ \mbox{MeV} }

\makeatletter
\def\section{\@startsection{section}{0}{\z@}{5.5ex plus .5ex minus
 1.5ex}{2.3ex plus .2ex}{\large\bf}}
\def\subsection{\@startsection{subsection}{1}{\z@}{3.5ex plus .5ex minus
 1.5ex}{1.3ex plus .2ex}{\normalsize\bf}}
\def\subsubsection{\@startsection{subsubsection}{2}{\z@}{-3.5ex plus
-1ex minus  -.2ex}{2.3ex plus .2ex}{\normalsize\sl}}

\renewcommand{\@makecaption}[2]{%
   \vskip 10pt
   \setbox\@tempboxa\hbox{\small #1: #2}
   \ifdim \wd\@tempboxa >\hsize     
       \small #1: #2\par          
     \else                        
       \hbox to\hsize{\hfil\box\@tempboxa\hfil}
   \fi}

\makeatother

\def\matheuro{{ \mbox{\euro}}}

\def\sid{SiD}

\def\micron{\ensuremath{\mu\mathrm{m}}}

\def\toprule{\hline}
\def\midrule{\hline}
\def\bottomrule{\hline}
\def\pT{\ensuremath{p_T}}
\newcommand {\sub}[1]{\ensuremath{_{\mathrm{#1}}}}
\newcommand {\siunit}[2]{\ensuremath{#1\,{\mathrm{#2}}}}
\newcommand {\num}[1]{\ensuremath{#1}}

\def\ALR{\ensuremath{A_{\mathrm{LR}}}}
\def\Leff{\ensuremath{\mathcal{L}_{\mathrm{eff}}}}
\def\Peff{\ensuremath{P_{\mathrm{eff}}}}
\def\Pem{\ensuremath{P_{e^-}}}
\def\Pep{\ensuremath{P_{e^+}}}
\def\Pmp{\ensuremath{P(e^-,e^+)}}
\def\sigmaLL{\ensuremath{\sigma_{\mathrm{LL}}}}
\def\sigmaLR{\ensuremath{\sigma_{\mathrm{LR}}}}
\def\sigmaRL{\ensuremath{\sigma_{\mathrm{RL}}}}
\def\sigmaRR{\ensuremath{\sigma_{\mathrm{RR}}}}

\begin{document}

\rightline{\begin{tabular}{l} 
     DESY 19-037, 
     FERMILAB-FN-1067-PPD,  
     IFIC/19-10\\  IRFU-19-10, 
    JLAB-PHY-19-2854, 
     KEK Preprint 2018-92 \\
       LAL/RT 19-001,   PNNL-SA-142168, 
     SLAC-PUB-17412 \\
        March 2019 \end{tabular}}

\title{\vskip 0.4in The International Linear Collider \\ A Global Project}

\author{\textbf{Prepared by:}
Philip Bambade$^1$, Tim Barklow$^2$,
Ties Behnke$^3$, Mikael Berggren$^3$, James Brau$^4$, Philip
Burrows$^5$, Dmitri Denisov$^{6,7}$, 
Angeles Faus-Golfe$^{1}$, Brian
Foster$^{3,5}$, Keisuke Fujii$^{8}$, Juan Fuster$^{9}$, Frank
Gaede$^{3}$, Paul Grannis$^{10}$, Christophe Grojean$^{3}$, Andrew
Hutton$^{11}$, Benno List$^{3}$, Jenny List$^{3}$, Shinichiro
Michizono$^{8}$, Akiya Miyamoto$^{8}$, Olivier Napoly$^{12}$, Michael
Peskin$^{2}$, Roman P\"oschl$^{1}$, Frank Simon$^{13}$,  Jan
Strube$^{4,14}$, Junping Tian$^{15}$, Maksym Titov$^{12}$, Marcel
Vos$^{8}$, Andrew White$^{16}$, Graham Wilson$^{17}$, 
Akira Yamamoto$^{8}$, Hitoshi Yamamoto$^{18}$,  Kaoru Yokoya$^8$  }
\affiliation{\vspace{.2 cm}
$^1$LAL-Orsay/CNRS, $^2$SLAC,
$^3$DESY, $^4$U. Oregon, $^5$Oxford U., $^6$BNL, $^7$Fermilab, \\ $^{8}$KEK, $^{9}$IFIC,
U. Valencia-CSIC,  $^{10}$Stony Brook U., $^{11}$Jefferson Lab,\\
$^{12}$IRFU, CEA Saclay, $^{13}$Max Planck Inst., Munich, 
$^{14}$PNNL, $^{15}$U. Tokyo, \\  $^{16}$U. Texas, Arlington, $^{17}$U. Kansas, $^{18}$U. Tohoku }

%

\collaboration{Representing the Linear Collider Collaboration and the
  global ILC community.}

\begin{abstract}
The International Linear Collider (ILC) is now under consideration as
the next global project in particle physics.   In this report, we
review of all aspects of the ILC program:  the physics motivation, the
accelerator design, the run plan, the proposed detectors, the
experimental measurements on the Higgs boson, the top quark, the
couplings of the $W$ and $Z$ bosons, and searches for new
particles. We review the important role that polarized beams play in
the ILC program.  The first stage of the ILC is planned to be a Higgs
factory at 250~GeV in the centre of mass.  Energy upgrades can
naturally be implemented based on the concept of a linear collider. We
discuss in detail the ILC program of Higgs boson measurements and the
expected precision in the determination of Higgs couplings.  We
compare the ILC capabilities to those of the HL-LHC and to those of
other proposed $\ee$ Higgs factories.  We emphasize throughout that
the readiness of the accelerator and the estimates of ILC performance
are based on detailed simulations backed 
by extensive R\&D and, for the accelerator technology, operational experience.
\end{abstract}

\maketitle

\tableofcontents

\clearpage
\newpage

\section{\label{sec:intro}Introduction}


While the Standard Model (SM) is a highly successful theory of
the fundamental interactions, it has serious shortcomings. New
fundamental interactions are 
required to address them.   A central focus of particle physics
now involves
searching for these new interactions and associated new particles.  
The SM is theoretically
 self-consistent, but it does not answer many obvious questions about
 particle physics.  It has no explanation for the 
dark matter or dark energy
that is observed in the cosmos,
or for  the cosmic excess of matter over antimatter.   It does
not address the mass scale of quarks, leptons, and gauge 
bosons, which is significantly lower than the Planck scale.   
It does not explain
the large mass ratios among the SM particles or the values of the quark and neutrino 
mixing angles.    
These and other considerations provide a compelling motivation
for new interactions beyond the SM.   On the other hand,
the current success of the SM indicates  that further search will be
very challenging and, most likely, requires new approaches and new methods.

The discovery of the Higgs boson in 2012 revealed the final particle 
predicted in the SM.   Within that theory,
the Higgs boson is the agent for  electroweak symmetry breaking
and the generation of the masses of all elementary
particles.  Thus, it occupies a central role in the SM and,
specifically, in many of the unresolved issues that we have listed above. 
 The properties of the Higgs boson are precisely specified
in the SM, while models of new interactions that address these issues
lead to
significant corrections to those predictions.  Thus, high-precision
measurement of the Higgs boson offers a new and promising 
 avenue for searches for
new physics beyond the Standard Model.   The discovery of deviations
of the Higgs boson properties from the SM predictions could well
provide the first evidence for new physics beyond the SM. 
 
This study of the Higgs boson properties is the most prominent goal of
the International Linear Collider (ILC).   The ILC has been designed
with this goal in mind, to provide a complete, high-precision picture
of the Higgs boson and its interactions.  Though the properties of the
Higgs boson are already being studied at the LHC, the ILC offers
significant advantages.  It will bring the measurements
to a new, qualitatively superior, level of precision, and it will
remove the many model-dependent assumptions required for the analysis
of the Higgs boson measurements at hadron colliders.   The ILC will be highly
sensitive to Higgs boson decays that yield invisible or
other exotic final states, giving unique tests of models of new weakly
interacting particles and dark matter. The ILC can also probe for
direct pair-production of particles with very weak interactions.  Since direct searches
at high-energy hadron colliders have not discovered new particles, it
is urgent and compelling to open this new path to the search for
physics beyond the SM.

As an $\ee$ linear  collider, the ILC brings a number of very powerful
experimental tools to bear on the challenge of producing a precise,
model-independent accounting of the Higgs boson properties.
The ILC has a well-defined, adjustable centre-of-mass energy.   It
produces conventional SM events at a level that is comparable to,
rather than overwhelmingly larger than, Higgs signal processes,
allowing easy selection of Higgs boson events.   At its initial stage
of 250~GeV, Higgs boson events are explicitly tagged by a recoil $Z$
boson.    At a linear collider, both the electron and positron beams
can be polarized, introducing additional observables.  Because all
electroweak  reactions at energies above the $Z$ resonance have
order-1  parity violation, beam polarization effects are large  and
provide access to critical physics information. 

After operation of a linear collider at the starting energy, it is
straightforward to upgrade the centre-of-mass energy.  This is the
natural path of evolution  for a new high-energy physics laboratory.
An upgrade in energy
systematically expands the list of physics processes that can be
studied with high precision and polarized beams.   An upgrade to
500~GeV
accesses the Higgs boson coupling to the top quark and the Higgs boson
self-coupling.  Together with the 250~GeV results, this  will give
 a complete accounting of the Higgs boson
profile.
An energy upgrade to 350~GeV begins the use of the ILC as a top quark
factory,
offering precision measurements of the top quark mass and electroweak
couplings.  At the same time, the ILC will study the reactions $\ee\to
f\bar f$ and $\ee\to W^+W^-$ with high precision.   Here also,
deviations from the SM predictions can indicate new physics. 
Finally, the ILC will
search directly for pair production of  weakly coupled particles  with
masses up to half
the centre-of-mass energy, without the requirement of special
signatures needed for searches at hadron colliders. 
Because of its upgrade capability and the unique access that $\ee$ beams give
to many important reactions, the ILC will 
continue to be a leading discovery machine in the 
world of particle physics for decades.

The ILC is mature in its design and ready for construction.   The technology
of the ILC has been advanced through a global program coordinated by
the International Committee for Future Accelerators
(ICFA).
In the mid-1990's, various technology options to
realise a high-energy linear collider were emerging. 
ICFA asked the 
Linear Collider Technical Review Committee to develop a standardised
way to  compare  these  technologies based on their parameters, such as
power consumption and luminosity. A second
review panel was organised by ICFA in 2002;
it concluded that both warm and cold technologies had
developed to the point where either could be the basis for a
high energy linear
collider. In 2004, the  International Technology Review Panel
(ITRP) was charged by ICFA to recommend an option that could focus the
worldwide R\&D effort.  This panel chose the  superconducting
radiofrequency technology (SCRF), in a large part due to its
energy efficiency and potential for broader applications. 

Today's design of the ILC accelerator is the result of 
nearly twenty years of
R\&D that has involved a broad, global community. 
The heart of the ILC, the SCRF cavities, is based on
pioneering work of the TESLA Technology Collaboration. Other aspects of the 
technology
emerged from the R\&D carried out for the competing linear collider projects JLC/GLC and NLC,
which were based on room-temperature accelerating structures. 
The ILC proposal
is supported by extensive R\&D and prototyping. The successful construction and
operation 
of the European XFEL (E-XFEL) at DESY provides
confidence both in the high reliability of the basic
technology and in the reliability of its performance and cost in 
industrial realisation.  Other communities acknowledge this; the SCRF 
 technology has also been chosen for new
free electron laser projects now under construction in the US and China.
Some specific optimisations and technological choices remain.
But the ILC is now ready to move forward to construction. 

The effort to design and
establish the technology for the linear collider culminated in the
publication of the Technical Design Report (TDR) for the International
Linear Collider (ILC) in 2013~\cite{Behnke:2013xla}. 
Twenty-four hundred (2400) scientists, from 48 countries and 
392 institutes and university
groups,
 signed the TDR.  This document
presented optimised collider and detector designs, and associated 
physics analyses based on their  expected performance.
From
2005 to the publication of the TDR, the
design of the ILC accelerator was conducted under the mandate of ICFA
as a worldwide
international collaboration, the Global Design Effort (GDE). 
Since 2013, ICFA has placed the  international activities for both the ILC and CLIC
projects under a single organisation, 
the Linear Collider Collaboration (LCC).

With knowledge of the mass of the Higgs boson, it became clear that
the 
linear collider could start its ambitious physics program
 with an initial centre-of-mass energy of 250~GeV at a cost
reduced from the TDR. A revised design of the ILC, the ILC250, was
thus  presented~\cite{Evans:2017rvt}.  This design  retains the final-focus and
beam-dump capability to extend the centre-of-mass energy to energies
as high as 1~TeV. Advances in the theoretical understanding of the impact of precision
measurements at the 
 ILC250 have justified that the 250~GeV operating point already gives
 substantial 
sensitivity to physics beyond the SM~\cite{Barklow:2017suo,Fujii:2017vwa}. 
 The cost estimate for ILC250 
  is similar in scale to that of the LHC.

In its current
form, the ILC250 is a $250\,{\mathrm{GeV}}$ centre-of-mass energy
(extendable up to a 1~TeV) linear $e^+e^-$ collider, based
on $1.3\,{\mathrm{GHz}}$ SCRF
cavities. It is designed to achieve a luminosity of $1.35\cdot
10^{34}~{\mathrm{cm}}^{-2}{\mathrm{s}}^{-1}$ and provide an integrated
luminosity of $400\,{\mathrm{fb}}^{-1}$ in the first four years of
running.  The scenario described in Section III gives a complete
program of $2\,{\mathrm{ab}}^{-1}$  of data at 250~GeV over 12 years.
The electron beam will be polarised to $80\,\%$, and the baseline plan includes an 
undulator-based
positron source which will  deliver
$30\,\%$ positron  polarisation.

The experimental community has developed
designs for two complementary detectors, ILD and SiD.  These detectors
are described in 
 \cite{Behnke:2013lya}. They are designed to 
optimally address the
ILC physics goals, with complementary approaches. One detector is based on
TPC tracking (ILD) and one on silicon tracking (SiD).
Both employ particle flow calorimetry based on
calorimeters with unprecedented fine segmentation.
Extensive R\&D and prototyping gives confidence that the
unprecedented levels of performance in calorimetry, tracking, and
 particle identification 
required to achieve the 
physics programme can be realised.
The  extensive course of
prototyping justifies our estimates of full-detector performance 
and cost.  
The detector R\&D program leading to these designs
has  contributed a number of advances in 
detector capabilities with applications well beyond the linear
collider program.    Similarly to the situation for the collider, some 
final optimizations and technology choices 
will need to be completed in the next few years. 

There is broad interest in Japan to host the
international effort to realise the ILC project.  
This interest has been growing over many years.
Political entities promoting the plan to host the ILC in Japan include 
the Federation of Diet Members for ILC and the Advanced Accelerator
Association, a consortium of industrial representatives that includes
most of the large high-tech companies in Japan.   The ILC has been
endorsed by the community of Japanese particle physicists (JAHEP)~\cite{Asai:2017pwp}. 
Detailed review in Japan of the many aspects of the
project is nearing a conclusion.
Since 2013 the MEXT ministry has been examining the ILC project in
great detail, including the aspect of risk minimisation.
This review concluded when
MEXT's ILC Advisory Panel released 
its report~\cite{AdvPanel} on July 4, 2018, summarising the
studies of the several working groups (WG) that
reviewed 
a broad range of aspects of the ILC.  The most recent studies include
a specific review of the scientific merit and the technical design for the ILC250. 
The  Physics WG scrutinised the scientific merit of the ILC250,
leading to their strong and positive statement on the importance of
the ILC250 to 
measure precisely the couplings of the Higgs boson \cite{AdvPanel}.
The TDR WG reviewed issues addressed in the Technical Design Report
and the ILC250 design, including the  cost estimate and technical feasibility.  
Other working groups of the MEXT review commented on manpower needs, 
organisational aspects, and the experience of previous large projects.
The report of the ILC Advisory Panel was followed by the beginning of
deliberations in a committee and technical working group 
established by the Science Council of Japan (SCJ),
the second stage of the review process.   The SCJ released its review
on December 19, 2018~\cite{SJCreport}.   The review
acknowledges a consensus in the particle physics community that
``the research topic
of precise measurement of Higgs couplings is extremely important'' but
expresses doubts about the cost of the project, which is well beyond
the scale of other proposals that have come before this committee.
The financing of the project will depend on negotiations with
international partners, led by the Japanese government after a clear
statement of interest.   The Japanese government is now preparing for
this step,  which can be followed by a move to the next phase of international
negotiations.  A new  independent committee (LDP Coordination Council
for the Realization of ILC),
led by high-ranking members of the Liberal Democratic 
Party, the majority party in the Diet,  has now 
convened to encourage the national government along this path.

Given a positive signal by the Japanese government, the ILC could move 
forward rapidly.
The potential timeline would have an initial period of 
about 4 years to obtain
 international agreements, prepare for the  construction, and form 
 the international laboratory and its governance structure.
 The construction phase would then need 9 years.

It is an important aspect of the discussions of the ILC in Japan that the
ILC has been organized from the beginning as a 
global project that will foster exchange between Japan
and other nations.   Thus, the  
scientific interest and political engagement of partner countries is of
major importance
for the Japanese authorities.  

The purpose of this report is to set out in detail the current status
of the ILC project, expanding on a recent paper
prepared as input to the Update of the
European Strategy for Particle Physics~\cite{Aihara:2019gcq}.  We discuss 
the physics reach of the ILC, the technological maturity of the accelerator,
detector, and software/computing designs,
and the further steps 
 needed to concretely realise the project.  Section~\ref{sec:ilc}
 describes the accelerator design and technology, reviewing both 
 current status of SCRF development and the general layout of the
 machine.  This section also presents luminosity and energy upgrade
 options,  as well as civil engineering plans, including site
 specific details, and cost and schedule estimates.
 Section \ref{sec:runscenarios}  presents the current
 thinking about the operations of the ILC, with estimates of the plan
 and schedule for the 
 collection of integrated luminosity.
 Section \ref{sec:physics} gives an overview of the physics case for
 the ILC as a 250~GeV collider.
 This includes a more detailed discussion of the significance of the Higgs boson 
 as a tool for searching for physics beyond the SM, the qualitative
 comparison of the ILC to the LHC as a facility for precision Higgs
 studies, and the theoretical approach for extracting Higgs boson
 couplings from $\ee$ data.  This section also discusses the physics
 opportunities of searches for exotic Higgs decays and studies of
 other processes of interest including SM fermion pair-production and
 searches for
 new particle pair production. Section~\ref{sec:highenergy} described
 the additional opportunities that the energy extension to 500~GeV
 will make available. 
 
 Section \ref{sec:detectors}
 provides detailed descriptions of the ILC detector designs
 that have been developed by the community,
 through detector R\&D and prototyping, and used as detector
 models to show the simulated performance on the various
 physics channels. Section \ref{sec:software}
 summarises the computing needs of the ILC program,
 including software.  These two sections provide the basis for a
 discussion of the experimental measurements of reactions crucial to
 the ILC program. All of the projections of experimental
 uncertainties given in this paper are based on full-simulation
 studies using the model detectors described in
 Sec.~\ref{sec:detectors},
 with capabilities justified by
 extensive R\&D programs. 

Building on this discussion, 
Sec.~\ref{sec:higgs}
 gives a description of physics simulations involving Higgs
 boson  reactions.  Section~\ref{sec:ew} describes physics simulations
 carried out for the reactions $\ee\to W^+W^-$ and $\ee\to f\bar f$.
 Section~\ref{sec:top} discusses simulations of measurements of top
 quark properties at the energy-upgraded ILC.
 These studies lead to  concrete
 quantitative estimates for the expected uncertainties in Higgs boson
 coupling determinations.   Based on the results of these studies, we
present in Sec.~\ref{sec:global}  what we feel are conservative
estimates for the precision that the ILC will attain in a highly
model-independent analsys for the determination of the Higgs boson
width and absolutely normalized couplings.   We compare these
estimates to those presented in the CDRs for other  $\ee$ Higgs
factories and those expected from the high-luminosity phase of the LHC.

Section~\ref{sec:searches} describes the capability of the ILC for
direct searches for pair-production of new particles, covering  a number of scenarios
that are difficult for the LHC but which can be investigated in detail
at $\ee$ colliders.  Section~\ref{sec:conclusion} gives our conclusions.

\section{\label{sec:ilc}ILC Machine Design}


The International Linear Collider (ILC) is a $250\,{\mathrm{GeV}}$ (extendable up to $1\,{\mathrm{TeV}}$) linear $e^+e^-$ collider, based on $1.3\,{\mathrm{GHz}}$ superconducting radio-frequency (SCRF) cavities.
It is designed to  achieve a luminosity of $1.35\cdot 10^{34}~{\mathrm{cm}}^{-1}{\mathrm{s}}^{-1}$ and provide an integrated luminosity of $400\,{\mathrm{fb}}^{-1}$ in the first four years of running.
The electron beam will be polarised to $80\,\%$, and positrons with $30\,\%$ polarization will be provided if the undulator based positron source concept is employed. 

Its parameters have been set by physics requirements first outlined in 2003,
updated in 2006, and thoroughly discussed over many years with the physics user community. 
After the discovery of the Higgs boson it was decided that an initial energy of $250\,{\mathrm{GeV}}$ provides the opportunity for a precision Standard Model and Higgs physics programme at a reduced initial cost~\cite{Evans:2017rvt}.
Some relevant parameters are given in Tab.~\ref{tab:ilc-params}.
This design evolved from two decades of R\&D, described in Sec.~\ref{sec:intro}, 
an international effort coordinated first by the GDE under ICFA mandate and
since 2013 by the LCC.

\begin{table*}[tbhp]
\begin{tabular}{lccccccc}
Quantity & Symbol & Unit & Initial & ${\cal{L}}$ Upgrade & TDR &  \multicolumn{2}{c}{Upgrades} \\
\hline
Centre of mass energy & $\sqrt{s}$ & GeV & $250$ & $250$ & $250$ & $500$ & $1000$ \\
Luminosity & \multicolumn{2}{c}{${\mathcal{L}}$ ~~~~$10^{34}{\mathrm{cm^{-2}s^{-1}}}$} & $1.35$ & $2.7$ & $0.82$ & $1.8 / 3.6$ & $4.9$ \\
Polarisation for $e^- (e^+)$ & $P\sub{-} (P\sub{+})$ & & ~$80\,\% (30\,\%)$~ &  ~$80\,\% (30\,\%)$~ &  ~$80\,\% (30\,\%)$~ &~$80\,\% (30\,\%)$~ &  ~$80\,\% (20\,\%)$~  \\
Repetition frequency &$f\sub{{rep}}$ & ${\mathrm{Hz}}$  & $5$ & $5$ & $5$ & $5$ & $4$ \\
Bunches per pulse  &$n\sub{{bunch}}$ & 1  & $1312$ & $2625$ & $1312$ & $1312 / 2625$ & $2450$ \\
Bunch population  &$N\sub{{e}}$ & $10^{10}$ & $2$ &  $2$ & $2$ & $2$ & $1.74$ \\
Linac bunch interval & $\Delta t\sub{{b}}$ & ${\mathrm{ns}}$ & $554$ & $366$ & $554$ & $554 / 366$ & $366$ \\
Beam current in pulse & $I\sub{{pulse}}$ & ${\mathrm{mA}}$& $5.8$ & $5.8$& $8.8$ & $5.8$ & $7.6$  \\
Beam pulse duration  & $t\sub{{pulse}}$ & ${\mathrm{\mu s}}$ & $727$ & $961$ & $727$ & $727 / 961$ & $897$ \\
Average beam power  & $P\sub{{ave}}$   & ${\mathrm{MW}}$ & $5.3$ & $10.5$ & $10.5$ & $10.5 / 21$  & $27.2$ \\  
Norm. hor. emitt. at IP & $\gamma\epsilon\sub{{x}}$ & ${\mathrm{\mu m}}$& $5$ & $5$ & $10$ & $10$ & $10$  \\ 
Norm. vert. emitt. at IP & $\gamma\epsilon\sub{{y}}$ & ${\mathrm{nm}}$ & $35$ & $35$ & $35$ & $35$ & $30$ \\ 
RMS hor. beam size at IP  & $\sigma^*\sub{{x}}$ & ${\mathrm{nm}}$  & $516$ & $516$ & $729$ & $474$ & $335$ \\
RMS vert. beam size at IP &$\sigma^*\sub{{y}}$ & ${\mathrm{nm}}$ & $7.7$  & $7.7$  & $7.7$  & $5.9$ & $2.7$ \\
Luminosity in top $1\,\%$ & ${\mathcal{L}}\sub{0.01} / {\mathcal{L}}$ &  & $73\,\%$  &  $73\,\%$ & $87.1\,\%$  & $58.3\,\%$ & $44.5\,\%$\\
Energy loss from beamstrahlung  & $\delta\sub{BS}$ &  & $2.6\,\%$  & $2.6\,\%$  & $0.97\,\%$  & $4.5\,\%$ & $10.5\,\%$ \\
Site AC power  & $P\sub{{site}}$ &  ${\mathrm{MW}}$ & $129$ & & $122$ & $163$ & $300$ \\
Site length & $L\sub{{site}}$ &  ${\mathrm{km}}$ & $20.5$ & $20.5$ & $31$ & $31$ & $40$ \\
\end{tabular}
\caption{Summary table of the ILC accelerator parameters in the initial 250~GeV staged configuration (with TDR parameters at 250~GeV given for comparison) and possible upgrades.  A 500~GeV machine could also be operated at 250~GeV with 10~Hz repetition rate, bringing the maximum luminosity to 
\siunit{5.4\cdot 10^{34}}{cm^{-2} s^{-1}}~\cite{Harrison:2013nvaxxx}.}
\label{tab:ilc-params}
\end{table*}


The fundamental goal of the design of the ILC accelerator is a high energy-efficiency.  The ILC design
 limits the overall power consumption of the accelerator complex during operation to $129\,{\mathrm{MW}}$ at  $250\,{\mathrm{GeV}}$ and $300\,{\mathrm{MW}}$ at  $1\,{\mathrm{TeV}}$, which is comparable to the power consumption of CERN.
This is achieved by the use of SCRF technology for the main accelerator, which offers a high RF-to-beam efficiency through the use of superconducting cavities, operating at $1.3\,{\mathrm{GHz}}$, where high-efficiency klystrons are commercially available.
At accelerating gradients of $31.5$ to $35\,{\mathrm{MV/m}}$ this technology offers high overall efficiency and reasonable investment costs, even considering the cryogenic infrastructure needed for the operation at $2\,{\mathrm{K}}$.

The underlying TESLA technology is mature, with a broad industrial base throughout the world, and is in use at a number of free electron laser facilities that are in operation (E-XFEL at DESY, Hamburg), under construction (LCLS-II at SLAC, Stanford) or in preparation (SCLF in Shanghai) in the three regions Asia, Americas, and Europe that contribute to the ILC project.
In preparation for the ILC, Japan and the U.S. have founded a collaboration for further cost optimisation of the TESLA technology.
In recent years, new surface treatment technologies utilising nitrogen during the cavity preparation process, such as the so-called   nitrogen infusion technique, have been developed at Fermilab, with the prospect of  achieving higher gradients and lower loss rates with a less expensive surface preparation scheme than assumed in the TDR (see Sec.~\ref{subsubsec:highgrad}).

When the Higgs boson was discovered in 2012, the Japan Association of High Energy Physicists (JAHEP) made a proposal to host the ILC in Japan~\cite{JAHEP:2012a,JAHEP:2012b}. 
Subsequently, the Japanese ILC Strategy Council conducted a survey of possible sites for the ILC in Japan, looking for  suitable geological conditions for a tunnel up to $50\,{\mathrm{km}}$ in length (as required for a $1\,{\mathrm{TeV}}$  machine), and the possibility to establish a laboratory where several thousand international scientists can work and live. 
As a result, the candidate site in the Kitakami region in northern Japan, close to the larger cities of Sendai and Morioka, was found to be the best option. 
The site offers a large, uniform granite formation with no currently active faults and a geology that is well suited for tunnelling.
Even in the great Tohoku earthquake in 2011, underground installations in this rock formation were essentially unaffected~\cite{bib:sanuki:desy2017}, which underlines the suitability of this candidate site. 

This section starts with a short overview over the changes of the ILC design between the publication of the TDR in $2013$ and today, followed by a description of the SCRF technology, and an description of the overall accelerator design and its subsystems. 
Thereafter, possible upgrade options are laid out, the Japanese candidate site in the Kitakami region is presented, and costs and schedule of the accelerator construction project are shown.

 \begin{figure*}[tbhp]
 \begin{center}
 \includegraphics[width=\hsize]{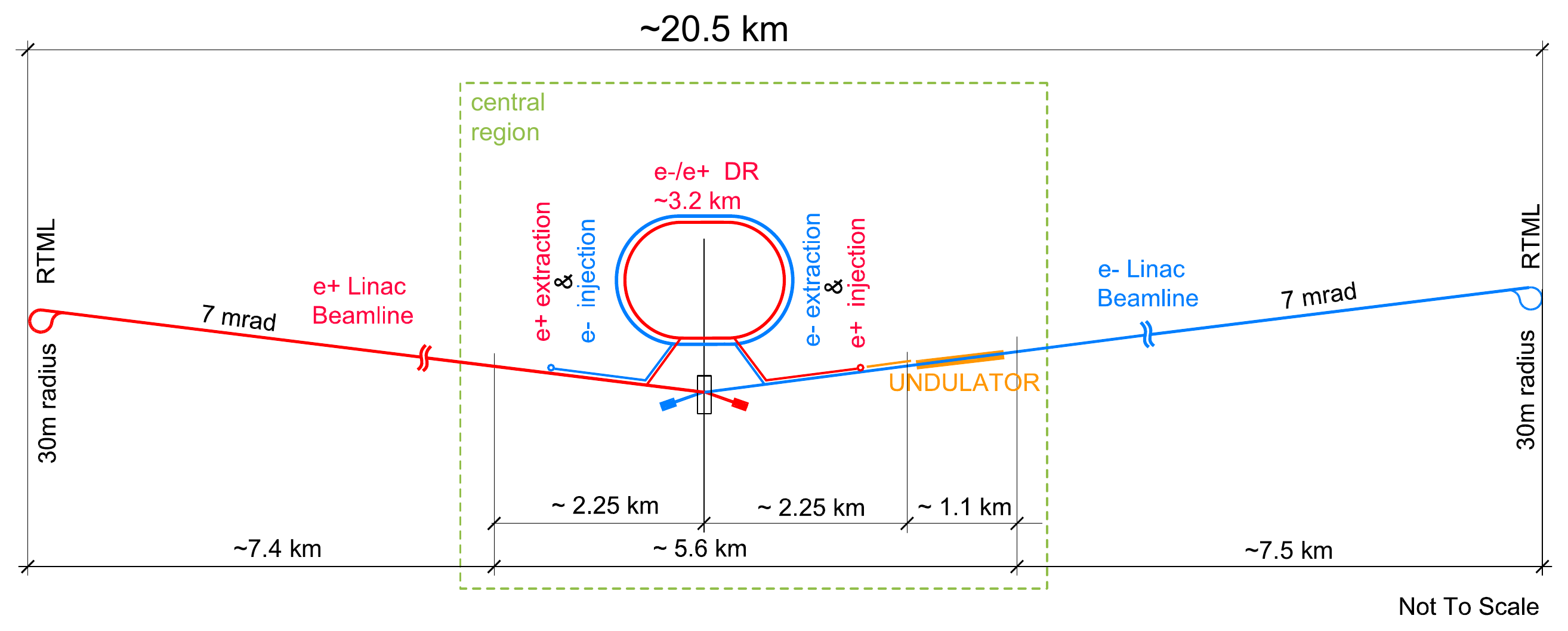}
\caption{Schematic layout of the ILC in the $250\,{\mathrm{GeV}}$ staged configuration.
\label{fig:ilc-schematic}}
 \end{center}
 \end{figure*}

\subsection{Design evolution since the TDR}
\label{sec:design_evo}

Soon  after the discovery of the Higgs boson, the Technical Design Report (TDR) for the ILC accelerator was published in 2013~\cite{Adolphsen:2013jya,Adolphsen:2013kya} after 8 years of work by the Global Design Effort (GDE).
The TDR design was based on the requirements set forth by the ICFA mandated parameters committee~\cite{Heuer:2006}:
\begin{itemize}
\item a centre-of-mass energy of up to \siunit{500}{GeV},
\item tunability of the centre-of-mass energy between $\sqrt{s} = \siunit{200}{GeV}$
 and \siunit{500}{GeV},
\item  a luminosity sufficient to collect \siunit{500}{fb^{-1}} within four years of operation, taking into account a three-year a ramp up.  This corresponds to a final luminosity of \siunit{250}{fb^{-1}} per year and an instantaneous luminosity of ${\mathcal{L}} = \siunit{2 \cdot 10^{34}}{cm^{-2}\,s^{-1}}$,
\item an electron polarisation of at least $80\,\%$,
\item  the option for a later upgrade to energies  up to \siunit{1}{TeV}.
\end{itemize}

The accelerator design presented in the TDR met these requirements (see Tab.~\ref{tab:ilc-params}), at an estimated construction cost of \siunit{7,982}{MILCU} for a Japanese site, plus \siunit{22.9}{Mh} (million hours) of labour in participating institutes~\cite[Sec. 15.8.4]{Adolphsen:2013kya}. 
Costs were expressed in ILC Currency Units ILCU, where \siunit{1}{ILCU} corresponds to \siunit{1}{US\$} at 2012 prices.

In the wake of the Higgs discovery, and the proposal by the Japan Association of High Energy Physicists (JAHEP) to host the ILC in Japan\cite{JAHEP:2012a} with its recommendation to start with a \siunit{250}{GeV} machine~\cite{JAHEP:2012b}, plans were made for a less expensive machine configuration with a centre--of--mass energy of $\sqrt{s} = \siunit{250}{GeV}$, around the maximum of the $Zh$ production cross section, half the TDR value.
Various options were studied in the TDR~\cite[Sect. 12.5]{Adolphsen:2013kya} and later~\cite{Dugan:2014}.
This resulted in a revised proposal~\cite{Evans:2017rvt} for an accelerator with an energy of \siunit{250}{GeV} and a luminosity of ${\mathcal{L}} = \siunit{1.35 \cdot 10^{34}}{cm^{-2}\,s^{-1}}$, capable of delivering about \siunit{200}{fb^{-1}} per year, or \siunit{400}{fb^{-1}} within the first four years of operation, taking into account the ramp-up.

Several other changes of the accelerator design have been approved by the ILC Change Management Board since 2013, in particular:
\begin{itemize}
\item The free space between the interaction point and the edge of the final focus quadrupoles ($L^*$) was unified between the ILD and SiD detectors~\cite{bib:cr-0002}, facilitating a machine layout with the best possible luminosity for both detectors.
\item A vertical access shaft to the experimental cavern was foreseen~\cite{bib:cr-0003}, allowing a CMS-style assembly concept for the detectors, where large detector parts are built in an above-ground hall while the underground cavern is still being prepared. 
\item The shield wall thickness in the Main Linac tunnel was reduced from $3.5$ to \siunit{1.5}{m}~\cite{bib:cr-0012}, leading to a significant cost reduction. This was made possible by dropping the requirement for personnel access during beam operation of the main linac.
\item Power ratings for the main beam dumps, and intermediate beam dumps for beam aborts and machine tuning, were reduced to save costs~\cite{bib:cr-0013}.
\item A revision of the expected horizontal beam emittance at the interaction point at \siunit{125}{GeV} beam energy, based on improved performance expectations for the damping rings and a more thorough scrutiny of beam transport effects at lower beam energies, lead to an increase of the luminosity expectation from $0.82$ to \siunit{1.35 \cdot 10^{34}}{cm^{-2}\,s^{-1}}~\cite{bib:cr-0016}.
\item The active length of the positron source undulator has been increased from $147$ to \siunit{231}{m} to provide sufficient intensity at \siunit{125}{GeV} beam energy~\cite{PWG:2018a}.
\end{itemize}

These changes contributed to an overall cost reduction, risk mitigation, and improved performance expectation.

Several possibilities were evaluated for the length of the initial tunnel. 
Options that include building tunnels with the length required for a machine with $\sqrt{s} = \siunit{350}{GeV}$ or \siunit{500}{GeV}, were considered.
In these scenarios, an energy upgrade would require the installation of additional cryomodules (with RF and cryogenic supplies), but little or no civil engineering activities.
In order to be as cost effective as possible, the final proposal (see Figure 1), endorsed by ICFA~\cite{ICFA:2017}, does not include these empty tunnel options. 

While the length of the main linac tunnel was reduced, the beam delivery system and the main dumps are still designed to allow for an energy upgrade up to  $\sqrt{s} = \siunit{1}{TeV}$.

\begin{figure}[thbp]
   \includegraphics[width=\hsize]{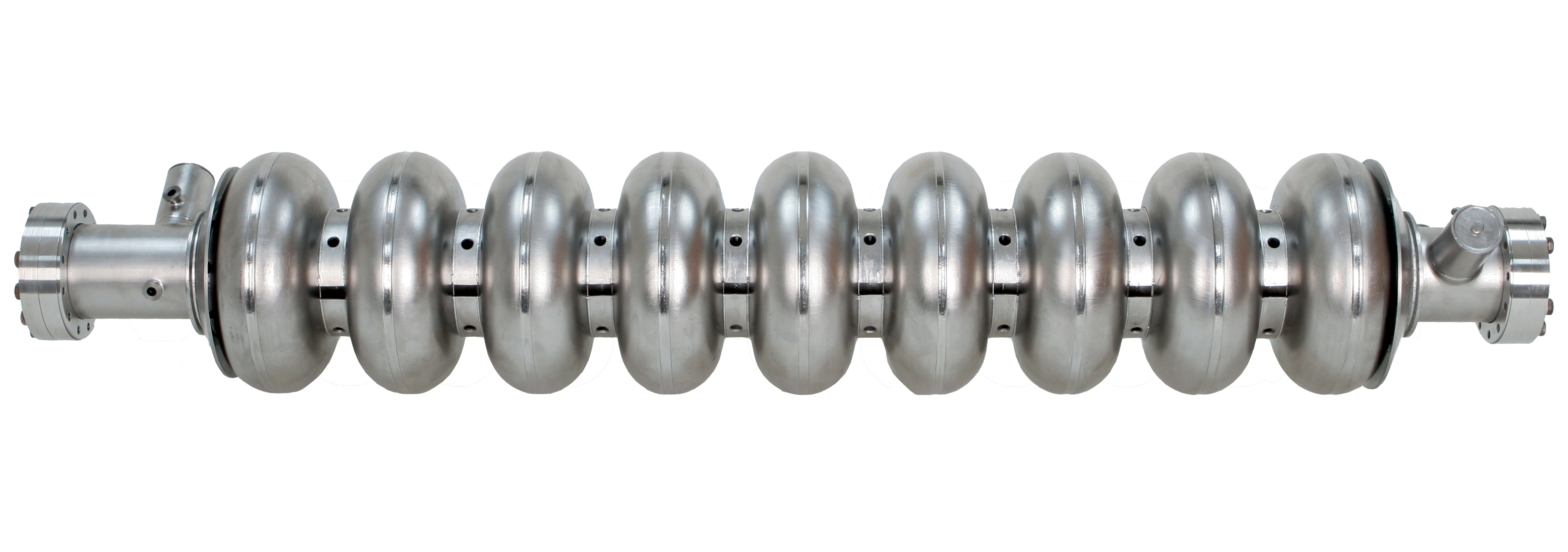}
\caption{A $1.3\,{\mathrm{GHz}}$ superconducting niobium nine-cell cavity.
}
\label{fig:tesla-cavity}
\end{figure}

\subsection{Superconducting RF Technology}

%
%
%
%

\begin{figure}[bthp]
   \includegraphics[width=\hsize]{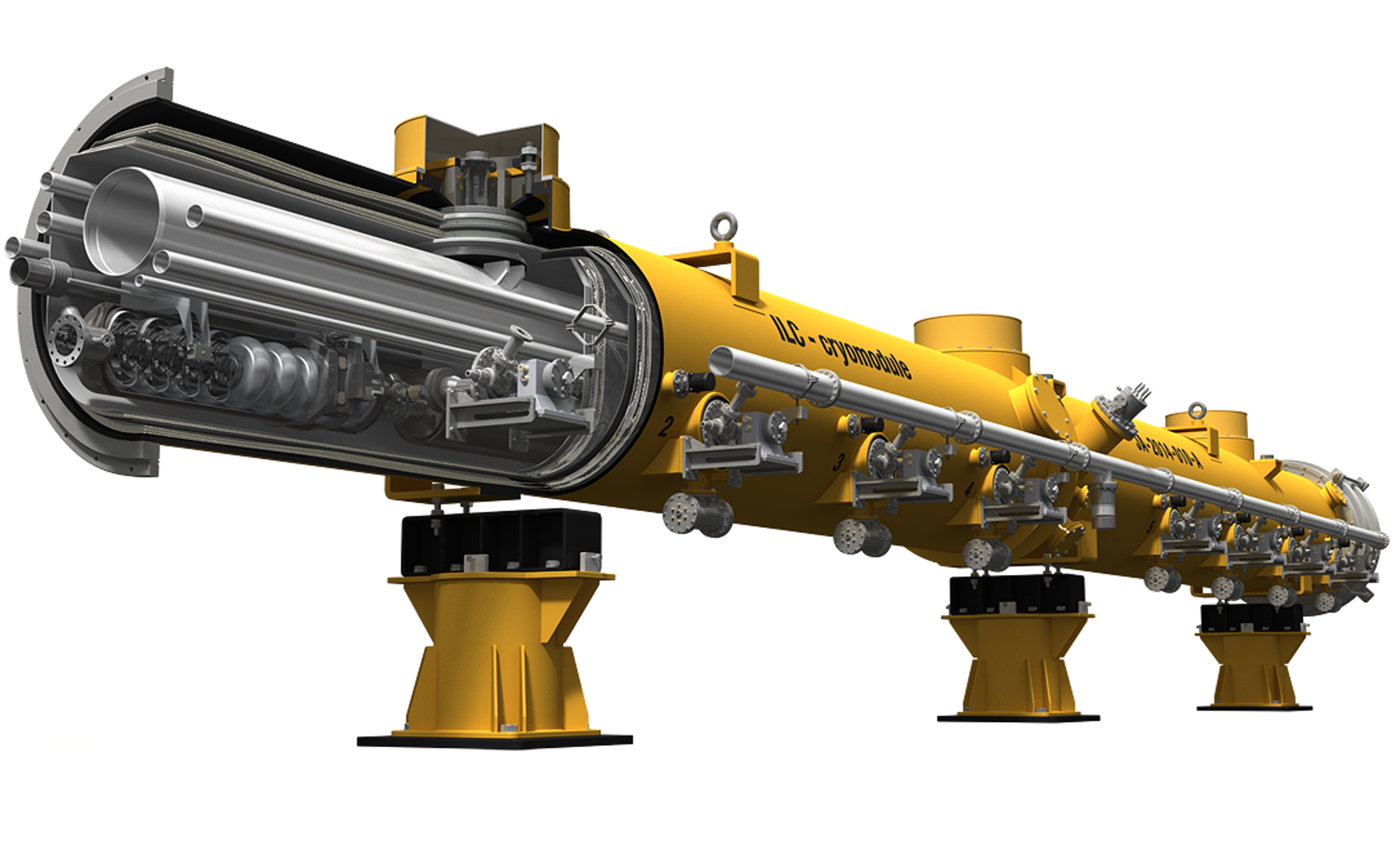}
\caption{An ILC type cryomodule. \copyright Rey.Hori/KEK.}
\label{fig:cryomodule}
\end{figure}

The heart of the ILC accelerator consists of the two superconducting Main Linacs that accelerate both beams from \num{5} to \siunit{125}{GeV}.
These linacs are based on the TESLA technology:
beams are accelerated in \siunit{1.3}{GHz} nine-cell superconducting cavities made of niobium and operated at \siunit{2}{K} (Fig.~\ref{fig:tesla-cavity}). These 
 are assembled into cryomodules comprising nine cavities or eight cavities plus a quadrupole/corrector/beam position monitor unit, and all necessary cryogenic supply lines (Fig.~\ref{fig:cryomodule}). 
Pulsed klystrons supply the necessary radio frequency power (High-Level RF HLRF) to the cavities by means of a waveguide power distribution system and one input coupler per cavity.

This technology was primarily developed at DESY for the TESLA accelerator project that was proposed in 2001.
Since then, the TESLA technology collaboration~\cite{bib:ttc} has been improving this technology, which is now being used in several accelerators in operation (FLASH at DESY~\cite{schreiber_faatz_2015,Vogt:2018wvy}, E-XFEL in Hamburg~\cite{bib:xfel}), under construction (LCLS-II at SLAC, Stanford, CA~\cite{bib:lcls-ii}) or planned (SHINE in Shanghai~\cite{Zhao:2018lcl}).

%

\begin{figure*}[tbhp]
   \includegraphics[width=0.8 \hsize]{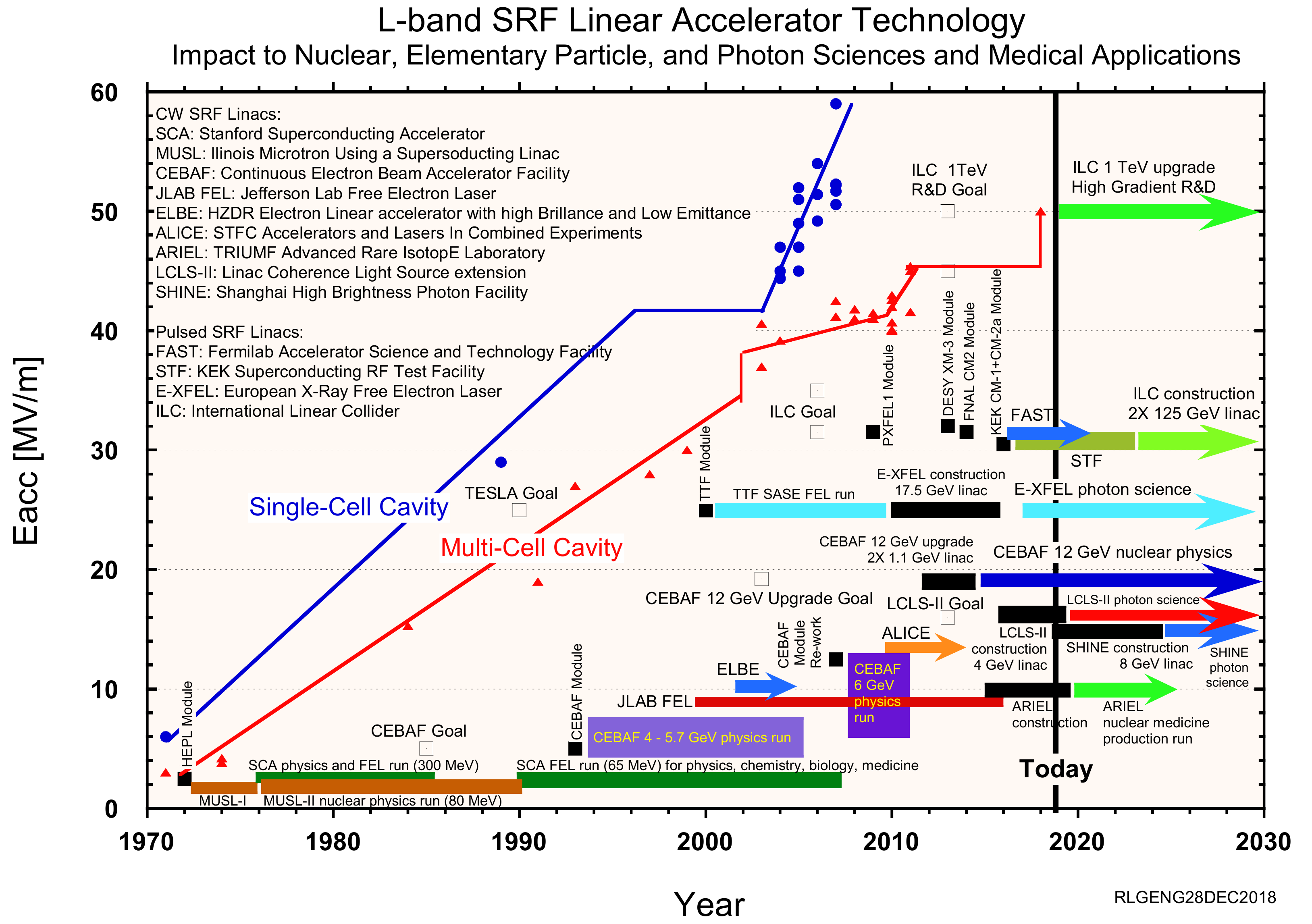}
\caption{Development of the gradient of SRF cavities since 1970
\cite[updated]{Geng:2015glc}.
}
\label{fig:gradients}
\end{figure*}

\subsubsection{The quest for high gradients}
\label{subsubsec:highgrad}

The single most important parameter for the cost and performance of the ILC is the accelerating gradient $g$.
The TDR baseline value is an average gradient $g = \siunit{31.5}{MV/m}$ for beam operation, with a $\pm 20\,\%$ gradient spread between individual cavities.
Recent progress in R\&D for high gradient cavities raises the hope to increase
the gradient by  $10\,\%$  to  $g = \siunit{35}{MV/m}$, which would reduce the total cost of the \siunit{250}{GeV} accelerator by about  $6\,\%$.
To achieve the desired gradient in beam operation, the gradient achieved in the low-power vertical test (mass production acceptance test) is specified $10\,\%$ higher to allow for operational gradient overhead for low-level
RF (LLRF) controls, as well as some degradation during cryomodule assembly (few ${\mathrm{MV/m}}$).
Figure~\ref{fig:gradients} shows how the achievable gradients have evolved over the past 50 years~\cite{Geng:2015glc}.

\paragraph{Gradient impact on costs:}
To the extent that the cost of cavities, cryomodules and tunnel infrastructure is independent of the achievable gradient, the investment cost per GeV of beam energy is inversely proportional to the average gradient achieved. This is the reason for the enormous cost saving potential from higher gradients.
This effect is partially offset by two factors:  the energy stored in the electromagnetic field of the cavity, and the dynamic heat load to the cavity from the electromagnetic field.  These grow quadratically with the gradient for one cavity, and therefore linearly for a given beam energy.
The electromagnetic energy stored in the cavity must be replenished by the RF source during the filling time that precedes the time when the RF is used to accelerate the beam passing through the cavity; this energy is lost after each pulse and thus reduces the overall efficiency and requires more or more powerful modulators and klystrons.
The overall cryogenic load is dominated by the dynamic heat load from the cavities, and thus operation at higher gradient requires larger cryogenic capacity.
Cost models that parametrise these effects indicate that the minimum of the investment cost per GeV beam energy lies at \num{50} or more GeV, depending on the relative costs of tunnel, SCRF infrastructure and cryo plants, and depending on the achievable $Q_0$~\cite{Adolphsen:2011a}. 
Thus,  the optimal gradient is significantly higher than the value of approximately \siunit{35}{MV/m} that is currently realistic; this emphasises  the relevance of achieving higher gradients.

It should be noted that in contrast to the initial investment, 
the operating costs rise when the gradient is increased, and this must be factored into 
the cost model.

\begin{figure}[htbp]
   \includegraphics[width=\hsize]{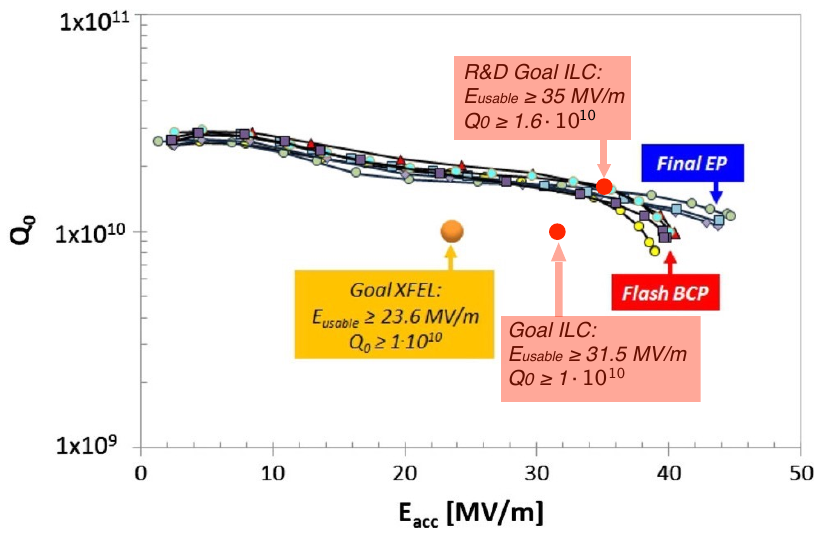}
\caption{Examples of the $Q_0\,(E\sub{acc})$ curves of some of the best
cavities, either treated at RI using ``EP final'', or at EZ using
``BCP flash.''
\cite[Fig. 19]{Singer:2016fbf}. 
Vendor ``RI'' employs a production process that closely follows the ILC specifications, with a final electropolishing step.
The ILC gradient / $Q_0$ goals are overlaid.}
\label{fig:cavity-gradient}
\end{figure}

\paragraph{Gradient limitations:}
Fundamentally, the achievable gradient of a SC cavity is limited when the magnetic field at the cavity walls surpasses the critical field $H\sub{crit,RF}$ of the superconductor.
This gradient depends on the material, operating temperature, and the cavity geometry. 
For the TESLA type cavities employed at the ILC, this limit is about \siunit{48}{MV/m} at \siunit{2}{K}.
The best E-XFEL production cavity reached \siunit{44.6}{MV/m} (Fig.~\ref{fig:cavity-gradient}).
The record for single cell cavities operating at \siunit{1.3}{GHz} is \siunit{59}{MV/m}~\cite{Eremeev:2007zza}.
%

Niobium is a type-II superconductor, and so it has two distinct superconducting phases, the Meissner state, with complete magnetic flux expulsion, which exists up to a field strength $H\sub{c1} \approx \siunit{180}{mT}/\mu_0$ ($\mu_0 = 4\pi \siunit{10^{-7}}{T\,m/A}$ being the vacuum permeability), and a mixed state in which flux vortices penetrate the material, up to a higher field strength $H\sub{c1}$, at which superconductivity breaks down completely.
In time-dependent fields, the penetrating vortices move due to the changing fields and thus dissipate energy, causing a thermal breakdown. 
However, for RF fields, the Meissner state may persist metastably up to the superheating field strength $H\sub{sh} \approx \siunit{240}{mT}/\mu_0$, which is expected to be the critical RF field critical field $H\sub{crit,RF}$~\cite{Padamsee:1998vf}.
Experimentally, niobium RF cavities have been operated at field strengths as high as $H=\siunit{206}{mT}/\mu_0$~\cite{Eremeev:2007zza}, and the best E-XFEL production cavities reach about $\siunit{190}{mT}$.
Recently, even \siunit{210}{mT} has been achieved at FNAL~\cite{Grassellino:2018tqg}.
In recent years, theoretical understanding of the nature of this metastable state and the mechanisms at the surface that prevent flux penetration has significantly improved~\cite{Gurevich:2017vnn,Kubo:2017cww}.
It appears that a thin layer of ``dirty'' niobium, \ie,  with interstitial impurities, on top of a clean bulk with good thermal conductivity, is favourable for high field operation.  

The gradient at which a SC cavity can be operated in practice is limited by three factors   in addition to those just listed~\cite{Padamsee:1998vf}:
\begin{itemize}
\item the thermal breakdown of superconductivity, when local power dissipation causes a local quench of the superconductor,
\item the decrease of the quality factor $Q_0$ at high gradients that leads to increased power dissipation,
\item the onset of field emission that causes the breakdown of the field in the cavity.
\end{itemize}
The onset of these adverse effects is mostly caused by micro-metre sized surface defects of various kinds. 
Producing a sufficiently defect-free surface in an economic way is thus the central challenge in cavity production.

More than 20 years of industrial production of TESLA type cavities have resulted in a good understanding which production steps  and quality controls are necessary to produce cavities with high-quality, nearly defect-free surfaces that are capable of achieving the desired high field strengths at a reasonable production yield.

\paragraph{Results from E-XFEL cavity production:}

The production and testing of $831$ cavities for the E-XFEL~\cite{Singer:2016fbf,Reschke:2017gjp} provides the biggest sample of cavity production data so far. 
Cavities were acquired from two different vendors, RI and EZ.
Vendor RI employed a production process with a final surface treatment closely following the ILC specifications, including a final electropolishing (EP) step,
while the second vendor EZ used buffered chemical polishing (BCP).
The E-XFEL specifications asked for a usable gradient of \siunit{23.6}{MV/m} with a $Q_0 \ge 1  \cdot 10^{10}$ for operation in the cryomodule;
with a $10\,\%$ margin this corresponds to a target value of \siunit{26}{MV/m} for the performance in the vertical test stand for single cavities.
Figure~\ref{fig:cavity-gradient} shows the $Q_0$ data versus accelerating gradient of the best cavities received, with several cavities reaching more than \siunit{40}{MV/m}, significantly beyond the ILC goal, already with $Q_0$ values that approach the target value $1.6\cdot10^{10}$ that is the goal of future high-gradient R\&D.

\begin{figure}[tbhp]
   \includegraphics[width=\hsize]{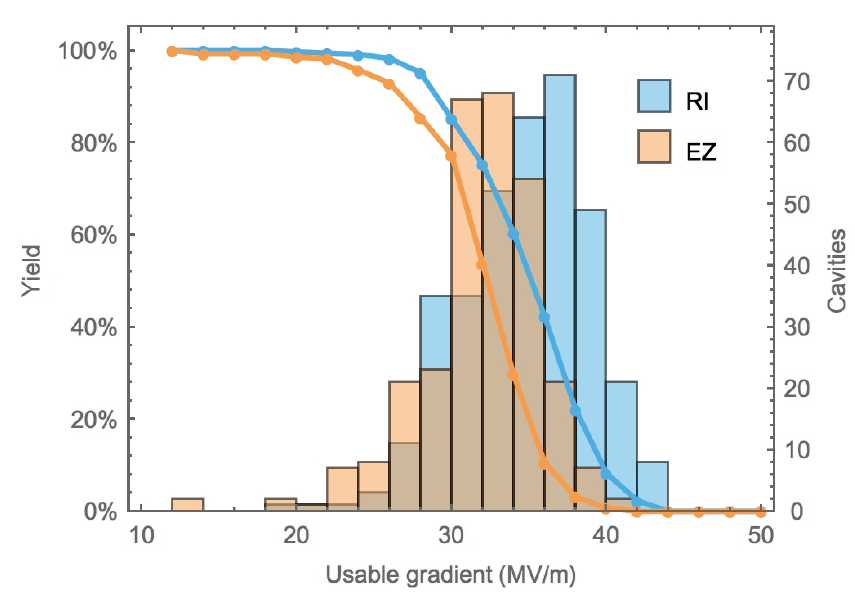}
\caption{Distribution and yield of the ``as received'' maximum
gradient of cavities produced for the E-XFEL, separated by vendor \cite[Fig. 33]{Reschke:2017gjp}. 
Vendor RI employs a production process that closely follows the ILC specifications, with a final electro polishing step.}
\label{fig:cavity-yield}
\end{figure}

E-XFEL production data, in particular from vendor RI, provide excellent statistics for the cavity performance as received from the vendors, as shown in Fig.~\ref{fig:cavity-yield}.
For vendor RI, the yield for cavities with a maximum gradient above \siunit{28}{MV/m} is $85\,\%$, with an average of \siunit{35.2}{MV/m} for the cavities that pass the cut.

Since the E-XFEL performance goal was substantially lower than the ILC specifications, cavities with gradient below \siunit{28}{MV/m}, which would not meet ILC specifications, were not generally re-treated for higher gradients, limiting our knowledge of the effectiveness of re-treatment for large gradients.
Still, with some extrapolation it is possible to extract yield numbers applicable to the ILC specifications ~\cite{bib:Walker:2017.lcws}.

The E-XFEL data indicate that after re-treating cavities with gradients outside the ILC specification of $\siunit{35}{MV/m} \pm 20\,\%$, \ie, below \siunit{28}{MV/m}, a yield of $94\,\%$ for a maximum gradient above \siunit{28}{MV/m} can be achieved, with an average value of \siunit{35}{MV/m}, meeting the ILC specification.
Taking into account limitations from $Q_0$ and the onset of field emission, the usable gradient is lower.  
This gives a $82\,(91)\,\%$ yield and an average usable gradient of \siunit{33.4}{MV/m} after up to one (two) re-treatments.
The re-treatment and testing rate is significantly higher than assumed in the TDR, but the E-XFEL experience shows that re-treatment can mostly be limited to a simple high-pressure rinse (HPR) rather than an expensive electropolishing step.  

Overall, the E-XFEL cavity production data prove that it is possible to mass-produce cavities meeting the ILC specifications as laid out in the TDR with the required performance and yield.

\paragraph{High-gradient R\&D -- nitrogen infusion:}

\begin{figure}[htbp]
   \includegraphics[width=\hsize]{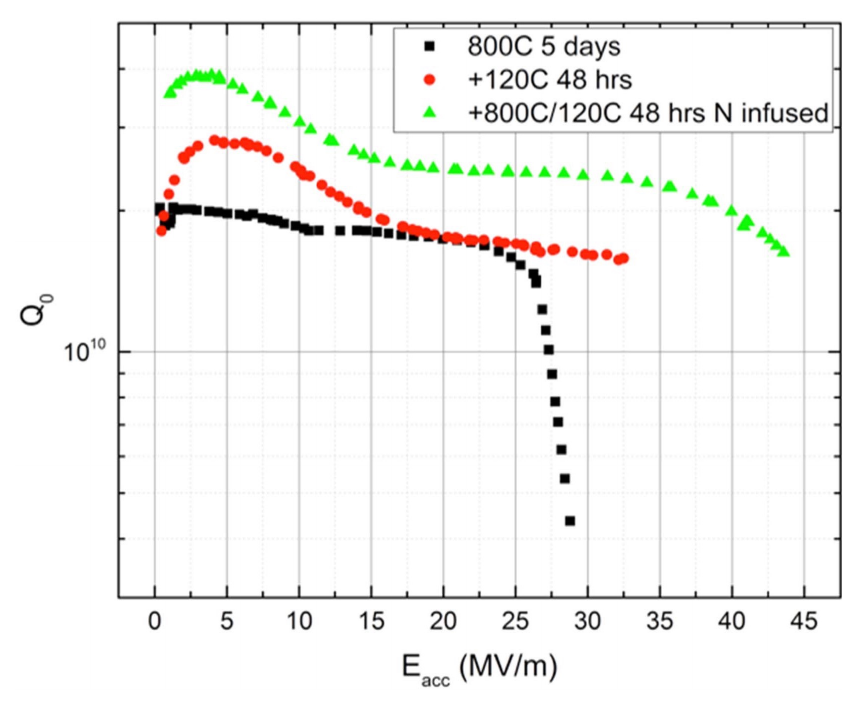}
\caption{Effect of successive cavity treatments on a single cavity: $\siunit{800}{^\circ C}$ bake for five days (black, lowest curve), followed by $48$ hours baking at $\siunit{120}{^\circ C}$ (red, middle curve). 
A third heat treatment including nitrogen infusion (green, top curve) significantly  raises the breakdown gradient and the quality factor of the cavity
\cite[Fig. 5]{Grassellino:2017bod}.
}
\label{fig:n2infusion}
\end{figure}

In recent years, new techniques have emerged that seem to indicate that higher gradients combined with higher quality factors are attainable in bulk niobium cavities.

In the early 2010s, nitrogen doping was developed as a method to substantially increase $Q_0$ by adding nitrogen during the \siunit{800}{^\circ C} baking, which leads to interstitial nitrogen close to the niobium surface~\cite{Grassellino:2013nza}.
This technique has been employed successfully in the production of the cavities for LCLS-II, with an average $Q_0$ of $3.0\cdot 10^{10}$ achieved in a prototype cryomodule~\cite{Wu:2018qyl}.
However, nitrogen doping reduces the critical RF field of the material and thus limits the achievable gradients to values below \siunit{30}{MV/m}, rendering doped material useless for high gradient applications.

By contrast, in nitrogen infusion the nitrogen is added during the low temperature baking at \siunit{120}{^\circ C}.
Experimental results seem to indicate that nitrogen infusion may offer a combination of three advantages:
\begin{itemize}
\item Reaching higher accelerating gradients,
\item higher $Q_0$ values, resulting in a reduced cryogenic load, 
\item a simplified and less expensive production process that does away with the final electropolishing step.
\end{itemize}

Figure~\ref{fig:n2infusion} \cite[Fig. 19]{Grassellino:2017bod} shows how the addition of nitrogen during the final \siunit{48}{h} long \siunit{120}{^\circ C} bake of a one--cell cavity drastically improves the cavity quality factor as well as the maximum gradient, which comes close to the best E-XFEL cavity results, but at higher $Q_0$.


Up to now, it has been difficult to reproduce these exciting results in other laboratories.
Success has been reported by groups at JLAB~\cite{Dhakal:2017xxq}, and  Cornell~\cite{Koufalis:2017blg}, but KEK has reported mixed results \cite{bib:Umemori:2018.lcws}, and DESY has so far not been able to reproduce these results~\cite{Wenskat:2018zco}.
These difficulties seem to indicate that the recipe for a successful application of nitrogen infusion is not yet fully understood, and that further research and development will be necessary before this process can be transferred to industry. 

Nevertheless, the infusion results have triggered a renewed interest in the research on highest gradients in niobium cavities, with a host of new experimental results, increased activity to achieve a more thorough theoretical understanding~\cite{Kubo:2017cww,Gurevich:2017vnn}, and application of state-of-the-art analytical methods such as   muon spin rotation (muSR) \cite{Romanenko:2013saa}.
Recently, a record gradient for TESLA shape cavities of \siunit{49}{MV/m} was reported~\cite{Grassellino:2018tqg}  with a low temperature treatment at \siunit{75}{^\circ C} after \siunit{120}{^\circ C} baking without nitrogen.
All these results provide reason for optimism that an improved understanding of the mechanisms that stabilise superconductivity in the presence of high fields will result in improved performance of industrially produced cavities for the ILC.

\paragraph{High-gradient R\&D -- alternative cavity shapes:}

Fundamentally, the achievable gradient in a niobium cavity is limited by the maximum magnetic field at the cavity surface, not the electrical field strengths.
The ratio between peak surface field $B\sub{pk}$ and gradient $g$ depends on the cavity geometry and is $B\sub{pk}/g = \siunit{4.26}{mT / (MV/m)}$ for TESLA type cavities.
A number of alternative cavity shapes have been investigated with lower ratios~\cite{Geng:2006wf},
resulting in single cells gradients up to \siunit{59}{MV/m}~\cite{Eremeev:2007zza}.
The reduced magnetic field, however, has to be balanced with other factors that favour the TESLA cavity shape, namely: a reasonable peak electrical field to limit the risk of field emission, sufficient iris width and cell-to--cell RF coupling, and a mechanical shape that can be efficiently fabricated.

Recently, new five-cell cavities with a new ``low surface field'' (LSF) shape~\cite{Li:2008a} have been produced at JLAB and have achieved gradient of up to~\siunit{50}{MV/m} in three of the five cells, which is a new record for multi-cell cavities~\cite{bib:Geng:2018.lcws}. 
The LSF shape aims to achieve a good compromise between the goal of a low magnetic field and the other criteria, and demonstrates that further improvements in gradient may be realised  in the future.

\subsubsection{Further cost reduction R\&D}

Additional strategies for cost reduction and improved cavity performance are also being investigated.

\paragraph{Low $RRR$ material:}

The niobium raw material and preparation of sheets  are a significant cost driver; R\&D is underway to re-evaluate the stringent limits on impurities, especially of tantalum, and the demand for a high residual resistivity ratio $RRR > 300$~\footnote{$RRR$ is the ratio of the material's room temperature resistivity to the normal conducting resistivity close to \siunit{0}{K}; heat conductivity from electrons is proportial to $RRR$. $RRR$ is reduced by impurities, in particular interstitial ones  from hydrogen, nitrogen and oxygen.}, to reduce the raw material cost.  The electrical 
conductivity and heat transport by electrons are proportional.  This implies that 
large $RRR$ values, indicative of low impurity content, make the cavities also 
less susceptible to thermal breakdown from surface defects.
However, when defect sizes can be successfully controlled to the extent necessary to achieve gradients above \siunit{35}{MV/m} routinely, the influence of heat conductivity and $RRR$ may be diminished, permitting the use of lower $RRR$ material~\cite{bib:Kubo:2018.ttc}.

\paragraph{Ingot and large-grain niobium:}

Together with direct slicing of discs from large niobium ingots, without rolling, forging and grinding or polishing steps, the cost for niobium sheets has the potential to be reduced by $50\,\%$~\cite{Evans:2017rvt,Kneisel:2014uqa}.
Without the mechanical deformation during rolling and forging, the grains from the initial crystallisation stay large, which makes later production steps, in particular deep--drawing of half cells, more challenging.
Nevertheless, if these challenges are overcome, tests with large--grain and ingot niobium show promising results~\cite{Reschke:2011a, Dhakal:2015xac}.

\subsubsection{Basic parameters}

The choice of operating frequency is a balance between the higher cost of larger, lower-frequency cavities and the increased cost at higher frequency associated with the lower sustainable gradient from the increased surface resistivity. 
The optimum frequency is in the region of \siunit{1.5}{GHz}, but during the early R\&D on the technology, \siunit{1.3}{GHz} was chosen due to the commercial availability of high-power klystrons at that frequency.

\subsubsection{Cavities}

The superconducting accelerating cavities for the ILC are nine-cell structures made out of high-purity niobium (Fig.~\ref{fig:tesla-cavity}), with an overall length of \siunit{1.25}{m}.
Cavity production starts from niobium ingots which are forged and rolled into \siunit{2.8}{mm} thick niobium sheets that are individually checked for defects by an eddy current scan and optical inspection~\cite{Adolphsen:2013jya}.
Cavity cells are produced by deep-drawing the sheets into half cells, \num{18} of which are joined by electron beam welding with two end groups to form the whole structure.
This welding process is one of the most critical and cost-intensive steps of the cavity manufacturing procedure. 
Utmost care must be taken to avoid irregularities, impurities and inclusions in the weld itself, and deposition of molten material at the inner surface of the cavity that can lead to field emission.

After welding, the inner surface of the cavity must be prepared.
The process is designed to remove material damage incurred by chemical procedures during the fabrication process, chemical residues from earlier production steps, hydrogen in the bulk niobium from earlier chemical processing, and contamination from particles.
In a last step, the cavity is closed to form a hermetically sealed structure ready for transport.
The treatment steps involve a series of rinses with ethanol or high pressure water, annealing in a high purity vacuum furnace at \siunit{800^\circ}{C} and \siunit{120^\circ}{C}, and electropolishing or buffered chemical polishing.
The recipe for the surface preparation has been developed over a long time.  Still, it remains subject to optimisation, since it is a major cost driver for the cavity production and largely determines the overall performance and yield of the cavities.
In particular the electropolishing steps are complicated and costly, as they require complex infrastructure and highly toxic chemicals.
One advantage of nitrogen infusion (see Sec.~\ref{subsubsec:highgrad}) 
is that the final electropolishing step is omitted.

Careful quality control during the production process is of high importance.
At the E-XFEL, several quality controls were conducted by the manufacturer during production, with nonconformities reported to the institute responsible for the procurement, where a decision was made whether to accept or reject a part~\cite{Singer:2016fbf}. 
With this ``build to print'' approach, in which the manufacturer guarantees that a precise production process will be followed but does not guarantee a specific performance, procurement costs are reduced, because the manufacturer does not carry, and does not charge for,  the performance risk.

Upon reception from the manufacturer, cavities are tested in a vertical cryostat (``vertical test''), where $Q_0$ is measured as a function of the gradient.
Cavities that fall below the specified gradient goal are re-treated by an additional (expensive) electropolishing step or a comparatively simple high-pressure rinse. 
After retreatment, the vertical test is repeated.

Re-treatment and tests constitute a major cost driver in cavity production. 
For the ILC TDR, it was assumed that $25\,\%$ of the cavities would fall below the \siunit{28}{MV/m} gradient threshold and undergo re-treatment and a second vertical test.
E-XFEL data from the vendor ``RI'' that followed the ILC production recipe indicate that $15\,\%$ to $37\,\%$ of the cavities fall below \siunit{28}{MV/m}, depending  on whether the maximum or the ``usable'' achieved gradient is considered~\cite{bib:Walker:2017.lcws}.
However, E-XFEL experience also shows that, in most of the cases, a high-pressure rinse is sufficient as re-treatment to remove surface defects, which is a cost saving compared to the electropolishing assumed in the TDR.

After successful testing, prior to installation in the cryomodule, cavities are equipped with a magnetic shield and the frequency tuner, which exerts mechanical force on the cavity to adjust the resonant frequency to the frequency of the external RF field~\cite[Sect. 3.3]{Adolphsen:2013kya}.

\subsubsection{Power coupler}

The power coupler transfers the radio frequency (RF) power from the waveguide system to the cavity. 
In the ILC, a coupler with a variable coupling is employed; this is realised using  a movable antenna.  Another role of the coupler is to 
separate the cavity vacuum from the atmospheric pressure in the waveguide, and to  insulate the cavity at \siunit{2}{K} from the surrounding room temperature.
Thus,  the coupler has to fulfill a number of demanding requirements: transmission of high RF power with minimal losses and no sparking, vacuum tightness and robustness against window breaking, and minimal heat conductivity.  
As a consequence, the coupler design is highly complex, with a large number of components and several critical high-tech manufacturing steps.

\begin{figure}[htbp]
   \includegraphics[width=\hsize]{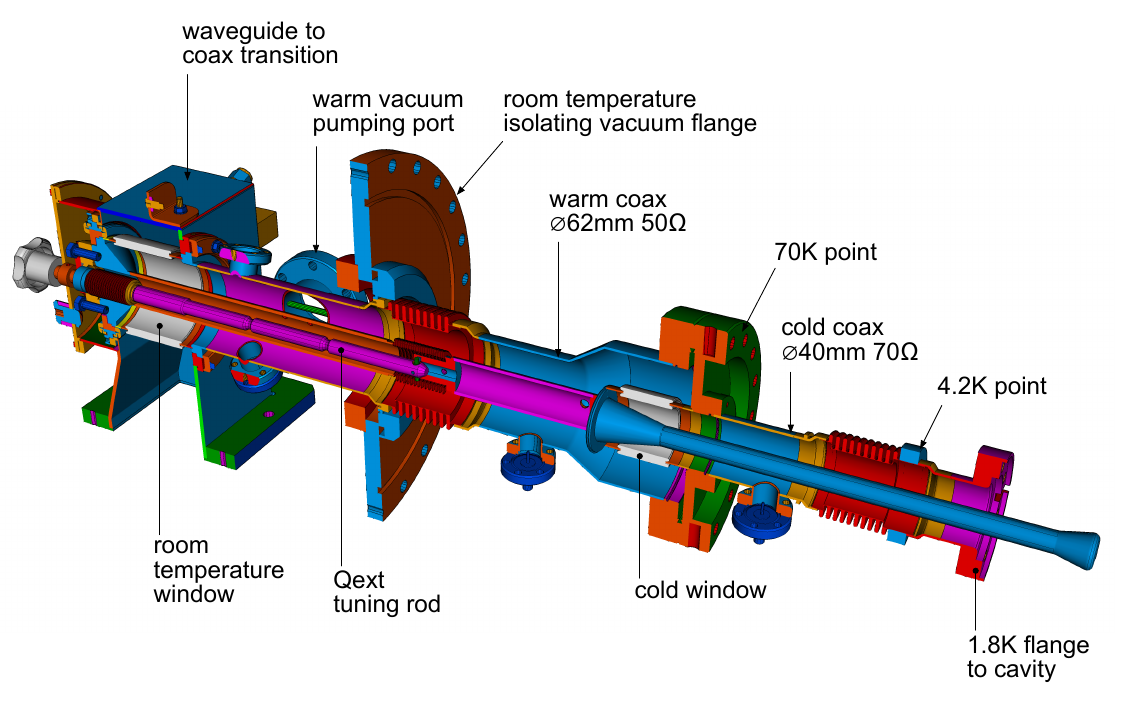}
\caption{An E-XFEL type coupler.
}
\label{fig:xfelcoupler}
\end{figure}

The baseline coupler design was originally developed in the 1990s for the TESLA Test Facility (TTF, now FLASH) at DESY,
and has since been modified by a collaboration of LAL and DESY for use in the E-XFEL.
About 840 of these couplers (depicted in Fig. \ref{fig:xfelcoupler}) were fabricated by three different companies for the  E-XFEL~\cite{Kaabi:2013wna},  where 800 are now in operation.
A lot of experience has been gained from this production~\cite{Sierra:2017wyc}.

\subsubsection{Cryomodules}

To facilitate transportation, installation and operation, 8 or 9 cavities are integrated into a \siunit{12.6}{m} long cryomodule~(Fig.~\ref{fig:cryomodule}), which houses the cavities, thermal insulation, and all necessary supply tubes for liquid and gaseous helium at \siunit{2-80}{K} temperature.

\begin{figure}[htbp]
   \includegraphics[width=\hsize]{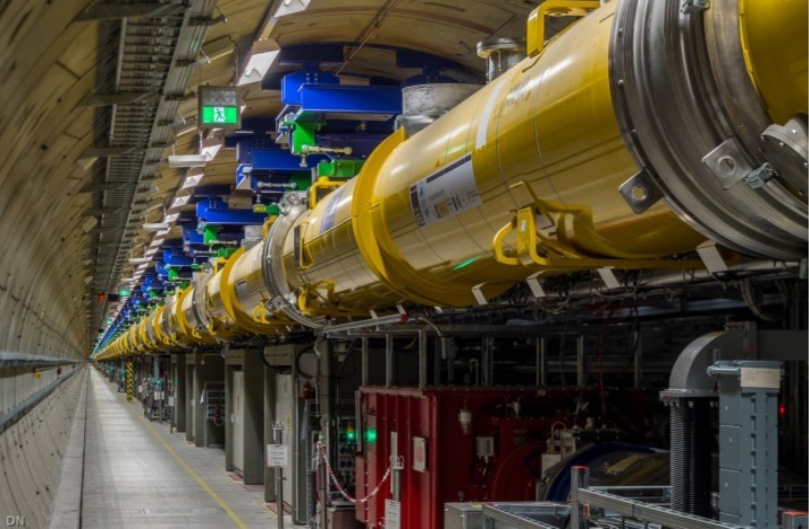}
\caption{View of installed cryomodules in the tunnel of the E-XFEL~\cite{Reschke:2018ywk}.
}
\label{fig:xfel-tunnel}
\end{figure}

Nine of these cryomodules are connected in the tunnel to form a cryostring with a common liquid helium supply.  RF for one such string is provided by two klystrons.
No separate helium transfer line is necessary, as all helium transport lines are integrated within the modules.  A  quadrupole / beam position monitor / corrector magnet unit  is mounted instead of the 9th cavity in every third module.
Figure~\ref{fig:xfel-tunnel} shows installed cryomodules in the tunnel of the  E-XFEL~\cite{Reschke:2018ywk}.

Cryomodule assembly requires a dedicated facility with large clean rooms, especially trained, experienced personnel, and thorough quality control~\cite{Berry:2017gpt}.
The cryomodules are certified for liquid helium pressure of up to \siunit{2}{bar}.  Thus they must  conform to the applicable pressure vessel codes, which brings with it very stringent documentation requirements for all pressure bearing parts~\cite{Peterson:2011zz}.

For the E-XFEL project, 103 cryomodules were assembled in a facility built and operated by CEA~\cite{Weise:2014zqa,Berry:2017gpt} and industrial partners, demonstrating the successful industrialization of the assembly process, with a final throughput of one cryomodule every four working days.
This production rate is close to the rate envisaged for a possible European contribution of 300 cryomodules to a \siunit{250}{GeV} ILC in Japan.

While the design gradient for E-XFEL accelerator modules of \siunit{23.6}{MV/m} is significantly lower than the aim of \siunit{31.5-35}{MV/m} for the ILC, a number of cryomodules have been built around the world that come close or reach the ILC TDR specification of \siunit{31.5}{MV/m}: An E-XFEL prototype module at DESY reached \siunit{30}{MV/m}~\cite{Kostin:2009a}, Fermilab has demonstrated cryomodule operation at the ILC specification of \siunit{31.5}{MV/m}~\cite{Broemmelsiek:2018iqr}, and KEK has reported stable pulsed operation of a cryomodule at \siunit{36}{MV/m}~\cite{Yamamoto:2018kml}.

\begin{figure}[htbp]
   \includegraphics[width=\hsize]{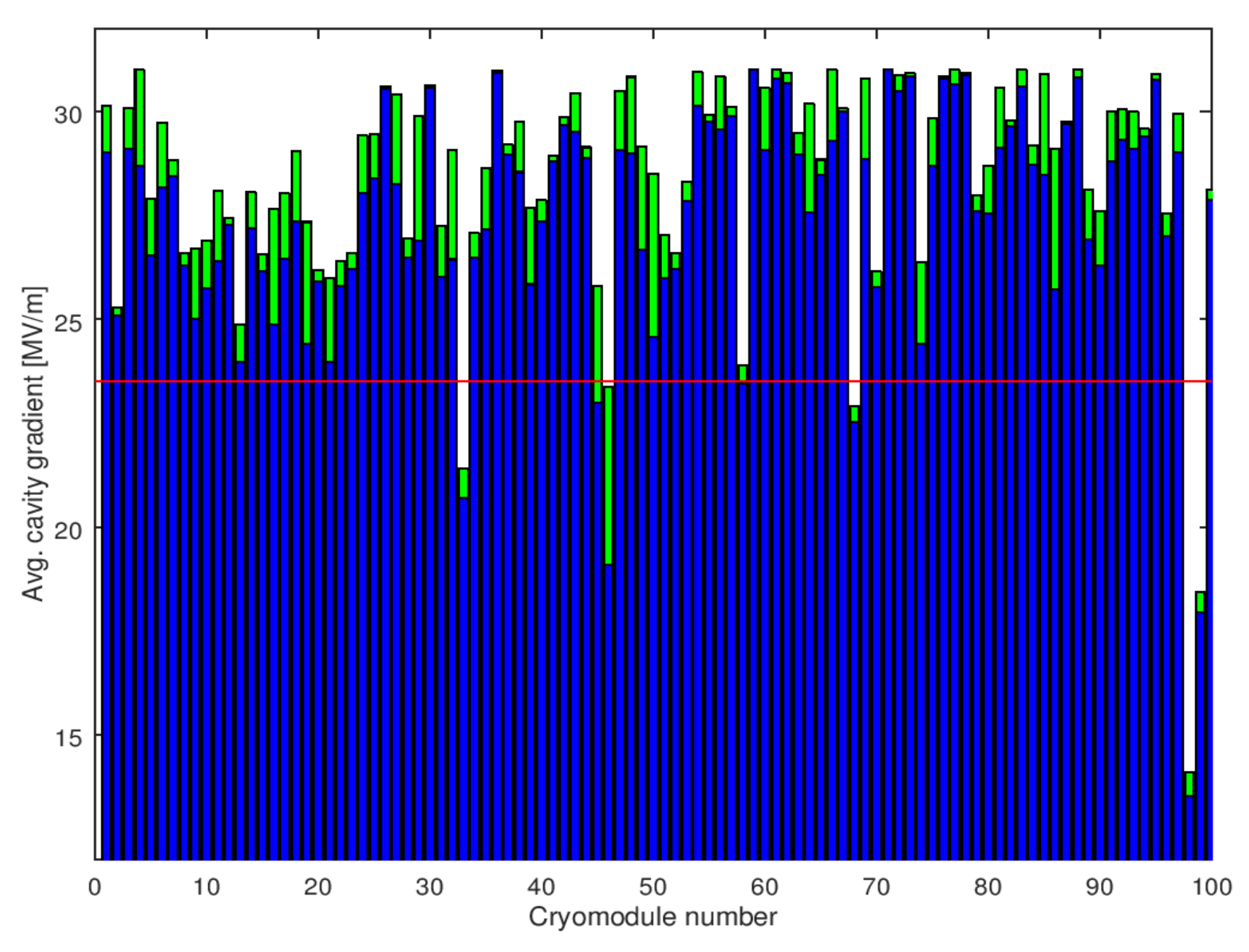}
\caption{Average of the operating (blue) and maximum
(green) gradient for cavities in each E-XFEL serial-production cryomodule.
The specification of \siunit{23.6}{MV/m} is marked by a red line
\cite{Kasprzak:2018kkr}.  Modules 98 and 99 were assembled from the lowest-performing cavities.}
\label{fig:cryomodules-performance}
\end{figure}

Figure~\ref{fig:cryomodules-performance} shows the average cavity gradients per cryomodule for the E-XFEL serial-production cryomodules~\cite{Kasprzak:2018kkr}. 
In the tests, the gradients were limited administratively to \siunit{31}{MV/m}; the true maxima might be higher.
For almost all of the modules, the cavity gradients are significantly above the E-XFEL specification of \siunit{23.6}{MV/m}.

\subsubsection{Plug-compatible design}

In order to allow various designs of sub-components from different countries and vendors to work together in the same cryomodule, a set of interface definitions has been internationally agreed upon.
This ``plug-compatible'' design ensures that components are interchangeable between modules from different regions and thus reduces the cost risk.
Corresponding interface definitions exist for the cavity, the fundamental-mode power coupler, the mechanical tuner and the helium tank.
The ``S1Global'' project~\cite{bib:s1g} has successfully built a single cryomodule from several cavities equipped with different couplers and tuners, demonstrating the viability of this concept.

\subsubsection{High-level radio-frequency}

The high-level radio-frequency (HLRF) system provides the RF power that drives the accelerating cavities.
The system comprises modulators, pulsed klystrons, and a waveguide power distribution system.

\paragraph{Modulators:}
The modulators provide the short, high-power electrical pulses required by the pulsed klystrons from a continuous supply of electricity. 
The ILC design foresees the use of novel, solid state Marx modulators.
These modulators are based on a solid-state switched capacitor network, where capacitors are charged in parallel over the long time between pulses, and discharged in series during the short pulse duration,
transforming continuous low-current, low voltage electricity into short high-power pulses of the required high voltage of \siunit{120}{kV} at a current of \siunit{140}{A}, over \siunit{1.65}{ms}.
Such Marx modulators have been developed at SLAC~\cite{Kemp:2011zz} 
and successfully tested at KEK~\cite{Gaudreau:2014pza}.
However, long-term data about the required large mean time between failures (MTFB) are not yet available.

\paragraph{Klystrons:}
The RF power to drive the accelerating cavities is provided by \siunit{10}{MW} L-band multi-beam klystrons. 
Devices meeting the ILC specifications were initially developed for the TESLA project, and later for the E-XFEL.
They are now commercially available from two vendors (Thales and Toshiba), both of which provided klystrons for the E-XFEL.
The ILC specifications ask for a $65\,\%$ efficiency (drive beam to output RF power), which are met by the existing devices.

\begin{figure}[htbp]
   \includegraphics[width=\hsize]{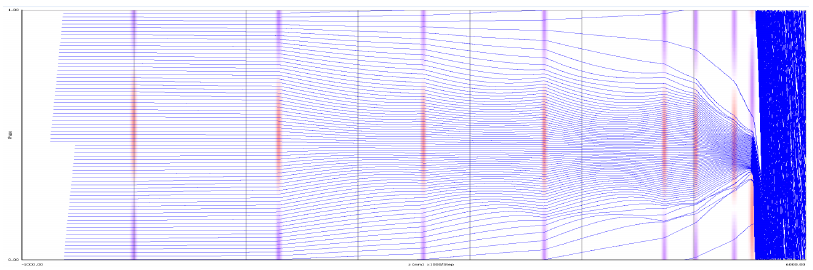}
\caption{Electron phase  profile of an \siunit{800}{MHz} klystron employing the Core Oscillation Method (COM)~\cite{Constable:2017hha}.
}
\label{fig:com}
\end{figure}

Recently, the High Efficiency International Klystron Activity (HEIKA) collaboration~\cite{Syratchev:2015a, Gerigk:2018ebm} has been formed that investigates novel techniques for high--efficiency klystrons.
Taking advantage of modern beam dynamic tools, methods such as the Bunching, Alignment and Collecting (BAC) method~\cite{Guzilov:2014a} and the Core Oscillation Method (COM)~\cite{Constable:2017hha} (Fig.~\ref{fig:com})
 have been developed that promise increased efficiencies up to $90\,\%$~\cite{Baikov:2015bif}.  
One advantage of these methods is that it is possible to increase the efficiency of existing klystrons by equipping them with a new electron optics, as was demonstrated retrofitting an existing tube from VDBT, Moscow. 
This increased the output power by almost 50\,\% and its efficiency from 42\,\% to 66\,\%~\cite{Jensen:2016a}.

To operate the ILC at an increased gradient of \siunit{35}{MV/m} would require that the maximum klystron output power is increased from $10$ to \siunit{11}{MW}. 
It is assumed that this will be possible by applying the results from this R\&D effort to high-efficiency klystrons.
 
\paragraph{Local Power--Distribution System (LPDS):}

In the baseline design, a single RF station with one modulator and klystron supplies RF to $39$ cavities, which corresponds to $4.5$ cryomodules~\cite[Sec. 3.6.4]{Adolphsen:2013kya}. Then  $2$ klystrons drive a $9$ cryomodule cryo-string unit.
The power is distributed by the LPDS, a system of waveguides, power dividers and loads. 
All cavities from a $9$-cavity module and half of a $8$--cavity module are connected in one LPDS, and three such LPDS units are connected to one klystron.
This arrangement allows an easy refurbishment such that a third klystron can be added to a cryo-string, increasing the available power per cavity by $50\,\%$ for a luminosity upgrade (cf.\ Sec.~\ref{subsec:upg-opt}).

The LPDS design must provide a cost--effective solution for the  distribution of 
the RF power with minimal losses, and at the same time provide the 
flexibility to adjust the power delivered to each cavity by at least $\pm20\,\%$ to allow for 
the specified spread in maximum gradient. 
The LPDS design therefore contains remotely controlled, motor-driven Variable Power Dividers (VPD), phase shifters, and H--hybrids that can distribute the power with the required flexibility.
This design allows one to optimise the power distribution during operation, based on the cavity performance in the installed cryomodule, and thus to get the optimum performance out of the system.
It does not require a measurement of the individual cavity gradients after the module assembly, and is thus compatible with the ILC production scheme, where only a fraction of the cryomodules are tested.
This is a notable difference from  the scheme employed at the E-XFEL, where $100\,\%$ of the modules were tested, and the the power distribution for each module was tailored to the measured cavity gradients, saving investment costs for the LPDS but making the system less flexible.

\subsubsection{Cryogenics}

The operation of the large number of superconducting cryomodules for the main linacs and the linacs associated with the sources requires a large--scale supply of liquid helium.
The cyomodules operate at \siunit{2}{K} and are cooled with superfluid helium, which at \siunit{2}{K} has a vapour pressure of about \siunit{32}{mbar}.

The accelerator is supplied with liquid helium by several cryogenic plants~\cite[Sec. 3.5]{Adolphsen:2013kya} of a size similar to those in operation at CERN for the LHC, at Fermilab, and DESY,
with a cooling capacity equivalent to about \siunit{19}{kW} at \siunit{4.5}{K}.
The \siunit{2}{K} and \siunit{4.5}{K} helium refrigerators are located in an underground access hall~\cite{bib:cr-0014} that is connected to the surface, where the helium compressors, gas tanks and further cryogenic infrastructure are located.
The total helium inventory is approximately $310 000$ liquid litres or about $41$ metric tonnes, about one third of the LHC's helium inventory.  A factor 2 more helium is needed for 
500~GeV operation.



\subsubsection{Series production and industrialisation, worldwide and in Europe}

Due to the construction of the E-XFEL, the industrial basis for the key SCRF components is broad and mature, in particular in Europe.
Europe has a leading supplier for raw material. 
In all three regions (Europe, America, Asia), several vendors for cavities have been qualified for ILC type cavities, and provided cost estimates in the past.
Two leading cavity vendors are European companies that have profited from large scale production of cavities for E-XFEL; 
both have won contracts for LCLS-II as a consequence.
RF couplers have also been successfully produced by European and  American vendors for the E-XFEL and LCLS-II projects.

ILC/TESLA type cryomodules have been built in laboratories around the world (DESY, CEA in Europe, FNAL and JLAB in America, KEK in Asia).
Series production has been established in America at Fermilab and JLAB for LCLS-II.
The largest series production was conducted by CEA in France, again for the E-XFEL, with the assembly of \num{103} cryomodules in total by an industrial partner under the supervision of CEA personnel, with a final throughput of one cryomodule produced every four working days.

ILC type, pulsed \siunit{10}{MW} klystrons are commercially available from two vendors in Japan and Europe.

For E-XFEL, China has been a supplier for niobium raw material and cryomodule cold masses (the cryostat with internal insulation and tubing).
For the planned SCLF project in Shanghai, China has started to develop cavity and cryomodule production capabilities, which will further broaden the worldwide production capabilities for SCRF components.
This reduces the risk that prices are pushed up by a monopoly of manufacturers for a large scale order of components as required for the ILC.

Overall, European industry is well prepared to produce the high-tech, high-value SCRF components needed for the ILC, which would likely constitute the largest fraction of any European in-kind contribution (IKC) to the ILC, at very competitive prices.
Thus, expenditure for the European IKC will likely stay in Europe, with an excellent chance to stay within the price range assumed in the value estimate.
Moreover, European companies are well positioned to win additional contracts from other regions, increasing the economic benefit for Europe from an ILC project.


\subsection{Accelerator design}

\subsubsection{Electron and positron sources}
\label{par:beampol}

The electron and positron sources are designed to produce \siunit{5}{GeV} beam pulses with a bunch charge that is $50\,\%$ higher than the design bunch charge of \siunit{3.2}{nC} ($\siunit{2\cdot 10^{10}}{e}$), in order to have sufficient reserve to compensate  any unforeseen inefficiencies in the beam transport.
In the baseline design, both sources produce polarized beams with the same time structure as the main beam, \ie, $1312$ bunches in a $\siunit{727}{\mu s}$ long pulse.

The electron source design~\cite{Adolphsen:2013kya} is based on the SLC polarized electron source, which has demonstarted that the bunch charge, polarisation and cathode lifetime parameters are feasible.
The long bunch trains of the ILC do require a newly developed laser system and powerful preaccelerator structures, for which preliminary designs are available.
The design calls for  a Ti:sapphire laser impinging on a photocathode based on a strained GaAs/GaAsP superlattice
structure, which will produce electron bunches with an expected polarisation of \siunit{85}{\%},
sufficient for \siunit{80}{\%} beam polarization at the interaction point, as demonstrated at SLAC~\cite{Alley:1995ia}.

The positron source poses a larger challenge. 

In the baseline design, hard gamma rays are produced in a helical undulator driven by the main electron beam, which are converted to positrons in a rotating target.
Positrons are captured in a flux concentrator or a quarter wave transformer, accelerated to \siunit{400}{MeV} in two normal conducting preaccelerators followed by a superconducting accelerator very similar to the main linac, before they are injected into the damping rings at \siunit{5}{GeV}.
The helical undulators produce photons with circular polarisation, which is transferred to the positrons produced in the target, which are longitudinally polarised as a result.
The positron polarisation thus achieved is $30\,\%$.
The E-166 experiment at SLAC has successfully demonstrated this concept  \cite{Alexander:2009nb}, albeit at intensities much lower than foreseen for the ILC. 
Technological challenges of the undulator source concept are the target heat load, the radiation load in the flux concentrator device, and the dumping of 
the high intensity photon beam remnant.

As an alternative, an electron-driven positron source concept has been developed.
In the electron-driven scheme, a \siunit{3}{GeV} electron beam from a dedicated normal conducting linac produces positrons in a rotating target.
The electron drive beam, being independent from the main linac, has a completely different time structure. 
Positrons are produced in $20$ pulses at \siunit{300}{Hz} with $66$ bunches each.  With this scheme, it takes about \siunit{67}{ms} to produce the  positrons needed for a single Main Linac pulse with its $1312$ bunches, compared to \siunit{0.8}{ms} for the undulator source.
This different time structure spreads the heat load on the target over a longer time, allowing a target rotation speed of only \siunit{5}{m/s} rather than \siunit{100}{m/s}, which reduces the engineering complexity of the target design, in particular the vacuum seals of the rotating parts.
Although not free from its own engineering challenges, such as the high beam loading in the normal conducting cavities, the electron driven design is currently considered to be a low risk design that is sure to work.

Aside from the low technical risk, the main advantage of the electron driven design is the independence of positron production and electron main linac operation, which is an advantage for accelerator commissioning and operation in general.
In particular, electron beam energies below \siunit{120}{GeV} for operation at the $Z$ resonance or the $WW$ threshold would be no problem.
The undulator source, on the other hand, offers the possibility to provide beams at the maximum repetition rate of \siunit{10}{Hz} given by the damping time in the damping rings of \siunit{100}{ms}, whereas the electron driven scheme is limited to \siunit{6}{Hz} due to the additional \siunit{66}{ms} for positron production.
The main difference between the concepts is the positron polarisation offered by the undulator source, which adds significantly to the physics capabilities of the machine.  The physics implications of positron polarization is discussed later in the report, in Secs.~\ref{subsec:beampol} and ~\ref{subsec:polarisation}. 

Both concepts have been reviewed recently \cite{PWG:2018a} inside the ILC community, with the result that both source concepts appear viable, with no known show stoppers, but they
require some more engineering work. 
The decision on the choice will be taken once the project has been approved, based on the physics requirements, operational aspects, and technological maturity and risks. 

\paragraph{Beam polarisation and spin reversal}

At the ILC, the electron beam and potentially the positron beam are longitudinally polarised at the source, \ie,  the polarisation vector is oriented parallel or antiparallel 
to the beam direction.
Whenever a longitudinally polarised beam of energy $E\sub{beam}$ is deflected by an angle $\theta\sub{bend}$, the polarisation vector undergoes a precession through an angle $\theta\sub{pol} =  \gamma a \theta\sub{bend}$~\cite{Moffeit:2005pb}, 
with the Lorentz factor $\gamma = E\sub{beam}/m\sub{e}$ and the electron's anomalous magnetic moment $a = (g-2)/2$. 
To preserve the longitudinal beam polarisation during the long transport from the source through the damping rings to the start of the main linac, which involves many horizontal bends, the beam polarisation vector is rotated into the transverse plane, perpendicular to the damping ring plane, before the beam is transferred to the damping rings, and rotated back to a longitudinal direction by a set of spin rotators at the end of the RTML (see Sec.~\ref{sec:rtml}).
Through the use of two rotators, it is possible to bring the polarisation vector into any desired direction, and compensate any remaining net precession between these spin rotators and the interaction point, so that any desired longitudinal or transverse polarisation at the IP can be provided.

To control systematic effects, fast helicity reversal is required.  This is helicity reversal of each beam independently, on a pulse to pulse basis, which must be achieved without a change of the magnetic fields of the spin rotator magnets.
For the electron beam, a fast helicity reversal is possible through a flip of the cathode laser polarisation.  For the undulator-based positron source, the photon polarisation is given by the undulator field.  Two parallel sets of spin rotators in front of the damping rings are used that rotate the polarisation vector either to the $+y$ or $-y$ direction.  With this scheme,
fast kickers can select a path through either of the two spin rotators and thus provide a fast spin reversal capability~\cite{Moffeit:2005pb,Malysheva:2016jdr}.

\subsubsection{Damping rings}

The ILC includes two oval damping rings of \siunit{3.2}{km} circumference, sharing a common tunnel in the central accelerator complex.
The damping rings reduce the horizontal and vertical emittance of the beams by almost six orders of magnitude\footnote{The vertical emittance of the positrons is reduced from $\epsilon_{\mathrm{y}} \approx 0.8\,{\mathrm{\mu m}}$ to $2\,{\mathrm{pm}}$.} within a time span of only \siunit{100}{ms}, to provide the low emittance beams required at the interaction point. 
Both damping rings operate at an energy of \siunit{5}{GeV}.

The damping rings' main objectives are
\begin{itemize} 
\item to accept electron and positron beams at large emittance and produce the low-emittance beams required for high-luminosity production.
\item to dampen the incoming beam jitter to provide highly stable beams.
\item to 
delay bunches from the source and allow feed-forward systems to compensate for pulse-to-pulse variations in parameters such as the bunch charge.
\end{itemize}

Compared to today's fourth generation light sources, the target value for the normalized beam emittance ($\siunit{4}{\mu m}$/\siunit{20}{nm} for the normalised horizontal / vertical beam emittance) is low, but not a record value, and it is thus considered to be a realistic goal.

The main challenges for the damping ring design are to provide
\begin{itemize} 
\item a sufficient dynamic aperture to cope with the large injected emittance of the positrons.
\item a low equilibrium emittance in the horizontal plane.
\item a very low emittance in the vertical plane.
\item a small damping time constant.
\item damping of instabilities from electron clouds (for the positron DR) and fast ions (for the electron DR).
\item a small (\siunit{3.2-6.4}{ns}) bunch spacing, requiring very fast kickers for injection and ejection.
\end{itemize}

Careful optimization has resulted in a TME (Theoretical Minimum Emittance) style lattice for the arcs that balances a low horizontal emittance with the required large dynamic aperture~\cite[Chap. 6]{Adolphsen:2013kya}. 
Recently, the horizontal emittance has been reduced further by lowering the dispersion in the arcs through the use of longer dipoles~\cite{bib:cr-0016}.
The emittance in the vertical plane is minimised by careful alignment of the magnets and tuning of the closed orbit to compensate for misalignments and field errors, as demonstrated at the CESR-TA facility~\cite{Billing:2011zc}.

The required small damping time constant requires large synchrotron radiation damping, which is provided by the insertion of $54$ wigglers in each ring.
This results in an energy loss of up to $7.7\,{\mathrm{MV}}$ per turn and up to $3.3\,{\mathrm{MW}}$ RF power to 
store the positron beam at the design current of $390\,{\mathrm{mA}}$.  This
actually exceeds the average beam power of the accelerated positron beam, $2.6\,{\mathrm{MW}}$ at 
a $250\,{\mathrm{GeV}}$.

Electron cloud (EC) and fast ion (FI) instabilities limit the overall current in the damping rings to about \siunit{400-800}{mA}, where the EC limit that affects the positrons is assumed to be more stringent. 
These instabilities arise from electrons and ions being attracted by the circulating beam towards the beam axis. 
A low base vacuum pressure of \siunit{10^{-7}}{Pa} is required to limit these effects to the required level.
In addition, gaps between bunch trains of around $50$ bunches are required in the DR filling pattern, which permits the use of clearing electrodes to mitigate EC formation.
These techniques have been developed and tested at the CESR-TA facility~\cite{Conway:2012zza}

In the damping rings, the bunch separation is only \siunit{6.4}{ns} (\siunit{3.2}{ns} for a luminosity upgrade to $2625$ bunches). 
Extracting individual bunches without affecting their emittance requires kickers with rise/fall times of \siunit{3}{ns} or less.
Such systems have been tested at ATF~\cite{Naito:2010zzb}.

The damping ring RF system will employ superconducting cavities operating at half the Main Linac frequency (\siunit{650}{MHz}).
Klystrons and accelerator modules can be scaled from existing \siunit{500}{MHz} units in operation at CESR and KEK~\cite[Sec. 6.6]{Adolphsen:2013kya}.

\subsubsection{Low emittance beam transport: ring to Main Linac (RTML)}
\label{sec:rtml}

The Ring to Main Linac (RTML) system~\cite[Chap. 7]{Adolphsen:2013kya} is responsible for transporting and matching the beam from the Damping Ring to the entrance of the Main Linac.
Its main objectives are
\begin{itemize} 
\item transport of the beams from the Damping Rings at the center of the accelerator complex to the upstream ends of the Main Linacs,
\item collimation of the beam halo generated in the Damping Rings,
\item rotation of the spin polarisation vector from the vertical to the desired angle at the IP (typically, in longitudinal direction).
\end{itemize}

The RTML consists of two arms for the positrons and the electrons. 
Each arm comprises a damping ring extraction line transferring the beams from the damping ring extraction into the main linac tunnel, a long low emittance transfer line (LTL), the turnaround section at the upstream end of each accelerator arm, and a spin rotation and diagnostics section.

The long transport line is the largest, most costly part of the RTML.
The main challenge is to transport the low emittance beam at \siunit{5}{GeV} with minimal emittance increase, and in a cost-effective manner, considering that its total length is about \siunit{14}{km} for the \siunit{250}{GeV} machine.

In order to preserve the polarisation of the particles generated in the sources, their spins are rotated into a vertical direction (perpendicular to the Damping Ring plane) before injection into the Damping Rings. 
A set of two rotators~\cite{Emma:1995kf} employing superconducting solenoids allows to rotate the spin into any direction required.

At the end of the RTML, after the spin rotation section and before injection into the bunch compressors (which are considered part of the Main Linac, not the RTML~\cite{bib:cr-0010}), a diagnostics section allows measurement of  the emittance and the  coupling between the  horizontal and vertical plane.
A skew quadrupole system is included to correct for any such coupling.

A number of circular fixed-aperture and rectangular variable-aperture collimators in the RTML provide betatron collimation at the beginning of the LTL, in the turn around and before the bunch compressors.

\subsubsection{Bunch compressors and Main Linac}

\begin{figure}[htbp]
   \includegraphics[width=\hsize]{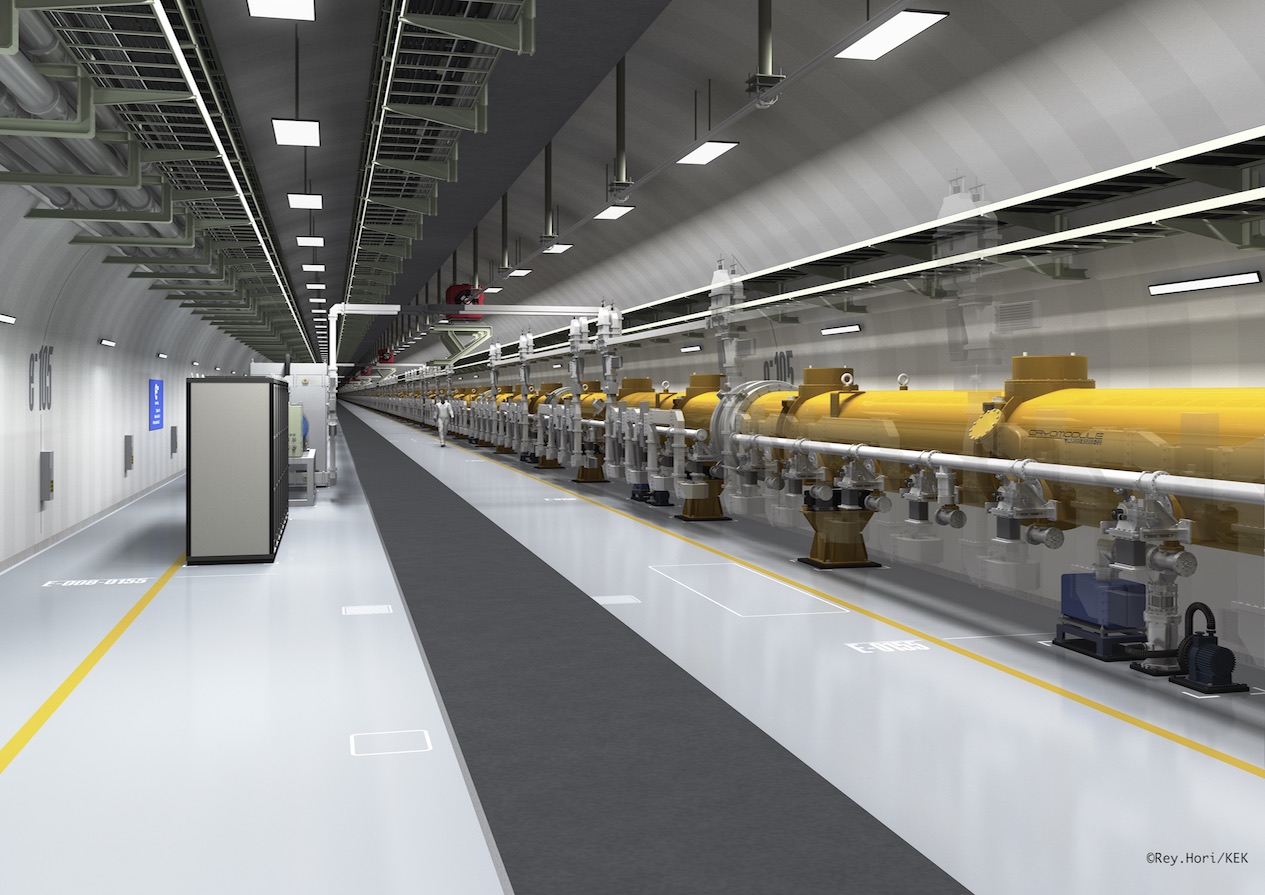}
\caption{Artist's rendition of the ILC Main Linac tunnel. The shield wall in the middle has been removed.
\copyright Rey.Hori/KEK.}
\label{fig:ilc-tunnel}
\end{figure}

\begin{figure}[htbp]
   \includegraphics[width=\hsize]{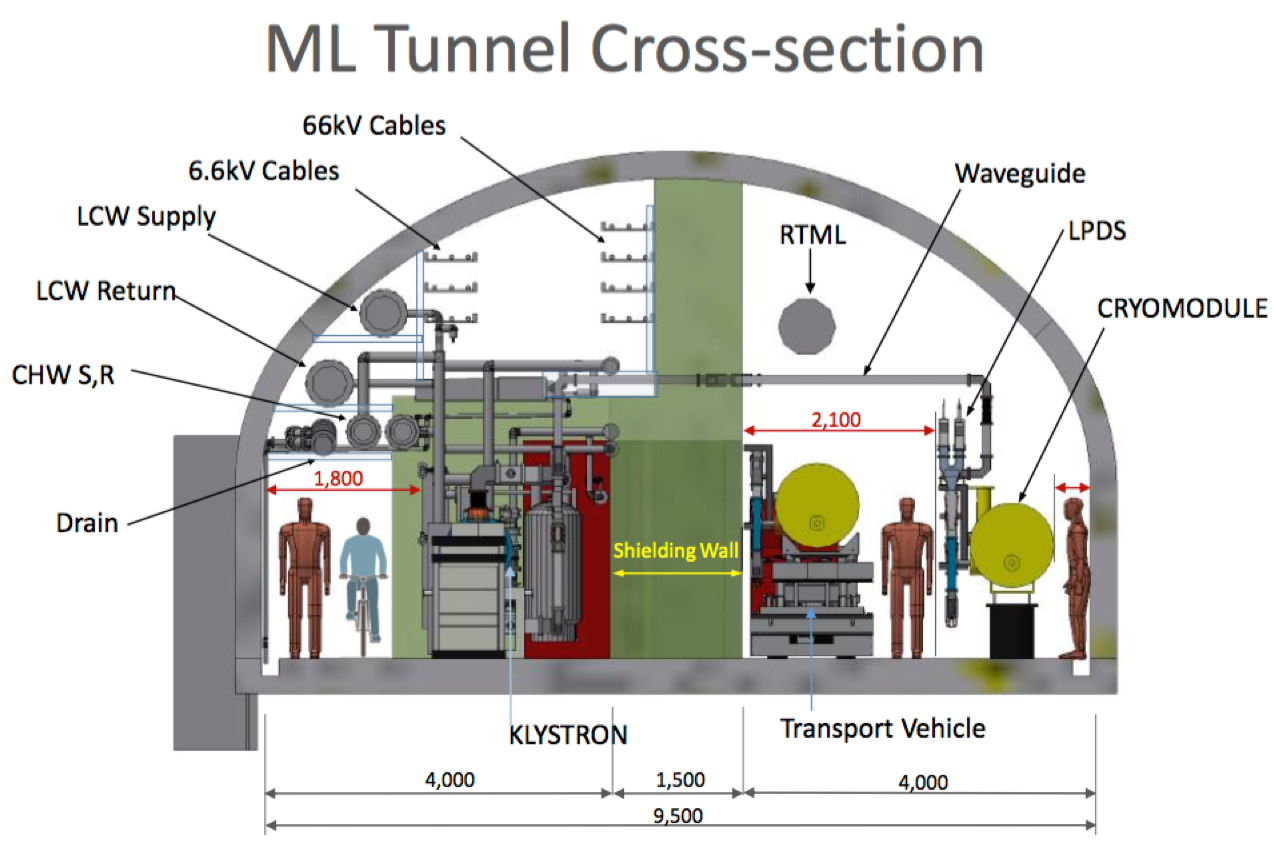}
\caption{Cross section through the Main Linac tunnel.}
\label{fig:ml-tunnel}
\end{figure}

The heart of the ILC are the two Main Linacs, which accelerate the beams from $5$ to \siunit{125}{GeV}.
The linac tunnel, as depicted in Figs.~\ref{fig:ilc-tunnel} and \ref{fig:ml-tunnel}, has two parts, separated by a shield wall. 
One side (on the right in Fig.~\ref{fig:ilc-tunnel}) houses the beamline with the accelerating cryomodules as well as the RTML beamline hanging on the ceiling.
The other side contains power supplies, control electronics, and the modulators and klystrons of the High-Level RF system.
The concrete shield wall (indicated as a dark-grey strip in in Fig.~\ref{fig:ilc-tunnel}) has a thickness of \siunit{1.5}{m}~\cite{bib:cr-0012}.
The shield wall allows access to the electronics, klystrons and modulators during operation of the klystrons with cold cryomodules, protecting personnel from X-ray radiation emanating from the cavities caused by dark currents.
Access during beam operation, which would require a wall thickness of \siunit{3.5}{m}, is not possible.

The first part of the Main Linac is a two-stage bunch compressor system~\cite[Sec. 7.3.3.5]{Adolphsen:2013kya}, each consisting of an accelerating section followed by a wiggler. 
The first stage operates at \siunit{5}{GeV}, with no net acceleration, the second stage accelerates the beam to \siunit{15}{GeV}.
The bunch compressors reduce the bunch length from $6$ to \siunit{0.3}{mm}.

After the bunch compressors, the Main Linac continues for about \siunit{6}{km} with a long section consisting entirely of cryomodules, bringing the beam to \siunit{125}{GeV}. 

\paragraph{RF distribution:}

Each cryomodule contains $9$ cavities, or for every third module, $8$ cavities and a package with a superconducting quadrupole, corrector magnets, and beam position monitor.
Nine such modules, with a total of $117$ cavities, are powered by $2$ klystrons and provide \siunit{3.83 (4.29)}{GeV} at a gradient of \siunit{31.5 (35)}{MV/m}.
Table~\ref{tab:ml-units} gives an overview over the units that form the linacs.
The waveguide distribution system allows an easy refurbishment to connect a third klystron for a luminosity upgrade.
The $50\,\%$ RF power increase would allow $50\,\%$ higher current through smaller bunch separation, and longer beam pulses because of a reduced filling time, so that the number of bunches per pulse and hence the luminosity can be doubled, while the RF pulse duration of \siunit{1.65}{ms} stays constant.

\paragraph{Cryogenic supply:}

A $9$ module unit forms a cryo string, which is connected to the helium supply line with a Joule-Thomson valve.
All helium lines are part of the cryomodule, obviating the need for a separate helium transfer line. 
Up to $21$ strings with $189$ modules and \siunit{2.4}{km} total length can be connected to a single plant; 
this is limited by practical plant sizes and the gas--return header pressure drop.  

\begin{table}[htbp]
\begin{tabular}{llrr}
Unit & Comprises & Length & Voltage \\
\hline
Cavity & \siunit{1.038}{m} active length & \siunit{1.25}{m} & \siunit{32.6 / 36.2}{MV} \\
Cryomodule & $8\,^2/_3$ cavities & \siunit{12.65}{m} & \siunit{282 / 314}{MV} \\
RF Unit & $4.5$ cryomodules & \siunit{58.2}{m} & \siunit{1.27 / 1.41}{GV} \\
Cryostring & 2 RF units & \siunit{116.4}{m} & \siunit{2.54 / 2.82}{GV} \\
Cryounit & up to 21 cryostrings & \siunit{2454}{m} & \siunit{53.4 / 39.3}{GV} \\
\end{tabular}
\caption{
  \label{tab:ml-units}
  Units that make up the main linacs. 
  The voltage takes into account that the beam is $5^\circ$ shifted in phase (``off crest'') for longitudinal stability, and is given for an average gradient of \siunit{31.5 / 35}{MV/m}.
  A RF unit is powered by one klystron, each cryostring is connected by a valve box to the liquid helium supply, and a cryounit is supplied by one cryogenic plant.
  Total lengths include additional space between components.  
}
\end{table}

\paragraph{Cost reduction from larger gradients:}

\begin{figure*}[htbp]
   \includegraphics[width=0.8\hsize]{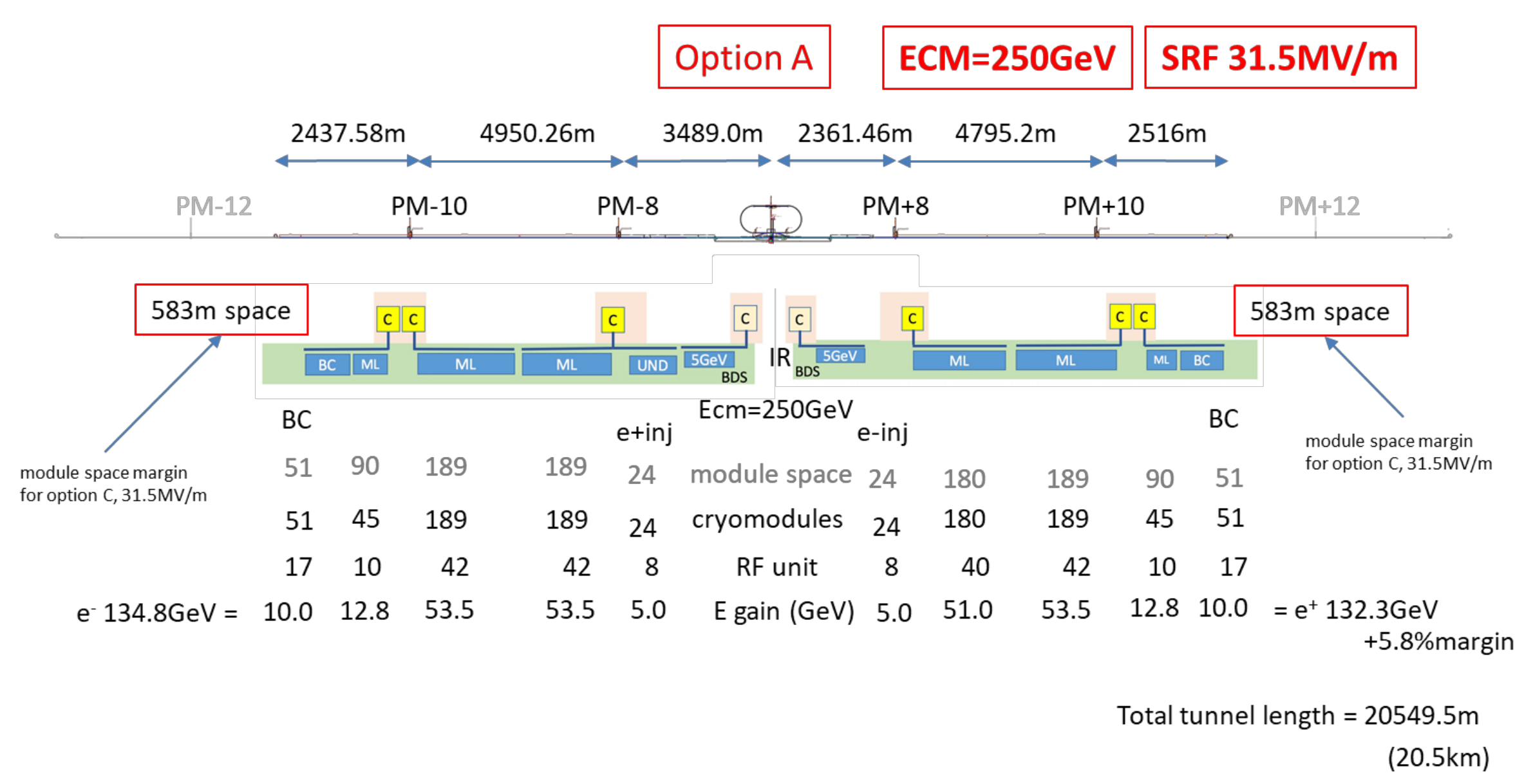}
   \includegraphics[width=0.8\hsize]{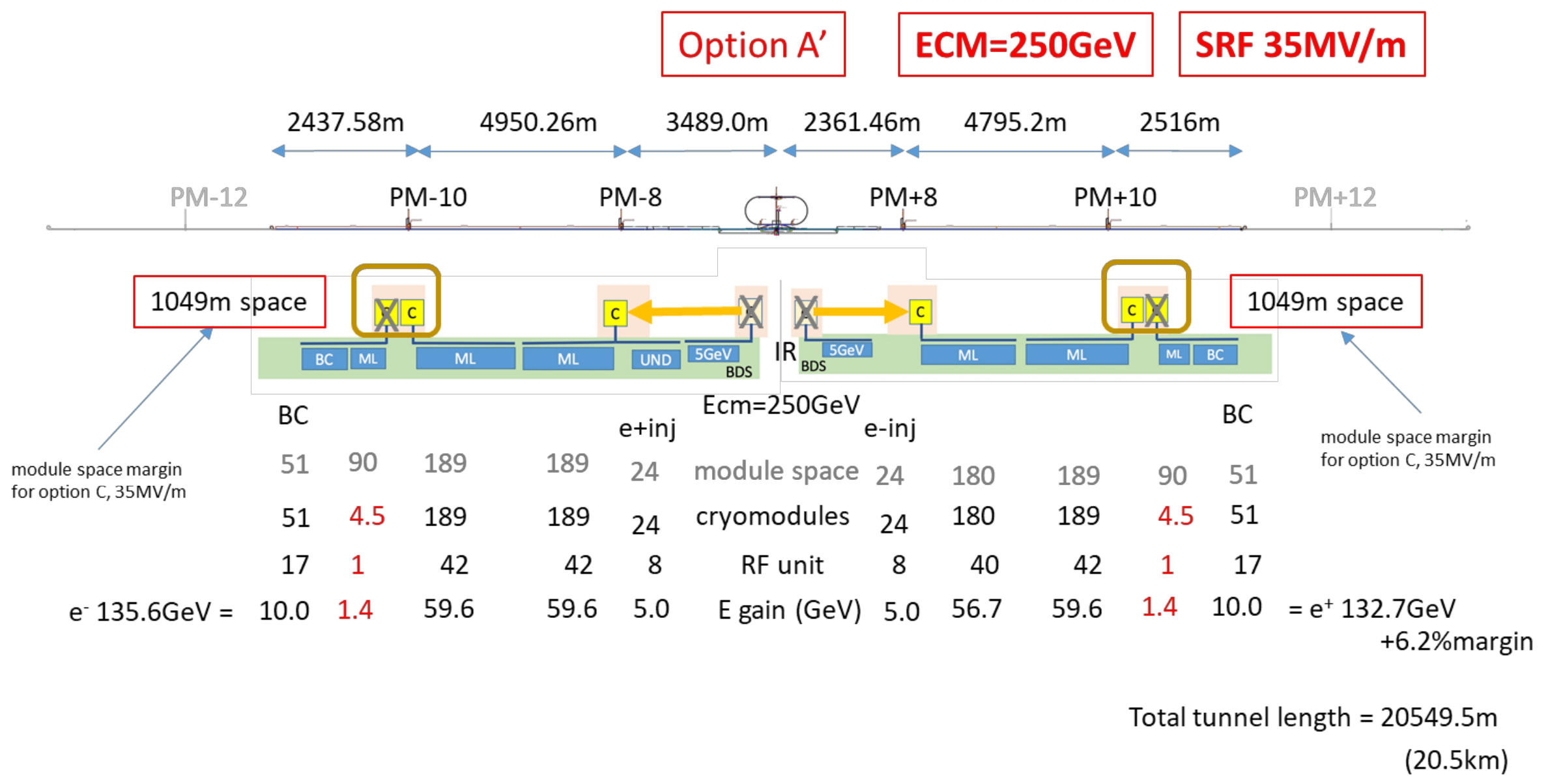}
\caption{Cryogenic layout for a gradient of \siunit{31.5}{MV/m} (top) and \siunit{35}{MV/m} (bottom)~\cite{Evans:2017rvt}.
``Module space'' indicates how many cryomodules can be physically installed, ``cryomodules'' and ``RF unit'' indicates the number of actually installed modules and klystrons (one klystron per 4.5 cryomodules). ``E gain'' indicates the energy gain in GeV. ``BC'', ``ML'', ``e+ inj'', ``e- inj'' and ``UND'' refer to the sections with need for liquid helium: bunch compressor, main linac, 5GeV boosters in the positron and electron source, and the positron source undulator section, respectively. PM$\pm8, 10, 12$ refer to access hall locations, ``C'' to cryo plants; meter numbers on top indicate the length of the corresponding section.}
\label{fig:ml-cryo-opta}
\end{figure*}

Figure~\ref{fig:ml-cryo-opta} shows the layout of the cryogenic supply system for the \siunit{250}{GeV} machine.
At the top, the situation is depicted for the gradient of \siunit{31.5}{MV/m} with a quality factor of $Q\sub{0}=1.0\cdot 10^{10}$, as assumed in the TDR~\cite{Adolphsen:2013kya}. 
In this case, the access points PM$\pm 10$ would house two cryogenic plants, each supplying up to $189$ cryomodules or an equivalent cryogenic load.   In this configuration $6$ large plants in the access halls plus $2$ smaller plants in the central region would be needed.
The bottom picture shows the situation for a gradient of \siunit{35}{MV/m} with $Q\sub{0}=1.6\cdot 10^{10}$, as could be expected from successful R\&D. 
The increased gradient would  allow reduction of the total number of cryomodules by roughly $10\,\%$ from $987$ to $906$. The increased quality factor would reduce the dynamic losses such that $4$ cryo plants would provide sufficient helium.

In general, the accelerator is designed to make good use of any anticipated performance gain from continued high gradient R\&D,  in the case that raising the gradient is seen to be  beneficial from an economical point of view, without incurring unwanted technology risk.

\subsubsection{Beam delivery system and machine detector interface}
\label{subsubsec:bds_mdi}

The Beam Delivery System (BDS) transports the $e^+/e^-$ beams from the end of the main linacs, focuses them to the required small beam spot at the Interaction Point (IP), brings them into collision, and transports the spent beams to the main dumps~\cite[Chap. 8]{Adolphsen:2013kya}.
The main functions of the BDS are
\begin{itemize}
\item measuring the main linac beam parameters and matching it into the final focus.
\item protecting beamline and detector from mis-steered beams~\footnote{On the electron side, the protective fast beam abort system is actually located upstream of the positron source undulator.}.
\item removing large amplitude (beam--halo) and off--momentum particles from the beam to minimize background in the detector.
\item accurately measuring the key parameters energy and polarisation before and after the collisions.
\end{itemize}
The BDS must provide sufficient diagnostic and feedback systems to achieve these goals.

The BDS is designed such that it can be upgraded to a maximum beam energy of \siunit{500}{GeV}; components such as the beam dumps, that are not cost drivers for the overall project but would be cumbersome to replace later, are dimensioned for the maximum beam energy from the beginning.
In other places, such as the energy collimation dogleg, those components necessary for \siunit{125}{GeV} beam operation are installed and space for a later upgrade is reserved.

Overall, the BDS is \siunit{2254}{m} long from the end of the main linac (or the undulator and target bypass insert of the positron source on the electron side, respectively) to the IP.

\paragraph{Diagnostics and collimation section:}
The BDS starts with a diagnostics section, where emittance, energy and polarisation are measured and any coupling between the vertical and horizontal planes is corrected by a set of skew quadrupoles.
The energy measurement is incorporated into the machine protection system and can, \eg,  extract off-momentum bunches caused by a klystron failure in the main linac that would otherwise damage the machine or detector.
An emergency dump~\cite{bib:cr-0013} is dimensioned such that it can absorb a full beam pulse at \siunit{500}{GeV}, sufficient for \siunit{1}{TeV} operation.

The diagnostics section is followed by a collimation system, which first removes beam halo particles (betatron collimation). 
Then, off-momentum particles are removed.
In this energy collimation section, sufficient dispersion must be generated by bending the beam in a dogleg, while avoiding excessive synchrotron radiation generation in dispersive regions that leads to an increase of the horizontal emittance.
This emittance dilution effect grows as $E\sub{beam}^6$ at constant bending radius for the normalised emittance, and determines the overall length of the energy collimation section for a maximum \siunit{500}{GeV} beam energy to about \siunit{400}{m}.

\paragraph {Final focus with feedback system and crab cavities:}

The final focus system demagnifies the beam to the required spot size of \siunit{516 \times 7.7}{nm^2} by means of a final quadrupole doublet.
Even the relatively small energy spread of $\approx 0.1\,\%$ leads to a significant spread of the focal length of the doublet and requires a correction to achieve the desired beam size, which is realised by a local chromaticity correction scheme~\cite{Raimondi:2000cx}.

To bring the beams to collision with the neccessary nanometre accuracy requires a continuous compensation of drift and vibration effects.
Along the ILC, the pulse length and bunch separation (\siunit{727}{\mu s} and \siunit{554}{ns}, respectively) are large enough to allow corrections between pulses as well as within a bunch train (intratrain feedback).
Beam-beam offsets of a fraction of the beam size lead to a measurable deflection of the outgoing beams,and these measurements are used to feed fast stripline kickers that stabilize the beam. 
Finally, the \siunit{3.9}{GHz} crab cavities close to the interaction point are incorporated that rotate the bunches to compensate for the \siunit{14}{mrad} beam crossing angle~\cite[Sect. 8.9]{Adolphsen:2013kya}.

\begin{figure}[htbp]
   \includegraphics[width=\hsize]{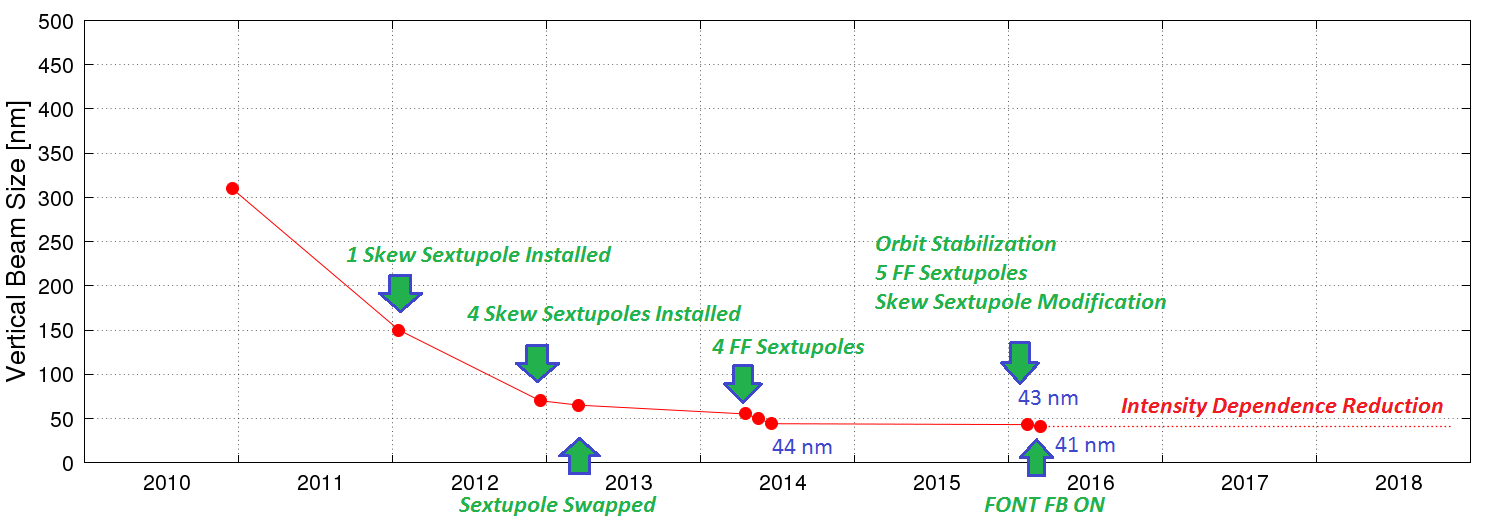}
\caption{Beamsizes achieved at the Accelerator Test Facility 2 (ATF2) as a function of time~\cite{bib:atf2esu}. The latest result (\siunit{41}{nm}~\cite{Okugi:2017jji}) is within $10\,\%$ of the goal beam size of \siunit{37}{nm}.}
\label{fig:atf-results}
\end{figure}

\paragraph {Test results from ATF2:}
The Accelerator Test Facility 2 (ATF2) was built at KEK in 2008 as a test bench for the ILC final focus scheme~\cite[Sec. 3.6]{Adolphsen:2013jya}.
Its primary goals were to achieve a \siunit{37}{nm} vertical beam size at the interaction point (IP), and to demonstrate beam stabilisation at the nanometre level~\cite{Grishanov:2005ek,Grishanov:2006kx}.
After scaling for the different beam energies (ATF2 operates at $E\sub{beam}=\siunit{1.3}{GeV}$), the \siunit{37}{nm} beam size corresponds to the TDR design value of $\sigma\sub{y}^* = \siunit{5.7}{nm}$ at \siunit{250}{GeV} beam energy.
As Fig.~\ref{fig:atf-results} shows, this goal has been reached within $10\,\%$~\cite{Okugi:2017jji} by the successive application of various correction and stabilisation techniques, 
validating the final focus design, in particular the local chromaticity correction~\cite{White:2014vwa}.

The fifth generation FONT5 feedback system~\cite{Apsimon:2018bpq} for the ILC and CLIC has also been tested at the ATF2, where a beam stabilisation to \siunit{41}{nm} has been demonstrated~\cite{Ramjiawan:2018egu}.

Since November 2016, intensity-dependence effects on the ATF2 beam size have been studied extensively.  
They show a degradation of the beam size with increasing intensity that is compatible with the effect of  wakefields.
Simulations and experiments in ATF2 show that the effect is not important when scaled to ILC. 
Also, it  could be mitigated by including a dedicated ``wakefield knob'' in the routine tuning procedure.

\paragraph {Machine detector interface (MDI):}



The ILC is configured to have two detectors that share one interaction point, with one detector in data taking position at any time, in a so--called ``push--pull'' operation~\cite[Sec. 8.4]{Adolphsen:2013jya}.
Both detectors are mounted on movable platforms that allow an exchange of the detectors within approximately \siunit{24}{hours}.

In the push--pull scheme, the innermost final focus quadrupole ``QD0'', a slim, superconducting magnet package combined with a sextupole for local chromaticity correction, is installed within the detectors. 
The other part of the final focus doublet (``QF1'') is located outside the detector on a bridge, and does not move with the detector.
Since the TDR, the free space $L^*$ between interaction point and the QD0 edge has been harmonised to a common value of $L^*=\siunit{4.1}{m}$~\cite{bib:cr-0002}, which facilitates the design of a final focus optics that delivers optimal and equal performance to both detectors.

The detectors are located in an underground cavern. 
In contrast to the TDR design, it is foreseen to have a large vertical access shaft~\cite{bib:cr-0003}, which permits a CMS--style detector installation concept, in which the detectors are assembled in large modules in a surface hall and lowered into the hall by means of a gantry crane capable of lowering pieces up to \siunit{4000}{t}.
As the CMS experience shows, this concept significantly reduces the schedule risk associated with the experimental hall, since the cavern needs to be available for detector installation only one or two years prior to commissioning.



\paragraph {Main dump:}

The main beam dumps~\cite[Sect. 8.8]{Adolphsen:2013kya} are rated for a maximum beam power of \siunit{17}{MW}~\cite{bib:cr-0013}, enough for a \siunit{1}{TeV} upgrade of the accelerator.
The main dump design is based on the successful SLAC \siunit{2.2}{MW} beam dump~\cite{Walz:1967nz}.
It  utilises water at \siunit{10}{bar} pressure (to prevent boiling) as absorber medium. 
The main engineering challenges lie in the safe recombination of the produced oxyhydrogen gas and in the safe containment and disposal of radioisotopes, in particular tritium and $^7{\mathrm{Be}}$ produced from spallation processes.
The entry window is another component that has to be carefully designed. 


\paragraph {Measurement of beam energy, luminosity, and beam polarisation:}

Two energy spectrometers, one located \siunit{700}{m} upstream of the IP, the other \siunit{55}{m} downstream, provide independent and complementary measurements of the beam energy with an accuracy of \siunit{100}{ppm}~\cite{Boogert:2009ir}.

The luminosity is measured to $10^{-3}$ accuracy from low angle Bhabha scattering in the so--called LumiCal (see Sect.~\ref{subsub:det:forward}) at polar angles from $30$ to \siunit{90}{mrad}.
Additional calorimeters (BeamCal) in the region $5$ to \siunit{30}{mrad} provide a fast signal that is sensitive to the beam sizes and offsets of the colliding beam, and that can thus be used for their tuning, as part of an intra-beam feedback system (see Sec.~\ref{subsubsec:bds_mdi}).

Beam polarisation is measured with \siunit{0.25}{\%} accuracy by means of Compton scattering: electrons that scatter off green or infrared light laser photons lose enough energy that they can be detected in a spectrometer; their momentum spectrum is used to fit the beam polarisation~ \cite{Vormwald:2015hla}.
Two such polarimeters are located \siunit{1800}{m} upstream and \siunit{150}{m} downstream of the IP, which allows to interpolate the precise polarisation at the IP and control the systematics, including effects from precession of the polarisation vector by transverse fields and depolarising effects in the interaction, which lead to a sizeable variation of the polarisation within the bunch during the collision (see Sect.~\ref{subsubsec:sysuncert}).



\subsection{Upgrade options \label{subsec:upg-opt}}

Given the high initial investment for a facility as large as the ILC, it is mandatory to have an interesting physics programme for several decades, with the possibility to adapt the programme to the needs arising from the knowledge obtained by the LHC, the ILC itself, all other particle physics experiments, and other domains of physics such as cosmology.
Several options exist for upgrades of the ILC in terms of energy, luminosity, and beam polarisation.

\subsubsection{Energy upgrade}
\label{subsubsec:upg-optE}

The obvious advantage of a linear collider is its upgradeability in energy.
Basically, the main linacs can be extended as far as desired, at constant cost per added beam energy, with some added cost for the relocation of the turn arounds and bunch compressors.
Additional costs arise when the beam delivery system (BDS), including the beam dumps, has to be extended to handle the increased beam energy. 
The current ILC BDS is designed to be easily upgradeable for centre of mass energies up to \siunit{1}{TeV} at minimal cost.

Depending on the actual gradient achieved for the construction of the ILC, up to $171$  cryomodules could be installed in addition to those needed to reach \siunit{250}{GeV}, which would increase the centre-of-mass energy by about \siunit{54}{GeV} to around \siunit{304}{GeV}, as Fig.~\ref{fig:ml-cryo-opta} shows, 
and possibly require the installation of two additional cryo plants.

A further energy upgrade would require extension of the tunnel.
The Kitakami site can accommodate a total accelerator length of at least \siunit{50}{km}, more than enough for a \siunit{1}{TeV} centre--of--mass energy.
Any extension of the accelerator would proceed by adding new cryomodules at the low energy (upstream) ends of the accelerator. There is no need to move modules already installed. 

An upgrade would likely proceed in two phases: a preparation phase while the accelerator is still operated and produces data, and a refurbishment phase where the accelerator is shut down.

During the preparation phase, the necessary components---in particular the cryomodules, klystrons, and modulators---would be acquired and built.
At the same time, civil engineering would proceed with the excavation of new access tunnels, underground halls, and the main tunnel.
Recent studies conducted during road tunnel construction in the Kitakami area, in the same rock formation as foreseen for the ILC, indicate that the level of vibrations caused by tunnelling activities would allow to bring the new tunnels quite close to the existing ones before machine operation would be affected~\cite{bib:sanuki:lcws2018}, minimising the shutdown time necessary.

During the installation phase, the newly built tunnels would be connected to the existing ones, the beam lines at the turn-around and the wiggler sections of the bunch compressors would be dismantled, and the new cryomodules would be installed as well as the new turn-around and bunch compressors. 
At the same time, any necessary modifications to the positron source and the final focus can be made.
With the cryomodules ready for installation at the beginning of the shut down period, it is estimated that the shutdown could be limited to about a year for an energy upgrade.


\subsubsection{Luminosity upgrade}
\label{subsubsec:upg-optL}

The luminosity of the ILC can be increased by increasing the luminosity per bunch (or per bunch charge), or increasing the number of bunches per second~\cite{Harrison:2013nva}.

Increasing the luminosity per bunch requires a smaller vertical beam spot size, which may be achieved by tighter focusing and/or smaller beam emittance.
Studies indicate that with enough operating experience, there is potential for a further luminosity increase. 
This route to increased luminosity is, however, invariably linked to higher beam disruption,  which brings a risk of a luminosity loss due to mis-steering the beam. 
Thus, a very accurate feedback system is required.

The ILC design also has the potential to increase the number of colliding bunches per second, by doubling the number of bunches per pulse, and possibly by increasing the pulse repetition frequency.

Doubling the number of bunches per pulse to $2625$ would require a smaller bunch spacing, requiring  the installation of $50\,\%$ more klystrons and modulators. 
Since  the RF pulse length of \siunit{1.65}{ms} is unchanged, the cryogenic load is essentially unchanged.
Doubling the number of bunches would double the beam current in the damping rings.
For the positron damping ring, this may surpass the limitations from electron cloud (EC) instabilities. 
To mitigate this risk, the damping ring tunnel is large enough to house a third damping ring, so that the positron current could be distributed over two rings.

The pulse repetition rate (\siunit{5}{Hz} in the baseline configuration) is limited by the available cryogenic capacity, the damping time in the damping rings, and the target heat load in the positron source target.
The damping rings are designed for a \siunit{100}{ms} damping time and thus capable of a repetition rate of up to \siunit{10}{Hz}, twice the nominal rate.
Operation at an increased repetition rate would be possible if after an energy upgrade the machine is operated below its maximum energy (e.g., \siunit{250}{GeV} operation of a \siunit{500}{GeV} machine for a larger low-energy data set), or if additional cryogenic capacity is installed.

\subsubsection{Polarisation upgrade}
\label{subsubsec:upg-optP}

The baseline design foresees at least $80\,\%$ electron polarisation at the IP, combined with $30\,\%$ positron polarisation for the undulator positron source.
At beam energies above \siunit{125}{GeV}, the undulator photon flux increases rapidly. 
Photon polarisation is maximal at zero emission angle; it is decreased and even inverted at larger angles.
Thus, collimating the surplus photon flux at larger emission angles increases the net polarisation. 
Studies indicate that $60\,\%$ positron polarisation at the IP may be possible at \siunit{500}{GeV} centre--of--mass energy with the addition of a photon collimator.


\subsection{Civil engineering and site}

In 2014, the ILC Strategy Council announced the result of its candidate site evaluation for the best possible ILC site in Japan~\cite{ILCSC:2014a}.
The evaluation was conducted by a number of Japanese experts from universities and industry, and reviewed by an international commitee. 
It considered technical as well as socio-environmental aspects, and concluded that the candidate site in the Kitakami region is best suited for the ILC.

\begin{figure}[htbp]
   \includegraphics[width=\hsize]{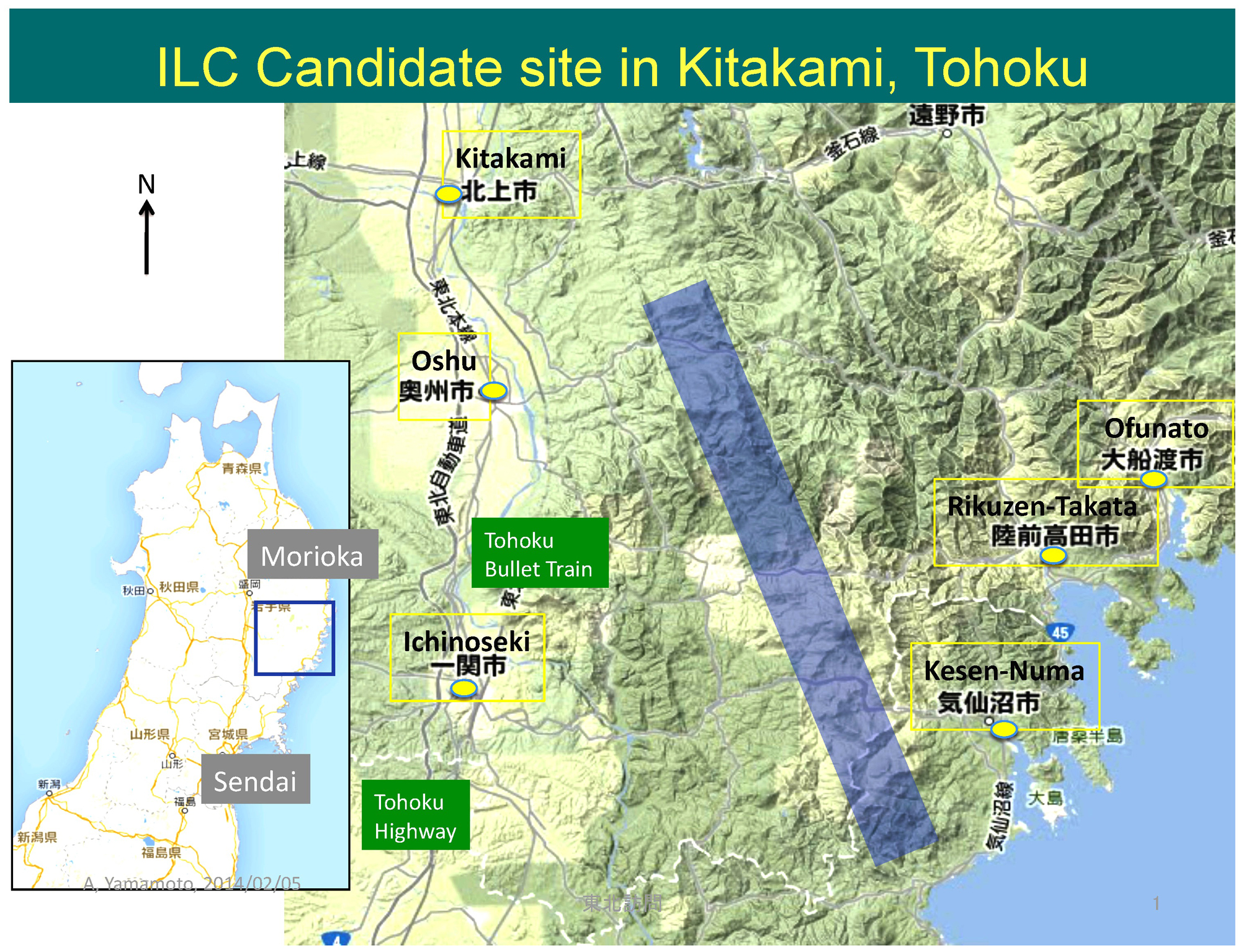}
\caption{The Kitakami candidate site for the ILC~\cite{Warmbein:2014a}.}
\label{fig:kitakami-site}
\end{figure}

The site (Fig.~\ref{fig:kitakami-site}) is located in the Japan's northern Tohoku region, not far from Sendai with its international airport, in the prefectures of Iwate and Miyagi.
The closest cities are Ichinoseki, Oshu, and Kitakami, which all offer Shinkansen (bullet train) access to Sendai and Tokyo.
The closest harbour is in the city of Kesen-Numa.
The coastal region in this area was severely hit by the great Tohoku earthquake in 2011. 
Both prefectures are supportive of the ILC project and view it as an important part of their strategy to recover from the earthquake disaster.

\begin{figure}[htbp]
   \includegraphics[width=\hsize]{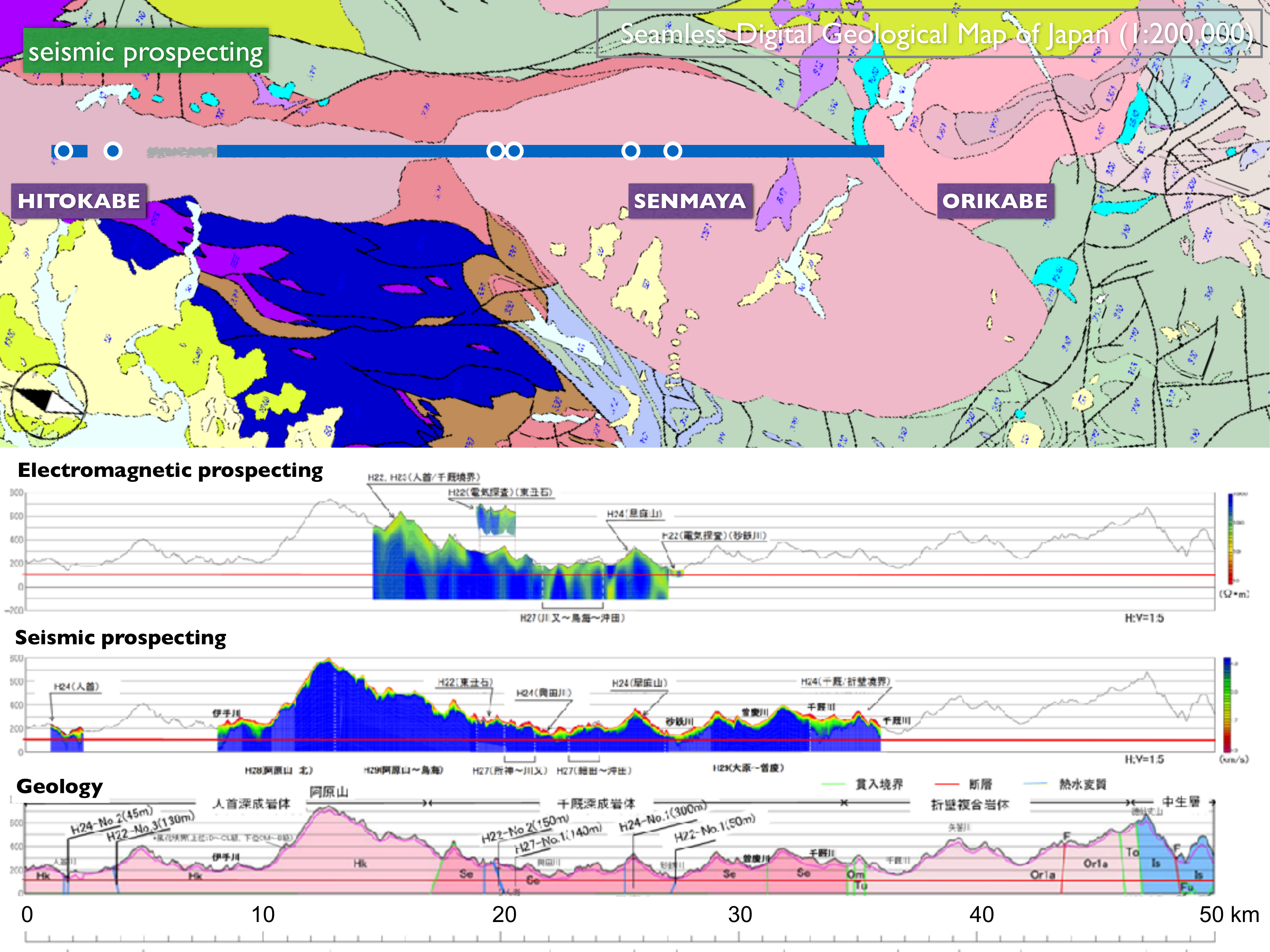}
\caption{Geological situation at the Kitakami site.}
\label{fig:kitakami-geology}
\end{figure}

The Kitakami site was largely selected because of its excellent geological condition. 
The proposed ILC trajectory lies in two large, homogeneous granite formations, the Hitokabe granite in the north and Senmaya granite to the south.
The site provides up to \siunit{50}{km} of space, enough for a possible \siunit{1}{TeV} upgrade or more, depending on the achievable accelerating gradient.  
Extensive geological surveys have been conducted in the area, including boring, seismic measurements, and electrical measurements~\cite{Sanuki:2015a}, as shown in Fig.~\ref{fig:kitakami-geology}.
The surveys show that the rock is of good quality, with no active seismic faults in the area.

Earthquakes are frequent throughout Japan, and the accelerator and detectors need  proper supports that isolate them from vibrations during earthquakes and micro tremors~\cite{Sanuki:2018b}. 
Proven technologies exist to cope with all seismic events, including magnitude 9 earthquakes such as the great Tohoku earthquake. 


Vibration measurements taken during the construction of a road tunnel show that accelerator operation would be possible during the excavation of a tunnel for an energy upgrade~\cite{Sanuki:2018a}.


\begin{figure*}[htbp]
 \begin{center}
 \includegraphics[width=0.36\hsize]{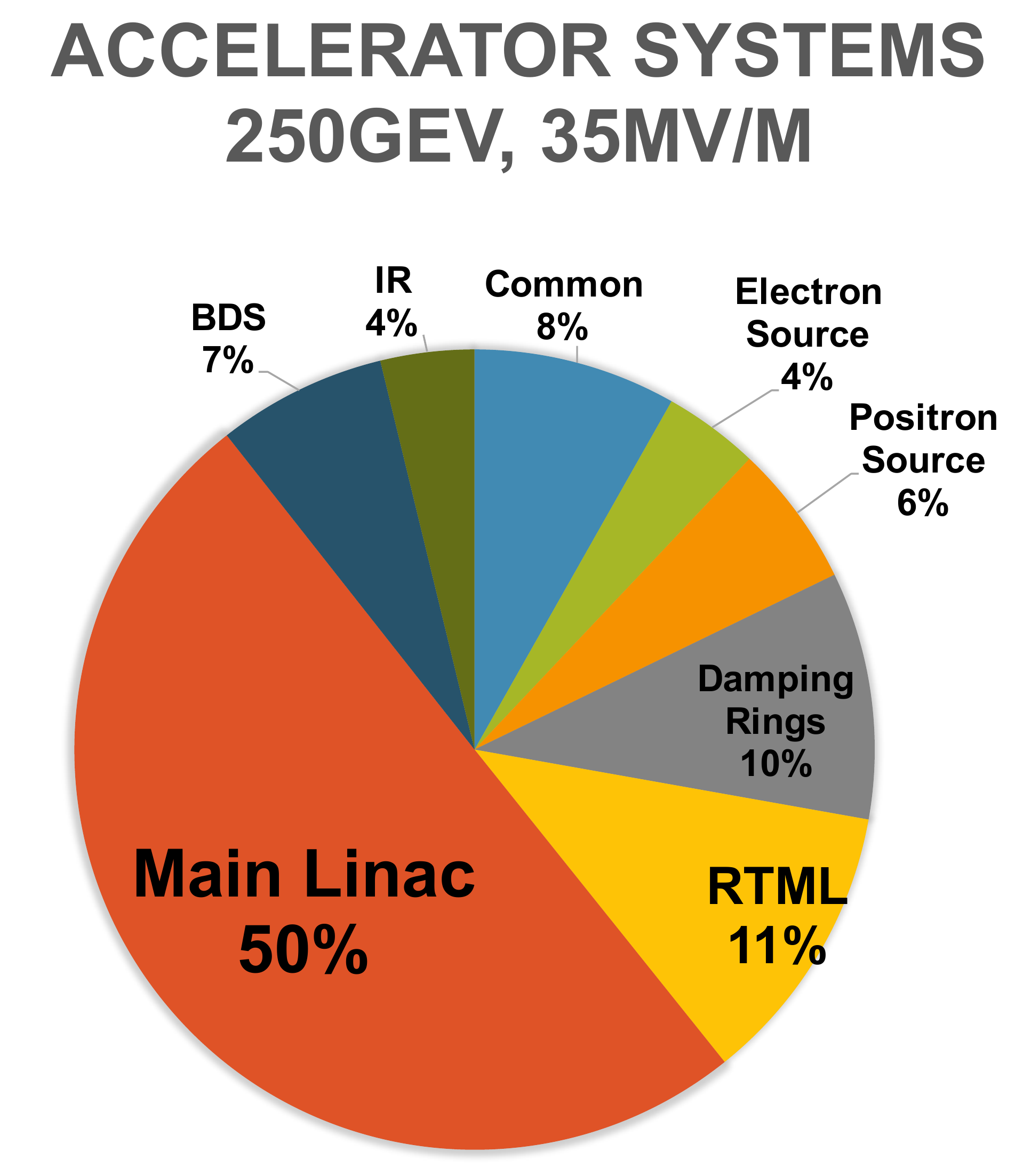}
 \includegraphics[width=0.4\hsize]{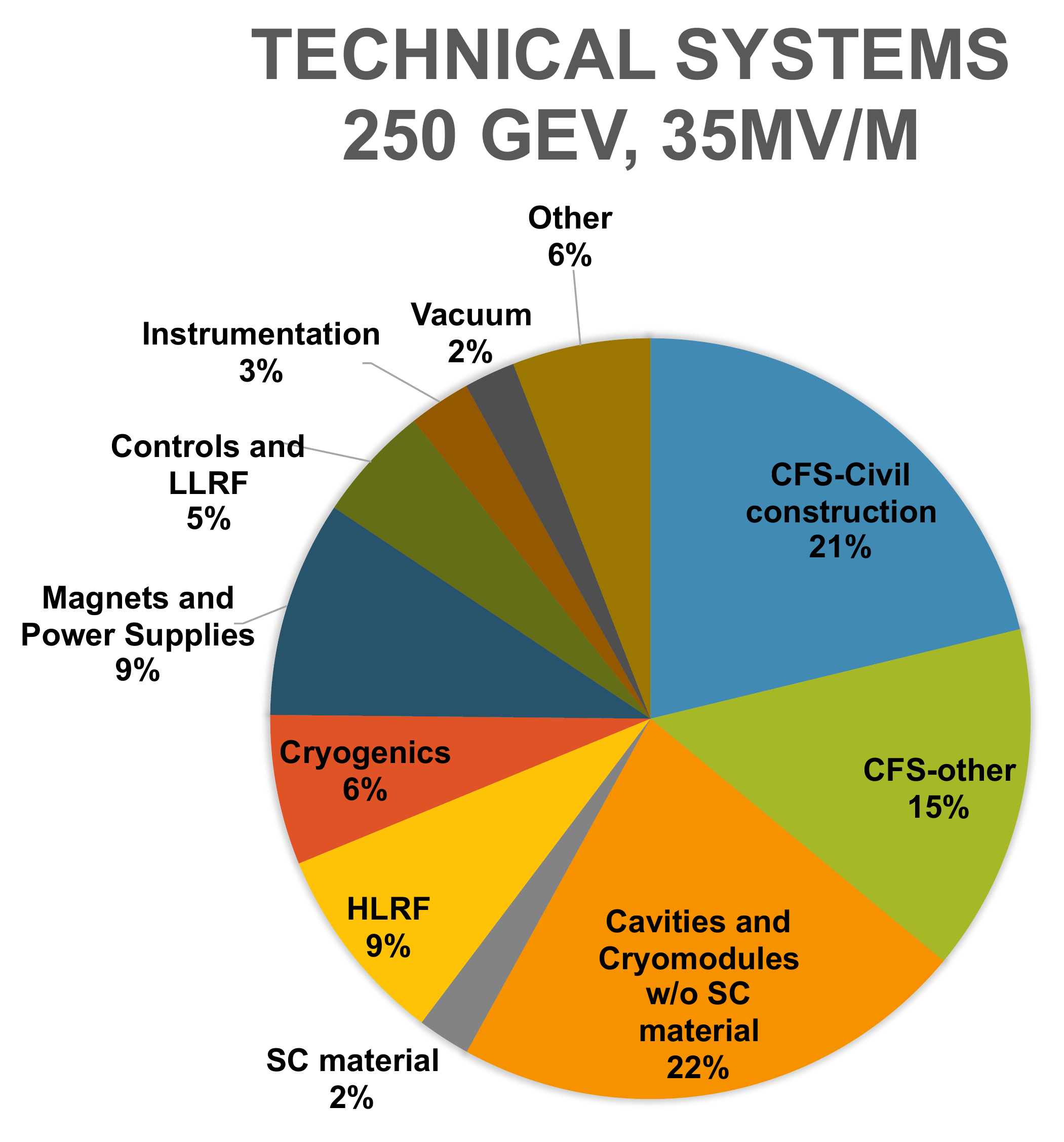}
\caption{Breakdown of Value costs into accelerator systems (left) and technical systems (right) for the \siunit{250}{GeV} ILC accelerator, assuming that cost reduction measures are successful and a gradient of \siunit{35}{MV/m} can be reached.
\label{fig:costs}}
 \end{center}
 \end{figure*}

\subsection{Cost and schedule}

%

For the Technical Design Report, the construction cost of the ILC accelerator was carefully evaluated from a detailed, bottom--up, WBS (Work Breakdown Structure)-based cost estimation~\cite[Sect. 15]{Adolphsen:2013kya}.
The TDR estimate distinguishes two cost categories: Value accounts for materials and supplies procured from industry and is given in ILCU (ILC Currency Unit, where $\siunit{1}{ILCU} = \siunit{1}{US\$}$ in 2012 prices), and Labour accounts for work performed in the participating institutions and is given in person--hours or person--years\footnote{One person--year corresponds to $1700$ working hours.}.

The Value of acquired goods reflects its worth in the local currency of the purchasing institution. 
Therefore, conversion of Value between currencies is performed based on Purchasing Power Parities (PPP), which are regularly evaluated and published by the OECD~\cite{OECD:2018,Eurostat:2012}, rather than currency exchange rates. 
The PPP values reflect local price levels and thus depend on the type of goods and the country, but fluctuate significantly less than currency exchange rates.
Therefore, conversions from ILCU to other currencies cannot be made on the basis of exchange rates to the U.S. dollar, but on PPP values.

The TDR estimate covers the cost of the accelerator construction, assumed to last 9 years plus one year of commissioning. 
It includes the cost for the fabrication, procurement, testing, installation, and commissioning of the whole accelerator, its components, and the tunnels, buildings \etc, and the operation of a central laboratory at the site over the construction period. 
It does not, however, cover costs during the preparation phase preceding the start of construction work (``ground breaking''), such as design work, land acquisition, infrastructure (roads, electricity, water) for the site.

Based on the TDR cost estimate, an updated cost estimate was produced for the \siunit{250}{GeV} accelerator. 
This updated cost estimate includes the cumulative effect of the changes to the design since the TDR (see Sect.~\ref{sec:design_evo}), and evaluates the cost for the reduced machine by applying appropriate scaling factors to the individual cost contributions of the TDR cost estimate.

The resulting Value estimate for the ILC accelerator at \siunit{250}{GeV} is 
\siunit{4,780-5,260}{MILCU}~\cite{Evans:2017rvt} in 2012 prices, where the lower number assumes a cavity gradient of \siunit{35}{MV/m}, while the higher number is based on the TDR number of \siunit{31.5}{MV/m}. 
In addition, \siunit{17,165}{kh} (thousand person-hours) are required of institutional Labour.

In 2018, the ILC Advisory Panel of the Japanese Ministry of Education, Culture, Sports, Science and Technoloy (MEXT) concluded its review of the ILC~\cite{ILCAP:2018}. 
For this review, costs were evaluated in Japanese Yen in 2017 prices, taking into account the local inflation for goods and construction costs.
For the purpose of this estimate, also the Labour costs were converted to Yen to yield \siunit{119.8}{G\yen}, resulting in a total range of the accelerator construction cost of \siunit{635.0 - 702.8}{G\yen}, where the range covers uncertainties in the civil construction costs (\siunit{18}{G\yen}) and of the gradient (\siunit{49.8}{G\yen}).
For the this estimate, conversion rates of $\siunit{1}{US\$} = \siunit{100}{JP\yen}$ and $\siunit{1}{\matheuro} = \siunit{1.15}{US\$}$ were assumed.

Operation costs of the accelerator and the central laboratory are estimated to be \siunit{36.6-39.2}{G\yen} (about \siunit{318-341}{M\matheuro}) per year.




\section{\label{sec:runscenarios}ILC Running Scenarios  }



One of the key advantages of $e^+e^-$ colliders is the ability to collect individual datasets at a series of different center-of-mass energies and beam polarisation settings. While each measurement one might wish to make has its own prefered data-taking mode, the combination with datasets collected at other beam energies and/or beam polarisations provides a unique robustness against systematic uncertainties. 
For example, a recent PhD thesis~\cite{Habermehl:417605} studied Dark Matter searches with consideration of non-neglibigle systematic uncertainties and showed that one obtains better results by sharing a given amount of total integrated luminosity between datasets with different beam polarisations rather than by investing the same total amount of luminosity into the (statistically) most favourable polarisation configuration.

Any physics projection will therefore depend on the exact running scenario, i.e.\ the ensemble of the integrated luminosities collected at the individual center-of-mass energies with the various polarisation settings. The ILC as currently under political consideration in Japan will be limited to a center-of-mass energy of 250\,GeV. Already at this energy, the ILC offers a formidable physics programme, which is described in detail in the following Sec.~\ref{sec:physics}. The intrinsic upgradability to higher energies, however, is a key feature of a linear collider, which clearly sets it apart from any circular $e^+e^-$ collider. In order to illustrate the full potential of the ILC, the upgrade options introduced in Sec.~\ref{subsec:upg-opt} are therefore included in the running scenarios. The timelines presented here are based on technological possibilities and physics requirements only, and do not include funding considerations.

\begin{figure}[tb]
\begin{center}
\includegraphics[width=0.90\hsize]{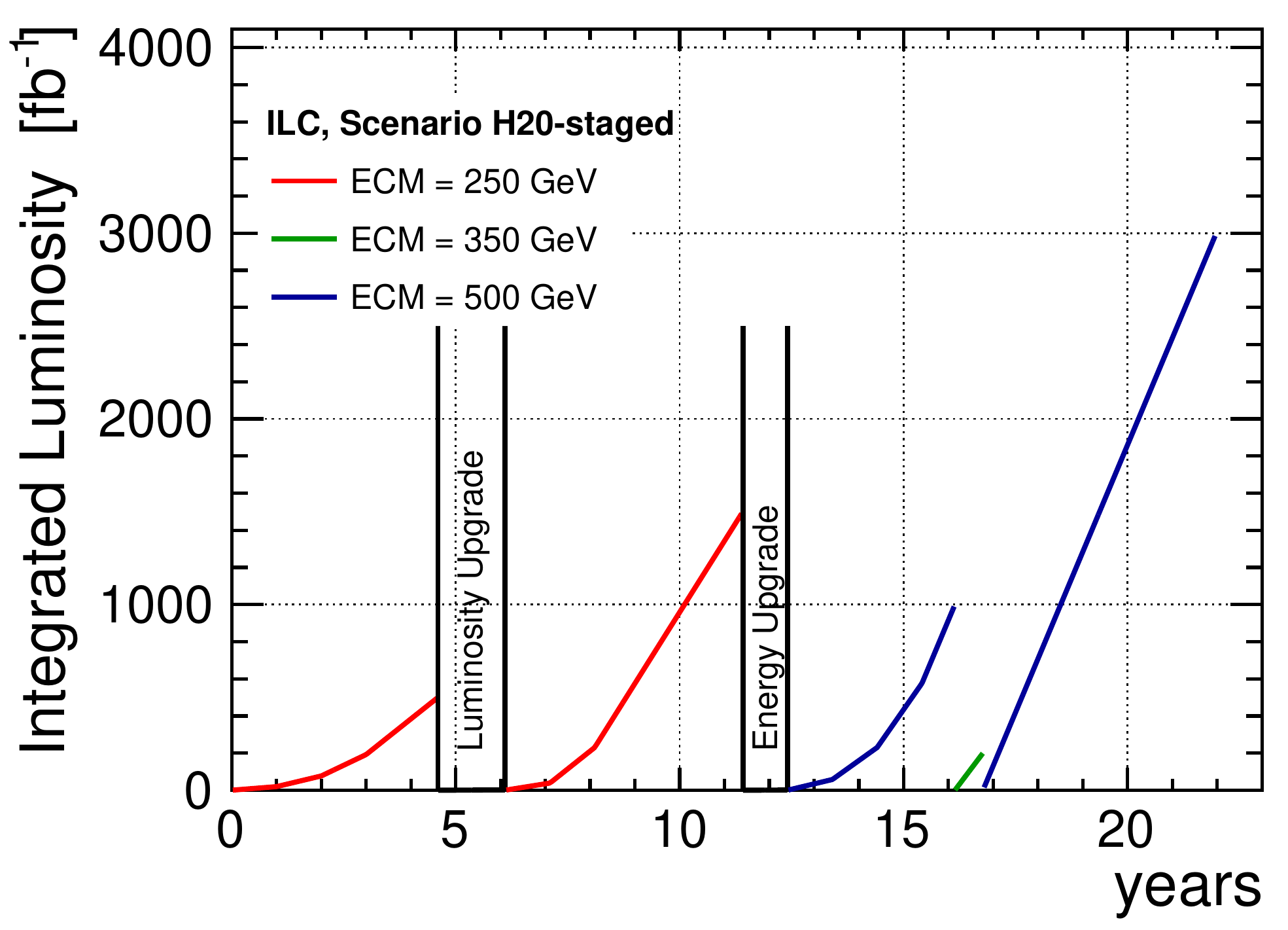}
\end{center}
\caption{The nominal 22-year running program for the staged ILC, starting operation at 250\,\GeV with the current baseline beam parameters for the 250\,GeV runs~\cite{Fujii:2017vwa}. }
\label{fig:H20staged}
\end{figure}

For the physics conclusions given in this paper, we have assumed the energy and luminosity evolution of the ILC shown in Fig.~\ref{fig:H20staged}. At each energy, the time is shared among the various choices for beam polarization in the manner explained in Sec.~\ref{subsec:runscen_pol}.
The full physics program is projected to take 22 years, including a realistic learning 
curve for the establishment of luminosity and scheduled downtimes for luminosity and energy upgrades. In this schedule, the ILC would accumulate 2\,ab$^{-1}$ at 250\,GeV by year 11. It woud then add datasets of 0.2\,ab$^{-1}$ at 350\,GeV and 4\,ab$^{-1}$ at 500\,GeV by year 22. 

The interplay between different datasets has been studied in detail in~\cite{Barklow:2015tja}, with a special focus on the optimisation of the Higgs precision measurements, resulting in a standard running scenario for ILC physics projections. The time evolution of this running scenario has been adapted to the staged construction of the ILC as first presented in~\cite{Fujii:2017vwa}. 

In this section, we will discuss the considerations that have led to the choice of this running scenario, the evolution of this scenario in accord with the design of the 
ILC accelerator, and the flexibility of the plan to respond to changes in machine
specifications or physics discoveries.

\subsection{Center-of-mass energies and integrated luminosities}
The three center-of-mass energies for ILC best motivated by our current knowledge are:
\begin{itemize}
\item $\sqrt{s}=250$\,GeV for collecting data near the threshold of the Higgsstrahlungs process, 
\item $\sqrt{s}=350$\,GeV for scanning the threshold for top quark pair production, and 
\item $\sqrt{s}=500$\,GeV or somewhat above for studying $t\bar{t}$ production in the continuum and enabling $t\bar{t}H$ and $ZHH$ production. 
\end{itemize}
Table~\ref{tab:lumiabstot} gives the total integrated luminosities foreseen at these energies for three alternative running scenarios. These scenarios are described in~\cite{Barklow:2015tja}, which presented a detailed evaluation of these and other possibilities. For comparison, the integrated luminosities assumed in the Snowmass community study~\cite{Asner:2013psa} is given in the last column. Since 2015, the scenario H20 has been the reference scenario for ILC physics projections.

\begin{table}[h]
\centering
  \renewcommand{\arraystretch}{1.10}
\begin{tabularx}{\columnwidth}{*{4}{>{\centering\arraybackslash}X} || *{1}{>{\centering\arraybackslash}X}} 
\hline
            &  \multicolumn{4}{c}{$\int{\mathcal{L} dt}$ [fb$^{-1}$]} \\
\hline
$\sqrt{s}$  & G20      &   H20   &  I20   & Snow   \\
\hline
250\,GeV    &  500      &  2000    &   500   & 1150   \\
350\,GeV    &  200      &   200    &  1700   &  200  \\
500\,GeV    & 5000      &  4000    &  4000   & 1600  \\
\hline
\end{tabularx}
\caption{Proposed total target integrated luminosities for $\sqrt{s}=250$,  $350$, $500$\,GeV based on $20$ ``real-time'' years of ILC operation under scenarios G20, H20 and I20. The total integrated luminosities assumed for Snowmass
are listed for comparison based on 13.7 ``real-time'' years. From~\cite{Barklow:2015tja}.}
\label{tab:lumiabstot} 
\end{table}

It must be stressed, however, that flexibility in the run plan remains one of the key assets of the ILC.  This plan can be adjusted whenever new insights,  discoveries either from the (HL-)LHC or from the ILC itself, require us to do so. In particular, the center-of-mass energy of the ILC can always be lowered from the nominal maximum energy without loss of efficiency, as long as the electron beam energy remains sufficiently high for positron production. In fact, the operation of the SCRF cavities below the maximum gradient
saves significant cryogenic and RF power, which in turn can be invested into higher instantaneous luminosity.

Future $e^+e^-$ colliders could also provide important physics measurements at other center-of-mass energies. Physics goals that motivate other choices are the high-statistics study of $Z$ and $W$, the exploration of the thresholds for any new color-singlet particles that might appear in the ILC energy region, and data-taking at additional center of mass energies to optimize the determination of Effective Field Theory parameters. The lower center-of-mass energies could be realized by doubling the repetition rate of the electron linac to 10\,Hz and adding a by-pass around the positron source for every second bunch train. Today, however, the priority of these issues seems lower than that  for the abovementioned three energies. Therefore they are not explicitly included in the current run plan of the ILC or in the current set of machine parameters.  Over a longer term, we plan to extend the linac to reach 
energies of 1\,TeV or higher.   Table~\ref{tab:lumiabstot1TeV} lists target integrated luminosities approriate to physics studies at these additional energies.

\begin{table}[h]
\centering
  \renewcommand{\arraystretch}{1.10}
\begin{tabularx}{\columnwidth}{l *{4}{>{\centering\arraybackslash}X}} 
\hline
$\sqrt{s}$     &  90\,GeV & 160\,GeV  &    1\,TeV \\
\hline
 $\int{\mathcal{L} dt}$ [fb$^{-1}$]         	   &  100       &  500    & 8000 \\
\hline
\end{tabularx}
\caption{Proposed total target integrated luminosities for other $\sqrt{s}$.  From~\cite{Barklow:2015tja}.}
\label{tab:lumiabstot1TeV} 
\end{table}

\subsection{Beam polarisation}
\label{subsec:runscen_pol}
At center-of-mass energies of up to 500\,GeV, the ILC beams are foreseen to be polarised  with absolute values of at least 80\% for the electrons and at least 30\% for the positrons. At 1\,TeV, the positron polarisation will reach at least 20\%. As an upgrade option, the positron polarisation can be increased to 60\% for center-of-mass energies around 500\,GeV; this is discussed in Sec.~\ref{subsubsec:upg-optP}. The accelerator design comprises sets of spin rotators which in principle allow one to prepare any desired direction of the polarisation vectors at the IP. However in the detailed running scenarios, we consider only longitudinal polarisation. The sign of the beam polarisations can be flipped 
on a train-by-train basis. This allows us to collect datasets with different helicity configurations quasi-concurrently compared to the typical time scales of changes in the accelerator or detector configuration, calibration and alignment. In a joint analysis of these datasets, large parts of the experimental systematic uncertainties cancel.  This is particularly important to minimize the 
systematic errors in the measurement of the left-right polarization asymmetry, a quantity that carries a great deal of information for every process that will be studied at the ILC.  But this idea has many other applications. The joint interpretation of the different datasets allows us to treat many systematic effects as nuisance parameters in global fits, and thereby to measure and subtract these effects~\cite{bib:PhDRobert}. 

The role of positron polarisation specifically at an initial 250-GeV stage of the ILC has been discussed in detail in a recent document~\cite{Fujii:2018mli}. In the case of a global fit to polarised total and differential cross-sections of various electroweak processes, it is shown there  that the uncertainties on some observables increase by up to a factor of 10 in the absence of positron polarisation due to the lack of redundancies required for ultimate control of systematic uncertainties (see Sec.~\ref{subsec:polarisation}). As we will see in Sec.~\ref{subsec:phys_eft}, the left-right asymmetry $A_{LR}(HZ)$ of the Higgsstrahlungs cross section plays an important role in our technique for obtaining a  model-independent fit to Higgs couplings.  Although the measurement of the absolute normalization of the Higgsstrahlungs cross section was not explicitly included in the study summarized in~\cite{Fujii:2018mli}, analogous deteriorations would also be expected for this quantity.

A part of the power of positron polarisation is that it allows one to collect four independent data sets with different mixtures of the physics reactions under study.   Tables~\ref{tab:pollumirel} through~\ref{tab:pollumiabs1TeV} give our
standard assumptions for the sharing of the total integrated luminosity (c.f.\ Tab.~\ref{tab:lumiabstot} and~\ref{tab:lumiabstot1TeV}) between the four possible beam helicity combinations. Due to the importance of  $A_{LR}(HZ)$~\cite{Durieux:2017rsg, Barklow:2017suo} noted above, the sharing for 250\,GeV, which was originally foreseen~\cite{Barklow:2015tja} to emphasize the $\operatorname{sgn}(P(e^-),P(e^+))=(-,+)$ configuration, is now  adjusted to provide equal amounts of luminosity for $(-,+)$ and $(+,-)$~\cite{Barklow:2017suo, Fujii:2017vwa}.

These integrated luminosities and polarisation configurations, especially as specified in Tab.~\ref{tab:pollumiabs} for the H20 running scenario, define the reference scenario for all ILC physics projections.  The order in which the various energies are surveyed will of course depend on the machine evolution and staging plan.

\begin{table}[h]
\centering
  \renewcommand{\arraystretch}{1.10}
\begin{tabularx}{\columnwidth}{l *{4}{>{\centering\arraybackslash}X}} 
\hline
        & \multicolumn{4}{c}{fraction with $\operatorname{sgn}(P(e^-),P(e^+))= $ } \\
           & (-,+) & (+,-) & (-,-) & (+,+) \\
\hline
$\sqrt{s}$ & [\%]  &  [\%] & [\%]  & [\%]  \\ 
\hline
250\,GeV (2015)   & 67.5 &  22.5 &  5    &   5   \\
250\,GeV (update) & \bf{45} &  \bf{45} &  5    &   5   \\
350\,GeV   & 67.5 &  22.5 &  5    &   5   \\
500\,GeV   &  40  &  40   &  10   &  10   \\
\hline
\end{tabularx}
\caption{Relative sharing between beam helicity configurations proposed for the various center-of-mass energies. The update of the luminosity
sharing for 250\,GeV originates from the importance of the left-right asymmetry of the Higgsstrahlung cross section in the EFT-based Higgs coupling fit.}
\label{tab:pollumirel} 
\end{table}

\begin{table}[h]
\centering
  \renewcommand{\arraystretch}{1.10}
\begin{tabularx}{\columnwidth}{l *{4}{>{\centering\arraybackslash}X}}    
\hline
        &  \multicolumn{4}{c}{int. luminosity with $\operatorname{sgn}(P(e^-),P(e^+))= $ } \\
           & (-,+)       & (+,-)       & (-,-)       &  (+,+)     \\
\hline
$\sqrt{s}$ & [fb$^{-1}$] & [fb$^{-1}$] &  [fb$^{-1}$] & [fb$^{-1}$] \\ 
\hline
250\,GeV (2015)   &  1350      &  450        &  100	      &   100  \\
250\,GeV (update) &  \bf{900}  &  \bf{900}   &  100	      &   100  \\
350\,GeV          &   135      &   45	     &   10	      &    10  \\
500\,GeV          &  1600      & 1600        &  400	      &   400  \\
\hline
\end{tabularx}
\caption{Integrated luminosities per beam helicity configuration resulting from the fractions in table~\ref{tab:pollumirel} in scenario H20. The update of the luminosity
sharing for 250\,GeV originates from the importance of the left-right asymmetry of the Higgsstrahlung cross section in the EFT-based Higgs coupling fit. 
}
\label{tab:pollumiabs} 
\end{table}

\begin{table}[h]
\centering
  \renewcommand{\arraystretch}{1.10}
\begin{tabularx}{\columnwidth}{l *{4}{>{\centering\arraybackslash}X}} 
\hline
        & \multicolumn{4}{c}{fraction with $\operatorname{sgn}(P(e^-),P(e^+))= $ } \\
           & (-,+) & (+,-) & (-,-) & (+,+) \\
\hline
$\sqrt{s}$ & [\%]  &  [\%] & [\%]  & [\%]  \\ 
\hline
90\,GeV    &  40  &  40   &  10   &  10   \\
160\,GeV   & 67.5 &  22.5 &  5    &   5   \\
1\,TeV     &  40  &  40   &  10   &  10   \\
\hline
\end{tabularx}
\caption{Relative sharing between beam helicity configurations proposed for low energy and $1$\,TeV running. From~\cite{Barklow:2015tja}.}
\label{tab:pollumirel1TeV} 
\end{table}

\begin{table}[h]
\centering
  \renewcommand{\arraystretch}{1.10}
\begin{tabularx}{\columnwidth}{l *{4}{>{\centering\arraybackslash}X}}    
\hline
        &  \multicolumn{4}{c}{integrated luminosity with $\operatorname{sgn}(P(e^-),P(e^+))= $ } \\
           & (-,+)       & (+,-)       & (-,-)       &  (+,+)     \\
\hline
$\sqrt{s}$ & [fb$^{-1}$] & [fb$^{-1}$] &  [fb$^{-1}$] & [fb$^{-1}$] \\ 
\hline
90\,GeV     &    40   	 &   40        &   10	      &    10  \\
160\,GeV    &   340   	 &  110        &   25	      &    25  \\
1\,TeV      &  3200   	 & 3200        &  800	      &   800  \\
\hline
\end{tabularx}
\caption{Integrated luminosities per beam helicity configuration resulting from the fractions in table~\ref{tab:pollumirel1TeV}. From~\cite{Barklow:2015tja}.}
\label{tab:pollumiabs1TeV} 
\end{table}

\subsection{Time evolution and upgrade options}

The possible real-time evolution of the integrated luminosity was studied in detail in~\cite{Barklow:2015tja}.  It is important to note that the plans given in that study assumed that the full 500\,GeV machine would be available from the beginning. With the introduction of a staged construction plan for the ILC, the time ordering of 
different runs needed to be adjusted. However, the details of trade-offs between scenarios is most fully documented in 
\cite{Barklow:2015tja}, so we will first review that study and the logic of its conclusions.   After this, we will describe our
current plan for the run scenario including the constraints from staging. 

\subsubsection{Running scenarios for the 500-GeV Machine}
\label{subsubsec:runscen_ilc500}

In the study \cite{Barklow:2015tja}, the peak luminosities used for each centre-of-mass energy are based on the numbers published in the ILC TDR~\cite{Adolphsen:2013jya}. But then, the plans took advantage of the reduced linac  electrical power and cryogenic loads when operating the full 500\,GeV machine at lower gradients.  This in particular allows 10-Hz and 7-Hz running, respectively, at the 250\,GeV  and 350\,GeV centre-of-mass energies. In addition, a luminosity upgrade (from 1312 to 2625 bunches per pulse) was been considered; this could require the installation of an additional positron damping ring, as described in  Sec.~\ref{subsubsec:upg-optL}.

More specifically, the study \cite{Barklow:2015tja} made the following assumptions:

\begin{itemize} 
\item A full calendar year is assumed to represent eight months running at an efficiency of 75\% as assumed in the ILC RDR~\cite{Phinney:2007gp}. This corresponds approximately to $Y= 1.6 \times 10^7$ seconds of integrated
running, thus 60\% more than a ``Snowmass year'' of $10^7$ seconds.
\item The start of ``Year 1'' is the start of running for physics. After the end of construction, there is one year foreseen for machine commissioning only, which is not shown on the plots.
\item A ramp-up of luminosity performance, defined as a set of yearly ramp factors $f \le 1$, is in general assumed after: (a) initial construction and `year 0' commissioning; (b) a downtime for a luminosity upgrade; (c) a change in operational mode which may require some learning curve (\eg,  going to 10-Hz collisions).
\item If the peak instantaneous luminosity is $L$, then the nominal integrated luminosity for any given
calendar year is $\int \mathcal{L} dt = f \times L \times Y$, where $f$ is the ramp factor associated with that year.
\item The peak instantaneous luminosities are those corresponding to the TDR beam parameters at 250, 350 and 500\,GeV, as shown in Tab.~\ref{tab:ilc-params}.
\item For the initial physics run after construction and year 0 commissioning, the RDR ramp of 10\%,
30\%, 60\% and 100\% over the first four calendar years is always assumed.
\item  The ramp after the shutdown for installation of the luminosity upgrade is assumed to be slightly shorter (10\%, 50\%, 100\%) with no year 0.
\item Going down in centre of mass energy from $500$\,GeV to $350$\,GeV or $250$\,GeV is assumed to have no ramp
associated with it, since there is no modification (shutdown) of  the machine. 
\item Going to 10-Hz operation at 50\% gradient does assume a ramp however (25\%, 75\%, 100\%),
since 10-Hz affects the entire machine including the damping rings and sources.

\end{itemize}

\begin{figure}
\begin{center}
\includegraphics[width=0.85\hsize]{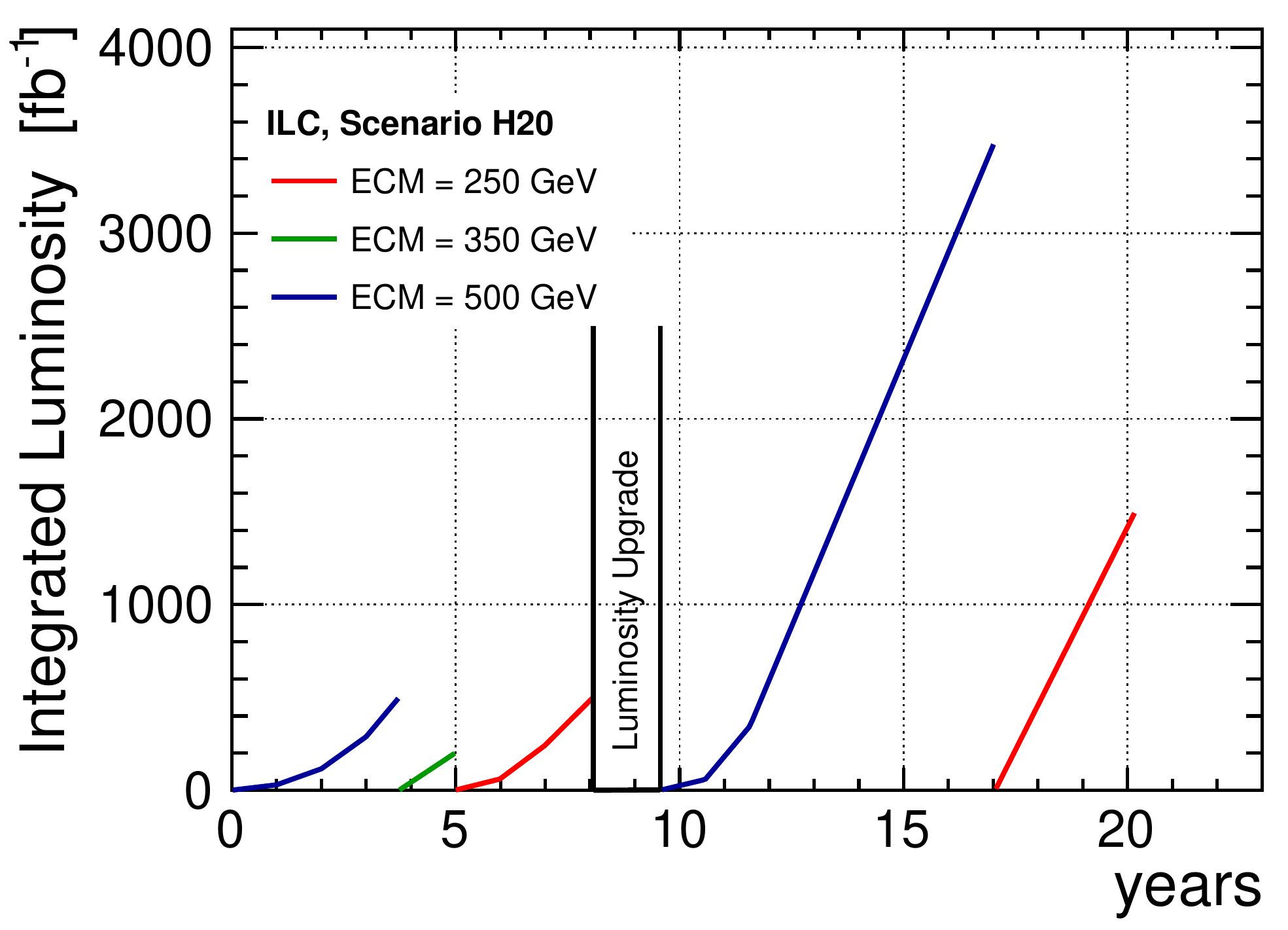}
\end{center}
\caption{The nominal 20-year running program for the 500-GeV ILC~\cite{Barklow:2015tja}.}
\label{fig:H20}
\end{figure}

Under these assumption, a possible real-time scenario for collecting
the integrated luminosities of the H20 scenario (c.f. Tab~\ref{tab:lumiabstot}) is shown in Fig.~\ref{fig:H20}. Since 
it was assumed that the full 500-GeV machine would be available from the start, the first foreseen run was intended to collect a dataset of 500\,fb$^{-1}$ at $\sqrt{s}=$500\,GeV in order to observe for the first time ever $t\bar{t}$ production via the electroweak force,
to survey the full kinematic reach for possible new particles and, last but not least, to collect a comprehensive set of Higgs precision data, with similar contributions from the Higgsstrahlung and $WW$ fusion processes (see \ Fig.~\ref{fig:HiggsProdILC}). 

After this general-purpose survey at the maximum energy, it was planned to collect dedicated datasets at lower energies, at the $t\bar{t}$ production threshold, for a precision determination of a theoretically well-defined top mass, and somewhat above the $ZH$ production threshold, near the maximum of the cross section.  The $ZH$ measurements at 250\,GeV would be a very important component of the program  even under the assumption that energies of  500\,GeV are immediately available.  This is true for two reasons.  First, in Higgsstrahlung production, each Higgs boson  is tagged by the recoil $Z$.  There are many measurements that 
rely on this tag to identify Higgs bosons or to measure absolute rates without the need to make assumptions about the Higgs decay modes.  These include 
the measurement of the normalised total cross section for the Higgsstrahlung process, the measurement  of  absolute branching ratios of the Higgs boson and the search for invisible and exotic decays.  At 500\,GeV, far above the threshold, recoil measurements become less characteristic, due to the more substantial ISR and increased amount of beamstrahlung with respect to \ 250\,GeV, and are subject to additional backgrounds.   Other measurements depend on precise reconstruction of the kinematics of the $\ee\to ZH$ process.
For example, the ultimate precision on 
the Higgs mass will be obtained using the kinematics of $Z$ recoil.  The search for deviations from the SM predictions for 
 Higgs decays requires as input a very precise value of this mass; see Sec.~\ref{sec:higgs:sigmazh}.
Another reaction that depends crucially on precise kinematic measurements is the $CP$ analysis of the $H \to \tau^+ \tau^-$ decay, discussed in Sec.~\ref{subsubsec:higgstautauCP}.

For a 500\,GeV machine running at 250\,GeV, the luminosity can be straightforwardly increased by a factor of 2 from the TDR value 
by the increase of the repetition rate for bunch trains  from 5 to 10\,Hz.  This improvement was incorporated in the plan H20 shown in Fig.~\ref{fig:H20}  even
at the initial stage of 250\,GeV running.

The H20 plan also included provision for an additional luminosity upgrade by doubling the number of bunches in each bunch 
train. This upgrade requires machine improvements as described in Sec.~\ref{subsubsec:upg-optL}, and after these improvements all further 
data would be taken in this mode.
This would give a total  4\,ab$^{-1}$ data sample at 500\,GeV.   A sample of this size is required for meaningful precisions on the top Yukawa coupling and on the Higgs self-coupling. These measurements remain by far statistically limited and  thus would profit from any further increase of the luminosity. In case of the top Yukawa coupling, it was noted that it is absolutely crucial to reach 500\,GeV, since already at 490\,GeV, thus when falling short of the target energy by only 2\%, the precision of the measurement would worsen by nearly a factor of 2. On the other hand, a moderate increase of the center-of-mass energy by 6\% to 530\,GeV would improve the precision on the top-Yukawa coupling by a factor of 2. This should be considered in the planning of the energy upgrade of an inital 250\,GeV machine, see also discussion in Sec.~\ref{subsubsec:upg-optE}.

Finally the H20 scenario planned a run at 250\,GeV, now with 4 times the TDR luminosity, to finish the collection of a 2\,ab$^{-1}$ data set.   This run would provide the ultimate precision on the Higgs boson mass and the total $ZH$ cross section. It should be stressed again that the current focus on three fixed center-of-mass energies does not preclude running at any other desired intermediate energy, \eg\ for scanning the production threshold of newly discovered particles.

At the end of this 20 year program, we envision  a further doubling of the energy to 1\,TeV.   This upgrade was presented already 
in the ILC TDR and is reviewed in Sec.~\ref{subsubsec:upg-optE}.  This energy upgrade could  possibly be preceeded by a run at the $Z$ pole if it is  required by the  physics.

\subsubsection{Running scenarios for the staged machine}

\begin{figure}
\begin{center}
\includegraphics[width=0.85\hsize]{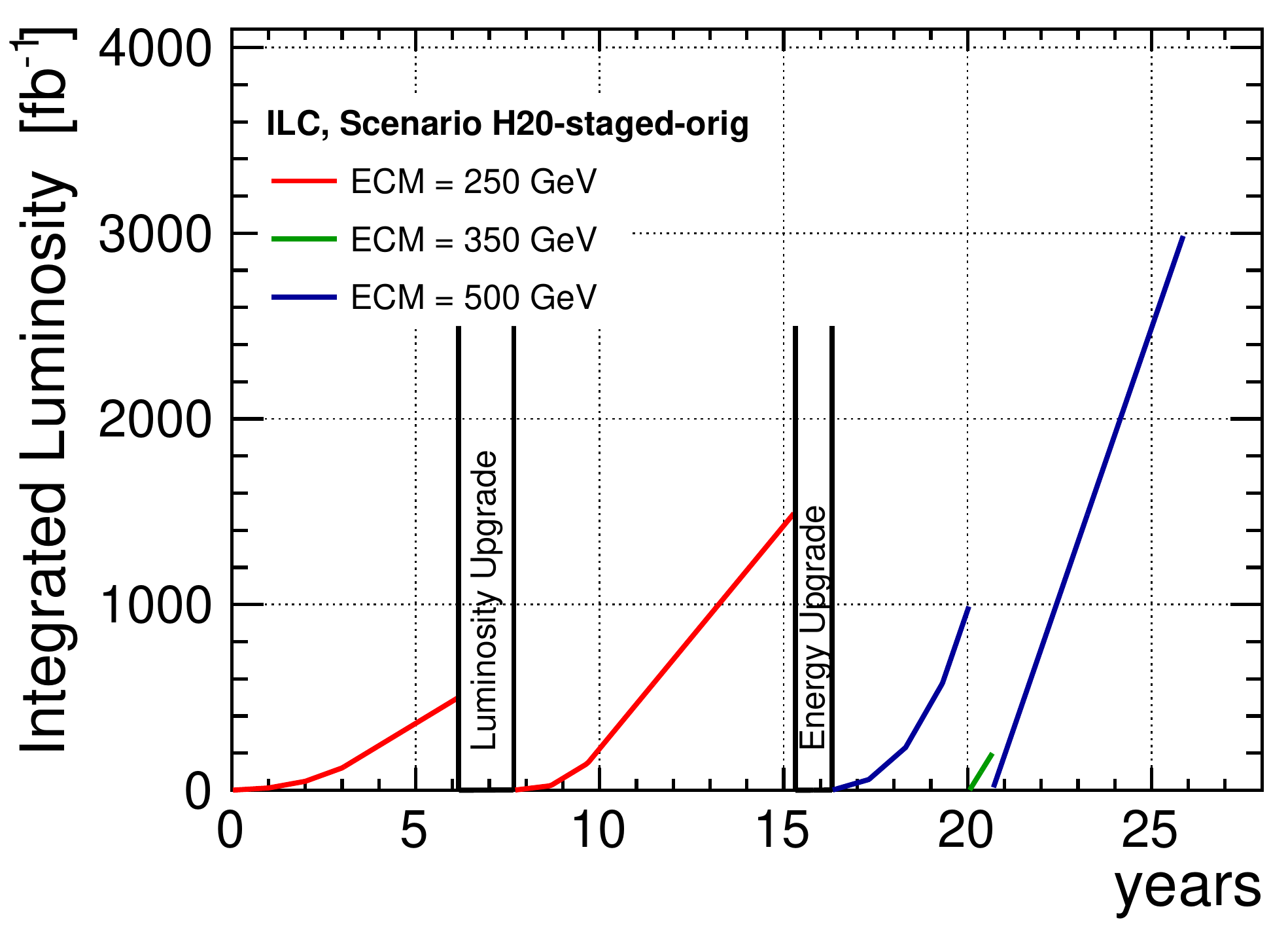}
\end{center}
\caption{The nominal 26-year running program for the staged ILC, starting operation at 250\,\GeV without the possibility to operate at 10\,Hz~\cite{Fujii:2017vwa}. The integrated luminosities are the same of for the original H20 scenario.}
\label{fig:H20staged-orig}
\end{figure}

With the introduction of the staging plan for the ILC machine construction, it was necessary to change the time ordering of the 
various energy steps in the program described in the previous subsection. However, the total integrated luminosities to be collected at each center-of-mass energy, which were already optimized for the physics goals,  were left untouched. Thus,  all physics projections based on the H20 scenario remained valid - albeit the results will arrive in a different time order. Figure~\ref{fig:H20staged-orig} shows the original plan for the time evolution of the staged H20 scenario. The assumptions differ from those listed in the previous subsection in the following points:
\begin{itemize} 
\item No 10\,Hz operation is assumed since in the 250\,GeV machine the cryomodules will be operated at full gradient and thus no spare cryo- and RF-power is available. Technically, it would be possible to increase the repetition rate (and thus the luminosity) at any time provided that resources for installing the additional cryo- and RF-power and for covering the higher operation costs could be found. This option is {\em not} included in the staging scenario.
\item The luminosity upgrade by doubling the number of bunches per train (c.f.\ Sec.~\ref{subsubsec:upg-optL}) is a smaller investment than the energy upgrade and will therefore happen first. In this plan, the second positron damping ring and the additional cryo- and RF-power needed for the luminosity doubling would  already be installed at the start of 500\,GeV operation.  Then the entire 500\,GeV run would be done at 2 times the TDR luminosity. 
\item The energy upgrade (described in\ Sec.~\ref{subsubsec:upg-optE}) requires only a relatively short machine shutdown of about one year, since major parts of the new tunnel can be constructed and the new parts of the machine can be installed without disturbing the operation of the 250-GeV machine. A shutdown is necessary only during the construction of the connections of the new parts of the  machine to the older ones.
\item After the energy upgrade the same ramp fractions as for a completely new machine are assumed, thus 10\%, 30\%, 60\% and 100\% over the first four calendar years.
\end{itemize}
With these assumptions, the real-time for realization of  the full H20 program increases from 20 to 26 years, 
mostly due to the much longer time to collect the 2\,ab$^{-1}$ at 250\,GeV without 10\,Hz operation.

In order to mitigate the absence of the 10\,Hz operation, which would require additional investments beyond the minimal 250-GeV machine, cost neutral ways to increase the luminosity at 250\,GeV have been studied, as discussed in Sec.~\ref{sec:ilc}. In 2017, a new set of beam parameters for the 250-GeV ILC was officially adopted~\cite{bib:cr-0016}.  It is this parameter set that is shown in the column ``initial'' of Tab.~\ref{tab:ilc-params}. The 65\% increase in instantaneous luminosity w.r.t.\ the TDR parameters is achieved by reducing the horizontal emittance by a factor of 2. This leads to a larger luminosity in each bunch crossing and thus to an increase of beamstrahlung, background from $e^+e^-$ pairs and pile-up from low-$p_t$ $\gamma\gamma \to$\,hadrons events. Neither these effects, nor the slightly wider luminosity spectrum which results from the increased beamstrahlung are included in the physics case studies presented in the following sections, since no new Monte-Carlo samples could be produced (and analysed) since the new beam parameters became available. However, even with the new beam parameters the background conditions at 250\,GeV do not become worse than what is expected at 500\,GeV, a case already studied in detail.  The ILC detectors have actually been designed for high performance in the more difficult  beam conditions at 1\,TeV. Therefore, the impact of the new beam parameters on the majority of the physics analyses is expected to be minor. The analysis most strongly affected is the mass measurement of the Higgs boson via the leptonic recoil method, described in Sec.~\ref{sec:higgs:sigmazh}. For this analysis, the new beam parameters have been estimated to result in a relative degradation of the ultimate precision on the Higgs mass by about 25\%~\cite{bib:jeans_awlc17} compared to the same amount of total luminosity collected with the TDR beam parameters.  This still corresponds to an impressive 
Higgs mass measurement to better than 20\,MeV. 

We have already shown in Fig.~\ref{fig:H20staged} the default running scenario for the staged ILC based on the new beam parameters for 250\,GeV~\cite{bib:cr-0016}. Compared to Fig.~\ref{fig:H20staged-orig}, the total run time shortens from 26 years to 22 years, thus recovering about 2/3 of the original increase in running time. A full-scale Monte-Carlo production with the new beam parameters and based on the ILD detector concept is planned for 2019.

None of the running scenarios explicitly includes the option to increase the positron polarisation to 60\% when operating at a center-of-mass energy of 500\,GeV. Numerous studies~\cite{Fujii:2018mli, Habermehl:417605, Moortgat-Picka:2015yla, Aurand:2009kp, MoortgatPick:2005cw} have shown that all physics measurements will profit from the corresponding increases in effective luminosity and effective polarisation. In this respect, all physics projections for 500\,GeV are still quite conservative.

\section{\label{sec:physics}Physics Case -- 250 GeV }


The core of the physics case for the ILC is to make high-precision
 measurements of the properties of the Higgs boson.    The Higgs
 field has a central role in 
 the SM.  It is responsible for the masses of all known
 elementary particles.
  It is also responsible for those aspects of the SM that are
 hardest to  understand----the
presence of spontaneous gauge symmetry breaking, and the  hierarchy of quark and lepton masses.  Also, within the SM, the  the  flavor mixing and $CP$ violation in weak 
interactions arise from the quark-Higgs Yukawa couplings, and neutrino masses, whatever their origin, require a coupling 
of neutrinos to the Higgs field.
If we wish to learn more about these features of the fundamental laws of nature, an obvious course is to measure the Higgs boson as well as we are able.  We will argue in this section and the succeeding ones that ILC will be able to determine the mass of the Higgs boson to a part in $10^4$ and the major couplings of the Higgs boson to better than 1\% accuracy.   This will qualitatively sharpen the picture of the Higgs boson that we will obtain even from the high-luminosity stage of the LHC. 

This set of measurements, and other measurements available for the first time at the ILC, will open new paths in the search for new fundamental interactions beyond the SM. 
Though the SM seems to account for all elementary particle phenomena observed up to now, it is manifestly incomplete.   It not only does not answer but actually is incapable of answering the questions posed in the previous paragraph.  It also cannot address basic facts about the universe in the large, in particular, the excess of matter over antimatter and the origin of the cosmic dark matter.  To make progress, we need observational evidence from particle physics of violations of the SM.  These will provide clues that can show the way forward.

Up to now, we have sought evidence for new interactions from direct searches for new particles at LEP, the Tevatron, and the LHC, from measurements of the $W$ and $Z$ bosons, and from searches for anomalies in flavor physics.  We are now approaching the limits of these techniques with current particle physics facilities.  The ILC will 
extend our search capabilities in precision measurements of $W$ boson couplings and fermion pair production, and will provide new opportunites for the direct discovery of new particles.  But, most of all, it will open a completely new road through the  high-precision study of the Higgs boson.

\subsection{Mysteries of the Higgs boson}

It is often said that the Higgs boson, as observed at the LHC, is an uninteresting particle, since it conforms so well to the expectations from the SM.   In fact, aside from our knowledge of the Higgs boson mass, the measurements make so far at the LHC tell us almost nothing about the true nature of this particle.   We now 
explain this statement, and, in the process, clarify the requirements for measurements of the Higgs boson couplings that can give insight into physics beyond the SM. 

New physics can correct the Higgs boson couplings in many ways.  However, in all cases, the size of the corrections is limited by the Decoupling Theorem, enunciated by Haber in \cite{Haber:1994mt}:    If the new particles that modify the SM have minimum mass $M$, then  the corrections to the SM predictions for the Higgs boson couplings are of size 
\beq 
              a \,     m_H^2 / M^2   \  .
\eeqn
where the coefficient $a$ is of order 1.   The exclusions of new particles through searches at the LHC suggest that $M$ is at least close to 1~TeV.   Then the effects of  new physics are limited to levels below 10\%.    We will illustrate this result with explicit models in the next subsection.

The proof of the theorem is simple and illustrative.   It can be shown that the SM is actually the most general renomalizable quantum field theory with $SU(3)\times SU(2)\times U(1)$ gauge symmetry and the known particle content.   If we add new particles with masses of $M$ and above, we can assess their influence on the Higgs boson by integrating them out of the theory.  This adds to the Lagrangian a set of new terms with the SM symmetries.   The terms in the new Lagrangian  can be organized by their operator dimension 
as 
\beq 
   {\L}  =  {\L}_{SM} + {1\over M^2} \sum_i\,  c_i \O_{6i} + {1\over M^4 } \sum_j\,  d_j\O_{8j} + \cdots
\eeq{geneffL}
where $\L_{SM}$ is the SM Lagrangian, $\O_{6i}$  are operators of dimension 6, $\O_{8j}$ are operators of dimension 8, \etc\ \   Shifts in the SM parameters due to new physics are not observable, since these parameters are in any case fit to experiment.  Then the leading observable corrections are of order $M^{-2}$. 

This theorem has a striking consequence.  Instead of a model with a single Higgs doublet, as we have in the SM, nature could be providing  a model with two or more Higgs fields, composite Higgs fields, even a whole Higgs sector.  All of this possible complexity is hidden from us by the Decoupling Theorem. 

The theorem has an appealing corollary, though.   Since the SM is the most general renormalisable model, once its parameters are known, its predictions for the Higgs couplings are determined precisely.   These predictions do depend on measured SM parameters such as $m_b$, $m_c$, and $\alpha_s$, but  it is argued in Ref.~\cite{Lepage:2014fla} that  
lattice QCD will determine these well enough to fix the SM predictions for Higgs to part-per-mil accuracy.   Then, if we can observe corrections to the SM predictions at the 1\% level, these corrections and the evidence that they give for new physics cannot hide.

\subsection{Examples of  new physics influence on the Higgs boson}
\label{subsec:newphysicofH}

Many models of physics beyond the SM illustrate the points made in the previous section. 
These examples point to  a goal of 1\% accuracy
for the measurement of Higgs boson couplings in the major decay modes.

Models with two Higgs doublets contain 5 physical Higgs particles: two  neutral $CP$-even states $h$, $H$,  a neutral $CP$-odd state $A$, and a pair of charged scalars $H^\pm$.  These states are mixed by two angles $\alpha, \beta$. 
The lighter
$CP$-even state $h$ is identified with the observed Higgs boson.  Its couplings to 
fermions depend on the mixing angles.   For example, in the ``Type II'' case, 
\beq
   g(Hb\bar b) = - {\sin\alpha\over \cos\beta}{m_b\over v}  \quad   g(Hc\bar c) =  {\cos\alpha\over \sin\beta}{m_c\over v}   \ .
\eeq{typeIIbcshift}
However, the mixing angles are connected to the masses in such a way that when the additional bosons become heavy, their effects in \leqn{typeIIbcshift} also become small,
\beq
     - {\sin\alpha\over \cos\beta} =  1 + \O({m_Z^2\over m_A^2}) \ ,
\eeqn
conforming to the Decoupling Theorem.   In Type II models, the $b$ and $\tau$ Yukawa couplngs are shifted together by about 5\% for $m_A = 500$~GeV, and by decreasing amounts as all of the additional bosons become heavier.

\begin{figure}
\begin{center}
\includegraphics[width=0.90\hsize]{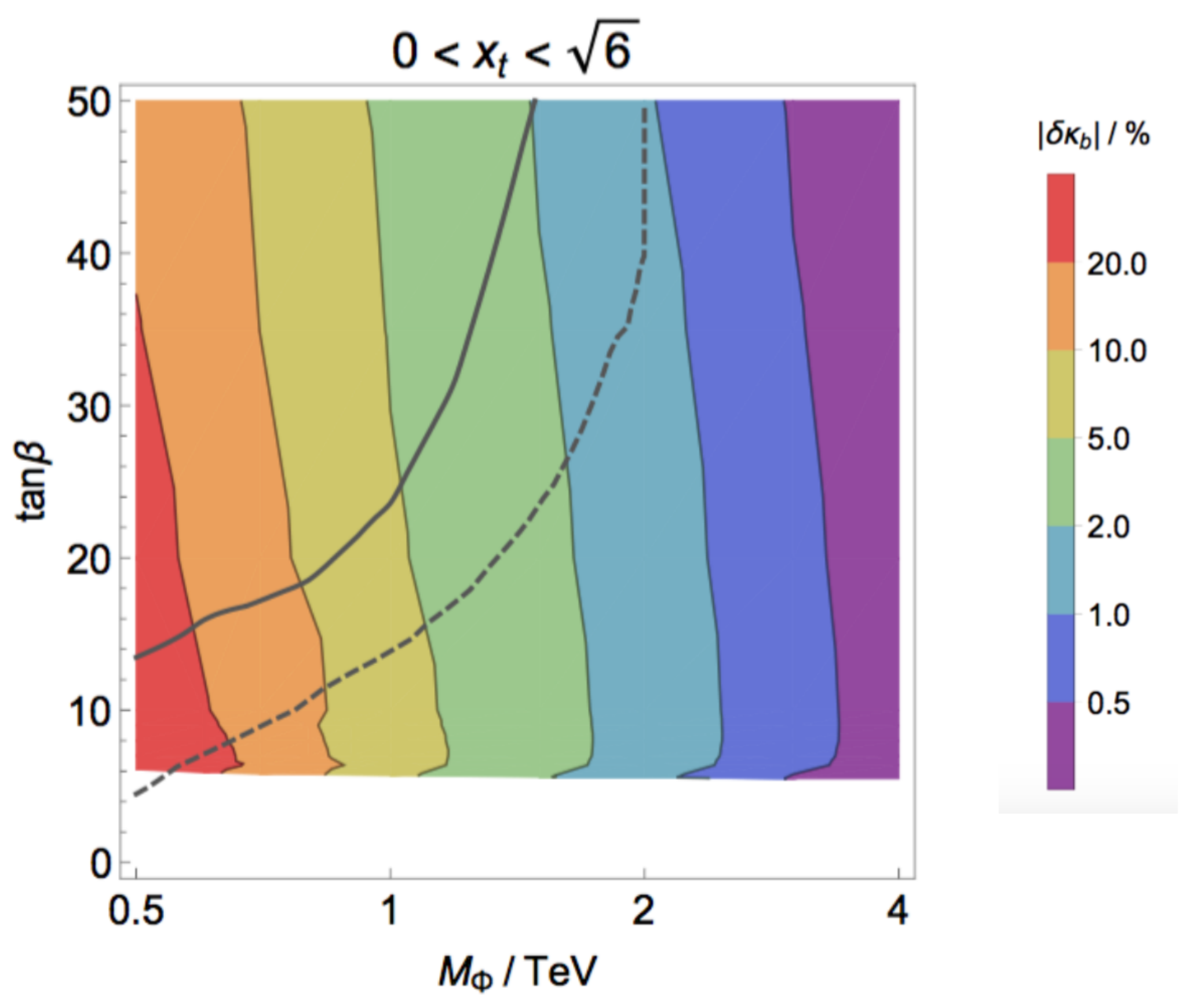}
\end{center}
\caption{Deviation from the SM prediction for the $Hb\bar b$ 
coupling over a parameter space of grand-unified SUSY models, from \cite{Wells:2017vla}.   
Models in the upper left-hand corner are excluded by current LHC searches.  
Models above the dashed line are expected to be excluded at the  HL-LHC. 
The color-code indicates the magnitude of the coupling deviation, in \%. }
\label{fig:WellsZhang}
\end{figure}

Supersymmetry (SUSY) models contain Type II  two-Higgs-double sectors, but they also contain other effects that modify the Higgs boson couplings.   Mixing between the scalar partners of $b_L$ and $b_R$ can generate a shift of the $Hb\bar b$ coupling through loop diagrams.  The magnitude of this effect in 
grand-unified SUSY models is shown in 
Fig.~\ref{fig:WellsZhang}~\cite{Wells:2017vla}.   
Note that it is possible to have a large shift of the Higgs boson coupling for 
parameter values at which the SUSY particles are too
heavy to be discovered at the LHC.   Thus, the search for shifts in the Higgs couplings away from the SM predictions provides a method of searching for this new physics that is independent of, and largely orthogonal to, the direct search for SUSY particles.   Other surveys of this effect in \cite{Cahill-Rowley:2014wba,Kanemura:2015mxa}  confirm this idea.

SUSY models typically predict very small shifts of the $HWW$ and $HZZ$ couplings, but other types of models can affect these couplings directly.   Models in which the electroweak phase transition becomes first-order and allows baryogenesis at the weak scale often involve mixing of the Higgs field with a heavy singlet field.  This gives
\beq
           g(HWW) = {2m_W^2\over v} \cos^2\phi\approx  {2m_W^2\over v} (1 -\half \phi^2) \ ,
\eeq{scalarmixing}
where $\phi \sim  m_H/m_S$, and similarly for the $HZZ$ coupling~\cite{DiVita:2017eyz}.   In composite models of the Higgs
 field, the Higgs boson often appears as a Goldstone boson of a new strong interaction theory, giving a coupling modification by $(1- v^2/f^2)^{1/2}$, where $f$ is the Goldstone boson decay constant~\cite{Contino:2013kra}.  This effect is similar to that in \leqn{scalarmixing}.

Models of Higgs compositeness, ``Little Higgs'' models, and models with extra space dimensions all contain new heavy vectorlike fermions $T$.   Typically, these fermions obtain a fraction of their mass from the Higgs mechanism (perhaps by mixing with the top quark) that is of order $m_t^2/m_T^2$.   Then they induce corrections of this order to the loop-generated Higgs couplings $g(Hgg)$ and $g(H\gamma\gamma)$.  Corrections as large as 10\% can be generated in specific models~\cite{Han:2003gf}.   The same mixing and compositeness effects modify the $Ht\bar t$ coupling~\cite{Agashe:2004rs}.

An interesting picture emerges.   Almost all models of new physics generate corrections to the Higgs boson couplings.   Almost always, these corrections are small, below the 10\% level, in accord with the Decoupling Theorem.  However, in precision experiments that make these coupling deviations  visible, each type of new physics affects the Higgs couplings in different ways.  In general,
\begin{itemize}
\item The $Hb\bar b$ and $H\tau\tau$ couplings are sensitive to models with additional Higgs doublets.
\item The $Hb\bar b$ coupling is sensitive to heavy SUSY particles with left-right mixing.
\item The $HWW$ and $HZZ$ couplings are sensitive to mixing of the Higgs field with singlet fields, and to composite structure of the Higgs boson.
\item The $Hgg$ and $H\gamma\gamma$ are sensitive to models with new vectorlike fermions.
\item The $Ht\bar t$ coupling is sensitive to models with composite Higgs bosons and top quarks.
\end{itemize}

In each new physics model, the  deviations from the SM predictions for the Higgs couplings form a pattern.  With precision experiments, it is possible not only to discover the existence of new physics but also to read the pattern and gain clues as to the way forward.   A worked example of such model discrimination at the level of precision expected at the ILC  is presented in
Section~7 of  Ref.~\cite{Barklow:2017suo}.

\subsection{Limitations of the LHC measurements on the Higgs boson}

Today, the LHC experiments are achieving 20\% uncertainties in their measurements of Higgs boson couplings.   Over the lifetime of the LHC, including its high-luminosity stage, these experiments will acquire a factor of 30 more data.   Shouldn't this lead to Higgs coupling measurements of the required high precision?  We believe that the answer to this question is no.   We give a high-level argument here. A detailed comparison of the expected ILC capabilities with those of the high-luminosity LHC will be presented in Sec.~\ref{subsec:higgs:ilclhc}.

We find three points relevant to this comparison.  First, the measurement of Higgs boson 
decays at the LHC is extremely challenging because of the difficulty of distinguishing 
signal from background.
In the two decay modes in which the Higgs boson was discovered, $H\to \gamma\gamma$ and $H\to 4\ell$, Higgs events are apparent, because all products of the Higgs are observed and the Higgs mass can be reconstructed with high accuracy.  Unfortunately, these modes correspond to tiny branching ratios,  0.2\% and 0.02\% of Higgs decays, respectively.  For more typical decay  modes, Higgs boson decay events have no obvious differences in appearance from SM background reactions with larger rates.   For example, $H\to WW\to e\nu \mu \nu $ events differ from $q\bar q\to WW \to  e\nu \mu \nu $ events only in subtle features of the final state.
To discover the Higgs boson in one of the major channels, the LHC experiments start from samples that are 10:1 background to signal in the highest significance regions.  (For $H\to b\bar b$, the ratio is 20:1.) They then  extract the signal by multivariate analysis and the use of machine-learning classifiers.  It is already a
 triumph that ATLAS and CMS have been able to obtain significant observations. 

Measuring  the Higgs couplings with high precision is even more of a challenge.   It is currently beyond the state of the art to determine the efficiency  for the rejection of SM background events from these signal regions to 1\% accuracy.   The residual background events must be subtracted from the Higgs signal, and so this 1\% would translate to a 10\% accuracy on the Higgs $\sigma\times BR$ or a 5\% error on the coupling. To go beyond this level is truly daunting.  Nevertheless, the studies reported in the HL-LHC Yellow Book~\cite{Cepeda:2019klc}  demonstrate that the HL-LHC can be expected to push beyond this level and reach accuracies on Higgs boson couplings of 2--4\%.

This brings us to the second point.  As we have emphasized already, the modifications of the Higgs boson couplings from new  physics are expected to be small.  In the previous section, we have argued that new physics interactions typically  affect specific
Higgs boson couplings at a level of 5\% or smaller.  A 2\% measurement of such a coupling  
would not meet even the $3\sigma$ criterion for positive evidence of new physics.

Finally, one must take into account that the HL-LHC measurements will ultimately be limited by the systematic understanding of backgrounds.  Any deviation in Higgs couplings observed at the LHC is  likely to be questioned (as, for example, the $t\bar t$ forward-backward asymmetry from the Tevatron was) without a clear means of confirming the result. 
One sometimes hears that the LHC can measure ratios of branching ratios with improved accuracy, but this statement is not borne out by results presented in Ref.~\cite{Cepeda:2019klc}, since each mode has different backgrounds and requires its own dedicated analysis. 

 In contrast, as we will argue below, the observation of Higgs coupling deviations at the ILC at  250~GeV will be very robust.  It will be be statistics-limited,  and it can be confirmed by experiments at 500~GeV that bring in a new production reaction with an independent  data set.

\subsection{$\ee\to ZH$}

The arguments just presented call out for a different way to measure Higgs boson couplings.   In this method, Higgs boson events should be apparent with a simple discriminator that can then be refined for high-accuracy $\sigma\times BR$ measurements.  Ideally, this method should identify Higgs boson events independently of the decay mode, allowing the measurement of the total cross section for Higgs production and the discovery of exotic and unanticipated Higgs decays.

This new method will be provided by the ILC.  It is the measurement of the reaction $\ee\to ZH$ at 250~GeV.   At an $\ee$ collider at this energy, it is true to a first approximation that any $Z$ boson observed with a lab energy of  110~GeV is recoiling against a Higgs boson.   The backgrounds to this signature (present at about 30\% of the signal level before cuts)  come from radiative $\ee\to Z\gamma$ and $\ee\to ZZ$, reactions that are well-understood and computed from electroweak theory at the 0.1\% level. 

\begin{figure}
\begin{center}
\includegraphics[width=0.5\hsize]{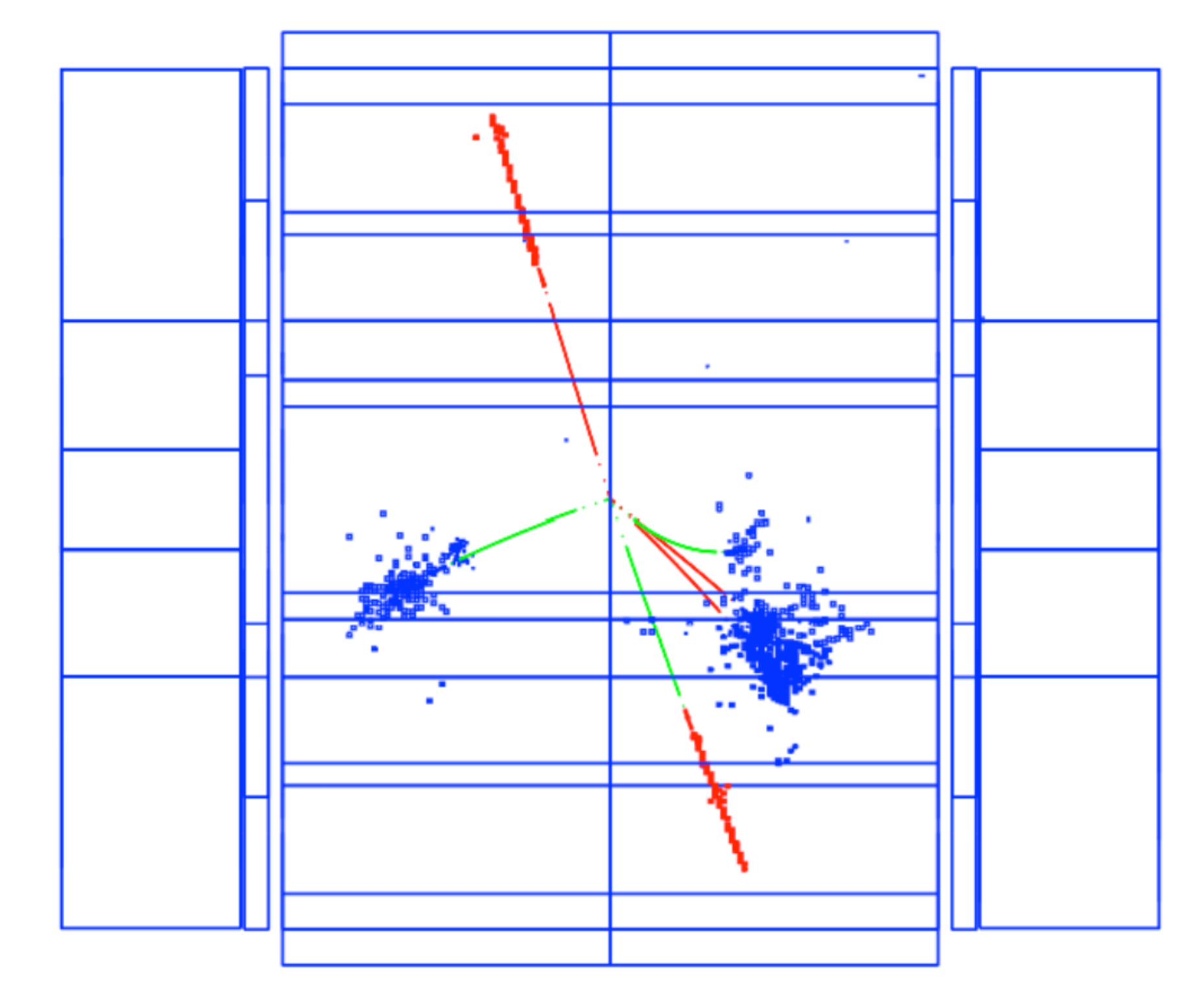}\ \ 
\includegraphics[width=0.40\hsize]{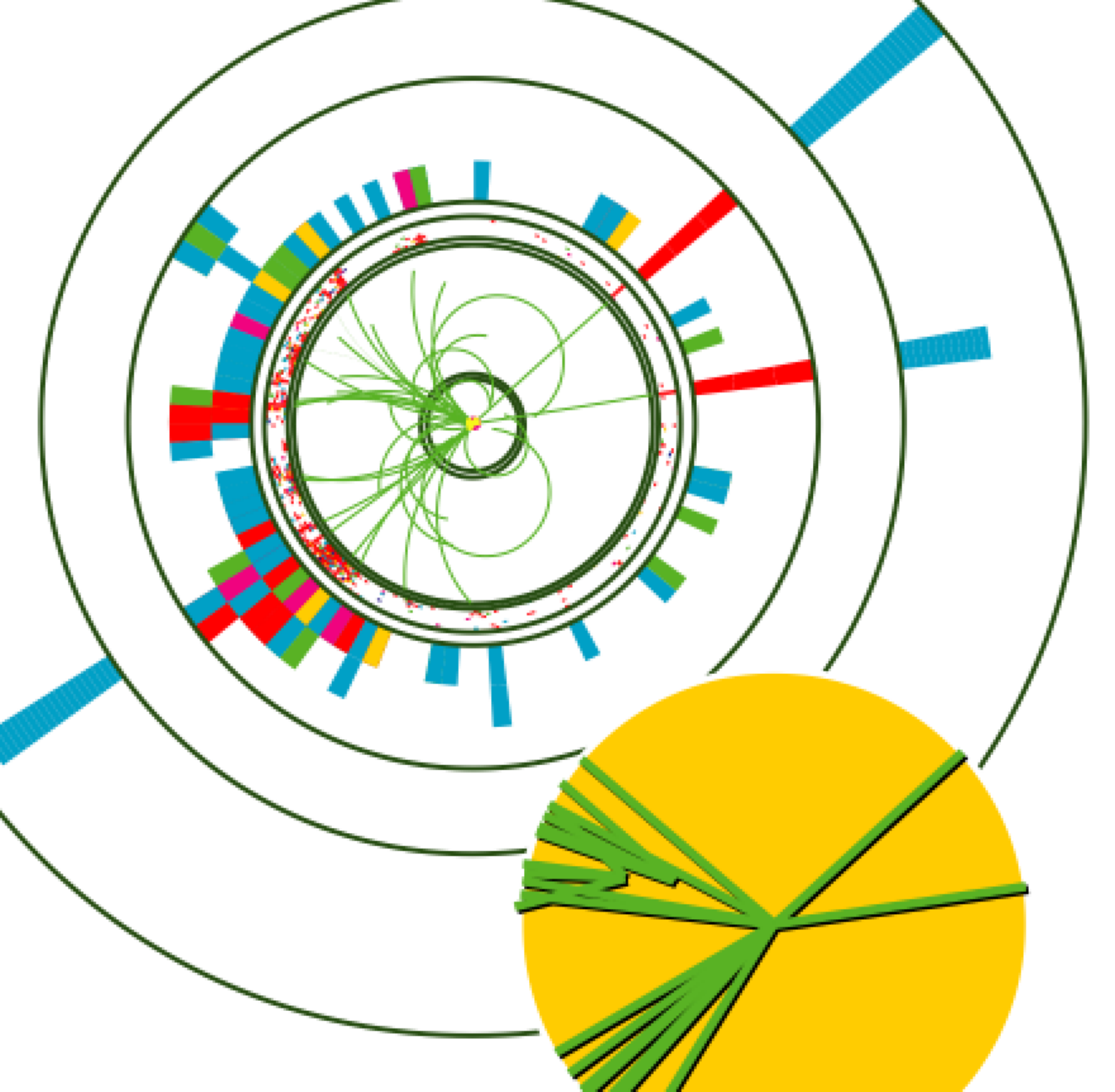}
\end{center}
\caption{Event displays of $\ee\to ZH$ events  with $Z\to \mu^+\mu^-$, from full simulation: Left: $H\to \tau^+\tau^-$, ILD detector model;  Right: $H\to b\bar b$, SiD detector model.}
\label{fig:HiggsEvents}
\end{figure}

The reaction $\ee\to ZH$ provides {\it tagged} Higgs decays.   Thus, events can be selected independently of the Higgs decay mode.  Then (1) the  total cross section for this reaction can be measured, giving a means of absolutely normalizing Higgs boson couplings; (2)  Higgs branching ratios can be measured  by counting, independently of the production cross section; and, (3)  exotic decay modes of the Higgs boson can be observed as products recoiling against the $Z$ tag.   Some event displays, from full 
simulation, are shown in Fig.~\ref{fig:HiggsEvents}. 

\subsection{Search for exotic Higgs decays}

The fact that the reaction $\ee\to ZH$ yields {\it tagged} Higgs decays opens the 
possibility of another type of search for new physics.
The Higgs field is unique among SM fields in that it can form a dimension-2 operator 
$\Phi^\dagger \Phi$ with zero SM quantum numbers.  If there is any sector of fields that contains its own scalar field $\Sigma$, there will in general be a renormalizable coupling 
\beq
          \Delta \L   =   \eta\    (\Phi^\dagger \Phi) \, (\Sigma^\dagger \Sigma)  \ .
\eeq{Higgsportal}
The coupling constant $\eta$ is dimensionless, so the connection can be made at any (high) mass scale.   Thus it is possible for the Higgs boson to access a sector of elementary particles that have no other connection to the fields of the SM. 

If there is a new sector of particles with zero SM quantum numbers such that some of those particles have pair-production thresholds below 125 GeV, those particles should appear in Higgs boson decays.   If the particle that makes up cosmic dark matter is light enough to be produced in this way, the Higgs boson will have decays to invisible final states.   It is also possible that the new particles are unstable with respect to decay back to SM models.   Such decays could produce  a number of different exotic
final states, including $4b$,
$4\tau$, $b\bar b$ + invisible states, and new particles with long lifetimes~\cite{Curtin:2013fra}.   With the data sample of the 250~GeV ILC, it is possible to search for all of these decay modes.   For  invisible Higgs decays, the expected 95\% CL exclusion limit on the  branching ratio is  $3\times 10^{-3}$, 
and for 
visible or partially visible modes the limits are in the range $10^{-3}$--$10^{-4}$~\cite{Liu:2016zki}. 

\subsection{Effective Field Theory framework for Higgs coupling determinations}
\label{subsec:phys_eft}

Though the goals of measuring the SM and exotic branching ratios of the Higgs boson are already very important, experiments at the ILC allow a further step.   The theory predictions described in Sec.~\ref{subsec:newphysicofH} refer to absolute partial decay widths.   These are 
related to the Higgs branching ratios by 
\beq
           \Gamma(H\to A\bar A) =   \Gamma_{tot} \cdot BR(H\to A\bar A)  \  .
\eeqn
The Higgs boson total width is very small---4~MeV in the SM---and it is not expected that any proposed collider can measure this value directly  with percent-level precision.
However, as we will now show, the ILC can determine the total width of the Higgs in a 
way that is highly model-independent and allows a 1\% absolute normalization of Higgs coupling constants.

A possible method to determine the Higgs width is to multiply each $HAA$ coupling by a parameter $\kappa_A$, and then fit these prarameters using data from $\ee\to ZH$.   In this simple method, the total cross section for $\ee\to ZH$ is proportional to $\kappa_Z^2$ and so the $\kappa_A$ parameters can be absolutely normalized.   This method has been used in much of the literature on Higgs coupling determinations, including \cite{Fujii:2015jha}.  In that paper, invisible and visible but exotic decay modes were treated by including these two partial widths as two additional parameters in the fit.  Using as inputs the measureable $\sigma\times BR$s for SM Higgs decay channels and Higgs decays to invisible final states, plus the total cross section for $\ee\to ZH$,  the ILC data would give a well-defined fit to the $\kappa_A$ parameters.

There are two problems with this method.  First, the method is not completely model-independent.  Modelling the effects of new physics as a general set of dimension-6 operators as in \leqn{geneffL}, we find two different Lorentz structures for the modifications of the $HZZ$ vertex,
\beq
   \Delta \L =    (1 + \eta_Z) {m_Z^2\over v}  h Z_\mu Z^\mu + \zeta_Z {1\over 2v} h
   Z_{\mu\nu} Z^{\mu\nu} \ ,
\eeq{etazeta}
where $Z_\mu$ is the gauge field and $Z_{\mu\nu}$ is the field strength, 
and a similar pair of structures for the $HWW$ vertex. The $\zeta_V$ coefficients have interesting phenomenological significance.    In weakly coupled models such as supersymmetry, these couplings are generated only by loop diagrams and have very small values; however, in composite Higgs models these coefficients 
can be comparable to the $\eta_V$ coefficients.   This makes it important to be able to determine the two couplings independently determined from data. The second problem with the method given in the previous paragraph is that it does not make the most effective use of the data set from $\ee$ colliders.  The total width of the Higgs boson is determined from the ratio  $\sigma(\ee\to ZH)/\Gamma(H\to ZZ^*)$.  Since branching ratio for $H\to ZZ^*$ is only 3\%, this strategy sacrifices a factor 30 in 
statistics.

A much more effective method for fitting Higgs boson couplings is described in \cite{Barklow:2017suo}.   In this method, we model the effects of new physics by the most general set of dimension-6 operators that can appear in \leqn{geneffL}.  The complete set of $SU(3)\times SU(2)\times U(1)$-invariant lepton- and baryon-number conserving dimension-6 operators  includes 59 terms~\cite{Grzadkowski:2010es}.  However, for 
the analysis of $\ee$ collider data, we can restrict ourselves to electron reactions 
producing the Higgs boson and other color-singlet states.  Since there is a single SM effective
 Lagrangian that should apply to all processes, this method allows us to combine data on 
Higgs production with additional data sets from $\ee\to W^+W^-$ and precision electroweak
measurements.   It is also possible to make use of additional observables for Higgs production
beyond simple rates.  In particular, the angular distribution and polarization asymmetry in 
$\ee\to ZH$ play important roles.  These considerations give the method based on 
Effective Field Theory much more power in extracting the most accurate estimates 
of the 
Higgs boson couplings from the data.

It is sometimes considered a restriction that the EFT model contains only operators of dimension 6 without considering operators of dimension 8 and higher.  However, there is a useful analogy to precision electroweak measurements.   There, 
the effects of the top quark and the Higgs boson are well-modeled by the $S$ and $T$ parameters~\cite{Peskin:1990zt}---which
are part of the dimension-6 effective field theory description---even though the masses of these particles are not far above the $Z$ mass.  Only when new particles are discovered and one wishes to compute their effects in detail is it necessary to go beyond the leading corrections.  Very light new particles can have a different effect that is not accounted for by Effective Field Theory, since they can provide new Higgs decay channels that give additional contributions to the Higgs width.  We take these possible effects into account explicitly as additional parameters in our global fit, as we will explain in Sec.~\ref{subsec:global:elements}.

After we describe the experimental methods and the expected measurement uncertainties for Higgs production in  Sec.~\ref{sec:higgs} and for $W$ pair production in Sec.~\ref{sec:ew},  we will present formal  projections for uncertainties in 
Higgs couplings in Sec.\ref{sec:global}, making use of the EFT  method.  We will show that the data set expected for the ILC at 250~GeV will  measure the $Hb\bar b$ couplings to 1\% accuracy, the $HWW$ and $HZZ$ couplings to better than 0.7\%, and the other 
major SM Higgs couplings to accuracies close to 1\%.    These measurements should be 
statistics-limited.  Above 250~GeV, a second Higgs production reaction, $\ee\to \nu\bar\nu h$ through $W$ boson fusion becomes important. We will show that,  
using the additional data on $\ee\to ZH$ and the independent measurements from the $W$ 
fusion reaction,  the uncertainties on Higgs couplings will decrease by another factor of 2.

\subsection{$\ee \to W^+W^-$}
\label{subsec:phys_WW}
\begin{figure}
\begin{center}
\includegraphics[width=0.80\hsize]{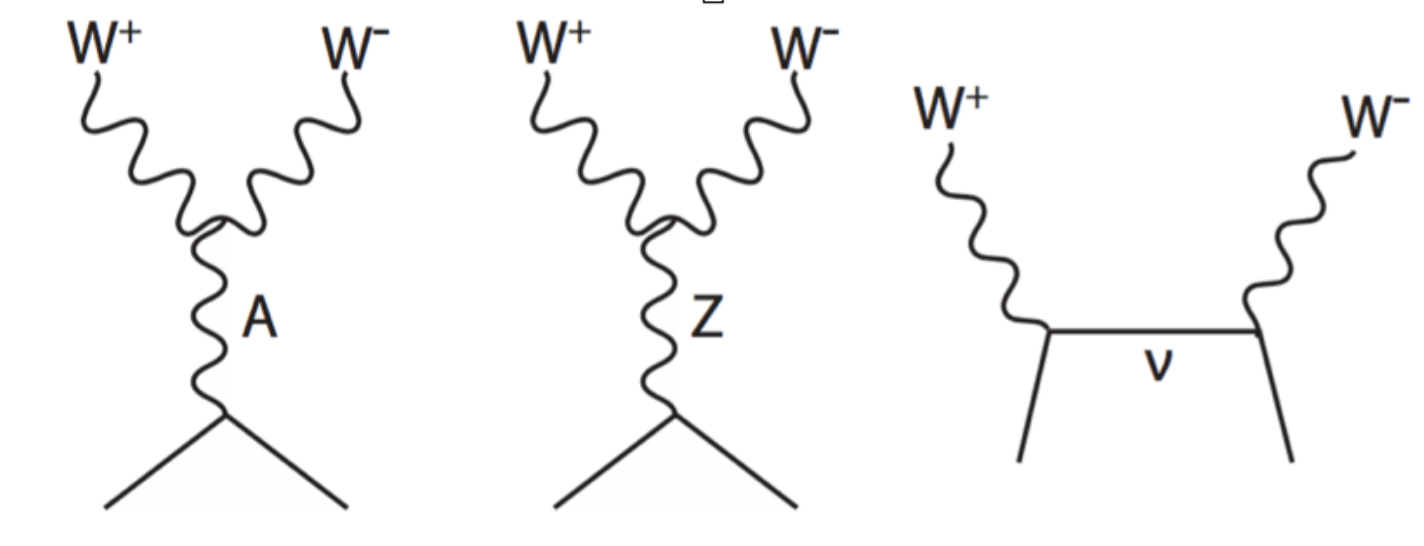}
\end{center}
\caption{Feynman diagrams contributing to the process $\ee\to W^+W^-$ when contributing  dimension-6 operators are included.}
\label{fig:eeWWdiagrams}
\end{figure}

The reaction $\ee\to W^+W^-$ contributes to the analysis described above, but it has its own independent interest.  This reaction provides an excellent way to test for the presence of dimension-6 
operators that involve the $W$ and $Z$ fields.   The Feynman diagrams contributing to the reaction are shown in Fig.~\ref{fig:eeWWdiagrams}.   The process involves interference between $s$-channel diagrams with $\gamma$ and $Z$ exchanges and a $t$-channel diagram with neutrino exchange.  In the SM, there are large cancellations among these diagrams, but these are not respected by the dimension-6 contributions.   Thus, the dimension-6 coefficients appear in the cross section formula enhanced by a factor $s/m_W^2$. 

The largest part of the dimension-6 effect can be described by shifts of the form 
factors for the $WW\gamma$ and $WWZ$ vertices.  These vertices can be parametrized as \cite{Hagiwara:1986vm}
\beqa
\Delta \L  &=& i g_V \biggl\{    V^\mu \bigl( { \hat W}^-_{\mu\nu}
W^{+\nu} -  {\hat W}^+_{\mu\nu}
W^{-\nu} \bigr)  +  \kappa_V  W^+_\mu W^-_\nu {\hat V}^{\mu\nu} \CR
& & \hskip 0.7in + {\lambda_V \over m_W^2}{\hat W}^-_\mu{}^\rho {\hat W}^+_{\rho \nu}
 {\hat  V}^{\mu\nu}  \biggr\} \ ,
\eeqa{WWgZvertex}
where $V = A$ or $Z$.
 In the SM, $g_{1A} = e$, $g_{1Z} = e s_w/c_w$ and the other coeficients are zero at the tree level.   The result $g_{1A} = e$ is exact due to the QED Ward identity.  The dimension-6 operator corrections generate a 3-parameter shift of the other 5 coefficients. These shifts can in principle be measured both at electron and at hadron colliders.  However, measurements in $\ee$ have  definite advantages.   First, it is possible to use final states with hadronic $W$ decays to determine the complete kinematics of each event and, using this information, to separate the production of 
transverse and longitudinal $W$ bosons.   Then, using beam polarization and $W$ final-state polarization, the 3 possible shifts of the form factors can be measured separately.   Second,  the greater intrinsic accuracy of measurements in $\ee$ give excellent results at a center of mass energies of 250--500~GeV.  At hadron colliders, 
the factor $s/m_W^2$ can be much larger, and one can take advantage of this by observing the reaction at $WW$ pair masses above 1~TeV.  However, this can lead to ambiguities due to the possible influence of dimension-8 operators, whose effects grow as $(s/m_W^2)^2$~\cite{Falkowski:2016cxu}. 

In Sec.~\ref{sec:ew}  below, after describing the experimental study of $\ee\to W^+W^-$, we will show that the ILC at 250~GeV is expected to improve the precision of $W$ form factor measurements by an order of magnitude over expected results from the HL-LHC.

\subsection{$\ee\to f\bar f$}
\label{subsec:phys_ff}

Fermion pair production provides a search for new forces that couple directly to the electron.   At LEP and LHC, $\ee$ and $q\bar q$ annihilation are used as probes for new gauge bosons appearing in the $s$-channel and for signals of fermion compositeness. 

The comparison with LEP 2 is instructive.   The ILC will operate at an energy not so far above that of LEP 2  (250~GeV {\it vs.}  180--208~GeV)  but with much higher luminosity (2 ab$^{-1}$  {\it vs.} a total of 1~fb$^{-1}$ over 4 experiments).   For statistics-limited measurements, this gives a factor 
\beq
   \biggl[{  s \cdot \int \L |_{ILC}\over s \cdot \int \L|_{LEP}} \biggr]^{1/2} \approx 
                    60 
\eeqn
improvement in sensitivity to deviations from the SM, or an improvement of  7.5 in the mass scale that can be accessed.   Though the comparison depends on the particular model, this corresponds to the ability to observed new vector bosons at 5--6 TeV and 
contact interaction scales of 70~TeV, comparable to  the projected reach of the HL-LHC. 

 In addition, the information from ILC is very specific.   Measuring the cross section for $\ee\to f\bar f$ in the forward and backward directions for $e^-_L$ and $e^-_R$ beams gives four  different observables, each of which corresponds to a different dimension-6 effective  interaction.   Discovery of an anomaly points directly to a model interpretation, either with an $s$-channel resonance or with new interactions at higher energy.   This information can be put together with results of resonance searches at the LHC. 

The reaction $\ee\to b\bar b$ deserves special consideration. In models in which the Higgs boson is composite, typically either the $t_L$ or the $t_R$ must be composite also to generate a large enough $t$ quark mass.    The $b_L$ is the $SU(2)$ partner of the $t_L$ and so must have the same admixture of composite structure.    If it is the $t_R$ that is more composite, it is not required that the $b_R$ is composite, but this often happens in models.  This can generate few-percent corrections to the rates for $e^-_{L,R} e^+\to b_R\bar b_L$ at the ILC~\cite{Funatsu:2017nfm,Yoon:2018xud}.   It is possible that this effect, rather than effects in Higgs decays, would be the first indication of a composite Higgs sector.

\subsection{Search for pair-production of new particles}

Despite the wide range of direct searches for new particle pair production at the LHC, 
those searches have blind spots corresponding to physically interesting models.  The two most important of these are:
\begin{enumerate}
\item  Insensitivity to new particles with electroweak interactions only that decay to an invisible partner with a mass gap of less than 5~GeV.   Though this case seems quite special, this is exactly the set of properties predicted for the charged Higgsino of SUSY models.  Dark matter scenarios involving coannihilation can also fall into this blind spot, since in those models there is an electroweak partner separated from the dark matter particle by $k_B T$ at the thermal dark matter freezeout temperature of 5-10~GeV.
\item Insensitivity to production of pairs of invisible particles observed through radiation of an initial-state gluon.   At the LHC, such ``mono-gluon'' events have as a background initial state radiation in the Drell-Yan process, and the sensitivity to such events is limited by the precision of our knowledge of the Drell-Yan cross section.
\end{enumerate}
In both cases, the ILC can detect these new physics events for particle masses almost up to half of  the collider center of mass energy.

The experimental aspects of these particle seaches are discussed in Sec.~\ref{sec:searches}.  A broader review of the opportunities for new particle discovery at $\ee$ colliders can be found in \cite{Fujii:2017ekh}.

\subsection{The central role of beam polarisation}
\label{subsec:beampol}

One theme that runs through all of the analyses discussed in the following sections is the  important role of beam polarisation.   The use of beam polarisation may be unfamiliar to many readers, since all recently operated colliders -- the Tevatron, PEP-II and KEKB, and the LHC -- have had unpolarised beams.   In hadronic 
collisions, the effects of polarisation are relatively small, first, because the dominant QCD interactions conserve parity and, second, because the proton is a composite particle, so high proton polarisation translates to much smaller polarisation for the constitutent quarks and gluons.   At a high-energy $\ee$ collider, the situation is very different.  The dominant interaction is the electroweak interaction, which has 
order-1 parity asymmetries in its couplings.  The beam particles are elementary, so that 80\% beam polarisation translates to 80\% polarisation in the colliding partons.
This implies  that polarisation effects are large at $\ee$  colliders and can be used to great advantage.   

It is very challenging to achieve high beam polarisation in circular colliders, especially for longitudinal polarisation.   Transverse beam polarisation was observed at LEP in single- and separated-beam operation but not for 
colliding beams~\cite{Assmann:1998qb}.  However,  a linear electron or positron
collider naturally preserves longitudinal beam polarisation.    The design of the 
ILC has been thought through  to produce, maintain, and control 
beam polarisation for both electrons and positrons, as has been explained in 
Sec.~\ref{par:beampol}.  This brings advantages for physics that we now discuss.

There are three major uses of beam polarisation in the experiments planned for ILC:
\begin{enumerate}
\item  Measurement of helicity-dependent electroweak couplings.
\item Suppression of backgrounds and enhancement of signals.
\item  Control of systematic uncertainties.
\end{enumerate}
We discuss the first two of these points here.  The third, which is particularly 
important to claim a discovery from  precision measurements, is discussed in Sec.~\ref{subsec:polarisation}.  A comprehensive review of the role of polarisation with many more examples can be found in~\cite{MoortgatPick:2005cw}, and, for  positron polarisation in particular, in~\cite{Fujii:2018mli}. 

To begin, we need some notation. Let $P_{e^-}$ and $P_{e^+}$ be, respectively, the longitudinal polarisations of the electron and positron colliding beams, equal to $+1$ for completely polarised right-handed beams and $-1$ for completely polarised left-handed beams.   Let $\sigma_{RR}$, $\sigma_{RL}$, $\sigma_{LR}$, $\sigma_{LL}$ be the cross sections for a given process 
with completely polarised beams of the four possible orientations.  Since the electron has only two spin  states, the 
cross section for general beam  polarisations is given by 
\begin{eqnarray}
\sigma_{\Pem\Pep} &=& \frac{1}{4}\bigl\{
     (1+\Pem)(1+\Pep) \sigmaRR  \nonumber \\
&& + (1-\Pem)(1-\Pep) \sigmaLL \nonumber \\
&& + (1+\Pem)(1-\Pep) \sigmaRL \nonumber \\ 
&& + (1-\Pem)(1+\Pep) \sigmaLR \bigr\},
\label{eq:pol:xsec}
\end{eqnarray}

For $s$-channel $\ee$ annihilation processes, helicity conservation implies that only $\sigma_{RL}$ and $\sigma_{LR}$ are nonzero.
In this case Eq.~\leqn{eq:pol:xsec} reduces to the simpler form
\begin{equation}
 \sigma_{\Pem\Pep} = 2 \sigma_0 (\Leff/\mathcal{L}) \left[1 - \ALR \Peff \right]
\label{eq:pol:xsecschan}
\end{equation}
where $\sigma_0$ is the unpolarised cross section, and \Leff\ and \Peff\  
are the effective luminosity and polarisation, defined, respectively, as 
\begin{equation}
\Peff= \frac{\Pem - \Pep}{1 - \Pep\Pem}
\label{eq:def-leff-peff}
\quad\mbox{\rm and }\quad
\Leff=\frac{1}{2}(1 -\Pep\Pem)\L \ . 
\end{equation}
The coefficient $\ALR$ is the intrinsic left-right asymmetry of the reaction cross section,
\beq
  \ALR =    {\sigmaLR - \sigmaRL\over \sigmaLR + \sigmaRL } \ . 
\eeq{eq:def-ALR}

In the electroweak interactions, it is typical that left-handed fermions have larger coupling constants than right-handed fermions.   Then, choosing $\Pem$ to be left-handed (negative) and $\Pep$ to be right-handed (positive) can confer important advantages.
Consider, for example, the typical ILC values $\Pem = -80\%$, $\Pep = +30\%$.    Then the effective polarization 
Eq. \leqn{eq:def-leff-peff} for the measurement of $\ALR$ values is  $\Peff = 90\%$.   The Higgsstrahlung process has a rather small polarisation asymmetry, $\ALR = 0.151$. Still, the luminosity is enhanced from the unpolarised case by  $ 40\%$. 

Such substantial values of the beam polarisations can be applied to physics measurements in the following ways:
\begin{itemize}
\item In $\ee\to f\bar f$, the $e^-_L$ and $e^-_R$ couple to different linear combinations of the $s$-channel $\gamma$ and $Z$
propagators.  Beam polarisation allows us to measure the couplings to these vector bosons independently.  In addition, an $s$-dependent change in the polarisation asymmetry can signal the presence of a new $s$-channel resonance.  
\item Similarly, in $\ee\to W^+W^-$, the separation of $\gamma$ and $Z$ couplings can be combined with information from the $W$ production angle and polarisations to completely disentangle the 14 complex parameters (28 real parameters) in the most general 
Lagrangian for triple gauge vertices.
\item In $\ee\to ZH$, measurement of  the polarisation asymmetry plays an important role in disentangling the various of  parameters 
that enter the EFT analysis of Higgs boson couplings.
\item If new particles are discovered in pair-production at the ILC, measurement of the production cross section with different beam polarisation settings allows their electroweak quantum numbers to be determined unambiguously.
\end{itemize}
We will illustrate all of these points in the sections to follow.

\begin{figure}
\centering
\includegraphics[width=0.95\linewidth]{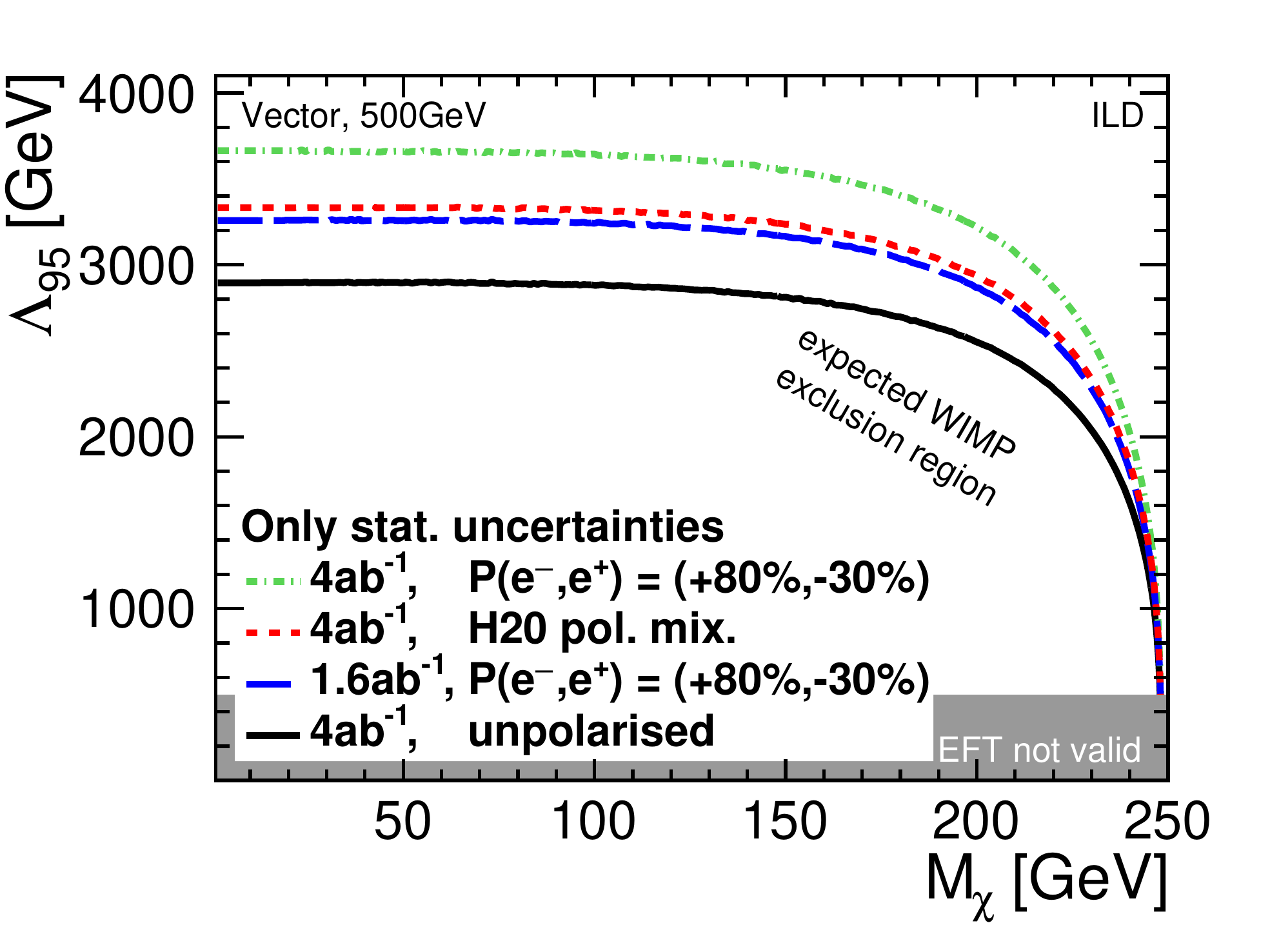}
		
\caption{Comparison of the 95\% confidence lower limit on the mediator scale for dark matter production using  the mono-photon channel,  for different assumptions on luminosity and polarisation.  See Sec.~\ref{sec:searches} for a description of the analysis~\cite{Habermehl:417605}. Note that this plot considers {\em statistical uncertainties only}. The corresponding comparison {\em including systematic uncertainties} is shown in Fig.~\ref{fig:polWIMPsys}.}
\label{fig:polWIMPstat}
\end{figure}


It is also possible to take advantage of differences in the polarisation effects on signal and background cross sections to enhance signals and control backgrounds.  Unlike annihilation processes, radiative Bhabha scattering and 2-photon processes have nonzero $\sigma_{LL}$ and $\sigma_{RR}$, so it is possible to test strategies for the suppression of these backgrounds using  data sets with
in which $\Pem$ and $\Pep$ have the same sign.   The reaction $\ee\to W^+W^-$ has a relatively large cross section among 
annihilation processes and  is often the dominant background to studies of or searches for other processes.   However, this process also has a large polarisation asymmetry, with $\sigma_{LR}/\sigma_{RL} \approx 30$.  Then running with $\Pem = +80\%$, $\Pep = -30\%$ essentially turns off this background.   

As an example of the effectiveness of background suppression, we show in Fig.~\ref{fig:polWIMPstat} a comparison of searches  for invisible dark matter particles $\chi$  in the mono-photon  mode $\ee\to \gamma+ \chi\chi$ under different assumptions on luminosity and polarisation.   The predicted 95\% confidence lower limit on the mediator scale $\Lambda_{95}$ is shown as a function of 
the $\chi$ mass. The expected limit for an unpolarized collider is shown as the black solid curve.  For this analysis, the statistically optimal choice is that of $(+80\%,-30\%)$ polarisation; this gives the projected limit shown in the green dot-dash curve in the figure. This mode might not be chosen in practice, however, because the optimal polarisation setting will be different for other studies (\eg, Higgs measurements) that might be done in the same run.  But the figure also
shows that a data set of 1.6~\iab\ with optimal polarisation is considerably more powerful than a data set of 4~\iab\ with unpolarised
beams.  The red dotted curve shows the result for the H20 scenario with polarisations given in Tab.~\ref{tab:pollumirel}.   For clarity, the  figure includes statistical errors only. 

The influence of polarisation on systematic errors is equally important.  Where experiments with unpolarised beams give one measurement, experiments with both electron and positron beams polarised give 4 independent measurements.  These can be used as cross-checks for  the understanding of systematics, and also to form combinations in which the 
dominant systematic errors cancel.  We will 
discuss this point in more detail in Sec.~\ref{subsec:polarisation}.

\section{\label{sec:highenergy}Physics Case --Beyond 250 GeV }


A key advantage of linear colliders is the possibility to upgrade
the center-of-mass energy.  The energy reach of circular electron-positron
colliders of a given circumference is limited by synchrotron
radiation, and this is difficult to overcome because of the steep
growth of synchrotron losses with energy.  Linear colliders, however,
can be upgraded to reach higher center of mass energies either by
increasing the length of the main linacs or by installing linac components
that support higher accelerating gradients.

In the major particle physics laboratories, the lifetime of collider elements
and infrastructure has rarely been  limited to the scope of the project they were
designed and built for. A famous example is the Proton Synchrotron at CERN.
Initially commissioned in 1959, it is still in operation as
part of the accelerator complex that prepares protons for injection in the
Large Hadron Collider. In this accelerator complex, the  expensive civil engineering
efforts to construct each component are reused, so that their cost is effectively shared. The tunnel
that was constructed for LEP now hosts the Large Hadron Collider and its
luminosity upgrade. In very much the same way, we
expect  that the ILC will  form the seed for a facility that contributes to the
cutting edge of particle physics for decades.  For electron-positron
collisions, 
any facility at energies much higher than
those already realised must be a linear collider in a long, straight
tunnel.   The ILC naturally offers a setting for this program.

The ILC project  today would be an $\ee$ colllider of
250~GeV in the center of mass. But, already,  considerable thought and
planning has already gone into the extension of this machine to higher energies.
In this section, we will briefly describe the further physics
opportunities that the ILC will offer at 350~GeV, 500~GeV, and beyond.
The physics goals for higher-energy $\ee$ colliders has already been 
discussed extensively in the literature. Particularly useful
references are the volumes presenting  detailed studies carried out for 
the ILC~\cite{Fujii:2015jha,Baer:2013cma}
and CLIC~\cite{Linssen:2012hp,deBlas:2018mhx,Roloff:2018dqu} design reports.

\subsection{Scope of ILC energy upgrades}
\label{subsec:highE:tech}

As explained 
 in 
Sec.~\ref{subsubsec:upg-optE},  the ILC TDR includes provisions for
running of the  ILC at  500~GeV and 1~TeV. 
The most obvious energy upgrade path is an extension of the linear
accelerator sections of the colliders, which provides an increase
in center-of-mass energy that is proportional to the length of the linacs.
The design of the ILC presented 
in the TDR~\cite{Behnke:2013xla,Adolphsen:2013jya,Adolphsen:2013kya} envisaged a
center-of-mass energy of 500~GeV{} in a facility with a total length
of 31~km.
The ILC TDR also documents
a possible extension to 1~TeV based on current superconducting RF
technology.   Space for a tunnel of  50~km length  is available at the Kitakami site in
Japan, enough to accomodate a 1~TeV machine based on  the TDR
technology. 

An even larger increase in center-of-mass energy may be achieved by
exploiting advances in accelerator technology. The development of
cavities with higher accelerating gradient can drive a significant
increase in the energy while maintaining a compact infrastructure.
Superconducting RF technology is evolving rapidly. Important
progress has been made toward developing cavities with a gradient well
beyond the 35 MV/m required for 
the ILC~\cite{Grassellino:2018tqg,Grassellino:2017bod}
and even beyond the 45 MV/m envisaged for
 the 1~Tev{} ILC.     In the longer term, alternate-shape
or thin-film-coated Nb$_3$Sn cavities
 or multi-layer coated cavities offer the potential of
significantly increased cavity performance~\cite{Adolphsen:2013jya}.
Novel acceleration schemes may achieve even higher gradients. 
The CLIC drive beam concept has
achieved accelerating gradients of up to 100 MV/m~\cite{Aicheler:2012bya}.
Finally, the advent of acceleration schemes based on plasma
wakefield 
acceleration or another advanced concept could 
open up the energy regime up to 30~TeV. A report of the status of accelerator R\&D and remaining
challenges is found in Refs.~\cite{Cros:2017jxp,Cros:2019tns}, with further details and
a brief description of the potential of such a machine in the
addendum~\cite{ALEGRO:2019alc}.

Thus, the ILC laboratory
has  paths to evolve into a laboratory for electron-positron
collisions at higher energies, and possibly even to a laboratory that
can offer the highest parton-parton center of mass energies
achievable at any collider.

\subsection{Improvement of ILC precision at higher energy}
\label{subsec:highE:physics}

Operation of the ILC at higher energies will produce new data sets
that will substantially improve the capabilites of the ILC for all of
the physics topics presented in Sec.~\ref{sec:physics}.    The ILC
simulation studies included extensive studies at 500~GeV 
 and also studies at 1~TeV in the center of mass.   We will present
 the results of these studies together with our studies from 250~GeV in
 the following sections.

Data-taking at higher energies will improve the results from 250~GeV
and give access to new SM reactions.   Let us first summarize the
expected
improvements in the areas that we have discussed so far:
\begin{itemize}
\item  For Higgs production, running   at 500~GeV will add a new data set
 of  $\ee\to ZH$ events.  It will also provide   a substantial data set of $\ee\to
 \nu\bar\nu H$ events, corresponding to $WW$ fusion production of the
 Higgs boson.   With these new samples, it will be possible to confirm
 any anomalies in the Higgs boson coupling seen at 250~GeV and to
 provide an independent comparison of the $ZZ$ and $WW$ couplings.
Though the backgrounds to the Higgs production processes are
relatively small  for both reactions, they are different in the two
cases, providing a nontrivial check of some systematics. In global
analysis, as we will see in Sec.~\ref{sec:global}, the addition of
500~GeV data leads to a decrease in the uncertainties on Higgs boson
couplings by about a factor of 2 and a decrease in the uncertainty on the Higgs total width by a factor 1.6.  

\item For $WW$ production, running at 500~GeV will give a data set of
  roughly the same size as that obtained at 250~GeV.  Further, since
  the effects of anomalous $W$ couplings, or the corresponding
  dimension-6 operators, increase as $s/m_W^2$, the new data set will
  provide much more sensitive constraints on their effects.

\item For $f\bar f$ production, similarly, the possible new effects   
  due to heavy gauge bosons or contact interactions scale as $s/M^2$,
  where $M$ is the new mass scale.   The discovery potential for $M$,
  or new limits, will increase by a factor close to 2. 

\item For new particle searches, the reach in $\ee$ pair production is
  close to half the $\ee$  centre-of-mass energy. The  improvement in
  reach is particularly relevant for color-singlet particles such as
  heavy Higgs bosons, electroweakinos, Higgsinos, and dark matter
  particles.  
\end{itemize}

All of these observations illustrate the more general point that
higher energy can be an important tool in tests of the SM
using Effective Field Theory.   We have emphasized that an analysis
within  EFT allows a more incisive search for new physics effects on
Higgs boson couplings by bringing together a large number of
observables from different physical processes.
 In the EFT formulae, the various operator contributions have different
energy-dependence, with certain operators having an impact that grows 
strongly with energy.  Thus there is great advantage in combining a
data set taken at 250~GeV with one or more data sets taken at higher
energies.

\subsection{New Higgs physics at higher energy}

Beyond the improvement in areas that we have already discussed, the
operation of the ILC above 250~GeV can give access to new and
important SM reactions.   Among Higgs boson couplings, there are two
that are 
inaccessible at 250~GeV.  These are the top quark Yukawa coupling and
the Higgs boson self-coupling.   In Secs.~\ref{subsec:higgsself}
and \ref{subsec:top:topYukawa}, we will describe the measurement of
these couplings at the ILC at 500~GeV and above.   Since these
couplings can show large deviations from the SM expectation in certain
classes of new physics models, it is necessary to measure these
couplings accurately to complete the full picture of the Higgs boson
interactions.

The interest in the top quark Yukawa coupling is obvious.   The top
quark is the heaviest SM particle, and, within the SM, its mass is
proportional to this coupling constant.   If there are new
interactions that promote the large value of the top quark mass, the
Yukawa coupling will receive corrections, and so it is important to
probe for them.

The measurement of the trilinear Higgs coupling is an equally  important goal
of a complete program of study for the Higgs boson.  While the
measurement of the Higgs field vacuum expectation value and the mass
of the Higgs boson express the mass scale of the Higgs field potential
energy and its variation, the trilinear Higgs coupling gives
information on the shape of the potential energy function and brings
us closer to understanding its origin.

The trilinear Higgs coupling is sensitive to the nature of the phase
transition in the early universe that led to the present state of
broken electroweak symmetry.  
The  SM predicts a continuous phase transition.   This has
implications for models of the creation of the matter-antimatter
asymmetry that we observe in the unverse today. According to
Sakharov's classic analysis~\cite{Sakharov:1967dj}, the net baryon number of the universe
needed to be created in an epoch with substantial deviations from
thermal equilibrium in which $CP$- and baryon-number-violating
interactions were active.    The baryon number of the universe could
have been created at the electroweak phase transition, making use of new
$CP$-violating interactions in the Higgs sector, but only if the phase
transition was strongly first-order.   In explicit models, this
typically requires large deviations of this coupling, by a factor
1.5--3, from its SM value~\cite{Morrissey:2012db}.

\subsection{Study of the top quark in $\ee$ reactions}

In addition, the extension of the ILC to higher energies will allow the precision
study of the top quark.   This is an essential goal of precision
experiments on the SM, for two reasons.  First, similarly to the Higgs
boson, the top quark stands closer to the essential mysteries of the
SM than any other quark or lepton.   It is heavier than the next
lighter fermion, the $b$ quark, by a factor of 40 and heavier than the
lightest quark, the $u$ quark, by a factor of $10^5$.  The reasons for
this are unknown, but they must be related to other mysteries of the
Higgs sector and SM mass generation.  In fact, it is not understood
whether the top quark is a  ``heavy'' quark because of special
interactions that the other quarks do not share or, alternatively, whether the top
quark is an ``ordinary'' quark receiving an order-1 mass while the
masses of the other quarks are highly suppressed.  Competing extensions of the
SM such as supersymmetry and composite Higgs models differ in their
answers to this question.   

Second, the fact that the top quark has spin, couples to the
parity-violating weak interactions, and decays to nonzero spin
particles through  $t\to bW$ gives a large number of independent
observables for each $t\bar t$ production process. An $\ee$ collider
with beam polarization can take advantage of all of these observables,
especially if it can produce $t\bar t$ well above threshold at
500~GeV in the center of mass.  

Thus, top quark physics is a place in which we expect to find
deviations from the predictions of the SM, in a setting where we have
many handles to search for these new physics effects. The top quark
physics potential of the ILC is discussed in Section~\ref{sec:top}.

\subsection{Direct searches for physics beyond the Standard Model}

The energy upgrade of the ILC greatly significantly extends the reach of 
direct searches for signatures of extensions of the Standard Model. Searches
for new particles at the ILC provide robust, loophole-free
discovery potential. Once a significant signal is observed the properties and
interactions of the new particle can be characterized with excellent precision. 
The discovery reach for massive particles is 
primarily limited by the kinematics of the process, with mass limits for pair-produced particles typically reaching half the center of mass energy.
An energy upgrade to 500~\GeV{} or 1~\TeV{} therefore yields an immediate extension
of the mass reach. 

The possibility of an energy upgrade renders a linear collider facility a very 
flexible tool, allowing it to react to new discoveries at the LHC, at the ILC or
elsewhere. The potential of the higher-energy stages for searches is evaluated in 
more detail in Section~\ref{sec:searches}.

In the following sections, we will present the capabilities of the ILC
in all of these areas.  Our discussion will be based on 
explicit simulation studies using the accelerator properties
and run plan  described in Secs.~\ref{sec:ilc} and
\ref{sec:runscenarios} and the detector models to be presented
in  Sec.~\ref{sec:detectors}.

\section{\label{sec:detectors}Detectors }
\subsection{Introduction}

The ILC accelerator is planned with one interaction region, equipped
with two experiments. The two experiments are swapped into the
Interaction Point within the so-called ``push-pull" scheme. The
experiments have been designed to allow fast move-in and move-out from
the interaction region, on a timescale of a few hours to a day. In
2008 a call for letters of intent was issued to the
community. Following a detailed review by an international detector
advisory group, two experiments were selected in 2009 and invited to
prepare more detailed proposals.  These  are the SiD detector and the ILD
detector described in this section. Both prepared detailed and costed proposals which were
scrutinised by the international advisory group and included in the
2012 ILC
Technical Design Report~\cite{Behnke:2013lya}.  In this section the two proposals are
briefly   introduced.

The ILC detectors are designed to make precision measurements on the
Higgs boson,  $W$, $Z$, $t$, and other particles.    They are able to
meet the requirements for such measurements, first, because the
experimental conditions are naturally very much more benign than those
at the LHC, and second, because the detector collaborations have
developed technologies specifically to take advantage of these more
forgiving conditions. 

An $\ee$ collider gives much  lower collision
rates and events of much lower complexity than a hadron collider, 
and detectors can be adapted to take
advantage of this.  The radiation levels at the ILC will be modest compared with the
LHC, except for the special forward calorimters very close to the
beamline, where radiation exposure will be an issue. 
This  allows the consideration of a wide range of
materials and technologies for the tracking and calorimeter systems. 
The generally low
radiation levels allow the innermost vertex detector elements to
be located at very small  radii, significantly enhancing the efficiency
for short-lived particle identification. More generally,  the relatively
benign ILC experiment environment permits  the design of 
tracking detectors with minimal material budget (see
Sec.~\ref{sub:sw-sim}).  This allows the detectors to meet the
stringent requirement on the track momentum resolution
 which is driven by the need to  precisely reconstruct 
 the $Z$ mass in the Higgs recoil analysis.  This requirement translates into a momentum
 resolution nearly an order of magnitude better than achieved in the
 LHC experiments.

At the same time, although they are studying electroweak particle
production, it is essential that the ILC detectors have excellent
performance for jets.   At an $\ee$ collider, $W$ and $Z$ bosons are
readily observed in their hadronic decay modes, and the study of these
modes plays a major role in most analyses.    To meet the requirements
of precision measurements, the ILC detectors are optimized from the
beginning to enable jet reconstruction and measurement using 
the particle-flow algorithm (PFA). This drives the goal of $3\%$
 jet mass resolution at energies above 100~\GeV,  a resolution
about twice as good as has been achieved in  the LHC
 experiments.

Finally, while the LHC detectors depend crucially on multi-level
triggers that filter out only a small fraction of events for analysis,
the  rate of interactions at the ILC is sufficiently low to allow
running without a trigger.     The ILC accelerator design is based on
trains of electron and positron bunches, with a repetition rate of
5~Hz, and with 1312 bunches (and bunch
collisions) per train (see Sec.~\ref{sec:ilc},
Tab.~\ref{tab:ilc-params}). 
The 199 ms interval between bunch trains provides ample time for a full
readout of data
 from the  previous train.  While there are background processes arising
 from  beam-beam interactions, the detector occupancies arising from these 
have been shown to be manageable.

The combination of extremely precise tracking, excellent jet mass
resolution, and triggerless running gives the ILC, at 250 GeV and at
higher energies, a superb potential for discovery. 

To meet these goals an ambitious R\&D program has
 been pursued throughout the past 10 years or so to develop and
 demonstrate the needed technologies. The results of this program are
 described in some detail in Ref.~\cite{RDliaision}. 
 The two experiments proposed for
 the ILC, SiD and ILD, utilise and 
rely on the results from these R\&D efforts.

Since the goals of SiD and ILD in terms of material budget, tracking
performance, heavy-flavor tagging, and jet mass resolution are very 
demanding, we feel it important to provide information about the level
of detailed input that enters our performance estimates.   These are
best
discussed together with the event reconstruction and analysis
framework that we will present in Sec.~\ref{sec:software}.   In that
section, we will present estimates of detector performance as
illustrations at the successive stages of event analysis.

\subsection{The SiD detector}
\begin{figure}[tb]
 \begin{center}
 \includegraphics[width=0.9\hsize]{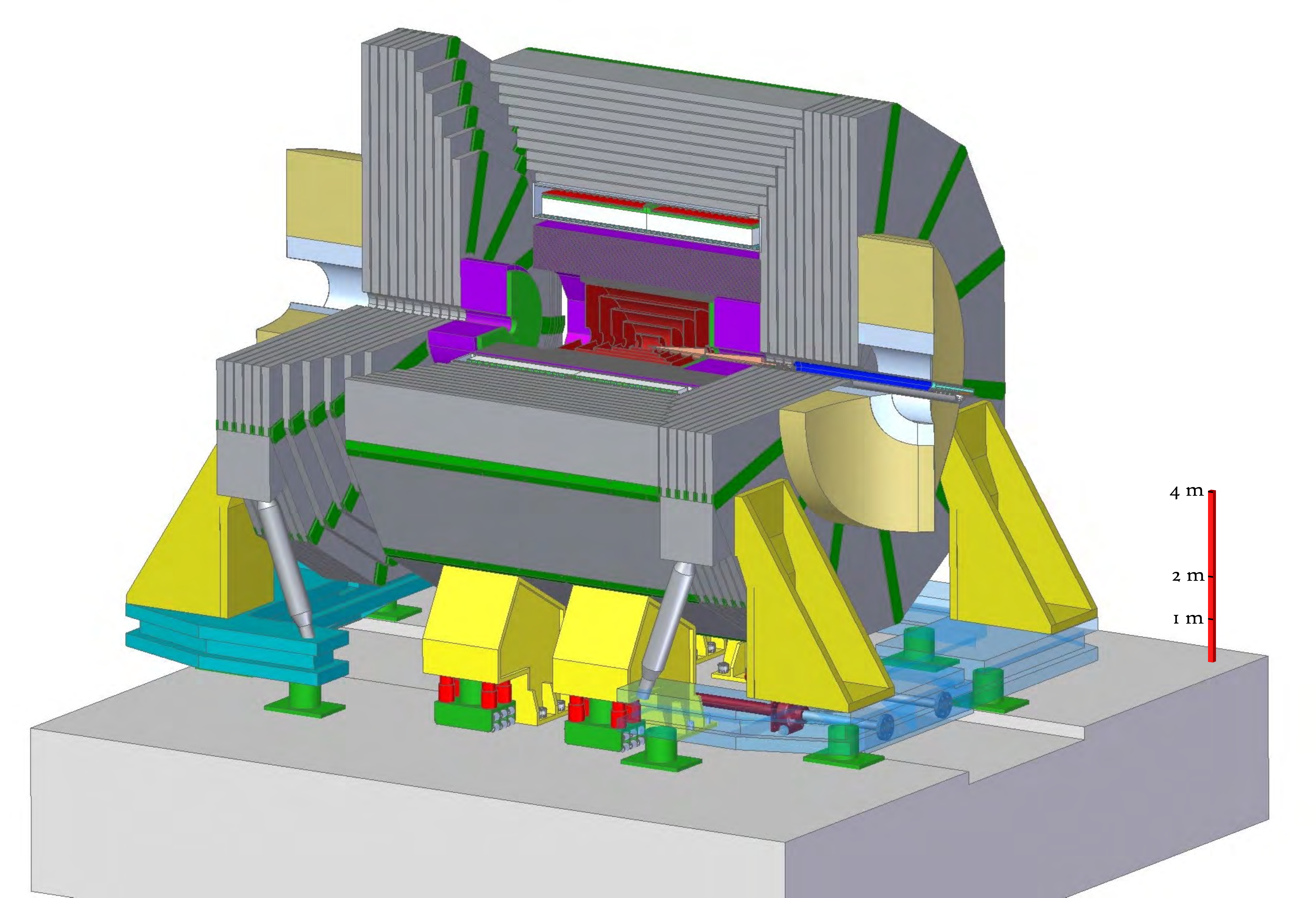}
\caption{The SiD detector concept.
\label{fig:fig_sid}}
 \end{center}
 \vspace{-0.7cm}
 \end{figure}

The SiD detector is a general-purpose experiment designed to perform
 precision measurements
at the ILC. It satisfies the challenging detector requirements resulting from the full range of 
ILC physics processes. SiD is based on the paradigm of particle flow, an algorithm by which
the reconstruction of both charged and neutral particles is accomplished by an optimised
combination of tracking and calorimetry. The net result is a significantly more precise jet
energy measurement which results in a di-jet mass resolution good enough to distinguish
between $W$s and $Z$s.
The SiD detector (Fig.~\ref{fig:fig_sid})  is a compact detector based on a powerful silicon
pixel vertex detector, silicon tracking, silicon-tungsten electromagnetic calorimetry, and
highly segmented hadronic calorimetry. 
SiD also incorporates a high-field solenoid, iron
flux return, and a muon identification system. The use of silicon 
sensors in the vertex, tracking,
and calorimetry enables a unique integrated tracking system ideally suited to particle
flow.

The choice of silicon detectors for tracking and vertexing ensures that SiD is robust
with respect to beam backgrounds or beam loss, provides superior charged particle momentum
resolution, and eliminates out-of-time tracks and backgrounds. The main tracking
detector and calorimeters are “live” only during a single bunch crossing, so beam-related
backgrounds and low-pT backgrounds from $\gamma\gamma$ processes will be reduced to the minimum
possible levels. The SiD calorimetry is optimised for excellent jet energy measurement
using the particle flow technique.
 The complete tracking and calorimeter systems are contained
within a superconducting solenoid, which has a 5 T field strength, enabling the overall
compact design. The coil is located within a layered iron structure
that returns the magnetic flux and is instrumented to allow the
identification of muons. 
All aspects of SiD are the result of intensive and leading-edge research aimed at achieving
performance at unprecedented levels. At the same time, the design represents a balance between cost
and physics performance. The key parameters of the SiD design are
listed in  
Table~\ref{sid:ConceptOverview:Table:Ovw_sidparams}.

\begin{table}[htbp]
\renewcommand{\arraystretch}{1.25}


\caption{\label{sid:ConceptOverview:Table:Ovw_sidparams}Key parameters of the baseline \sid design. (All dimension
are given in cm).}

{
\begin{tabular}{l l r r r}
    \toprule
    \sid Barrel& Technology& In rad& Out rad& z extent \\
    \midrule
    Vtx detector& Silicon pixels& 1.4& 6.0& $\pm \quad 6.25$ \\
    Tracker& Silicon strips& 21.7& 122.1& $\pm \quad 152.2$ \\
    ECAL& Silicon pixels-W& 126.5& 140.9& $\pm \quad 176.5$ \\
    HCAL& Scint-steel& 141.7& 249.3& $\pm \quad 301.8$ \\
    Solenoid& 5 Tesla SC & 259.1& 339.2& $\pm \quad 298.3$ \\
    Flux return& Scint-steel& 340.2 & 604.2& $\pm \quad 303.3$ \\
    \bottomrule

   \toprule
 \sid Endcap& Technology& In z& Out z& Out rad \\
    \midrule
Vtx detector& Silicon pixels& 7.3& 83.4& 16.6 \\
Tracker& Silicon strips& 77.0& 164.3& 125.5 \\
ECAL& Silicon pixel-W& 165.7& 180.0& 125.0 \\
HCAL& Scint-steel& 180.5& 302.8& 140.2 \\
Flux return& Scint/steel& 303.3& 567.3& 604.2 \\
LumiCal& Silicon-W& 155.7& 170.0& 20.0 \\
BeamCal& Semicond-W& 277.5& 300.7& 13.5 \\
    \bottomrule
\end{tabular}
}

\end{table}

\subsubsection{Silicon-based tracking}
\begin{figure}[tb]
 \begin{center}
 \includegraphics[width=0.9\hsize]{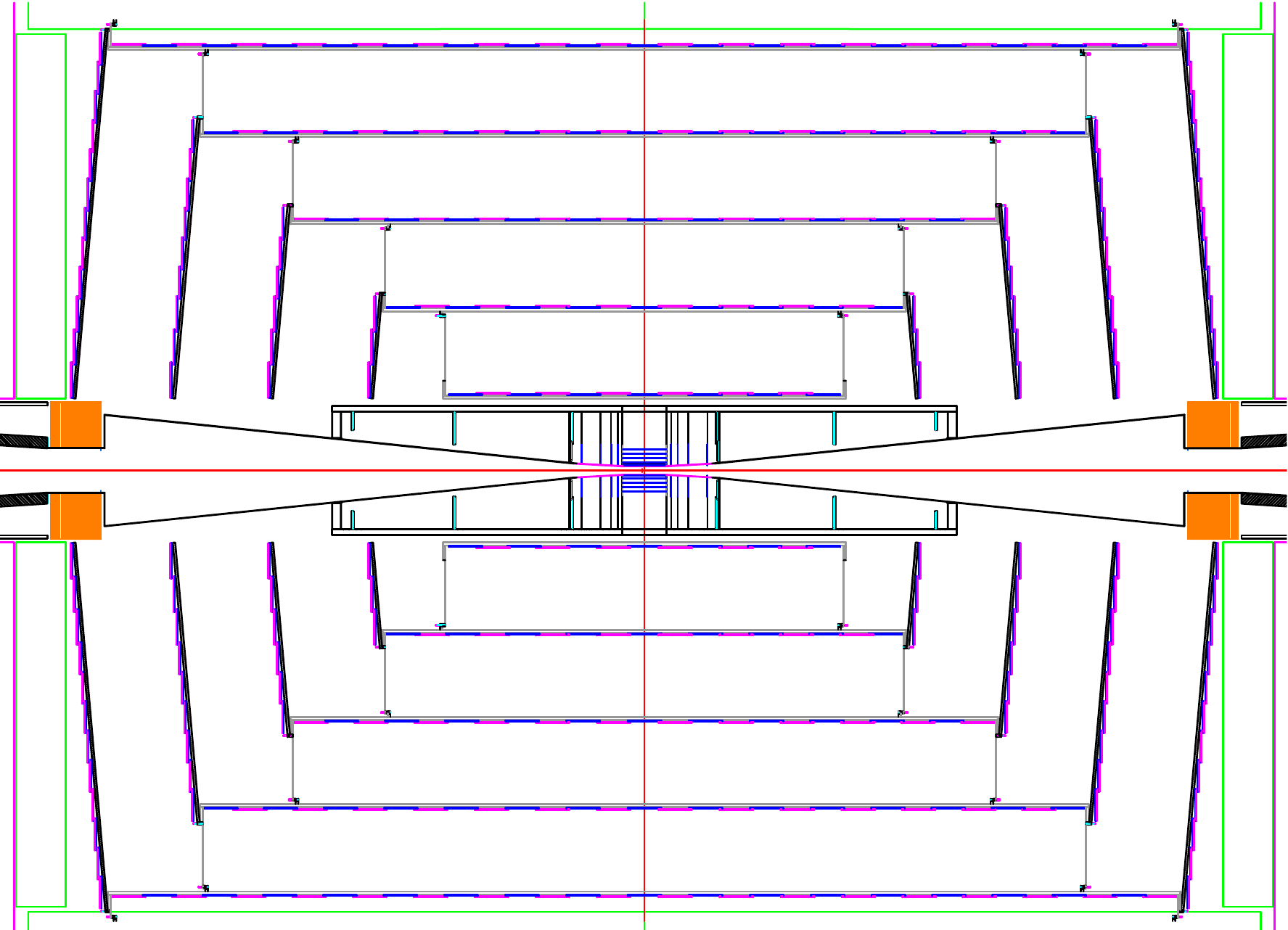}
\caption{r-z view of vertex detector and outer tracker.
\label{fig:fig_vxdtrk}}
 \end{center}
 \vspace{-0.7cm}
 \end{figure}

The tracking system (Fig.~\ref{fig:fig_vxdtrk}) is a key element of the SiD detector concept. The
particle flow algorithm requires excellent tracking with superb efficiency and
two-particle separation. The requirements for precision measurements, in
particular in the Higgs sector, place high demands on the momentum resolution at
the level of $\delta (1/\pT)  \sim 2-5 \times 10^{-5}/$GeV/$c$.

Highly efficient charged particle tracking is achieved using the pixel detector
and main tracker to recognise and measure prompt tracks, in conjunction with the ECAL, which can
identify short track stubs in its first few layers 
to catch tracks arising from secondary decays of long-lived particles. With
the choice of a 5~T solenoidal magnetic field, in part chosen to control the $\ee$-pair
background, the design allows for a compact tracker design. 

\subsubsection{Vertex detector}

To unravel the underlying physics mechanisms of new observed processes, the
identification of heavy flavours will play a critical role. One of the main
tools for heavy flavour identification is the vertex detector. The physics goals
dictate an unprecedented spatial three-dimensional point resolution and a very
low material budget to minimise multiple Coulomb scattering. The running 
conditions at the ILC impose the readout speed and radiation tolerance. 
These requirements are normally in tension. High
granularity and fast readout compete with each other and tend to increase the
power dissipation. Increased power dissipation in turn leads to an increased
material budget. The challenges on the vertex detector are considerable and
significant R\&D is being carried out on both the development of the sensors and
the mechanical support.
The SiD vertex detector uses a barrel and disk layout. The barrel section
consists of five silicon pixel layers with a pixel size of
$20~\times~20~\micron^2$. The forward and backward regions each have four
silicon pixel disks. In addition, there are three silicon pixel disks at a
larger distance from the interaction point to provide uniform coverage for the
transition region between the vertex detector and the outer tracker. This
configuration provides for very good hermeticity with uniform coverage and
guarantees excellent charged-track pattern recognition capability
 and impact parameter resolution 
over the full solid angle. 
This enhances the capability of the integrated tracking system and, 
in conjunction with the high magnetic field, makes for a very compact
system, thereby minimising the size and costs of the calorimetry.

To provide for a very robust track-finding performance the baseline 
choice for the vertex detector has a sensor technology that provides
time-stamping of each hit with sufficient precision to assign it to
a particular bunch crossing. This significantly suppresses
backgrounds. 


Several vertex detector sensor technologies are being developed.  One of these is a 
monolithic CMOS pixel detector with time-stamping capability (Chronopixel~\cite{Sinev:2015iwr}),
being developed in collaboration with SRI International. 
The pixel size is about  $10~\times~10~\micron^2$ with a design goal of 99\% charged-particle
 efficiency.
The time-stamping feature of the design means each hit is accompanied by a time tag with sufficient precision to assign it to a particular bunch crossing of
the ILC -- thus the name Chronopixel. This reduces the occupancy to negligible levels, even in the
innermost vertex detector layer, yielding a robust vertex detector which operates at background
levels significantly in excess of those currently foreseen for the ILC. Chronopixel differs from the
similar detectors developed by other groups by its capability to record time stamps for two hits in
each pixel while using standard CMOS processing for manufacturing. 
Following a series of prototypes, the Chronopixel has been proven to be
a feasible concept for the ILC. The three prototype versions
were fabricated in 2008, in 2012, and in 2014.
The main goal of the third prototype was to test possible solutions for a high capacitance problem
discovered in prototype 2. The problem was traced to the TSMC 90 nm technology design rules,
which led to an unacceptably large value of the sensor diode capacitance. Six different layouts
for the prototype 3 sensor diode were tested, and the tests demonstrated that the high capacitance
problem was solved.

With prototype 3 proving that a Chronopixel sensor can be successful with all known problems solved, optimal sensor design would be the focus of future tests.
The charge collection efficiency for different sensor diode options needs to be measured to determine
the option with the best signal-to-noise ratio. Also, sensor efficiency for charged particles with sufficient energy to penetrate the sensor thickness and ceramic package, along with a trigger telescope measurement, needs to be determined. Beyond these fundamental measurements, a prototype of a few cm$^2$ with a final readout scheme would
test the longer trace readout resistance, capacitance, and crosstalk.

A more challenging approach is the 3D vertical integrated silicon technology, for which a full 
demonstration is also close.

Minimising the support material is critical to the development of a high-performance 
vertex detector. An array of 
low-mass materials such as reticulated foams and silicon-carbide
materials are under consideration. An alternative approach that is being pursued very actively is the
embedding of thinned, active sensors in ultra low-mass media. This line of R\&D
explores thinning active silicon devices to such a thickness that the silicon
becomes flexible. The devices can then be embedded in, for example, Kapton
structures, providing extreme versatility in designing and constructing a vertex
detector.

Power delivery must be accomplished without exceeding the material budget and
overheating the detector.  The vertex detector 
design relies on power pulsing during bunch trains to minimise heating 
and uses forced air for cooling. 

\subsubsection{Main tracker}
\begin{figure}
\begin{center}
\includegraphics[width=0.80\hsize]{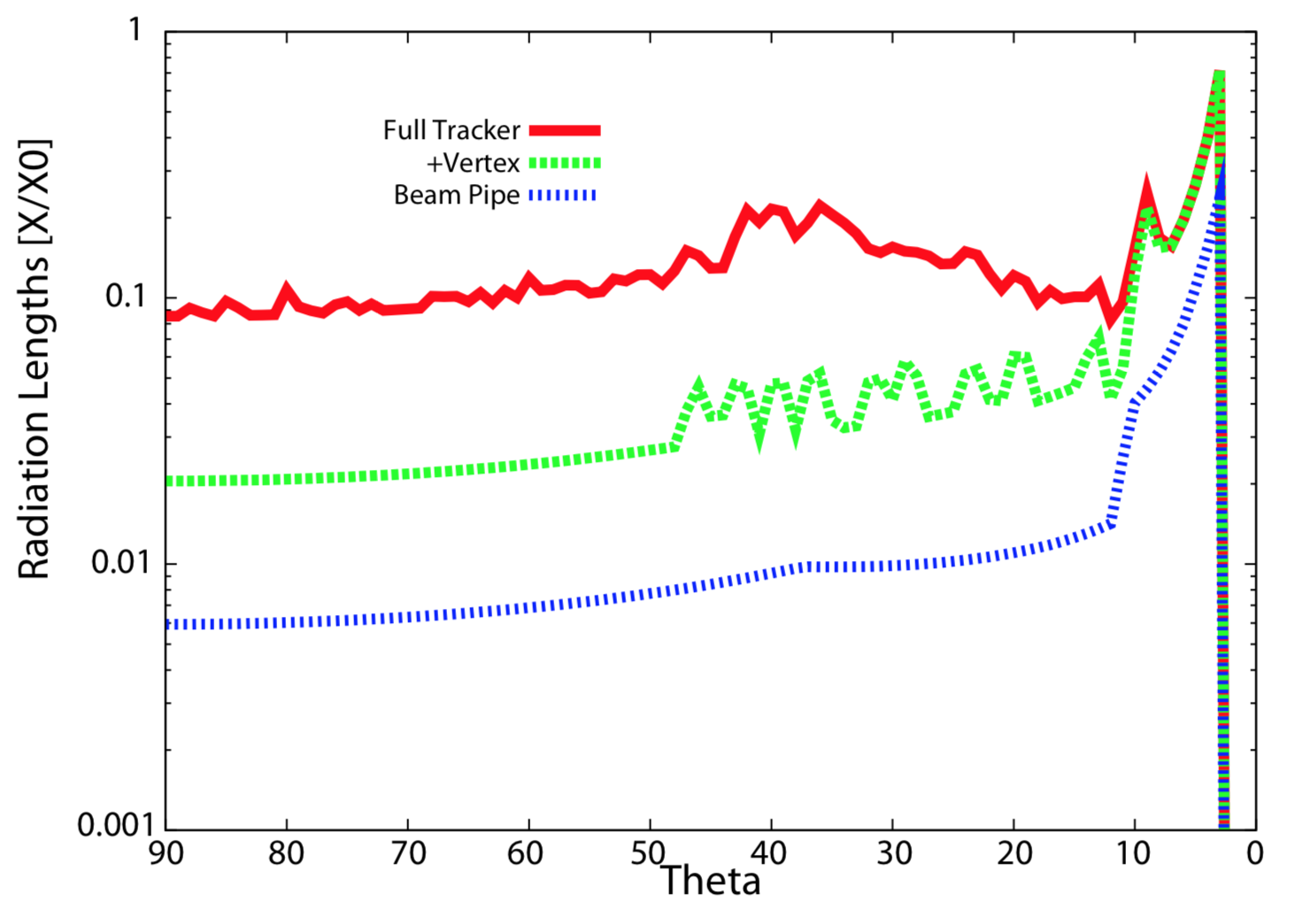}
\end{center}
\caption{Material in the SiD detector, in terms of fractions of a radiation length, as a function of the polar angle.}
\label{fig:sid_mat_budget}
\end{figure}

The main tracker technology of
choice is silicon strip sensors arrayed in five nested cylinders in the central
region and four disks following a conical surface with an angle of 5 degrees
with respect to the normal to the beamline in each of the end regions. The geometry of the endcaps
minimises the material budget to enhance forward tracking. The detectors are
single-sided silicon sensors, approximately 10 $\times$ 10 cm$^2$ with a readout
pitch of 50~\micron. The endcaps utilise two sensors bonded back-to-back for
small angle stereo measurements. With an outer cylinder radius of 1.25~m
and a 5~T field, the charged track momentum resolution will be better than
$\delta (1/\pT) = 5 \times 10^{-5} $/(GeV/$c$) for high momentum tracks with coverage down to polar angles of 10 degrees.
A plot of the momentum budget as a function of polar angle is shown in Fig.~\ref{fig:sid_mat_budget}.

The all-silicon tracking approach has been extensively tested using full Monte-Carlo
simulations including full beam backgrounds. Besides having an excellent momentum resolution
it provides robust pattern recognition even in the presence of backgrounds and has a
real safety margin, if the machine backgrounds will be worse than expected.

\subsubsection{Main calorimeters}

The SiD  baseline design incorporates the elements needed to
successfully implement the PFA approach. This imposes a number of
basic requirements on the calorimetry. The central calorimeter
system must be contained within the solenoid in order to reliably associate
tracks to energy deposits. The electromagnetic and hadronic sections
must have imaging capabilities that allow both efficient
track-following and correct assignment of energy clusters to tracks. These
requirements imply that the calorimeters must be finely segmented both
longitudinally and transversely. In order to ensure that no significant amount
of energy can escape detection, the calorimetry must extend down to small
angles with respect to the beampipe and must be sufficiently deep to prevent
significant energy leakage. Since the average penetration depth of a hadronic
shower grows with its energy, the calorimeter system must be designed for the
highest-energy collisions envisaged.

In order to ease detector construction the calorimeter mechanical design consists of a series of modules of
manageable size and weight. The boundaries between
modules are kept as small as possible to prevent significant non-instrumented
regions. The detectors are designed to have excellent long-term stability and reliability,
since access during the data-taking period will be extremely limited, if not
impossible.

The combined ECAL and HCAL systems consist of a
central barrel part and two endcaps, nested inside the barrel. The entire barrel system is contained
within the volume of the cylindrical superconducting solenoid. 


SiD's reliance on particle flow calorimetry to obtain a jet energy resolution of  $\sim$3\% demands a highly segmented (longitudinally and laterally) electromagnetic calorimeter. It also calls for a minimized lateral electromagnetic shower size, by minimizing the Moliere radius to efficiently separate photons, electrons and charged hadrons~\cite{calor:2018}.

The SiD ECal design employs thirty longitudinal layers, the first twenty each with 2.50 mm tungsten alloy thickness and 1.25 mm readout gaps, and the last ten with 5.00 mm tungsten alloy.  The total depth is 26 radiation lengths, providing good containment of electromagnetic showers.

Simulations have shown the energy resolution for electrons or photons
to be well described by 0.17 / $\sqrt{E}$ $\oplus$ 0.009, degrading a
bit  at higher energies due to changes in sampling fraction and a small leakage.

The baseline design employs tiled, large, commercially produced silicon sensors (currently assuming 15 cm wafers). The sensors are segmented into pixels that are individually read out over the full range of charge depositions. The complete electronics for the pixels is contained in a single chip, the KPiX ASIC~\cite{Brau:2013yb}, which is bump bonded to the wafer. The low beam-crossing duty cycle ($10^{-3}$) allows reducing the heat load using power pulsing, thus allowing passive thermal management within the ECal modules.

Bench tests of the KPiX bonded sensor  with a cosmic ray telescope trigger yielded 
a Landau distribution with a peak of the signal at about 4 fC is consistent with our expectation for minimum-ionizing particles (MIP) passing through the fully-depleted 320 $\mu$m thick sensors. Crosstalk between channels has been managed and the 
 noise distribution shows an RMS of 0.2 fC, well below the 4 fC MIP signal, and exceeding the ECal requirement.

The overall mechanical structure of the ECal barrel has been designed for minimal uninstrumented gaps. Input power and signals are delivered with Kapton flex cables.
The KPiX chip has an average power less than 20 mW, resulting in a
total heat load  that is managed with a cold plate and water pipes routed 
into the calorimeter.

A first SiD ECal prototype stack of nine (of thirty) layers has been constructed and was exposed to a 12.1 GeV electron beam at the SLAC End Station Test Beam Facility. 
This data collection demonstrated good measurements of multiple particle overlap and reconstruction of overlapping showers~\cite{Steinhebel:2017qze}.  Comparison of the deposited energy distribution in each of the nine layers also agrees well with simulations.
An algorithm developed to count the number of incident electrons in each event was used to assess the ability of the calorimeter to separate two showers as a function of the separation of the showers, achieving 100\% for separations of $>$10 mm.

The hadronic
calorimeter has a depth of 4.5 nuclear interaction lengths, consisting of
alternating steel plates and active layers. The baseline choice for the active
layers is scintillator tiles read out via silicon photomultipliers. For this approach SiD is closely following the analog hadron calorimeter developments within the CALICE collaboration. In this context, the simulated HCAL energy resolution has been shown to reproduce well the results from the CALICE AHCAL prototype module exposed to pion beams.

\subsubsection{Forward calorimeters}
\label{subsub:det:forward}
Two special calorimeters are foreseen in the very forward region: LumiCal for a precise luminosity measurement, and BeamCal for the fast estimation of the collision parameters and tagging of forward-scattered beam particles. LumiCal and BeamCal are both compact cylindrical electromagnetic calorimeters centered on the outgoing beam, making use of semiconductor-tungsten technology. BeamCal is placed just in front of the final focus quadrupole and LumiCal is aligned with the electromagnetic calorimeter endcap. 

LumiCal makes use of conventional silicon diode sensor readout. It is a precision device with challenging requirements on the mechanics and position control, and must achieve a small Moliere radius to reach its precision targets. Substantial work has been done to thin the silicon sensor readout planes within the silicon-tungsten assembly. Dedicated electronics with an appropriately large dynamic range is under development.

BeamCal is exposed to a large flux of low-energy electron-positron pairs originating from beamstrahlung. These depositions, useful for a bunch-by-bunch luminosity estimate and the determination of beam parameters, require radiation hard sensors. The BeamCal has to cope with 100\% occupancies, requiring dedicated front-end electronics. A challenge for BeamCal is to identify sensors that will tolerate over one MGy of ionizing radiation per year. Sensor technologies under consideration include polycrystalline chemical vapor deposition (CVD) diamond (too expensive to be used for the full coverage), GaAs, SiC, Sapphire, and conventional silicon diode sensors. The radiation tolerance of all of these sensor technologies has been studied in a high-intensity electron beam. 

For SiD, the main activities are the study of these radiation-hard sensors, development of the first version of the so-called Bean readout chip, and the simulation of BeamCal tagging for physics studies. SiD coordinates these activities through its participation in the FCAL R\&D Collaboration.

\subsubsection{Magnet coil}

The SiD superconducting solenoid is based on the CMS solenoid
design philosophy and construction techniques, using a slightly modified CMS
conductor as its baseline design. Superconducting strand count in the coextruded
Rutherford cable was increased from 32 to 40 to accommodate the higher 5~T
central field. 

Many iron flux return configurations have been simulated in two
dimensions so as to reduce the fringe field. An Opera 3D calculation with the Detector
Integrated Dipole (DID) coil has been completed.
Calculations of magnetic field with a 3D ANSYS program
are in progress. These will have the capability to calculate forces and stress
on the DID as well as run transient cases to check the viability of using the
DID as a quench propagator for the solenoid. Field and force calculations with
an iron endcap HCAL were studied. The field homogeneity improvement was found
to be insufficient to pursue this option. 

Conceptual DID construction and
assembly methods have been studied. The solenoid electrical power system,
including a water-cooled dump resistor and grounding, was established.
Significant work has been expended on examining different conductor stabiliser
options and conductor fabrication methods. This work is pursued as a cost- and
time-saving effort for solenoid construction.

\subsubsection{Muon system}
The flux-return yoke is instrumented with position sensitive detectors to
serve as both a muon filter and a tail catcher. The total area to be
instrumented is very significant -- several thousand square meters. Technologies
that lend themselves to low-cost large-area detectors are therefore under
investigation. Particles arriving at the muon system have seen large amounts of
material in the calorimeters and encounter significant multiple scattering
inside the iron. Spatial resolution of a few centimetres is therefore
sufficient. Occupancies are low, so strip detectors are possible. The SiD 
baseline design uses scintillator technology, with RPCs as an alternative. 
The scintillator technology uses extruded scintillator readout with wavelength 
shifting fibre and SiPMs, and has been successfully demonstrated. 
Simulation studies have shown that nine or more layers of sensitive detectors 
yield adequate energy measurements and good muon detection efficiency and purity.
The flux-return yoke itself has been optimised with respect to the
uniformity of the central solenoidal field, the external fringe field,
and ease of the iron assembly. 
This was achieved by separating the  barrel and end sections of the
yoke along a 30 degree line.

\subsubsection{The machine-detector interface}
A time-efficient implementation of the push-pull model of
operation sets specific requirements and challenges for many detector and
machine systems, in particular the interaction region (IR) magnets, the
cryogenics, the alignment system, the beamline shielding, the detector design
and the overall integration. The minimal functional requirements and interface
specifications for the push-pull IR have been successfully developed and
published~\cite{Parker:2009zz,Buesser:2012et}.  All further IR design
work on both the detectors and machine sides are constrained by these 
specifications. 

\subsection{The ILD detector}
The ILD detector  has been developed by a proto-collaboration with the goal to develop and eventually propose a fully integrated detector for the ILC. 

\begin{figure}
    \centering
    \includegraphics[width=0.35\textwidth]{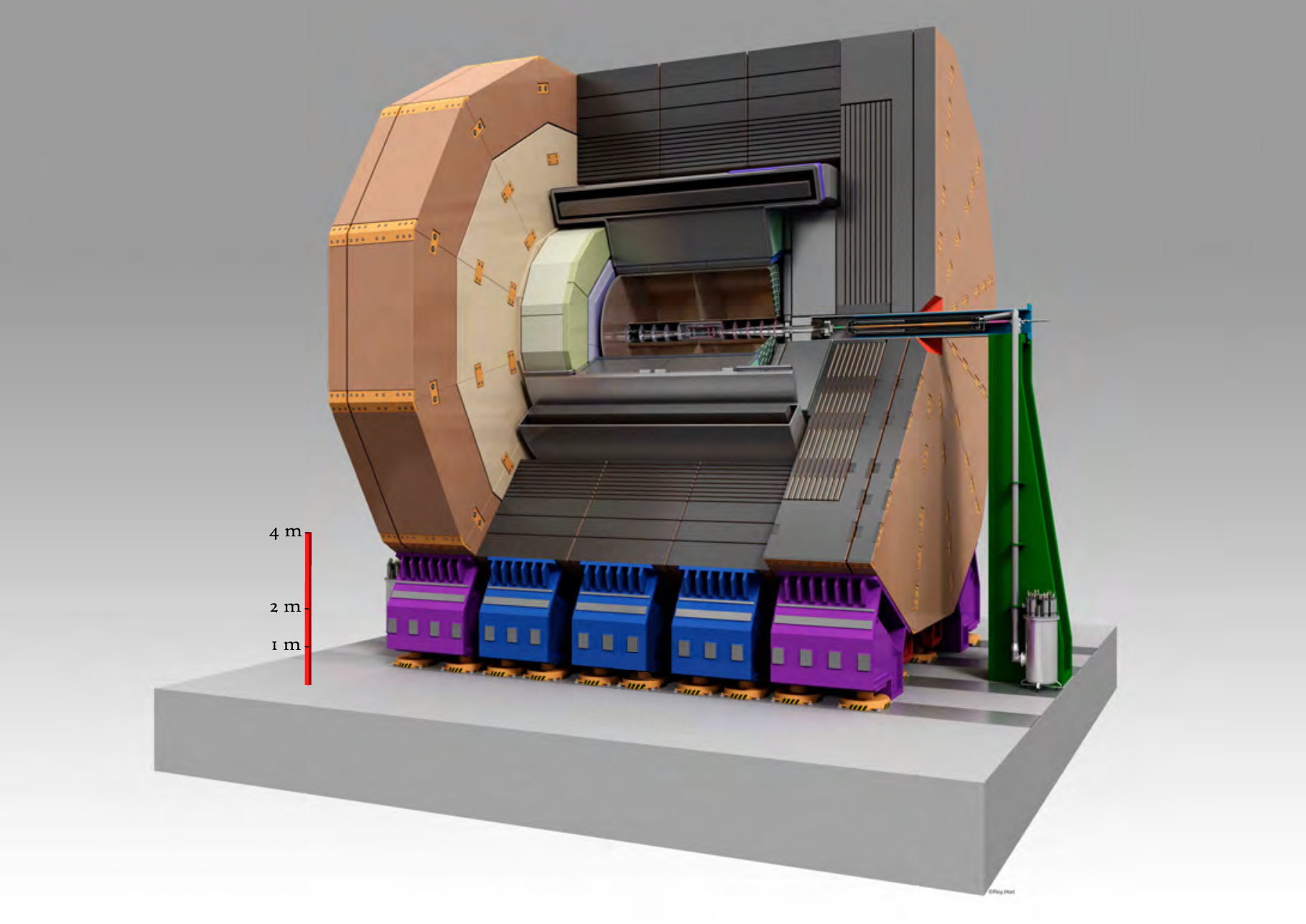}
    \caption{3D-picture of the ILD detector.}
    \label{fig:ild_3d}
\end{figure}

The ILD detector concept has been designed as a multi-purpose detector. It should deliver excellent physics performance for collision energies between 90~GeV and 1~TeV, the largest possible energy reach of the ILC. The ILD detector has been optimized to perform excellently at the initial ILC energy of 250 GeV (for more details see ~\cite{Abe:2010aa,Behnke:2013lya}). An artist's view of the ILD detector is shown in Fig.~\ref{fig:ild_3d}. 

The science which will be done at the ILC requires a detector that
truly covers all aspects of the $\ee$ events.  The tracking philosophy
is very different from that of SiD, as will be discussed in a moment.
However, similarly to SiD, the ILD detector has been designed to combine the traditional
precision detector 
elements such as as vertex detectors and trackers in an overall design
philosophy 
that optimizes jet reconstruction using particle flow.

\subsubsection{Vertexing and tracking}
\label{subsubsec:ILDtracker}
The high precision vertex detector positioned very closely to the
interaction point is followed by a hybrid tracking layout, realised as
a combination of silicon tracking with a time projection chamber, and
a calorimeter system. The complete system is located inside a large
solenoid providing a magnetic field of 3.5-4 T. On the outside of the
coil, 
the iron return yoke is instrumented as a muon system and as a tail catcher calorimeter. 

The vertex detector is realised as a multi-layer pixel-vertex detector
(VTX), with three super-layers, each comprising two layers. The
detector
 has a pure barrel geometry. To minimise the occupancy from background hits,
the first super-layer is only half as long as the outer two. Whilst the underlying detector technology has not yet been decided, 
the VTX is optimised for point resolution and
 minimum material thickness. 
	
A system of silicon strip and pixel detectors surrounds the VTX detector. In the barrel, two layers of silicon strip detectors (SIT) are arranged to bridge the gap between the VTX and the TPC. In the forward region, a system of two silicon-pixel disks and five silicon-strip disks (FTD) provides low angle tracking coverage.

A distinct feature of ILD is a large volume time projection chamber
(TPC) with up to 224 points per track. The TPC is optimised for
3-dimensional point resolution and minimum material in the field cage
and in the end-plate. It also allows d$E$/d$x$-based particle
identification. At the ILC,  a TPC has a number of specific strengths
 which make this type of detector attractive. A time projection
 chamber offers true three-dimensional points, and offers many of
 those along a charged particle trajectory. The intrinsic disadvantage
 of a TPC, its slow readout speed, does not harm the
 performance at the ILC, since the time between bunches is relatively
 long, around 300~ns. On the other hand the large number of
 points offer superb pattern recognition capabilities, and allows the
 detailed reconstruction of kinks or decays in flight within its
 volume. This can be achieved at a very low material budget, rather
 uniformly distributed over the sensitive volume. The excellent
 performance of the system is particularly striking at low momenta, at
 a few GeV and below, where the combination of three dimensional
 reconstruction and low material 
allows the efficient and precise reconstruction of tracks. 

Outside the TPC, a system of Si-strip detectors in between the TPC and
the ECAL (SET), provide additional high precision space points which
 improve the tracking performance and provide additional
    redundancy in the regions between the main tracking volume and the
    calorimeters. 

\begin{figure}
\centering
\includegraphics[width=0.85\hsize,viewport={0 -10 600 500},clip]{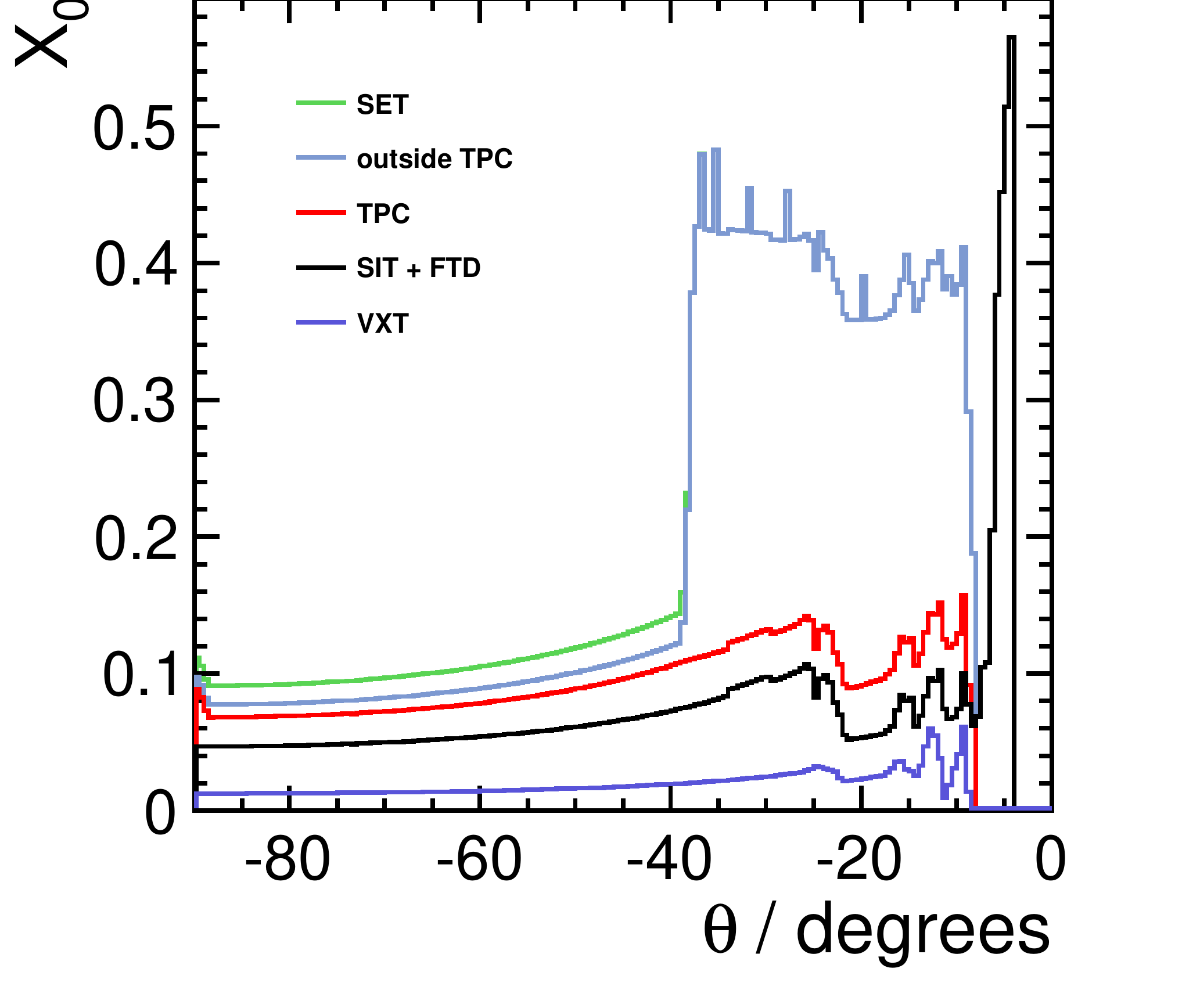}
\caption{Material in the ILD detector, in terms of fractions of a radiation length, as a function of the polar angle.}
\label{fig:ILD_mat_budget}
\end{figure}

A key aspect of the ILD detector design is the low mass of the tracking system. The total material as a function of angle, in radiation lengths, is shown in Fig.~\ref{fig:ILD_mat_budget}.

\subsubsection{Calorimetry}

A highly segmented electromagnetic calorimeter (ECAL) provides up to
30 samples in depth and small transverse cell size, split into a
barrel and an end cap system. For the absorber, Tungsten has been
chosen; for the sensitive area, silicon diodes 
or scintillator strips are considered.



This is followed by a segmented hadronic calorimeter (HCAL) with up to 48 longitudinal samples and small transverse cell size. Two 
options are considered, both based on a steel-absorber structure. One option uses scintillator tiles of $3 \times 3$\,cm$^2$, 
which are read out with an analogue system. The second uses a gas-based readout which allows a $1 \times 1$\,cm$^2$ 
cell geometry with a semi-digital readout of each cell. 

At very forward angles, below the coverage provided by the ECAL and
the HCAL, a system of high precision and radiation hard calorimetric
detectors (LumiCAL, BeamCAL, LHCAL)
 is foreseen. The LumiCAL and BeamCAL are based on technologies developed in the context of the FCAL collaboration. These detectors 
extend the calorimetric coverage to almost $4\pi$, measure the
luminosity, and  monitor
 the quality of the colliding beams. the LHCAL system bridges the electromagnetic endcap calorimeter with the forward systems. 

\subsubsection{Coil and yoke}
A large volume superconducting coil surrounds the calorimeters, creating an axial $B$-field of nominally 3.5-4\,Tesla.

An iron  yoke, instrumented with scintillator strips or resistive
plate chambers (RPCs), returns the magnetic flux of the solenoid, and,
at the same
 time, serves as a muon filter, muon detector and tail catcher calorimeter.

\subsubsection{Detector integration and performance}
The ILD detector is designed to operate in the ILC interaction region with a push-pull scheme, allowing the rapid interchange of ILD with SiD. Detailed studies have been done to understand the impact this scheme might have on the detector and its design. In addition the ILD detector is optimised for operation in the seismic active region in the north of Japan. Extensive simulation studies for the main components have shown that the detector is stable against seismic events.

Key plots to evaluate the projected performances of the ILD and SiD
detectors will be presented in the following chapter.    These plots
will also illustrate the successive stages of event reconstruction
from raw data and will describe the level of detail that we have
considered in making these estimates of performance.  
As we have already noted, the ILD and SiD detectors include many
technologies that have been developed in close cooperation with R\&D
collaborations 
and have been extensively tested. For both detectors, the performance numbers 
of key systems are based on results from prototypes, wherever 
possible, and extrapolated to the full detector performance. This
strong 
check against experimental results ensures that the performance 
numbers are reliable and are considered a realistic estimate of the ultimate detector performance.

\section{\label{sec:software} Computing, Event Reconstruction, and
  Detector Performance}

   %
%
\newcommand{\fix}[1]{\textcolor{red}{\texttt{#1}}} 

\newcommand{\CPP}{C\nolinebreak\hspace{-.05em}\raisebox{.4ex}{\tiny\bf
    +}\nolinebreak\hspace{-.10em}\raisebox{.4ex}{\tiny\bf +}}

This section will describe the software framework used for ILC event
analysis, working from raw data or digitized simulation data to
physics objects.   We will first describe the core software tools used
by the detector groups.  We will then follow the path by which this
software to used to provide detailed detector models and model data
sets,   to reconstruct the data including as much realism as possible,
and to produce the final physics objects for analysis.    At the
successive stages of this process, we will illustrate the intermediate results 
with  performance plots that also can be used to benchmark the
detector models.  Finally, we will discuss the computing concept and
costs for the ILC experiments. 


More than 15 years ago the linear collider community started to develop common software
tools to facilitate the development and optimization of detector concepts based on realistic
simulations of physics interactions. These software tools eventually led to the creation of
a common software ecosystem called \emph{iLCSoft}~\cite{bib:ilcsoft}.
The \emph{iLCSoft} tools are used by both ILC detector concepts as well as by CLIC
and partly by CEPC and FCC.

From the start, a strong emphasis has been placed on developing flexible and generic tools
that can easily be applied to other experiments or new detector concepts. 
This approach of developing common tools wherever possible has helped considerably in
leveraging the limited manpower and putting the focus on algorithm development that
is crucial for the physics performance. 


\subsection{\label{sub:sw-core-tools}Core software tools}

\begin{figure}
\begin{center}
\includegraphics[width=0.60\hsize]{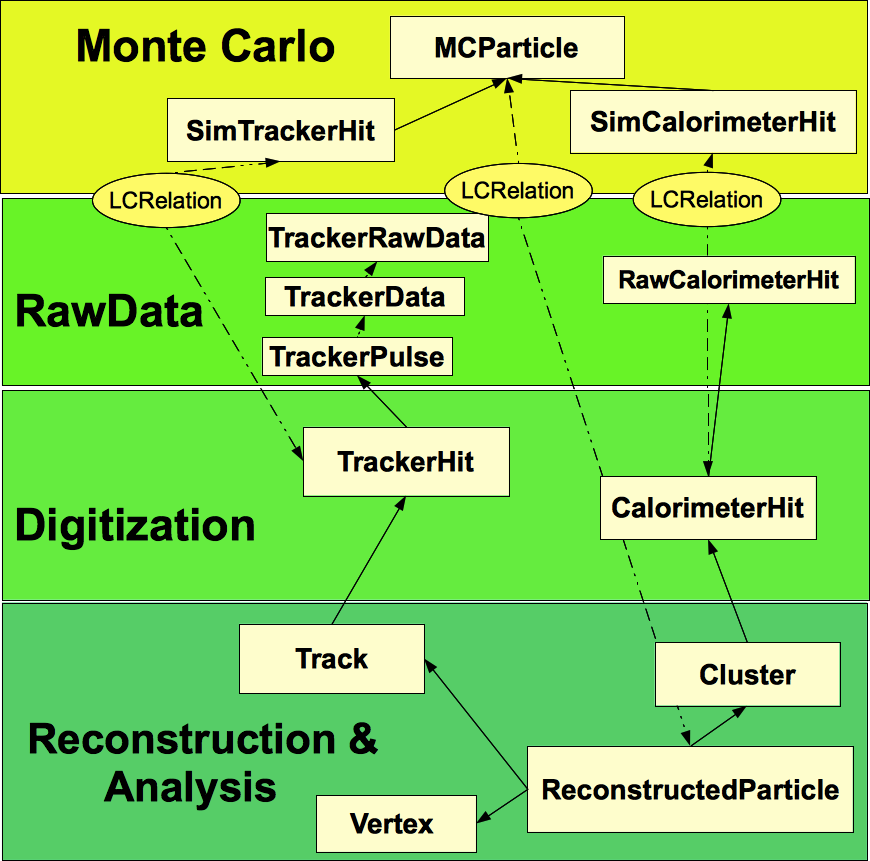}
\end{center}
\caption{Schematic view of the hierarchical EDM in LCIO.}
\label{fig:lcio_edm}
\end{figure}

The foundation for the development of common software was laid with LCIO~\cite{Gaede:2003ip}, the
event data model (EDM) and persistency tool for linear collider studies. At the core of LCIO is a hierarchical
EDM for any particle physics experiment, as shown in Fig.~\ref{fig:lcio_edm}.
It provides data classes for all phases of the event processing,
starting from Monte Carlo truth information, continuing to generate
raw data and digitization, and processing this  to the final reconstruction and analysis. Objects at higher levels of the processing
point back to the lower level constituting objects. As a specific design decision, there are no pointers back to the
Monte Carlo truth but these can be added if needed using dedicated generic LCRelation objects.
These relation objects can be used to create many-to-many relations between arbitrary types in the EDM.
A special class LCGenericObject holds user defined data in named vectors of types int, float and double.
This feature is used in many test beams for conditions data and raw data from the DAQ.
LCIO provides APIs in \CPP, Java and Fortran, but today \CPP\ is used almost exclusively.

The \CPP\ application framework Marlin~\cite{Gaede:2006pj} provides an easy to use environment for developing software
modules on all levels of processing and uses LCIO as its transient data format, i.e. all data that is read in or created
by a software module (called \emph{Processor}) are stored in the \emph{LCEvent} class from LCIO. Marlin processors
are self-documenting and controlled via xml-steering files. As processors have well defined input and output data, Marlin
provides a \emph{"Plug-And-Play"} environment, where any specific algorithm can easily be exchanged with another
equivalent implementation for direct comparisons and benchmarking.

\begin{figure}
\begin{center}
\includegraphics[width=0.90\hsize]{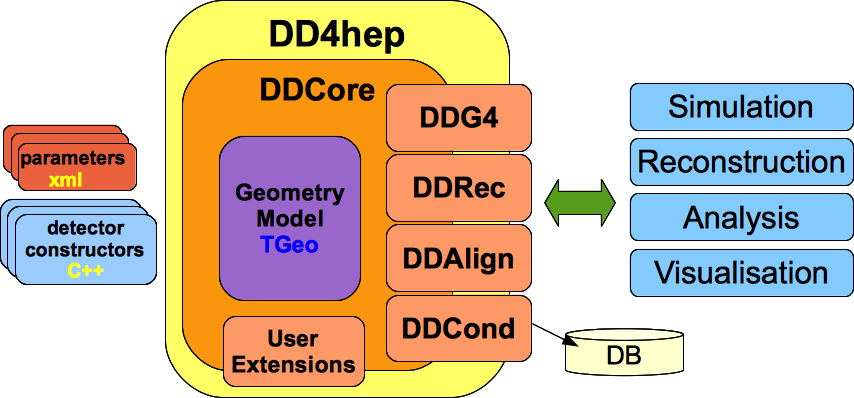}
\end{center}
\caption{Schematic view of DD4hep with its main components and interfaces.}
\label{fig:dd4hep}
\end{figure}

The generic detector description toolkit DD4hep~\cite{Frank:2014zya,Frank:2015ivo} provides a powerful tool for describing
the detector geometries, materials and readout properties. DD4hep follows a modular component based approach and provides
interfaces to full simulations with GEANT4~\cite{Agostinelli:2002hh} via DDG4, to reconstruction programs via DDRec and to
conditions data and alignment with DDCond and DDAlign respectively, see Fig.~\ref{fig:dd4hep}.
DD4hep is an excellent example for the development of generic software tools for the wider HEP community and was one of the
first incubator projects adopted by the Hep Software Foundation. While it was developed to address the needs of the linear
collider community, it is now used by several other projects and is under evaluation by LHC experiments.

\subsection{\label{sub:sw-generators}Event generators}

Both detector concepts have created large, realistic Monte Carlo samples with the full Standard Model physics as well as various
BSM scenarios that have been used for the physics analyses presented in the following sections.
In a first step, large generator samples with $e^+e^-$ 
events are created with the Whizard~\cite{Kilian:2007gr} event generator.
Whizard uses tree-level matrix elements and loop corrections to generate events with the final state partons and leptons
based on a realistic beam energy spectrum, the so called \emph{hard sub-process}. The hadronization into the visible final state
is performed with Pythia~\cite{Sjostrand:2006za} tuned to describe the LEP data.

\begin{figure}
\begin{center}
\includegraphics[width=1.\hsize]{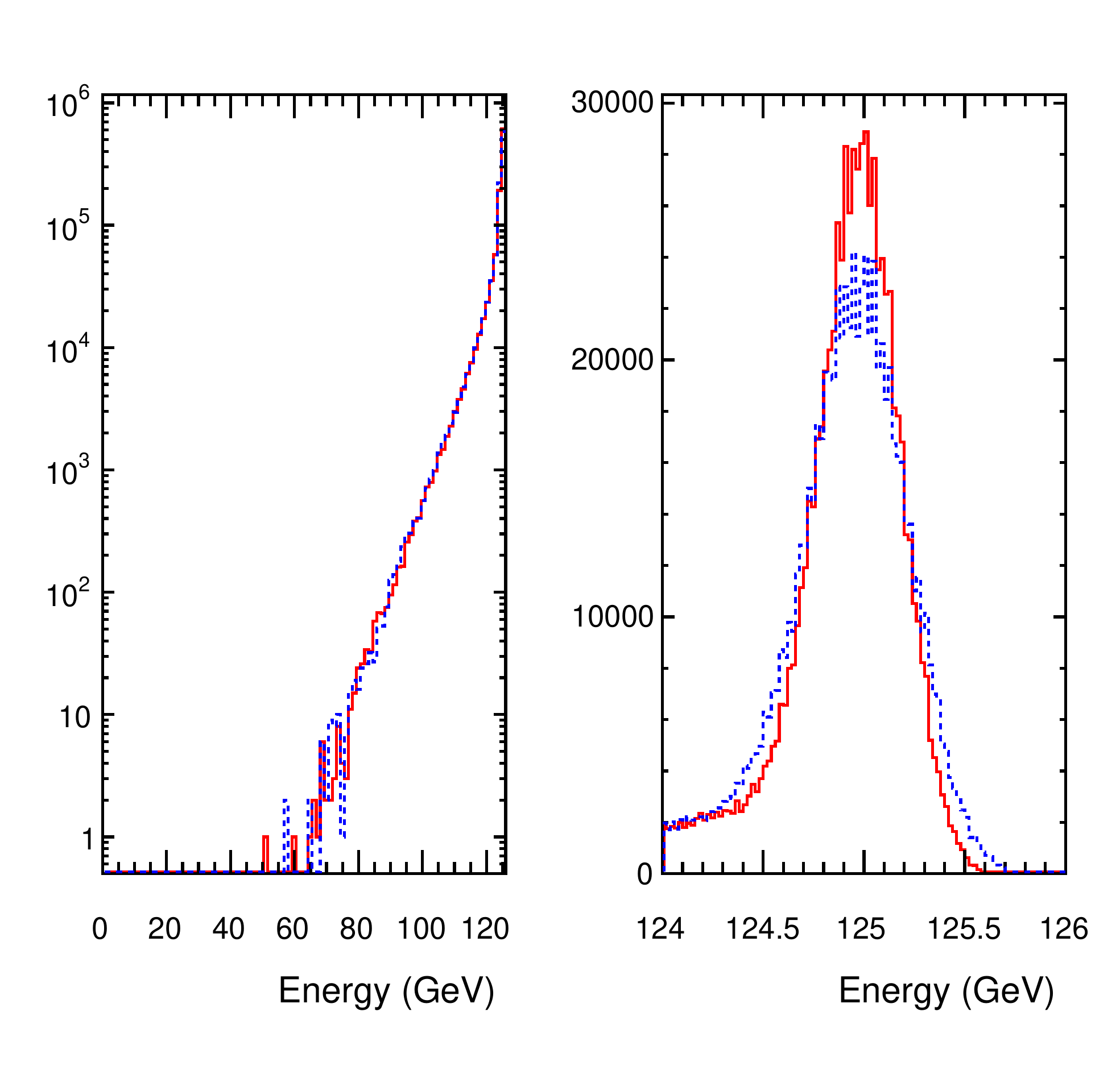}
\end{center}
\caption{Beam energy spectra for $\sqrt{s}=250~\rm{GeV}$ Set-A, created with GuineaPig (blue-dashed: $e^-$, red-solid $e^+$).}
\label{fig:lumi_spectrum}
\end{figure}

\begin{figure}
\begin{center}
\includegraphics[width=0.90\hsize]{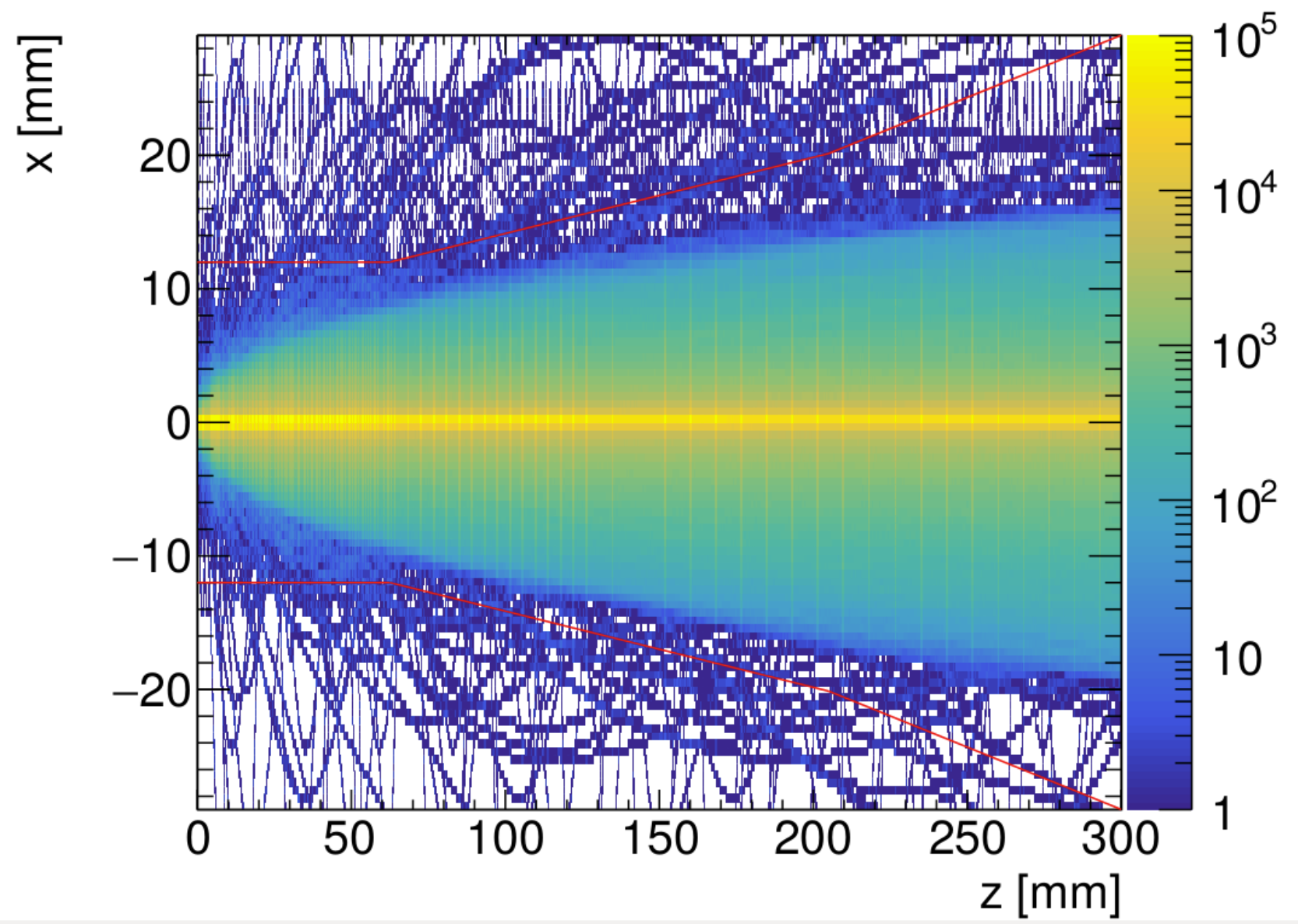}
\end{center}
\caption{Cone of background from incoherent $e^+e^-$-pairs, generated with Guinea-Pig and simulated in the 5 T B-field of the SiD
  detector (from~\cite{Schutz:2017ihd}).}
\label{fig:pair_bg}
\end{figure}

The input spectrum is created with Guinea-Pig~\cite{Schulte:1998au}, a dedicated simulation program for computing
beam-beam interactions at linear colliders. The two dominating effects of the strong beam-beam interactions are 
beamstrahlung, leading to the available luminosity spectrum (see Fig.~\ref{fig:lumi_spectrum}), and the creation of
incoherent $e^+e^-$-pairs that are the source of the dominant
background at the ILC. 
These electrons and positrons
are predominantly created in a forward cone as shown in Fig.~\ref{fig:pair_bg}. It is this cone that restricts the minimal
allowed radius of the innermost layer of the vertex detector.

Another source of background at the ILC are $\gamma \gamma \rightarrow hadrons$ events, due to bremsstrahlung and beamstrahlung photons.
These types of events are generated for $\gamma \gamma$ cms-energies from $300$~MeV to 2~GeV with a dedicated generator based
on Ref.~\cite{Chen:1993dba}; for higher energies Pythia is used.

\subsection{\label{sub:sw-sim}Simulation}

Both detector concepts have adopted DD4hep for describing their detector simulation models and use \emph{ddsim}, a python application that
is based on the DDG4 component, to provide a gateway to full simulations with GEANT4.
In DD4hep the detector geometry is implemented in dedicated \CPP\  modules for every subdetector and the actual parameters with dimensions
and materials are provided via compact xml-files. DD4hep contains a
 large palette of predefined sub-detector drivers, allowing for an
easy implementation of a new detector concept by providing suitable compact files.
A dedicated software package lcgeo~\cite{bib:lcgeo}, which is shared by SiD, ILD and CLICdp, contains all subdetector drivers for the
detector concepts under study by these groups, together with the corresponding compact parameter files.

\begin{figure}
\begin{center}
\includegraphics[width=0.80\hsize]{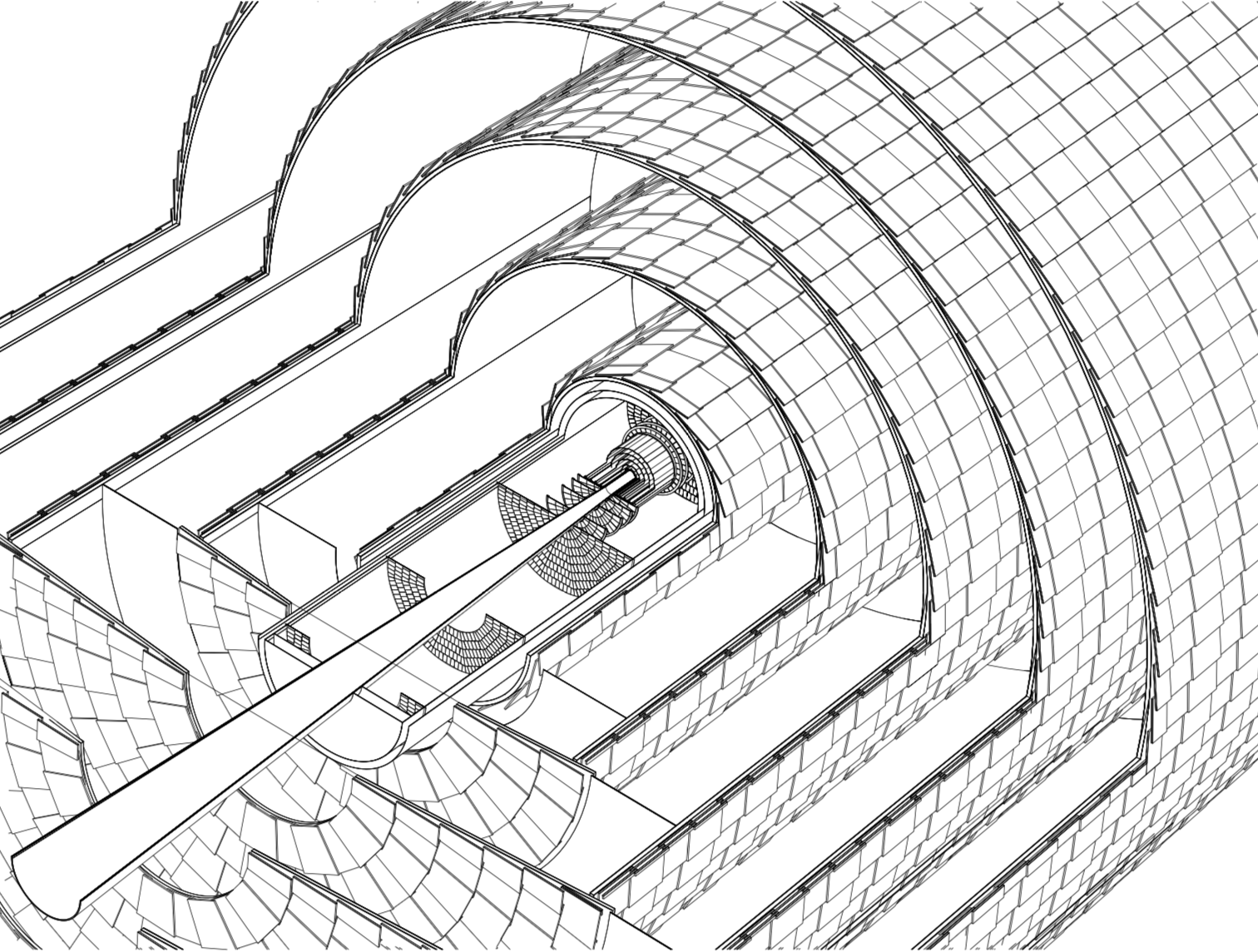}
\end{center}
\caption{Cut-away view of the tracking system as implemented in the \emph{SIDLOI3} simulation model (from~\cite{Behnke:2013lya}).}
\label{fig:sid_trk}
\end{figure}
Both detector concept groups have invested considerable effort into making their full-simulation models as realistic as possible, by
\begin{itemize}
\item following the exact dimensions and layout of detector elements from engineering models
\item implementing correct material properties
\item implementing precise descriptions of the actual detector technology
\item adding realistic amounts of dead material from supports and services, such as cables and cooling pipes
\item introducing realistic gaps and imperfections into the subdetectors
\end{itemize}
Care has been taken to include realistic material estimates in particular in the tracking region where
the material budget has a direct impact on the detector performance.
Figure~\ref{fig:sid_trk} shows the tracking detector as implemented for the SiD simulation model.
The average material budget in the tracking volume of the simulation models has already been shown in Sec.~\ref{sec:detectors}, Figs.~\ref{fig:sid_mat_budget}
and~\ref{fig:ILD_mat_budget}, for SiD and ILD respectively.

Before the two concepts had decided to move to the common geometry description and simulation with DD4hep, they had
implemented their detailed simulation models in Mokka~\cite{MoradeFreitas:2002kj} and slic~\cite{bib:slic}. These models
have been ported into DD4hep preserving all features and dimensions, thus resulting in equivalent simulation results.
Most of the physics analyses in the next sections are based on simulations using these older programs.

The high level of detail in the simulation models as described above is a key prerequisite for the
realistic understanding of the expected detector performance and the physics reach of the ILC for both detector concepts.

\subsection{\label{sub:sw-digi}Digitzation}

The output of the detailed full simulations with GEANT4 from 
ddsim are \emph{SimTrackerHit} and \emph{SimCalorimeterHit} objects.
These store the deposited energy in the sensitive detector elements, such as silicon wafers and calorimeter cells, together with
the position and pointers to the \emph{MCParticle} that created the energy deposition. In the digitization step, carried out in dedicated
Marlin processors, these hits are converted into \emph{TrackerHit} and \emph{CalorimeterHit} objects,
taking into account all relevant effects from the detector and the readout electronics.

The \emph{SimTrackerHits} contain the exact energy-weighted position of the individual energy depositions in a given sensitive
detector element. For silicon strip-and pixel detectors as well as the ILD-TPC, these positions are smeared according to
resolutions that have been established from test beam campaigns for the different sensor technologies, thereby including effects
from charge sharing, clustering and position reconstruction. Table~\ref{tab:ild_trk_res} shows the point resolution parameters used for ILD.

\begin{table}[tbp]
\renewcommand{\arraystretch}{1.25}

\centering\small
\begin{tabular}{llcl}
\hline
 Subdetector &  \multicolumn{3}{c}{ Point Resolution }  \\
\hline
        VTX    &  $ \sigma_{r\phi,z}  $ & $=$ & $ 2.8 \mu\mathrm{m}$   (layer 1)   \\
               &  $ \sigma_{r\phi,z}  $ & $=$ & $ 6.0 \mu\mathrm{m}$   (layer 2)   \\
               &  $ \sigma_{r\phi,z}  $ & $=$ & $ 4.0 \mu\mathrm{m}$   (layers 3-6)   \\

        SIT    &  $ \sigma_{\alpha_{z}}   $ & $=$ & $ 7.0 \mu\mathrm{m}$    \\
               &  $  \alpha_{z}         $ & $=$ & $ \pm 7.0^\circ $ (angle with z-axis)        \\

        SET    &  $ \sigma_{\alpha_{z}}   $ & $=$ & $ 7.0 \mu\mathrm{m}$    \\
               &  $  \alpha_{z}         $ & $=$ & $ \pm 7.0^\circ $ (angle with z-axis)        \\

       FTD     &  $\sigma_{r}$      & $=$ & $ 3.0 \mu\mathrm{m}$    \\
  \emph{Pixel} &  $ \sigma_{r_\perp}$  & $=$ & $ 3.0 \mu\mathrm{m}$    \\

     FTD       &  $ \sigma_{\alpha_r}   $ & $=$ & $ 7.0 \mu\mathrm{m}$    \\
  \emph{Strip} &  $ \alpha_{r}         $ & $=$ & $ \pm 5.0^\circ $ (angle with radial direction)        \\

       TPC    &  $ \sigma^2_{r\phi} $ & $=$ & $ \bigl( 50^2+900^2\sin^2\phi + \bigl( (25^2/22)\times$  \\
              &                      &     &   $(4T/B)^2\sin\theta\bigr) (z/\mathrm{cm}) \bigr)\,\mu\mathrm{m}^2$  \\
               &  $ \sigma^2_{z}    $ & $=$ & $ (400^2+80^2\times (z/\mathrm{cm})) \,\mu\mathrm{m}^2 $ \\
               &   \multicolumn{3}{c}{ where $\phi$ and $\theta$ are the azimuthal and} \\
               &   \multicolumn{3}{c}{ polar angle of the track direction } \\
\hline
\end{tabular}
\caption[Simulated ILD tracking point resolutions.]{Effective point resolutions as used in the digitization of the ILD tracking detectors.
  The parameterization for the TPC takes into account geometric effects due to the direction of the track with respect to the pad row and
  has been established from test beam data.
        \label{tab:ild_trk_res} }
\end{table}

In the TPC hit digitization, simulated hits that are closer than the established double-hit resolution of 2~mm in $r\phi$ and 5~mm
in $z$ are merged into one. For the silicon detectors this treatment is not necessary, due to the expected low occupancies.

The \emph{SimCalorimeterHits} contain the total energy deposited in each calorimeter cell, together with the individual depositions
from the individual Monte Carlo steps. For scintillating calorimeters
Birk's Law is already applied during the simulation, resulting in different light yields for different
particles.  Dedicated digitizers take into account effects of non-uniformity of the light yield
for scintillators as well as cross-talk between neighboring channels. The latter is important in particular for the simulation of
(semi)-digital calorimeters using RPCs and is possible due to the availability of the individual simulation steps, containing
the exact position of the energy deposition.

During the calorimeter digitization, a two step calibration is applied for every calorimeter type and sampling structure. In a first step
the hits are calibrated to a MIP signal and in a second step, the total energy is calibrated to an absolute value of the
cell energy in  GeV. This calibration is an iterative procedure, based on the application of the full
\emph{particle flow algorithm }  to single particle events with photons and $K^0$s and thereby
repeatedly adjusting the calibration constants.

\subsection{\label{sub:sw-reco}Reconstruction}

\subsubsection{Tracking} 

The first step of the event reconstruction consists of identifying the trajectories of charged particles based on the positions of their
energy depositions in the detector (\emph{SimTrackerHits}), typically referred to as \emph{pattern recognition}. In a second step the
kinematic parameters of these trajectories are fitted based on the known equations of motion in a magnetic field and the errors of the
hit positions. Often both steps are carried out together, \eg, by using a Kalman-Filter and simply referred to as \emph{Tracking}.

\begin{figure}
\begin{center}
\includegraphics[width=0.80\hsize]{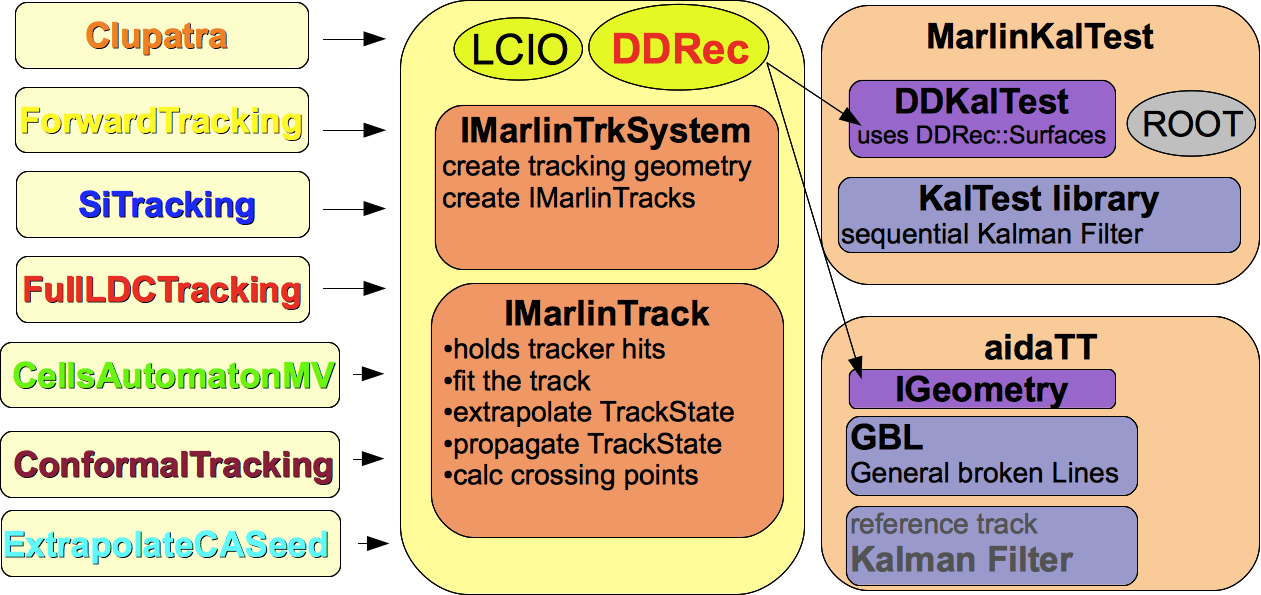}
\end{center}
\caption{Schematic view of the MarlinTrk tracking tools available in iLCSoft. They are based on the LCIO event data model and the
DDRec geometry description.}
\label{fig:imarlintrk}
\end{figure}

The tracking packages in iLCSoft is called MarlinTrk and provides a generic tracking-API \emph{IMarlinTrk} and underlying fitting code,
using the Kalman-Filter package \emph{KalTest}~\cite{Li:2013cxa}.
The \emph{IMarlinTrk} interface provides code to iteratively add hits to a track segment,
thereby updating the track parameters, extrapolation of the current track state to the next measurement surface or any given point
in space. It uses LCIO as data model for the Track and TrackState with a \emph{perigee} track parameterization with
track curvature $\omega$, impact parameters $d_0$ and $z_0$ and direction parameters $\phi_0$ and $\rm{tan}(\lambda)$.
A palette of different pattern recognition algorithms are programmed against \emph{IMarlinTrk} as shown in Fig.~\ref{fig:imarlintrk}.
ILD uses the following different algorithms in the different parts of the tracking region
(for more details see Ref.~\cite{Gaede:2014aza}):

\begin{itemize}
\item SiliconTracking\\
  Algorithm used in the innermost Si-tracking detector VXD and SIT,
  based on a brute-force triplet seeding followed by
  a road search using the extrapolation to the next layer provided in MarlinTrk.
\item ForwardTracking\\
  Stand alone pattern recognition in the FTD forward tracker using a
  Cellular-Automaton to find a 
(possibly large) set of
  track candidates that are reduced to a unique and consistent set through the use of a Hopfield Network.
\item Clupatra\\
  Pattern recognition algorithm for the TPC, based on topological clustering in the outer TPC pad row layers for seeding,
  followed by a Kalman-Filter based road search inwards.
\item FullLDCTracking\\
  A collection of algorithms for merging track segments from the previous algorithms and assignments of leftover hits followed
  by a final re-fit using a Kalman-Filter.
\end{itemize}

SiD had originally developed their stand-alone tracking software in the Java framework \emph{LCSim}~\cite{bib:LCSim}
using a triplet based seeding followed by a road search and a final track fit. More recently SiD has adopted the \emph{ConformalTracking}
algorithm originally developed for CLICdp. It uses a conformal mapping transforming circles going through the origin (IP)
into straight lines which are then identified using a Cellular-Automaton.

The correct reconstruction of the kinematics of charged particles requires a sufficiently detailed description of the material
the particles have traversed, in order to correctly account for effects of energy-loss and multiple-scattering in the fit.
The DD4hep component DDRec provides dedicated surface classes for track reconstruction and fitting. These surface classes provide the
geometric information of the corresponding measurement surfaces as well as material properties, averaged in a suitable way. Surfaces
are also used to account for effects from dead material layers, such as support structures or cables and services.

\begin{figure}
  \begin{tabular}[c]{c}
    \includegraphics[width=0.85\hsize]{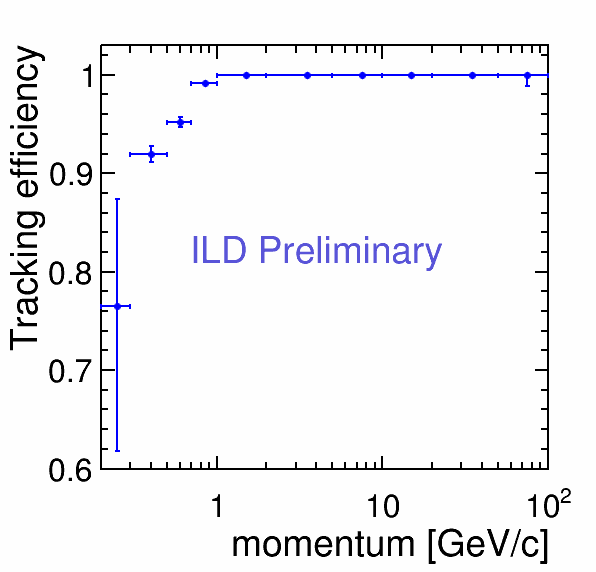} \\
    \includegraphics[width=0.85\hsize]{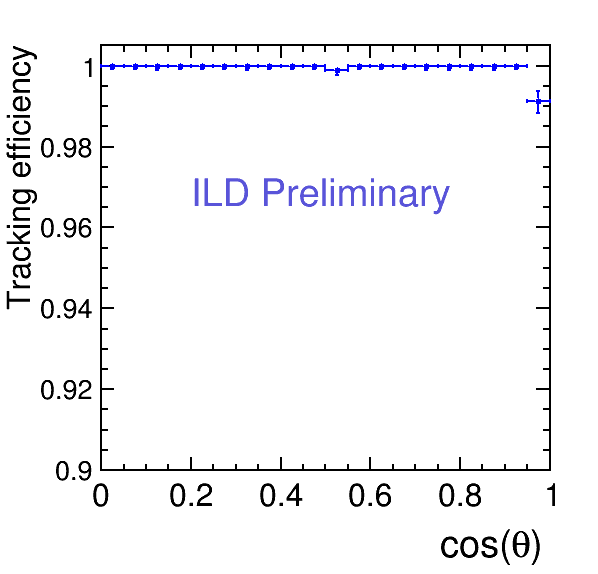}
\end{tabular}
  \caption{Tracking efficiency for $t\bar t$-events at $\sqrt{s}=500~\rm{GeV}$ in the ILD detector as a function of
    momentum ($\rm{cos}(\theta)>.99$) [upper] and $\rm{cos}(\theta)$ ($p>1~\rm{GeV}$) [lower] for prompt tracks
    ($d_{IP}<10~\rm{cm}$). Decays in flight are excluded and tracks are required to have left at least 4 hits in the detector.
  Background from $e^+e^-$-pairs for two bunch crossings is overlaid to the $t\bar t$-events.}

\label{fig:ild_trkeff}
\end{figure}

The resulting tracking efficiencies for the ILD detector are shown as a function of the momentum and
$\rm{cos}(\theta)$ in Fig.~\ref{fig:ild_trkeff}.

\begin{figure}
\begin{center}
\includegraphics[width=0.95\hsize]{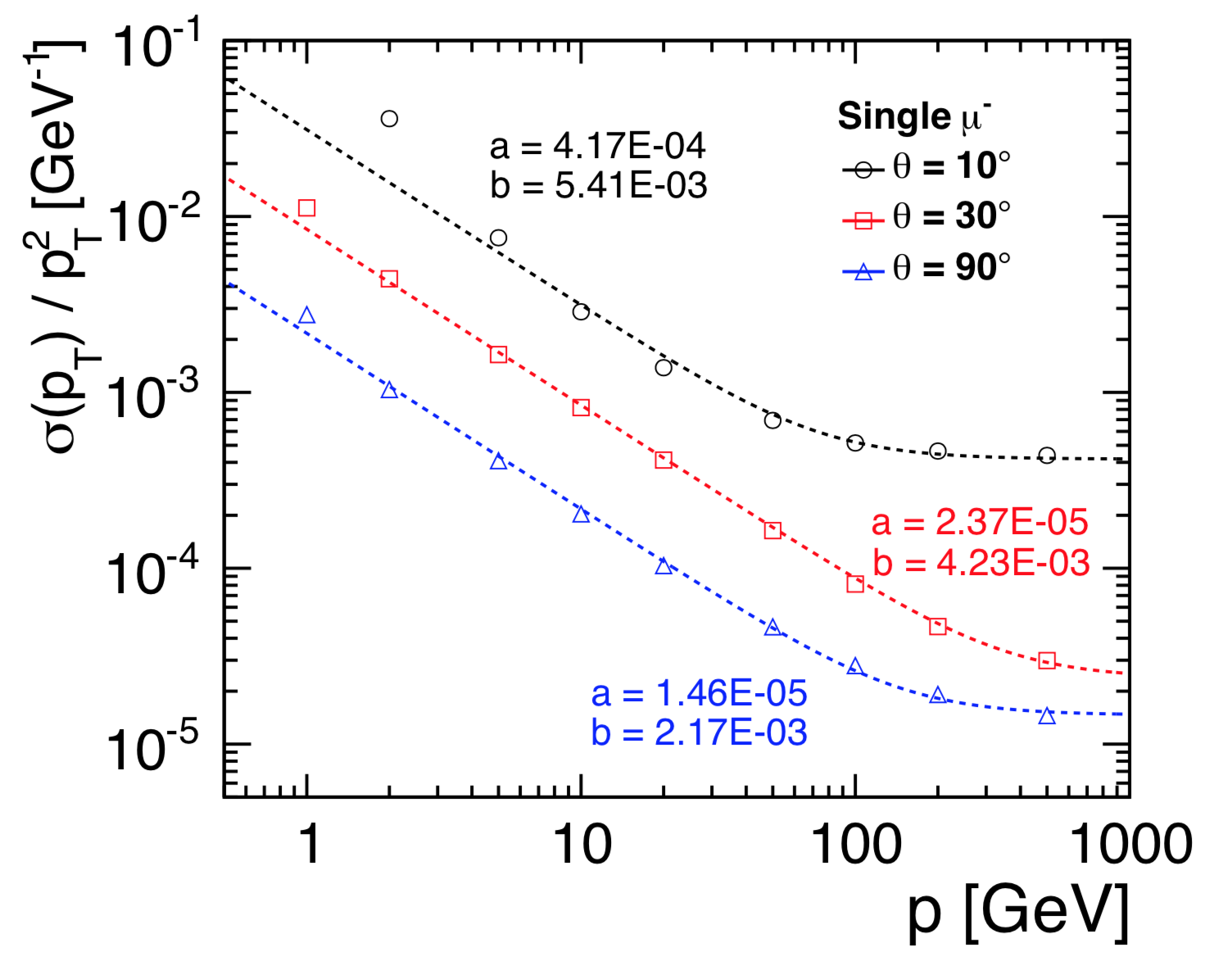}
\end{center}
\caption{Normalised transverse momentum resolution for single-muon events as function of momentum
  in the \emph{SIDLOI3} simulation model (from~\cite{Behnke:2013lya}). The dashed lines are fits to the data points according
  to eq.~\ref{eq:trk_res_param}.}
\label{fig:sid_mom_res}
\end{figure}

The normalised transverse momentum resolution $\sigma(1/p_T)$ for single-muon events the SiD detector model is shown in
Fig.~\ref{fig:sid_mom_res} together with fits using the parameterisation:

\begin{equation}
  \frac{\sigma(p_T)}{p^2_T} = a~\oplus~\frac{b}{p~\rm{sin}\theta}     \label{eq:trk_res_param}
\end{equation}

Comparable results are obtained for ILD, and both detector concepts achieve their design goals for the momentum resolution of
$\sigma(p_T)/P^2_T  < 2 \times 10^{-5} \rm{GeV}^{-1}$ for high momentum central tracks.

\begin{figure}
\begin{center}
\includegraphics[width=0.85\hsize]{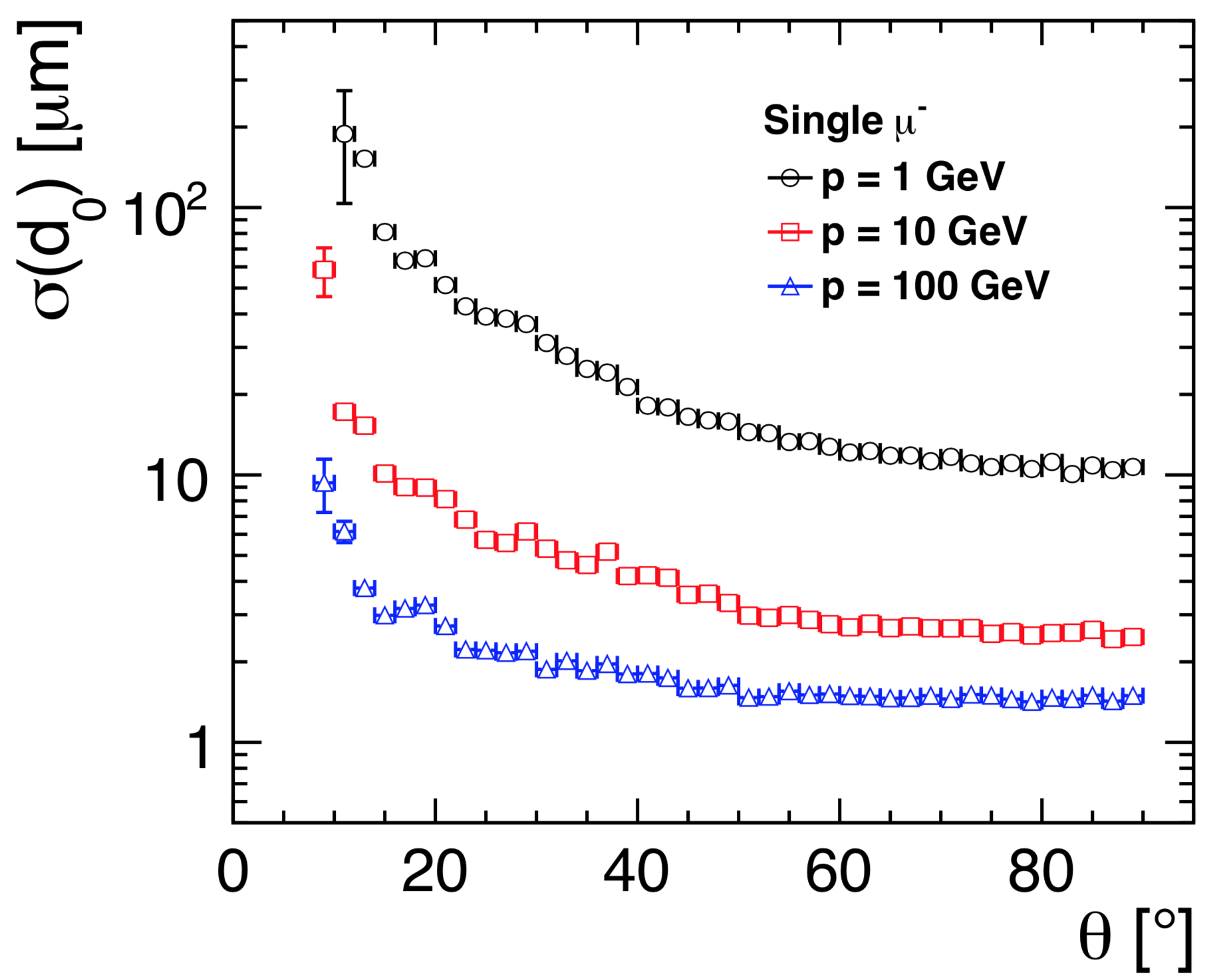}
\end{center}
\caption{Impact parameter resolution $\sigma(d_0)$ for single-muon events as function of polar angle
  in the \emph{SIDLOI3} simulation model (from~\cite{Behnke:2013lya}).} 
\label{fig:sid_d0_res}
\end{figure}

The impact parameter resolution as a function of polar angle for single-muon events in SiD is shown in Fig.~\ref{fig:sid_d0_res} for
different particle momenta. A resolution of a few $\mu\rm{m}$ is achieved for high mometum tracks over a large range of the polar
angle down to $\sim 20^{o}$.

The tracking software is completed with dedicated processors for the identification and reconstruction of kinks and $V^0$s.
Tracks with kinks can arise from bremsstrahlung, typically for electrons, or a large angle deflection due to multiple scattering.
$V^0$s are almost exclusively decays of $K^0_s$ and $\Lambda^0$ and gamma conversions.

\subsubsection{Particle Flow:} 

The \emph{particle flow algorithm} (PFA) aims at reconstructing every individual particle created in the event in order
to take the best available measurement for the given particle type:
\begin{itemize}
\item charged particles\\
  using the momentum measured in the tracking detectors with the excellent resolution described  above.
\item photons\\
  measured in the Ecal with an energy resolution of $\sigma(E)/E \sim  17\% / \sqrt{(E/\rm{GeV})}$. 
\item neutral hadrons\\
  measured predominantly in the HCAL\footnote{Hadronic showers often start in the ECAL and might extend into the Muon system. 
    This is taken into account in PandorPFA.} with an energy resolution of $\sigma(E)/E \sim  50\% / \sqrt{(E/\rm{GeV})}$. 
\end{itemize}

\begin{figure}
\begin{center}
\includegraphics[width=0.85\hsize]{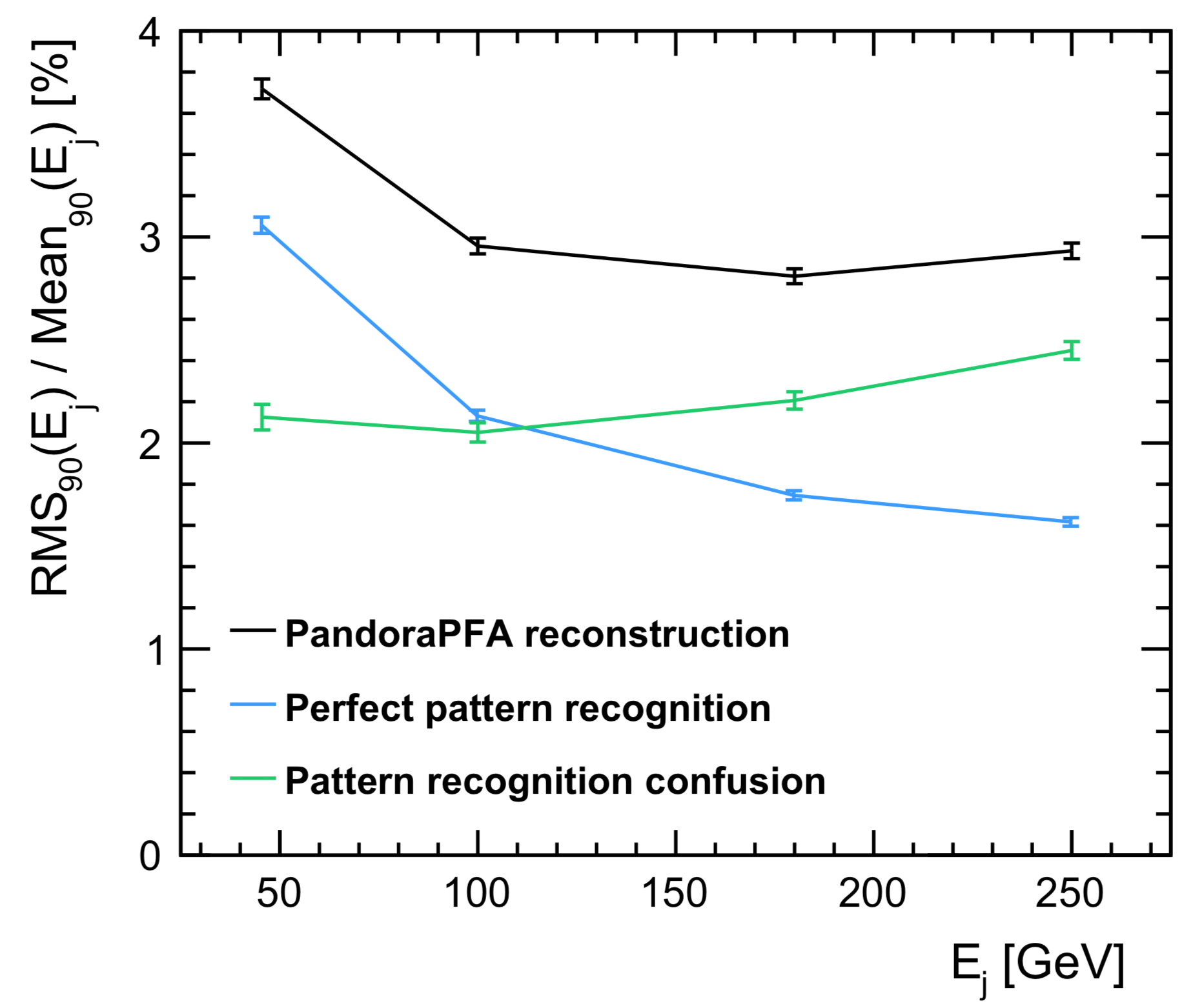}
\end{center}
\caption{Jet energy resolution (in \%) for $Z'$ events as a function of the jet energy in a realistic detector for PandoraPFA.
  Also shown are the effect of \emph{confusion} and the result assuming \emph{perfect PFA} (from~\cite{Marshall:2015rfa}).} 
\label{fig:pandorapfa_perfect}
\end{figure}

The best jet energy measurement in hadronic events would be achieved if the above algorithm would work perfectly. However in reality
there is always confusion in the assignment of individual \emph{CalorimeterHits} to Clusters and showers as well as in the assignment
of tracks to clusters. This effect is demonstrated in Fig.~\ref{fig:pandorapfa_perfect} for PandoraPFA~\cite{Marshall:2015rfa}, the
implementation of PFA available in iLCSoft that is used by both detector concepts.

The input to PandoraPFA are collections of Tracks, Kinks, $V^0$s and collections of all digitized \emph{CalorimeterHits} together with
some geometrical information retrieved from DDRec.
Following~\cite{Marshall:2015rfa} the main steps of the algorithm are:

\begin{itemize}
\item \emph{CalorimeterHits} are clustered using a simple cone-based algorithm, seeded either from isolated hits in the first calorimeter
  layers or by the projection of Tracks to the front face of the ECAL.

\item the clustering algorithm is configured to prefer splitting of clusters rather than risking to falsely merge particles into single
  clusters.

\item Clusters are associated to Tracks based on topological (position and direction) and kinematic (momentum and energy) consistency.
  In case of significant discrepancies a re-clustering is initiated.

\item Clusters without associated Tracks are transformed into neutral \emph{ReconstructedParticles} unless they can be more likely
  interpreted as fragments of charged particles.
  
\item consistent Track-Cluster combinations are transformed into charged \emph{ReconstructedParticles}.
  
\item particle identification plugins are applied to label specific particle types, such as photons, electrons and muons.

\item a dedicated weighting procedure known as \emph{software compensation} is applied to the hits inside a cluster in order
  to equalize hadronic and electromagnetic shower components.
  
\end{itemize}

The final output collection of PandoraPFA is a set of objects called
``PandoraPFO''s.  This represents the final output of the \emph{Reconstruction}
process. This collection is either directly used for physics analyses or serves as input to higher-level reconstruction
algorithms where necessary.

\begin{figure}
  \begin{tabular}[c]{c}
    \includegraphics[width=0.85\hsize]{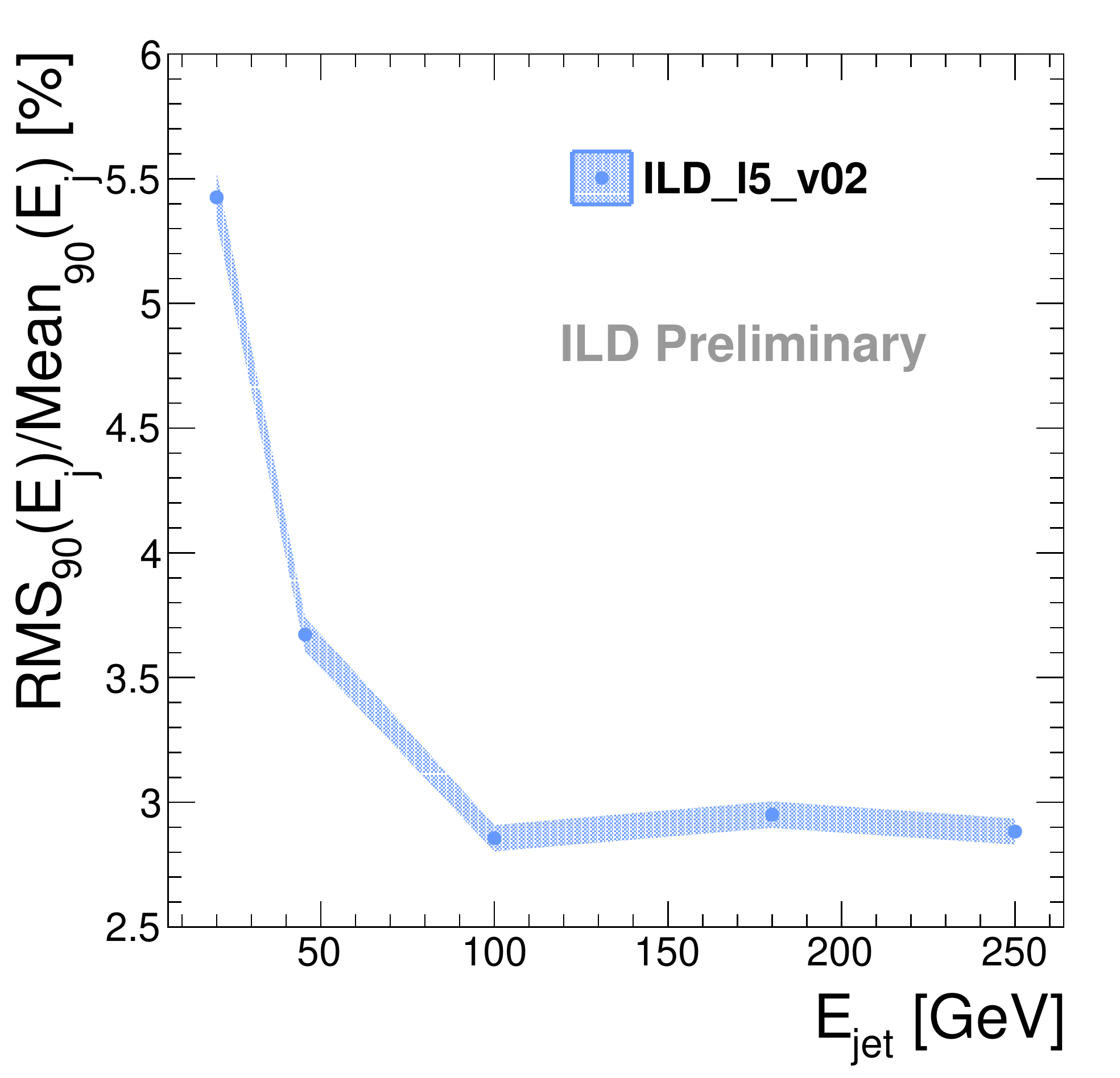} \\
    \includegraphics[width=0.85\hsize]{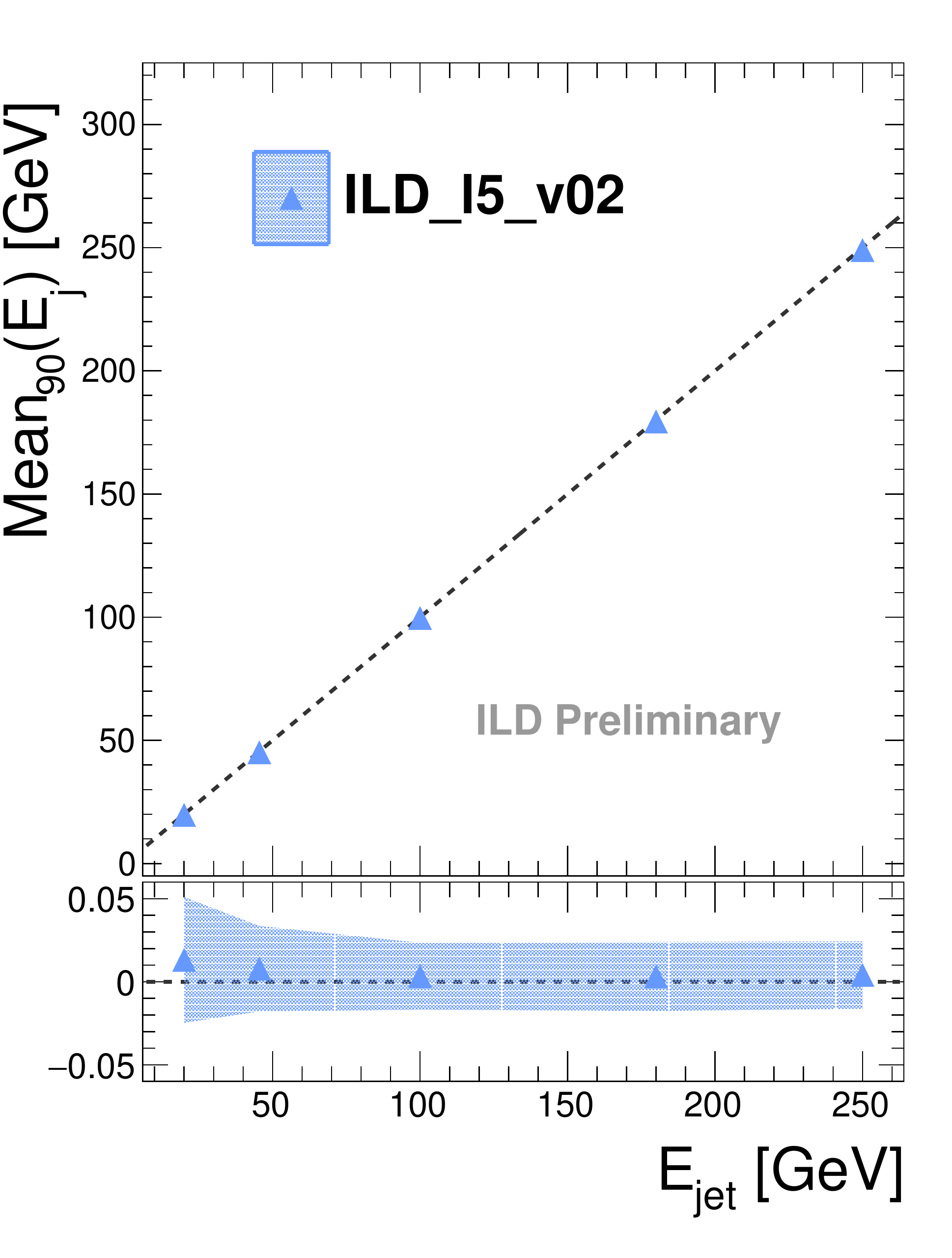}
  \end{tabular}
  \caption{Upper: Jet energy resolution for $Z\rightarrow u,d,s$ events as a function of the jet energy in the standard ILD simulation model.
    Lower: The resulting jet energy scale for the same events.}
\label{fig:ild_jer_jes}
\end{figure}

Fig.~\ref{fig:ild_jer_jes} shows the jet energy resolution and jet energy scale that is achieved for two variants of the ILD detector
for a dedicated event sample of hadronic $Z\rightarrow u,d,s$ events.
The jet energy resolution is evaluated using $RMS_{90}(E)$, the root mean square of the energy of the central 90\% of the events.
The restriction to $u,d,s$ quarks is chosen to focus on  the detector and PFA performance without the extra complication of missing
energy due to neutrinos.

\subsection{\label{sub:sw-HLR}High-level reconstruction}

After having reconstructed all of the individual particles in the event, the next step in the processing is the reconstruction of
primary and secondary vertices. This is carried out in iLCSoft with the LCFIPlus~\cite{Suehara:2015ura} package that is also used
for the tagging of heavy flavor jets.

The primary vertex of the event is found in a tear-down procedure. First an initial vertex is fitted by a $\chi^2$-minimization using
all charged tracks in the event and a constraint from the expected beam spot
$(\sigma_x=516~\rm{nm}, \sigma_y=7.7~\rm{nm},\sigma_z \sim 200~\mu\rm{m}~~\rm{at}~~~E_{cms}=250~\rm{GeV})$.
Then all tracks with a $\chi^2$-contribution larger than a given threshold value are removed.

In a second step LCFIPlus tries to identify secondary vertices, starting out from forming all possible track-pairs from tracks not
used in the primary vertex. The pairs have to fulfill suitable requirements with respect to their invariant mass, momentum direction
and $\chi^2$. $V^0$s are excluded from these initial pairs. Secondary vertices are then formed using so far leftover tracks in
an iterative procedure and eventually adding compatible tracks originally used in the primary vertex.

Secondary vertices and optionally isolated leptons can be used by LCFIPlus for jet clustering, aiming at high efficiency for correctly identifying
heavy flavor jets. The actual jet clustering is then performed by using a cone-based clustering with a Durham-like algorithm.
Alternatively users can use $k_T$ jet clustering algorithms from
Fastjet~\cite{Cacciari:2006sm},
which is interfaced to Marlin in a dedicated
package MarlinFastJet.

\begin{figure}
\begin{center}
\includegraphics[width=0.9\hsize]{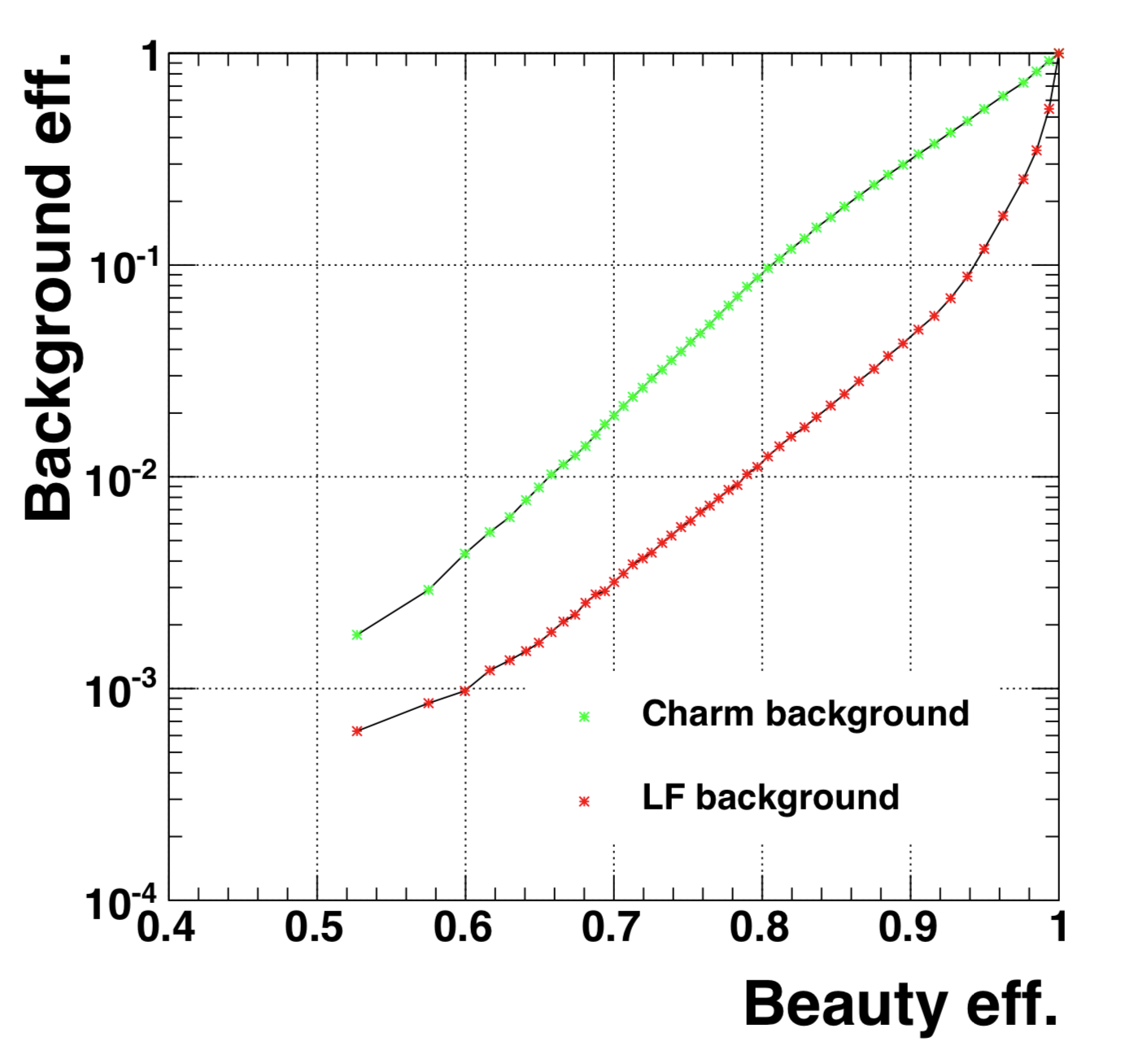}
\end{center}
\caption{Mis-identification efficiency of light quark jets (red points) and charm jets (green points) as beauty jets versus beauty identification efficiency in
  di-jets events at $\sqrt{s} = 91~\rm{GeV}$ (from~\cite{Behnke:2013lya}).} 
\label{fig:sid_flavor_tag}
\end{figure}

LCFIPlus also provides algorithms for jet flavor tagging using boosted decision trees (BDTs) based on suitable variables from tracks and vertices.
Fig.~\ref{fig:sid_flavor_tag} shows the mis-identification efficiency for jets from light quarks and c-quarks as a function of the b-tagging efficiency
for the SiD detector using LCFIPlus.

There is a large palette of additional high level reconstruction algorithms available in iLCSoft addressing the needs for physics analyses, e.g.
\begin{itemize}
\item particle identification using dE/dx, shower shapes and multi-variate methods
\item $\gamma\gamma$-finders for the identification of $\pi^0$s and $\eta$s
\item reconstructed particle to Monte-Carlo truth linker for cross checking analysis and reconstruction efficiencies
\item tools for jet clustering using Monte-Carlo truth information
\item processors for the computation of various event shapes 
\end{itemize}

\subsection{\label{sub:sw-fastsim}Fast simulation}

In addition to the full simulation and reconstruction 
outlined in the previous sections,
there is a need for simulation that can quickly generate
substantial samples of simulated and reconstructed events.
Situations where this is desirable include detector optimisation
and new physics searches. In these cases,
similar processes need to be simulated and reconstructed at
a, possibly very large, number of different conditions.
In the first case, one needs to  modifying various aspects
of the detector in steps, in the latter,
one needs to  explore the entire allowed parameter space
of a theory for new physics.
In addition to these cases,
fast simulation is also an asset for simulating high cross section
SM processes, such as $\gamma\gamma$ processes, where the investment 
in processor power and intermediate storage might be
prohibitively large to attain the goal that simulation statistics
should be a negligible source of systematic uncertainty.

To meet these needs,
a fast simulation program needs to be fast, flexible, and
accurate. 
The SGV program\cite{Berggren:2012ar} used at ILC meets these needs.
The time to simulate and reconstruct an event is similar to
the time it takes to generate it ($\sim 1-10$~ms).
The response of the detector is as far as possible calculated
from the detector design (so there is no need to parametrisise
pre-existing full simulation results).  SGV
has been shown to compare well  both with full simulation
and with  real data~\cite{Abdallah:2003xe}.

The program uses a simplified ``{\it cylinders-and-discs}'' description
of the detector,
which is used to calculate the Kalman-filtered track-helix covariance matrix
of each generated charged particle.
By Cholesky decomposition of the covariance matrix,
the track-parameters are simulated in a way such that all correlations
are respected.
The calorimetric response is calculated from the expected single-particle
performance of the different components of the calorimetric system,
for each particle impinging on it. Optionally,
the effects of shower-confusion can be included.
To reduce the needed storage for a Giga-event size sample,
event filtering can be applied at different steps of the processing,
directly after generation, after the detector response is known,
or after higher-level event analysis is done.
Events passing all filters are output in LCIO DST-format,
and can seamlessly be further analysed within the Marlin framework.


\subsection{\label{sub:sw-computing}Computing concept}

An initial computing concept for the ILC, including a first estimate of the required resources, has been developed by the LCC Software and Computing Group.

\begin{figure*}
\begin{center}
\includegraphics[width=0.6\hsize]{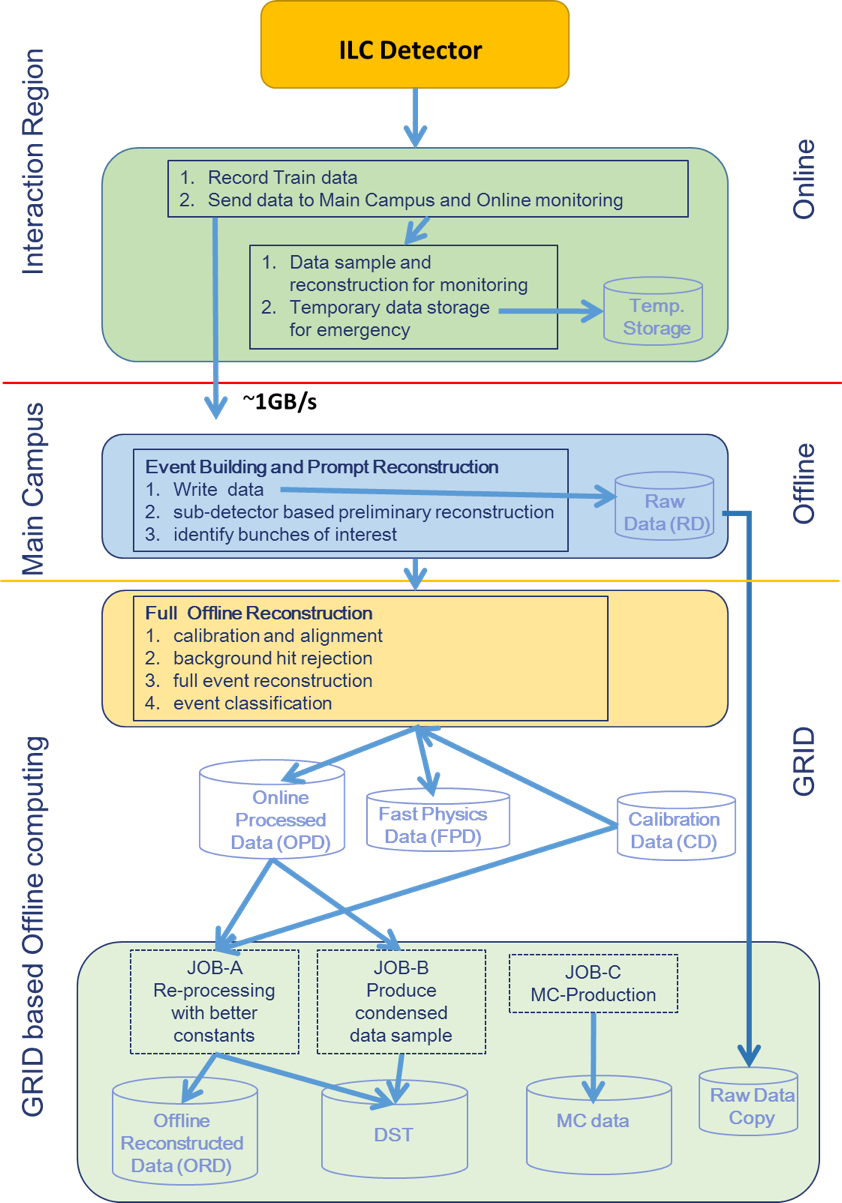}
\end{center}
\caption{Computing concept foreseen for the ILC, distributed over on-site computing at the interaction region, the main campus and Grid-like
  offline computing.}
\label{fig:computing_scheme}
\end{figure*}

The foreseen computing concept follows in general terms that of the current LHC experiments and Belle~II, with a strong on-site computing
center complemented by large Grid-based computing resources distributed around the world. This concept is schematically shown in
Fig.~\ref{fig:computing_scheme}.

Due to the much lower event rates at the ILC compared to the LHC, we will be
able to run in an un-triggered mode in which  collision data from every bunch crossing will be recorded. At the experimental site,
we require only limited computing resources for online monitoring, QA and data-buffering for a few days.

Prompt reconstruction, event building, and filtering of the interesting collisions will be performed at the main ILC campus.
A small fraction of the initial raw data will be distributed to major participating Grid sites in the world for further skimming and
final redistribution for physics analysis.
A copy of the raw data from all bunch crossings will be kept to allow  for future searches for new exotic signatures.

\subsection{\label{sub:sw-res-estimate}Computing resource estimate}

Based on our detailed physics and background simulations,  we estimate the total raw data rate of the ILC to be $\sim$1.5GB/s.

The total estimated storage needs will be a few tens of PB/y.
The computing power needed for simulation, reconstruction, and analysis will be a few hundred kHepSpec06.
Given that these numbers are already smaller than what is now
needed by the LHC experiments, and given an expected annual increase
of 15\% and 20\%, respectively, for storage and CPU
at flat budget, we expect the overall computing costs for the ILC
will be more than an order of magnitude smaller than those for the LHC.


\section{\label{sec:higgs}Physics Simulations: Higgs
}


\begin{figure}
\begin{center}
\includegraphics[width=0.85\hsize]{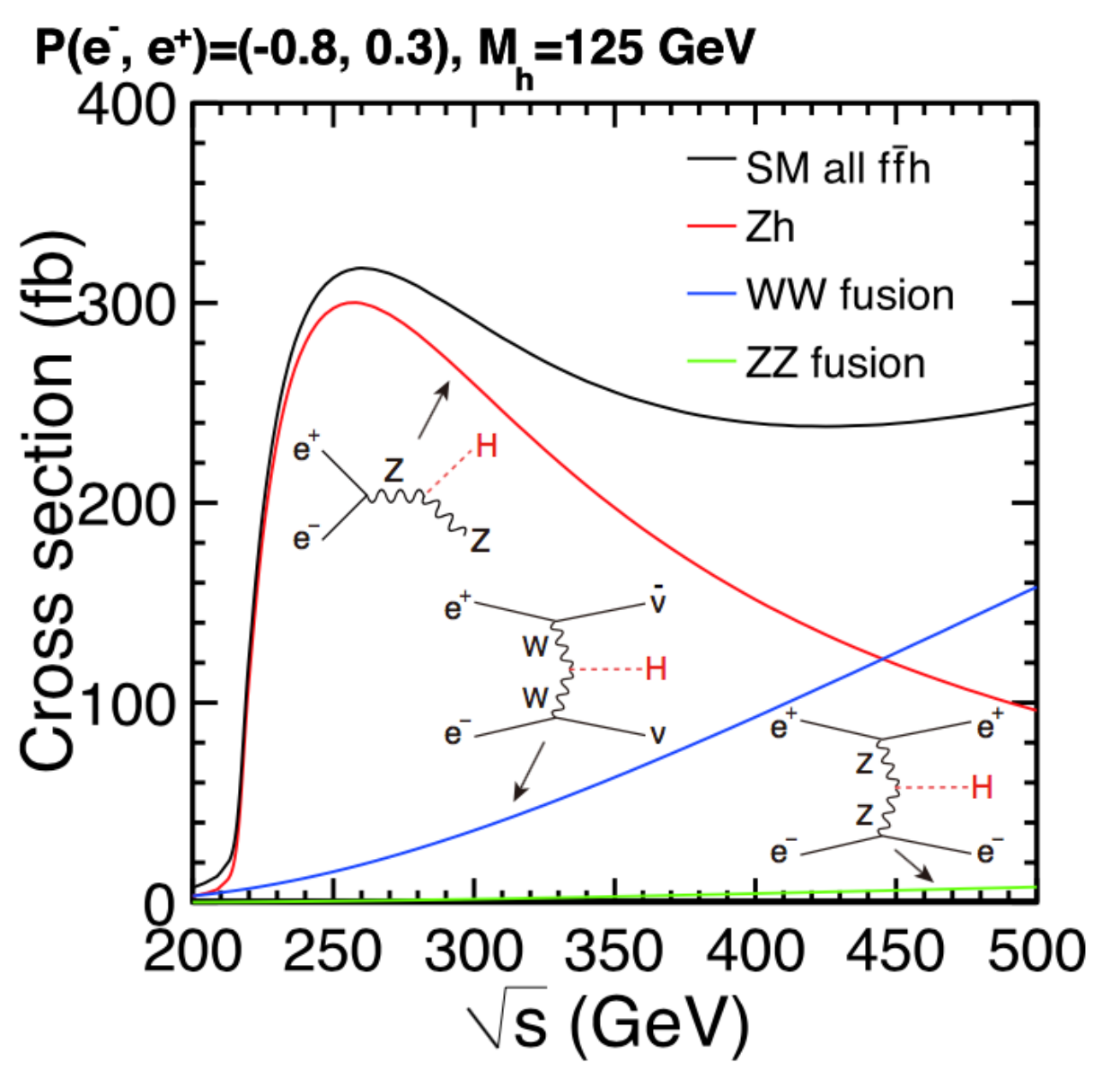}
\end{center}
\caption{Cross sections for the three major Higgs production processes
  as a function of 
center of mass energy, 
from~\cite{Baer:2013cma}.}
\label{fig:HiggsProdILC}
\end{figure}

The physics case for the precision study for the Higgs
boson
presented  in Section~\ref{sec:physics}
will be realized through the measurement of  total cross sections and 
$\sigma\cdot BR$  values for the various final-states.  The major
Higgs production cross sections at the ILC are shown in
Fig.~\ref{fig:HiggsProdILC} as a function of centre of mass energy for
the optimal choice (-80\%/+30\%) of  ILC beam polarisations.    In
Tab.~\ref{tab:higgserrors},  we present our estimates for the
statistical errors that will be obtained for the total cross section for $\ee\to ZH$ and for
the $\sigma\cdot BR$s for 
this process and the $WW$ fusion process, 
for a reference luminosity sample of 
250~\ifb\ and for three different ILC energies. There is a similar 
table for the opposite beam polarisation state (+80\%/-30\%).  In this case, the errors for $ZH$ observables are almost the same, due to a compensation of lower background and lower signal cross sections.   The $WW$ fusion process has a much reduced cross section and comparably lower precision\cite{Barklow:2017suo}. These estimates are
based on full-simulation analyses using the tools presented in
Sec.~\ref{sec:software}.     The purpose of this
section is explain how these numbers are obtained, what the factors
are  that limit them, and how these limitations might be relaxed.

\begin{table}[htb]
\begin{center}
\begin{tabular} {lcccccc}
\multicolumn{7}{l} {-80\% $e^-$, +30\% $e^+$ polarization:} \\  \hline
 & \multicolumn{2}{c}{250 GeV} & \multicolumn{2}{c}{350 GeV}  & \multicolumn{2}{c}{500 GeV} \\
 &  $Zh$ & $ \nu\bar\nu h$  & $ Zh$ & $\nu\bar\nu h$ & $Zh$ & $ \nu\bar\nu h$\\ 
\hline
$\sigma$  &    2.0     &    &1.8  &  &  4.2   &     \\  \hline
$h\to invis.$ &  0.86   &  &1.4 &   &   3.4 &     \\
\hline
$h\to b\bar b$  &   1.3 &  8.1 & 1.5  &  1.8  & 2.5  &  0.93  \\ 
$h\to c\bar c$ &  8.3 &   &11 & 19  &   18 & 8.8 \\ 
$h\to gg$ &  7.0 &  &8.4  & 7.7 & 15  &  5.8\\
$h\to WW$  &  4.6 &   &5.6$^*$ & 5.7$^*$ &  7.7   &  3.4\\
$h\to \tau\tau$  & 3.2 &   &4.0$^*$ & 16$^*$ &  6.1  &  9.8\\
$h\to ZZ$ & 18 &   & 25$^*$ & 20$^*$ & 35$^*$  & 12$^*$    \\ 
$h\to \gamma\gamma$  & 34$^*$ &   &39$^*$ &  45$^*$ &  47 &  27 \\ 
$h\to \mu\mu$ & 72 &   &87$^*$&  160$^*$ &  120 &  100 \\
\hline\hline
$a$  &  7.6   &     &2.7$^*$ &   &  4.0   &  \\
$b$ &  2.7  &    & 0.69$^*$ &    & 0.70   &   \\
$\rho(a,b)$  &  -99.17 &   & -95.6$^*$ &  & -84.8 &   \\
\end{tabular}
\caption{Projected statistical errors, in \%, for Higgs boson 
measurements. The errors are 
quoted for luminosity samples of 250~fb$^{-1}$
  for $\ee$ beams with -80\% electron polarization and +30\% positron
  polarization. 
  Except for the first and last segments of  each set, these are measurements
  of  $\sigma \cdot BR$, relative to the Standard Model
  expectation.
The top lines gives the error for the total cross section relative to
the Standard Model and  the 95\% confidence upper limit on the branching ratio for
Higgs to invisible decays.  The bottom lines in each half give the
expected errors on the $a$ and $b$ parameters and their correlation
(all in \%)  for $\ee\to Zh$ (see
\leqn{eqn:CPHZZ}.    All error estimates in this table are
based on full simulation, and the entries marked with a $^*$ are 
extrapolated from full simulation results. }  
\label{tab:higgserrors}
\end{center}
\end{table}


\begin{table}
\begin{center}
\begin{tabular} {lcccccc}
measurement  & efficiency &  S/B final. \\
\hline
$\sigma_{Zh}$ in $\mu^+\mu^-h$  & 88\% & 1/1.3 \\
$BR(h\to b\bar{b})$ in $q\bar{q}h$ & 33\% & 1/0.89 \\
$BR(h\to\tau\tau)$ in $q\bar{q}h$ & 37\% &  1/0.44 \\
$BR(h\to WW)$ in $\nu\bar{\nu}h$ & 20\% &  1/1.6
\end{tabular}
\caption{Typical signal efficiencies (second column) and signal over background ratio (S/B) 
after the final cuts (third column) for some of the 
representative Higgs measurements (first column) at the ILC.}
\label{tab:ILCEffSB}
\end{center}
\end{table}

\begin{figure}
\begin{tabular}[c]{c}
\includegraphics[width=0.85\hsize]{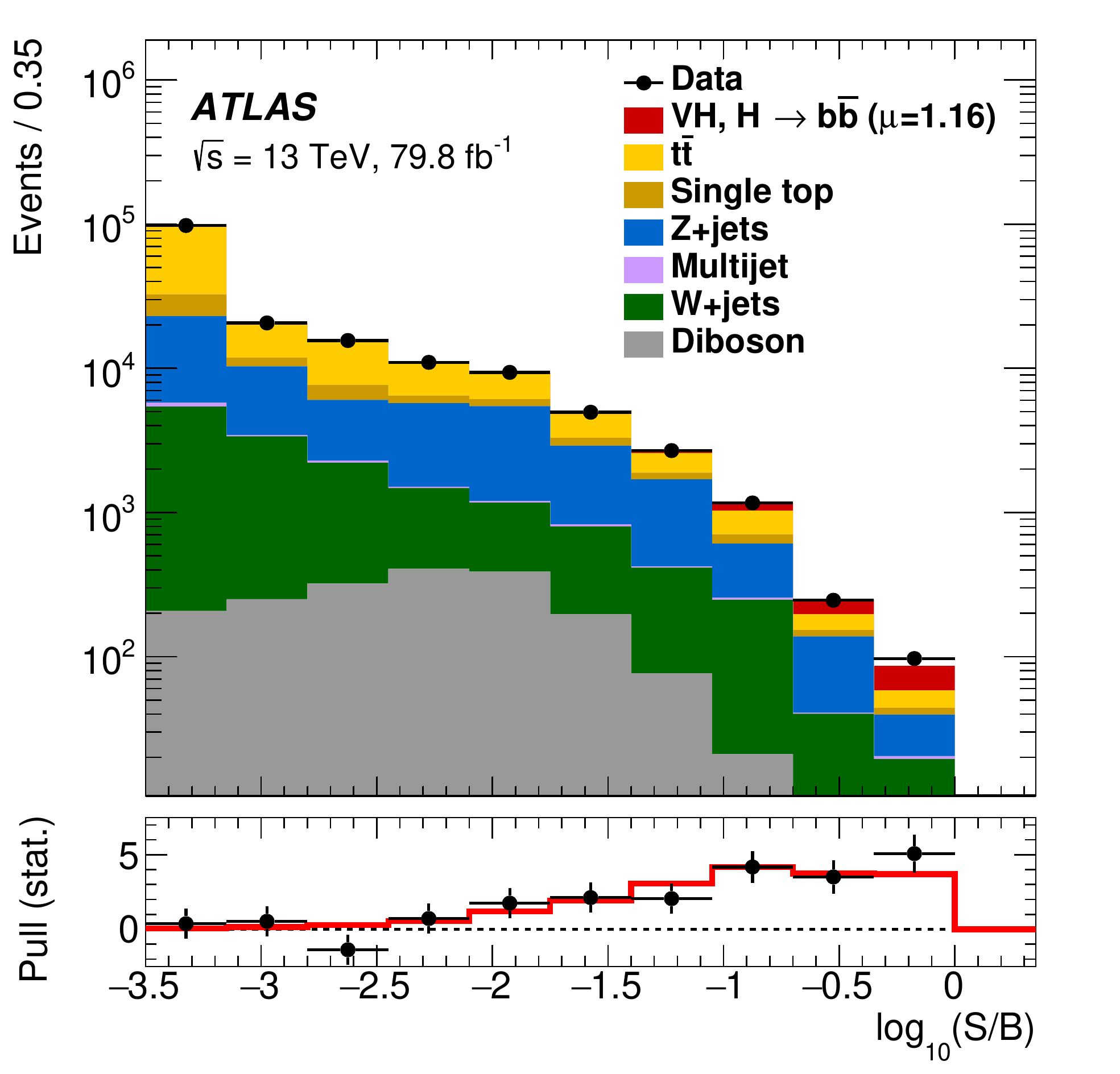} \\
\includegraphics[width=0.85\hsize]{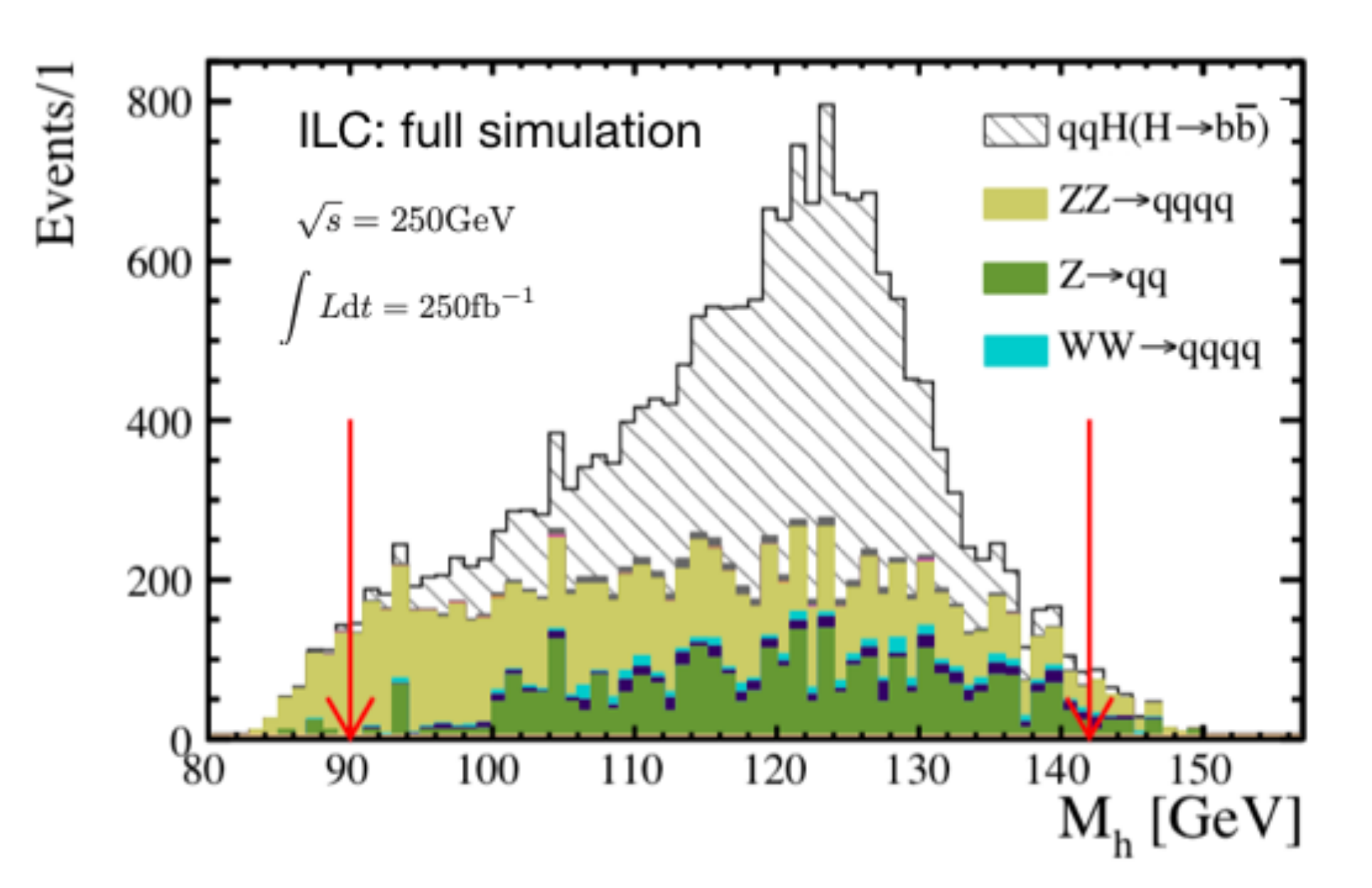}
\end{tabular}
  \caption{Upper: signal $H\to b\bar{b}$ and background events in different categories of S/B
  measured by ATLAS~\cite{Aaboud:2018zhk,Sirunyan:2018kst} using LHC Run 2 data; lower: signal $h\to b\bar{b}$ 
  and background events in the $b\bar{b}$ mass spectrum expected from the 
  ILC full simulation~\cite{Ogawa:2018}.}
  \label{fig:LHCILCHbb}
\end{figure}

We begin with the observation that 
precision Higgs measurements will be much easier to obtain 
at a lepton collider than  at a hadron collider.
Table~\ref{tab:ILCEffSB} gives the typical signal efficiencies for ILC
analyses and the corresponding signal to
background ratios (S/B) after final cuts. The difference with
LHC can be clearly seen using the example of $H\to b\bar{b}$ measurements.
The decay of $H\to b\bar{b}$
has been discovered by ATLAS and
CMS~\cite{Aaboud:2018zhk,Sirunyan:2018kst} 
with a significance of 5.4$\sigma$, 5.5$\sigma$, respectively, 
after producting  about  4 million Higgs events per experiment.  At
the ILC,  with only 400 Higgs events which will be produced with an
 integrated luminosity of 1.3 fb$^{-1}$ (corresponding to 2 days of
 running time),
the decay of $H\to b\bar{b}$ will be 
measured with a similar significance, around 5.2$\sigma$ according to the full 
simulation result~\cite{Ogawa:2018}.  The S/B ratios for these
analyses are illustrated in  Fig.~\ref{fig:LHCILCHbb}. Clearly, if one
wishes to measure the rate for $h\to b\bar b$,
there are strong advantages in starting from a situation in which the
signal stands well above any background process that would need to be
controlled.   The challenge of physics at a linear collider is to make use of this
advantage in the most optimal way and realize the potential to achieve 
very high precision.

A full simulation analysis contains two components.  The first is the 
detector simulation.  This provides  the 
realistic interactions between each final state particle and 
any part of the detector that the particle passes through, including creation of new
particles during the interaction; concrete algorithms for tracking,
particle flow analysis, vertex reconstruction and particle identification; 
the resulting performance of the various detector resolutions for track momentum, jet energy,
and impact parameters; and the  efficiencies for tracking, flavor tagging, 
and isolated lepton finding.  These aspects have already been
described in Section~\ref{sec:software}. The second component
 is the event seletion, that is, the algorithms
for discriminating
between signal and background events. That will be our main concern in
the discussion of this section.

First, however, we would like to emphasize to the reader a number
of effects that are included in the detector modelling and event
generation, and that must be included for a solid estimate of
detection efficiencies and signal-background discrimination:
\begin{itemize}
\item {\it beamstrahlung and ISR}
are implemented in the event generators for both signal and background processes.
These effects are important for estimation of 
signal and background contributions in the analyses that make use of the 
nominal value of the centre-of-mass energy. A representative example is seen in
Section~\ref{sec:higgs:sigmazh}, Fig.~\ref{fig:RecoilMassLep250}, for
the determination of the Higgs boson mass from $Z$ recoil. 
Both beamstrahlung and ISR effects will drag signal events
from the more sensitive peak region to the less sensitive tail region,
and at the same time will induce more background contribution in the signal region.
\item {\it overlay of beam background events} is implemented in every signal 
and background event sample.  This will affect the performance of 
reconstructed variables related to jets and  hence degrades the signal and background
discrimination. There is a method to partially remove the effect of
this events which will 
be introduced in next section. 
\item {\it full Standard Model background} is checked in all of  the
  analyses to be described,
in order not to miss any significant contribution. For example,  2-fermion 
events developed with a parton shower can   become background for the  4-fermion signal.
Another example is the background contribution to  Higgs observables
that 
comes from tail of the Breit-Wigner structure of a $Z$ boson in
$\ee\to ZZ$.   It is not correct neither to neglect the 
$Z$ natural width nor to ignore the  similar diagram with the $\gamma$ propagator.
\item {\it explicit jet clustering and jet paring} algorithms are used
  in all analyses.   These often become the 
limiting factors in the analyses with 4 or more jets  in the final state.
The confusion between two color singlets, for instance $Z$ and $h$ in $\ee\to Zh\to 4 jets$,
could produce a much wider spread of the reconstructed dijet invariant mass
than that due to  the pure detector resolution. Hence,  simply
smearing the  dijet mass variable 
at the parton level according to the detector resolution is often too optimistic. 
\item {\it control of systematics} is taken into account in the design of every
selection cut.
\end{itemize}

This section is organized as follows. In
Sec.~\ref{subsec:higgs_common},  we will introduce  the common
procedures for event selections. Section~\ref{subsec:higgs_ana} will discuss the 
analyses for the main Higgs observables. The analysis strategies and 
selection cuts in some representative channels will be  discussed in great detail. 
Section~\ref{subsec:higgs_improve}  presents some estimates for
improvement of the key algorithms in the future.
Section~~\ref{subsec:higgsself}  gives a dedicated discussion of the 
 measurement of Higgs self-coupling.

\subsection{Common procedures for event selections}
\label{subsec:higgs_common}
The full simulation analysis at the event selection level can be described 
in two steps: {\it pre-selection} and {\it final-selection}. At the pre-selection step
each signal event is characterized according to its final states at parton level by
numbers of isolate leptons (meaning electron or muon unless otherwise stated), 
isolated taus, isolated photons and jets, and nature of missing momentum.
Here isolated particle is meant to be not coming from a jet. The procedures
for the pre-selection are typically as follows:
\begin{itemize}
\item {\it isolated lepton finder}, which will try to reconstruct the isolated leptons
in each event. The main algorithms are implemented in the processor called
IsolatedLeptonTagging in the iLCSoft, based on a multivariate method.
It starts with selecting energetic electron/muon (momentum $P>5$ GeV), 
from the reconstructed particles collection, by requiring the particle has characteristic
energy fractions deposited in each sub-detector, namely $E_{ecal}/E_{tot}$,
$E_{tot}/P$, $E_{yoke}$, where $E_{ecal}$ ($E_{hcal}$) is the energy deposited 
in ECAL (HCAL), $E_{tot}$ is the sum of $E_{ecal}$ and $E_{hcal}$, and
$E_{yoke}$ is the energy deposited in Yoke. For electron, 
it is required that $E_{ecal}/E_{tot}>0.9$ and $0.5<E_{tot}/P<1.3$. For muon,
it is required that $E_{tot}/P<0.3$ and $E_{yoke}>1.2$ GeV. The selected
electron/muon is then further required to have impact parameters consistent with
that from primary vertex. Double cones are then defined around that electron/muon,
and variables such as the energies of the charged and neutral particles within a cone 
are utilized for isolation requirement. The exact criteria for isolation 
are realized by a MVA, trained using true isolated leptons as signal 
and leptons from jets as background.
A cut on the MVA output is then required as the last step of isolated lepton finder.
\item {\it isolated tau finder}, which will try to reconstruct the isolated taus. 
The main algorithms are implemented in TaFinder in the iLCSoft. It starts
with finding the most energetic charged particle as a tau candidate. Then the
remaining most energetic particle which is within a cone of $\cos\theta=0.99$ around
the tau candidate will be combined to the tau candidate if the invariant mass
of the combined tau candidate does not exceed 2 GeV. This combining step
will be iterated until there is no any more particle to combine. The resulting 
combined tau candidate is identified as an isolated tau.
\item{\it isolated photon selection}, which will try to reconstruct the isolated photon.
A photon is first identified based on its cluster properties by PandoraPFA. For
most of the signal processes with an isolated photon, it is usually sufficient to tag
the most energetic photon which has energy larger than several tens of GeV, as
the candidate isolated photon. If there are other photons within a very small cone of
$\cos\theta=0.999$ around the candidate photon, those other photons are most
probably split ones hence are merged into the candidate photon.
\item {\it overlay removal}, which will try to remove the pile-up beam background
events in every event. An exclusive jet clustering is performed using 
longitudinal invariant $k_t$ algorithm~\cite{Catani:1993hr}
for all the particles except the selected isolated lepton/tau/photon in above step. 
As a result, the particles from beam background events, which usually have very low-$p_t$, 
are clustered into beam jets and 
are effectively removed by the exclusive jet clustering process. The 
input parameters such as $R$ and number of required jets are carefully optimized 
for each signal process. Alternative algorithms include anti-$k_t$~\cite{Cacciari:2008gp} 
and Valencia~\cite{Boronat:2014hva}.
\item {\it jet clustering and flavor tagging}, are done using LCFIPlus 
as introduced in~\ref{sub:sw-HLR}. 
All the particles belonging to the
jets obtained in previous step are then re-clustered into a few jets using another inclusive 
jet clustering algorithm, Durham algorithm~\cite{Catani:1991hj}.
Each jet is flavor tagged using the reconstructed information of 
its secondary and tertiary vertices. 
\end{itemize}

At the final-selection step, the reconstructed leptons, taus, photons and jets 
will be first combined to reconstruct $W$, $Z$, $h$ or $top$ according to the signal. 
Then various cuts will be applied to further suppress background events.
Details are explained measurement by measurement in the following. 
Unless stated otherwise, the analysis is done at $\sqrt{s}=250$ GeV,
a nominal integrated luminosity of 250 fb$^{-1}$ is assumed, and
the cuts and results are illustrated with left-handed beam polarization
$e^-_Le^+_R: P(e^-,e^+)=(-0.8,+0.3)$. Additional comments will be given
when $\sqrt{s}$ or right-handed beam polarization 
$e^-_Re^+_L: P(e^-,e^+)=(+0.8,-0.3)$ has a significant impact on the results.
The results are straightforwardly extrapolated into that for the running scenario
introduced in Sec.~\ref{sec:runscenarios} and are then used as input for the 
Higgs coupling determination by a global fit introduced in Sec.~\ref{sec:physics}.
The full information for the uncertainties of Higgs observables for 
$e^-_Le^+_R$ and $e^-_Re^+_L$ can be found in Tab.6 of Ref.~\cite{Barklow:2017suo}.

\subsection{Analyses for Higgs observables}
\label{subsec:higgs_ana}

\subsubsection{$m_h$ and $\sigma_{Zh}$}
\label{sec:higgs:sigmazh}
The signal processes are $\ee\to Zh$, $Z\to l^+l^-$ or $q\bar{q}$ and 
$h\to$anything. Thanks to the known four momenta of initial states, 
the four momentum of final state $h$ can be reconstructed 
as the recoil against the four momentum of $Z$, which is directly
measured from its decay products $l^+l^-$ or $q\bar{q}$. 
The mass of $h$ ($m_X$) can therefore be reconstructed as
\beq
m_X^2=s+m_Z^2-2E_Z\sqrt{s},
\eeq{eqn:recoilmass}
where $m_Z$ and $E_Z$ are measured mass and energy of $Z$
respectively. The signal events can hence be tagged without looking
at the decay products of $h$. This technique is traditionally called
{\it recoil mass} technique, and the two types of signal processes 
($Z\to l^+l^-$ and $Z\to q\bar{q}$) are called
leptonic recoil and hadronic recoil channels. The recoil mass technique
makes possible the measurement of the inclusive cross section of $\ee\to Zh$ ($\sigma_{Zh}$),
that plays a unique role in the determination of the absolute values of
Higgs couplings as explained in Sec.~\ref{sec:physics}.
Meanwhile, the Higgs mass ($m_h$) can be straightforwardly determined by 
the $m_X$ spectrum. The detailed analyses for leptonic recoil channels
$\mu^+\mu^-h$ and $e^+e^-h$ and for hadronic recoil channel $q\bar{q}h$
can be found respectively in references~\cite{Yan:2016xyx} and 
~\cite{Tomita:2015,Thomson:2015jda,Miyamoto:2013zva}. 
For simplicity only the analysis for $\mu^+\mu^-h$ channel is illustrated in detail here.

The event pre-selection in $\mu^+\mu^-h$ channel starts with requiring at least 
two isolated muons with opposite charges and invariant mass ($m_{ll}$) consistent 
with the $Z$ mass (in the range $m_{ll}\in[50,130]$ GeV). 
It is quite possible that one or two muons in such a candidate 
muon pair are actually from Higgs decay, for instance from 
$h\to ZZ^*/WW^*/\tau^+\tau^-\to \mu^+\mu^-+X$. To minimize the possibility of
this case, or to maximize the possibility that the candidate muon pair is 
indeed from the primary $Z$ decay, the following strategy is taken:
when there are more than one such candidate muon pairs,
the pair which minimizes following $\chi^2$ 
\beq
\chi^2=(\frac{m_{ll}-M_Z}{\sigma_Z})^2+(\frac{m_{X}-M_h}{\sigma_h})^2
\eeq{eqn:chi2forleptonpair}
is identified as from the primary $Z$ decay. Here $M_Z$ is 91.2 GeV,
$M_h$ is 125 GeV, $\sigma_Z$ and $\sigma_h$ are resolutions for 
$Z$ mass and recoil mass reconstructions. 
After the pre-selection, the remaining background events are dominated by
leptonic and semi-leptonic decays of $\ee\to ZZ$, leptonic decay of $\ee\to WW$, 
and leptonic decay of $\ee\to \gamma Z$. 

In the final selection, the cuts $p_T^{ll}>10$ GeV and $|\cos\theta_{mis}|<0.98$,
where $p_T^{ll}$ is the transverse momentum of muon pair and $\theta_{mis}$
is the polar angle of missing four momentum, are applied to suppress
$\gamma Z$ background events. $E_{vis}>10$ GeV,
where $E_{vis}$ is the visible energy other than the muon pair,
and $m_{ll}\in[73,120]$ GeV
are applied to suppress $WW$ background events. 
$ZZ$ as well as $WW$ and $\gamma Z$ background events are
further suppressed by a dedicated BDT cut 
which is trained using distributions of polar angle of each muon,
angle between two muons, and polar angle of the muon pair. After a final cut that requires
$m_X\in[110, 155]$ GeV, the remaining signal and background events are 
shown in the $m_X$ spectrum in Fig.~\ref{fig:RecoilMassLep250} 
for the $Z\to\mu^+\mu^-$ channel, where the signal peak is clearly seen.
The overall signal efficiency is 88\%, with an
average signal over background ratio of 1/1.3.

The number of signal events and its statistical uncertainty are obtained by 
fitting $m_X$ spectrum with signal component modeled by a kernel function
and background component modeled by a third order polynomial, 
shown in Fig.~\ref{fig:RecoilMassLep250}. As shown by the green histogram in Fig.~\ref{fig:RecoilMassLep250},
the signal spectrum has a 
considerable non-Gaussian tail in the high mass end, which is due to
the overestimate of effective $\sqrt{s^\prime}$ in $\ee\to Zh$ reaction 
when beamstrahlung and ISR effects are included, recall Eqn.~\ref{eqn:recoilmass}.
It's worth noting that these effects become so significant at $\sqrt{s}=500$ GeV,
as shown in Fig.~\ref{fig:RecoilMassLep500},
that the measurement uncertainty could be underestimated by a factor of 2
if the effects are not properly included in the simulation. 

\begin{figure}
\begin{center}
\includegraphics[width=0.85\hsize]{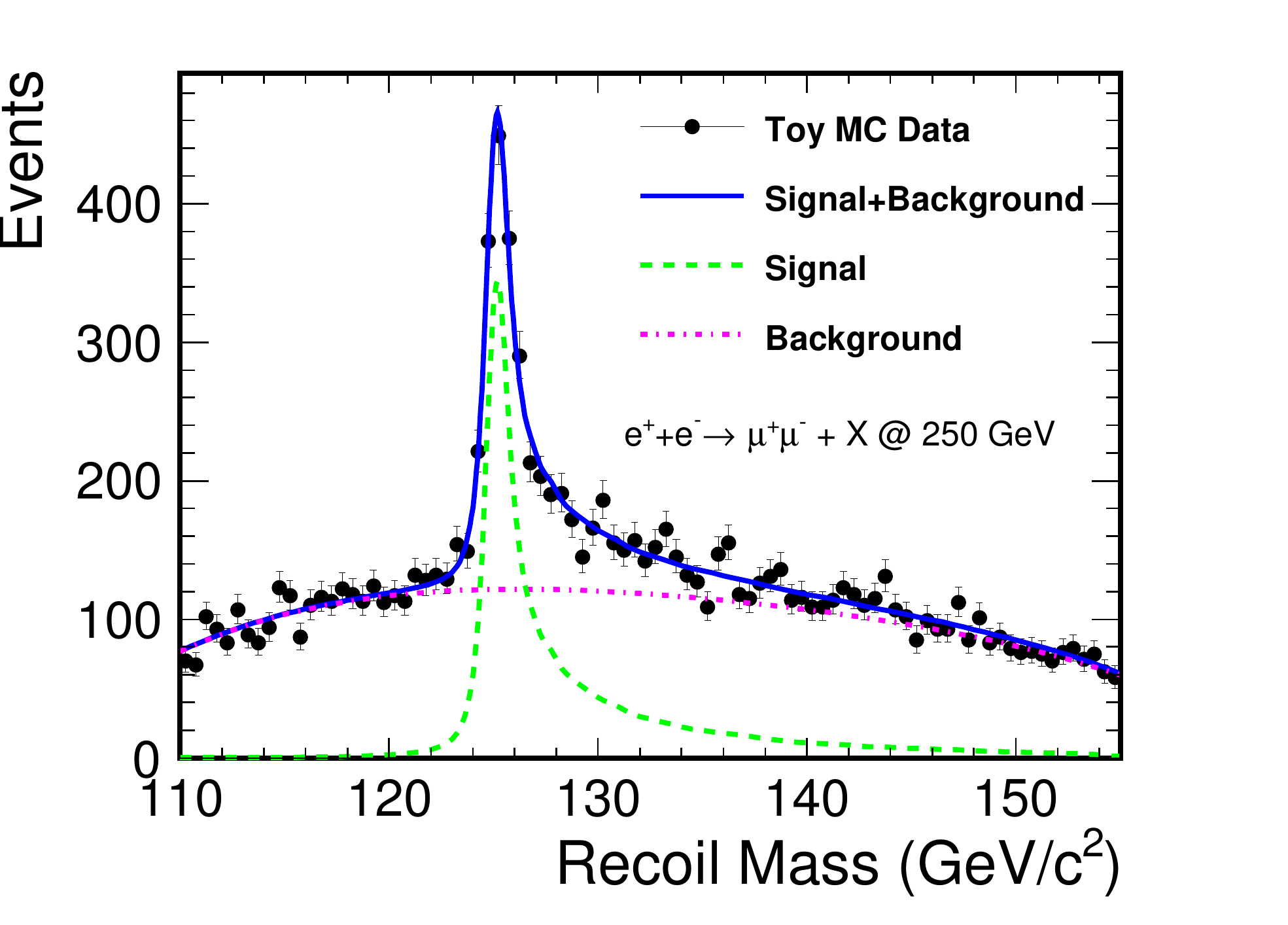}
\end{center}
  \caption{Recoil mass spectrum against
 $Z\to\mu^+\mu^-$ for signal $e^+e^-\to Zh$ and SM background 
  at 250 GeV \cite{Yan:2016xyx}.}
  \label{fig:RecoilMassLep250}
\end{figure}

\begin{figure}
\begin{center}
\includegraphics[width=0.85\hsize]{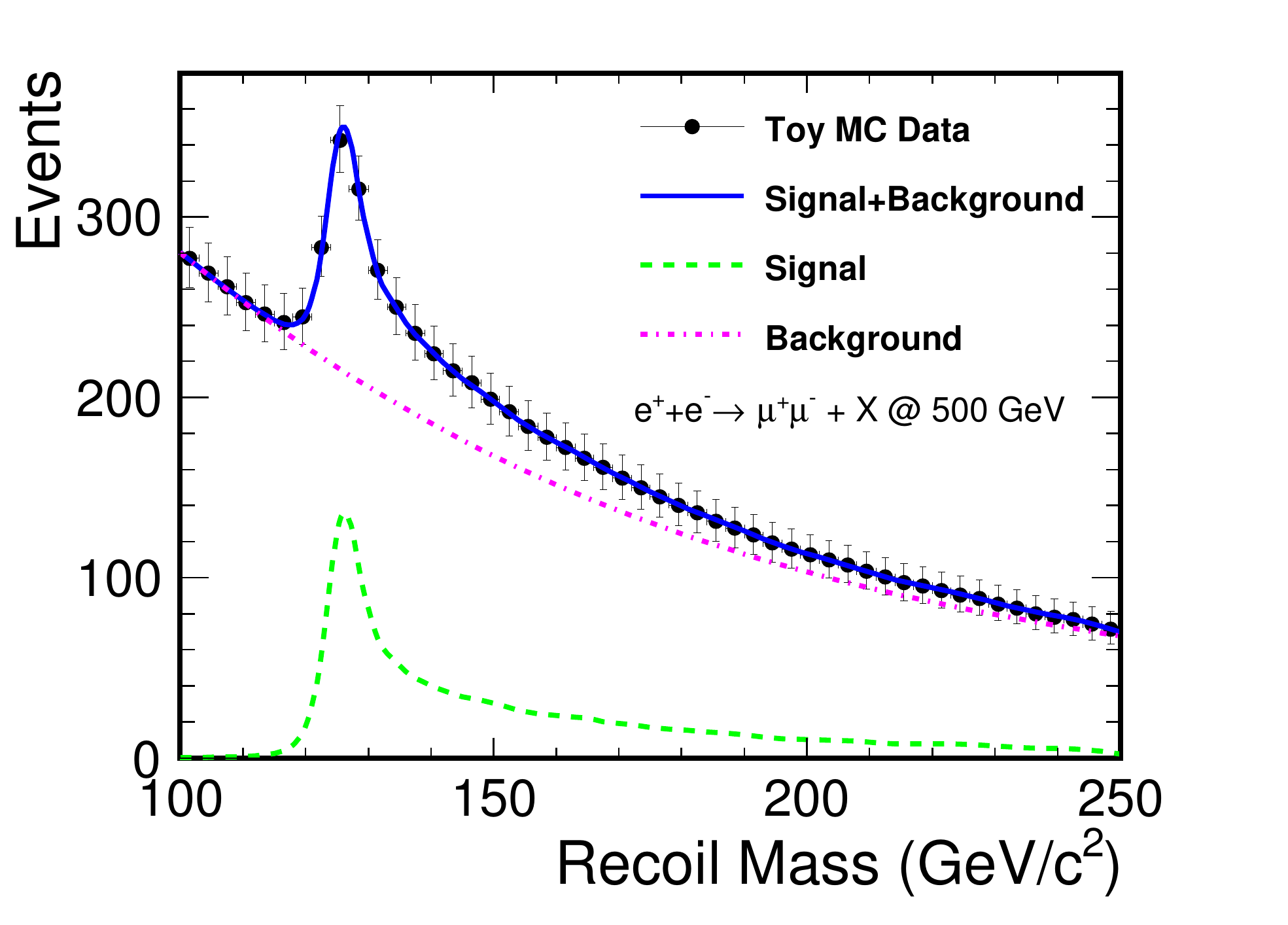}
\end{center}
  \caption{Recoil mass spectrum against
 $Z\to\mu^+\mu^-$ for signal $e^+e^-\to Zh$ and SM background 
  at 500 GeV \cite{Yan:2016xyx}.}
  \label{fig:RecoilMassLep500}
\end{figure}

For $e^-_Le^+_R$, the estimate of relative 
uncertainty on $\sigma_{Zh}$ measurement ($\delta\sigma_{Zh}$) 
is 2.5\% for the leptonic
recoil channel, where the contribution from $e^+e^-h$ channel is
slightly smaller than $\mu^+\mu^-h$ channel due to the higher 
electron bremsstrahlung. For $e^-_Re^+_L$, $\delta\sigma_{Zh}$
is estimated to be 2.9\%. By combining the hadronic recoil channel,
$\delta\sigma_{Zh}$  is estimated to be 2.0\% for both $e^-_Le^+_R$
and $e^-_Re^+_L$, as shown in Tab.~\ref{tab:higgserrors}. 
The enabled measurement of left-right asymmetry 
for $\sigma_{Zh}$ plays a very important role in the EFT fit as
explained in Sec.~\ref{sec:physics}. 
The Higgs mass $m_h$ is also measured from the fit shown
in Fig.~\ref{fig:RecoilMassLep250}. The estimate of $m_h$ uncertainty 
is 14 MeV for ILC250, with the dominant contribution 
from $\mu^+\mu^-h$ channel. 
The uncertainty in the Higgs boson mass ($\delta m_h$) does play a role 
as a source of systematic error for predictions of Higgs boson couplings. 
In most cases, $\Delta m_h\sim 100$ MeV 
would be already sufficient, but this is not true
for $h\to ZZ^*$ or $h\to WW^*$. 
It has been pointed out in  \cite{Lepage:2014fla} that 
\beq
\delta_W =6.9\cdot\delta m_h,~~~~\delta_Z =7.7\cdot \delta m_h,
\eeq{eqn:MassH}
where $\delta_W$ and $\delta_Z$ are the 
relative errors for $g(hWW)$ and $g(hZZ)$ respectively. 
At ILC250, the 14 MeV accuracy for Higgs boson mass results in 
systematic errors of 0.1\% for $\delta_W$ and
$\delta_Z$.

\begin{table}[htb]
\centering
\begin{tabular}{l|c|c|c|c|c|c|c|c}
\hline
${h\to }$ & bb & cc & gg & $\tau\tau$ & $\mathrm{WW^{*}}$ & 
$ZZ^{*}$ & $\gamma\gamma$ & $\gamma Z$ \\
\hline
eff. [\%] & 88.25 & 88.35 & 87.98 & 88.43 & 88.33 & 88.52 & 88.21 & 87.64 \\
\hline
\end{tabular}
\caption{The efficiencies of the major SM Higgs decay modes,
after all the event selection cuts, shown here for the case 
of the $\mathrm{\mu^{+}\mu^{-}h}$
channel and $e_{L}^{-}e_{R}^{+}$ at $\sqrt{s}$=250 GeV~\cite{Yan:2016xyx}.
The uncertainties due to finite MC statistics on these values are below 0.14\%.}
\label{tab:recoilmass}
\end{table}

As pointed out in the very beginning of this analysis, the key idea which enables
the inclusive $\sigma_{Zh}$ measurement is that the signal is tagged
independently of Higgs decay modes. Hence it is crucial to examine whether all
the pre-selection and final-selection cuts satisfy this criterion. This can be verified 
by checking the signal efficiency for each individual Higgs decay mode
and evaluating the efficiency uniformity among all the decay modes.
Table~\ref{tab:recoilmass} lists the efficiencies of major SM Higgs decay modes 
after all cuts in the $\mu^+\mu^-h$ channel. It is seen that there is no
discrepancy in efficiencies of SM decay modes beyond 1\%. 
This is not a surprise because the analysis strategies and selection cuts
are carefully designed to make it so. The cut $E_{vis}>10$ GeV may
deserve a few more words, since it apparently suppresses the 
$h\to invisible$ mode. The strategy behind is that $\sigma_{Zh}$ can be
measured as 
\beq
\sigma_{Zh}=\sigma_{Zh}^{vis}+\sigma_{Zh}^{inv},
\eeq{eqn:sigmazh}
where $\sigma_{Zh}^{vis}$ is the total cross section for all $h\to visible$
modes, which is measured here, and $\sigma_{Zh}^{inv}$ is the cross section
for $h\to invisible$ mode, which can be measured separately, described in
Sec.~\ref{sec:higgs:invisible}. A detailed and quantitative analysis taking into 
account the possibility of existing BSM decay modes is performed in~\cite{Yan:2016xyx}.
It concludes that the relative bias on $\sigma_{Zh}$,
induced by the Higgs decay modes dependence, can be
controlled at below 0.1\% (0.2\%) for the $\mu^+\mu^-h$ ($e^+e^-h$) channel,
which is much smaller than the
expected statistical uncertainty even at the full ILC250.

In the hadronic recoil channel, a more complicated strategy is applied in order to
keep the analysis still decay modes independent. Instead of the simple categorization
into visible and invisible modes in leptonic channel, the signal events in hadronic channel
are categorized according to number of taus, number of leptons, and number of jets
in the final state. In principle, as long as the categories are 
inclusive, we can design and 
optimize the selection cuts category by category. The studies in~\cite{Tomita:2015} show
that by varying the SM decay branching ratios by $\pm 5\%$ (absolute) in each decay mode,
the bias on measured $\sigma_{Zh}$ is at most around 0.5\% relatively.  
More efforts would be needed in future to further reduce the bias to a much lower level
in particular even under assumption that there would be other unknown exotic decay modes.
At higher $\sqrt{s}$, the hadronic recoil analysis generally becomes less challenging, because
the two jets from primary $Z$ are more boosted hence are easier to 
be identified from the Higgs decay products, as studied in~\cite{Thomson:2015jda,Miyamoto:2013zva}
for $\sqrt{s}=350$ and 500 GeV.

\subsubsection{$\sigma_{\nu\nu h}$ and $\sigma_{eeh}$}
\label{subsubsec:higgs:nunuee}

The second leading Higgs production process, 
$\ee\to\nu\bar{\nu}h$ via $W$-fusion, provides a direct measurement
for $hWW$ coupling. It plays a crucial role in the global fit based on $\kappa$ formalism,
and still helps improve the global fit results based on EFT formalism even though 
the cross section is not very large at $\sqrt{s}=250$ GeV, 
$\sigma_{\nu\nu h}=$ 14 fb for $e^-_Le^+_R$. 
The signal channel used is 
$e^+e^-\to\nu\bar{\nu}h,~h\to b\bar{b}$, in which direct observable is
$\sigma_{\nu\nu h}\cdot BR_{bb}$. Together with $BR_{bb}$ measurement
by $Zh$ process, $\sigma_{\nu\nu h}$ is then measured.
The analysis is briefly described here, and more details can be found 
in~\cite{Durig:2014lfa,Tian:2017}. 

\begin{figure}
\begin{tabular}[c]{c}
\includegraphics[width=0.85\hsize]{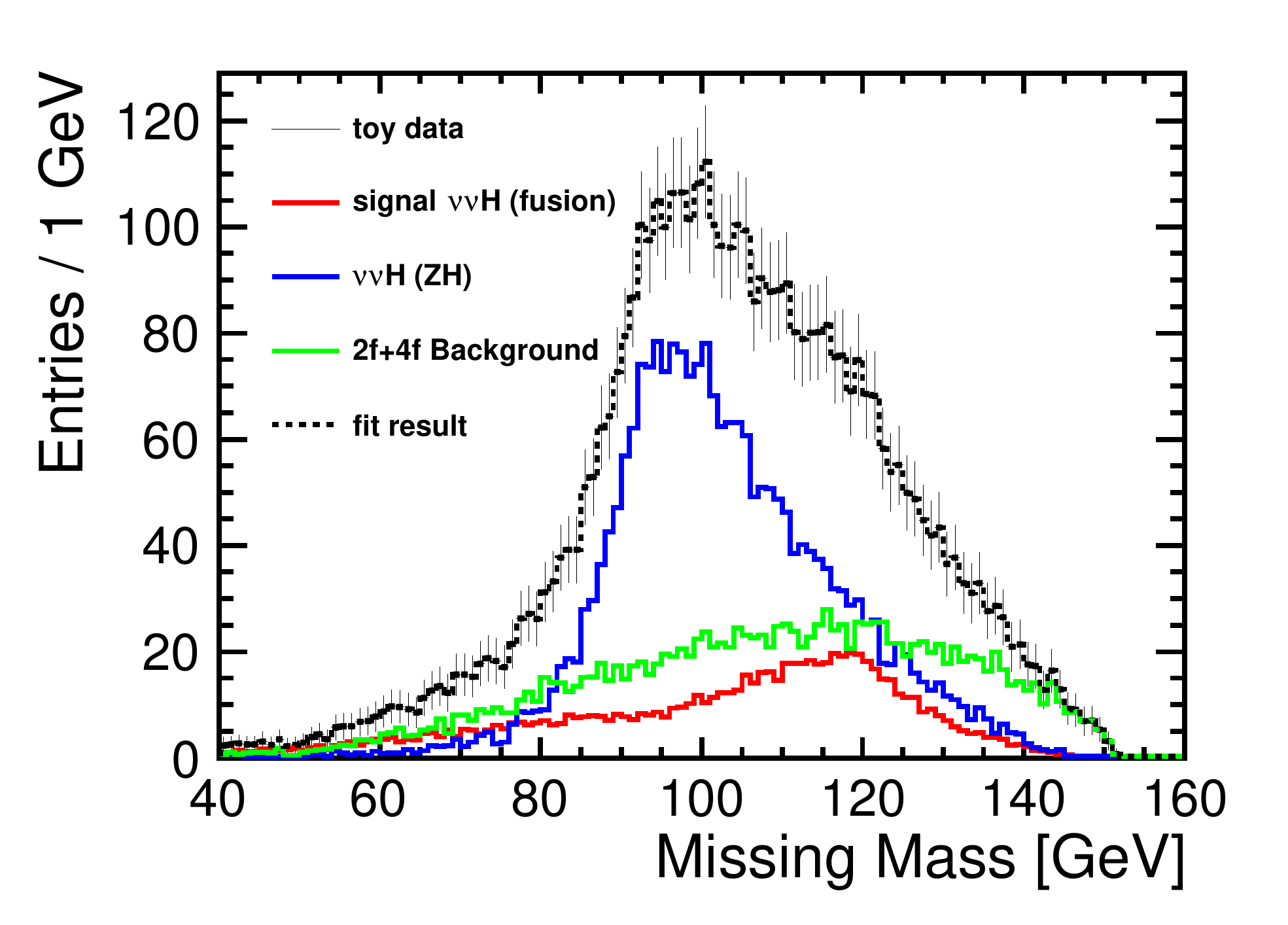} \\
\includegraphics[width=0.85\hsize]{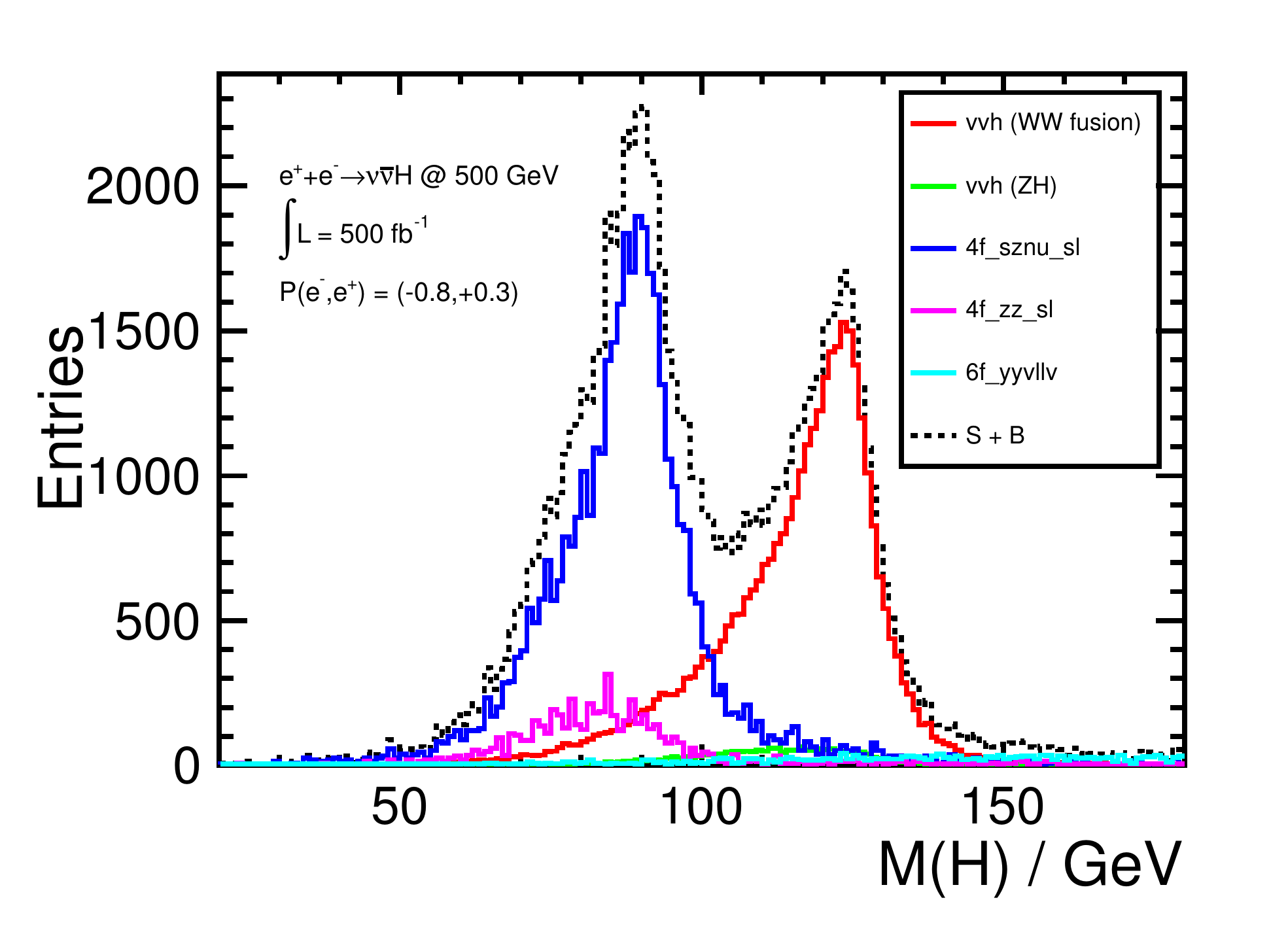}
\end{tabular}
  \caption{Missing mass spectrum (upper) and Higgs mass spectrum (lower) 
  for the signal $e^+e^-\to\nu\bar\nu h, h\to b \bar{b}$ and the SM background 
  at 250 GeV and 500 GeV respectively \cite{Durig:2014lfa,Tian:2017}.}
  \label{fig:vvHbb}
\end{figure}

\begin{figure*}[htb]
\begin{center}
\includegraphics[width=0.85\hsize]{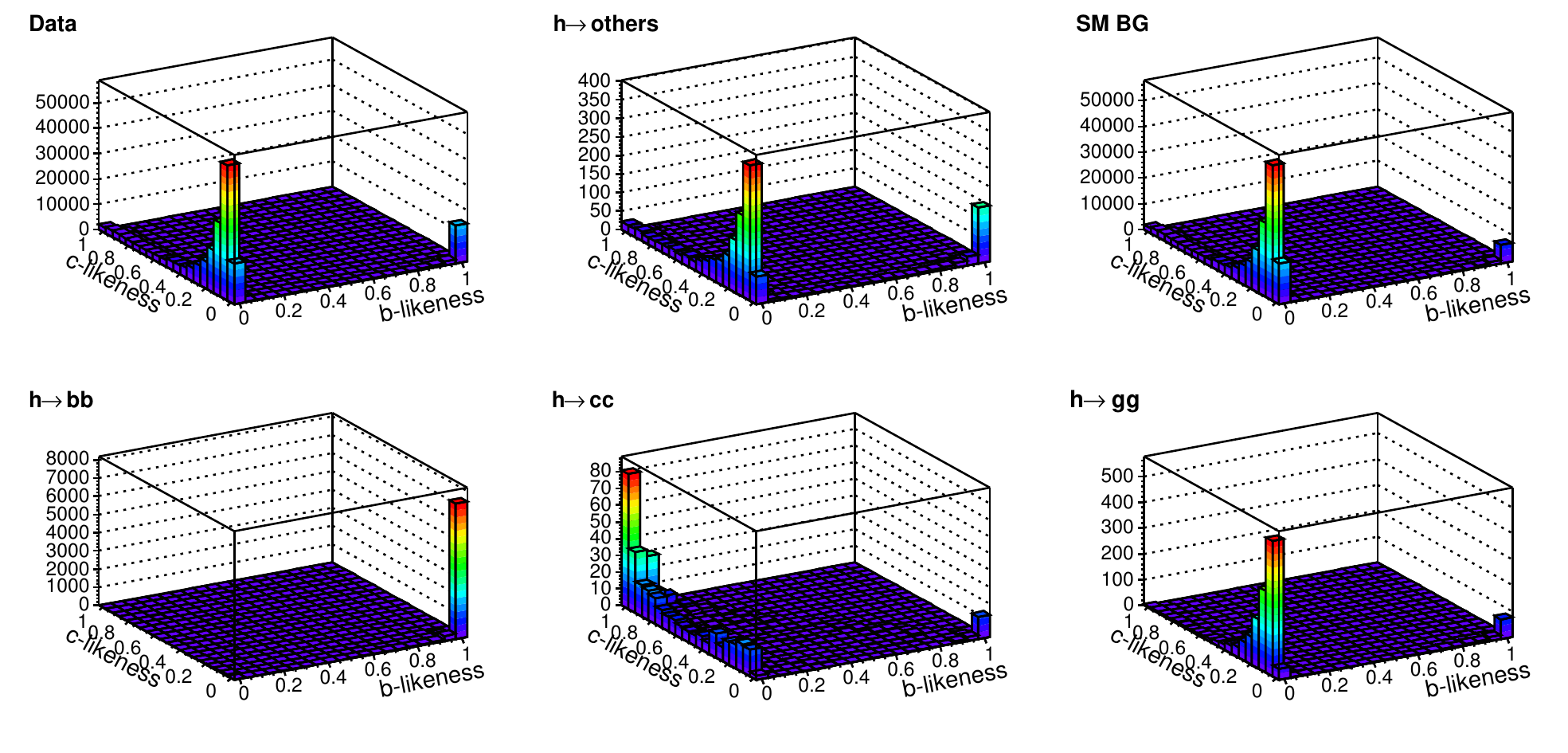}
\end{center}
  \caption{Template of b-likeliness versus c-likeness for signal $h\to b\bar{b}/c\bar{c}/gg$
  (bottom left/middle/right) events,
and for $h\to others$ / SM background (top middle/right) events, and distribution for 
all the events (top left), in $Z\to qq$ channel normalized to 250 fb$^{-1}$. The b-likeness
is defined as a combined function of the two b-tags (say $x_1$ and $x_2$) 
of the two jets from $h$ candidate: $\mathrm{b-likeness} = \frac{x_1x_2}{x_1x_2+(1-x_1)(1-x_2)}$.
The c-likeness is defined in a similar way.}
  \label{fig:qqHbbccgg250}
\end{figure*}

The signal final states consist of two b-jets and two missing neutrinos.
The pre-selection starts with vetoing events with one or more isolated 
leptons. Then jet-clustering and flavor tagging are performed with expected number of
jets equals 2. The two jets are required that in each jet there are at least 6
reconstructed particles and $Y_{3\to2}<0.1$, where $Y_{3\to 2}$ is the 
jet distance value from 3 jets to 2 jets step defined by Durham algorithm. The 
b-tagging of the two jets are required to be $btag1>0.8$ and $btag2>0.2$.
The di-jet invariant mass is required to be $m_{bb}\in[110,150]$ GeV.
The missing mass, defined as the recoil mass against the di-jet,
is required to be larger than 20 GeV.
After the pre-selection, the remaining dominant background events are
from $\gamma Z$ ($Z\to b\bar{b}$), $\nu\bar{\nu}Z$ ($Z\to b\bar{b}$) 
and $Zh$ ($Z\to\nu\bar{\nu}$, $h\to b\bar{b}$).

In the final selection, $\gamma Z$ and $\nu\bar{\nu}Z$ background 
events are further highly suppressed by a BDT cut, which is trained 
using input variables di-jet mass, polar angle of di-jet, angle between
two jets, and $Y_{2\to1}$. The remaining signal and background events
are plotted in the missing mass spectrum, shown in Fig.~\ref{fig:vvHbb} (upper).
The signal efficiency is 36\% and the average signal over background ratio is around 1/4.
The most dominant background events turn out to come from $Zh$ ($Z\to\nu\bar{\nu}$) and 
have significant overlap with signal events in the missing mass spectrum.
This is because the invariant mass of $\nu\bar{\nu}$ of signal events
can not be far away from $M_Z$, limited by available phase space at $\sqrt{s}=250$ GeV.
Therefore it is necessary to fit simultaneously $\sigma_{\nu\nu h}\cdot BR_{bb}$
and $\sigma_{Zh}\cdot BR_{bb}$. Note a useful constraint can be added into the fit
that $\sigma_{Zh}\cdot BR_{bb}$ is also measured 
using $Z\to l^+l^-$ and $Z\to q\bar{q}$ channels. 
As a result, the estimate of relative uncertainty on $\sigma_{\nu\nu h}\cdot BR_{bb}$
is 8.1\%, shown in Tab.~\ref{tab:higgserrors}, 
and the correlation between $\sigma_{\nu\nu h}\cdot BR_{bb}$ and
$\sigma_{Zh}\cdot BR_{bb}$ is -34\%.

The left-handed beam polarisation does help significantly the $\sigma_{\nu\nu h}$
measurement here, simply because it enhances the cross section by a factor of 2.34.
The $\sigma_{\nu\nu h}$ can be measured much better at $\sqrt{s}=500$ GeV, 
shown in~\ref{fig:vvHbb} (lower), thanks
to a fact of 10 increase on cross section and much easier separation with $Zh$ ($Z\to\nu\bar{\nu}$).

The third leading Higgs production process, $\ee\to e^+e^- h$ via $Z$-fusion, is not easy to measure
due to its very small cross section at $\sqrt{s}=250$ GeV, $\sigma_{eeh}=0.7$ fb. 
A full simulation analysis is performed and 
suggests that this process is already discoverable at $\sqrt{s}=250$ GeV with 2 ab$^{-1}$ and
$\sigma_{eeh}$ can be measured with a significance of 9$\sigma$~\cite{Ogawa:2018}. 
At $\sqrt{s}=500$ GeV, the significance will be significantly improved to 60$\sigma$.

\subsubsection{$\mathrm{BR}(h\to b\bar{b}/c\bar{c}/gg)$}
The capabilities of making precise measurements for $\mathrm{BR}(h\to c\bar{c}/gg)$ 
demonstrate  another unique advantage of a lepton collider, enabled by:
(1) clear separation between b-jets, c-jets and light quark/gluon jets thanks to the
excellent flavor tagging performance introduced in Sec.~\ref{sec:software}; and (2)
the democracy about cross sections between Higgs processes and
other SM background processes induced by electroweak interactions.
The branching ratios $\mathrm{BR}(h\to b\bar{b}/c\bar{c})$ offer important measurements of the 
Yukawa couplings between the Higgs boson and third/second generation quarks. 
$\mathrm{BR}(h\to gg)$ offers a direct measurement of the $hgg$ coupling,  
This is complementary to that at the LHC, where this coupling is obtained from the Higgs
production cross section, and has much smaller theoretical uncertainties. These measurements are performed
using the leading Higgs production process $\ee\to Zh$. All the major $Z$ decay
channels $Z\to l^+l^-$, $Z\to\nu\bar{\nu}$ and $Z\to q\bar{q}$ are used in the analyses;
see details in Ref.~\cite{Ono:2013sea}. 

We now discuss in more detail the analysis procedure for the 
 $Z\to q\bar{q}$ channel. 
The signal final states consist of four jets, common for $h\to b\bar{b}/c\bar{c}/gg$. 
In the pre-selection, all the particles
in each event are first clustered into four jets using Durham algorithm. The
four jets are paired into two di-jet pairs, $j_1j_2$ and $j_3j_4$, 
as for respectively $Z$ and $h$ candidates by minimizing the $\chi^2$ defined as 
$$\chi^2=(\frac{m_{j_1j_2}-M_Z}{\sigma_Z})^2+(\frac{m_{j_3j_4}-M_h}{\sigma_h})^2,$$
where $m_{j_1j_2}$ ($m_{j_3j_4}$) is the invariant mass of $j_1j_2$ ($j_3j_4$),
and $\sigma_Z$ ($=4.7$ GeV) and $\sigma_h$ ($=4.4$ GeV) are the widths
of invariant mass spectra of $Z$ and $h$ respectively determined using MC truth information. 
A cut $\chi^2<10$ is applied. In the final-selection, to suppress the leptonic or 
semi-leptonic background events, the number of
charged particles in each jet is required to be $> 4$. To suppress the $q\bar{q}$
background events, the jet clustering parameter $Y_{4\to3}$ 
is required to be consistent with 4-jet characteristic, that $\log Y_{4\to 3}>-2.7$.
In addition two cuts are applied on the event thrust and thrust angle,
that $\mathrm{thrust}<0.9$ and $|\cos\theta_{\mathrm{thrust}}|<0.9$. The remaining
background events are dominated by $q\bar{q}q\bar{q}$, mainly from hadronic decays of $WW$ and $ZZ$. 
A cut on the angle between $j_3$ and $j_4$ is applied, $105^\circ<\theta_{j_3j_4}<160^\circ$.
A kinematic fitting is performed, using four-momentum conservation constraints plus the constraint
$m_{j_1j_2}-m_{j_3j_4}=M_Z-M_h$. Then, two cuts are
applied on the fitted $Z$ and $h$ masses, 
that $m_{j_1j_2}\in[80,100]$ GeV and $m_{j_3j_4}-M_h\in[-15,10]$ GeV. As a final cut,
a multivariate likelihood is derived and required to be $\mathrm{Likelihood}>0.375$.
After all the cuts, the signal efficiency is 26\%, with an average $S/B$ ratio of around 1/10, 
including all events of $h\to b\bar{b}/c\bar{c}/gg$ (note the $S/B$ ratio for $h\to b\bar{b}$ events
is much higher). 

A template fit is then performed to extract the numbers of signal events $h\to b\bar{b}$,
$h\to c\bar{c}$ and $h\to gg$ respectively, 
for which it is crucial that the different signal events are distinguishable with themselves 
as well as with background events. 
The templates are constructed as 3-D histograms using
3 variables, namely b-likeness, c-likeness and bc-likeness defined for the two jets 
$j_3$ and $j_4$ (as from $h$ candidate). Five templates are made using separated MC samples:
signal $h\to b\bar{b}$, $h\to c\bar{c}$, $h\to gg$, SM background and $h\to\mathrm{others}$ background. 
The projected 2-D templates for b-likeness versus c-likeness are shown in Fig.~\ref{fig:qqHbbccgg250}, 
each of which has been normalised to an integrated luminosity of 250 fb$^{-1}$.
It demonstrates that the three types of signal events can indeed be clearly distinguished
with themselves and with background events, thanks to
the excellent flavor tagging performance and good signal over background ratio.

We just used $Z\to q\bar{q}$ channel to illustrate the analysis, it is worth commenting that 
$Z\to\nu\bar{\nu}$ channel is as powerful as $Z\to q\bar{q}$ channel despite its branching ratio
is a factor of 3 smaller. This is largely due to the factor that the signal and background 
discrimination in $Z\to q\bar{q}$  channel is much degraded by performance of 
the realistic jet clustering and jet pairing algorithms at now, 
as a result of which the S/B ratio in $Z\to q\bar{q}$ channel is a factor of 5
lower than that in $Z\to\nu\bar{\nu}$ channel. From the perspective of a better jet clustering
or jet pairing algorithm in future, the analysis in $Z\to q\bar{q}$ channel can be 
significantly improved.

By combining $Z\to q\bar{q}/\nu\bar{\nu}/l^+l^-$ channels, 
the estimates of statistical uncertainties for $\sigma_{Zh}\cdot BR_{bb}$,
$\sigma_{Zh}\cdot BR_{cc}$ and $\sigma_{Zh}\cdot BR_{gg}$ are respectively
1.3\%, 8.3\% and 7.0\%, shown in Table~\ref{tab:higgserrors}.

\subsubsection{$\mathrm{BR}(h\to WW^*/ZZ^*)$}
The measurements of branching ratios of $h\to WW^*/ZZ^*$ play an important role
in the global fit as the Higgs total width is determined by
$$\Gamma_h=\frac{\Gamma_{WW}}{BR_{WW}}=\frac{\Gamma_{ZZ}}{BR_{ZZ}}.$$
Depending on how each $W/Z$ decays and how Higgs is produced, there
are quite many signal channels that can be used. The analysis strategies as well as
the signal background discrimination also vary quite a lot channel by channel.
For $Zh$ production and $h\to WW^*$, 
the signal channels are listed in Table~\ref{tab:ZhWWchannels}, where the channels
with marks are studied based on full simulation and enter the combined estimate of 
statistical uncertainty. The details of event selections can be found in~\cite{Ono:2012,Barklow:2017,Liao:2017}.
One of the dominant background processes in all channels is $e^+e^-\to W^+W^-$, suppression of which 
can be helped by the right-handed beam polarisations.
Due to the multiple jets in the signal final states, the analysis could also benefit significantly 
from an improved jet clustering algorithm in future.
The estimate of statistical uncertainties for $\sigma_{Zh}\cdot BR_{WW}$ is
4.6\%, shown in Table~\ref{tab:higgserrors}. It's worth noting from Table~\ref{tab:ZhWWchannels}
that there are still many more channels which yet to be employed in full simulation in future 
to improve the $\sigma_{Zh}\cdot BR_{WW}$ measurement, in particular the fully hadronic channel 
 $Z\to q\bar{q}$ and $WW^*\to qqqq$ that has the largest branch ratio.

\begin{table}
\begin{center}
\begin{tabular} {lcccccc}
$h\to$ / $Z\to$  & $l^+l^-$ &  $\nu\bar{\nu}$ & $q\bar{q}$ \\
\hline
$WW^*\to qqqq$        & 3.0\%$^{**}$    & 9.0\%$^{*}$   & 31\% \\
$WW^*\to qql\nu$      & 2.0\%    & 5.8\%    & 20\%$^{*}$ \\
$WW^*\to l\nu l\nu$   & 0.3\%    & 1.0\%    & 3.3\%
\end{tabular}
\caption{Signal channels in $e^+e^-\to Zh,~h\to WW^*$ and their branching ratios. 
The entries marked with * or ** are currently studied by full simulation and enter the combined result. 
The entry marked with ** is based on CEPC studies.}
\label{tab:ZhWWchannels}
\end{center}
\end{table}

\begin{figure}
\begin{center}
\includegraphics[width=0.85\hsize]{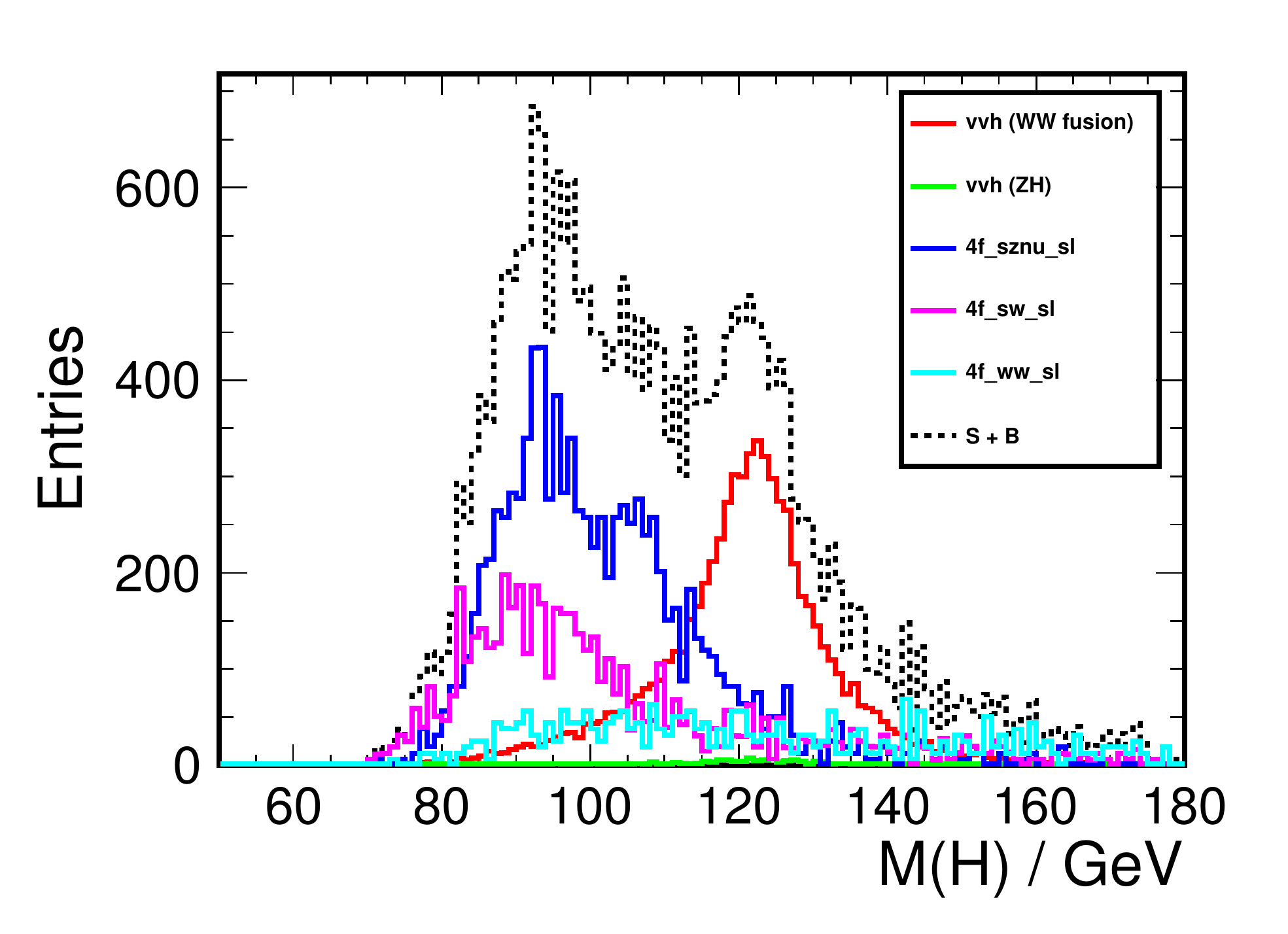}
\end{center}
  \caption{Higgs mass spectrum for the signal $e^+e^-\to\nu\bar\nu h, h\to WW^{*}\to qqqq$
  and the SM background events, normalized to 500 fb$^{-1}$
  at 500 GeV \cite{Durig:2014lfa}.}
  \label{fig:vvHWW500}
\end{figure}

\begin{figure}
\begin{center}
\includegraphics[width=0.85\hsize]{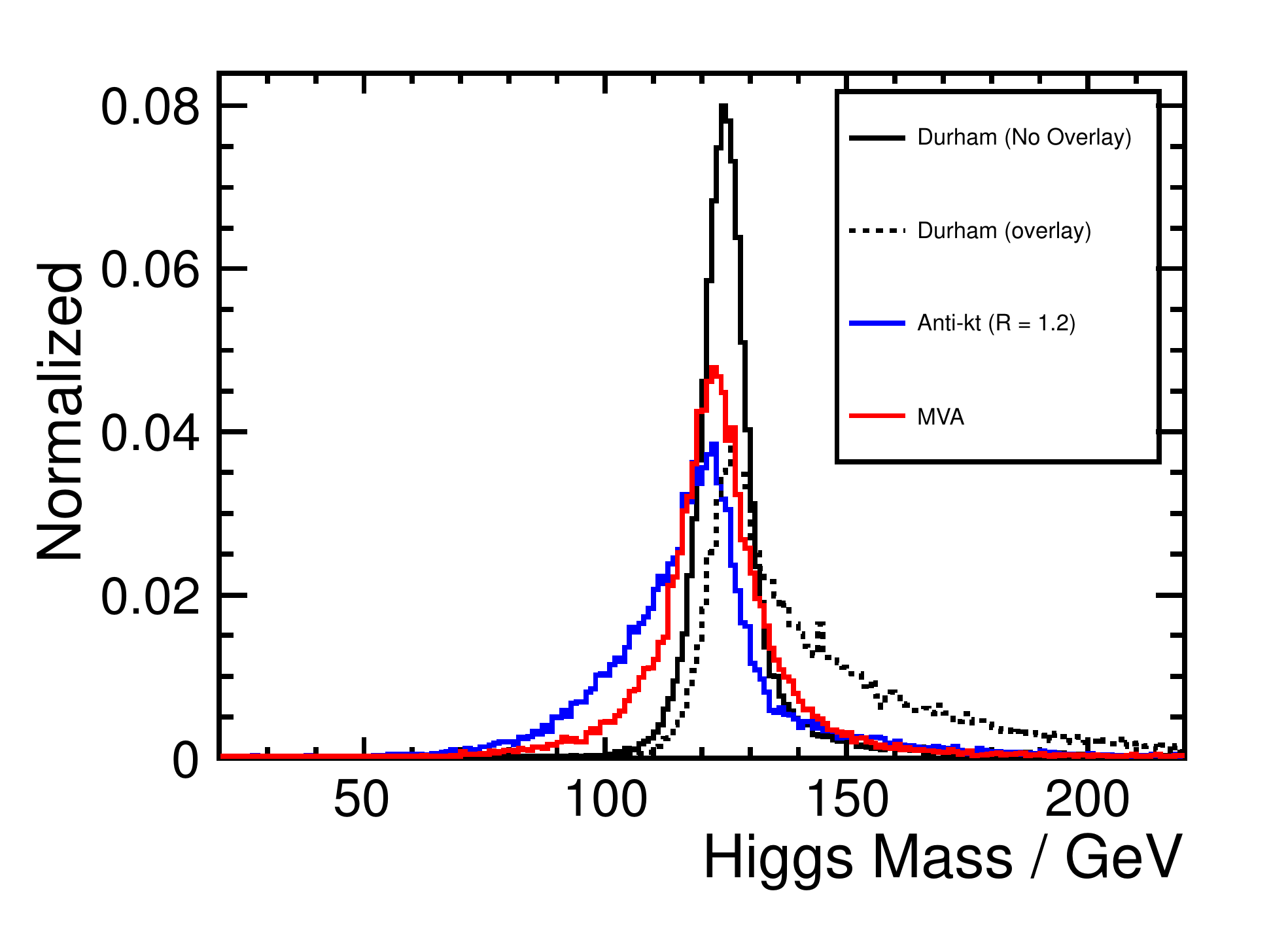}
\end{center}
  \caption{Higgs mass spectrum for the signal $e^+e^-\to\nu\bar\nu h, h\to WW^{*}\to qqqq$
   in different cases of overlay: no overlay (solid black) 
  illustrating the performance of perfect overlay removal; without overlay and without
  applying any removal algorithm (dashed black); overlay removal with traditional 
  exclusive jet clustering algorithm, anti-$k_T$ here (blue); overlay removal with a new
  method based on MVA, optimized for this particular signal channel (red). The histograms 
  are normalized to 500 fb$^{-1}$ at $\sqrt{s}=$500 GeV \cite{Durig:2014lfa}.}
  \label{fig:vvHWW500ovl}
\end{figure}

At $\sqrt{s}=500$ GeV, the $BR(h\to WW^*)$ measurement can also be improved significantly by including 
Higgs production via $W$-fusion. Two signal channels $e^+e^-\to\nu\bar{\nu}h$, $h\to WW^*\to qqqq / qql\nu$
have been studied based on full simulation. As an illustration, 
the remained signal and background events in $WW^*\to qqqq$ channel
after all cuts are plotted in Fig.~\ref{fig:vvHWW500} in the reconstructed $h$ mass spectra. The signal peak
can be clearly observed with the dominant background from $\nu\bar{\nu}h$ ($h\to others$), 
$\nu\bar{\nu}Z$ and $W^+W^-$. The average S/B ratio is around $1/1.6$ in the $M(h)\in(114,142)$ GeV 
region. The estimate of statistical uncertainty for $\sigma_{\nu\bar{\nu}h}\cdot BR_{WW}$ is
3.4\% with 250 fb$^{-1}$ at $\sqrt{s}=500$ GeV, shown in Table~\ref{tab:higgserrors}.
It is worth noting here the significant impact of overlay events. Figure~\ref{fig:vvHWW500ovl} 
shows the reconstructed $h$ mass spectra for signal events in cases of no overlay, with overlay but using
inclusive Durham jet clustering algorithm, overlay removal using anti-$k_T$ algorithm, and overlay
removal using a new MVA based algorithm; see details in~\cite{Durig:2014lfa}. 
It can be said that the performance of $h$ mass reconstruction,
in the realistic case even with an optimized overlay removal algorithm to date, is still far away from the 
perfect case of no overlay. Therefore the $\sigma_{\nu\bar{\nu}h}\cdot BR_{WW}$ measurement will 
benefit a lot from a better overlay removal algorithm in future.

For $BR(h\to ZZ^*)$ measurement, in general it is more challenging due to its small branching ratio.
Though the analysis can be done similarly by combining analyses optimized in many individual channels, 
a different strategy was used in~\cite{Asner:2013psa}.
All the signal events are selected against background with a single multivariate method using many
variables as input. As a result, the estimate of statistical uncertainty for
$\sigma_{Zh}\cdot BR_{ZZ}$ is 18\%, shown in Tab.~\ref{tab:higgserrors}.

\subsubsection{$\mathrm{BR}(h\to \tau^+\tau^-)$}
The measurement of $BR_{\tau\tau}$ provides a very important probe of the Higgs couplings to
third generation fermions. And it is going to be one of the most precise Higgs measurements at the ILC, 
thanks to the relatively large branching ratio and very clean signal and background separation. 
The full simulation is performed using the leading Higgs production process $e^+e^-\to Zh$ and 
all the decay channels from $Z\to q\bar{q} / \nu\bar{\nu}/ l^+l^-$; see details in~\cite{Kawada:2015wea}.
The $\tau$ is reconstructed using TaFinder and the four momenta of missing neutrinos are 
calculated using collinear approximation. The remained signal and
 background events in $Z\to q\bar{q}$ channel are shown in 
 Fig.~\ref{fig:qqHtautau250}. The S/B ratio is higher than 2/1. The signal efficiency is 36\% and
 the dominant background is from $e^+e^-\to ZZ\to q\bar{q}\tau^+\tau^-$.
 The estimate of statistical uncertainty for $\sigma_{Zh}\cdot BR_{\tau\tau}$ is 
3.2\%, shown in Table~\ref{tab:higgserrors}.

\begin{figure}
\begin{center}
\includegraphics[width=0.85\hsize]{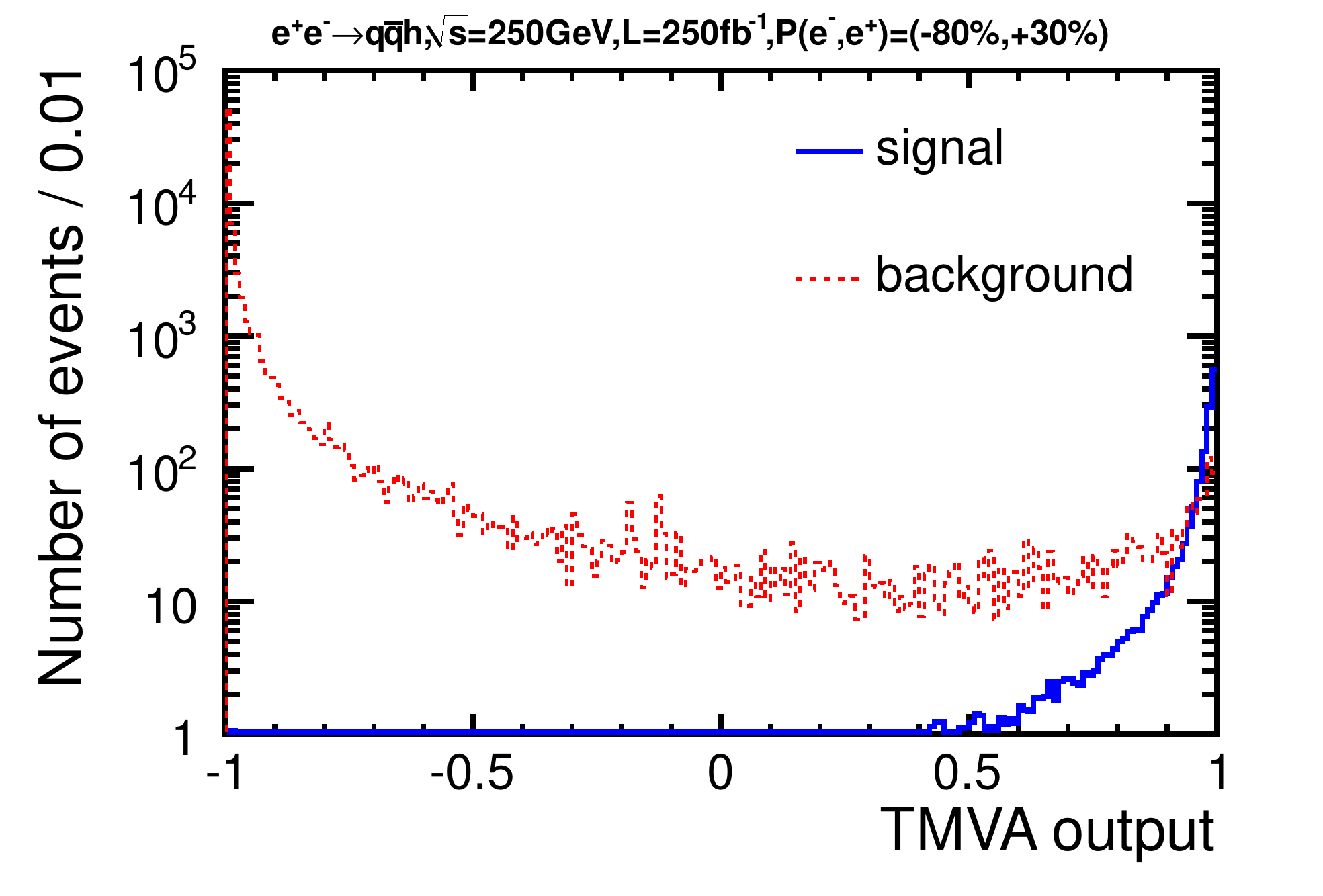}
\end{center}
  \caption{MVA output for the signal $e^+e^-\to q\bar{q} h, h\to\tau^+\tau^-$
 and the SM background at 250 GeV \cite{Kawada:2015wea}.}
  \label{fig:qqHtautau250}
\end{figure}

\subsubsection{$\mathrm{BR}(h\to \mathrm{invisible/exotic})$}
\label{sec:higgs:invisible}
As introduced in~\ref{sec:higgs:sigmazh}, the recoil technique enables that 
 Higgs events can be tagged without looking into the Higgs decay products. 
This feature is extremely useful for probing Higgs to invisible or other exotic decays.
Full simulation studies are performed for $e^+e^-\to Zh$, $h\to\mathrm{invisible}$
using two signal channels $Z\to q\bar{q}$ and $Z\to l^+l^-$; 
see details in~\cite{Ishikawa:2014,Tian:2015,Kato:2016}. The dominant contribution comes
from $Z\to q\bar{q}$ channel. After all the cuts,  the recoil mass spectrum for the 
remaining signal and background events 
is plotted in Fig.~\ref{fig:qqHinv250}. The main background events come from $ZZ/\nu\bar{\nu}Z\to\nu\bar{\nu}q\bar{q}$ 
and $WW\to qql\nu$. The signal
peak would be seen clearly for the value  $BR(h\to\mathrm{invisible})=10\%$ assumed 
in the figure.  The actual sensitivity is much greater.  We estimate the 
95\% C.L. upper limit for $BR(h\to invisible)$
to be 0.86\% for the left-handed beam polarisations, as shown in Tab.~\ref{tab:higgserrors}. 
An upper limit factor of 1.5 lower can be obtained for the right-handed beam polarisations,
thanks to the much reduced background level.

Other exotic decays have not been studied based on full simulation. Nevertheless
according to the fast simulation results in Ref.~\cite{Liu:2016zki}, at ILC250 we would be able 
to probe partially visible exotic decays with branching ratios of $10^{-3}$ or below.

\begin{figure}
\begin{center}
\includegraphics[width=0.85\hsize]{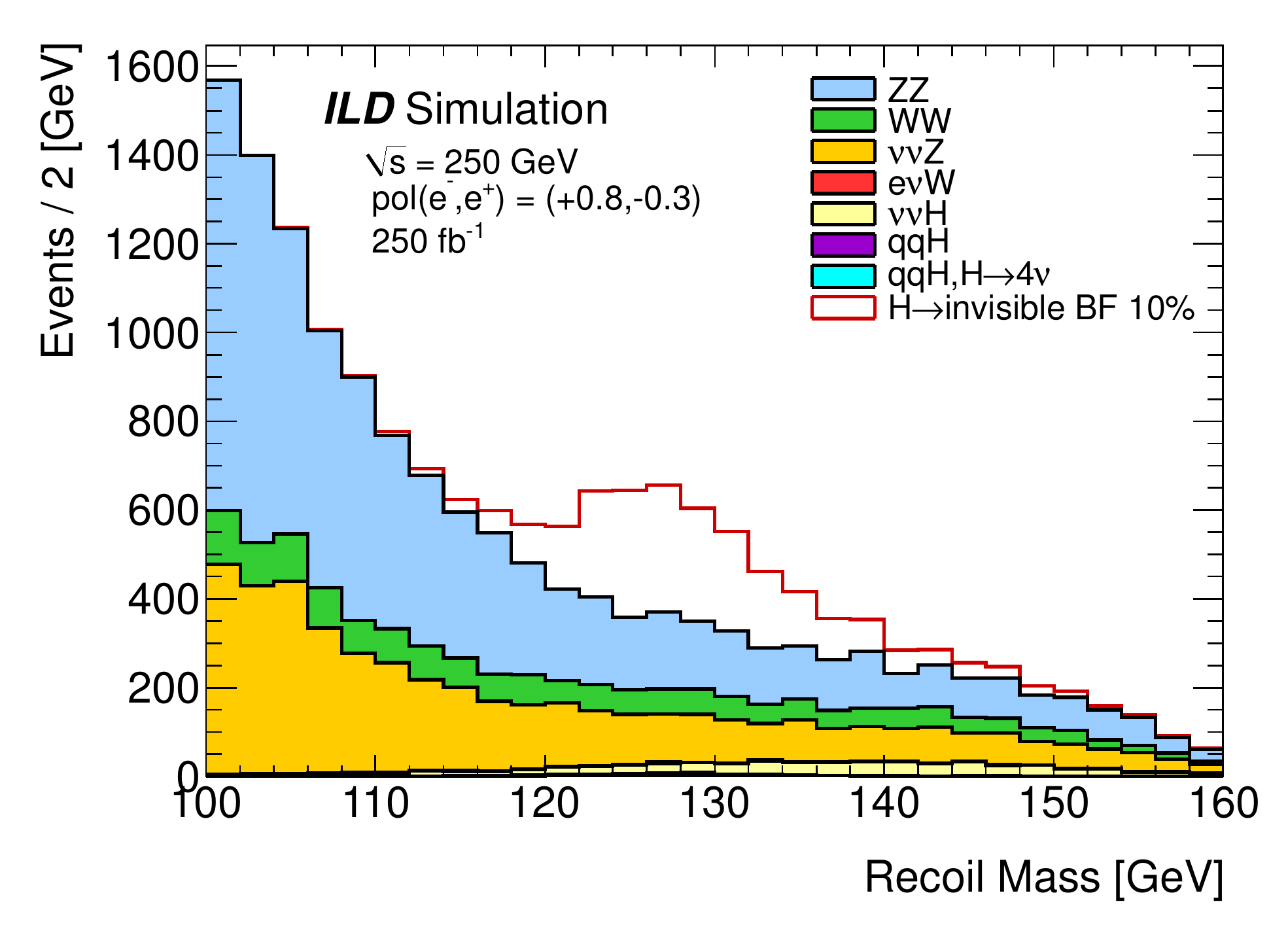}
\end{center}
  \caption{Recoil mass spectrum against
 $Z\to q\bar{q}$ for signal events $e^+e^-\to Zh, h\to invisible$ assuming $BR(h\to invisible)=10\%$
  and SM background events at 250 GeV for the right-handed beam polarisation \cite{Ishikawa:2014}.}
  \label{fig:qqHinv250}
\end{figure}

\subsubsection{$\mathrm{BR}(h\to \mu^+\mu^- / \gamma\gamma / \gamma Z)$}
The measurement of the SM rare decay branching ratios
$\mathrm{BR}(h\to \mu^+\mu^- / \gamma\gamma / \gamma Z)$
are a bit challenging at the ILC, mainly due to the limited number of signal events. 
We expect significant contributions from HL-LHC for these measurements.
Full simulations are performed in~\cite{Kawada:2018wyz,Calancha:2013}, 
and the estimates of statistical uncertainties for
$\sigma_{Zh}\cdot BR_{\mu\mu}$ and $\sigma_{Zh}\cdot BR_{\gamma\gamma}$
are respectively 72\% and 34\%, shown in Table~\ref{tab:higgserrors}. 
$BR(h\to\gamma Z)$ is also studied based on full simulation~\cite{Fujii:2018}, 
a significance of 2$\sigma$ would be expected with full ILC250.

\subsubsection{Higgs CP properties}
\label{subsubsec:higgstautauCP}
Higgs CP properties can be measured via the
$h\tau\tau$ coupling at the tree level, 
\beq
\Delta{\cal L}_{h\tau\tau}=-\frac{\kappa_\tau y_\tau}{\sqrt{2}}h{\tau^+}
(\cos\Psi_{CP}+i\sin\Psi_{CP}\gamma_5)\tau^-,
\eeq{eqn:CPHtautau}
where the CP phase angle $\Psi_{CP}$ is determined using the transverse spin
correlation between the two $\tau$, as shown in Fig.~\ref{fig:qqHtautauCP} (upper)
in the $\Delta\phi$ (angle between transverse spins of two $\tau$)
distribution for different values of $\Psi_{CP}$. 
The spin of each $\tau$ is estimated using
the polarimeter vector which can be fully reconstructed in some of $\tau$ decay modes,
such as $\tau\to\pi\nu/\rho\nu$,
taking advantage of precise measurements for impact parameters; see the method detail
in~\cite{Jeans:2015vaa}. Full simulation studies are performed using signal channels 
$Z\to q\bar{q}/l^+l^-$ and $h\to \tau^+\tau^-$ in~\cite{Jeans:2018anq}. 
Figure~\ref{fig:qqHtautauCP} shows the distribution of reconstructed $\Delta\phi$
for the remained signal and background events in one of the golden event categories. 
The estimate of statistical uncertainties for CP phase angle is $4.3^\circ$ with full ILC250.
Note that the Higgs CP violating effects can also be probed in $hZZ$ coupling using the
$\tilde{b}$ parameter shown in next section.

\begin{figure}
\begin{tabular}[c]{c}
\includegraphics[width=0.85\hsize]{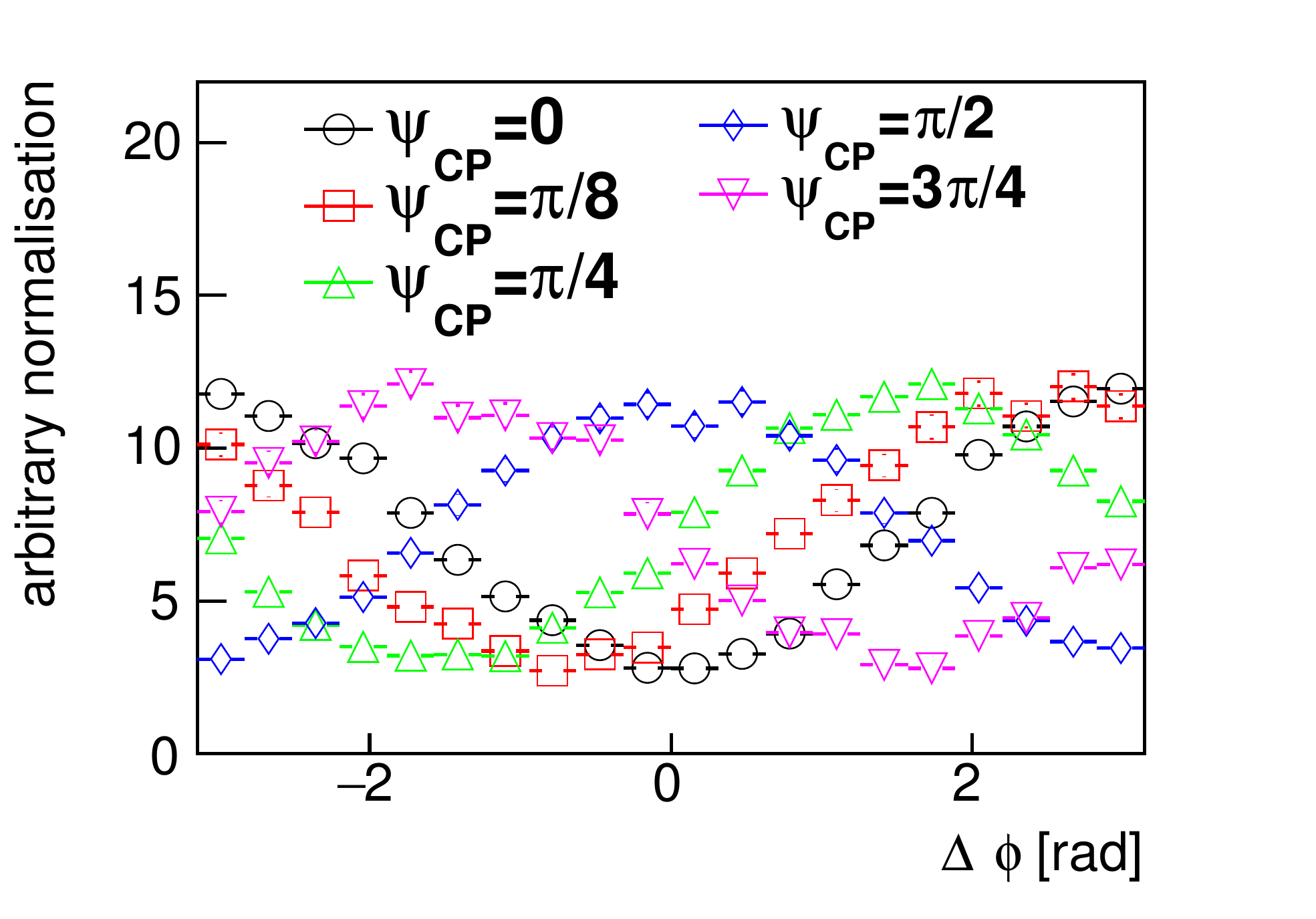} \\
\includegraphics[width=0.85\hsize]{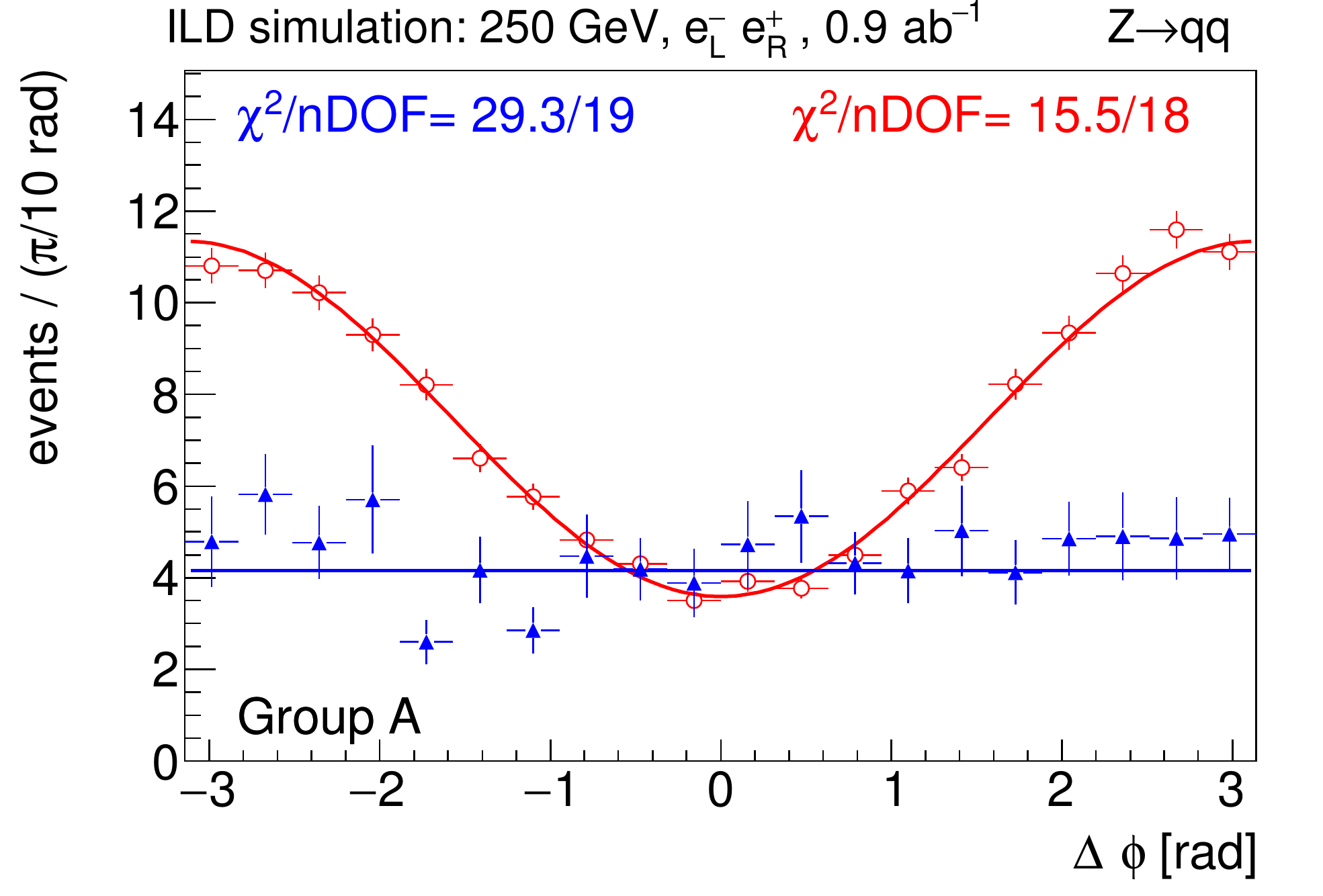}
\end{tabular}
  \caption{Upper: $\Delta\phi$ distributions at MC Truth level for different 
  values of $\Psi_{CP}$ for the signal $e^+e^-\to q\bar{q} h, h\to\tau^+\tau^-$ at 250 GeV;
  Lower: reconstructed $\Delta\phi$ distributions after all cuts in the dominant category
  for the SM signal (in red) and the SM background (blue) respectively \cite{Jeans:2018anq}.}
  \label{fig:qqHtautauCP}
\end{figure}

\subsubsection{Angular analyses for anomalous $HVV$ couplings}
The $hZZ$ coupling can be deviated from SM not only in total strength but also
in Lorentz structures, which can be detected by measuring differential cross sections.
Full simulation studies are performed using $e^+e^-\to Zh$ events 
for measuring following effective $hZZ$ couplings:
\beq
\Delta{\cal L}_{hZZ}=(1+a)\frac{m_Z^2}{v}hZ_\mu Z^\mu+
\frac{1}{2}\frac{b}{v}hZ_{\mu\nu}{Z}^{\mu\nu}+
\frac{1}{2}\frac{\tilde{b}}{v}hZ_{\mu\nu}\tilde{Z}^{\mu\nu},
\eeq{eqn:CPHZZ}
where the first $a$-term is a rescaling of SM $hZZ$ coupling, the second $b$-term and 
the third $\tilde{b}$-term represent respectively anomalous CP-even and CP-odd 
$hZZ$ couplings. The total cross section $\sigma_{Zh}$ is sensitive to both $a$ and $b$
parameters, but $b$ is distinguished from $a$ in the differential cross sections; see
Fig.~\ref{fig:ZHanomHVV1} (upper) how $Z$ production angle depends on values of $b$.
$\sigma_{Zh}$ depends on $\tilde{b}$ rather weakly, only quadratically. But the angle 
between $Zh$ production plane and $Z$ decay plane, namely $\Delta\Phi$, is very
sensitive to $\tilde{b}$; see Fig.~\ref{fig:ZHanomHVV1} (lower) for $\Delta\Phi$ distributions for
different values of $\tilde{b}$. The analysis details can be found in~\cite{Ogawa:2017bmg}.
The estimate of statistical uncertainties for $a$ and $b$ are 
0.076 and 0.027 respectively, with a large correlation $\rho=-99.17\%$, 
shown in Table~\ref{tab:higgserrors}. This large correlation can be significantly reduced
by measurements at $\sqrt{s}=500$ GeV, as shown in Fig.~\ref{fig:ZHanomHVV2},
because the effect of $b$-term is momentum dependent.
The CP violating parameter $\tilde{b}$ can be determined with a statistical uncertainty of 
0.004 for the full ILC250, with almost no correlation with $a$ or $b$.
\begin{figure}
\begin{tabular}[c]{c}
\includegraphics[width=0.85\hsize]{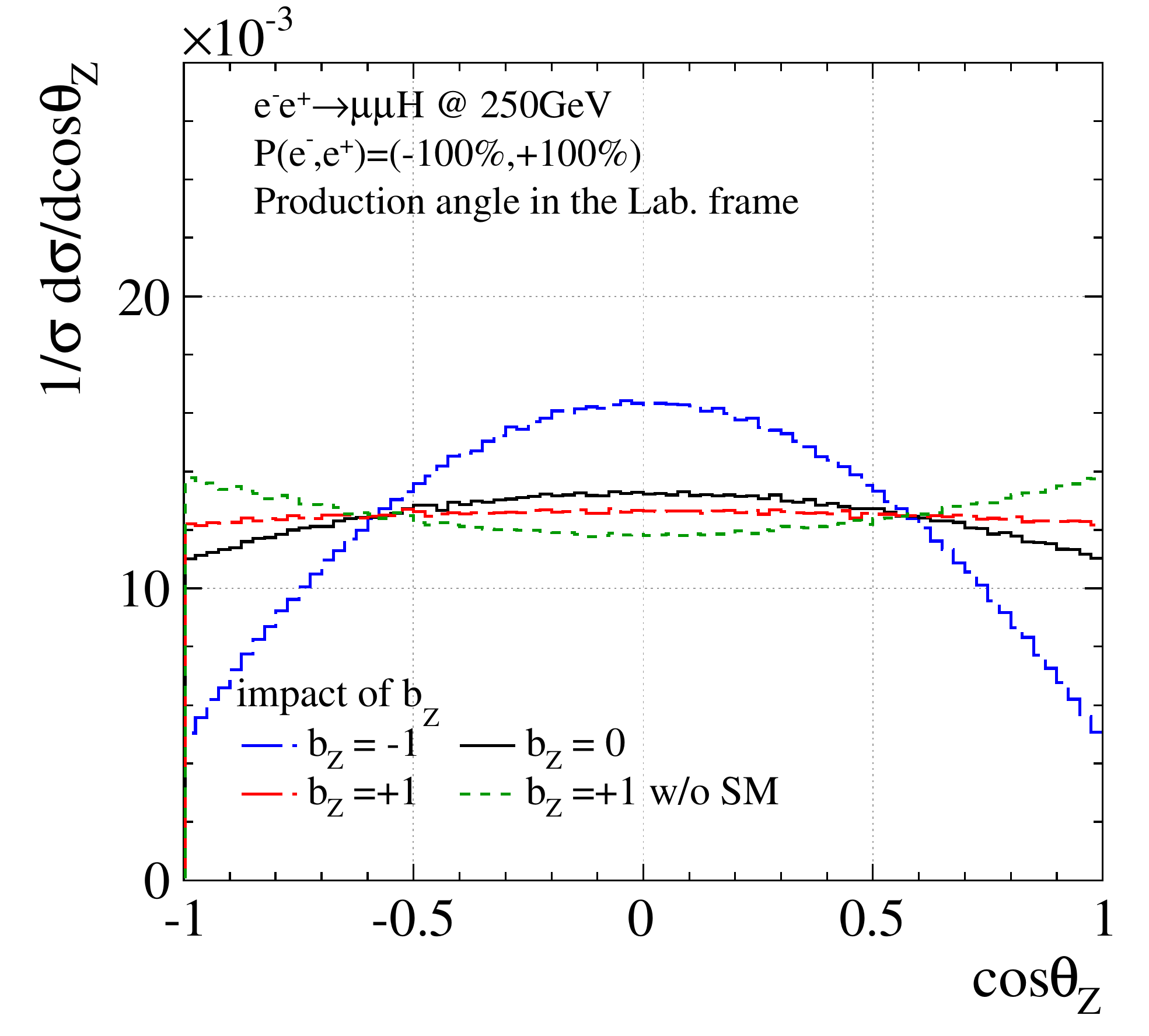} \\
\includegraphics[width=0.85\hsize]{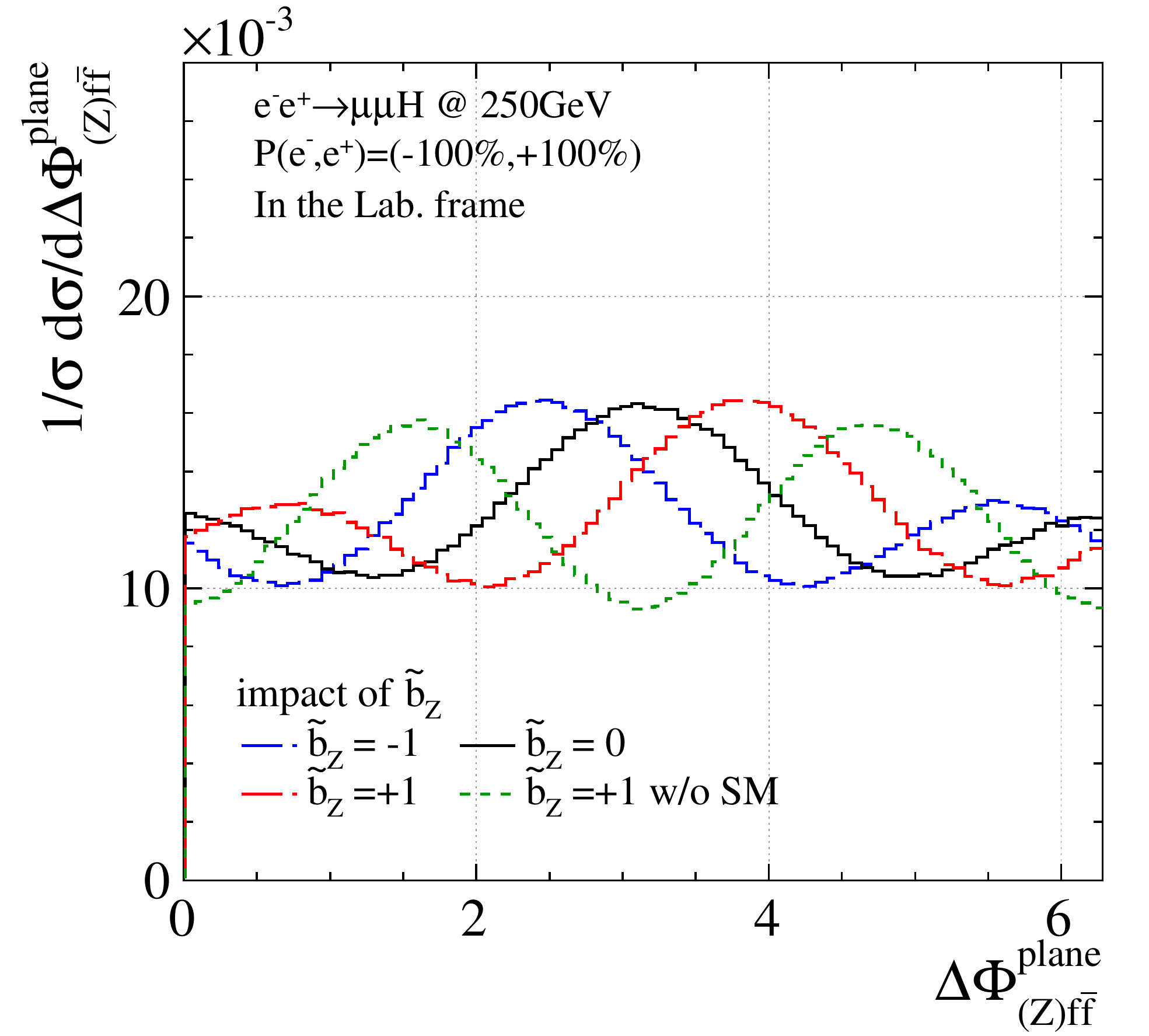} 
\end{tabular}
  \caption{Upper (lower): $\cos\theta_Z$ ($\Delta\Phi$) distributions at MC Truth level for different 
  values of anomalous coupling $b$ ($\tilde{b}$) 
  for the signal $e^+e^-\to \mu^+\mu^- h, h\to everthing$ at 250 GeV;
  \cite{Ogawa:2017bmg}.}
  \label{fig:ZHanomHVV1}
\end{figure}

\begin{figure*}
\begin{center}
\includegraphics[width=0.85\hsize]{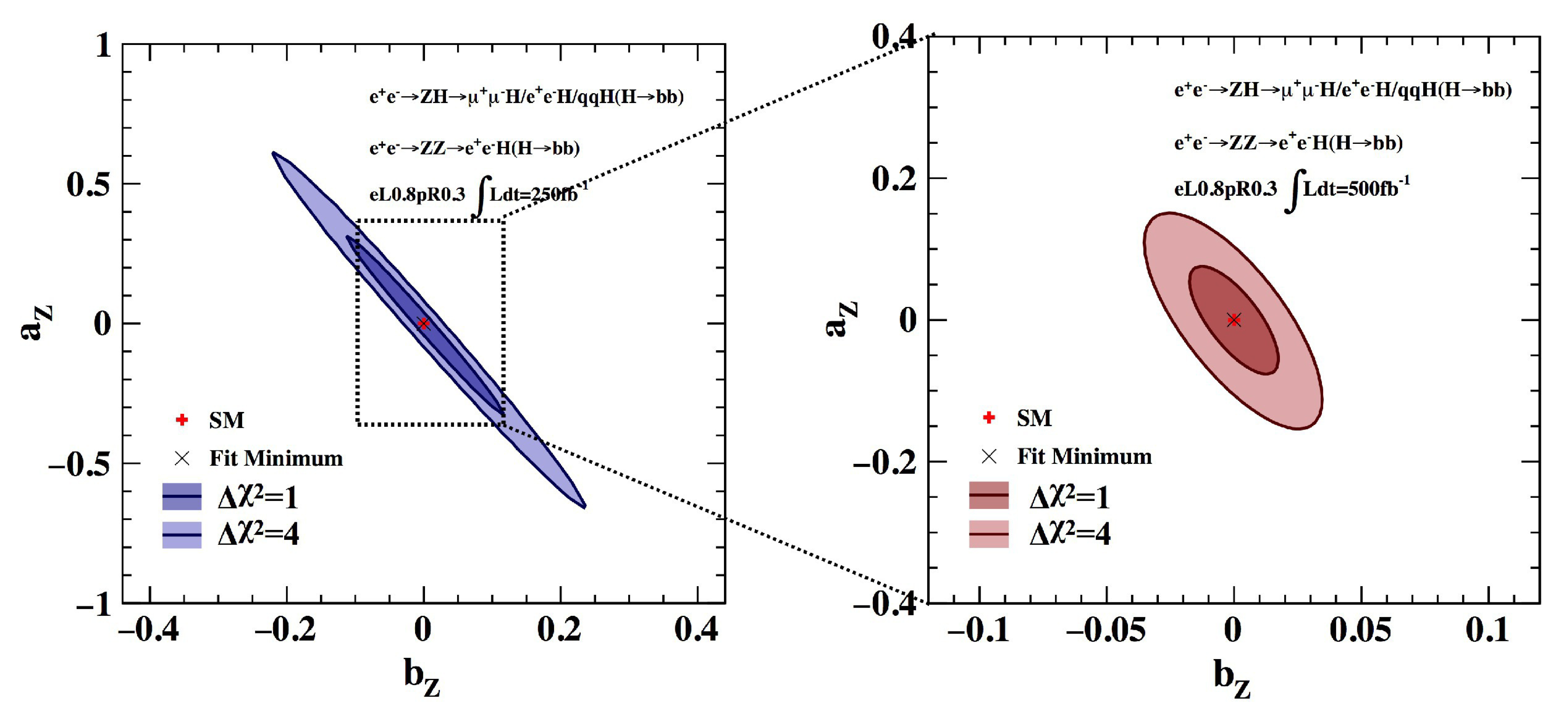}
\end{center}
  \caption{68\% and 95\% C.L. contour plots for fitted parameter $a$ versus $b$ at 250 GeV (left)
  and 500 GeV (right) \cite{Ogawa:2017bmg}.}
  \label{fig:ZHanomHVV2}
\end{figure*}



\subsection{Systematic uncertainties, and the importance of beam polarisation}
\label{subsec:polarisation}

  For the studies of the Higgs boson in which we wish to claim that precisely measured deviations from the SM can give a discovery of new physics, we must be certain that systematic errors are both small and well-constrained.  In this section, we will discuss 
the sources of systematic error that we consider in our Higgs coupling analysis.

We will also discuss the capability that the ILC gives to control systematic uncertainties using the availability of polarized beams. In precision measurement, it is useful, wherever possible, to  measure effects correlated to sources of systematic uncertainty.  For this, it is crucial to always have one more degree of freedom that (statistically) absolutely required.  In the ILC program, electron and positron polarisation
provide tools to validate estimates of systematic errors, and to reduce these 
sources of uncertainty.  In Sec.~\ref{subsec:beampol}, when we reviewed in general terms the importance of the
 use of beam polarisation to meet the physics goals of the ILC, we did not emphasize this 
aspect of the physics implication of polarisation.  But it is clear that, for each measurement that can be done at an unpolarized collider, a collider with control of the polarization for each beam can provide four independent data sets.  We will explain in this section how this tool 
can be used not only to estimate but also to reduce systematic errors.

\subsubsection{Systematic uncertainties considered in the Higgs coupling fit}
\label{subsubsec:sysuncert}
The evaluation of systematic uncertainties for experiments which have not yet been built is a difficult task and will to some extent always remain guess-work until real data have been taken. To some extent, we can rely on the experience from previous $e^+e^-$ experiments, especially at LEP, where many uncertainties could be controlled to a typical level of 1\%.
The ILC detector designs, which aim for higher precision, make use of this experience,
as explained in Sec.~\ref{sec:detectors}.  Assuming this basic level of performance, detailed studies of systematic uncertainties at the ILC have concentrated on cases where the statistical uncertainties are expected to be significantly below 1\%, and on searches in channels with large irreducible backgrounds. An example for the first case is a global analysis of total rates and differential distributions of various 2-fermion and 4-fermion SM processes, extracting simultaneously the total unpolarised cross sections, the relevant left-right asymmetries, the beam polarisations and the charged triple gauge couplings, see Sec.~\ref{subsec:ew_WWana} and Ref.~\cite{bib:PhDRobert}. An illustrative example for the second category, though not directly connected to Higgs physics, is the WIMP search in the mono-photon channel, see Sec.~\ref{subsec:searches_monophoton} and Ref.~\cite{Habermehl:417605}. 

Studies of this type lead us to the following estimates of the dominant systematic uncertainties.  These sources of systematic uncertainty are also applied to the measurements of triple gauge boson couplings described 
in Sec.~\ref{subsec:ew_WWana}.
\begin{itemize}
\item The luminosity at the ILC will be measured from low-angle Bhabha scattering with the help of a dedicated forward calorimeters, the LumiCals (see Sec.~\ref{subsub:det:forward} and Ref.~\cite{Abramowicz:2010bg}). This measurement is extremely sensitive to the exact alignment of the LumiCals on the two sides of the detector, as well as to beam backgrounds and has been studied in detailed simulations both for the ILC and for CLIC~\cite{Bozovic-Jelisavcic:2014aza, Lukic:2013fw}. Based on these studies, the resulting systematic uncertainty on all Higgs cross section and cross-section-times-braching-ratio measurements is assumed to be 0.1\%
\item Another 0.1\% is assumed for the net systematic effect of the finite knowledge of luminosity-weighted long-term average values of the beam polarisations at the $e^+e^-$ interaction point.    Compton polarimeters in the Beam Delivery System will provide 
time-stable measurements of the beam polarisations at their locations at the level of 0.25\%~\cite{Vormwald:2015hla, List:2015lsa}.  To obtain the polarisations relevant for 
the experiments, one must also consider also the effects of spin transport, misalignment of beam line magnets as well as depolarisation during the beam-beam interaction~\cite{Beckmann:2014mka}. The absolute scale of the luminosity-weighted average polarisation at the IP is finally calibrated from collision data, \eg, from a global fit to SM processes with strong polarisation dependence~\cite{bib:PhDRobert}. 
\item Theoretical uncertainties are also assumed to have reached  the level of 0.1\% by the time of ILC operation.   This requires the computation of all relevant processes to 2 loops in the electroweak interactions, a task feasible within the current state of the art~\cite{Blondel:2019qlh}.  Another question is the availability of high-precision values for the most important input parameters---$m_b$, $m_c$, $\alpha_s$, and $m_h$.  We expect to obtain the first three  of these to sufficient accuracy from lattice QCD~\cite{Lepage:2014fla}.
For $m_h$, the  ILC recoil measurement described in Sec.~\ref{sec:higgs:sigmazh} will provide 
the high precision needed.
\item For flavor tagging, systematic errors of 1\% have already been reached at LEP. With the advances in detector technology and the larger integrated luminosity, we assume that for each data set at the ILC this can be reduced  and also improved as  a function of integrated luminosity by probe-and-tag measurements.   We expect an uncertainty of 
  $0.3\% \sqrt{0.250/L}$, where $L$ is the integrated luminosity in ab$^{-1}$.
  This is an error of 0.1\% for the ILC250.


\end{itemize}

\subsubsection{Control of systematic uncertainties using beam polarisation} 
\label{subsubsec:pol:systematics}

In the remainder of this section we will highlight the impact of the beam polarisation on the control of systematic uncertainties using these two studies as examples.   We have already pointed out that beam polarisation provides subsets of the data that can be used as cross-checks of systematic errors on 
efficiencies for signal identification and background suppression.  However, the studies
in Refs.~ \cite{Habermehl:417605} and \cite{bib:PhDRobert} go beyond this to illustrate the use of beam polarisation to actually reduce systematic errors beyond what is possible at 
an unpolarised collider.   Both of these studies were carried out for measurements at the 
500~GeV ILC, but the same principles apply to the 250~GeV data.

Several principles combine to produce this  result.   The first is that different polarization
settings produce event samples with different mixtures of signal and background 
processes.   The differences in these  mixtures arise from order-1 polarization asymmetries that vary from 
process to process, to first approximation, in the manner predicted by the SM.   In the SM, for example, lepton pair production has a small polarization asymmetry while the   polarization asymmetry for $b\bar b $ production is large and that for $W$ pair production is almost maximal.   For certain modes, for example, lepton pair production and di-jet production, the detection  efficiency is naturally very high and therefore has a small uncertainty, while for other processes, for example, $b\bar b$ production, this efficiency is smaller and also more complicated to estimate.  If we introduce nuisance factors for the more uncertain efficiencies
and determine these from data, the correlation of the relative compositions with polarization
allows us to determine these parameters in terms of the efficiencies that are better known. 

The second principle is that systematic uncertainties that are correlated with polarization can be cancelled locally in the data set using fast reversal of the beam helicities.  This principle 
was essential to the excellent measurement of $\sin^2\theta_w$ by the SLD experiment 
from a very small data sets; almost all systematic errors were cancelled by flipping the $e^-$ beam  polarization in a pseudo-random fashion~\cite{Abe:1996nj}.   The principle of rapid helicity reversal is built into the ILC design, which gives the capability to flip the sign of 
polarisation for each of the two beams independently on a train-by-train basis (see Sec.~\ref{par:beampol}.  This helicity reversal is fast compared to typical time-scales of changes in the configuration, calibration, and alignment of the detector and the accelerator.
It implies that data sets with the same beam energy but different beam helicities can be considered as being collected  essentially concurrently.  

\begin{figure}
\centering
\includegraphics[width=0.95\linewidth]{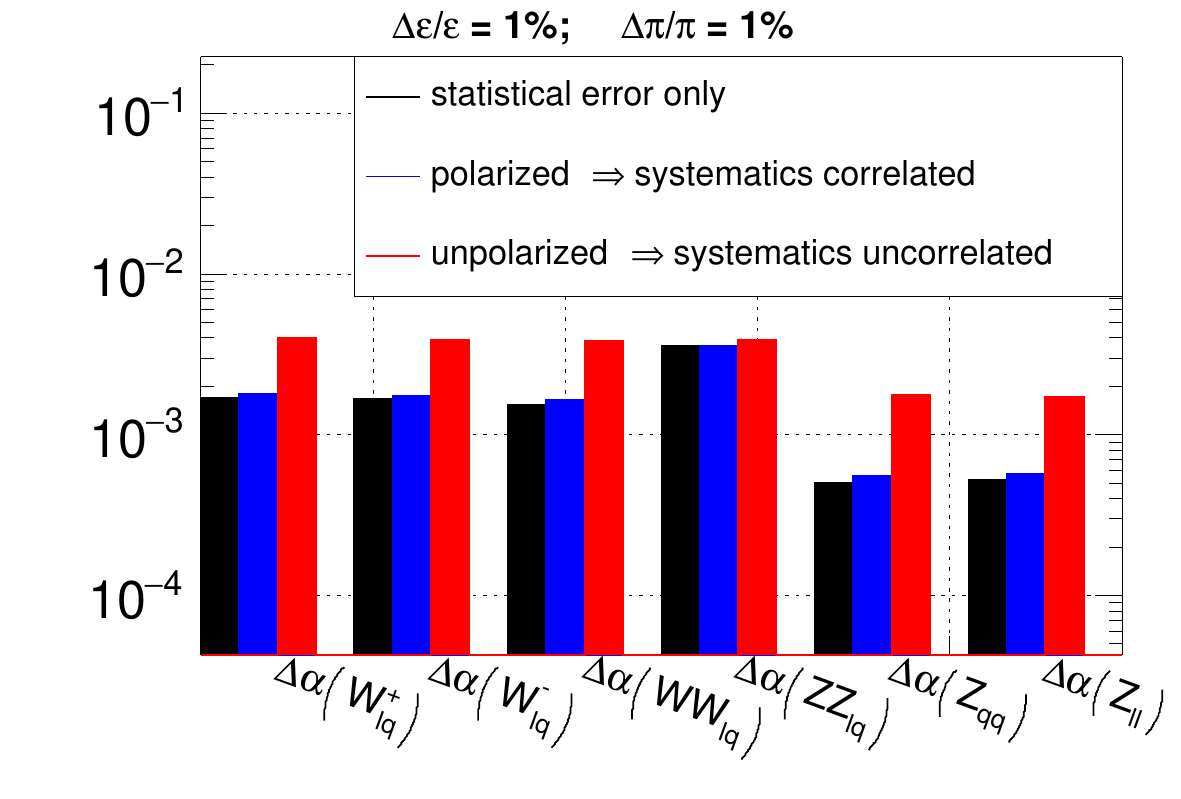}
		
\caption{Uncertainties on the unpolarised cross sections of various 2-fermion and 4-fermion processes as obtained from the global fit introduced in the text~\cite{bib:PhDRobert}, assuming a systematic uncertainty of 1\% on the selection efficiencies and purities, each. In the case of polarised beams, it is estimated that only 10\% of the uncertainty is uncorrelated between data sets.  Applying that estimate to the analysis of data sets taken ``quasi-concurrently", the impact of the systematic uncertainties is minimal.  Without the redundancies provided by data sets with correlated systematic uncertainties, the total uncertainties increase by a factor 2 for $WW$ and single-$W$ processes and a factor of 5 for 2-fermion processes.}
\label{fig:alpha_error_corr_uncorr}
\end{figure}

The improvement in the measurement of the absolute normalization of cross sections can be very significant.  The study of Ref.~\cite{bib:PhDRobert} considered the full set of 2-fermion production processes and 4-fermion production processes (including $\ee\to WW / ZZ \to 4$~fermions and well as single-$W$ production) at 250~GeV.   Each channel was assigned a 1\% systematic uncertainty in its selection efficiency and signal purity. Based on the correlations of experimental effects between ``quasi-concurrently'' taken data sets, discussed in section~\ref{subsubsec:pol:systematics}, it was estimated that only 10\% of this uncertainty is uncorrelated between data sets with different  beam polarisation configurations. Thus, a global fit using all four
polarization settings allows one to determine the relative efficiencies and remove most of the systematic uncertainty.  The result for the final normalization uncertainties are shown in 
Fig.~\ref{fig:alpha_error_corr_uncorr}.  For each of several 2- and 4-fermion channels, the black bars show the statistical uncertainties, the red bars show the full uncertainties for unpolarised data, and the blue bars show the uncertainties for polarised data samples.   The final uncertainties are larger in unpolarized case by a factor of 2 for $WW$ and single-$W$ processes and by a factor of 5 for lepton pair production.

\begin{figure}
\centering
\includegraphics[width=0.95\linewidth]{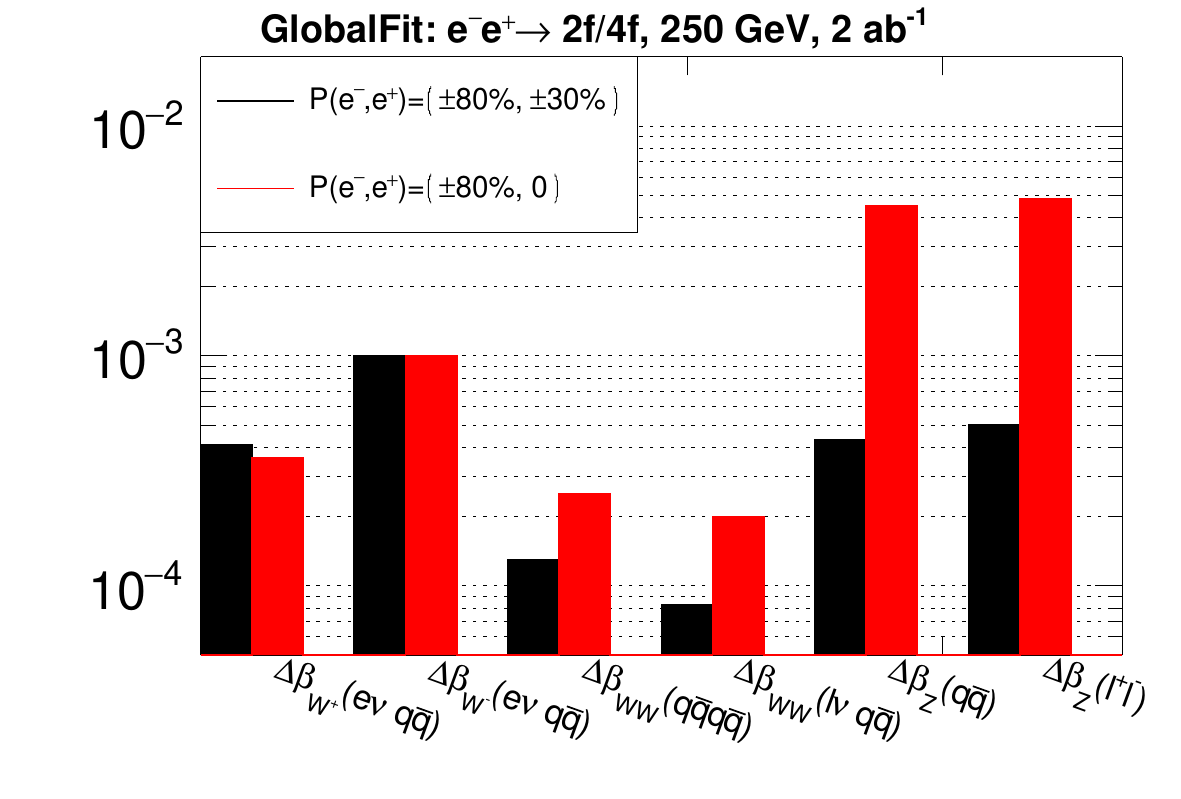}
		
\caption{Uncertainties $\Delta \beta$ on \ALR\ of various 2-fermion and 4-fermion processes as obtained from the global fit introduced in the text~\cite{bib:PhDRobert} with both beams polarised (with the standard 45\%/45\%/5\%/5\% sharing between the four helicity configurations) and in the absence of positron polarisation (with a 50\%/50\% sharing between the two remaining helicity configurations). In the absence of positron polarisation, the  uncertainties on \ALR\ increase by a factor 2 for $WW$ and by about a factor of 10 for 2-fermion processes. Alone the single-$W$ processes remain unaffected.}
\label{fig:beta_error_noposipol}
\end{figure}

The same principles can be applied to the measurement of polarisation asymmetries $A_{LR}$, which, as we have seen, play a large role in the ILC program.   Though many systematic 
errors automatically cancel in $A_{LR}$, there are new sources of systematic uncertainty, for example, the possibility of a correlation between the helicity orientation and the luminosity delivered per bunch train.   This is effectively controlled if both the electron and positron 
bunches can be polarised.  Roughly, the polarization asymmetry in $W$ pair production is almost maximal, and the small uncertainty in this quantity can be transferred to the value of $A_{LR}$ for other processes. The point is illustrated in Fig.~\ref{fig:beta_error_noposipol}, again from Ref.~\cite{bib:PhDRobert}, which shows a comparison of the final uncertainty on 
$A_{LR}$ in a global fit between a collider with $e^-$ and $e^+$ polarization (black bars) and a collider with only $e^-$ polarization (red bars). The improvement is a factor of 10 for 
fermion pair production.  (In this illustrative study, the systematic errors from detector efficiency and theory are set equal to zero.) 

\begin{figure}
\centering
\includegraphics[width=0.95\linewidth]{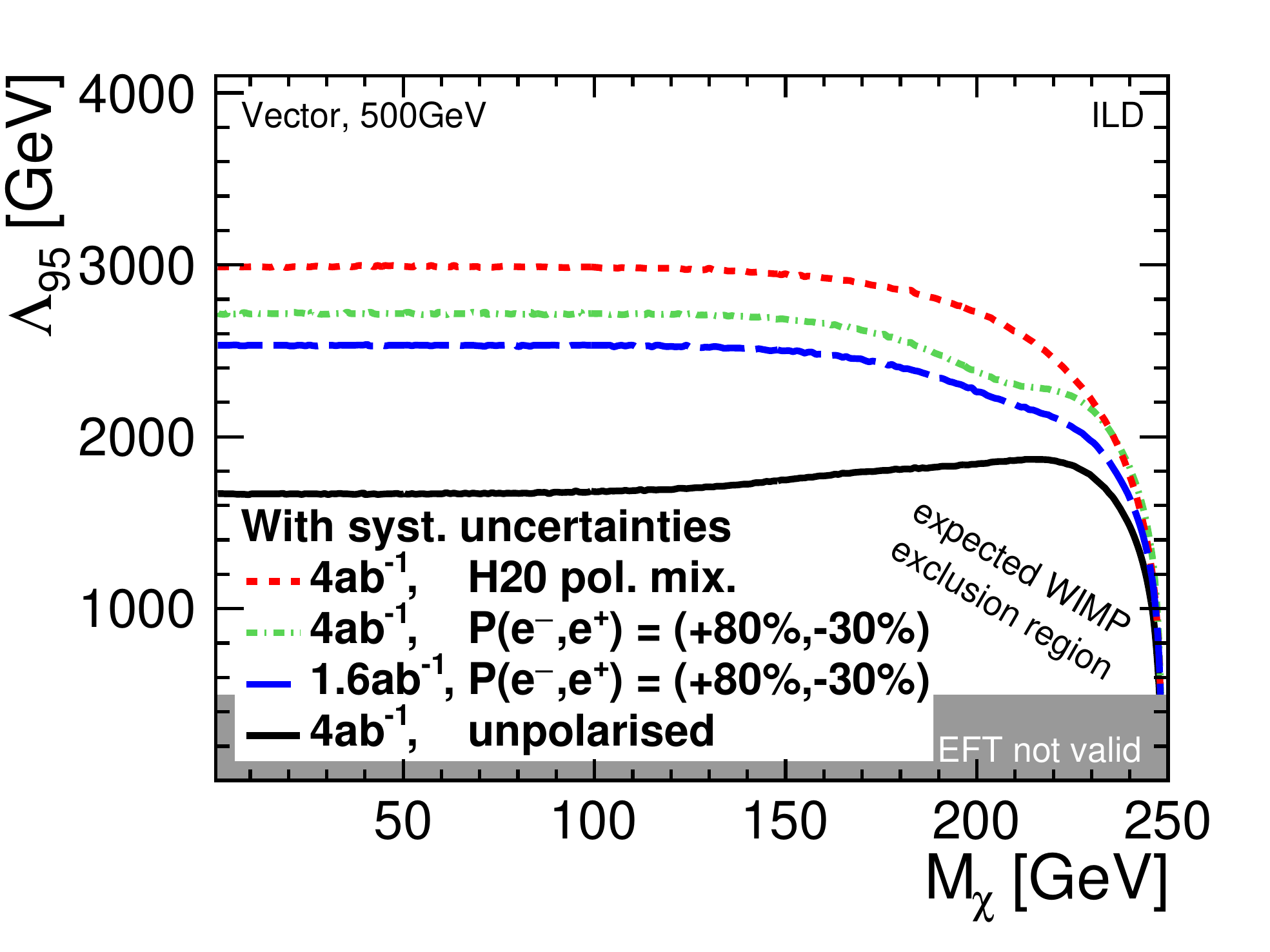}
		
\caption{Comparison of the reach of the search for WIMP production in the mono-photon channel for different assumptions on luminosity and polarization, {\em including} systematic uncertainties (see Sec.~\ref{sec:searches} for a description of the analysis)~\cite{Habermehl:417605}. }
\label{fig:polWIMPsys}
\end{figure}

Similar large effects from polarisation are seen in cases in which the signal is detected in 
the shape of  a  distribution.    An illustration here is given by a study of the search for 
dark matter particles $\chi$ using the mono-photon signature~\cite{Habermehl:417605}, already discussed in Sec.~\ref{subsec:beampol}.   In our earlier discussion, we pointed out that the signal from 
$\ee\to \gamma \chi\chi$ sits on top of a large irreducible background from 
$\ee\to \gamma \nu\bar \nu $.  The study includes a careful evaluation of the systematic uncertainties, including those on selection efficiencies, luminosity, beam energy (spectrum) and polarization as well as on the theoretical modelling of the background. 
The limit calculation uses fractional event counting based on the observed energy spectrum of the selected photon candidates and considers normalisation and shape-dependent uncertainties as well as the correlations between these. If the mass of the $\chi$ is relatively high, only low-energy photons can appear in the signal process.  Then the high-energy part of the photon spectrum can be used to determine nuisance parameters assigned to the 
polarization and efficiencies.  The results for the lower limit on the mediator scale, including 
systematic errors, are shown in Fig.~\ref{fig:polWIMPsys}. This figure should be compared to 
Fig.~\ref{fig:polWIMPstat}, in which systematic errors are set to zero.   Note that, in this case, the strongest limits are set using a mixture of beam polarizations (the dashed red curve in both cases) since this allows the systematic errors from beam polarisation to be better 
controlled.

\subsection{Estimation of future improvements}
\label{subsec:higgs_improve}

As elaborated in the previous section, all the estimates for Higgs measurements shown in
Tab.~\ref{tab:higgserrors} are based on available full simulation studies, 
which are performed using the analysis
techniques known at present. These estimates are clearly too
conservative in the sense that we have not been able to analyse all
useful signal channels.  In addition, analysts working closely with
the data are always able to invent algorithms to that are more cleverly
optimized to  the data that is actually collected.   The projected
uncertainties  quoted here and in Sec.~\ref{subsec:global:elements} do not take advantage of
these  likely improvements.

Since the formal estimates of the performance of HL-LHC given in the
HL-LHC Yellow Report~\cite{Cepeda:2019klc} do take into account
improvements in systematic uncertainties that are anticipated but not
yet realised, it seems to us reasonable to define also for ILC an
optimistic scenario with improved performance.   We use this scenario
to compare to the results of Ref.~\cite{Cepeda:2019klc} in the manner
explained in Sec.~\ref{subsec:higgs:ilclhc}.   This scenario, which we 
refer to as ``S2'' in that discussion, includes the following
improvements in the analysis just described.  In all cases, these
improvements are under study using our full simulation
tools
and are  suggested, if not yet validated, by our current results: 
\begin{itemize}
\item 10\% improvement in signal efficiency of the  jet clustering algorithm.
\item 20\% improvement in the performance of the  flavor tagging algorithm. 
\item 20\% improvement in statistics by including more signal channels
  in  $\sigma_{Zh}\cdot BR(h\to WW^*)$. 
\item a factor of 10 improvement in the precision electroweak input
  $A_{\ell}$ thorugh the measurment of $e^+e^-\to\gamma Z$ with
  polarized beams at 250~GeV.
\item 30\% improvement in the precision of Higgs self-coupling and 
top-Yukawa coupling at 500 GeV, which
is a consequence of the improvements in jet clustering algorithm, flavor tagging
algorithms and statistics by including more signal channels. 
\end{itemize}

\subsection{Measurement of the Higgs boson self-coupling}
\label{subsec:higgsself}

\begin{figure*}[htb]
\centering
\includegraphics[width=130mm]{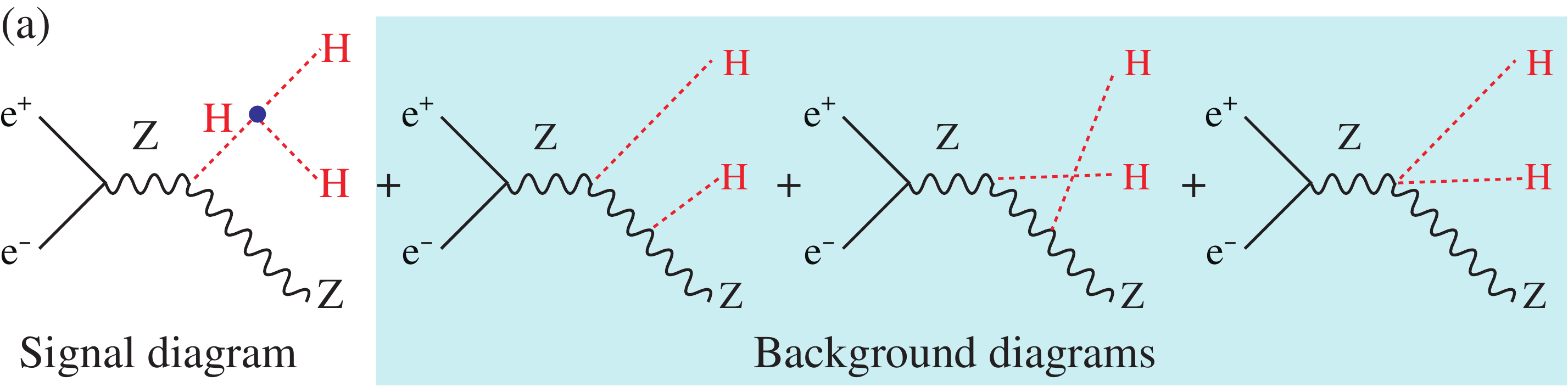}
\\~\\
\includegraphics[width=130mm]{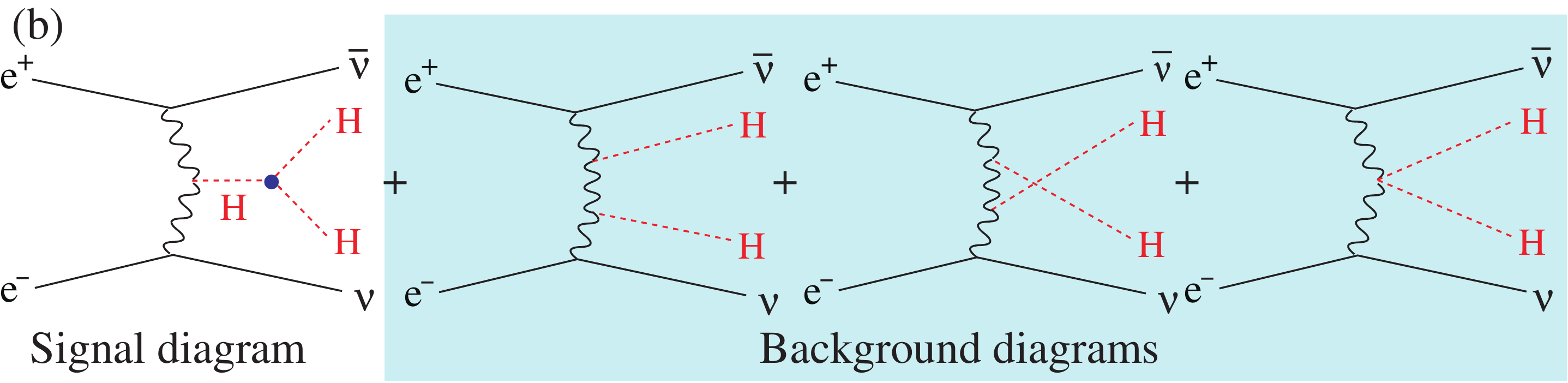}
\caption{Diagrams contributing to (a) $e^+e^- \to Zhh$ and (b) $e^+e^- \to \nu\bar{\nu}hh$.} 
\label{fig:hhdiagrams}
\end{figure*}
\begin{figure}[htb]
\centering
\includegraphics[width=0.85\hsize]{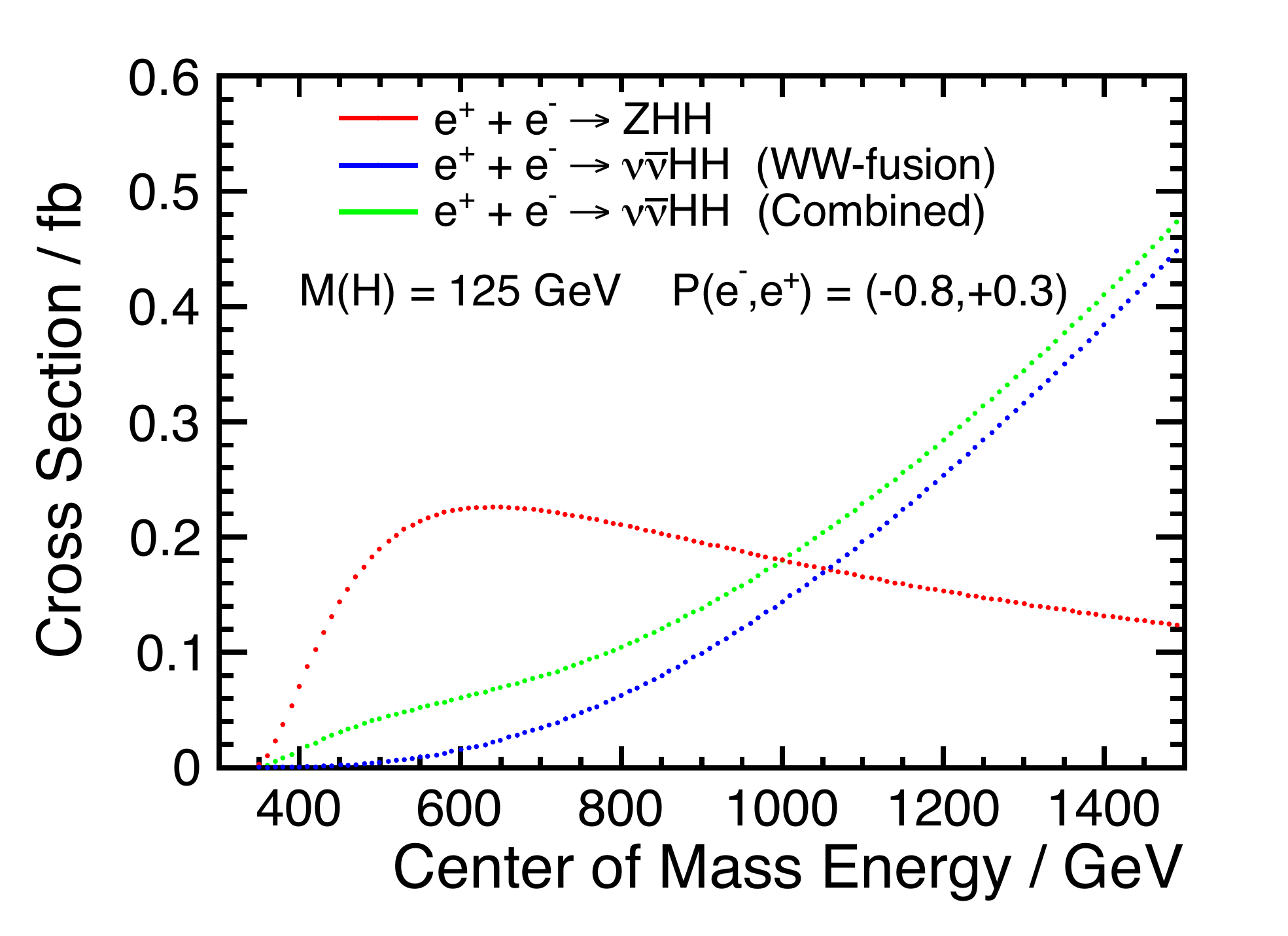}
\caption{Cross sections for the double Higgs production processes, 
$e^+e^- \to Zhh$ and $e^+e^- \to \nu\bar{\nu}hh$, as a function of $\sqrt{s}$ for $m_h=125\,$GeV.} 
\label{fig:sigzhh_vvhh}
\end{figure}

\begin{figure*}[htb]
  \centering
  \begin{tabular}[c]{ccc}
    \includegraphics[width=0.45\hsize]{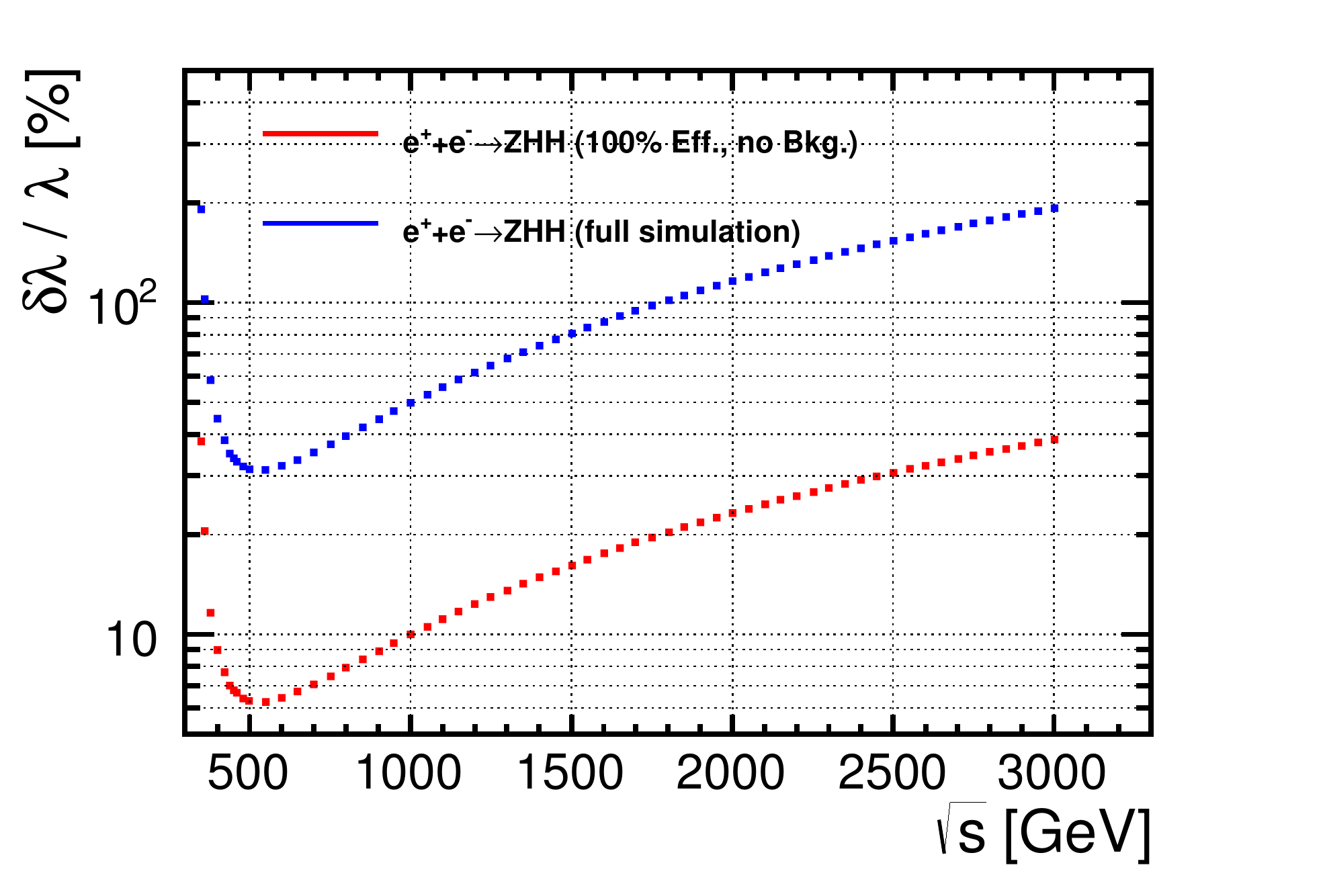} &
    \includegraphics[width=0.45\hsize]{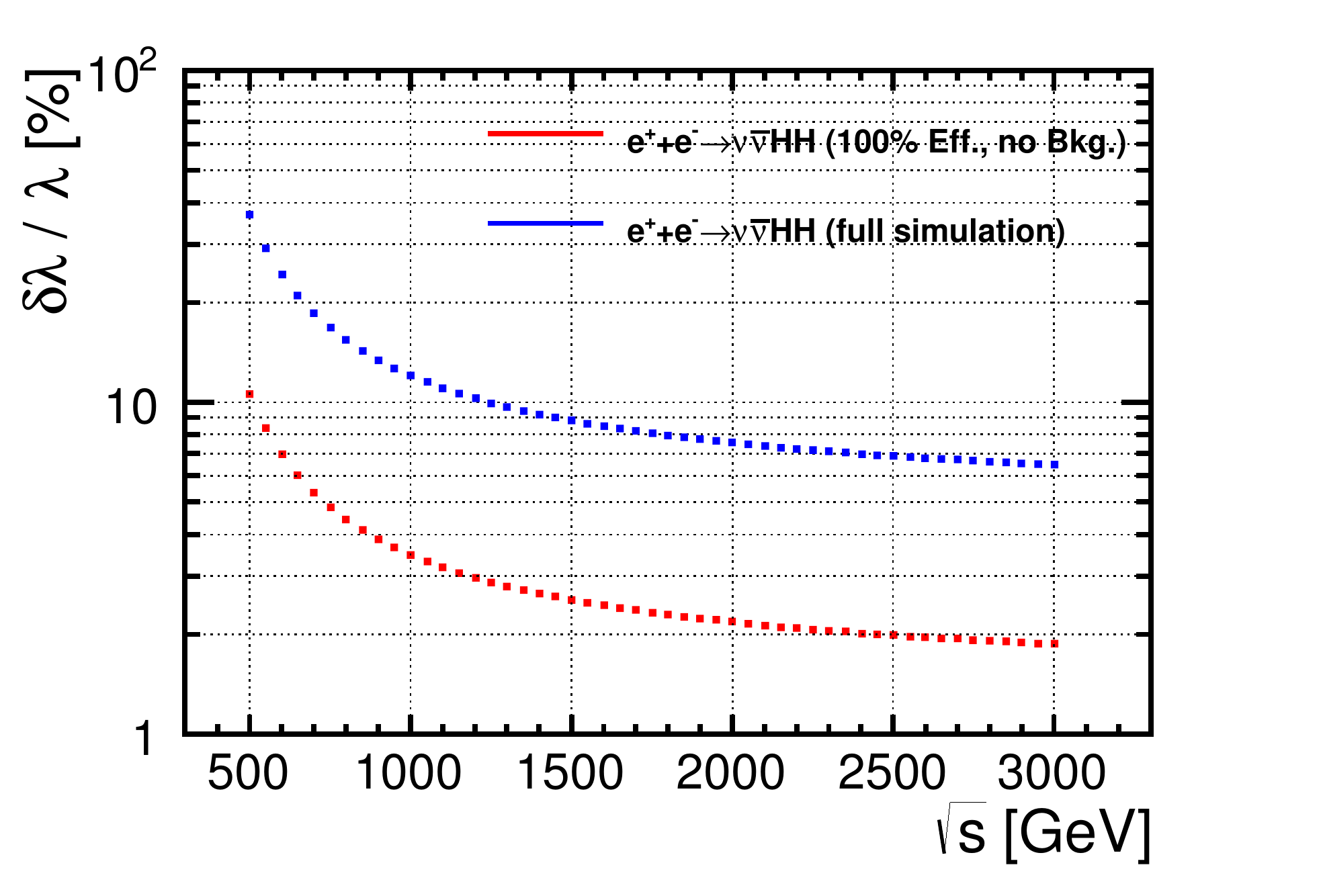} \\   
 \end{tabular}
  \caption{The expected precisions of $\lambda$ as a function of $\sqrt{s}$ for $e^+e^-\to Zhh$ ~(left) and
for $e^+e^-\to\nu\bar{\nu}hh$ ~(right). The two lines in each plot correspond to ideal situation (red) and 
realistic situation (blue) as described in the text. Same integrated luminosities of 4 $\mathrm{ab}^{-1}$ is 
assumed at all $\sqrt{s}$.}
  \label{fig:HHHSensitivity}
\end{figure*}

\begin{figure*}[htb]
  \centering
  \begin{tabular}[c]{ccc}
    \includegraphics[width=0.45\hsize]{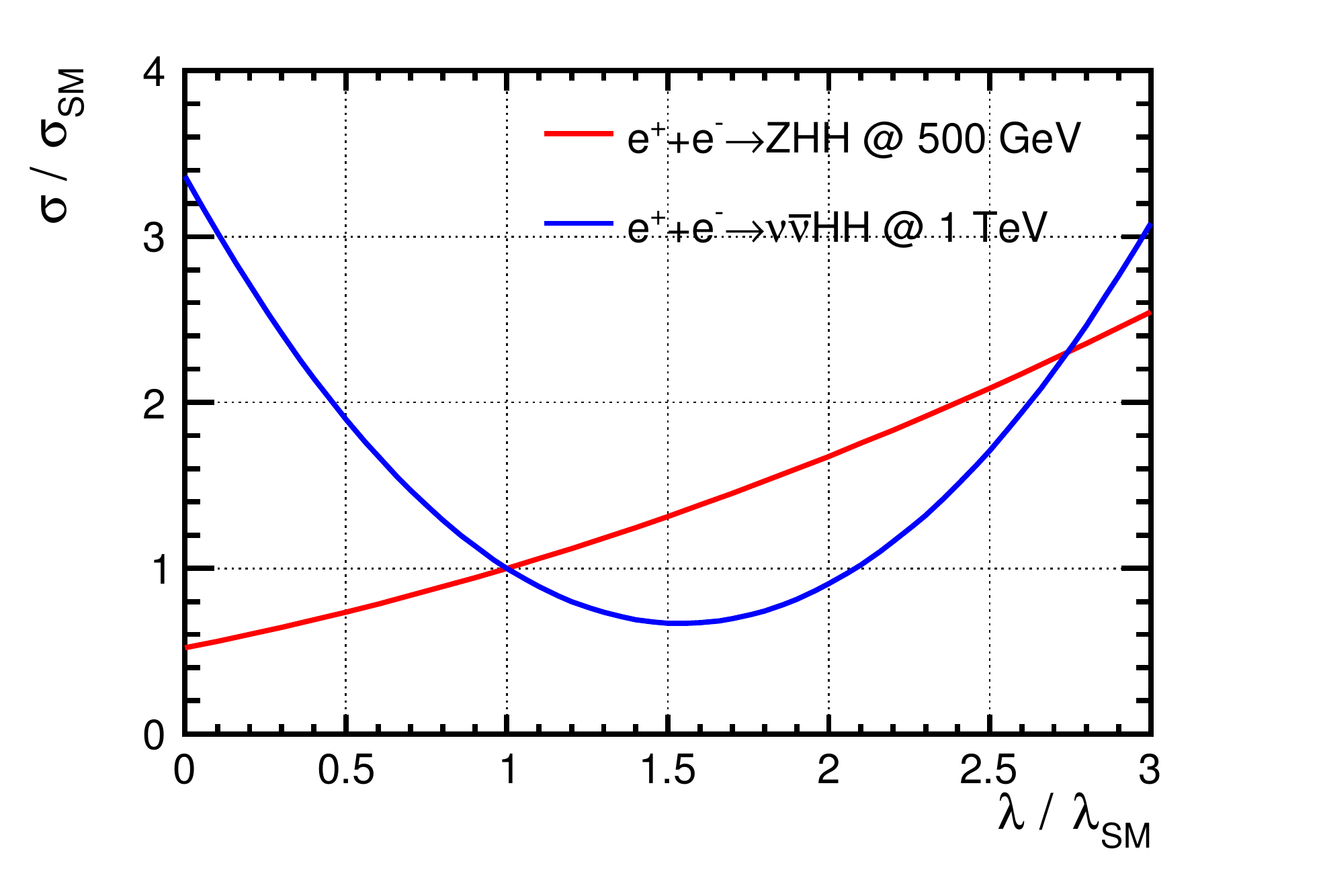} &
    \includegraphics[width=0.45\hsize]{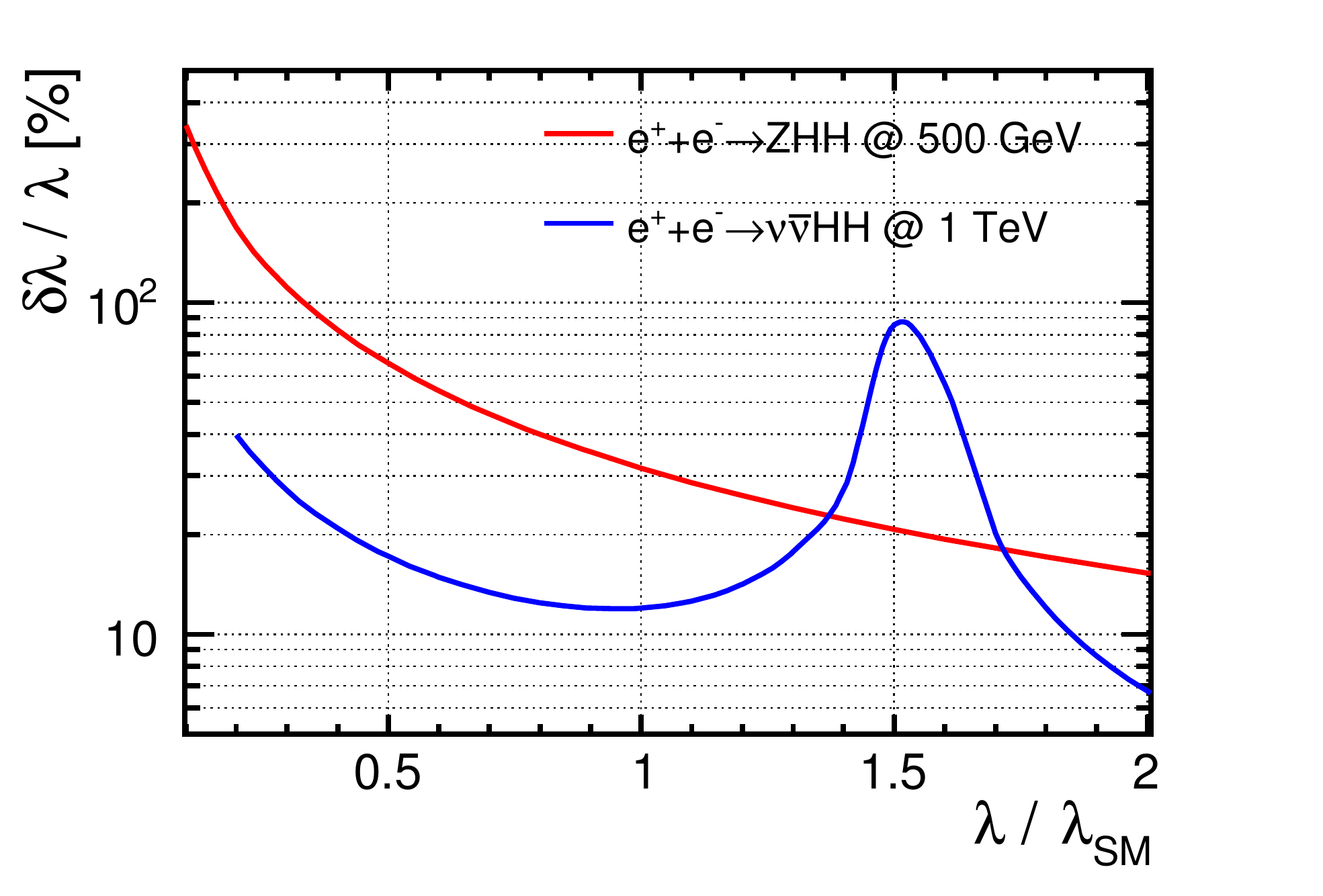} \\   
 \end{tabular}
  \caption{Left: the cross section as a function of $\lambda$ for $e^+e^-\to Zhh$ ~(red line) and for 
  $e^+e^-\to\nu\bar{\nu}hh$ ~(blue line), where values of both $\lambda$ and $\sigma$ are scaled to their 
  SM values. Right: expected precisions of $\lambda$ when $\lambda$ is deviated from its SM value.}
  \label{fig:HHHBSM}
\end{figure*}

The  trilinear Higgs coupling can be measured at colliders in two
different ways. First, the coupling can be measured directly, using 
processes with Higgs pair production
that  diagrams involve the triple Higgs coupling at the tree level. 
Second, the coupling can be measured indirectly, since
radiative corrections to single-Higgs processes can include effects
due to the tripple Higgs coupling.

The important Higgs pair production reactions at $\ee$ colliders are 
$\ee\to ZHH$ and $\ee\to \nu\bar\nu HH$, shown in Fig.~\ref{fig:hhdiagrams}. 
Note in both reactions there are diagrams that do not involve trilinear Higgs coupling.
The first of these processes can be studied already at 500~GeV; 
the second, which is a 4-body process, requires somewhat higher energy. 
The cross sections of these two processes as a function of $\sqrt{s}$ are shown in Fig.~\ref{fig:sigzhh_vvhh}.
Full simulation studies at a $\sqrt{s}=$500~GeV show that a discovery of the
double Higgs-strahlung process is possible within the H20 program, 
using $Z\to l^+l^-/\nu\bar{\nu}/q\bar{q}$ and $hh\to b\bar{b}b\bar{b}/b\bar{b}WW^*$ channels.
With 4~\iab\ at 500~GeV, a combination of those decay channels
would yield a precision of 16.8\% on the total cross section for
$\ee\to ZHH$~\cite{Duerig:2016dvi,Tian:2013qmi,KurataHHH}.  Assuming the SM with only the
trilinear Higgs coupling free, this corresponds to an uncertainty of
27\% on that coupling.

At still higher energy vector boson fusion becomes the dominant
production channel. Making use of this channel, with  8~\iab\ at
1~TeV, the studies of 
Refs.~\cite{Tian:2013qmi,KurataHHH,Roloff:2019crr} show that, in the
same context of varying the trilinear Higgs coupling only, this
coupling can be determined to 10\%. 

The impact of the center-of-mass energies on the trilinear Higgs coupling
measurement is studied by extrapolating the full simulation results done 
at 500 GeV and 1 TeV to other energies. Due to the existence of diagrams 
that do not involve the trilinear Higgs coupling in both reactions, 
to get the correct extrapolation a careful 
analysis taking into account the dependence on $\sqrt{s}$ for both the
total cross sections and interference contributions was performed in Ref.~\cite{TianHHH:2015}.
The results are shown in Fig.~\ref{fig:HHHSensitivity} as the blue lines for the two reactions. 
In addition to the results from realistic full simulations, the expectations for
the ideal case,  assuming no background and 100\% signal efficiency,
are shown as the red lines in the figure. The differences between the blue and the read lines, is
as large as a factor of 4-5.  This suggests that there is much room for improvement in the clustering algorithm used to identify 2-jet systems with the Higgs boson mass, which 
lead to improvements in the final results.  Improvements could also come from better 
flavor-tagging algorithms and inclusion of additional  signal channels such as  
 $Z\to\tau^+\tau^-$. The figure does imply that $\sqrt{s}=500$--600 GeV is
optimal for $\ee\to Zhh$ but that CM energies of  1 TeV or above would be needed for $\ee\to\nu\bar{\nu}hh$. 

Since large deviations of the  trilinear Higgs coupling are expected in some new physics 
models,  in particular in models of electroweak baryogenesis,
it is interesting to see how the expected precisions would change in that case.
Figure~\ref{fig:HHHBSM}(left) gives the cross sections of the two reactions as a function
of the actual triple Higgs coupling $\lambda$, and  Figure~\ref{fig:HHHBSM}(right)
 shows the expected precisions of the ILC 
 measurements. 
 The natures of interference between the triple Higgs coupling and the SM production
 amplitude is very different for the 
two reactions, constructive for $\ee\to Zhh$ but destructive 
for $\ee\to\nu\bar{\nu}hh$.  Therefore, 
the two reactions, useful at 500 GeV and 1 TeV
respectively, are complementary in determining the trilinear Higgs coupling. 
If the trilinear Higgs coupling is indeed a factor of 2 larger, as expected in some models,
the double Higgsstrahlung process at 500 GeV becomes very useful and 
would already provide 
a measurement of around 15\% precision for the trilinear Higgs coupling.


The indirect determination of the trilinear Higgs coupling is based on
the observation of McCullough~\cite{McCullough:2013rea} that the cross
section for $\ee\to ZH$ contains a radiative correction involving the
trilinear coupling that lower the cross section by about 1.5\% from
250~GeV to 500~GeV, with most of the decrease taking place below
350~GeV.  Taken a face value in the simple context with only the
trilinear coupling free, the ILC cross section measurements would determine the trilinear 
coupling to about 40\%.

It is important to note, however, that the determination of the
trilinear coupling involves two separate questions.  First, is the SM
violated?   The accuracies with which this question can be answered
are those given above.  Second, can the violation of the SM be
attributed to a change in the trilinear coupling or the Higgs
potential rather than being due to other possible new physics
effects?  A precise way to ask this question is: Can the shift of the
trilinear coupling be measured  independently of possible effects of
all 
other dimension-6 EFT operators?   To our knowledge, this latter
question has only been addressed for determinations of the trilinear
coupling at lepton colliders.   In Ref.~\cite{Barklow:2017awn} it is
shown that, after the ILC H20 program of single-Higgs measurements is
complete, the uncertainty in the measurement of the total cross
section for  $\ee\to ZHH$ receives a negligible 2.5\% uncertainty due
to variation of the other relevant dimension-6 EFT perturbations.   In
Ref.~\cite{DiVita:2017vrr}, it is shown that, when the cross section
for $\ee\to ZH$ is fit together with other relevant observables at
250~GeV and 500~GeV, the
uncertainty in the coupling is not substantially changed from the value of
40\%. This conclusion, however, might be sensitive to the precision of the inputs from precision electroweak observables.   A study of those effects is in progress.

\section{\label{sec:ew}Physics Simulations: Electroweak Production of 2- and 4-Fermion Final States }

The precision studies of the Higgs boson described in the previous section receive important and complementary support from analyses of 2- and 4-fermion final states which do not directly involve Higgs bosons, but our well-known SM gauge bosons --- or potentially their yet to be discovered siblings. In this section, some of the key examples introduced in Sec.~\ref{subsec:phys_WW}  will be discussed in more detail, highlighting the level of realism on which the projections are based.

\subsection{Analyses of $e^+e^- \to W^+W^-$} 
\label{subsec:ew_WWana}

The analysis of four-fermion processes, e.g.\ from $W$-pair production, but also 
 $Z$-boson pairs and single-boson processes, plays a key role in the ILC physics program.  As discussed in Sec.~\ref{subsec:phys_WW}, constraints on triple gauge couplings (TGCs) are an important input to the dim-6 EFT-based interpretation of Higgs precision measurements introduced in Sec.~\ref{subsec:phys_eft}. Prospects for this type of analysis have been intensely studied in full detector simulation (c.f.\ Sec.~\ref{sec:software}), however only at center-of-mass energies of at least 500\,GeV. In this section, we will start out by summarizing these detailed studies, and then proceed to their recent extrapolations to a center-of-mass energy of 250\,GeV.

\subsubsection{Full simulation analyses of TGCs}
\label{subsubsec:ew_fullsimww}
The prospects for probing charged TGCs at the ILC have been studied in full, GEANT4-based simulation of the ILD detector concept at $\sqrt{s}=$500\,GeV~\cite{Marchesini:94888} in the context of the ILD Letter of Intent~\cite{Abe:2010aa} and at $\sqrt{s}=$1\,TeV~\cite{Rosca:2016hcq} as a benchmark for the ILD Detailed Baseline Design (Vol 4 ILC TDR~\cite{Behnke:2013lya}). Both analyses focused on the channel $e^+e^- \to W^+ W^- \to qql\nu$, $l=e, \mu$ and followed a similar, cut-based selection approach. Thereby they exploit the known initial state for a full kinematic reconstruction of both $W$ bosons, under consideration of optional photon radiation collinear to the beam direction.
Neither the case $l=\tau$, nor the fully hadronic mode, nor contributions from single-$W$ production were included at the time. Especially the fully hadronic mode will profit substantially from the recent advances in reconstructing the jet charge with the ILD detector, c.f.\ Sec.~\ref{subsec:ew_ffana}, in order to determine the charges of the $W$ boson candidates.

\begin{figure*}[]
\centerline{
\includegraphics[width=0.5\linewidth]{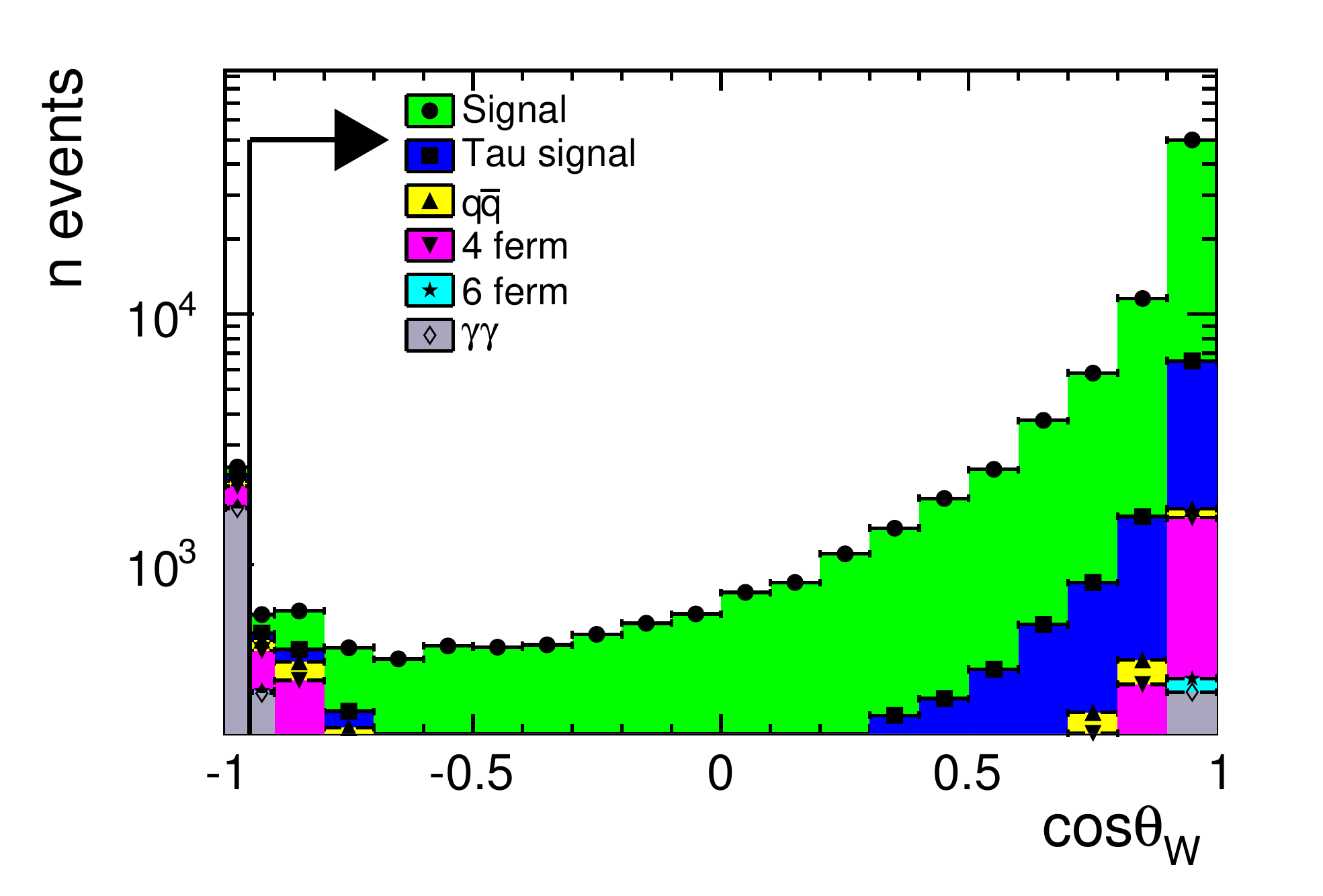}
\includegraphics[width=0.5\linewidth]{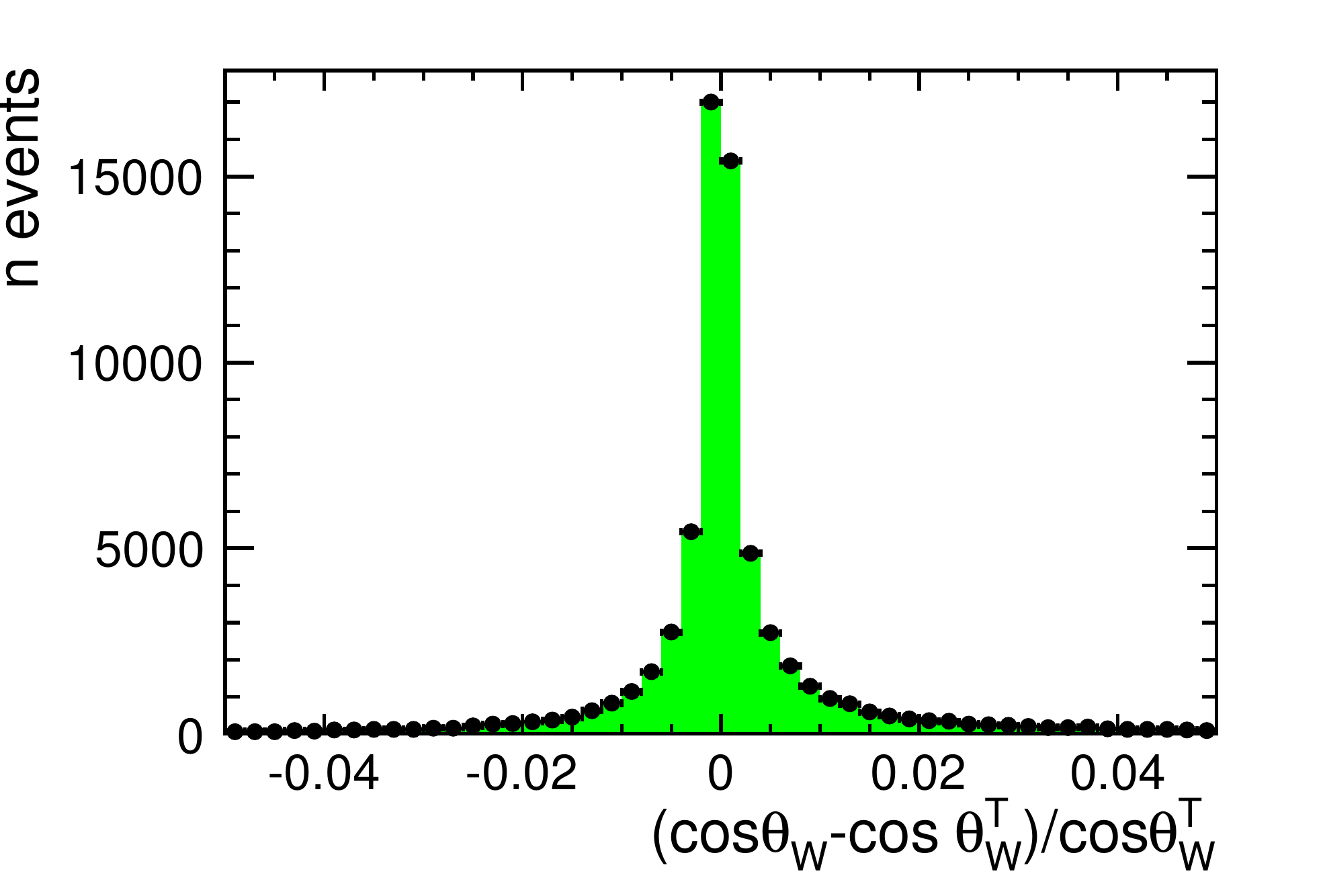}
}
\caption{Left: Reconstructed polar angle distribution $\cos\theta_{W}$ of the $W^-$ boson candidates for signal and SM background events before the application of the final selection cut $\cos\theta_{W} >$ -0.95. Right: relative deviation of the reconstructed $\cos\theta_{W}$ from the MC truth value $\cos\theta_{W}^\mathrm{T}$. Both from~\cite{Marchesini:94888}.}\label{fig:WW_costheta}
\end{figure*}

Figure~\ref{fig:WW_costheta} shows one of the final observables sensitive to anomalous TGCs,
namely the cosine of the polar angle of the $W^-$ boson, $\cos\theta_{W}$, which is reconstructed from the hadronically decaying $W$ boson. The left part of the figure illustrates the high purity of the selection which ranges between 85\% and 95\%, depending on whether  $WW \to qq\tau \nu$ is considered as background or not. The right panel shows relative deviation of the reconstructed $\cos\theta_{W}$ from MC truth, indicating a resolution of better than 0.5\%.
 
In the 500\,GeV analysis, no pile-up from $\gamma \gamma \to$ low $p_t$ hadrons was considered, which has, at $\sqrt{s}=$500\,GeV, an expectation value of 1.2 events per bunch crossing, c.f.\ Sec.~\ref{sec:software}.
This type of background was considered, however, in the TGC study at 1\,TeV, where its expectation value increases to 4.1 pile-up events per bunch crossing. Figure~\ref{fig:WW_overlayremoval} shows the impact of these pile-up events on the reconstructed hadronic $W$-boson mass without (``Durham'') and with (``Kt'') application of suitable suppression algorithms (see Sec~\ref{subsec:higgs_common} for a detailed description of the algorithm). It can be seen that the residual effect is small even at 1\,TeV, where the number of pile-up events is expected to be nearly four times higher than at 500\,GeV.

\begin{figure}
	\centering
		\includegraphics[width=0.95\linewidth]{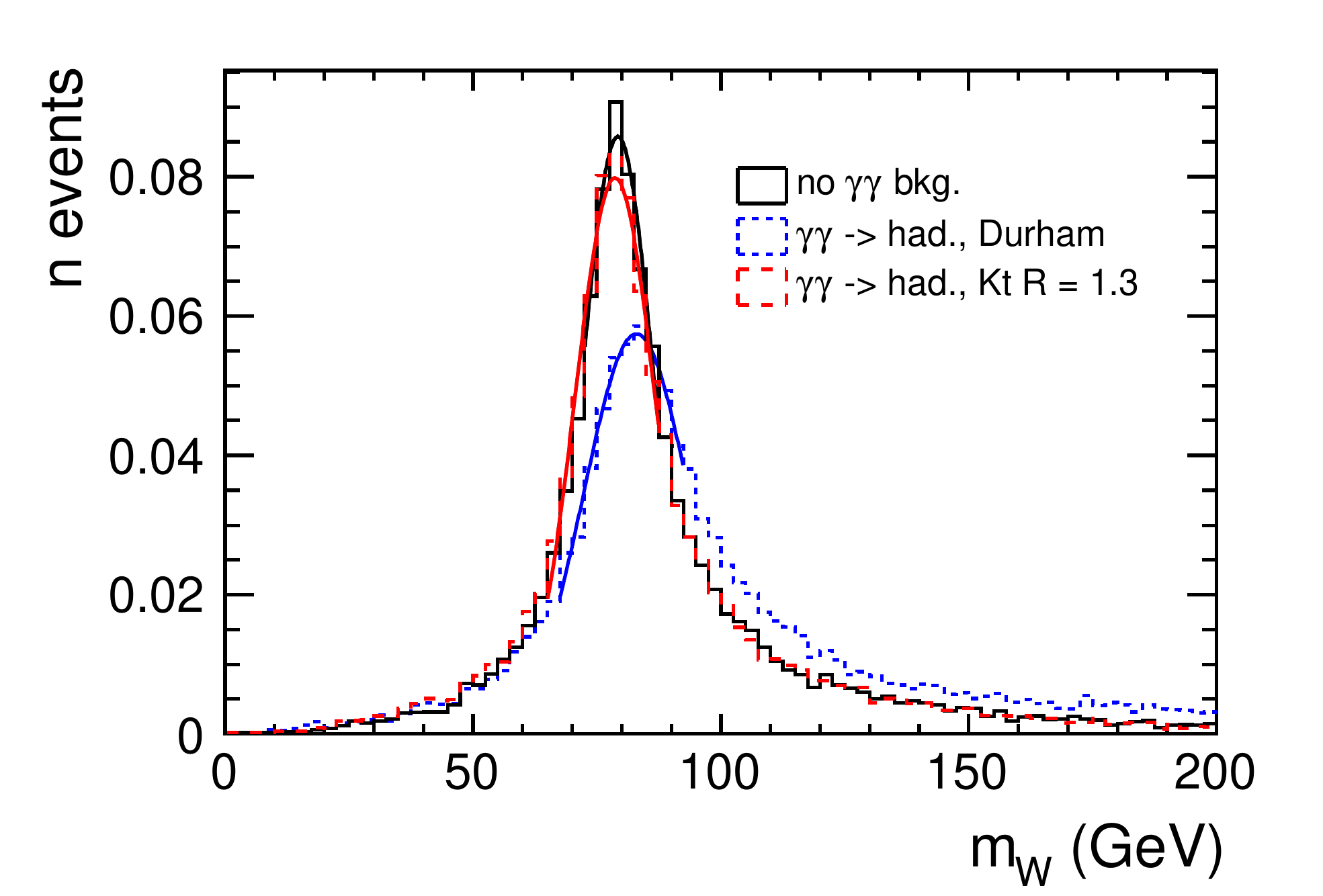}
		
	\caption{Reconstructed mass of the hadronically decaying $W$ boson at $\sqrt{s}=$1\,TeV without pile-up compared to the situation with pile-up in absence (``Durham'') or presence (``Kt'') of a mitigation strategy in the analysis. From~\cite{Rosca:2016hcq}. 
	}
	\label{fig:WW_overlayremoval}
\end{figure}

Both full-simulation studies used only three out of five possible angular distributions which could provide sensitivity to TGCs: besides the production angle of the $W^-$, the two angles describing the direction of the decay lepton in the restframe of its mother $W$ were considered in a five-parameter fit based on MC templates. The free parameters of the fit were the three TGCs $ g^Z_1$, $\kappa_\gamma$ and $\lambda_\gamma$ as well as the absolute values of the electron and positron beam polarisations. With this approach, statistical uncertainties of about $6$-$7 \cdot 10^{-4}$ were obtained for all three couplings for an integrated luminosity of 500\,fb$^{-1}$ at $\sqrt{s}=$500\,GeV, shared equally between all four beam polarisation configurations.
In the context of the 500\,GeV analysis a thorough evaluation of the systematic errors was performed. It was found at the time that uncertainties on the selection efficiencies of signal and background of 0.2\% and 1\%, respectively, would lead to systematic effects in the same order of magnitude as the statistical uncertainty for 500\,fb$^{-1}$.

A recent study dedicated to a global fit of total and differential cross sections of various SM processes sensitive to TGCs and/or beam polarisation has shown, however, that a much better control of systematic effects can be achieved --- provided that {\em both} beams are polarised~\cite{bib:PhDRobert, Fujii:2018mli}.  

The full simulation study at $\sqrt{s}=$1\,TeV, following the same fitting approach, found statistical uncertainties of about $2$-$3 \cdot 10^{-4}$ for all three TGCs for an integrated luminosity of 1000\,fb$^{-1}$. The effect of different sharings of the luminosity between the four polarisation sign combinations on the TGC precisions were found to be minor. 

\subsubsection{Extrapolation of TGC prospects to 250\,GeV}

While extensive studies of $WW$ production exist at higher center-of-mass energies, no complete analysis based on full detector simulation is available yet for $\sqrt{s}=$250\,GeV. Nevertheless substantial progress has been made in various important aspects which have been incorporated in
an extrapolation~\cite{Fujii:2017vwa} based on a) the previously discussed full-simulation studies at 500\,GeV~\cite{Marchesini:94888} and 1\,TeV~\cite{Rosca:2016hcq} and b) the actual LEP results at $\sim$ 200\,GeV~\cite{Schael:2004tq}:
\begin{itemize}
\item As discussed in Sec.~\ref{subsec:phys_WW}, the sensitivity of measured cross sections to the TGCs depends on the center-of-mass energy as $s/m_W^2$.
\item Naively, the statistical uncertainties on measured cross sections scale as $1/\sqrt{\sigma \L}$. However at higher center-of-mass energies, the $W$ bosons are more and more boosted into the forward direction due to increasing amount of ISR and beamstrahlung. Therefore the experimental acceptance decreases for higher $\sqrt{s}$. A correction factor for this effect has been derived~\cite{Karl:2017let} from a comparison of the full detector simulation studies at 500\,GeV and 1\,TeV.
\item The dependence on the sharing of luminosity between the four different polarisation configurations was found minor~\cite{Rosca:2016hcq} and therefore no corrections for differences in the assumptions of the full simuation studies w.r.t.\ the H20 running scenario (c.f.\ Sec.~\ref{subsec:runscen_pol}) were applied.
\item The improved treatment of systematic uncertainties based on a nuisance parameter technique in a global fit to many observables and datasets explored in~\cite{bib:PhDRobert} was assumed, which leads to a constant ratio between systematic and statistical uncertainties up to luminosities of at least 2\,ab$^{-1}$.
\item The full simulation studies were found to be limited their MC-based, binned fit of 3D-template histograms. The relative improvement expected when including the fully hadronic channel and when exploiting all five sensitive angles (production angle of one of the $W$ bosons plus decay angles of both $W$ bosons, see e.g.\ Fig. 5.16 in~\cite{Marchesini:94888} for an ilustration) in an unbinned fit~\cite{Barklow:1995sk}, or, equivalently, when applying an optimal observable technique~\cite{Diehl:1997ft, Diehl:2002nj}, was estimated in a parton-level study to be a factor 2.4 in the case of $ g^Z_1$, and a factor of 1.9 for the other two couplings.
\item Since none of the ILC full-simulation studies evaluated the precisions for single-coupling fits, i.e.\ when fixing the other two anomalous couplings to 0 as done in hadron collider studies,
the corresponding LEP2 results~\cite{Schael:2004tq} were extrapolated up in center-of-mass energy,
and then the minimum of this extrapolation and of the 3-coupling extrapolation from ILC studies was taken.
\end{itemize}

The results of this procedure are displayed in Tab.~\ref{table:ew_tgc} and Figs~\ref{fig:TGC_1par} and~\ref{fig:TGC_3par} in comparison with the LEP2 and LHC results as well as HL-LHC projections, where applicable. In case of the single-parameter fits, the 250\,GeV stage of the ILC will improve the precision on $ g^Z_1$ and $\kappa_\gamma$ by factors of 5 and 30 w.r.t\ to HL-LHC, while the projections for $\lambda_\gamma$ are comparable. The loss in precision when fitting all three couplings simultaneously to ILC data is minor, and the resulting precisions are used as input for the EFT-based Higgs coupling fit discussed in Sec.~\ref{subsec:global:elements}. Actually,
it has been shown it is possible even to determine simultaneously the 14 
complex couplings in the most general parametrisation of triple gauge boson 
vertices, including e.g.\ $CP$ violating contributions,  at $e^+e^-$ linear 
colliders when both beams are polarised and all polarisation configurations, 
including transverse polarisation, are exploited~\cite{Diehl:2002nj}.

\begin{table}
\begin{center}
  \begin{tabular} {|l|c||c|c|c||c|c|c||}
    \hline
 &   &  \multicolumn{3}{|c||}{total error ($\times 10^{-4}$) } & \multicolumn{3}{c||}{correlation} \\
    \hline
    Exp & $N_{par}$ & $ g^Z_1$  & $\kappa_\gamma$ & $\lambda_\gamma$ & $g^Z_1\ \kappa_\gamma$ &  $g^Z_1\ \lambda_\gamma$  & $\kappa_\gamma\ \lambda_\gamma$  \\
    \hline
    LEP~2     & 3     &  $516$  & $618$  & $376$  & -0.17 & -0.62 & -0.15 \\
    ILC~250   & 3        & $4.4$ & $5.7$ & $4.2$ & 0.63 & 0.48 & 0.35 \\
    \hline
    LEP~2     & 1     & $300$ & $626$ & $292$ & -- & -- & -- \\
    LHC      & 1     & $319$ & $1077$ & $198$ & -- & -- & -- \\
    HL-LHC   & 1      & $19$ & $160$ & $4$ & -- & -- & -- \\
    ILC~250   & 1       & $3.7$ & $5.7$ & $3.7$ & -- & -- & -- \\
    \hline

\end{tabular}
  \caption{TGC precisions for LEP~2, Run1 at LHC, HL-LHC and the ILC at $\sqrt{s}=250$~GeV with 2000~fb$^{-1}$ luminosity (ILC~250). The LEP~2 result is from ALEPH~\cite{Schael:2004tq} at  $\sqrt{s}\approx 200$~ GeV with
    0.68~fb$^{-1}$.  The LHC result is from ATLAS\cite{Aad:2014mda} at $\sqrt{s}=7$~TeV with 4.6~fb$^{-1}$.  The HL-LHC estimate is from a 2013 overview of HL-LHC physics~\cite{bib:HLLHCMoenig}. From~\cite{Fujii:2017vwa}.}
\label{table:ew_tgc}
\end{center}
\end{table}

\begin{figure}
	\centering
		\includegraphics[width=0.95\linewidth]{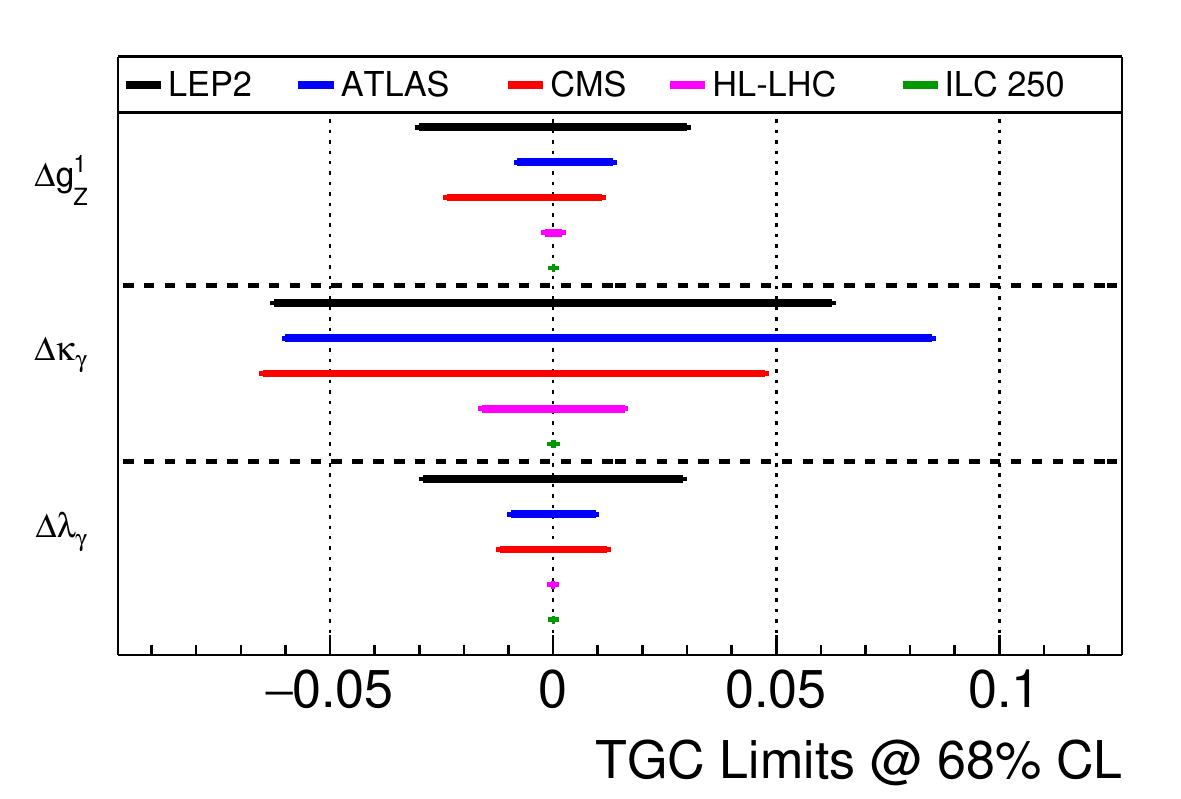}
		
	\caption{Comparison of the reachable TGC precision from single parameter fits: ILC~\cite{bib:TGC_EPS17}, final results from LEP combined from ALEPH, L3 and OPAL results~\cite{bib:LEPTGC} and the LHC TGC limits for $\sqrt{s} = 8$\,TeV data and an integrated luminosity of $\mathcal{L} = 20.3$\ifb and  $\mathcal{L} = 19.4$\ifb for ATLAS and CMS, respectively~\cite{bib:LHCTGC}. 
	}
	\label{fig:TGC_1par}
\end{figure}

\begin{figure}
	\centering
		\includegraphics[width=0.95\linewidth]{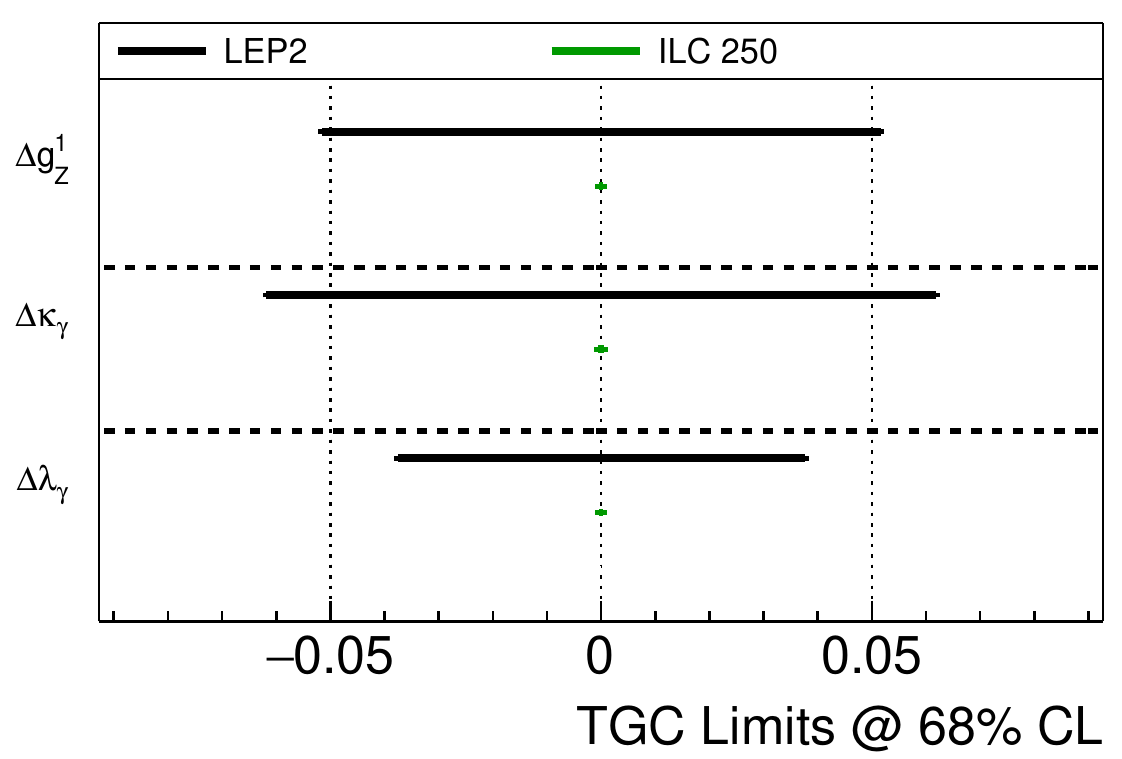}
		
	\caption{Comparison of the reachable TGC precision from a simultaneous fit of all three parameters: ILC~\cite{bib:TGC_EPS17} and final results from LEP combined from ALEPH, L3 and OPAL results~\cite{bib:LEPTGC}. No comparable hadron collider results are available.
	}
	\label{fig:TGC_3par}
\end{figure}

\subsubsection{$W$ mass measurement at 250\,GeV}
\label{subsubsec:ew_mw}
The analysis of $W^+W^- \to qql\nu$ discussed in the previous sections, as well as the study of single-$W$ events also offer an excellent setting for the measurement of the $W$ mass. As discussed in Sec.~\ref{subsec:phys_ff}, the available statistics at the 250\,GeV ILC will be about a factor of 2000 larger than at LEP2, which makes it obvious that a pure consideration of statistical uncertainties is meaningless. While the studies discussed in Sec.~\ref{subsubsec:ew_fullsimww} showed that this kind of events can be selected with high efficiency and purity, a careful extrapolation of the systematic uncertainties of previous measurements is therefore much more instructive than the evaluation of the statistical uncertainty from full detector simulation.
Apart from a scan of the production threshold, kinematic reconstruction of $W$ pair events 
and calorimetric comparison of hadronic $W$ and  $Z$ decays in single-boson events are the most promising techniques, which are described in detail in~\cite{Freitas:2013xga, Wilson:2016tto}.   With a combination of methods and considering advances in theory as well as in the performance of the detectors the systematic limit has been estimated as 2.4\,MeV. This is expected to be reached already at the 250\,GeV stage of the ILC. Additional datasets at higher center-of-mass energies 
could then provide independent information in order to cross-check and constrain systematic effects.

\subsection{Analyses of $e^+e^- \to f \bar{f}$}
\label{subsec:ew_ffana}
Another important class of processes at $e^+e^-$ colliders is fermion-antifermion production, which is highly sensitive to various new physics models, as discussed in Sec.~\ref{subsec:phys_ff}. Thereby, the important observables are the polarised total cross sections, in particular in form of the left-right asymmetry \ALR, as well as the differential cross section as a function of the polar angle, $d\sigma/d \cos{\theta}$, which contains even more information than the forward-backward asymmetry $A_{\mathrm{FB}}$.

\subsubsection{General experimental aspects}

At center-of-mass energies above the $Z$ pole, di-fermion production will be accompanied frequently by a significant amounts of ISR. For example,  at $\sqrt{s}=$250\,GeV, about half of the di-fermion events return to the $Z$ pole. The ISR photons may escape undetected through the beam pipe, or they can be produced at a sufficiently large angle to be measured in the detector.  The forward acceptance of the ILC detectors is assisted by dedicated forward calorimeters described in  Sec.~\ref{subsub:det:forward}.

In the latter case, energy and momentum constraints can be employed to reconstruct the full event kinematics from the angles of the fermions and the photon, {\em without relying on their calorimetrically measured energies or momenta}. This technique offers an excellent opportunity to cross calibrate the energy scales of various subsystems, \eg, to calibrate the photon energy scale against the momentum scale of the tracking systems in $e^+e^- \to \mu^+\mu^-\gamma $ events. While in principle also the beam energy spectrum can be obtained from this method, it suffers from large event-by-event statistical fluctuations due to the relatively large width of the $Z$ resonance~\cite{Wilson:2016hne}.

But also in the case that there is no photon detected, the amount of collinear beamstrahlung or ISR energy can be reconstructed from kinematic constraints on an event-by-event basis. In this case, however, the measured momenta of the fermions have to be used. The previously mentioned case of $e^+e^- \to \mu^+\mu^-\gamma $, then provides an excellent method for an in-situ determination of the beam energy spectrum, since the muon momentum scale can be calibrated to 10\,ppm from $J/\psi \to \mu^+\mu^-$ decays~\cite{Wilson:2016hne}.

In presence of beam polarisation, another important observable becomes accessible, namely the left-right asymmetry of 2-fermion processes:
\begin{equation}
\ALR = \frac{\sigmaLR-\sigmaRL}{\sigmaLR+\sigmaRL}
\label{eq:defALR}
\end{equation}
The parameters \sigmaLR, \etc,  are the chiral cross sections for fully polarised beams, defined in Sec.~\ref{subsec:beampol}.  Their 
relation to the cross section for partial polarisation is described in that section.  Here, we will focus on the measurement of
the polarised cross sections and of angular distributions for the various 2-fermion processes.

\begin{figure*}[hbt]
       \includegraphics[width=0.45\linewidth]{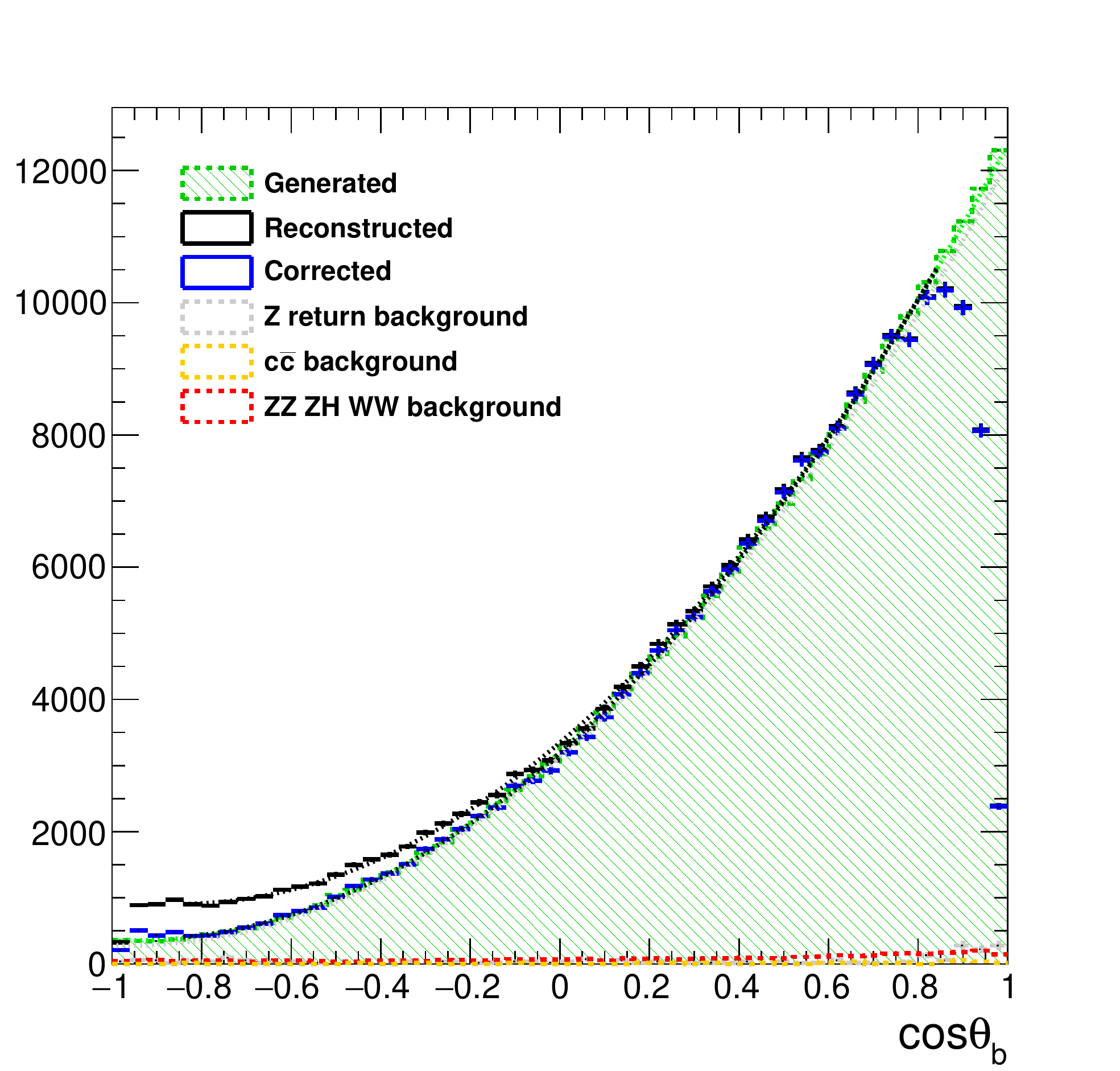}
       \llap{\shortstack{%
                       \includegraphics[clip, trim=0cm 0cm 1.8cm 1.7cm, scale=.15]{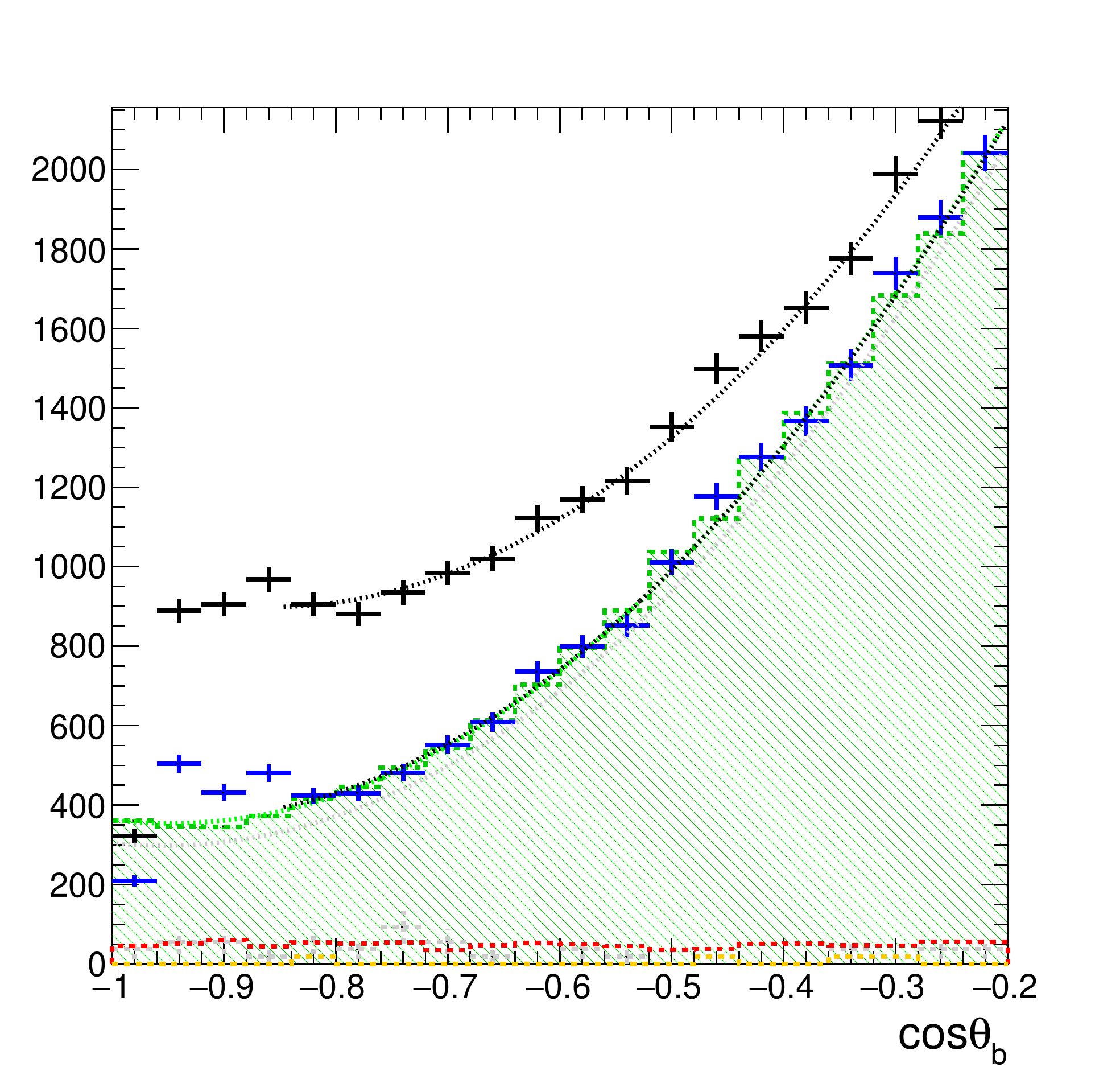}\\
                       \rule{0ex}{0.67in}%
               }
               \rule{2.2in}{0ex}}
        \hspace{0.2cm}       
        \includegraphics[width=0.45\linewidth]{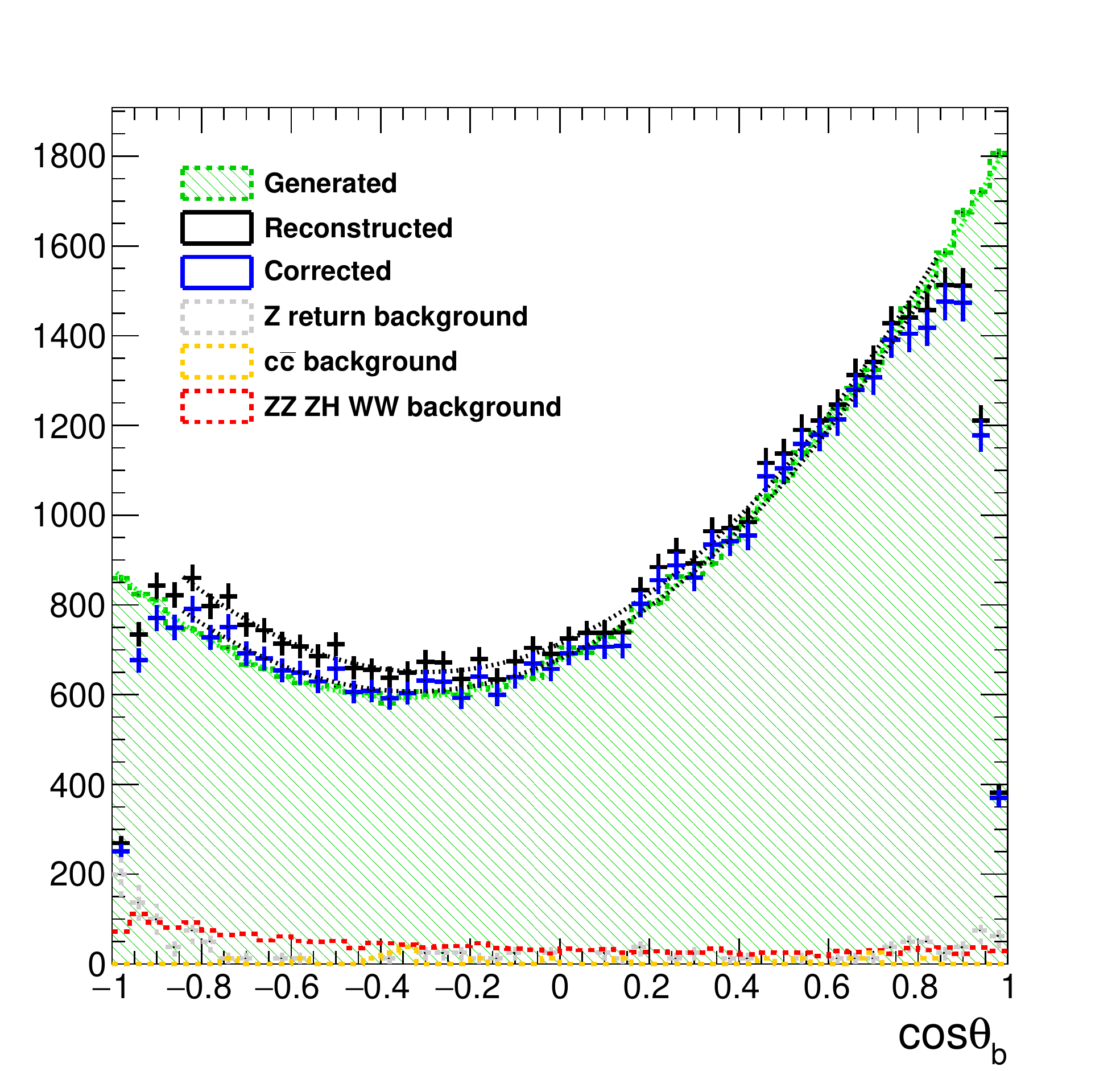}
	\caption{Polar angle distribution $\cos{\theta_b}$ of generated $b$-quarks and final reconstructed 
         $b$-jets including any SM  background remaining after event selection. 
         Left:  $P(e^+,e^-)=(+100\%,-100\%)$ with a zoom of the region with negative 
         $\cos{\theta_b}$.
         Right: $P(e^+,e^-)=(-100\%,+100\%)$. From~\cite{Bilokin:2017lco}.  }
	\label{fig:ffbar_basym}
\end{figure*}

\subsubsection{Inclusive $e^+e^- \to f\bar{f}$ analyses}
Di-fermion production at a center-of-mass energy of 250\,GeV has recently been studied both by ILD and SiD, albeit with complementary goals. ILD has performed a study of all di-lepton channels in full detector simulation, including all leptonic 2-fermion and 4-fermion backgrounds and focussing on events with $\sqrt{s'}>230$\,GeV~\cite{bib:deguchi_lcws18, Yamashiro:2018ant}. After a simple cut-based event selection, purities of 97-99\% can be obtained, while retaining a signal of 26 million events in the $e^+e^-$ case and of about 0.75 million events in the $\mu^+\mu^-$ case and 0.6 million events in the $\tau^+\tau^-$ case. The polar angle distributions of the selected events are then compared to the predictions of various BSM models, which fall into two classes: tree-level exchange of additional, $E6$-insprired $Z'$ bosons and loop-effects from dark matter candidates on the $\gamma/Z$ propagator. In the case of the $Z'$ models, the reach for a 3$\sigma$ observation ranges between 1.6 and 4.8\,TeV, depending on the exact model, while it is between 165 and 460\,GeV in case of the dark matter models, again depending on the exact type of model. These numbers so far combine only the electron and muon channels, therefore further improvement is expected once the $\tau$-channel has been included in the combination.

SiD on the other hand has performed a study focussing on the measurement of \ALR\ from radiative returns to the $Z$ pole. They studied inclusively all di-lepton and di-jet channels in fast detector simulation~\cite{bib:ueno_lcws18}. Thereby they make use of the method mentioned above in order to obtain the boost between the $Z$ rest frame and the lab frame from the angles of the two leptons or jets. After a simple cut-based event selection, about 4.5 million hadronic $Z$ events and about 0.5 million leptonic $Z$ events remain over a background of 1.2 million events for 250\,\ifb\ with \Pmp=(-80\%,+30\%). Exploiting the modified Blondel scheme~\cite{Blondel:1987wr, Monig:2001db} in order to extract \ALR\ directly from the polarised cross sections measured in the four different beam helicity configuraions, the estimated uncertainty on \ALR\ for the full 2\,\iab\ is $\Delta \ALR = 0.00039$. This number is strictly speaking a statistical uncertainty only. However, due to the redundancy offered by the presence of positron polarisation in combination with the ``quasi-concurrent'' collection of the four data sets with different beam helicity configurations, the impact of systematic uncertainties is expected to be very small, especially if \ALR\ is extracted from a global fit to several physics processes, which is, in contrast to the modified Bondel scheme, fully robust against unequal absolute polarisation values when flipping the sign of the polarisation.
For more details on the discussion of the impact of polarisation on the control of systematic uncertainties see Sec.~\ref{subsec:polarisation}.

\subsubsection{$e^+e^- \to \tau^+\tau^-$}
In the special case of $e^+e^- \to \tau^+\tau^-$, the decays of the $\tau$ lepton can be used to determine their polarisation, which adds extra information about the $Z \tau \tau$ vertex. For the polarisation measurement, the individual $\tau$ decay modes have to be identified and treated separately.
It has been shown in full simulation of the ILD detector at $\sqrt{s}=500$\,GeV~\cite{Behnke:2013lya, Suehara:2009nj} that the leptonic decay modes can be identified with efficiencies and purities of about 99\%. For the $\pi$ and $\rho$  decay modes as well as for the three-prong $\tau \to a_1 \nu_{\tau}$ decay, the same study reached purities of 90\% at efficiencies between 96\% and 91\%, while for the one-prong $\tau \to a_1 \nu_{\tau}$ decay efficiencies and purities of about 70\% were achieved. In a more recent study~\cite{Tran:2015nxa}, covering only the separation of the hadronic decay modes, efficiencies and purities of about 90\% were demonstrated also for the three-prong decay. 

After a polarisation analysis of the leptonic channels and the $\pi$ and $\rho$ channels via an optimal observable technique, the polarisation of the $\tau$ leptons can be measured with a relative uncertainty of about 1\% already with an integrated luminosity of 500\,\ifb.  

\subsubsection{$e^+e^- \to b\bar{b}$}

The couplings of the $b$-quark are of particular interest because as a lighter sister of the top-quark, it could be partially composite in extensions of the SM. Also, a rather large discrepancy in the $\sin^2{\theta_{eff}}$ values extracted from the measurement of the foward-backward asymmetry of $b\bar{b}$ production at LEP and the \ALR\ measurement at SLD still persists today and could be easily 
confirmed or rejected by remeasuring $b\bar{b}$ production at the ILC 250. The prospects for this measurement have recently been evaluated in full, GEANT4-based simulation of the ILD detector~\cite{Bilokin:2017lco, bilokin:tel-01946099}.

Thereby the special difficulty is to distinguish the $b$-quark from the $\bar{b}$-quark. This can either be achieved by reconstructing the charge sum of the tracks from the $b$/$\bar{b}$ decay vertices, which requires highest efficiency and precision for the reconstruction of tracks, also at small momenta and in the forward region --- or via the charge of Kaons identified via their specific energy loss in the Time Projection Chamber of ILD, see Sec.~\ref{subsubsec:ILDtracker}.

Figure~\ref{fig:ffbar_basym} compares the $\cos{\theta}$ distribution of the $b$-quark at the generator-level, at reconstruction-level and after all corrections applied, for both opposite-sign polarisation configurations. Both techniques for identifying the $b$-direction, vertex charge and Kaon ID, have been combined here and contribute about equally to the final event sample. These distributions are then used to extract the left- and right-handed couplings of the $b$-quark to the $Z$-boson and the photon --- or alternatively form factors $F^{\gamma /Z}_{1A/V}$. The expected precisions on a subset of couplings and form factors is compared to the corresponding LEP results in Fig.~\ref{fig:LEPILCResult_3}. The ILC projections in this plot are based on only 500\,\ifb\ at 250\,GeV, corresponding to the data collected in the first couple of years before the luminosity upgrade. For the full data set, further improvement by about a factor of 2 is expected.

\begin{figure}[tbh]
	\centering
	\includegraphics[width=0.8\columnwidth]{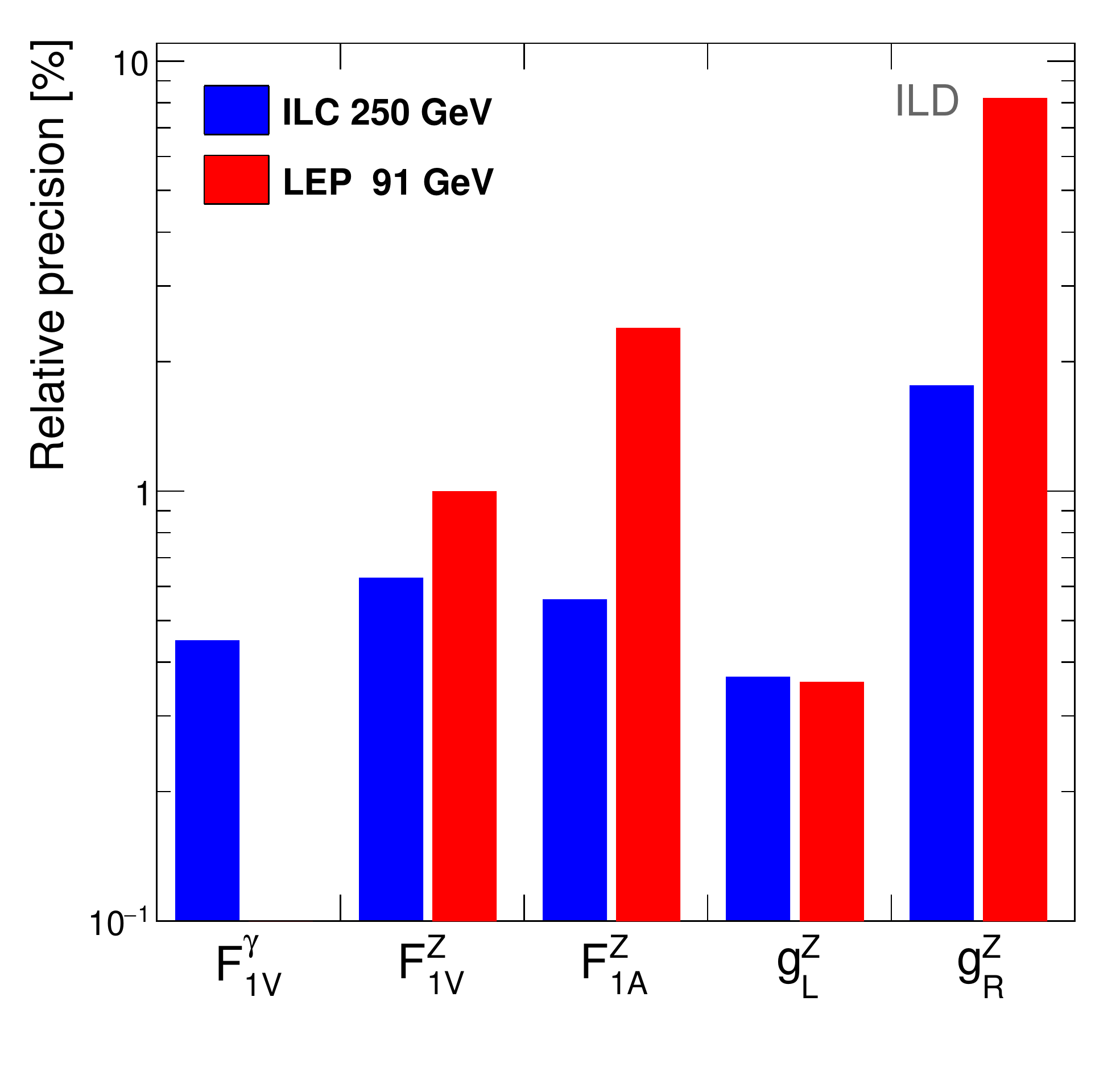}
	\caption{\sl  Comparison of the LEP measurements to the expected 
                  precision at the ILC. The results of the ILC correspond to the 
                  integrated luminosity of $\mathcal{L}_I = 500$\,fb$^{-1}$ to be collected 
                   at $\sqrt{s} = 250$\,GeV before the luminosity upgrade. Final results for the
                  full 250\,GeV dataset would improve the precision further by about a factor of 2.
                   From~\cite{Bilokin:2017lco}.}
	\label{fig:LEPILCResult_3}
\end{figure}


\section{\label{sec:top}Physics Simulations: Top quark}


In this section, we will review the  highlights of the top quark
program of linear colliders.     Since the top quark has not yet been
studied in $\ee$  reactions,  its study at a linear collider gives the
opportunity to dramatically improve the precision with which we know
its properties.   As we have explained above, such precision
measurement can reveal clues to the origin of the large mass of the
top quark and possibly, through this, to the nature of the Higgs
interactions that give mass to all fermions.
The potential of linear $e^+e^-$ colliders for top quark physics is
discussed in more detail in the 
ILC design reports~\cite{Baer:2013cma,Behnke:2013lya}
and in Refs.~\cite{Agashe:2013hma,Vos:2016til,Abramowicz:2018rjq}.

\subsection{Selection and reconstruction of top-quark pairs}
\label{subsec:top:reco}

The $e^+e^- \rightarrow t\bar{t}$ production process has a sizeable cross section above
the top quark pair production threshold. With a left-handed electron beam and
right-handed positron beam the cross section reaches approximately~1 pb at 
$\sqrt{s}=$ 500~\GeV{}. Top quark pair production, with the top quarks decaying to a
$W$-boson and a bottom quark, is the leading six-fermion process. Most recent 
simulation studies~\cite{Amjad:2015mma,Bernreuther:2017cyi,Abramowicz:2018rjq}
have focused on the the lepton+jets final state, where one of the $W$-bosons decays 
to a charged lepton and a neutrino and the other $W$-boson to jets. Compared to the 
fully hadronic final state~\cite{Devetak:2010na}, the lepton + jets final state offers 
the advantage of the presence of an energetic charged lepton, that helps to tag top 
and anti-top quarks and is an efficient polarimeter.

The selection of top quark pair events at the ILC is straightforward. For lepton+jets
events the requirement of a charged lepton and two b-tagged jets is sufficient to reduce
the Standard Model background. The efficiency of the selection can be very high, 
between 50 and 80\% depending on the purity requirement. 

The complete reconstruction of the $t\bar{t}$ system is more
challenging. The assignment of the six fermions to top and anti-top quark candidates
suffers from combinatorics that can lead to significant migrations in differential
measurements. Their effect on observables such as the forward-backward asymmetry are 
kept under control by a rigorous selection on the reconstruction quality. 
The size of potential systematic effects due to the selection and reconstruction 
of the complex six-fermion final state is then expected to be
sub-dominant~\cite{Amjad:2015mma}. Later studies have extended this conclusion 
to a broader set of 
observables~\cite{Bernreuther:2017cyi,Durieux:2018tev, Abramowicz:2018rjq}. 

To take full advantage of the large integrated luminosity envisaged in the ILC 
operating scenario, a rigorous control of experimental and theoretical uncertainties 
is required.  Ultimately, we expect that the data-driven techniques developed for the bottom-quark
analysis of Ref.~\cite{Bilokin:2017lco, bilokin:tel-01946099} will supply an 
in-situ measurement of the rate of wrong combinations. This will allow one to correct
differential measurements using the statistical power of the entire data set. 

A complete and quantitative analysis of systematic limitations is currently 
ongoing. The results presented in this section are based on the
prospects of Ref.~\cite{Amjad:2015mma,Bernreuther:2017cyi,Abramowicz:2018rjq}.
Where needed, results are extrapolated to the full integrated luminosity of
4~\iab.

\subsection{Measurement of the top quark mass}
\label{subsec:top:topmass}

The top quark mass is a fundamental parameter of the Standard Model that must be
determined experimentally. Precise measurements are essential for precise tests of
the internal consistency of the Standard Model, through the electro-weak
fit~\cite{Baak:2014ora} or the extrapolation of the Higgs potential to
very high energy
scales~\cite{Degrassi:2012ry}.   The precise value of the top quark is
also needed as input to the theory of flavor-changing weak
decays~\cite{Buras:2009if} and models of the grand unification of the
fundamental interactions~\cite{Langacker:1994vf}.

The top quark pair production threshold was identified long ago~\cite{Gusken:1985nf} as
an ideal laboratory to measure the top quark mass, and other properties such as the top quark
width and the Yukawa coupling and the strong coupling constant~\cite{Strassler:1990nw}.
The large natural width of the top quark acts as an infrared cut-off,
rendering the threshold cross section insensitive to the non-perturbative confining part
of the QCD potential and allowing a well-defined  cross section
calculation within  perturbative QCD.  This calculation has now been
carried out the N$^3$LO order~\cite{Beneke:2015kwa} with  NNLL resummation~\cite{Hoang:2013uda}. Fully differential results are available in WHIZARD~\cite{Bach:2017ggt}.

Given this precise theoretical understanding of the shape of the
$t\bar t$ threshold cross section as a function of center of mass
energy, it is possible to extract the value of the top quark mass by
scanning the values of  this cross section near threshold.  We
emphasize that the top quark mass determined in this way is, directly,
a short-distance quantity that is not subject to significant
nonperturbative corrections.
It is also closely related to the $\msb$ top quark mass, the input to
the theory calculations listed above. The uncertainty in the
conversion is less than 10~MeV~\cite{Marquard:2015qpa}.  This
contrasts with the situation at hadron colliders, where the
conversion uncertainties, the nonperturbative corrections, and the
experimental systematics in the measured top quark mass
contribute  independent uncertainties, each of which is about
200~MeV.

\begin{figure}[tb]
 \begin{center}
 \includegraphics[width=0.8\hsize]{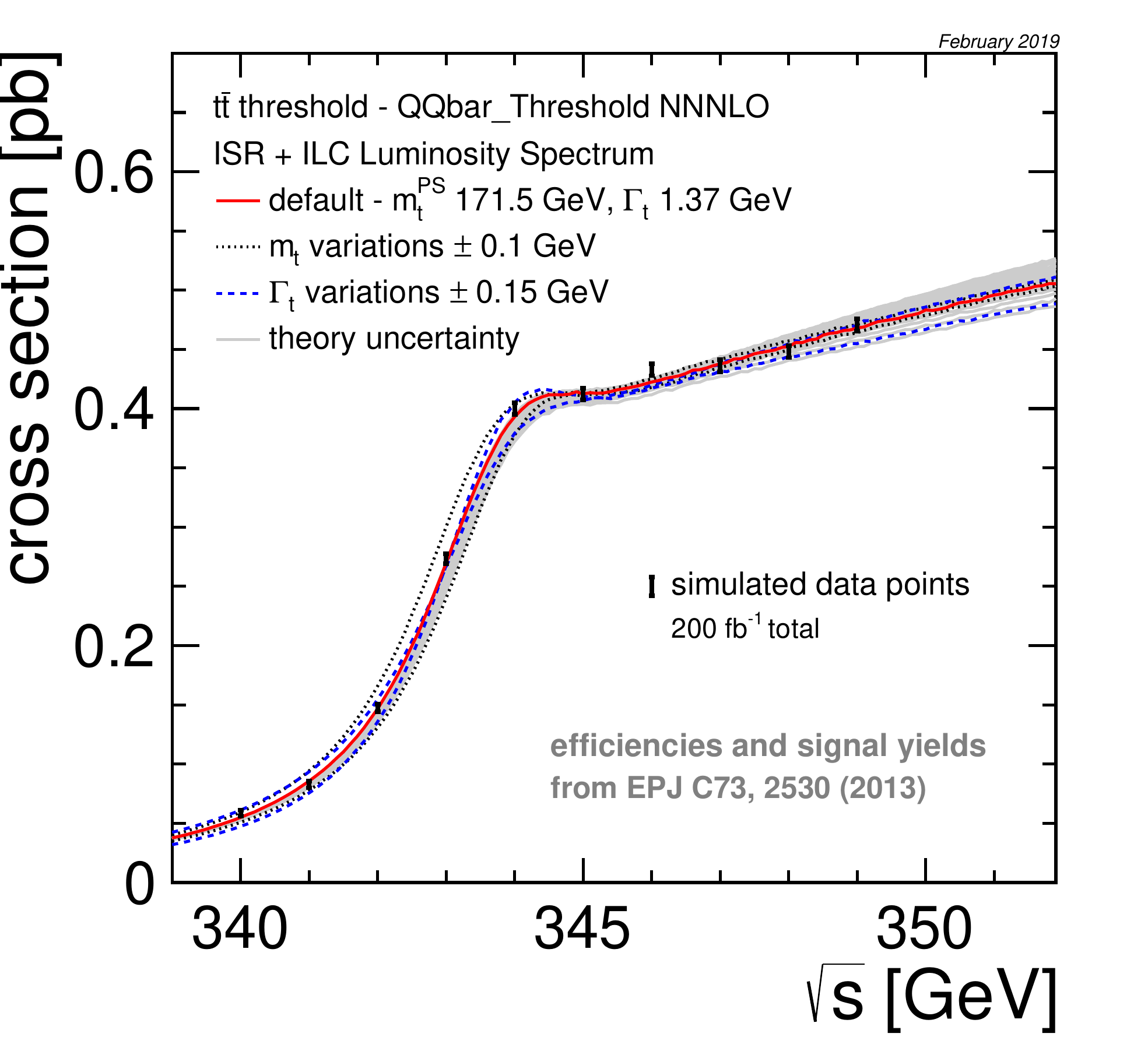}
 \includegraphics[width=0.8\hsize]{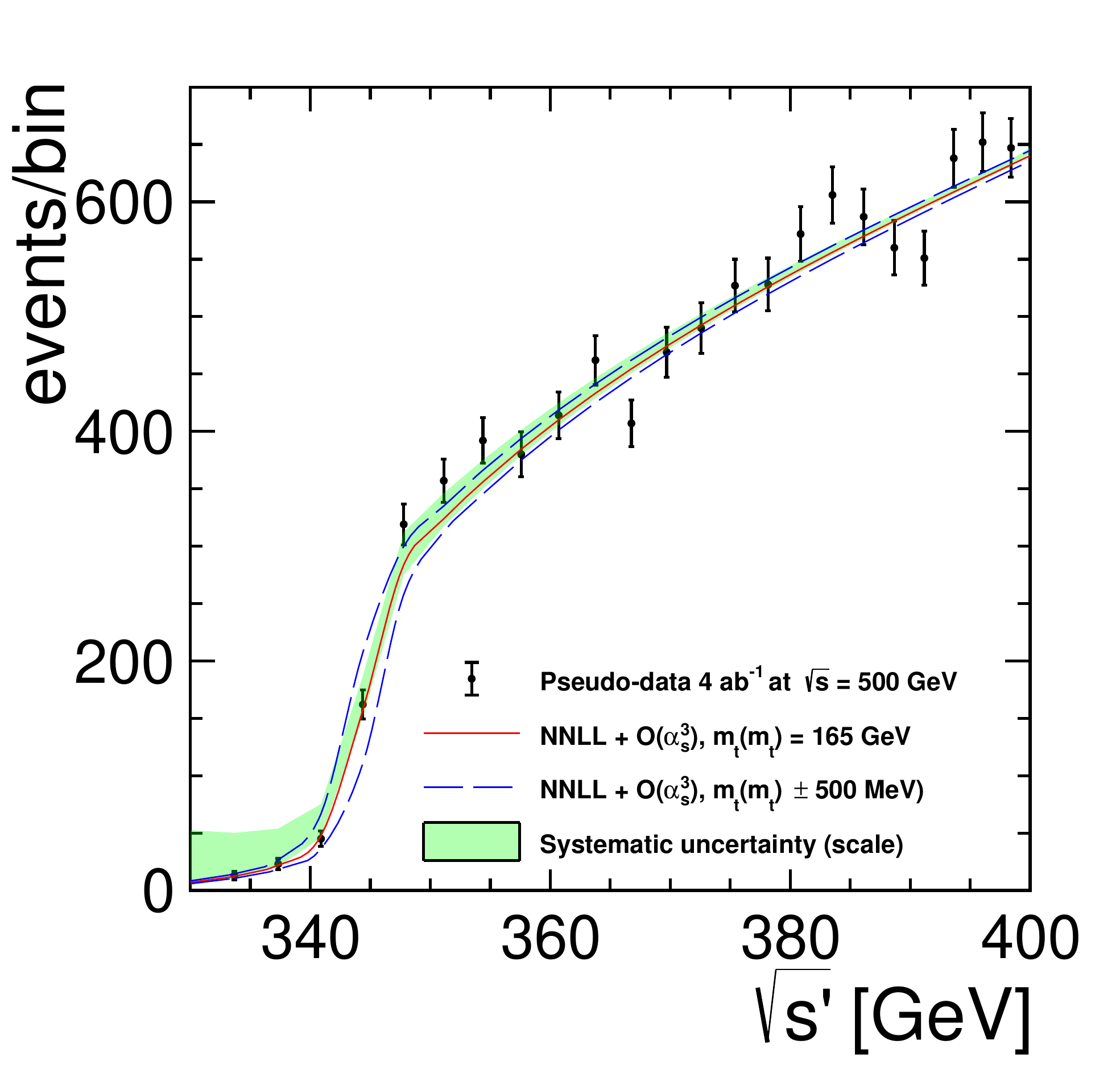}
\caption{Two methods to extract the top quark mass: (upper panel) a scan of the center 
of mass energy through the top quark pair production threshold, and (lower panel) a reconstruction of the differential $t\bar{t}\gamma$ cross section as a function of 
$s'$ during operation at $\sqrt{s}=$ 500~\GeV{}.
\label{fig:top_mass}}
 \end{center}
 \vspace{-0.7cm}
 \end{figure}

A simulation of the threshold scan is presented in
Fig.~\ref{fig:top_mass}. The scan of the $t\bar t$ threshold measures the
top quark pair production cross section at ten $e^+e^-$ center-of-mass energy points. 
The error bars on the pseudo-data point represent the statistical uncertainty of
the measurement, the uncertainty band indicates the theory (scale) uncertainty 
of the calculation. A fit of the line shape will give
a precise extraction of the top quark mass~\cite{Martinez:2002st,Horiguchi:2013wra,Seidel:2013sqa}.
The statistical uncertainty on the threshold mass is reduced to 
below 20~MeV with a scan of ten times 20~\ifb. The total uncertainty 
on the $\msb$ mass can be controlled to the level
of 50~\MeV. These systematic uncertainties
include a rigorous evaluation of theory uncertainties in the 
threshold calculation and in the conversion
to the $\msb$ scheme~\cite{Simon:2016pwp}.

The top quark mass can also be measured precisely in operation above the $t\bar t$
threshold. The top quark mass can be extracted from the differential cross section of $t\bar t \gamma$  events as a function of the center of mass energy of the $t \bar t$ system 
$\sqrt{s'} = \sqrt{s} ( 1 - E_{\gamma}/\sqrt{s})$, as shown in 
 Fig.~\ref{fig:top_mass}. A fit with a calculation that matches the 
NNLL prediction for the threshold region with an $\mathcal{O} (\alpha_{s}^3)$ calculation for 
the continuum yields a statistical uncertainty with 100~\MeV{} for 4~\iab{} at 500~\GeV{}~\cite{Abramowicz:2018rjq}. Including the theory uncertainty due to scale variations
and experimental systematic uncertainties the total uncertainty is estimated to be below
200~\MeV. 

A direct mass 
measurement can reach a statistical precision below 100~\MeV{}~\cite{Seidel:2013sqa} and will 
be helpful to clarify the interpretation of such measurements.

A linear $e^+e^-$ collider
can thus achieve a precision that goes well beyond
even the most optimistic scenarios for the evolution 
of the top quark mass measurement at the LHC.

\subsection{Searches for flavour changing neutral current interactions of the top quark}
\label{subsec:top:fcnc}

Among the direct searches for physics beyond the Standard Model with top quarks 
in the final state, the searches for flavour changing neutral current interactions of 
the top quark have been studied in most detail. Thanks to the excellent charm
tagging performance and the clean experimental environment such searches at the ILC
can compete with the sensitivity of the LHC to anomalous $ t Z c$, $t H c$ and 
$t\gamma c$ couplings.

Searches for $e^+e^- \rightarrow t c$ production can
already be performed during the 250~\GeV{} stage~\cite{Hesari:2014eua}. Greater
sensitivity can be achieved in searches for $t \rightarrow Hc$ and 
$t\rightarrow \gamma c$ decays above the $t\bar{t}$ production threshold.
Based on full-simulation studies~\cite{Zarnecki:2018wsw, Abramowicz:2018rjq} 
scaled to an integrated luminosity of 4~\iab{} 
at a center of mass energy of 500~\GeV{}{}, the 95\% C.L. limits on FCNC branching
ratios are expected to reach $BR(t \rightarrow Hc) \sim 3 \times 10^{-5}$ and 
$BR(t \rightarrow \gamma c) \sim 10^{-5}$, well in excess of the 
limits expected after 3~\iab{} at the HL-LHC.

\subsection{Measurement of the top quark electroweak couplings}
\label{subsec:top:topelectroweak}

Composite Higgs models and models with extra dimensions naturally
predict  large corrections to the top quark couplings to the $Z$ and
$W$ bosons~\cite{Richard:2014upa,Barducci:2015aoa,Durieux:2018ekg}.
The study of top quark pair production at an $e^+e^⁻$ collider therefore provide
a stringent test of such extensions of the SM.

The potential of the 500~GeV ILC for the measurement of the cross section and forward-backward asymmetry in
$t\bar{t}$ is characterized in detail in Ref.~\cite{Amjad:2015mma}. It
is important to note that these measurements search for deviations
from the SM in the main production mechanism of the $t\bar t$ system
through $s$-channel $\gamma$ and $Z$ exchange.   With two
configurations of the beam polarization, measurement of the angular
distribution, and measurement variables sensitive to the $t$ and $\bar
t$ polarizations, all 6 possible $CP$-conserving form factors can be
disentangled and constrained at the 1\% level. 
Especially designed $CP$-odd observables can also provide precise and
specific constraints on the $CP$-violating form factors~\cite{Bernreuther:2017cyi}. 
The expected 68\% C.L. limits on the form factors with 500~\ifb{} at
a center of mass energy of 500~\GeV{} are compared to the HL-LHC
expectation of Ref.~\cite{Baur:2004uw,Baur:2005wi} in Fig.~\ref{fig:top_ew_couplings}.

\begin{figure}[tb]
 \begin{center}
 \includegraphics[width=0.8\hsize]{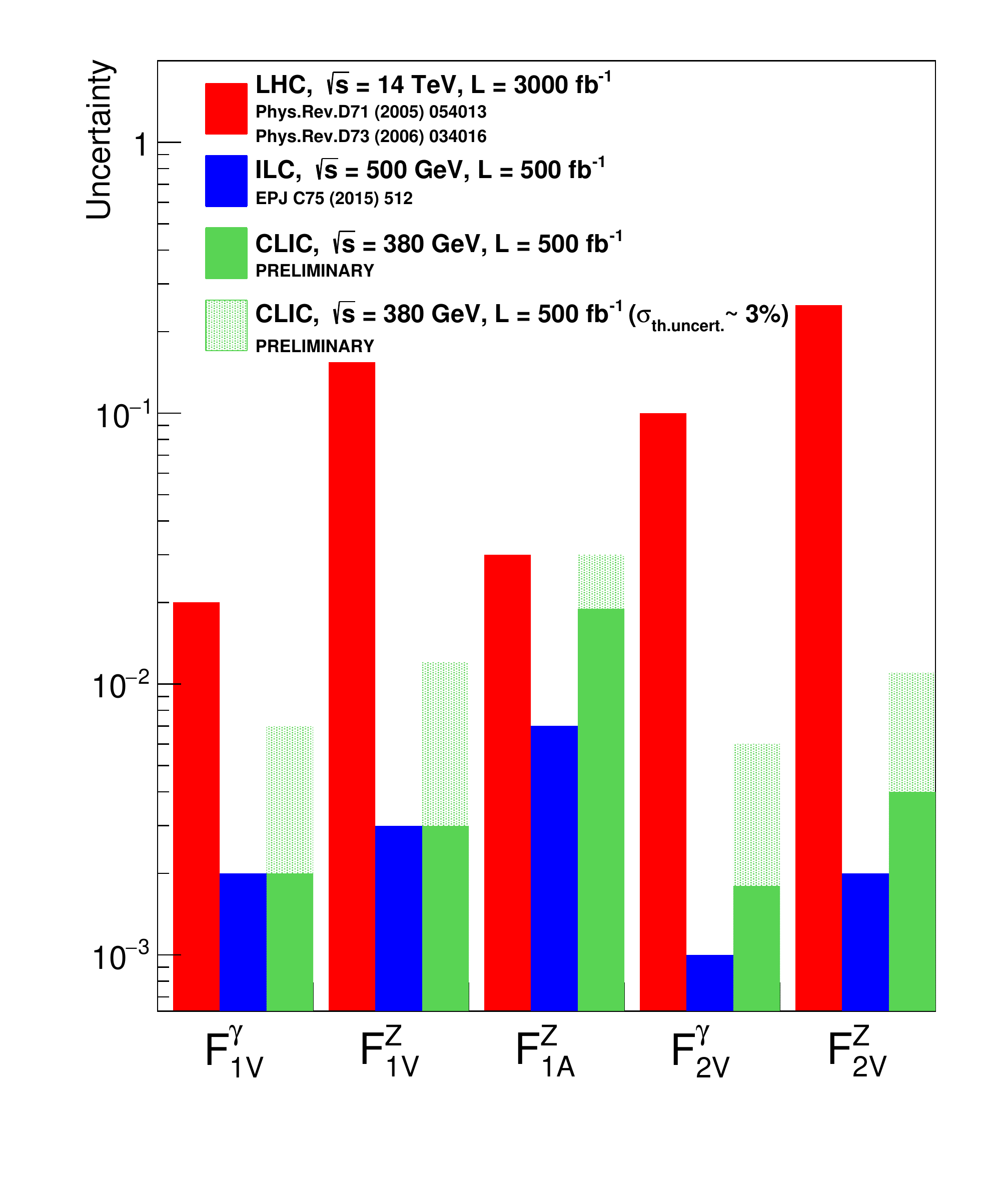}
 \includegraphics[width=0.8\hsize]{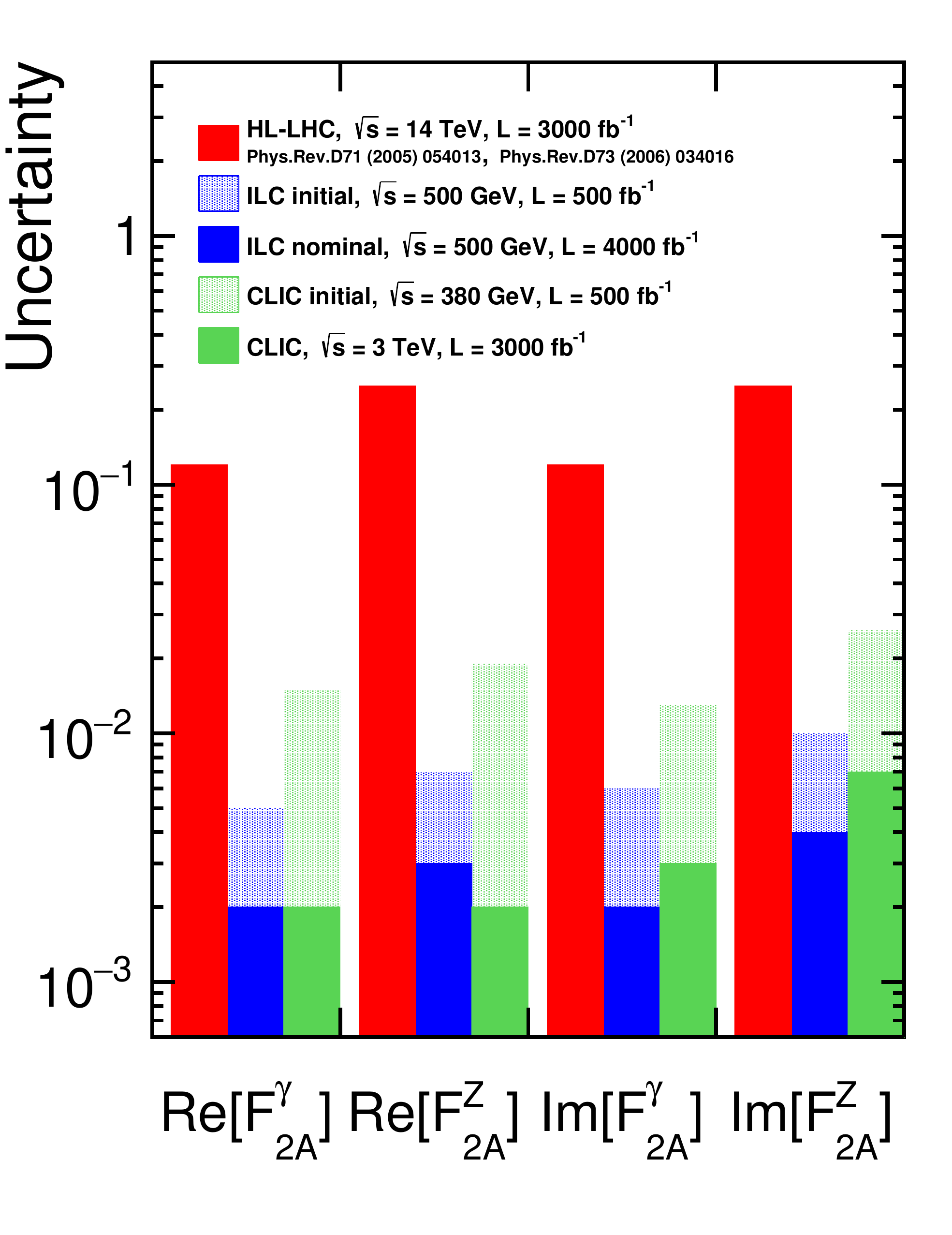}
\caption{The ILC prospects for the measurement of the electroweak couplings of the top quark, expressed as 68\% C.L. bounds on the form factors in a general expression 
for the Lagrangian that describes the $t\bar{t}Z$ and $t\bar{t}\gamma$ vertices:
(upper panel) CP-conserving form factors from Ref.~\cite{Abramowicz:2016zbo}, 
and (right) CP-violating form factors, from Ref.~\cite{Bernreuther:2017cyi}.
\label{fig:top_ew_couplings}}
 \end{center}
 \vspace{-0.7cm}
 \end{figure}

The corrections to the top quark electroweak couplings can 
be parametrized by set of 
dimension-6 operators of the SM EFT that contain the top quark as a
field. There is a large number of such operators, although the
restriction to $\ee$ reactions allows us to concentrate on a limited
set of operators that appear at the tree level in electroweak pair
production.
Because the top quark is massive, helicity conserving operators such
as chiral 4-fermion operators and helicity violating operators that
give corrections to the top quark magnetic moments must be considered
on an equal footing. 

In 
Ref.~\cite{Durieux:2018tev}, the authors consider the
perturbation of the reaction $\ee\to t\bar t$ by the 10 dimension-6
operators that contribute to the cross section at the tree level.
They show that a combination of
 the 500~GeV run, with excellent sensitivity to two-fermion operators,
with 1~TeV{} data, with increased sensitivity to four-fermion
operators, yields  tight constraints independently on 
all operator coefficients.  This study demonstrates the feasibility of a global EFT
analysis of the top sector at the ILC.  It also gives an  expected sensitivity 
of the ILC to top electroweak couplings that  exceeds that of the HL-LHC programme 
by one to two orders of magnitude. Translated into discovery potential for concrete 
BSM scenarios, a linear collider operated above the top quark pair production 
threshold can probe for compositeness of the Higgs sector to very high scales, up 
to 10~TeV and beyond~\cite{Durieux:2018ekg}. 

Figure~\ref{fig:top_bottom_ew_fit} presents the results of a combined fit of 
the Wilson coefficients dimension-six operators that affect the electroweak 
interactions of bottom and top quarks. 
For each operator, limits are extracted from existing LEP I and LHC run 2 data and 
from prospects for the high-luminosity stage of the LHC and for ILC runs at 
$\sqrt{s}=$ 250~\GeV{} and 500~\GeV.

The LHC measurements in the top quark sector are extrapolated to the complete program,
including the high-luminosity phase of the LHC. The detailed analyses presented in
Ref.~\cite{Azzi:2019yne} predict that significant progress can be made, especially
in the measurements on rare associated processes.
The HL-LHC scenarios S1 and S2 are defined in analogy to the two scenarios
defined for Higgs coupling measurements in Ref.~\cite{Cepeda:2019klc}. 
Both contemplate 3~ab$^{-1}${} at $\sqrt{s}=$ 14~\TeV, but assume a very different
scaling of the systematic uncertainties. In scenario S1 systematic uncertainties 
are fixed to today's values;
in S2 the experimental systematic uncertainties scale with integrated luminosity
like the statistical uncertainties and theory uncertainties are reduced by a factor of 2 with respect 
to the current state of the art.

The ILC250 scenario includes
measurements of the cross section and forward-backward asymmetry of bottom quark 
pair production, with a total integrated luminosity of 2~ab$^{-1}$ divided between the
left-right and right-left beam polarizations, following Ref.~\cite{Bilokin:2017lco}.
The ILC500 prospects includes in addition the projections at $\sqrt{s}=$ 500~\GeV{} of
Ref.~\cite{Durieux:2018tev} for top-quark pair production, scaled to an integrated
luminosity of 4~ab$^{-1}${} at $\sqrt{s}=$ 500~\GeV{}.

\begin{figure*}[tb]
 \begin{center}
 \includegraphics[width=1.0\hsize]{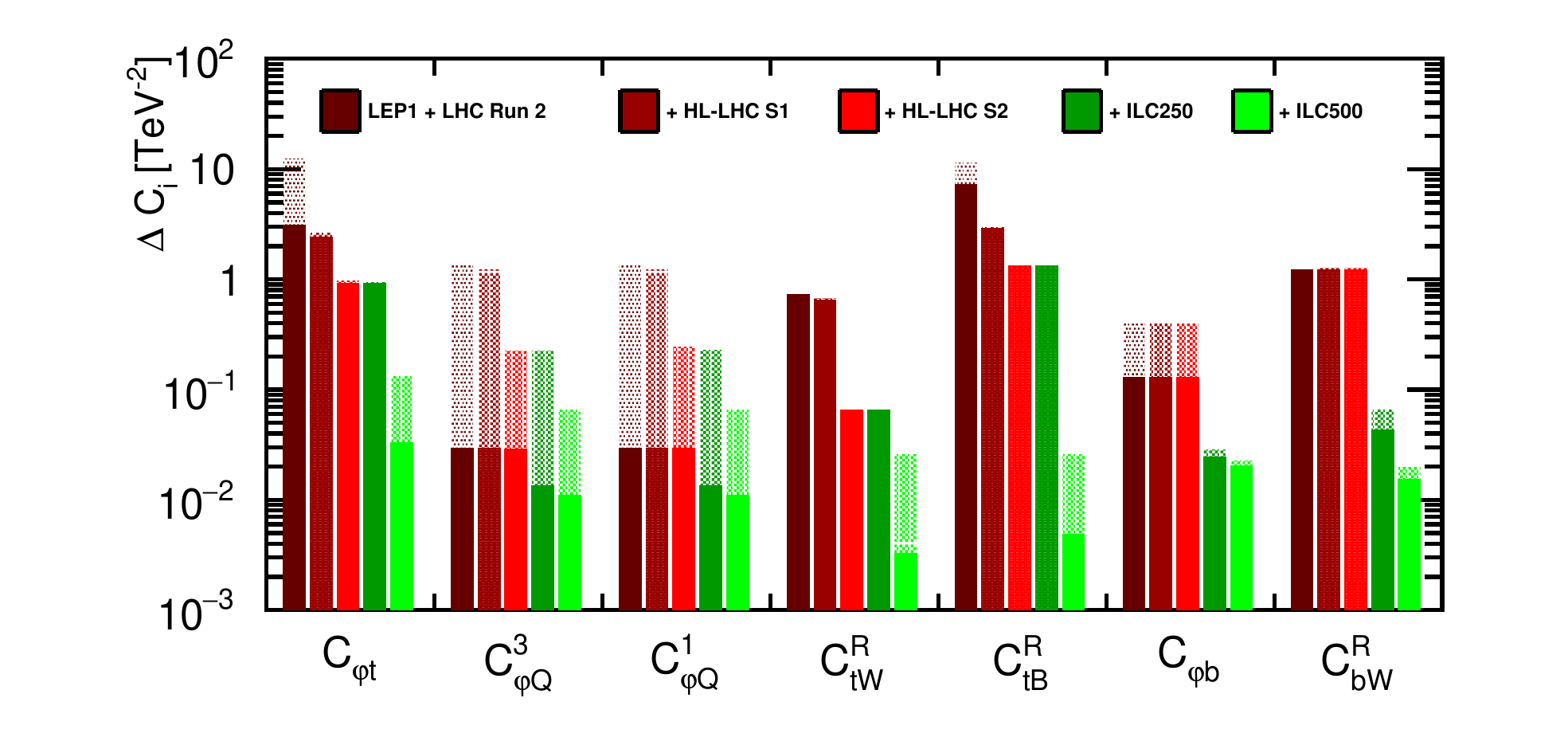}
\caption{The 68\% C.L. limits on the Wilson coefficients of the dimension-six 
operators that affect the electroweak interactions of the top and bottom quark. 
The filled bars indicate {\em individual} limits, from a 
single-parameter fit, while the dashed bars present the {\em marginalized} limits of
a seven-parameter fit. 
The first bar indicates the current limits, obtained in a fit to LEP~I and LHC run 2 
measurements. These results are compared with prospects for the HL-LHC, with
3~\iab{} at $\sqrt{s}=$ 14~\TeV{}. The S1 and S2 scenarios are defined in analogy 
with the 
Higgs scenarios of Ref.~\cite{Cepeda:2019klc}: S1 envisages no improvement of the systematic 
uncertainties, while S2 foresees that theory uncertainties can be reduced by a
factor 2 and experimental systematic uncertainties are expected to scale with
$1\sqrt{s}$. The green bars indicate the 250~\GeV{} and
500~\GeV{} runs of the ILC, with 2~\iab{}. 
\label{fig:top_bottom_ew_fit}}
 \end{center}
 \vspace{-0.7cm}
 \end{figure*}

The $Z$-pole data yield tight constraints on coefficients of operators that are specific 
to the bottom quark $C_{\varphi b}$ and $C_{dW}$. The current LHC constraints on the
top quark operators, from single top production, top quark decay and associated
$t\bar{t}X$ production are relatively weak. The operator coefficients
$C^{1/3}_{\varphi}$, that affect both bottom-quark and top-quark interactions 
have tight individual limits from LEP, but are poorly constrained in a global fit. 
The S2 scenario for the HL-LHC foresees a significant
increase of the precision of the measurements. The 250~GeV run at the ILC considerably
sharpens the limits on the operators that affect the bottom-quark interactions with 
$Z$-bosons and photons. Finally, with the 500~\GeV{} data, the operators specific to 
the top quark, the dipole operators $C_{tB}$ and $C_{tW}$ and the operator that
$C_{\varphi t}$ that modifies the right-handed coupling of the top quark. 
With respect to the current precision the constraints on all operator coefficients 
are expected to improve by one to two orders of magnitude. The increase in precision 
is very significant even with respect to the most aggressive scenario for the HL-LHC.

Similar analyses, now requiring  only 4 relevant dimension-6 operator coefficients, can
improve the constraints on four-fermion operators involving $b$, $c$,
and light-fermion sectors beyond the results projected in
Sec.~\ref{subsec:ew_ffana}.

\subsection{Measurement of the top quark Yukawa coupling}
\label{subsec:top:topYukawa}

As with the trilinear Higgs coupling, the top quark Yukawa coupling can be
measured either directly or indirectly.  In the literature, most
estimates of the accuracy of determination of the top quark Yukawa
coupling are done within the simple context of the SM with only this
one parameter varied. We will first quote uncertainties
within this model in this section and then explore the implications of a general 
EFT analysis.

Consider first the indirect determination of the top quark Yukawa coupling.
For the Higgs boson decays $H\to gg$, $H\to \gamma\gamma$, and $H\to Z \gamma$, 
there are no SM tree diagrams and so diagrams with top quark loops give leading contributions.  
For $h\to gg$, the top quark loop diagram gives the single largest
contribution. In Tab.~\ref{tab:ILCLHC}, it is shown that the ILC
program up to 500~GeV will determine the effective coupling in this process to bettter
than 1\%. Even higher precision can be obtained in a joint fit
including also the top quark radiative corrections to the cross
sections for $e^+e^- \rightarrow ZH$, $e^+e^- \rightarrow \nu\bar\nu
H$, and $e^+e^- \to  \gamma H$~\cite{Boselli:2018zxr}. However, one
should be uncomfortable that, for this determination, the simple model is
too simple, since new heavy colored particles can also contribute to
these processes at the 1-loop level. 

An indirect determination that calls out the top quark more
specifically is the measurement of the influence of the top quark
Yukawa coupling on the shape of the  $t\bar{t}$ pair production 
cross section very close to the $t\bar{t}$ threshold, due to the Higgs
boson-exchange contribution to the $t\bar t$ potential. In principle,
this effect could give a 4\% determination of the Yukawa coupling if
the QCD  theory of the top quark threshold region were precisely
known.   However, the Higgs-exchange effect is of the same size as
the N$^3$LO QCD corrections.  At this time, the threshold shape is
calculated only to this N$^3$LO order, by the use of a very sophisticated
NRQCD framework~\cite{Beneke:2015kwa}, combined with NNLL resummation of
large logarithms~\cite{Hoang:2013uda}.   Propagating the QCD
uncertainties gives an uncertainty of 20\% on the top quark Yukawa
coupling~\cite{Vos:2016til}, and there is no clear path at this time to improve the
accuracy of the QCD result.

A more direct -- and more robust -- extraction of the top quark Yukawa coupling 
is possible from measurements of the cross section of the associated production 
process of a Higgs boson with a top quark pair. In the SM the $pp \rightarrow t\bar{t}h$ 
and $e^+e^- \rightarrow t\bar{t}h$ production rates are simply
proportional to the square of the Yukawa coupling. The first measurements of the 
$t\bar{t}h$ rate at the LHC~\cite{Sirunyan:2018hoz,Aaboud:2018urx} 
have an uncertainty of approximately 30\% and are expected to improve considerably 
during the remainder of the LHC program~\cite{Cepeda:2019klc}.

At an electron-positron collider, the cross section for $t\bar{t}h$ production 
increases rapidly above $\sqrt{s} \sim 500 $~GeV, reaching several fb for 
$\sqrt{s} = 550$~GeV. Detailed studies of selection
 and reconstruction of these complex multi-jet events
have been performed by the ILC at 500~GeV~\cite{Yonamine:2011jg} and
 1~TeV~\cite{Behnke:2013lya,Price:2014oca} and by CLIC
at 1.5~TeV~\cite{Abramowicz:2018rjq}. The direct measurement of the
top quark Yukawa coupling at the ILC reaches 2.8\%
precision~\cite{Fujii:2015jha}, with 4~\iab{} at 550~GeV.  With a sample of
2.5~\iab\  at 1~TeV, this precision would improve to 2\%~\cite{Asner:2013psa}.  
From the energy-dependence of the cross section and the top polarizations,
this reaction can also be used to probe for non-standard forms of the
$tth$ coupling~\cite{Han:1999xd}.

In principle, the corrections to the top quark electroweak couplings
and to the top quark Yukawa coupling should be parametrized by
dimension-6 operators of the SM EFT. The studies of
Refs.~\cite{Azatov:2016xik,Durieux:2018ggn} 
show that the indirect extraction of the Yukawa coupling from the $hgg$ or
$ h \gamma \gamma$ vertices do not provide robust measurements of the
top quark Yukawa coupling in a general multi-parameter fit. The indirect determination 
is certainly very sensitive to new physics in the Higgs sector, but in case a 
deviation from the Standard Model predictions is observed, further measurements
are needed to unambiguously identify the operator that gave rise to 
the effect.  

The direct determination from the $t\bar{t}h$ rate, be it at the LHC
or at the ILC, is more robust. However, even in this case
vertices arising from dimension-6 operators
that do not directly involve the Higgs boson can affect the cross
section for $\ee\to t\bar t h$ and thus create ambiguity in the
extraction of the top quark Yukawa coupling. In $\ee \rightarrow t\bar t h$,
the EFT corrections arise from 4-fermion $eett$ operators and from
operators that correct the $\gamma$ and $Z$ anomalous moments of the
$t$ quark. Similarly, in hadron-hadron collisions, the cross section
for $gg\to t\bar t H$ is corrected by dimension-6 operator that alter
the top quark vector coupling to gluons and those which create a
possible axial vector coupling to gluons and a gluonic magnetic
moment. The 34-parameter fit on current LHC data of Ref.~\cite{Hartland:2019bjb} 
indeed finds that the marginalized limits on the operator $C_{t \varphi}$
that shifts the top quark Yukawa coupling are considerably weaker 
than the individual constraints and the results of fits with fewer parameters.

The LHC and a linear $e^+e^-$ collider offer excellent opportunities
to probe the interaction between the top quark and the Higgs boson, both
directly and indirectly. In a global EFT fit the coefficients of all operators 
affecting the Higgs branching ratios and $t\bar{t}h$ production rate must be 
constrained to sufficient precision, such that the coefficient of the operator that 
shifts the Yukawa coupling can be extracted unambiguously. Precise measurements at
a linear collider operating above the $t\bar{t}$ threshold provides powerful
constraints to such a fit.



\section{\label{sec:global}Global Fit to Higgs Boson Couplings, and Comparisons of ILC to Other Colliders}

In this section, we make use of the simulation results presented in
Secs.~\ref{sec:higgs} and \ref{sec:ew} to present projections for the
uncertainties in Higgs boson couplings that will be obtained from the
ILC.  We will present projections both for the 250~GeV stage and for
the stage that includes at 500~GeV in the centre of mass, following
 the plan presented in Sec.~\ref{sec:runscenarios}.

\subsection{Elements of the fit to Higgs couplings from Effective Field Theory}
\label{subsec:global:elements}

To extract
Higgs boson couplings from measurements, we will use the method of
Effective Field Theory (EFT) sketched in Sec.~\ref{subsec:phys_eft}.
This method has been explained in full technical  detail in
\cite{Barklow:2017suo,Barklow:2017awn}.
Here we will present an overview of the EFT analysis, supplying those 
technical details that are relevant to the evaluation of our fitting procedure.

In the EFT method, we represent the effects of new physics on the
Higgs boson and other SM observables by the most general linear
combination of dimension-6 operators invariant under $SU(2)\times
U(1)$.  In the most general settting, this formalism contains a very large
number of parameters.  However, in the special case of $\ee$
collisions, there are some simplifications.   First, for the purpose
of computing deviations from the SM due to dimension-6 operators, it
suffices to work at the electroweak tree level.   (The basic SM
predictions must of course be computed as accurately as possible,
typically to 2-loop order in electroweak couplings.)  Second, it
suffices to consider only CP-even observables, since the 
 contributions of $CP$-odd operators can be bounded
 by independent measurements.   With these simplifications, a total of
 16 operator coefficients appear in the analysis.  One additional
 parameter $c_6$ appears in double Higgs production, and 10 additional
 parameters appear in analyses of top quark production, but these do
 not enter the extraction of the Higgs couplings we will discuss here.

To determine
 these operator coefficients, we can use precision electroweak measurements and
 data on $\ee\to W^+W^-$ in addition to data 
from Higgs processs.
    The analysis also makes use of  specific constraints from the LHC
that should be available when the ILC runs and have a clear
model-independent interpretation.  These are the ratios of 
branching ratios of the Higgs boson to the final states 
$\gamma\gamma$, $ZZ^*$, $Z\gamma$, $\mu^+\mu^-$.  The measurements of these 
four channels are all based on Higgs bosons produced centrally through the dominant gluon fusion process. 
The ratios of rates should be extracted  from LHC data in
a way that most systematic errors are common and cancel in the ratios. 
 It is shown in 
\cite{Barklow:2017suo,Barklow:2017awn} that this data suffices to determine these
coefficients independently and without important 
degeneracies.\footnote{An 17th operator contributes to $G_F$
  but is controlled by constraints from the measurement of  $\ee\to
  \mu^+\mu^-$ at high energy.  The bound from LEP 2 is already very
  strong.}

\begin{figure}
\begin{center}
\includegraphics[width=0.80\hsize]{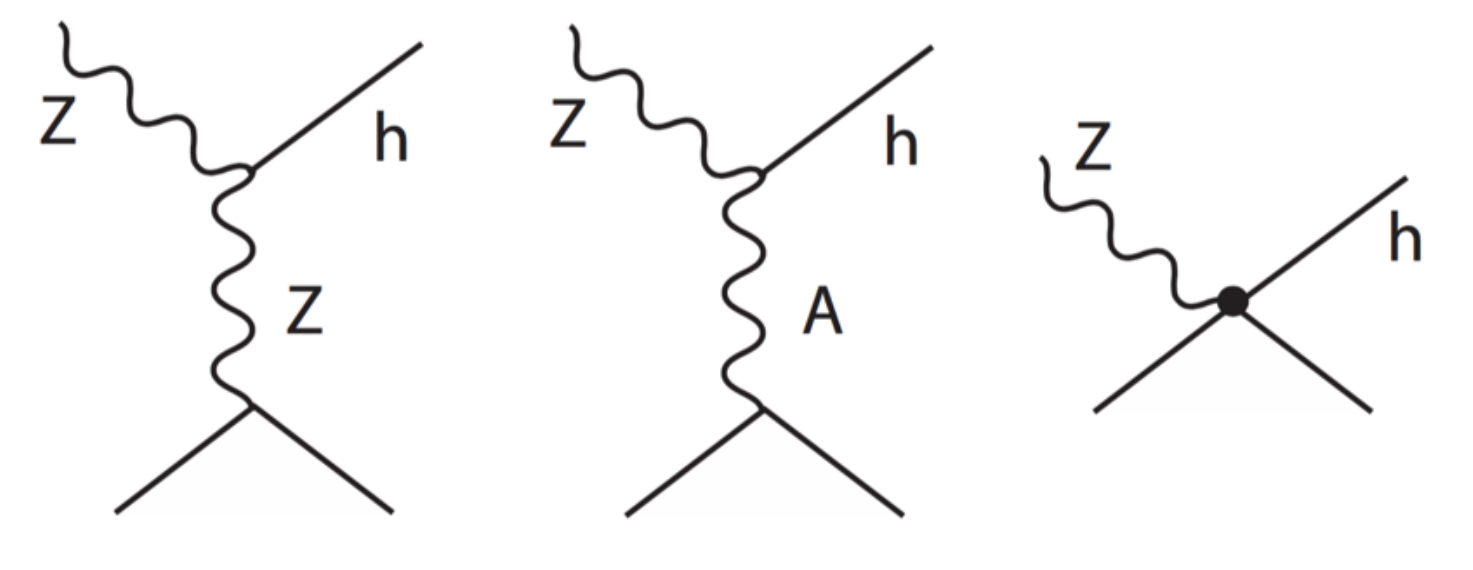}
\end{center}
\caption{Feynman diagrams contributing to the process $\ee\to Zh$ when 
contributing  dimension-6 operators are included. }
\label{fig:eeZhdiagrams}
\end{figure}

\begin{figure}
\begin{center}
\includegraphics[width=1.0\hsize]{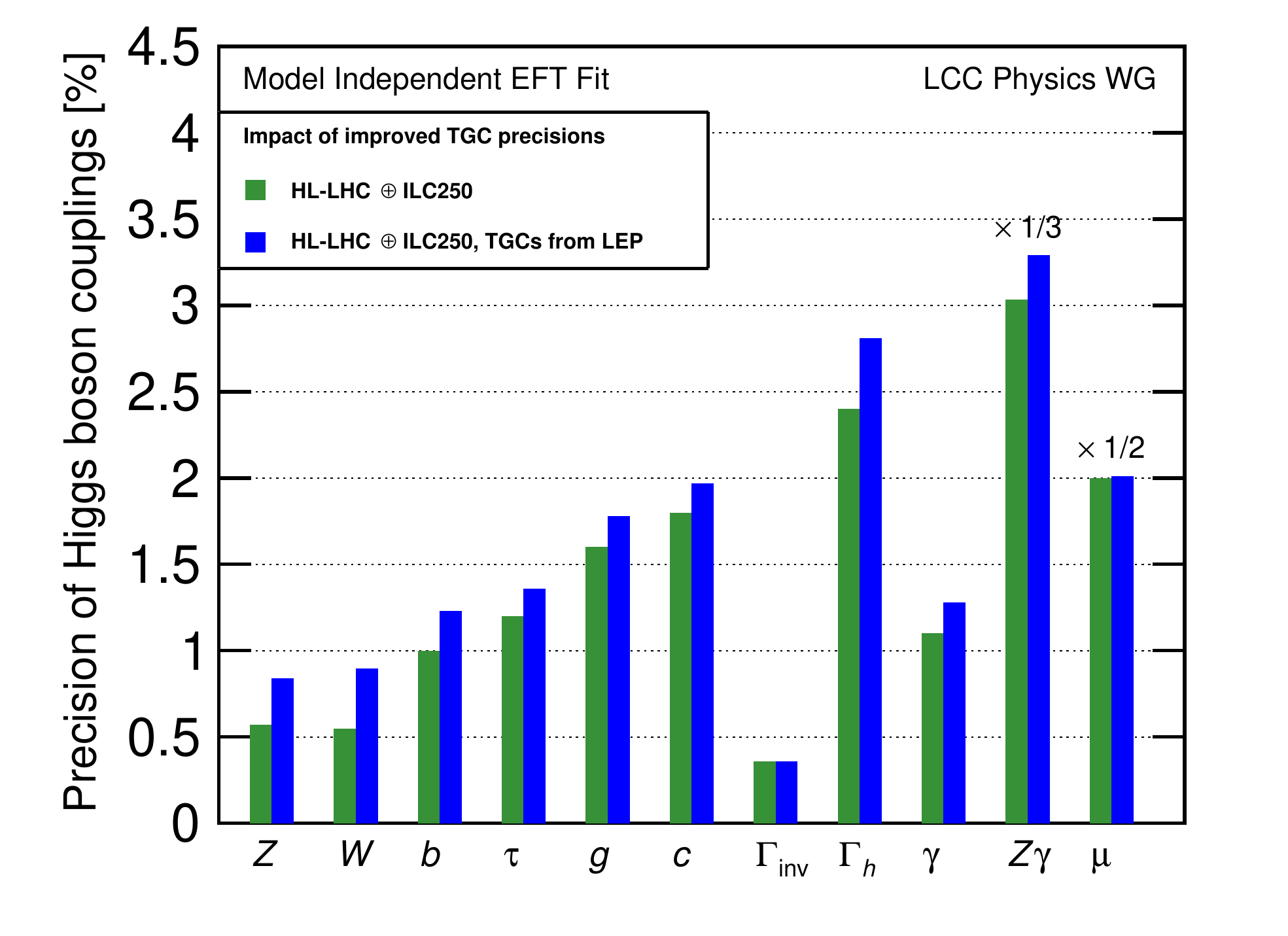}
\caption{Projected Higgs boson coupling uncertainties when including the charged 
triple gauge coupling precisions as expected from ILC250, compared to 
the case of using the corresponding LEP results instead.}
\label{fig:WWeffectonH}
\end{center}
\end{figure}

\begin{figure*}
\begin{center}
\includegraphics[width=0.7\hsize]{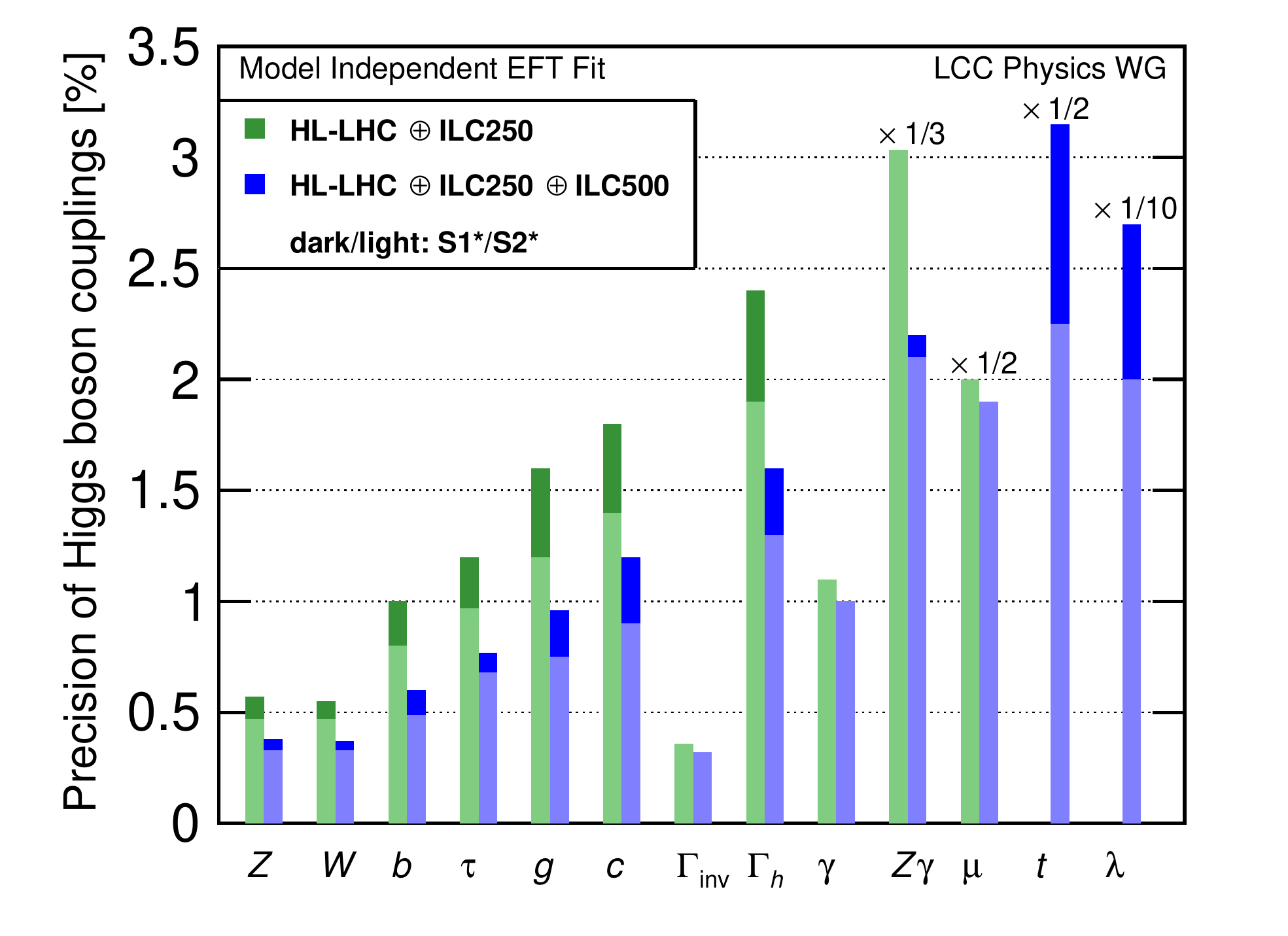}
\caption{Projected Higgs boson coupling uncertainties for the ILC
  program at 250~GeV and an energy upgrade to 500~GeV, using the
  highly model-independent analysis presented in \cite{Fujii:2017vwa}. This
  analysis makes use of  data on $\ee\to W^+W^-$ in addition to Higgs
  boson observables and also incorporates projected LHC results, as described
  in the text.  The darker bands correspond to the values given in Tab.~\ref{tab:ILCHiggs}.  The lighter bands
  correspond to the  scenario S2* in Table~\ref{tab:ILCLHC}, which is defined in the discussion of Sec.~\ref{subsec:higgs:ilclhc}.   The column $\lambda$ refers to the $HHH$ coupling.  In the last four columns, 
  all bars are rescaled by the indicated factor.}
\label{fig:ILCmodelindep}
\end{center}
\end{figure*}

\begin{table}[!htbp]
\begin{center}
\begin{tabular}{lcc}
   coupling     &   2~\iab\ at 250      &   + 4~\iab\ at 500   \\ \hline 
$HZZ$            &            0.57&            0.38                  \\ 
$HWW$            &         0.55               &   0.37    \\ 
 $Hbb$            &              1.0  &                0.60   \\ 
$H\tau\tau$    &          1.2  &                   0.77   \\ 
$Hgg$ &  1.6  &       0.96          \\ 
$Hcc$         &   1.8  &  1.2  \\ 
$H\gamma\gamma$ &  1.1 &   1.0  \\ 
$H\gamma Z$     & 9.1&   6.6   \\
$H\mu\mu$ &  4.0  &  3.8  \\ 
$Htt$  &   -     &      6.3   \\ 
\hline 
$HHH$  &  -    &   27   \\ \hline 
$\Gamma_{tot}$ & 2.4  & 1.6  \\  
$\Gamma_{inv}$ &   0.36  & 0.32 \\ 
$\Gamma_{other}$ &   1.6  & 1.2 \\  \hline
\end{tabular}
\end{center}
\caption{ \label{tab:ILCHiggs}    Projected uncertainties in the Higgs
  boson couplings for the ILC at 250~GeV and 500~GeV, with
  precision LHC input, assuming the
  integrated luminosities in the H20 program.   All values
  are given in percent (\%). The definition of a Higgs coupling
  uncertainty, for the purpose of this
  table,  is half the fractional uncertainty in the corresponding
  Higgs boson partial width.  (See the text
  for further explanation.)  The ILC at 250~GeV only does 
not have direct sensitivity to the $Htt$ and $HHH$ couplings; 
thus no  model-independent  values are given in these lines. The
  bottom lines give, for reference, the projected uncertainties in the
  Higgs boson total width, the 95\% confidence limits on the Higgs
  boson invisible width, and the 95\% confidence limits on possible
  exotic Higgs decay processes that are not explicitly recognized, as
  described in the text.
   The analysis, which based on SM Effective Field Theory,  is highly model-independent}
\end{table}

  The way that we make use of LHC inputs illustrates the complementarity of LHC and ILC results 
  on the Higgs boson.   The LHC has special strength in gathering statistics on rare modes  of 
  Higgs decay, especially those with leptons in the final state.  On the other hand, results from 
  the ILC are needed to determined the Higgs total width and the absolute normalization of branching
  ratios and partial widths with minimal model assumptions.   In a similar way, the HL-LHC will 
  probe deeply for specific exotic decays of the Higgs boson, especially those involving 
  muons, while ILC is needed to survey the most general possibilities for exotic decays.

Here is an outline of the analysis. We need to fit 16 operator
coefficients plus 4 SM parameters which are shifted by dimension-6
effects.  The 16
EFT coefficients arise in the following way:  2 from Higgs boson
operators ($c_H$, $c_T$), 4 from operators involving the
squares or cubes of SM gauge field strengths ($c_{WW}$, $c_{WB}$,
$c_{BB}$, and $c_{3W}$), 3 from Higgs current couplings to leptons
($c_{HL}$, $c_{HL}^\prime$, $c_{HE}$), 5 from the operators that shift
the Higgs coupling to $b$, $c$, $g$, $\tau$, $\mu$, and two more from
Higgs current couplings to quarks ($c_{W}$ and $c_Z$).  

The parameters are
constrained rather specifically, in a way that we can outline.  Measurements of
$\alpha$ and 
$G_F$ and the $W$, $Z$ and $H$ masses constrain the SM parameters plus
one additional parameter ($c_T$).  Purely leptonic precision electroweak 
measurements ($\Gamma(Z\to \ell^+\ell^-)$ and $A_{\ell}$) constrain
two of the three $c_{H\ell}$ parameters, and measurements of the $W$
and $Z$ total widths fix $c_W$ and $c_Z$.   Measurements on $\ee\to
W^+W^-$ constrain  the third $c_{H\ell}$  parameter, plus $c_{WB}$ and
$c_{3W}$.    The LHC measurement of the ratio of branching ratios 
$\Gamma(h\to \gamma\gamma)/\Gamma(h\to ZZ^*)$ will put a strong
constraint on $c_{BB}$.   The Higgs branching ratios to fermion and
gluon states constrain those 5 parameters.  At this point, only the
parameters $c_H$ and $c_{WW}$ remain.  These are constrained, respectively, by the
normalized cross section for $\ee\to ZH$ and the polarization
asymmetry or angular distribution in this reaction.  

To account for the possibility of non-standard Higgs boson decays, we
add two more parameters to our global fit.  The first is the Higgs branching
ratio to invisible decay products.   This is independently measurable
at an $\ee$ collider using the $Z$ tag in $\ee\to ZH$.  The second is
the branching ratio to exotic modes that somehow do not correspond to
any category that has been previously defined.   Though it might be
argued that any Higgs decay mode above the $10^{-3}$ level of
branching ratio should be directly observed, we add this parameter 
as insurance against modes not yet thought of.  It is determined by the constraint that the 
Higgs branching ratios, including this one, sum to 1.   We call this
possible contribution to the Higgs width $\Gamma_{other}$.

The ratio of the Higgs couplings to $W$ and $Z$ plays an
important role in the extraction of Higgs boson couplings and the
Higgs boson total width.   In the
$\kappa$ formalism, one parameter is assigned for each of these
couplings, and these parameters are determined independently  from Higgs production
cross sections.   This typically leads to very small errors on the $Z$
coupling and large  errors on the $W$ coupling.  In the EFT
formalism, as we have shown in Eq.~\ref{etazeta}, two
parameters are needed to describe each of these couplings, making the
$\kappa$ description oversimplified.   The
corresponding $W$ and $Z$ parameters are linked by
not-so-simple formulae involving other EFT parameters.  However, these
formula can be evaluated with the help of data from precision
electroweak and $WW$ reactions, leading to  constraints that are at
the same time tight and highly
model-independent~\cite{Barklow:2017awn}.  This is one illustration of
the synergies between different measurements that the EFT method
brings into play. 

It is remarkable that, though the EFT analysis introduces a large
number of free parameters, each one has a direct counterpart in a
physical observable that can be measured in the $\ee$ environment. 
In particular, beam polarisation is very powerful in providing needed
information.  For example, in the EFT framework, the process $\ee\to
ZH$ involves three diagrams, shown in Fig.~\ref{fig:eeZhdiagrams}.
Only the first diagram appears in the SM.   The third diagram is
required to be small by precision electroweak constraints.  The second
diagram, with $s$-channel $\gamma$ exchange,  is generated by the
operator 
corresponding to the coefficient $c_{WW}$.  Under a spin reversal $e^-_L
\leftrightarrow e^-_R$, the $Z$ diagram flips sign while the 
$\gamma$ diagram keeps the same sign.  Thus,  measurement of the 
polarisation asymmetry  in the total cross section for $\ee\to ZH$
directly measures the $c_{WW}$ parameter.   Beam polarisation plays
another important role.  With beam polarisation, the branching ratios
of the   Higgs boson are measured for  two different   polarisation
settings.   The statement that the same branching ratio must appear in
each pair of measurements helps to sharpen the global fit.  At the
same time, this comparison provides a check of assigned systematic
errors.   In Sec.~\ref{subsec:lincirc}, we will assess the importance of 
polarisation quantitatively and present results on the trade-off between polarisation and increased luminosity.

The precise measurement of the triple gauge bosons couplings expected
at the ILC also plays an
important role in global fit.   We have described the measurement of
these couplings through analysis of $\ee\to W^+W^-$ in
Sec.~\ref{subsec:ew_WWana}.
The ILC is expected to improve the precision of our knowledge of these
couplings by a factor of 10 over results from LEP and by a similar large
factor over results from LHC.   Figure~\ref{fig:WWeffectonH} shows the
significance of this set of inputs.  In the figure, the results of our
global fit, in green, are compared to the same fit using as inputs the
LEP constraints on the triple gauge boson couplings.

From this analysis, we derive the projected uncertainties on Higgs
couplings shown in Tab.~\ref{tab:ILCHiggs}. Here and in the rest of this section, the 
uncertainty presented in the tables for each $HA\bar A$ coupling is defined to be 
half of the fractional uncertainty in the corresponding partial width.  In cases such as 
$HZZ$ in which multiple EFT coefficients contribute to a given partial width (see Eq.~\ref{etazeta}), the
quoted uncertainty includes the uncertainties in these EFT parameters and their correlation.

\begin{table*}[!htbp]
\begin{center}
\begin{tabular}{l|cc|c|c|ccc}
    &  ILC250      &   ILC500   & CEPC &  FCC-ee &
   CLIC350 & 
   CLIC1.4 & CLIC3 \\ 
coupling &   EFT fit & EFT fit & $\kappa$ fit &  $\kappa$ fit &
              $\kappa$   fit &   $\kappa$   fit &
   $\kappa$ fit  \\ \hline 
$HZZ$            &            0.57 &   0.38    &   0.25   & 0.17     &  0.6     &     0.6    &   0.6     \\ 
$HWW$            &         0.55  &   0.37   &   1.4   &   0.43   &   1.0    &   0.6  &  0.6 \\ 
 $Hbb$            &              1.0  &  0.60   &  1.3    &    0.61  &   2.1    &  0.7  &  0.7 \\ 
$H\tau\tau$    &          1.2  &   0.77   &   1.5   &  0.74    &   3.1    &  1.4  &  1.0 \\ 
$Hgg$ &                      1.6  & 0.96       &  1.5    &  1.01    &   2.6    &   1.4   &  1.0  \\ 
$Hcc$                       &   1.8  &  1.2   &   2.2   &   1.21   &    4.4   &  1.9  &  1.4 \\ 
$H\gamma\gamma$ &  1.1 &   1.0     &  1.6    &   1.3   &    -   &  4.8  &  2.3 \\ 
$H\mu\mu$                &  4.0  &  3.8     &  5.0    &  3.8    &    -
& 12.1 &  5.7 \\ 
$Htt$  &                       -     &      6.3     &  -    &  -    &
-    & 3.0  &  3.0  \\ 
$HHH$                         &  -    &   27     &   -   &   -   &   -
& 35 &  9 \\ \hline 
$\Gamma_{tot}$             & 2.4  & 1.6    &   2.8    &  1.3     &
4.7    & 2.6  & 2.5 \\  
$\Gamma_{inv}$          &   0.36  & 0.32    &  $<$ 0.3    &   -   &   -
& -  & - \\  \hline
\end{tabular}
\end{center}
\caption{ \label{tab:askthem}    Projected uncertainties in the Higgs
  boson couplings quoted in the CDRs presented to the European
  Strategy Study.  The methodology of the fit is indicated.  The precise definition of a Higgs coupling uncertainty for the ILC EFT analysis
  is given at the end of Sec.~\ref{subsec:global:elements} and in the caption of Tab.~\ref{tab:ILCHiggs}.
 For the ILC, the  values are taken from
  Tab.~\ref{tab:ILCHiggs}.    For the CEPC, the values are taken from
  Ref.~\cite{CEPCStudyGroup:2018ghi}, Tab.~11.4.  For the FCC-ee, the values are taken
  from Ref.~\cite{Benedikt:2018qee}, Tab.~1.2. As discussed later,
  the capability for $\Gamma_{inv}$  is similar to that for CEPC. 
 For CLIC, the values are taken from 
Ref.~\cite{Charles:2018vfv}, Tab.~2, with $HHH$ values from
Ref.~\cite{Roloff:2019crr}.
  All values
  are given in percent (\%). The
  bottom lines give, for reference, the projected uncertainties in the
  Higgs boson total width and the 95\% confidence limits on the Higgs
  boson invisible width.  For  ILC, CEPC, and FCC-ee, the 
 values given for the $\gamma\gamma$ and $\mu\mu$ modes are those
 combined with expected
LHC results.}
\end{table*}

\begin{table*}[!htbp]
\begin{center}
\begin{tabular}{l|cc|cc|c}
 &  2/ab-250 & +4/ab-500 &  5/ab-250 &  + 1.5/ab-350 &  2/ab-350 \\
coupling &  pol.  &   pol.  &   unpol.  &  unpol &  $e^-$  pol. 
  \\  \hline 
$HZZ$            &            0.57&   0.38    &   0.69   & 0.40     &  0.51            \\ 
$HWW$            &         0.55  &   0.37   &   0.67  &   0.40   &   0.50    \\ 
 $Hbb$            &            1.0  &  0.60   &  0.88    &   0.65  &   1.0   \\ 
$H\tau\tau$    &          1.2  &   0.77   &   0.96  & 0.74   &   1.3     \\ 
$Hgg$ &                      1.6  & 0.96       & 1.2    &  0.98     &   1.6    \\ 
$Hcc$                       &   1.8  &  1.2   &   1.4  &   1.1    &    2.2  \\ 
$H\gamma\gamma$ &  1.1 &   1.0     &  1.2    &   1.0   &    1.1    \\ 
$H\gamma Z$           &  9.1  &  6.6   & 9.6   &  9.1  &   8.9  \\
$H\mu\mu$                &  4.0  &  3.8     &  3.8    &  3.7    &    4.0
 \\ 
$Htt$  &                       -     &      6.3     &  -    &  -    &
-     \\ 
$HHH$                         &  -    &   27     &   -   &   -   &   -
 \\ \hline 
$\Gamma_{tot}$             & 2.4  & 1.6    &  1.9    &  1.5     & 2.4    \\  
$\Gamma_{inv}$          &   0.36  & 0.32    &  0.34    &  0.30   &   0.58\\  
$\Gamma_{other}$       &   1.6  & 1.2    &  1.1    &  0.95   &   1.6\\
\hline
\end{tabular}
\end{center}
\caption{ \label{tab:oursimple}    Projected uncertainties in the Higgs
  boson couplings computed using the EFT method, with  ILC uncertainties per unit of luminosity and assuming a run with the quoted integrated luminosity
  and energy. The notation is that of Tab.~\ref{tab:ILCHiggs}.  The first two columns are identical to the ILC values from 
  Tab.~\ref{tab:ILCHiggs}.   In the last column, we assume $\pm$
  80\% electron polarisation and zero positron polarisation.  }
\end{table*}
\begin{figure*}
\begin{center}
\includegraphics[width=0.7\hsize]{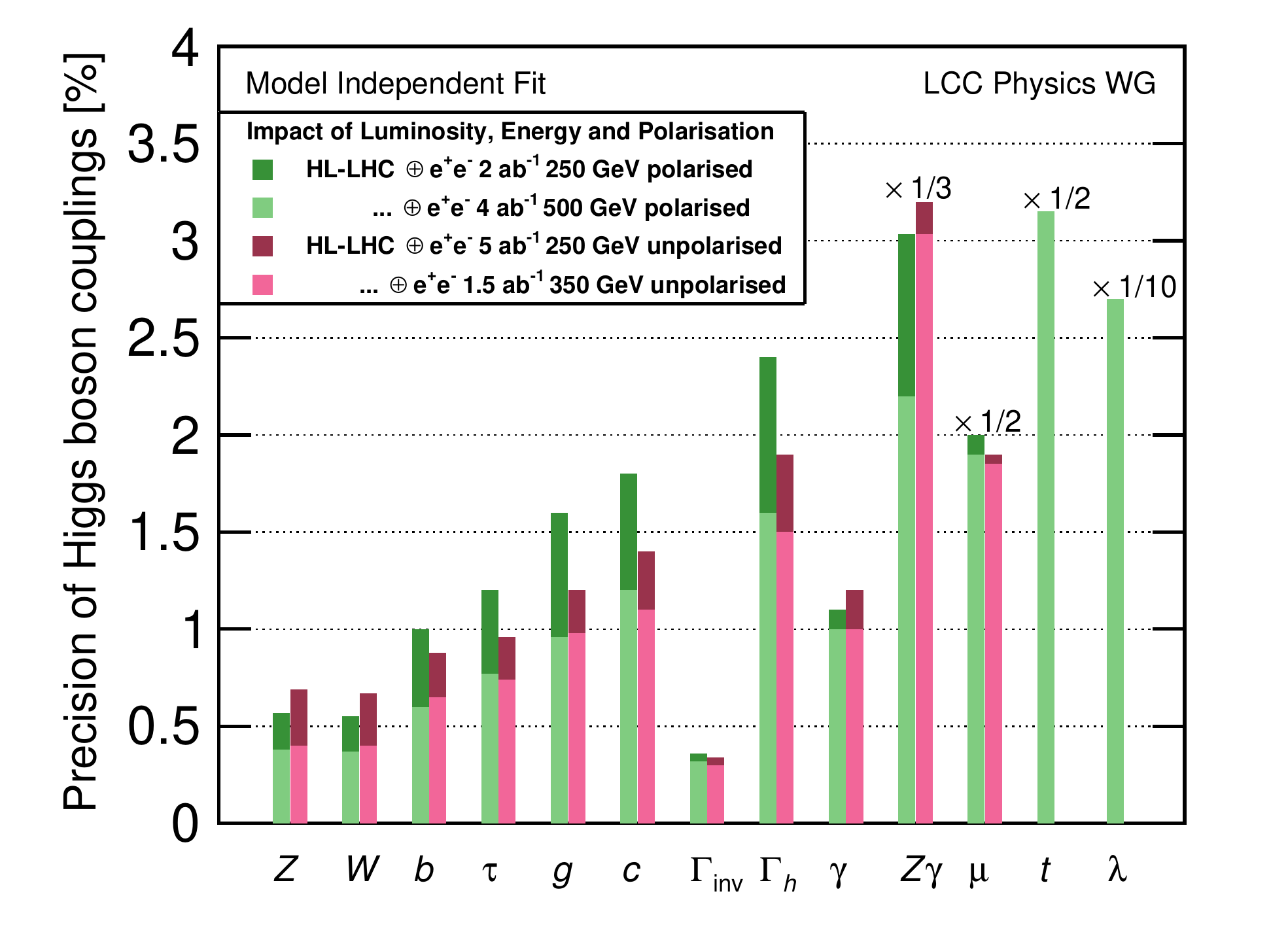}
\caption{Projected Higgs boson coupling uncertainties for selected scenarios from Table~\ref{tab:oursimple}.
In particular it shows that at $\sqrt{s}=250$\,GeV, 2\,\iab  with polarised beams yield comparable results to a much larger data set of 5\,\iab with unpolarised beams.}
\label{fig:oursimple}
\end{center}
\end{figure*}

\begin{table*}[!htbp]
\begin{center}
\begin{tabular}{l|cc|cc}
 &  2/ab-250 & +4/ab-500 &  5/ab-250 &  + 1.5/ab-350\\
coupling &  pol.  &   pol.  &   unpol.  &  unpol  
  \\  \hline 
$HZZ$            &             0.50&   0.35  &   0.41   & 0.34           \\ 
$HWW$            &         0.50  &   0.35  &   0.42   &   0.35     \\ 
 $Hbb$            &           0.99 &  0.59   &  0.72   &    0.62       \\ 
$H\tau\tau$    &          1.1  &   0.75   &   0.81   &  0.71      \\ 
$Hgg$ &                      1.6  & 0.96       &  1.1   &  0.96     \\ 
$Hcc$                       &   1.8  &  1.2   &   1.2   &   1.1    \\ 
$H\gamma\gamma$ &  1.1 &   1.0     &  1.0    &   1.0   \\ 
$H\gamma Z$           &  9.1  &   6.6    &  9.5     &   8.1     \\
$H\mu\mu$                &  4.0  &  3.8     &  3.8    &  3.7   
 \\ 
$Htt$  &                       -     &      6.3     &  -    &  -     \\ 
$HHH$                         &  -    &   27     &   -   &   -  \\ \hline 
$\Gamma_{tot}$             & 2.3  & 1.6    &   1.6    &  1.4       \\  
$\Gamma_{inv}$          &   0.36  & 0.32    &  0.34    &   0.30  \\  
$\Gamma_{other}$       &   1.6  & 1.2    &  1.1    &  0.94\\
\hline
\end{tabular}
\end{center}
\caption{ \label{tab:ournotsosimple}    Projected uncertainties in the Higgs
  boson couplings computed using the same methodology as
 in Tab.~\ref{tab:oursimple} but including projected improvements in
 precision electroweak measurements.  In the first two columns, the polarised collider projections from Tab.~\ref{tab:oursimple} are modified to include 
 an improvement by a factor 10 in $A_\ell$, as discussed in Sec.~\ref{subsec:higgs_improve}. In the second two 
 columns, the unpolarised collider projections from  from Tab.~\ref{tab:oursimple} are modified to include 
 the improvement of the uncertainties on precision electroweak observables described
 in the 
 FCC-ee CDR~\cite{Benedikt:2018qee}.}
\end{table*}

\begin{figure*}
\begin{center}
\includegraphics[width=0.7\hsize]{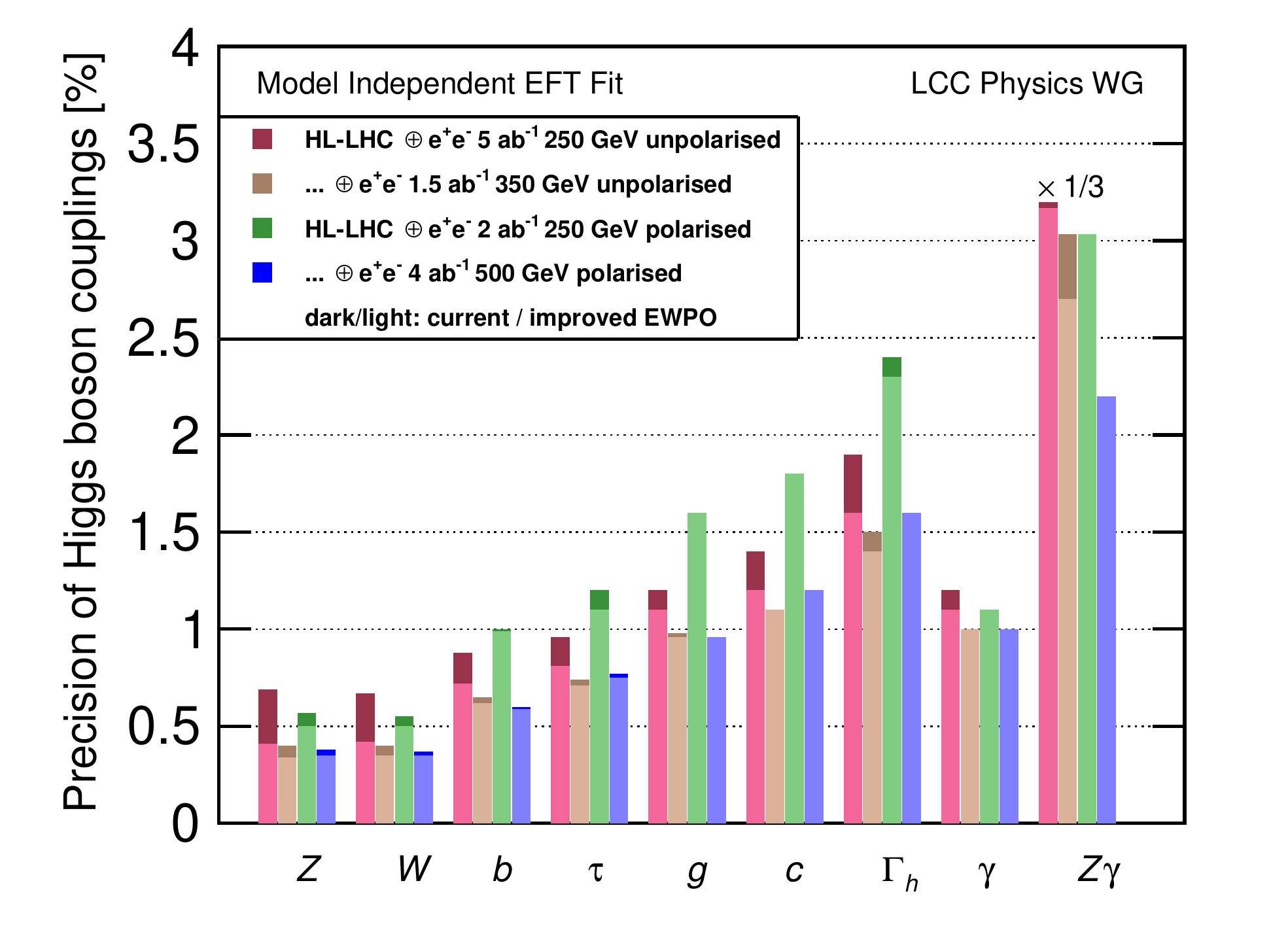}
\caption{Impact of improved electroweak precision observables on the projected precisions for various Higgs couplings for the combinations of luminosity, energy and polarisation from Tab.~\ref{tab:ournotsosimple}.
For the unpolarised cases, EWPO projections from the FCC-ee CDR~\cite{Benedikt:2018qee} have been assumed, while for the polarised case only an improved precision for $A_\ell$ is assumed. Couplings for which there is no improvement due to improved EWPO have been omitted from the figure.   The notation of the figure is the same as that in Fig.~\ref{fig:ILCmodelindep}.}
\label{fig:ournotsosimple}
\end{center}
\end{figure*}

Table~\ref{tab:ILCHiggs} is the main result of this report
in relation to the ILC capabilities for Higgs boson coupling measurements.
This table gives  the
current state of our understanding of the ILC capabilities. We emphasise that the
analysis leading to these projections is completely model-independent,
in the sense that all models of new physics describable either by the
addition  of local operators to the SM EFT (for heavy new particles)
or by the addition of invisible and exotic Higgs decays (for light new
particles) are included.   Given the
run plan and  detector designs described above, we have a high degree
of confidence that these estimated uncertainties will be achieved --
and, probably, surpassed -- in the realisation of the ILC program.
The projections in the table are summarised in Fig.~\ref{fig:ILCmodelindep}.

\subsection{Comparison of $\ee$ Higgs factory proposals}
\label{subsec:lincirc}

In this section, we will present a comparison of the capabilities of
the ILC for precision Higgs measurement with those of other proposed
linear and circular colliders, including CLIC, CEPC, and FCC-ee.  We
will present three  sets of quantitative comparisons.

To begin, each collider proposal has presented its own set of
projections in its documentation for the European strategy study.   We
have copied the relevant numbers for projected Higgs boson coupling
uncertainties into Tab.~\ref{tab:askthem}. 
For colliders other than the ILC, those estimates have been 
 made using the more model-dependent
 $\kappa$ fit.   The small values of the $ZZ$ coupling uncertainty
 relative to the
 $WW$ coupling uncertainty reflects the model-dependence of the $\kappa$ formalism
 as discussed in Sec.~\ref{subsec:global:elements}.

It is interesting to ask how the proposals would compare if a common
fitting technique is used. In almost all cases, the measurement errors
are dominated by statistics and the efficiencies used in the analyses
are similar.  A direct way to make the comparison is  to use
the results of our ILC analyses to estimate efficiencies and
statistical errors for all of the colliders.  That is, we  assume
the luminosity samples in the collider proposals, assume the same
measurement errors per unit of luminosity used to generate
 Tab.~\ref{tab:ILCHiggs},  take account of differences in
the cross sections resulting from the use (or not) of polarized beams,
and rerun our fitting  program for those conditions.   This is the
method used to generate 
Tab.~3 of Ref.~\cite{Barklow:2017suo}.  As a proxy for  CEPC, we assume
 a sample of  5~\iab\ at 250~GeV without polarisation. As a proxy for  
FCC-ee, we use a sample of 5~\iab at 250~GeV plus 1.5~\iab\ at 350~GeV,
without polarisation.  The run plan for CLIC includes only 1~\iab\ at
380~GeV before the energy upgrade to 1~TeV.  Since we are
uncomfortable using the EFT formalism
 with dimension-6 operators only at 1~TeV and above, we represent CLIC by a sample of
2~ab$^{-1}$, similar to ILC, with 80\% $e^-$ polarisation only, at 350~GeV.  
The results are presented in Tab.~\ref{tab:oursimple} and visualised in Fig.~\ref{fig:oursimple}.

Though not all differences among the various proposals are included in
this table, the table does usefully show how increased luminosity trades off
against beam polarisation.   We see that beam polarisation is a very
powerful tool, essentially compensating the advantage of larger event
samples claimed by the circular machines.  Note that the advantage is not uniform;
increased luminosity is a generally greater benefit for smaller couplings such as $Hcc$, while
polarisation has special benefit for specific couplings such as $H\gamma Z$. The comparison of 2~\iab\
data samples at 250 and 350~GeV is also interesting, since the two
energy settings bring different advantages to the Higgs physics study.

To make the comparison in Tab.~\ref{tab:oursimple} more realistic, we should indicate how the 
improved precision electroweak
measurements that can be achieved at circular colliders affect these numbers. The answer to this question is
given in Tab.~\ref{tab:ournotsosimple}, which is compared 
 with Tab.~\ref{tab:oursimple} in Fig.~\ref{fig:ournotsosimple}.   For the ILC columns, we have 
assumed that the input measurement of $A_\ell$, the polarisation
asymmetry of the lepton-$Z$ coupling, will be improved by a factor 10
by measurement of the polarisation asymmetry in $\ee\to Z\gamma$ at
250~GeV.   This is one of the improvements to our Higgs analysis
currently under study listed in Sec.~\ref{subsec:higgs_improve}.   The
improvement in $A_\ell$ by a dedicated ``Giga-Z''  run at the $Z$ pole
would be comparable~\cite{Baer:2013cma}.  For the third and fourth columns, we assume the
improvements in measurements of precision electroweak observables
described in the FCC-ee CDR~\cite{Benedikt:2018qee}.

The ILC results with polarised beams are somewhat improved in the most precisely determined couplings
by the improvement in the input $A_\ell$. The changes for circular colliders, which make use of the large
improvements from $Z$ pole running, are more significant for the results at  250\,GeV.
 More surprisingly, though,
the improvement in precision electroweak inputs turns out to make only a small difference
 when the  dataset at 350\,GeV is added.  One
reason for this, pointed out in Ref.~\cite{Barklow:2017awn}, is that
the EFT coefficients responsible for precision electroweak corrections
also contribute terms of order $s/m_Z^2$ to the $\ee\to ZH$ cross
section.  Then these coefficients are powerfully constrained by comparing measurements 
of this cross section at different centre-of-mass energies.

\subsection{Comparison of the ILC and the HL-LHC Higgs capabilities}
\label{subsec:higgs:ilclhc}

\begin{table}[!htbp]
\begin{center}
\begin{tabular}{lccccc}
   coupling     &  current  &    S1*     &     S1     &    S2*   &   S2   \\ \hline 
$HZZ$ - LHC  &     11.      &        &       2.4  &        &  1.5 \\ 
\phantom{$HZZ$} - ILC 250 &      &   0.57  &  0.46   &   0.47   &  0.37 \\ 
\phantom{$HZZ$} - ILC 500&      &   0.38  &  0.20  &  0.33   & 0.18 \\ 
 \hline 
$HWW$ - LHC  &    15.       &        &    2.6    &        &  1.7 \\ 
\phantom{$HWW$} - ILC 250 &      &   0.55 &  0.44   &   0.47  &  0.36 \\ 
 \phantom{$HWW$} - ILC 500 &      &   0.37 &  0.19   &  0.33 & 0.18 \\ 
   \hline 
$Hbb$ - LHC  &    29.       &        &         6.0   &        & 3.7 \\ 
\phantom{$Hbb$} - ILC 250 &      &  1.0  & 0.83   &   0.80   &  0.69 \\ 
\phantom{$Hbb$} - ILC 500 &      &  0.60  & 0.43   &  0.49  &  0.37 \\ 
 \hline 
$H\tau\tau$ - LHC  &    17.       &        &             2.8  &        & 2.0 \\ 
\phantom{$H\tau\tau$} - ILC 250 &      &  1.2  &  0.98   &   0.97  &  0.86 \\ 
\phantom{$H\tau\tau$} - ILC 500 &      &  0.77  &  0.63   &  0.68   & 0.59 \\ 
    \hline 
$Hgg$ - LHC  &     15.      &        &            4.0   &        &
               2.5  \\ 
\phantom{$hgg$} - ILC 250 &      &  1.6  &  1.6   &   1.2   &  1.2 \\ 
 \phantom{$hgg$} - ILC 500 &      &  0.96  &  0.91   &  0.75  & 0.70 \\ 
\hline 
$Hcc$ - LHC  &    -       &        &           -  &        &  - \\ 
\phantom{$Hcc$} - ILC 250 &      &   1.8  &  1.8   &   1.4  &  1.3 \\ 
\phantom{$Hcc$} - ILC 500 &      &  1.2  &  1.1   &   0.90  &  0.85 \\ 
    \hline 
$H\gamma\gamma$ - LHC  &    15.       &        &        2.9  &        & 1.8 \\ 
\phantom{$H\gamma\gamma$} - ILC 250  &      &   1.1  &  1.1  &   1.1   &  1.0\\ 
 \phantom{$H\gamma\gamma$} - ILC 500  &      &   1.0  &  0.97 &   1.0  &  0.96\\ 
     \hline 
$H\gamma Z$ - LHC  &    15.       &        &          &        & 9.8 \\ 
\phantom{$H\gamma Z$} - ILC 250  &      &   9.1  &    &   9.1   &  \\ 
 \phantom{$H\gamma Z$} - ILC 500  &      &   6.6  &   &   6.3  &  \\ 
\hline 
$H\mu\mu$ - LHC  &   70.        &        &    6.7  &        & 4.3 \\ 
\phantom{$H\mu\mu$} - ILC 250 &      &  4.0  &  4.0  &  4.0  &  4.0 \\ 
\phantom{$H\mu\mu$} - ILC 500 &      &  3.8  &  3.7  &  3.8  & 3.7\\ 
     \hline 
$Htt$ - LHC  &   14.        &        &           5.5   &        &  3.4
  \\  
 \phantom{$Htt$} - ILC 500  &        &    6.3    &     4.1     &    4.5   & 2.8
\\ 
\hline 
$HHH$ - LHC  &         &        &  80  &        &    50
  \\  
 \phantom{$HHH$} - ILC 500  &        &    27  &    27      &   20  & 20
\\ 
\hline \hline
$\Gamma_{tot}$ - LHC            &       28       &      &   5   &       &  4  \\ 
\phantom{$\Gamma_{tot}$} - ILC 250 &  &   2.4   &   1.4    &    1.9   & 1.1   \\ 
\phantom{$\Gamma_{tot}$} - ILC 500 & & 1.6 & 0.70 & 1.3  &  0.60  \\  \hline
$\Gamma_{inv}$ - LHC                & 26  &     &   &    &  3.8  \\ 
\phantom{$\Gamma_{inv}$} - ILC 250 &  &   0.36   &    -     &   0.36   &   -  \\ 
\phantom{$\Gamma_{inv}$} - ILC 500 & & 0.32 &  - & 0.32  & - \\  \hline
$\Gamma_{other}$ - LHC                &  &     &   &    &    \\ 
\phantom{$\Gamma_{other}$} - ILC 250 &  &   1.6   &    -     &   1.4   &   -  \\ 
\phantom{$\Gamma_{other}$} - ILC 500 & & 1.2 &  - & 1.1  & - \\  \hline
\end{tabular}
\end{center}
\caption{ \label{tab:ILCLHC}   Projected uncertainties in the Higgs
  boson couplings for HL-LHC and  for ILC  with the
  specific LHC inputs described in the text, 
in various scenarios.   All values
  are given in percent (\%).  The precise definition of a Higgs coupling uncertainty for the ILC EFT analysis
  is given at the end of Sec.~\ref{subsec:global:elements}.  The values labelled ``current'' are taken mainly
  from Table 8 of the CMS publication Ref.~\cite{Sirunyan:2018koj}; the values for $\Gamma_{tot}$ and $\Gamma_{inv}$ are found in the text of 
Sec.~8.2 and Sec~3.9, respectively.    The
  LHC S1 values are those from the $\kappa$ fit to CMS projections,
  given in Tab.~36 of Ref.~\cite{Cepeda:2019klc}; the ATLAS projections are
  similar.    The S2 values are those from the ATLAS/CMS combination
  given in  Fig.~30 of Ref.~\cite{Cepeda:2019klc}.  Values for the HHH
  coupling are found in the text of Sec.~3.2 of 
Ref.~\cite{Cepeda:2019klc}. Values for $\Gamma_{tot}$ and $\Gamma_{inv}$ are found in the text of 
Sec.~2.7.1 and Sec~6.1, respectively, in Ref.~\cite{Cepeda:2019klc}; these are CMS results only.
For ILC, the S1*
  results are those presented in Sec.~\ref{subsec:global:elements} for
  ILC programs at 250~GeV and 500~GeV.   The scenario S1 includes the 
same values for ILC measurement uncertainties but also includes additional
model-dependent assumptions that are used in the LHC S1 analysis.
These are described in the text.  The
scenarios
S2* and S2 assume the improved performance in ILC measurements
presented in Sec.~\ref{subsec:higgs_improve}.}
\end{table}

Finally, we compare the capabilities of the ILC for precision Higgs
measurement to those of the HL-LHC.  

In Sec.~\ref{sec:physics}, we have presented 
qualitative comparisons of the approach to Higgs physics 
that is possible at the ILC to the approach that must be taken at the LHC.   
Here we will compare the quantitative projections given in 
Sec.~\ref{subsec:global:elements} to the projections presented in the HL-LHC
Yellow Report~\cite{Cepeda:2019klc}.  The comparison is not so straightforward because 
of the different frameworks used in the analyses. In 
Tab.~\ref{tab:ILCLHC}, we quote projected uncertainties in Higgs boson couplings
for HL-LHC given in Ref.~\cite{Cepeda:2019klc} and
present results from the ILC global fit in a number of scenarios.

The projections in Ref.~\cite{Cepeda:2019klc} are based on operational experience
with detectors that have successfully  made measurements on the Higgs
boson, have exceeded their expectations from the proposal stage, and,
based on that experience, expect further improvements beyond the level
of their current methodologies.  These estimates are based on
extrapolation of current results.   They do depend on the assumption that the
improvement of the ATLAS and CMS detectors by the Phase-II upgrades will 
fully compensate for the effects of the high-pileup environment expected at the 
HL-LHC.  With this understanding, these estimates give the 
expectations for the performance of the ATLAS and CMS experiments in
the HL-LHC program.

In addition to the formal HL-LHC projections, which are ATLAS/CMS combinations, 
the individual LHC experiments have actually
produced two sets of projections, a final one (S2) described above
and a maximally conservative one (S1) that includes the increase in
statistics from HL-LHC but uses only current methodologies and current
estimates of systematic errors.  It is interesting to us to compare the S1 and S2 
projections, since this comparison gives an idea of the improvements expected by the LHC 
experiments beyond the current state of the art to the end of the HL-LHC program.  
In Tab.~\ref{tab:ILCLHC} we have quoted  both the S1
numbers from CMS (those from ATLAS are similar) and the final (S2) projections from
Ref.~\cite{Cepeda:2019klc}.  At the S2 level, the systematic errors are estimated to be small enough that the 
projections benefit by about 20\% from making an ATLAS/CMS combination.

We call to the reader's attention the fact that the improvements
projected for HL-LHC from the current ATLAS and CMS uncertainties on 
the $HZZ$ and $HWW$ couplings are very
significant already in the S1 analysis.  This is because the high statistics
of the HL-LHC allows one to make use of the vector boson fusion 
production mode, which has a low cross section but  relatively small
theoretical and modelling uncertainties.
   On the other hand, the projected improvement in the $Hbb$
couplings is based mainly on a higher-precision understanding of
analyses such as that shown in the upper plot of
Fig.~\ref{fig:LHCILCHbb}.

\begin{figure*}
\begin{center}
\includegraphics[width=0.7\hsize]{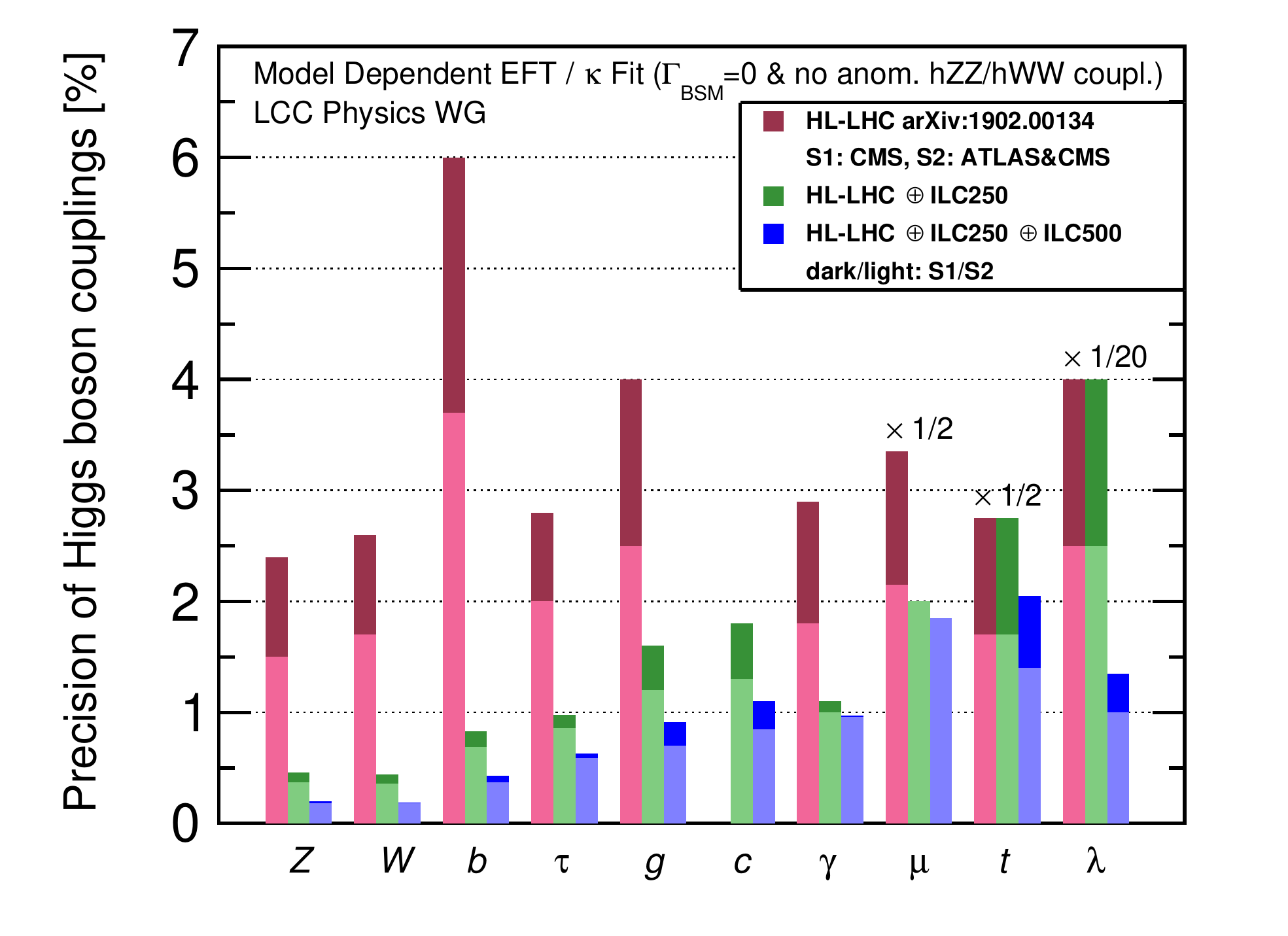}
\caption{Projected Higgs boson coupling uncertainties for the LHC and
  ILC
using the model-dependent assumptions appropriate to the LHC Higgs
coupling fit.   The
dark and light red bars represent the projections in the scenarios S1
and S2 presented in Ref.~\cite{Cepeda:2019klc}.    The dark and light green bars represent the
projections in the ILC scenarios S1 and S2 described in the
text.  The dark and light blue bars show the projections for scenarios S1 and S2
when
data from the 500~GeV run of the ILC is included. The notation of the figure is the same as that in Fig.~\ref{fig:ILCmodelindep}.}
 \label{fig:ILCLHC}
\end{center}
\end{figure*}

The ILC estimates have a very different basis. 
 It is
always risky to estimate errors for experiments that have not yet been
constructed or taken data.   We have designed the ILC detectors to
have the 
superb performance characteristics detailed in
Secs.~\ref{sec:detectors} and \ref{sec:software}.   As far as is
possible today, these projected performances are justified by R\&D and
test beam measurements.   However, from this point, we wished to be
quite conservative.  We then take the expected precision of our measurements to be those 
of our current analyses  
of fully simulated, digitised events.  This conservative choice is the basis of the estimates quoted in Sec.~\ref{subsec:global:elements}.  Experience at all other colliders 
has shown that final precision with real data exceeds such {\it a priori} 
estimates. 

To compare these projections with those for HL-LHC, we have defined four scenarios, called
S1*, S1, S2*, S2.
The projections in Tab.~\ref{tab:ILCLHC} labeled S1* are those
from  Tab.~\ref{tab:ILCHiggs}  in Sec.~\ref{subsec:global:elements}.   
While, as we have stressed, the ILC analysis is highly
model-independent, the LHC analysis relies on certain model 
assumptions that are difficult to
remove with only the constraints available at a hadron collider.   The
LHC results in Tab.~\ref{tab:ILCLHC} assume  that the Higgs boson has no decay
modes beyond those predicted in the SM, and they  assume that the Higgs
boson couplings to $WW$ and $ZZ$ are modified only by a rescaling.  In
the ILC EFT analysis, each of these these couplings depends on two additional 
independent constants $\zeta_W$ and $\zeta_Z$ defined
in Eq.~\ref{etazeta}.   For a sharper comparison, then, we have then recast the ILC EFT analysis
adding these two assumptions, that is, assuming no
Beyond-Standard-Model decays and assuming $\zeta_{W} = \zeta_Z = 0$.
This gives the set of values labelled S1.  We do foresee some improvements in our 
analyses, as described in Sec.~\ref{subsec:higgs_improve}.   These reflect improvements to our
methods that are under study  and seem promising but are not yet completely
validated.  Making these improvements gives the uncertainties S2* and S2 (for model-independent and model-dependent
EFT fits) quoted in Tab.~\ref{tab:ILCLHC}.  These estimates are intended 
give an indication that the ILC capabilites are not fixed
but rather are improvable with further experimental effort.
 We remind the reader that all
estimates quoted for ILC require certain specific inputs from HL-LHC,
as explained in Sec.~\ref{subsec:global:elements}.  Our use of HL-LHC results nicely 
illustrates the complementarity of the two machines, as is discussed in that section.

It is subtle to directly compare the projections for HL-LHC and ILC taking into account their 
two different philosophies.  On the ILC side, since we 
have no experience with the actual operation of the detectors and the accelerator, we have been very 
cautious in making extrapolations beyond our current full-simulation 
results to the actual performance that we might 
eventually achieve.   We therefore regard our 
scenario S1, and even our  scenario S2, to be more conservative than the final (S2) HL-LHC projections. 
In any event, we hope that we have 
described the various estimates given in Tab.~\ref{tab:ILCLHC} clearly enough that the reader
can make his or her own judgement as to the most appropriate comparison of the ILC to the HL-LHC.

In all cases, however, it is only the ILC results that cross a
boundary into the region in which  we can robustly claim discovery of
deviations from the SM of the size generally expected in new physics models. 

In summary, Figs.~\ref{fig:ILCmodelindep} and \ref{fig:ILCLHC} illustrate the
capabilities of the ILC and the comparison of the ILC and LHC
projections.  Figure~\ref{fig:ILCmodelindep} shows the uncertainty
projections for the 250~GeV stage of the ILC, in the highly
model-independent framework S1*.  These results are
compared to results obtained in the same framework with the addition
of data from an energy upgrade to 500~GeV.   This justifies the
statement made earlier that deviations from the SM seen at the 250~GeV
stage of the ILC can be confirmed with an independent data set after
the upgrade to higher energy.   Figure~\ref{fig:ILCLHC} shows the
comparison of the ILC projections in the S1 and S2 scenarios to the 
projections given for the S1 and final (S2)  HL-LHC projections given
in Ref.~\cite{Cepeda:2019klc}.
Note that, while the improvement from the S1 to S2 scenarios for ILC  is a
matter of conjecture, the improvement from the 250~GeV to the 500~GeV
values is based on completed full-simulation studies.

\section{\label{sec:searches}Physics Simulations: Direct Searches for
  New Particles}

\def\MXN#1{\mbox{$ M_{\widetilde{\chi}^0_#1}                                $}}
\def\MXC#1{\mbox{$ M_{\widetilde{\chi}^{\pm}_#1}                            $}}
\def\XP#1{\mbox{$ \widetilde{\chi}^+_#1                                     $}}
\def\XM#1{\mbox{$ \widetilde{\chi}^-_#1                                     $}}
\def\XPM#1{\mbox{$ \widetilde{\chi}^{\pm}_#1                                $}}
\def\XMP#1{\mbox{$ \widetilde{\chi}^{\mp}_#1                                $}}
\def\XN#1{\mbox{$ \widetilde{\chi}^0_#1                                     $}}
\def\XNN#1#2{\mbox{$ \widetilde{\chi}^0_{#1,#2}                             $}}
\def\MXn{\mbox{$ M_{\widetilde{\chi}^0}                                     $}}
\def\MXc{\mbox{$ M_{\widetilde{\chi}^{\pm}}                                 $}}
\def\Xp{\mbox{$ \widetilde{\chi}^+                                          $}}
\def\Xm{\mbox{$ \widetilde{\chi}^-                                          $}}
\def\Xpm{\mbox{$ \widetilde{\chi}^{\pm}                                     $}}
\def\Xn{\mbox{$ \widetilde{\chi}^0                                          $}}
\def\Xnn{\mbox{$ \widetilde{\chi}^0                                         $}}
\def\p#1{\mbox{$ \mbox{\bf p}_1                                         $}}
\def\Xgen{\mbox{$ \widetilde{\chi}                                          $}}
\def\Mlsp{\mbox{$ M_{\mathrm {LSP}}                                     $}}
\def\lsp{\mbox{$ {\mathrm {LSP}}                                     $}}
\newcommand{\Ptmis}   {\mbox{$/\mkern-11mu P_t \,                          $}}
\newcommand{\Tpmis}   {\mbox{$\theta_{/\mkern-11mu p}                      $}}
\newcommand{\grav}    {\mbox{$ \widetilde{\mathrm G}                           $}}
\newcommand{\Gino}    {\mbox{$ \widetilde{\mathrm G}                           $}}
\newcommand{\tanb}    {\mbox{$ \tan \beta                                  $}}
\newcommand{\smu}     {\mbox{$ \widetilde{\mu}                                 $}}
\newcommand{\smur}    {\mbox{$ \widetilde{\mu}_{\mathrm R}                     $}}
\newcommand{\smul}    {\mbox{$ \widetilde{\mu}_{\mathrm L}                     $}}
\newcommand{\msmu}    {\mbox{$ M_{\widetilde{\mu}}                             $}}
\newcommand{\msmur}   {\mbox{$ M_{\widetilde{\mu}_{\mathrm R}}                 $}}
\newcommand{\msmul}   {\mbox{$ M_{\widetilde{\mu}_{\mathrm L}}                 $}}
\newcommand{\sel}     {\mbox{$ \widetilde{\mathrm e}                           $}}
\newcommand{\sell}    {\mbox{$ \widetilde{\mathrm e}_{\mathrm L}               $}}
\newcommand{\selr}    {\mbox{$ \widetilde{\mathrm e}_{\mathrm R}               $}}
\newcommand{\msel}    {\mbox{$ M_{\widetilde{\mathrm e}}                       $}}
\newcommand{\snu}     {\mbox{$ \widetilde\nu                                   $}}
\newcommand{\msnu}    {\mbox{$ m_{\widetilde\nu}                               $}}
\newcommand{\msell}   {\mbox{$ M_{\widetilde{\mathrm e}_{\mathrm L}}           $}}
\newcommand{\mselr}   {\mbox{$ M_{\widetilde{\mathrm e}_{\mathrm R}}           $}}
\newcommand{\sfe}     {\mbox{$ \widetilde{\mathrm f}                           $}}
\newcommand{\sfeb}    {\mbox{$ \overline{\widetilde{\mathrm f}}                $}}
\newcommand{\sfel}    {\mbox{$ \widetilde{\mathrm f}_{\mathrm L}               $}}
\newcommand{\sfer}    {\mbox{$ \widetilde{\mathrm f}_{\mathrm R}               $}}
\newcommand{\sfelb}   {\mbox{$ \overline{\widetilde{\mathrm f}_{\mathrm L}}    $}}
\newcommand{\sferb}   {\mbox{$ \overline{\widetilde{\mathrm f}_{\mathrm R}}    $}}
\newcommand{\msfe}    {\mbox{$ M_{\widetilde{\mathrm f}}                       $}}
\newcommand{\sle}     {\mbox{$ \widetilde{\ell}                                $}}
\newcommand{\sq}     {\mbox{$ \widetilde{q}                                $}}
\newcommand{\sqr}     {\mbox{$ \widetilde{q}_{\mathrm R}                                $}}
\newcommand{\sql}     {\mbox{$ \widetilde{q}_{\mathrm L}                                $}}
\newcommand{\msle}    {\mbox{$ M_{\widetilde{\ell}}                            $}}
\newcommand{\stau}    {\mbox{$ \widetilde{\tau}                                $}}
\newcommand{\stone}   {\mbox{$ \widetilde{\tau}_1                              $}}
\newcommand{\sttwo}   {\mbox{$ \widetilde{\tau}_2                              $}}
\newcommand{\staur}   {\mbox{$ \widetilde{\tau}_{\mathrm R}                    $}}
\newcommand{\mstau}   {\mbox{$ M_{\widetilde{\tau}}                            $}}
\newcommand{\mstone}  {\mbox{$ M_{\widetilde{\tau}_1}                          $}}
\newcommand{\msttwo}  {\mbox{$ M_{\widetilde{\tau}_2}                          $}}
\newcommand{\stq}     {\mbox{$ \widetilde {\mathrm t}                          $}}
\newcommand{\stqone}  {\mbox{$ \widetilde {\mathrm t}_1                        $}}
\newcommand{\stqtwo}  {\mbox{$ \widetilde {\mathrm t}_2                        $}}
\newcommand{\msq}    {\mbox{$ M_{\widetilde {\mathrm q}}                      $}}
\newcommand{\mstq}    {\mbox{$ M_{\widetilde {\mathrm t}}                      $}}
\newcommand{\sbq}     {\mbox{$ \widetilde {\mathrm b}                          $}}
\newcommand{\sbqone}  {\mbox{$ \widetilde {\mathrm b}_1                        $}}
\newcommand{\sbqtwo}  {\mbox{$ \widetilde {\mathrm b}_2                        $}}
\newcommand{\msbq}    {\mbox{$ M_{\widetilde {\mathrm b}}                      $}}
\newcommand{\msbqone}    {\mbox{$ M_{\widetilde {\mathrm b}_1}                 $}}
\newcommand{\msbqtwo}    {\mbox{$ M_{\widetilde {\mathrm b}_2}                 $}}
\newcommand{\mstqone}    {\mbox{$ M_{\widetilde {\mathrm t}_1}                 $}}
\newcommand{\mstqtwo}    {\mbox{$ M_{\widetilde {\mathrm t}_2}                 $}}
\newcommand{\eeto}    {\mbox{$ {\, \mathrm e}^+ {\mathrm e}^- \to             $}}
\newcommand{\Ecms}    {\mbox{$ E_{\mathrm{\small cms}}$}}

In this section, we will discuss the prospects at the ILC for the
direct discovery of new particles.   Our discussion will of course be
given in the context in which the LHC experiments have carried out a
large number of new particle searches, some reaching deeply into the
mass region above 1~TeV.   Still, we will explain, experiments at
$\ee$ colliders can bring a new approach to new particle searches and
still have very interesting windows for discovery. 

In general, the new particle searches done at the LHC have focused on 
scenarios within each theory of new physics that give the 
{\it best} possible experimental prospects to observe new physics.
However,
a negative result will only make it possible to claim that 
new physics is absent in a specific region of the full theoretical 
parameter space.
There is  no guarantee that new physics would be discovered
even if it is within the kinematic reach of the experiment.
The actual parameters of the theory might be far from the
ones giving the searched-for signature.

It is a rather different perspective to concentrate on the 
{\it worst} possible
points in the theoretical parameter space.
This clearly cannot reach as far out as in the previous case,
but now  a negative result
would make it possible to claim that the new physics theory is
ruled out at {\it all} possible parameter values below the
kinematic reach of the experiment.
It would also make discovery of the new physics {\it guaranteed}
if it is indeed energetically reachable.

These two avenues in the search for new physics
are in fact the main difference between searches at
hadron colliders and lepton colliders.
Hadron colliders are well suited for the first approach,
with their large reach into unknown territory in
energy,
but are less well suited for the second one due to
huge background levels and to the initial state
being unknown.
Lepton colliders have a lower
reach in energy,
but excel in fully exploiting
all possible manifestations of new physics within
reach. 
When comparing exiting limits  on new physics from LHC or LEP,
commonly presented in the mass-plane
of a pair of new states,
one must note that the former are incomplete ones
showing
models than {\it might} be excluded (for some - but not all -
  other model parameters),
while the latter shows complete ones,
i.e. models that
{\it must} be excluded ( for any value of other parameters).

ILC---like LEP---will explore all corners of the parameter spaces of
theoretical models.   It offers a guaranteed discovery within the
kinematic reach of the machine and, in the case of no discovery, sets
immutable limits that can be the final word the models it considers.
In this section, we will concentrate on this aspect, explaining how
ILC will expand the region of fully-explored theory space beyond
that of LEP.

It is clear that an ILC operating at 500 GeV, or even at 1 TeV,
will vastly extend the fully explored region.
But, already at 250 GeV, ILC will significantly extend this region:
While it is true that 250 GeV is not much more than the maximum 
energy of 208 GeV that LEP reached,
there are other features that are
ameliorated by orders of magnitude: 
The luminosity is 1000 times higher,
and both beams are  polarised.
The beam-spot is sub-microscopic in size, allowing to find 
displaced vertices at much smaller distances, also in channels 
(like \stau~pair production),
where there is no reconstructable primary vertex.
Furthermore, 
many aspects of the detectors are better than the LEP ones by a 
factor ten or more.
Since computing power has been increased by orders of magnitude,
all interactions can be recorded and analysed, 
i.e.
no trigger will be needed for experiments at the ILC, 
unlike the conditions at LEP.
Taken together, this means that much more subtle effects
can be probed for at energies that in principle were reachable at LEP.

Many of these features also are relevant in exploiting LHC's blind-spots: 
namely any signal stemming from processes without QCD interactions,
or with only soft final states.
Here, trigger-less operation of almost fully hermetic detectors
is a great advantage.
Processes where only kinematic reconstruction of the full event
would reveal BSM physics,
can be studied at a lepton collider. 
In contrast, 
at a $pp$ collider, only partial reconstruction 
in the transverse plane is possible.
In addition,
ILC detectors will be more precise than their LHC counter-parts,
since the low background-rates means that it is not necessary
to compromise between performance and radiation-hardness.

We will discuss a few particular classes of signatures,
which have been studied in depth:
\begin{itemize}
\item Pair-production of new short-lived states decaying to visible
  SM particles and another lighter new state, the lighter
  state being invisible, the so-called {\it antler} signatures.
  R-parity conserving SUSY is an example.
\item Production of new invisible final states, where only the presence
  of initial state radiation could reveal new phenomena,
  the {\it mono-photon} signature.
  A prominent example is dark matter production.
\item Production of new scalars, similar to the SM Higgs boson, but with smaller
  coupling to the $Z$, and possibly very different decay
  branching ratios, the {\it new-scalar} signature.
  Here nMSSM and 2HDM models are typical examples.
\end{itemize}

In addition to these cases discussed in detail,
other extensions to the SM can be searched for at the ILC.
Compared to a hadron collider, a lepton collider is much less dependent 
on  missing energy signature to find new physics.
For example, in R-parity violating SUSY 
or models with visible signs of a dark sector, or in composite models,
new physics does not necessarily manifest itself
with a  missing energy signature, but rather by the presence of
new states.
New physics could also manifest itself as 
new couplings, rather than new particles, \eg, in unexpected
flavour signatures. 
ILC-250 would be able to probe such signatures,
in some cases with a sensitivity
equal to that of dedicated flavour experiments, like BELLE II 
or LHCB.
In general, due to the low background levels, 
the ILC can be used to search for any
new particle in nature with electromagnetic, hyper-charge or
electroweak quantum numbers and thus provides discovery potential
complementary to that of the LHC.
A comprehensive overview of the potential of
the full ILC program  to discover new particles and phenomena 
can be found in \cite{Fujii:2017ekh}.

\subsection{Pair-production signatures}
\label{subsec:searches_antlers}

A event-signature that often occurs in BSM theories is the
``antler'' topology.
In such processes,
a pair of (not necessarily identical) new states are produced.
These particles then decay into SM particles and a lighter
new state. 
The lighter state might further decay to other SM particles,
and an even lighter new state.
At the end of such a cascade of decays,
a detector-stable new state,  $\chi$, is produced, 
which is not directly detectable.
The properties of the visible decay-products not only reveals the
presence of physics beyond the standard model, 
but also contain a large amount of information about the properties
of the new states.

In the case of direct decays of a pair of new state(s) (denoted by
$X$ and $Y$) produced
in an $e^+e^-$ collision at $E_{lab} = E_{cms} = M_o$ , 
the endpoints
of the energies of the standard model particles $x$ and $y$ 
in the process
$\eeto X Y \rightarrow x y  \chi \chi$
can be found to be
\begin{align}
E_{i^{max}_{(min)}}=&
 \frac{M_0} { 4 } 
\left (  
\sqrt {  \frac{ \lambda_{0,X,Y} + 4  M^2_0 M^2_{i^\prime} } { M^4_0 }}
\frac{ \sqrt {   \lambda_{i^\prime,i,\chi} + 4  M^2_{i^\prime} M^2_{i} }}
 {M^2_{i^\prime} } \right . \nonumber \\ 
  & \left . \begin{smallmatrix}+ \\ (-)\end{smallmatrix} 
\sqrt {  \frac{ \lambda_{0,X,Y} } { M^4_0 }}
\frac{\sqrt { \lambda_{i^\prime,i,\chi} }  } {M^2_{i^\prime} } 
\right )  
\label{eq:searches_genspartendp} 
 \end{align}
where
the shorthand  $\lambda_{k,l,m} = \lambda(M^2_k,M^2_l,M^2_m)$ is used\footnote{
The K\"all\'en function $\lambda$ is defined as $\lambda(a,b,c) = a^2 + b^2 +c^2 -2ab - 2ac - 2bc$}, and 
$i^\prime$ is either $X$ or $Y$ (and similarly, $i$ is the corresponding
SM particle, either $x$ or $y$)\cite{Berggren:2015qua}.
By determining these endpoints of the energy-spectra of the two SM particles ($x$ and $y$),
and using the knowledge of $\Ecms, M_x$ and $M_y$, both $M_\chi, M_X$\ and $M_Y$ can 
be determined.
If  the two initially produced new particles have the same mass, 
so that $ M_{X} = M_{Y} = M_{i^\prime}$, 
then  $\lambda_{0,X,Y} = M^4_0 - 4 M^2_o M^2_{i^\prime}$.
If, in addition, the masses of the produced SM particles can be neglected,
$\lambda_{i^\prime,i,\chi} = (M^2_{i^\prime}-M^2_\chi)^2$.
Hence, in the important case of pair production,   
$\eeto Y \bar{Y} \rightarrow y \bar{y} \chi \chi$ with $M_y \approx 0$, one finds the
simpler relation
\begin{align}
E_{y^{max}_{(min)}} &=  
   { \Ecms \over 4}  \left ( 1 - \left( { M_\chi \over M_{Y} } \right )^2 \right ) \nonumber \\
   &\left ( 1  \begin{smallmatrix}+ \\ (-)\end{smallmatrix}  \sqrt{1 -  4 \left ( {  M_{Y} \over \Ecms   } \right )^2} \right )
\label{eq:searches_sleptpairendp}
\end{align}
from which $M_\chi$ and $M_Y$ can be determined from the end-points.

R-parity conserving SUSY is a model that predicts a variety of
antler-type signatures, 
in sfermion or bosino production.
The lightest SUSY particle (the LSP) would be the final,
undetectable, new state, and would usually be the lightest
neutralino, \XN{1}, even though other candidates also
would be possible.
Both pair-production and associated production can occur,
and the decays might be direct or in cascades.
The produced pair might be fermions (bosinos),
or scalars (sfermions),
and can carry a variety of different quantum-numbers.
Hence,
an extensive study of SUSY covers a wide range of possible
antler topologies.
The essential difference between SUSY and the general case,
is that SUSY predicts the couplings, 
by the fundamental principle of SUSY:
{\it spaticles couples as particles}.
From the experimental point of view,
the implication of this is rather in the interpretation in terms
of exclusion- or discovery-reach,
than in the actual analysis-methods required.
In the SUSY case, 
regions in the mass-plane can be fully exploited, 
since the production cross-section is predicted.
In the general case,
the conclusion would be that discovery or exclusion is
possible down to some minimal cross-section,
determined by the data.
Therefore,
the extensive study of SUSY at past and future lepton colliders
serves as a boiler-plate for any search for new physics with
antler signatures.
In the following we will concentrate on the SUSY case.

\subsubsection{Loop-hole free searches}
\label{subsec:searches_noloophole}

In \cite{Berggren:2013vna},
it is shown how experiments an $e^+e^-$ collider can systematically and exhaustively
search for {\it any} Next-to-lightest SUSY Particle (NLSP), and thereby guarantee discovery, 
or set immutable limits for SUSY within the kinematic reach of the accelerator.

LHC has set no limits on processes giving the weakest limits, such as 
sleptons in general, and staus in particular.
In Refs.~\cite{Aaboud:2017leg,Aad:2014vma,Sirunyan:2018nwe},
it is shown that limits on selectrons and smuons can be set in the best possible case 
- either requiring that not only that both selectrons and smuons 
have the same mass, but also that the left- and right-handed states
are degenerate, or that the mass difference to the LSP is very large.
No limits at all could be set for stau production.
One the other hand,
LHC does give stringent limits on a gluino or first- or second-generation
squarks \cite{Aaboud:2017vwy,*Sirunyan:2018vjp}.
Also for a stop,
the LHC coverage is increasing,
and excludes a stop with a mass of 1 TeV, 
if the LSP mass is below 250 GeV \cite{Aaboud:2017ayj,*Sirunyan:2018vjp}.
However,
both theoretically, and given these limits,
it is quite unlikely that a coloured sparticle would be the NLSP.

Instead,
the most stringent absolute limits on the NLSP comes from LEP.
There
limits on all SUSY particles has been set.
In
\cite{Abdallah:2003xe,Heister:2001nk,Achard:2003ge,Abbiendi:2003ji},
searches for sleptons are reported, in 
\cite{Abdallah:2003xe,Achard:2003ge,Heister:2002hp,Abbiendi:2002mp},
the results of the searches for squarks can be found, 
and in \cite{Abdallah:2003xe,Abbiendi:2003sc,Heister:2002mn,Acciarri:1999km},
the results for charginos and neutralinos are given.
In addition,
combined results can be found in \cite{
LEPSUSYWG/04-01.1,*LEPSUSYWG/04-02.1,*LEPSUSYWG/02-04.1,*LEPSUSYWG/01-03.1}.

A summary of the LEP results is that 
a chargino NLSP below
between 92 and 103 GeV
(depending on the mass-difference) is excluded,
whatever the nature of the chargino is.
For the second neutralino,
a general exclusion in the mass-plane is not possible,
due to the complicated structure of the neutralino mass-matrix,
which allows for situations where the cross-sections both for
$\eeto~\XN{2} \XN{2}$~ and $\eeto~\XN{1} \XN{2}$~ can be small
at the same time. 
For any {\it given} SUSY model, however, the combination of the
searches for $\eeto~\XN{2} \XN{2}$, $\eeto~\XN{1} \XN{2}$ and
$\eeto~\XPM{1} \XMP{1}$, 
as well as the searches for $\eeto~\sel \sel$ and $\eeto~\snu_e \snu_e$ 
(since the \sel~ and the $\snu_e$ contribute to t-channel production of \XN{2} and \XPM{1}, 
respectively) 
will be likely to yield constraints on \MXN{2}.
\begin{figure*}
   \centering
      \subfigure[]{\includegraphics[width=0.35\linewidth]{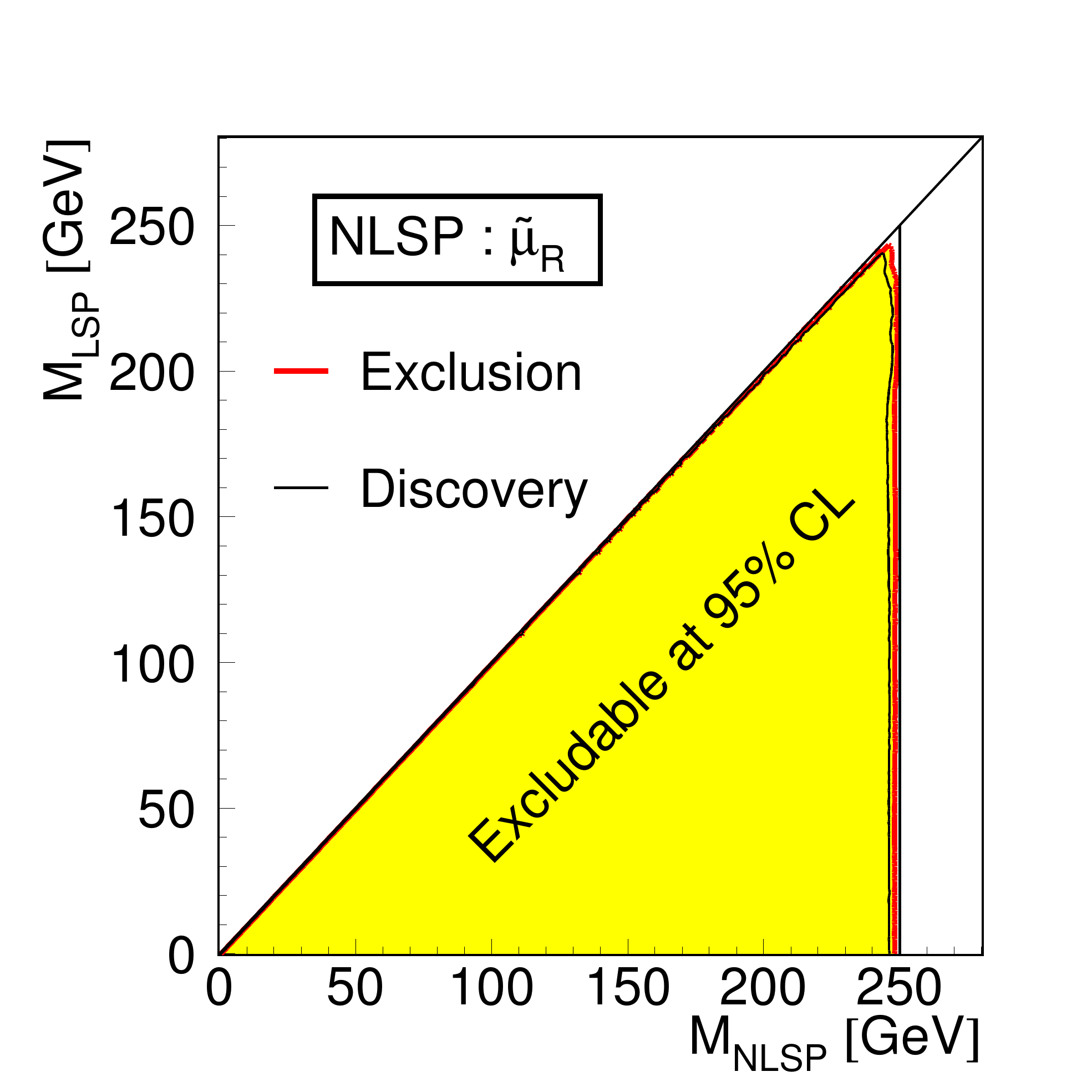}}
      \hspace{0.1\linewidth}
      \subfigure[]{\includegraphics[width=0.35\linewidth]{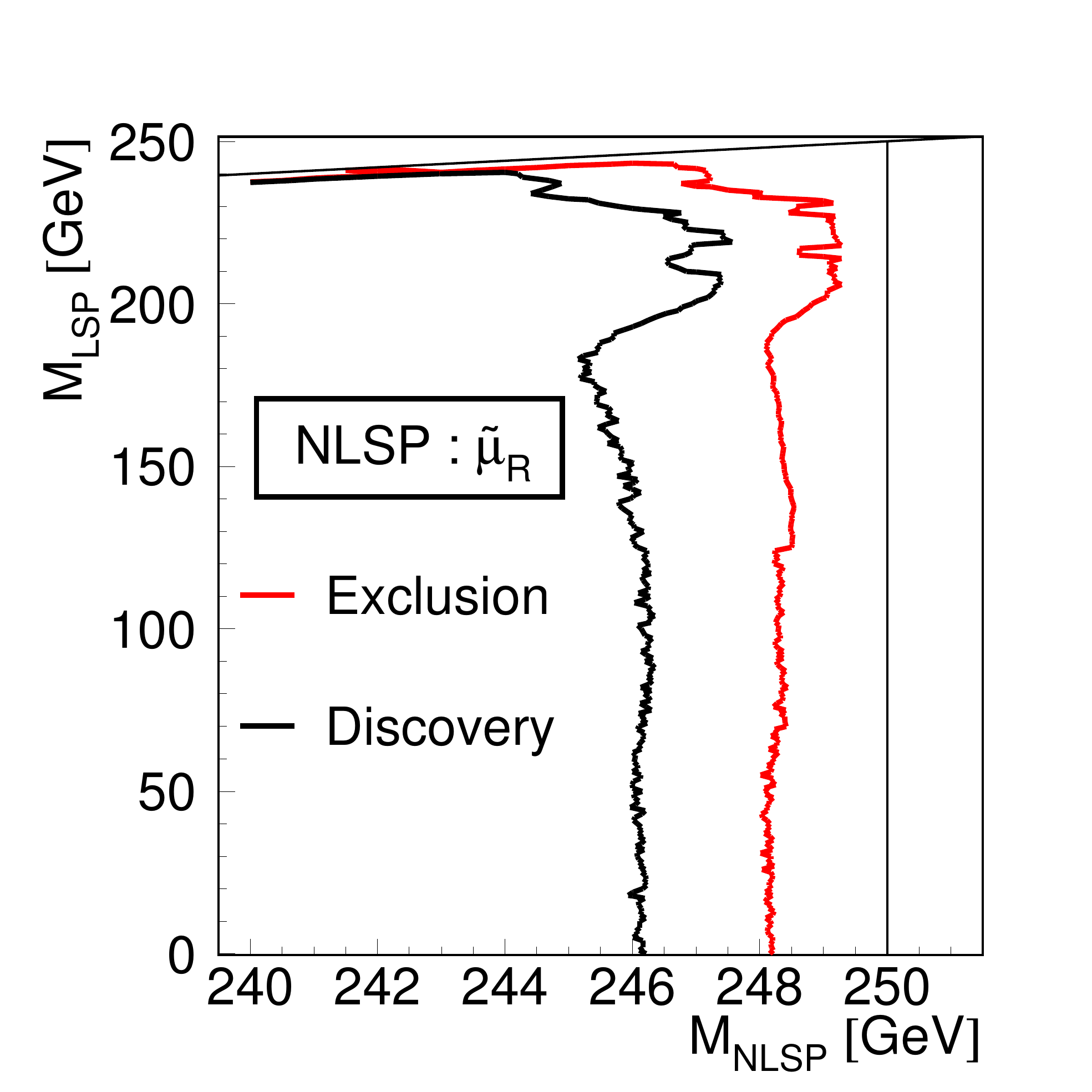}}

     \subfigure[]{\includegraphics[width=0.35\linewidth]{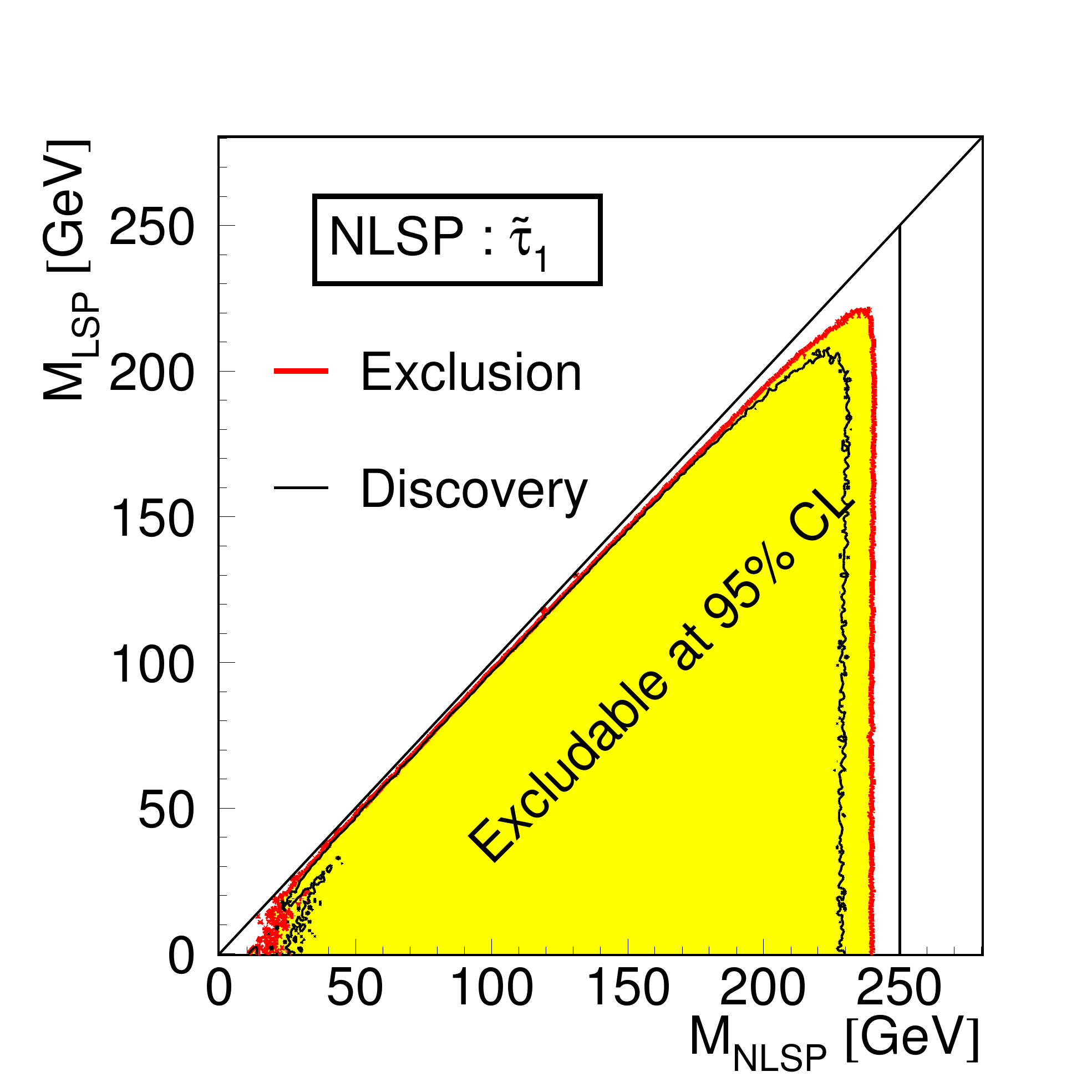}}
     \hspace{0.1\linewidth}
     \subfigure[]{\includegraphics[width=0.35\linewidth]{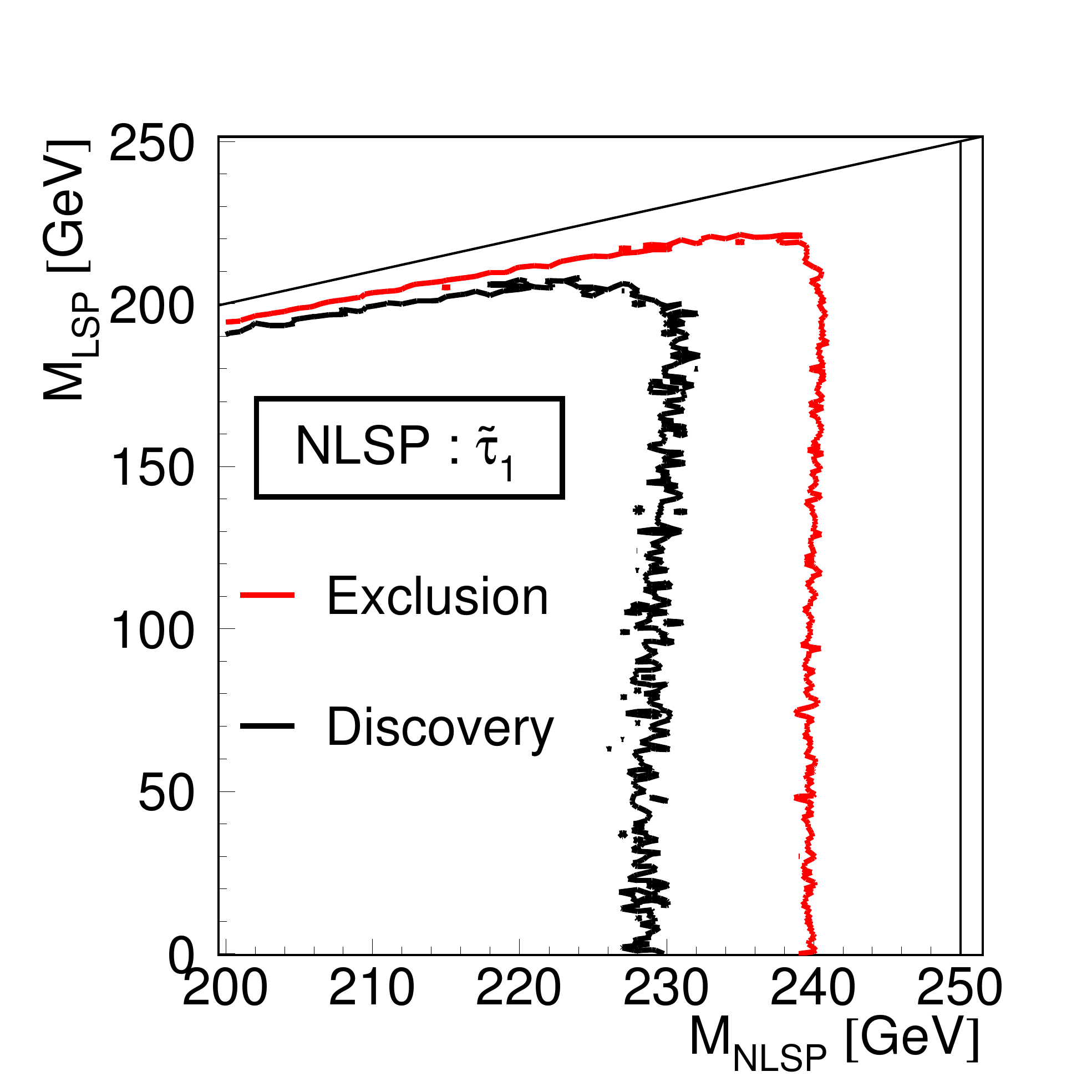}}
\caption{ILC discovery reach for a $\smur$ 
(top) $\stone$ (bottom) 
NLSP for $\int \mathcal{L} \, \mathrm{dt}$ = 
500 fb$^{-1}$ at $\sqrt{s}$ = 500 GeV. 
For the \stone, the mixing angle was chosen to give the lowest
possible production cross-section. (a,c) full scale, (b,d) zoom to last 
few GeV before the
kinematic limit~\cite{Berggren:2013vna}. \label{fig:searches_noloophole1}}
\end{figure*}

In the slepton sector,
smuons and selectrons are excluded
below 95 to 100 GeV, if the mass difference to the LSP is
above 4 GeV. 
Staus are excluded below 87 to 93 GeV, 
if the difference is above 8 GeV. 
Selectrons and
smuons are completely excluded below $M_Z/2$ (from the width of the $Z$),
while staus are excluded below 28 GeV for any mass-difference and
mixing. 
The weaker limit for the staus is due to the fact that
it is possible that the stau mixing is such that it does not couple at
all to the $Z$, 
only to the photon, 
and hence that the constraint from the width of the $Z$ cannot be
applied.
In fact, the limit from the stau at minimal cross-section is the weakest limit on any
NLSP candidate, 
and therefore represents the current absolute exclusion for any MSSM model.

In addition, LEP excludes third generation squarks below 94 to 98 GeV at
mass differences to the \XN{1}  larger than 8 GeV 
and the mixing angle giving the
minimal cross-section are excluded. 
For any mixing, mass-difference and dominant
decay mode, 
a stop with mass below 63 GeV is excluded.
However, these coloured sector limits are
essentially superseded by the LHC ones.

It can be noted that, except for the chargino,
the LEP limits fall short of the kinematic limit
by 10 to 20 \% even for large mass-differences,
and for small differences by 50 \% or more.
This is due to the fact that the size of the data-sets at the
highest energies were tiny - 500 pb$^{-1}$ at 206 GeV, and
only 33  pb$^{-1}$ at 208 GeV.
This low luminosity is particularly unfavorable for the sfermions,
because of the slow ($\beta^3$) rise of the cross-section close to threshold.
Also,
the LEP detectors all were triggered,
meaning that in the low mass-difference cases,
either some auxiliary activity was needed to provide
a trigger, 
or only a small fraction of the events - 
those where the detectable SM decay-products happened to be
almost aligned with the direction of the decaying sparticle -
would be registered.
This resulted in a quite low selection efficiency in these cases.
Furthermore,
due to the large size of the beam-spot at LEP,
using impact-parameters as a tool to separate signal
and background was not very effective.
Finally,
the beams at LEP were unpolarised,
which is a particular draw-back when searching for signs
of a chiral theory such as SUSY.

In contrast, the ILC has none of these problems,
as already mentioned,
which means that the ILC will largely extend the
territory explored by LEP.
The same features of the ILC allows to probe for signals in the
LHC blind areas for un-coloured states at lower mass differences.

In \cite{Berggren:2013vna},
the prospects at the ILC at 500 are evaluated.
Two cases were studied in more detail, 
the least and the most challenging ones, 
namely the cases where the NLSP is either the \smur~or the \stone.
The first case profits from a very clean and well measured
signal,
with no other parameters than the two masses involved,
while the second one has the most difficult signal 
(due to the partly invisible SM system),
and in addition has a further theory parameter,
namely the \stau~mixing angle.
For both these cases,
the full mass-plane was scanned over a 1-by-1 GeV grid,
using the detailed fast simulation SGV, described in  \ref{sub:sw-fastsim}.
In the \stone~case, the mixing-angle was chosen such that the
production cross-section was as small as possible.
The resulting exclusion/discovery reaches are shown in
Fig.~\ref{fig:searches_noloophole1}.
One can note that the exclusion limit,
even for the rather modest luminosity used in the study
is only 0.8 (4) \% from the kinematic limit for the \smur (\stone).
Hence, the area of assured discovery-potential or power of
model-independent exclusion
will increase by a factor of 6 to 7 with respect to current (LEP) results.

Even for an ILC operating at 250 GeV, 
a substantial increase in reach is expected.
The area of the excluded mass plane will increase
much more than the modest increase in energy might suggest at first glance.
Reasonably assuming that also at 250 GeV, 
reach will fall only a few percent short of the kinematic limit,
the area covered by ILC-250 will increase
by 70 \% to 80 \% at large mass-differences compared to the LEP results.
At the smallest mass-differences,
even larger improvements might be expected,
once dedicated analyses have been performed in this region.
Also at 250~GeV,
the reach into LHC's blind areas will be important.
\subsubsection{Sleptons}
\label{subsec:searches_sleptons}
In \cite{Berggren:2015qua} and \cite{Bechtle:2009em}
more in-depth analyses of specific models are presented.
The emphasis in these works is to estimate the precision
with which various parameters can be extracted.
The analyses were done with the full SM background simulated  
with full detector simulation at \Ecms = 500 GeV.
In \cite{Bechtle:2009em}, also the signal was simulated with
full simulation,
while the detailed fast simulation SGV was used in  \cite{Berggren:2015qua}
\footnote{The models studied allows for \selr, \smur. and \stone~production also
at \Ecms = 250 GeV}.
\begin{figure*}[]
  \begin{center}
    \subfigure[]{\includegraphics[width=0.3\linewidth] {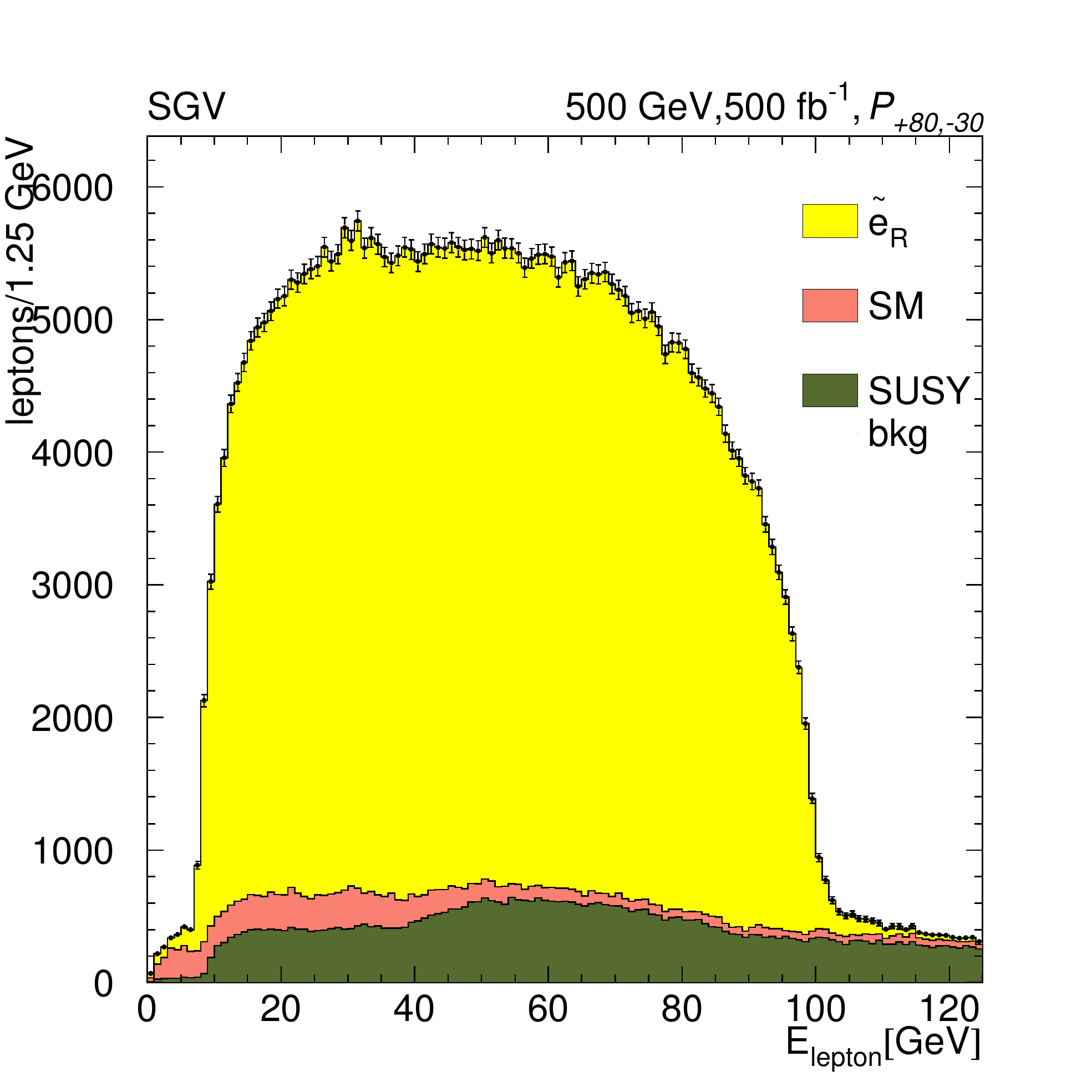}}
    \hspace{0.01\linewidth}
    \subfigure[]{\includegraphics[width=0.3\linewidth] {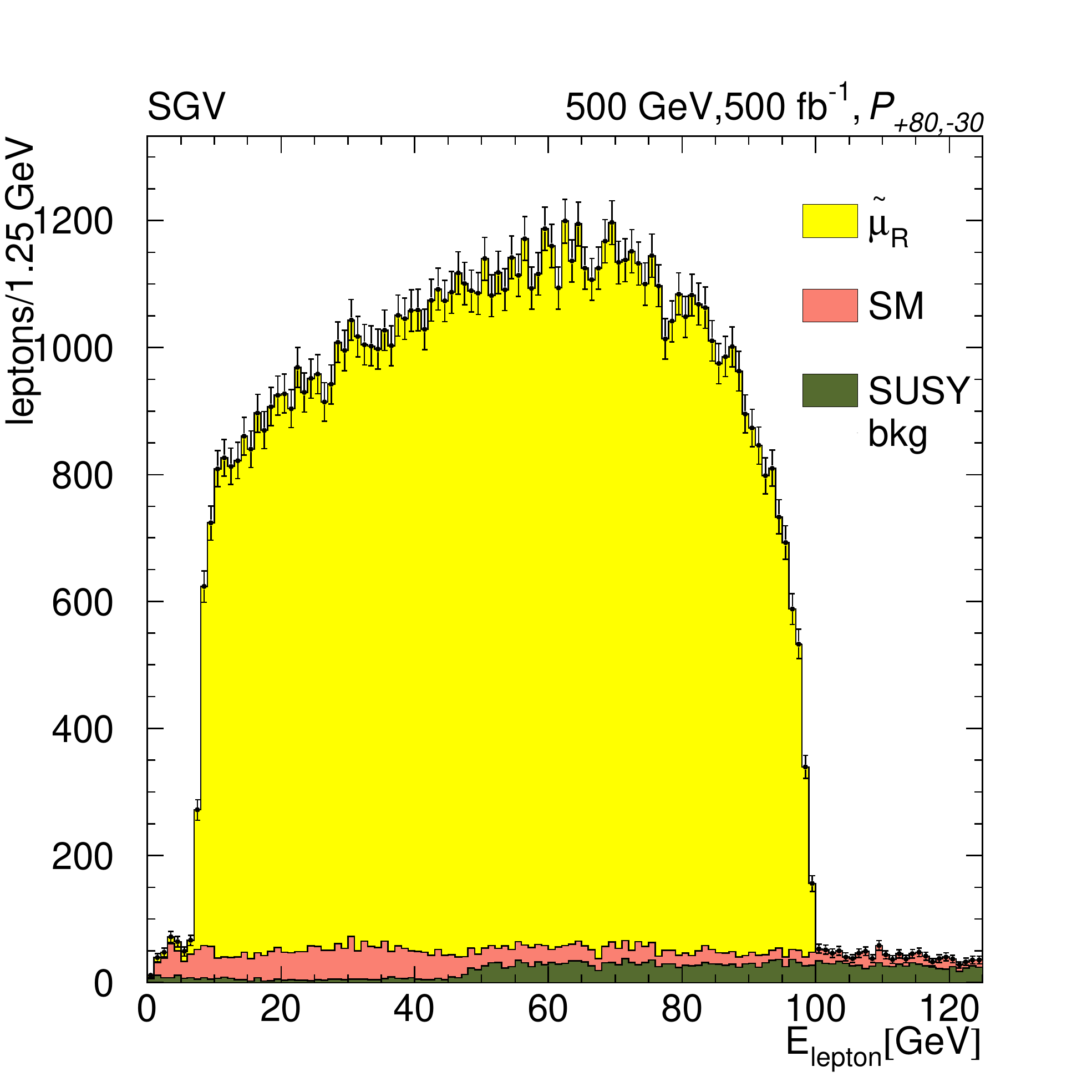}}
    \hspace{0.01\linewidth}
    \subfigure[]{\includegraphics [width=0.3\linewidth]{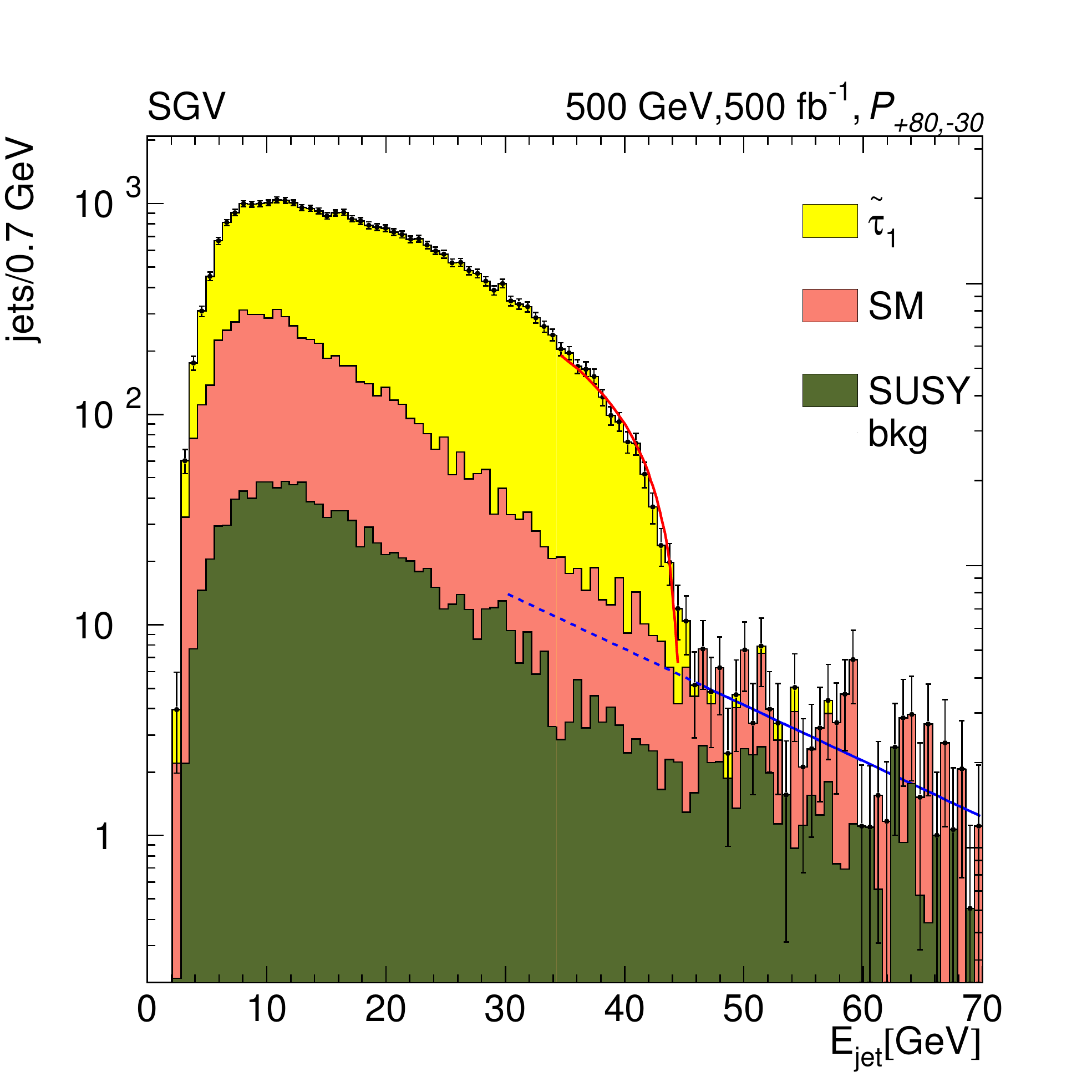}}
  \end{center}
  \caption{\label{fig:searches_sleptons} Property determination of SUSY  (a) selectron.  (b) muon and 
(c) $\tau$-jet energies in selected di-leptons events
after collecting
500 fb$^{-1}$ of data for beam-polarisation  $\mathcal{P}_{-80,+30}$~\cite{Berggren:2015qua}. }
\end{figure*}
In Fig.~\ref{fig:searches_sleptons},
the energy-spectra of the visible decay-products of 
\selr , \smu~and \stone~are shown.
A number of novel techniques were utilised
to extract the relevant edges from the distributions
(the truncated sub-sample method  \cite{Berggren:2015qua},
and finite impulse response method \cite{Caiazza:416980}),
both giving precisions a factor two or more better than
traditional methods.
Once the edges were determined,
applying Eq. \ref{eq:searches_genspartendp},
the masses of \selr~and  \smur~could 
be estimated with an error of 2~\permil~and
4~\permil, respectively.
Fitting for a single value of \MXN{1} in these two spectra,
an error of 1.5~\permil~ was obtained.
Using the value of  \MXN{1},
and fitting spectrum in Fig.~\ref{fig:searches_sleptons}c for $M_{\stone} $,
the  \stone~mass could be determined to 2~\permil.

Furthermore, 
as can also be seen from Eq. \ref{eq:searches_genspartendp},
close to the threshold,
the decay-products become mono-energetic which means
that an almost background-free threshold-scan can be done
at a collider -- such as ILC -- where \Ecms~ can be freely chosen.
The result of such a scan is shown in Fig.~\ref{fig:searches_smuselthreshold}.
The precision of the masses are comparable to those
obtained from the fit to the spectra,
but are independent of \MXN{1}.
In addition,
the fit to the shape of the threshold makes it possible to
exclude the hypothesis that the new states discovered are
fermions,
as can be see by the fits of either $\sigma \propto \beta^3$ (expected for
scalars) or
 $\sigma \propto \beta$ (expected for fermions). 
 
A further measurement possible in these models is the determination of the
polarisation of the $\tau$-lepton from the \stone~decay.
This is achieved by studying the spectrum of the $\pi$:s in
the $\tau \rightarrow \pi \nu_\tau$ mode, or the ratio of
$E_\pi^\pm$ to $E_\pi^0 + E_\pi^\pm$ in the  $\tau \rightarrow \rho \nu_\tau \rightarrow \pi^\pm \pi^0 \nu_\tau $ mode.
In \cite{Bechtle:2009em} it was found that
the degree of polarisation could be determined to $\sim$ 8 \%.
Also in \cite{Bechtle:2009em},
it was found that the cross-section for \stone~pair-production could be determined to 4~\%.
The difference in the cross-section when the beam-polarisations are
reversed can be used to determine the \stau~ mixing angle\footnote{The other
possibility to determine the mixing-angle, namely the \stone \sttwo associated production
have not yet been studied in detail},
which together with the determination of $\tau$-polarisation can be used to determine
the size of the chirality-conserving gauagino fraction of the \XN{1} relative to it's
chirality-flipping higgsino fraction.
\begin{figure*}[]
  \begin{center}
    \subfigure[]{\includegraphics[width=0.3\linewidth]{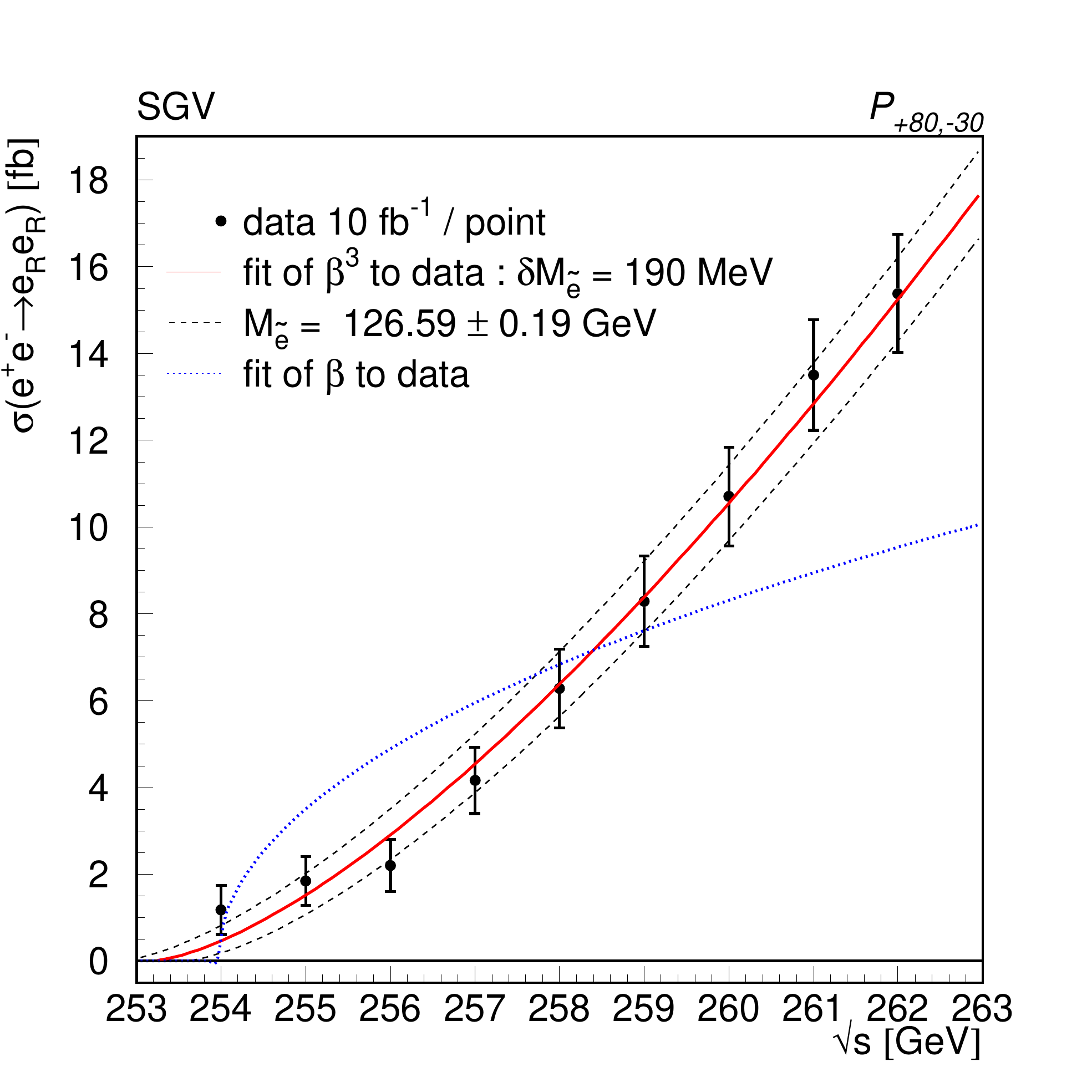}}
    \hspace{0.01\linewidth}
    \subfigure[]{\includegraphics[width=0.3\linewidth]{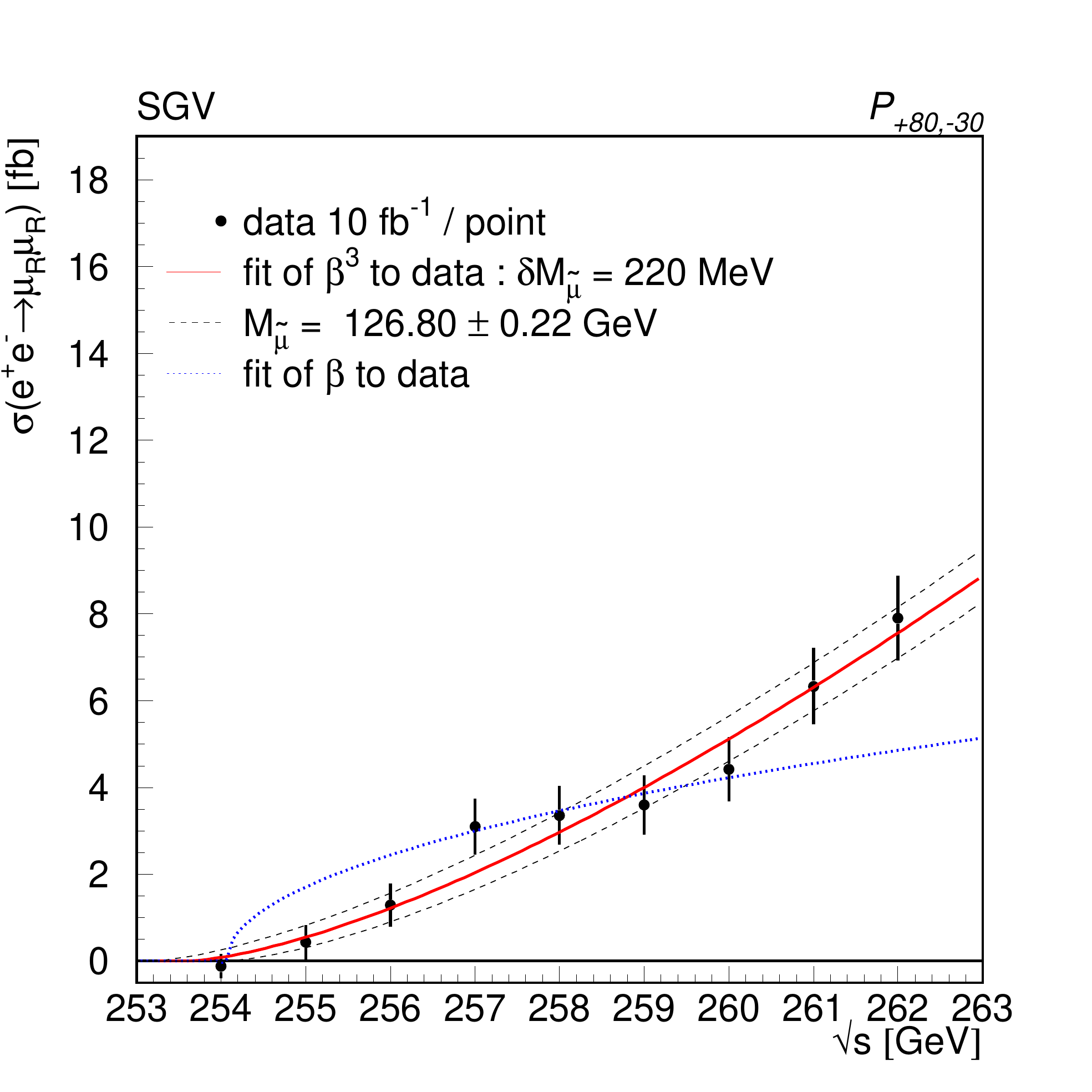}}
  \end{center}
  \caption{\label{fig:searches_smuselthreshold} : Scans over the threshold for slepton production
(a) scan of the $\eeto \selr \selr$ threshold  (b)  scan of the $\eeto \smur \smur$ \cite{Berggren:2015qua}. }
\end{figure*}

\subsubsection{Bosinos}
\label{subsec:searches_bosinos}

In \cite{Berggren:2015qua,Berggren:2013vfa,Baer:2016new,Chera:402736}
detailed studies of specific points where a bosino is the NLSP
are presented.
Many different topologies are covered by the analyses,
depending on the mass-difference.
The bosinos might decay to on-shell $Z$ or $W$ bosons,
undergo three-body decays (mediated by virtual
$Z$ or $W$ bosons), or decay radiatively.
In addition,
mixing in the bosino-sector will yield relations between the masses
of $\XN{2}$, $\XPM{1}$ and the LSP, relations that are different
for different models.
The same is true for the production cross-sections,
in particular the relation between $\XN{2}\XN{2}$
pair-production and
$\XN{1}\XN{2}$ associated production.
Also assumed relations at the high scale have implications on
the phenomenology at the EW-scale,
in particular on how large the LSP-NLSP mass difference can be.
For bosino production it is therefore needed to study
various cases in detail,
and avoid assumptions on other related processes.
The LEP experiments all carried out a comprehensive
search for a $\XPM{1}$ NLSP which were combined in \cite{LEPSUSYWG/02-04.1}.
For the   $\XN{2}$ NLSP case,
as mentioned is section \ref{subsec:searches_noloophole},
only cross-section limits can be given,
if no assumptions on the model is done.
Such limits were given by the experiments \cite{Abdallah:2003xe,Acciarri:1999km,Abbiendi:2003sc}.

At LHC, the reach of the search for the non-coloured bosinos
can be quite large, but always with strong model assumptions.
Even so, the limits tend to disappear for low mass-differences,
and are largely absent in the region allowed if GUT-scale unification
of the bino and wino mass-parameters ($M_1$ and $M_2$) is assumed
\cite{Aaboud:2018jiw,Aaboud:2017leg,Sirunyan:2018ubx}.

Once again, the conditions at the ILC will allow to extend the
model-independent LEP limits to higher masses.
Because the $\XPM{1}$ cross-section is quite large, and has
a sharp ($\propto \beta$) threshold dependence,
the increase in reach at ILC-250 with respect to LEP is
not as large as it is for the sfermions: already LEP could
exclude  an $\XPM{1}$ NLSP to only a few GeV below the kinematic limit
and at all mass-differences.
However,
the ILC potential
becomes very important at a 500 GeV,
and even more so after a future energy-upgrade to 1 TeV. 
Nevertheless,
the limit/discovery potential of ILC-250 is still sizeable
compared to current and future LHC limits,
in particular as the LHC limits for bosinos suffer
more from model-dependence than the sfermion ones.

This is illustrated by a specific example in figure~\ref{fig:searches_bosinoexcl}, 
which shows the current limits in the $\MXN{1}$ - $\MXC{1}$ 
plane from ATLAS~\cite{Aad:2014vma}, together with the
projected discovery reach at 14 TeV with $\int \mathcal{L} \, \mathrm{dt}$ = 3000 
fb$^{-1}$ \cite{ATL-PHYS-PUB-2018-048}
Here it is assumed that  $\MXN{2}=\MXC{1}$, that $\XPM{1}$ and $\XN{2}$ are 
pure Winos, and that Br($\widetilde{\chi}\rightarrow W^{(*)}/Z^{(*)}\XN{1}$)
=1
\footnote{Note that the more difficult case $\widetilde{\chi}\rightarrow h^{(*)}\XN{1}$ is 
not considered.}.
The brown-shaded area indicates the corresponding limit from LEP 
\cite{Heister:2002mn,Abdallah:2003xe,Abbiendi:2002vz},
which assumes only  $\XPM{1}$ pair production, with no assumption on the decay mode,
nor the nature of the $\XPM{1}$.
The expected limits for the ILC at $\sqrt{s}=500$ or $1000$\,GeV are also shown 
with the same 
assumptions as for the LEP exclusion.
As can be seen from the (loophole) region not covered by the LHC, there is a large
discovery potential for the ILC, even after the high luminosity LHC data has been 
fully exploited.
\begin{figure*}[]
   \centering
      \includegraphics[width=0.5\linewidth]{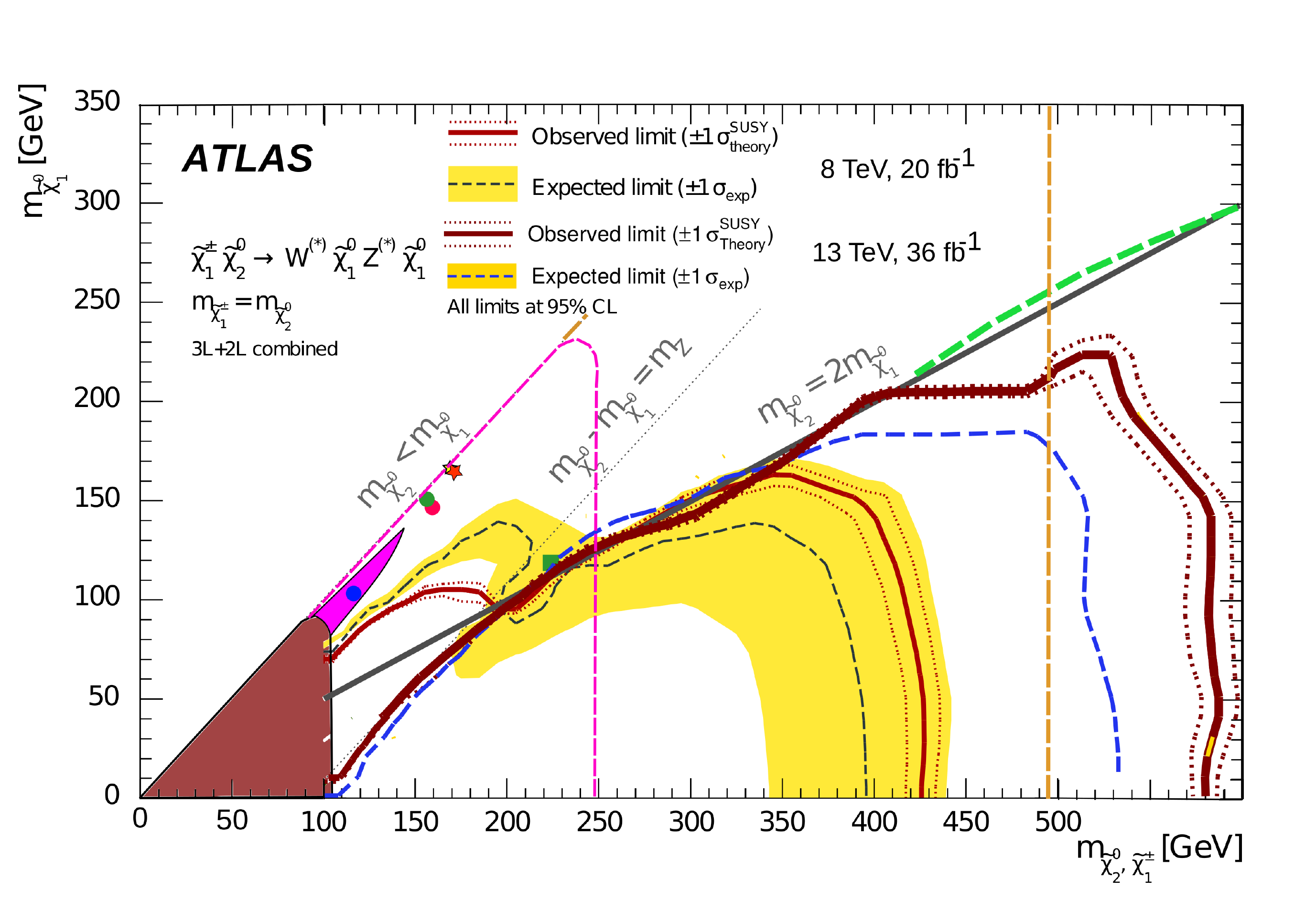}
\caption{Discovery or exclusion regions in the $M_{NLSP} - M_{LSP}$ plane
  for a $\XPM{1}$ or $\XN{2}$ NLSP. Solid brown area: LEP exclusion;
  Solid red and dashed grey/blue lines: ATLAS exclusion (observed and expected),
  for the 8 TeV data (thinner lines) and the 13 TeV data (thicker lines).
  Dashed green line: ATLAS 14 TeV discovery projections for $\int \mathcal{L} \, \mathrm{dt}$ = 3000 fb$^{-1}$;
  Dashed magenta (orange) lines: ILC discovery expectation for $E_{CMS}$ = 500 (1000) GeV;
  Solid black line: below line, no GUT scale gaugino mass unification.
  The symbols indicate the positions in the mass-plane of the analyses mentioned in the text.
  The magenta solid area is the ATLAS low $\Delta(M)$ search at 13 TeV, which however is within a different model.
 \label{fig:searches_bosinoexcl}}
\end{figure*}

\subsubsection{Small mass differences}
\label{subsec:searches_lowdm}
The case with antler topologies with small mass-differences
is particularly interesting for ILC, already at 250 GeV.
Partly because the experimental limits from LEP are much weaker then
for high mass differences,
and largely absent at LHC,
but also for theoretical reasons.

One reason to particularly search for SUSY with small
mass-differences is the possibility that the LSP is the (full) explanation
for Dark Matter:
Over a large region of SUSY parameter space, co-annihilation with the NLSP
is an attractive mechanism which acts to reduce the relic density of the LSP to its
cosmologically observed value~\cite{deVries:2015hva}. 
An example of such a model is the one presented in~\cite{Berggren:2015qua} and
discussed in the previous section.
In this model,
the NLSP is the \stone,
with a mass 10 GeV above the LSP.
Co-annihilation requires
a small mass difference between the NLSP and the LSP in order to be effective,
and thus the expected value of the relic density depends strongly on the exact
masses and mixings of the involved particles,
 requiring measurements at the permille and
percent-level, respectively.
This is discussed in~\cite{Lehtinen:415433}, 
where also a detailed analysis of the relic-density determination that
the measurements presented in~\cite{Berggren:2015qua} would imply.
Figure~\ref{fig:searches_STC10omega}a shows the precision
of the fitted relic density relative to the model value.
In the figure, 
the model value was chosen to be
the central value determined from cosmology using the observations 
of {\sc planck}~\cite{Ade:2015xua}.
It was also verified the other model-values were faithfully reproduced
by the fit, see Fig.~\ref{fig:searches_STC10omega}b.
\begin{figure*}[]
   \centering
      \subfigure[]{\includegraphics[width=0.35\linewidth]{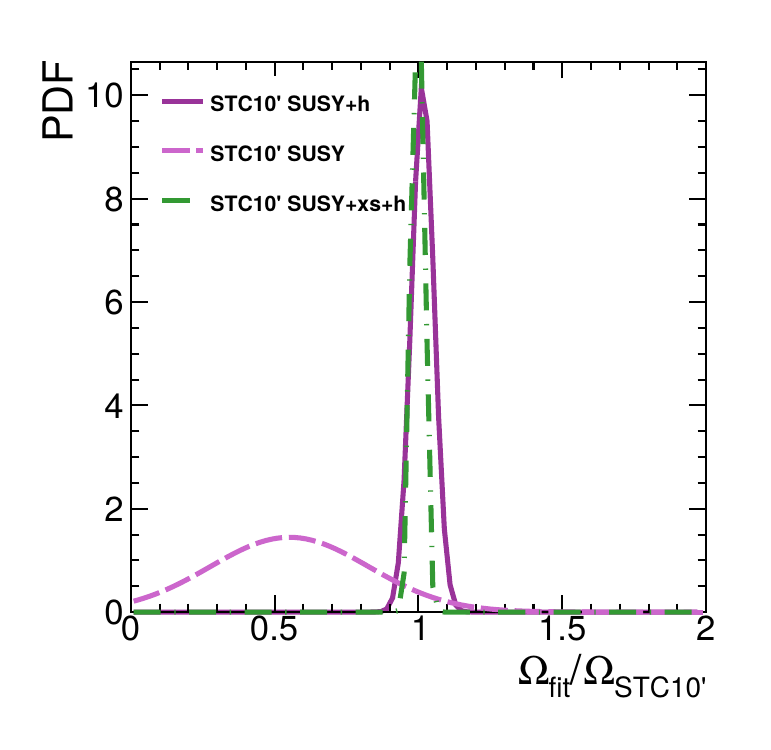}}
      \hspace{0.1\linewidth}
      \subfigure[]{\includegraphics[width=0.35\linewidth]{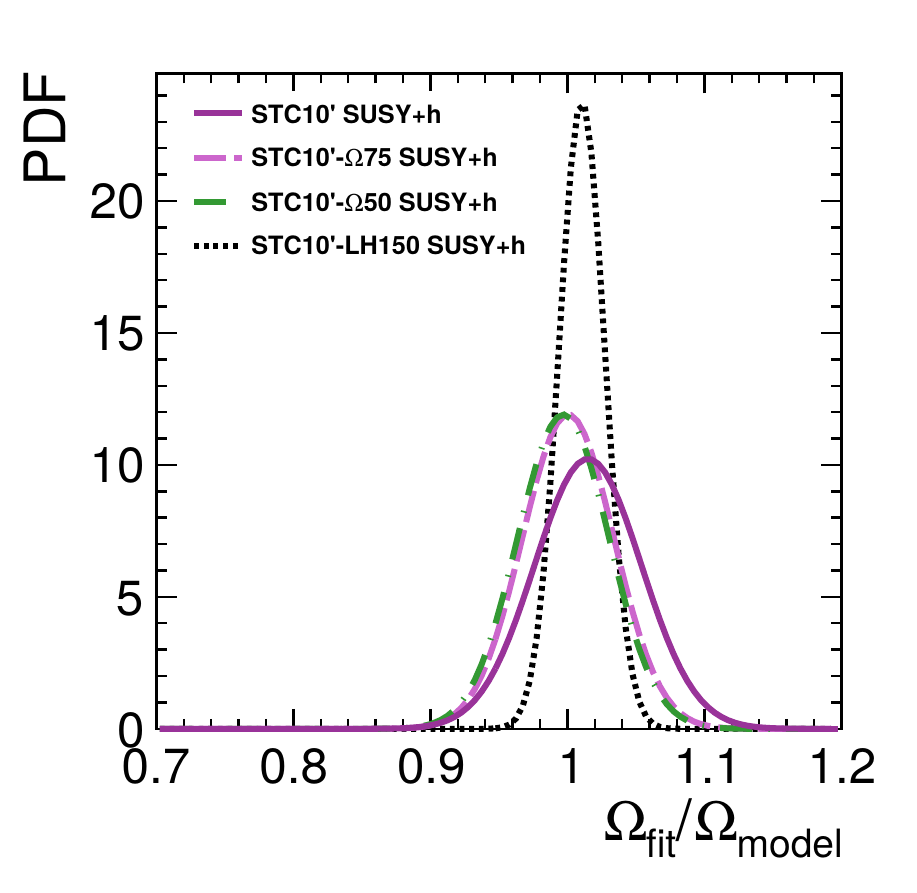}}
\caption{ (a) Comparison  of  relic   density  fitted to  the   measurements of the SUSY model
to the model value, with or without using input from ILC higgs-measurements and with, in addition, using 
measured cross-sections. (b) Comparison between fitted and model value, when the model value was by hand modified as indicated.
From~\cite{Lehtinen:415433}. \label{fig:searches_STC10omega}}
\end{figure*}

A second reason to search for such low mass-difference
processes, applying to SUSY
is that they tend to occur in many possible SUSY scenarios,
as shown in Fig.~\ref{fig:searches_tomohikoscan},
because of the mass-relations between different bosinos in the Wino- and
Higgsino-sectors,
the second lightest bosino will be close in mass to the LSP,
if the latter is dominantly Wino or Higgsino.
Only in the case of a large admixture of Bino in the LSP
can the mass-difference be arbitrarily large.
Furthermore, 
{\it if} GUT-scale unification of  the Bino and Wino mass 
parameters $M_1$ and $M_2$ holds,
the next-to-lightest bosino {\it cannot} be 
heavier than twice the LSP mass~\cite{Choi:2001ww}.
\begin{figure*}[]
  \begin{center}
    \subfigure[]{\includegraphics[width=0.3\linewidth] {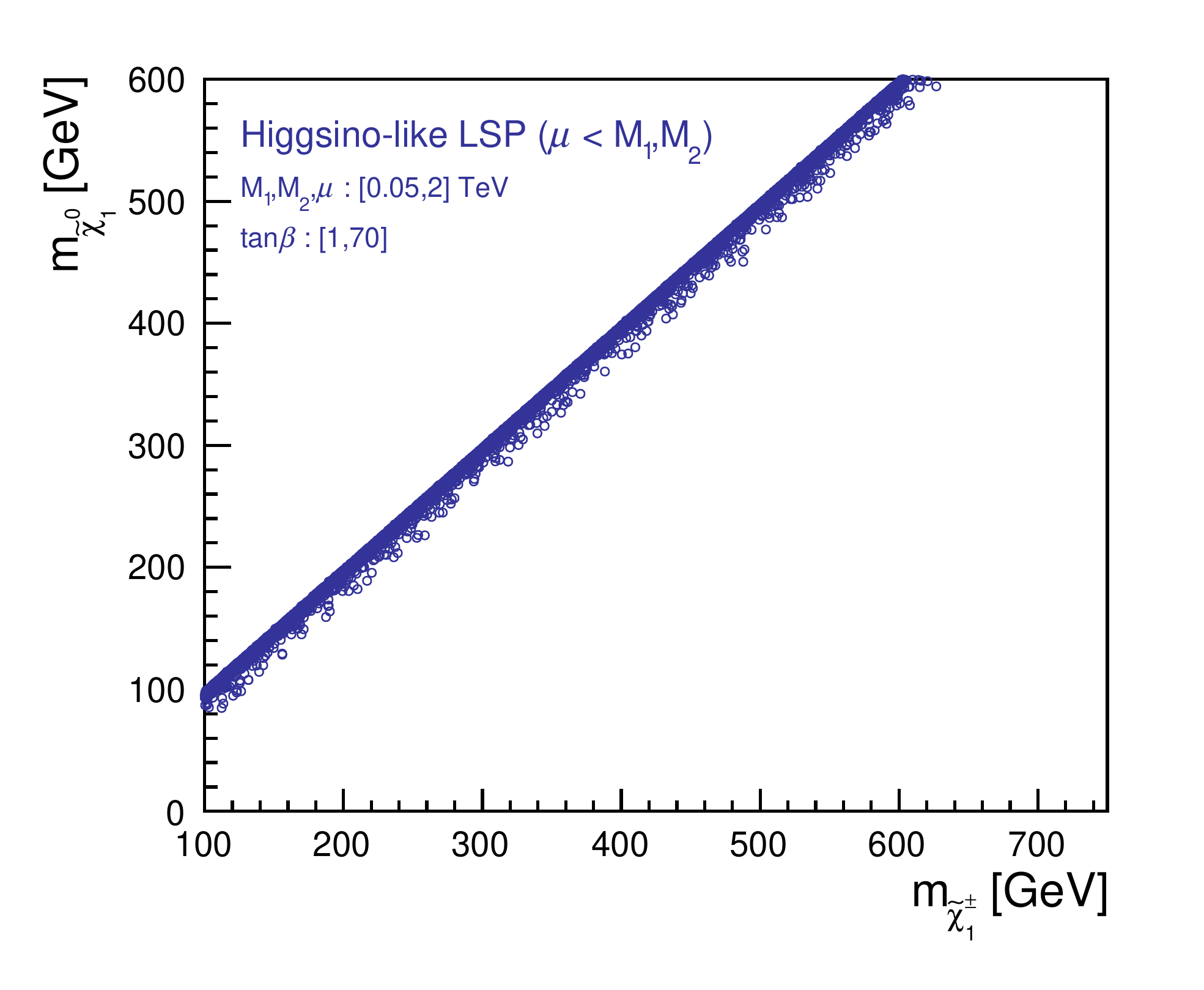}}
    \subfigure[]{\includegraphics[width=0.3\linewidth] {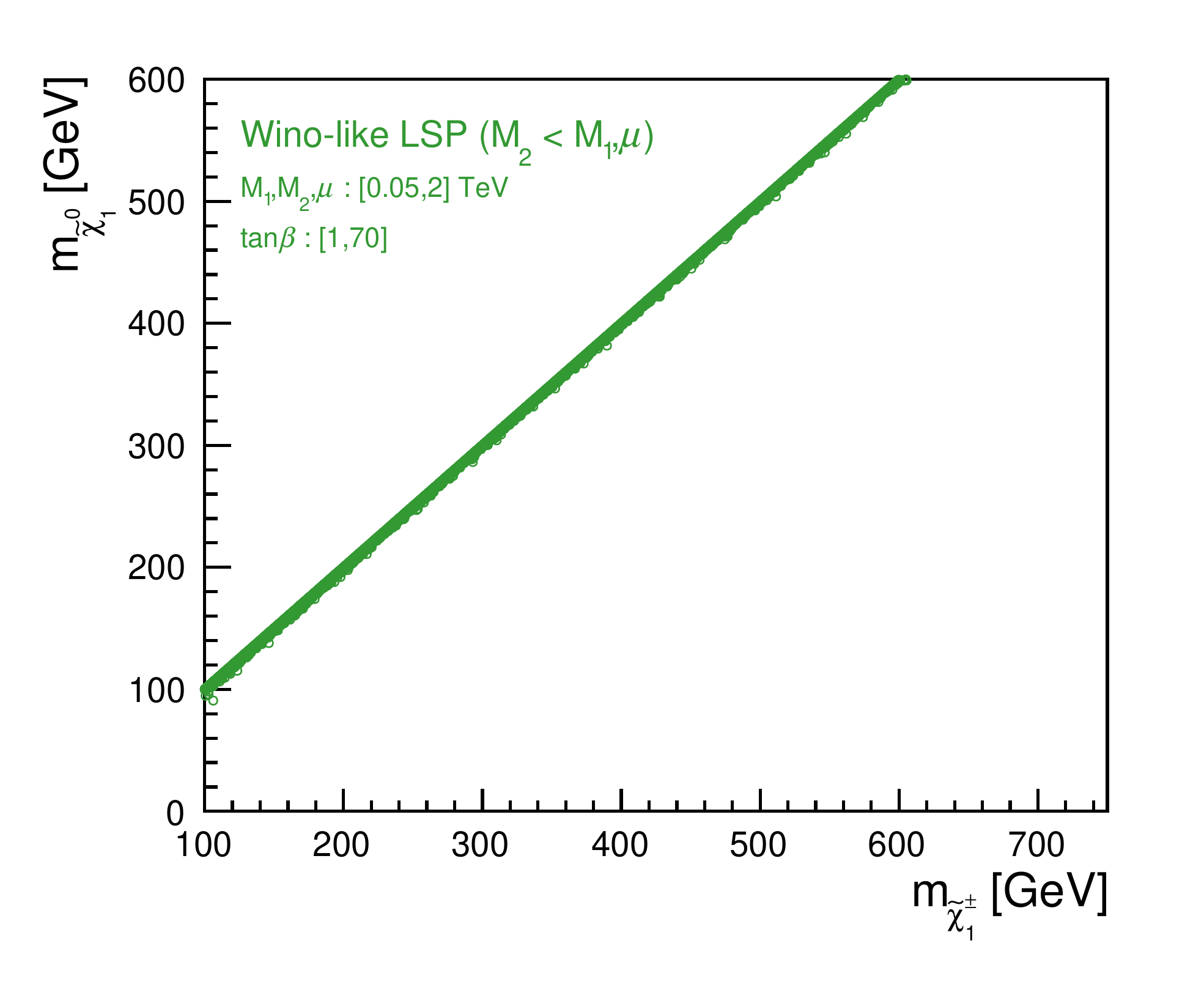}}
    \subfigure[]{\includegraphics[width=0.3\linewidth] {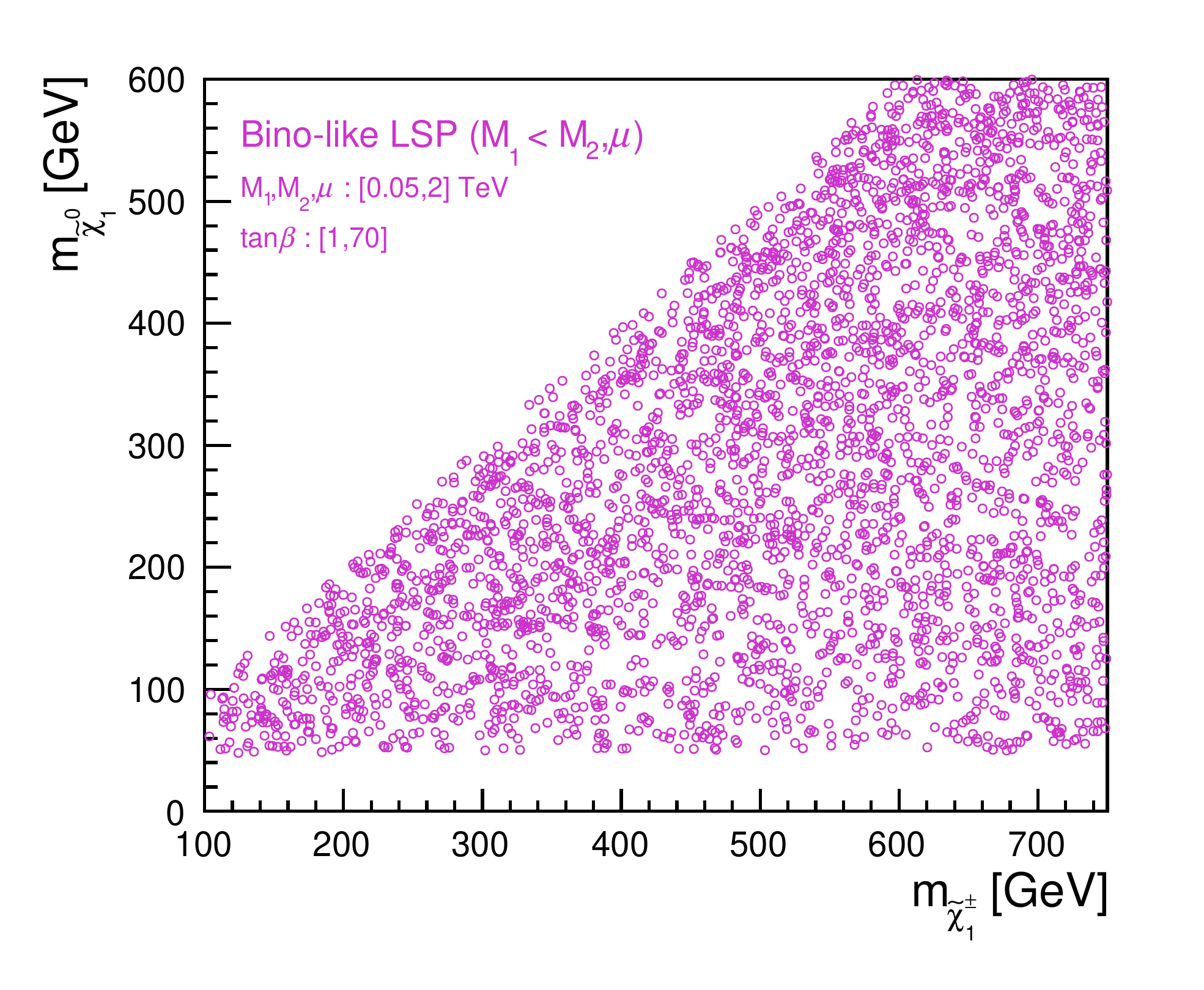}}
  \end{center}
  \caption{\label{fig:searches_tomohikoscan} \MXN{1} vs \MXC{1} when scanning over the bosino parameters
$M_1 , M_2, \tan{\beta}$ and $\mu$. (a) Higgsino-like LSP ($\mu < M_1 , M_2$), 
(b)  Wino-like LSP ($M_2 < \mu , M_1$), (c) Bino-like LSP ($M_1 < \mu , M_2$). }
\end{figure*}

In fact, 
light higgsinos are a fundamental requirement 
of natural SUSY models.
The generic formula relating $M_Z$ to SUSY parameters reads
~\cite{Bae:2014yta}:
\beqa
\frac{m_Z^2}{2} &=& \frac{(m_{H_d}^2+\Sigma_d^d)-(m_{H_u}^2+\Sigma_u^u)\tan^2\beta}{(\tan^2\beta -1)}
-\mu^2 \CR 
 & & \hskip 0.4in \simeq -m_{H_u}^2-\mu^2
\eeqa{eq:searches_mzs}
To avoid unnatural fine-tuning between the terms on the right-hand side in
this expression, each term should individually be of the order of the left-hand side,
\ie,  $M^2_Z$,
and in particular $\mu$ should be as close as possible to $M_Z$.
This leads to a dominantly higgsino LSP and that
also $\XN{2}$ and $\XPM{1}$ are mainly higgsino.
Mass differences within the higgsino sector are small, 
typically  below $20$\,GeV,
depending on the values of the other SUSY parameters, 
in particular on $M_1$ and $M_2$.
The other SUSY particles can be more heavy:
top squarks may range up to $\sim 3$ TeV and gluinos 
up to $\sim 4$ TeV with little cost to naturalness~\cite{Bae:2014yta}. 
Such heavy top squarks and gluinos may well lie beyond the reach of even HL-LHC.

In the clean environment of the ILC, the soft visible
decay products of $\XN{2}$ and $\XPM{1}$ can be easily detected 
--- without any need to rely on large-mass-gap decays
of heavier particles. 
The ILC capabilities 
have been studied in detector simulations performed for different 
benchmark points with mass 
differences ranging from $770$\,MeV~\cite{Berggren:2013vfa}
to $20$\,GeV~\cite{Baer:2016new}. 
Two examples of the striking signals and the extraction of 
kinematic endpoints are given in Fig.~\ref{fig:searches_higgsinos}. 
The resulting precisions on masses and polarised 
cross sections reach the percent level even in the experimentally most difficult 
cases and allow to determine other SUSY parameters.
They will also play an important role in unveiling the nature of dark matter: 
in this case with the result that the LSP only contributes a small fraction of the 
total abundance. Such a situation might call for additional, non-WIMP constituents of 
dark matter such as axions.
\begin{figure*}[]
  \begin{center}
    \subfigure[]{\includegraphics[width=0.45\linewidth] {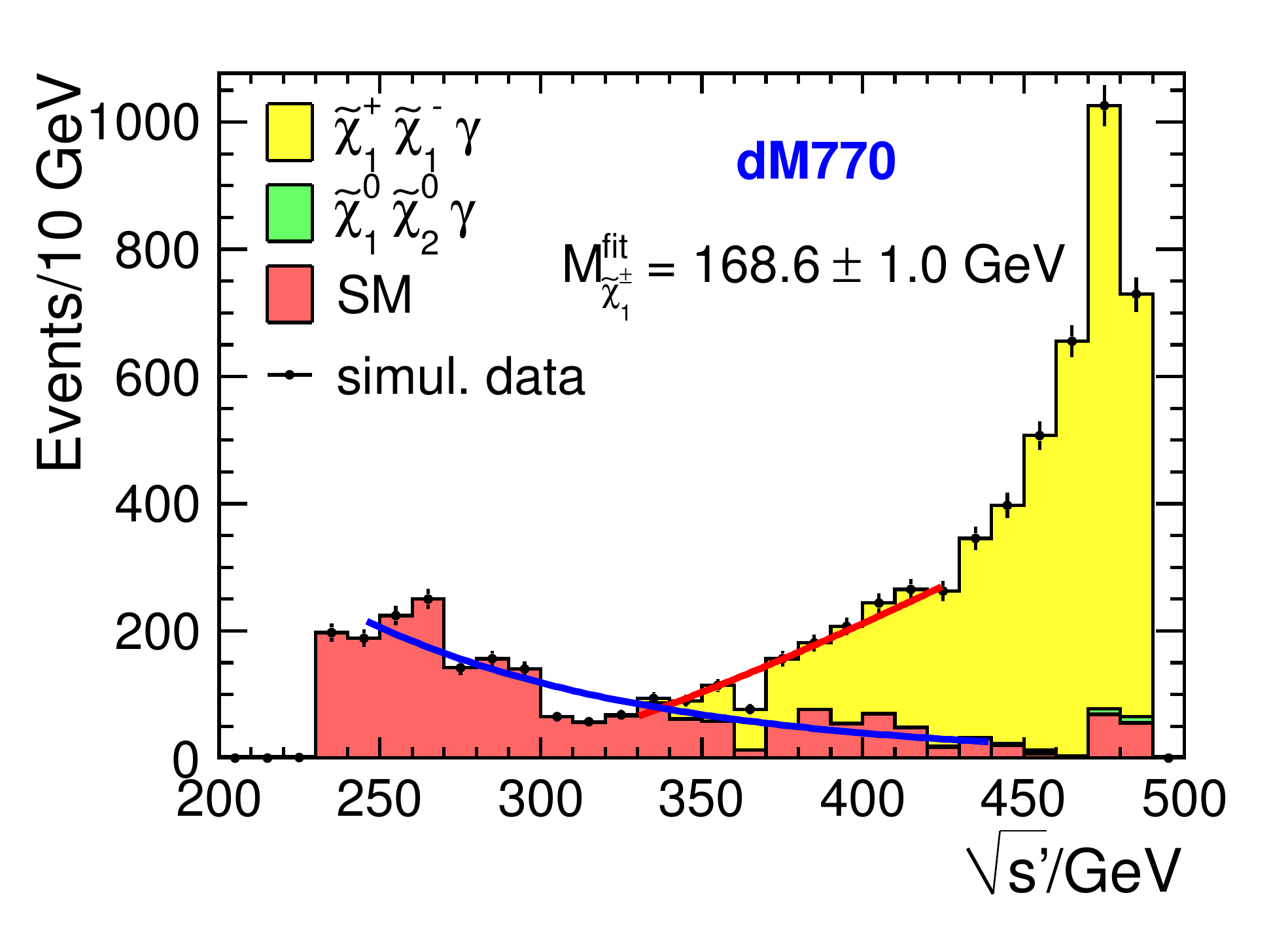}}
    \hspace{0.05\linewidth}
    \subfigure[]{\includegraphics [width=0.45\linewidth]{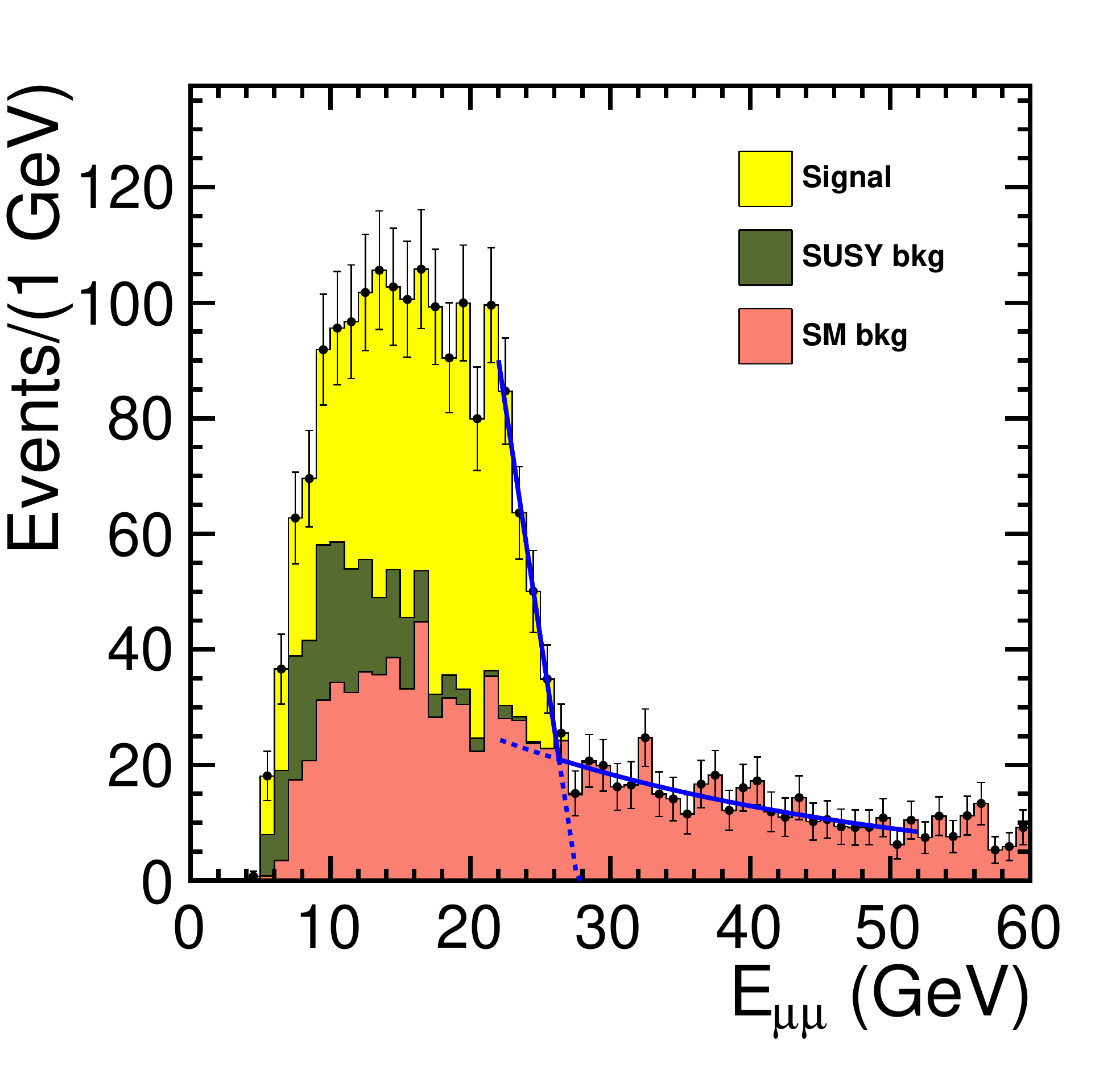}}
  \end{center}
  \caption{\label{fig:searches_higgsinos} Higgsino mass determination for (a) the charged higgsino from the recoil against an ISR photon in a scenario with a mass splitting of $770$\,MeV~\cite{Berggren:2013vfa}, using the SGV detector 
simulation. (b) the neutral higgsino from the energy of its visible decay products in a scenario with a mass splitting of  $10$\,GeV~\cite{Baer:2016new}, using full detector simulation.}
\end{figure*}
In \cite{Baer:2016usl},
it is shown that an ILC operating at 1 TeV
would have guaranteed discovery/exclusion reach over the
entire class of natural SUSY models, which is illustrated
in Fig. \ref{fig:searches_nuhm2_excl} for the example of the NUMH2 model
\cite{Baer:2005bu}.
Only highly fine-tuned, un-natural, models would still be allowed if ILC at 1 TeV
failed to discover SUSY.
No such statement can be expected to come out of HL-LHC:
even though discovery would be possible at HL-LHC,
no guarantee is possible.
\begin{figure*}[]
  \begin{center}
      \includegraphics[width=0.45\linewidth] {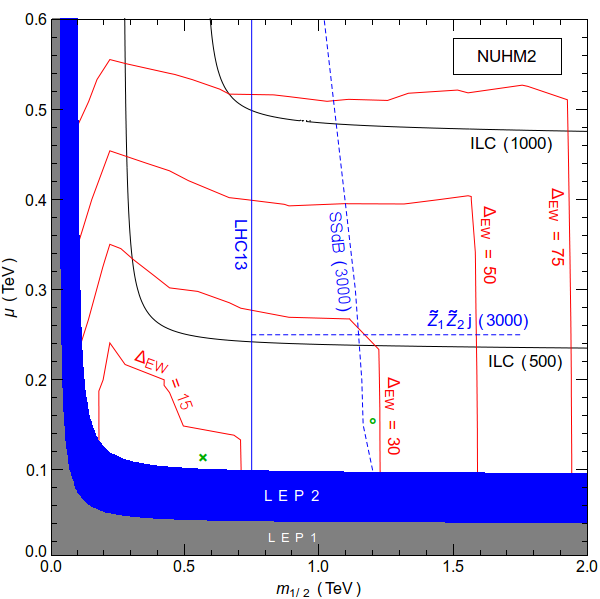}
  \end{center}
  \caption{\label{fig:searches_nuhm2_excl} The $m_{1/2}$ vs. $\mu$ plane in the 
NUHM2 model for $\tan\beta =15$, 
$m_0=5$\,TeV, $A_0=-8$\,TeV and $m_A=1$\,TeV. We show contours of the
naturalness measure $\Delta_{EW}$ \cite{Baer:2015rja}
along with current limits from LHC13 and future reach of HL-LHC and ILC.
Location of benchmark points is indicated in green.} 
\end{figure*}

\subsection{Mono-photon signature}
\label{subsec:searches_monophoton}

The primary probe at the ILC for the direct production of WIMP dark matter are photons
emitted as initial-state radiation in association with the pair production of dark matter.
Such a Mono-photon search is analogous  to Mono-$X$ searches at the LHC.
The main backgrounds to this search are the radiative neutrino production, which is irreducible,
and the radiative Bhabha scattering process, in which the outgoing electron and positron escape 
undetected in the beam pipe.
At LEP, searches for photon events with missing energy were performed~\cite{Abdallah:2003np,*Abdallah:2008aa},
and were later re-analysed within the  effective
operator framework~\cite{Fox:2011fx}\footnote{Note that under LEP or ILC conditions the 
effective field theory approximation is accurate, while it is questionable
in similar analyses at hadron colliders. 
}.

The prospects to detect WIMPs with such methods at the ILC and to determine their properties 
have been studied
for a centre-of-mass energy of $500$\,GeV
in detailed detector simulation~\cite{Bartels:2012ex,Habermehl:417605}. 
Also at the ILC, the experimental sensitivity have been interpreted 
in the framework of effective
operators.
Figure~\ref{fig:searches_WIMPs}a shows the exclusion reach found, and 
Fig.~\ref{fig:searches_WIMPs}b shows the extrapolation of these
results to a wide range of integrated luminosities and centre-of-mass energies 
~\cite{Habermehl:417605}.
For the full $500$\,GeV-program of the ILC, scales of new physics ($\Lambda$) 
of up to $3$\,TeV  can be probed,
while the $1$\,TeV-energy-upgrade of the ILC would extend this even 
to $4.5$\,TeV or more, 
depending on the integrated luminosity.
At 250 GeV, 
the full reach will be attained already at a modest integrated luminosity.

If a WIMP would be discovered, 
its properties could be determined precisely due to the known initial
state of a lepton collider~\cite{Bartels:2012ex}. 
In particular, 
its mass could be determined with a precision of about 1\%, and the type of operator 
(or the angular momentum of dominant partial wave) of the WIMP pair production 
process can be determined. 
By such detailed measurements of WIMP properties as offered at the ILC, 
it is often possible to constrain WIMP
production rates in the early universe along with WIMP scattering or annihilation
rates and the local WIMP abundance~\cite{Baltz:2006fm}. 
Such checks could verify or falsify the simple assumptions 
associated with thermal DM production within the WIMP miracle scenario, 
thus giving important insights into the nature of dark matter.

Searches for WIMP dark matter at the ILC are highly complementary to those
at hadron colliders and at direct detection experiments: 
as an electron-positron collider, 
ILC is sensitive to WIMP couplings to electrons, 
whereas hadron colliders and direct detection experiments are sensitive to WIMP 
couplings to quarks. 
Depending on the
type of particle mediating the WIMP-SM interaction, 
there is a priori no reason for these couplings to be of similar strength.
Thus, 
if the LHC does not discover a deviation from the SM expectation 
in its Mono-$X$ searches, 
it is essential to complement the picture by probing 
the WIMP-lepton couplings at an electron-positron collider.
Moreover, 
while LHC can probe larger WIMP masses due to its higher centre-of-mass energy, 
ILC can probe smaller couplings, thus higher energy scales for the 
WIMP-electron interaction due to its higher precision.

\begin{figure*}[]
\setlength{\unitlength}{1.0cm}
\subfigure[]{\includegraphics[width=0.45\linewidth]{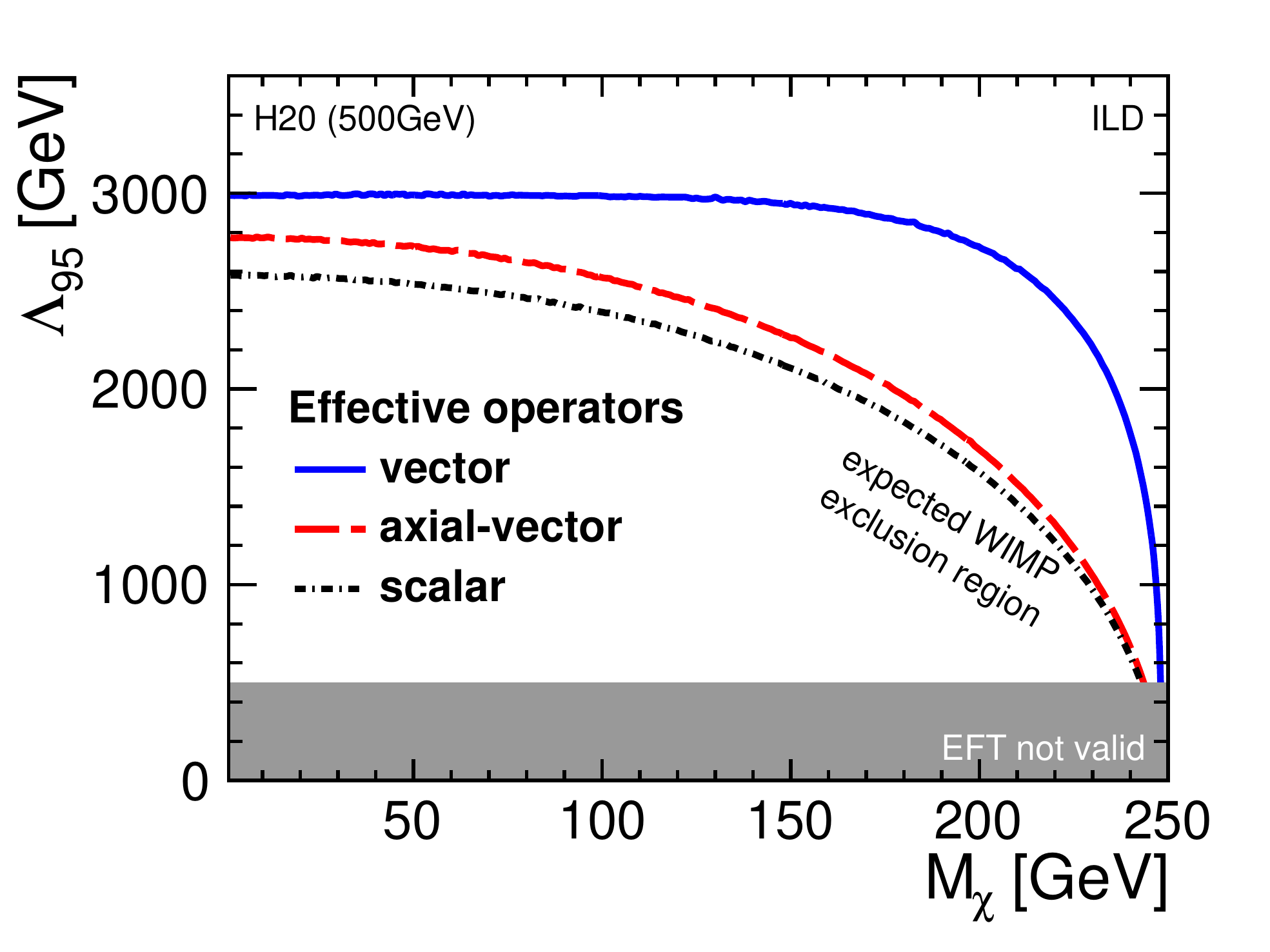}}
\hspace{0.05cm}
\subfigure[]{\includegraphics[width=0.45\linewidth]{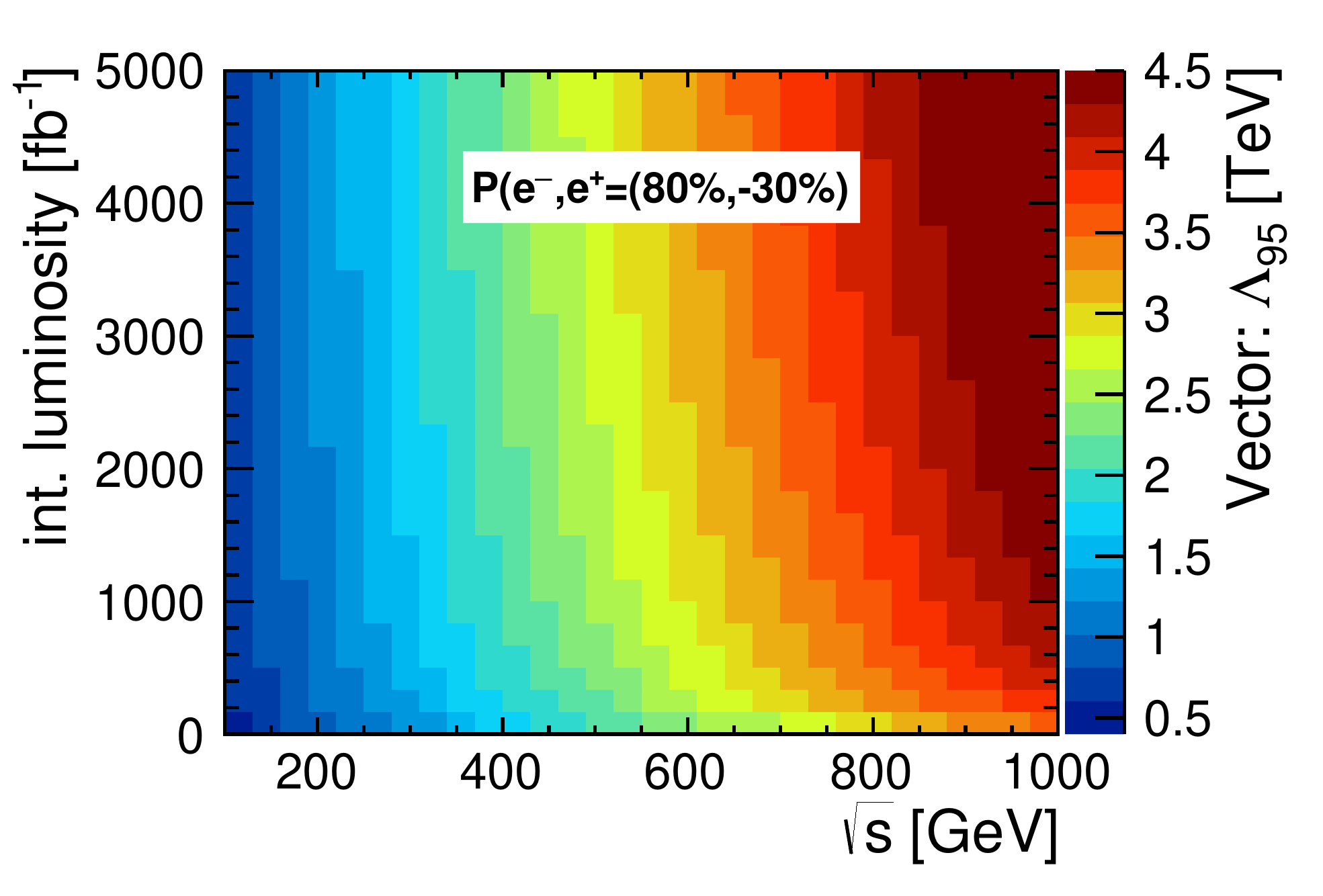}}
\caption{\label{fig:searches_WIMPs} Left: Observational reach ($3\sigma$) of the ILC for a Spin-1 
  WIMP in terms of WIMP
  mass and $\kappa_e$ for three different chiralities of the WIMP-fermion couplings
   Right: Expected sensitivity for a vector operator in an EFT-based interpretation as a function of integrated
  luminosity and centre-of-mass energy~\cite{Habermehl:417605}.}
\end{figure*}

\subsection{New-scalar signatures}
\label{subsec:searches_newscalars}

In many models with extended Higgs sectors, \eg,
Two Higgs Doublet Models, The Next-to-Minimal Supersymmetric  Standard  Model
and  Randall  Sundrum  models,  there  exists  a  light  scalar
$S^0$,
lighter than the Standard Model Higgs.
The coupling of the $S^0$ to the $Z$ can be very small,
compared to the coupling a standard model Higgs with the
same mass would have
to the $Z$.
Such a light scalar with suppressed couplings to the Z boson would
have escaped detection at LEP.
With a factor of 1000 higher luminosity and polarised beams,
the ILC is expected to have substantial discovery potential for
this kind of states.
Furthermore,
searches for additional scalars at LEP and LHC are usually dependent on the
model details,
in particular on the decay branching ratios of the new scalar.
Thus, to be able to search for such new states,
it is paramount to have a more general analysis without
model-dependent assumptions.
The recoil-mass technique,
in particular with the Z boson decaying into a pair of leptons,
offers the possibility to achieve this.

The OPAL collaboration at LEP searched
for light scalars with this method,
but the  results were limited due to the low luminosity~\cite{Abbiendi:2002in}.
The large luminosity offered by the ILC 
makes the recoil mass technique correspondingly more powerful~\cite{Asner:2013psa}
Therefore a search for a light scalar with a very weak interaction with
the Z boson using the model-independent analysis would become  viable
at the ILC-250.

A study was performed using the full GEANT4-based simulation of the
ILD concept.
As a preliminary result~\cite{yanichep},
exclusion cross-section limits for
masses of the new scalar between 10 and 120
GeV are given in terms of a scale factor $k$ with respect to the
cross-section of the Standard Model Higgsstrahlung process would
have had, would the Higgs-mass have been the one assumed for the new scalar.

Background events are rejected by considering kinematic variables
only relied on muons and the reconstructed $Z$:
The invariant mass, transverse momentum and polar angle of the muon pair,
as well as the polar angle of the missing momentum,
and the polar angle of each muon, and the angle between them.
Thus, no information on the decay of $S^0$ is used,
and the results will indeed be model-independent.
The recoil mass distributions obtained after applying the cuts
are shown in Fig.~\ref{fig:searches_newscalars}a, 
for a number of hypotheses
on the mass of $S^0$, and $k=1$.
\begin{figure*}[]
\setlength{\unitlength}{1.0cm}
\subfigure[]{\includegraphics[width=0.35\linewidth]{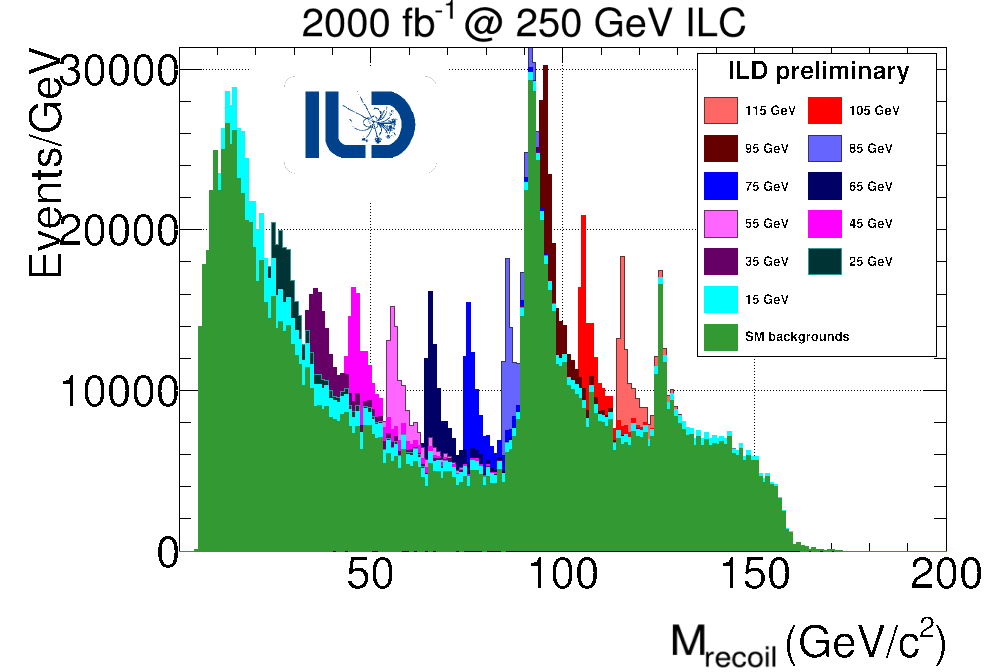}}
\hspace{0.0\linewidth}
\subfigure[]{\includegraphics[width=0.3\linewidth]{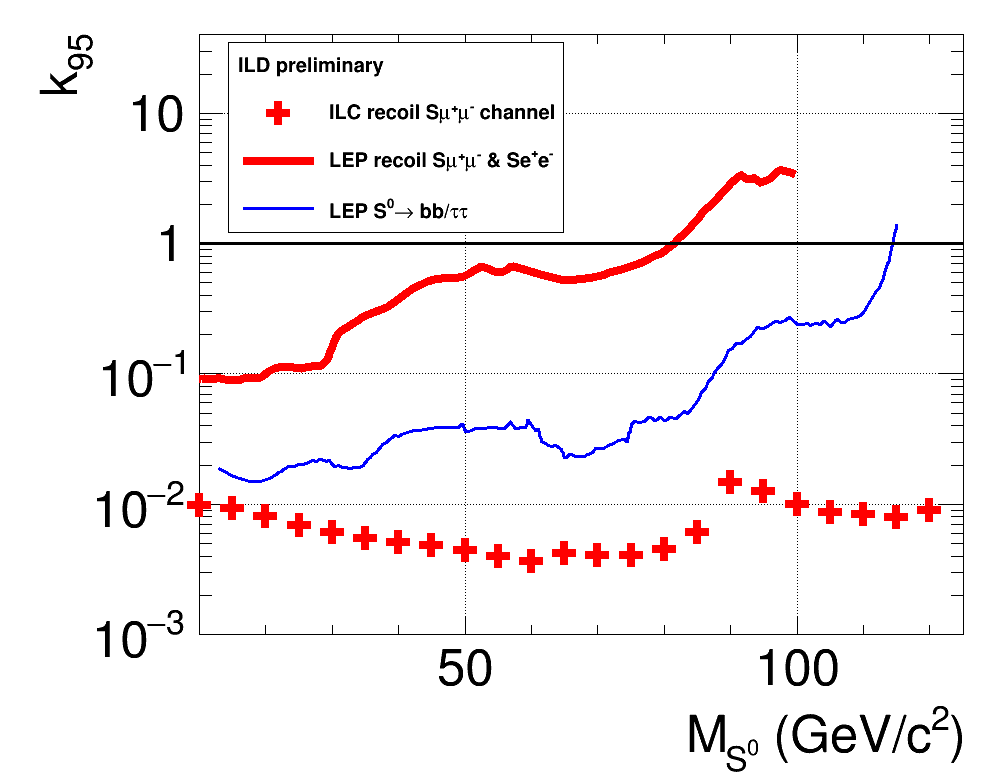}}
\hspace{0.0\linewidth}
\subfigure[]{\includegraphics[width=0.3\linewidth]{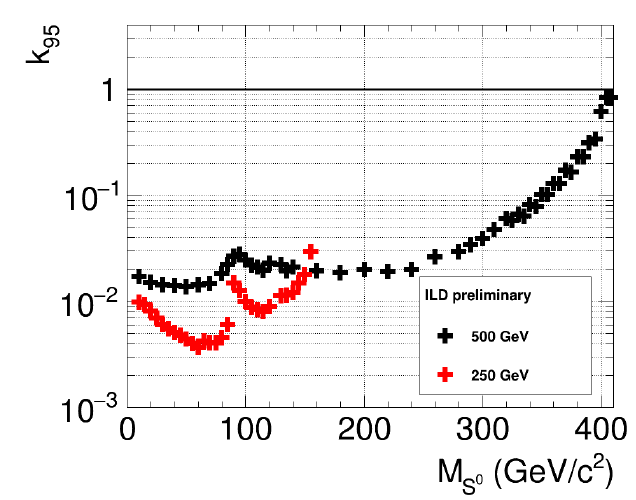}}
\caption{\label{fig:searches_newscalars} (a) The recoil mass distributions for various signals
and all backgrounds after the cuts From \cite{yanichep}.
(b) The 2 $\sigma$ exclusion limits for the cross section scale factor $k$ comparing the LEP and
ILC-250 results. From \cite{yanichep}.
(b) The 2 $\sigma$ exclusion limits for the cross section scale factor $k$ comparing for ILC 250 and ILC-500. From \cite{yanlcws18}.}
\end{figure*}

The main backgrounds depend on the scalar mass.
In
the small mass region, the two fermions background
$\eeto~\mu^+ \mu^-$ with an energetic ISR
photon is the overwhelming background;
while in the $Z$-pole region,
$\eeto~ZZ \rightarrow \mu^+ \mu^- + X$
is an irreducible background,
as is - obviously - $\eeto~Z^* \rightarrow ZH \rightarrow \mu^+\mu^- H$
at $M_{S^0} \sim M_H$.
The two fermion background can be
further rejected by taking into account ISR photon return effects.
The ISR photon veto cuts are applied to the ISR photons in the
centre region and forward region, separately.

The obtained 2 $\sigma$ expected exclusion limits for the
cross section scale factor
$k_{95}$ are shown for scalar mass from 10 \GeV to 120 \GeV
in  Fig.~\ref{fig:searches_newscalars}b.
It is one to two orders of magnitude more sensitive than LEP, and
covering substantial new phase space.
In particular,
at all studied points, a new scalar with a coupling to the $Z$ greater
than 1\% of that of a SM-higgs at the same mass would be excluded or
discovered at ILC-250.
Preliminary studies indicate that an ILC operating at 500 GeV
could discover or exclude such a scalar with a mass up to 350 GeV,
even if the coupling is only one tenth of the coupling of a
would-be SM Higgs at the same mass \cite{yanlcws18}


\section{\label{sec:conclusion}Conclusion}


In this report, we have reviewed the full panorama of the
International Linear  Collider project.   

This machine addresses
compelling physics questions.   In
our quest to discover new interactions beyond the Standard Model, the
couplings of the Higgs boson are the most obvious place to look, and
the one place where today we are not looking with sufficient power. 
The ILC  will supply the capabilities that
we need to study the Higgs boson and other particles with the degree
of precision that is actually required to learn their secrets. 

The ILC provides a fully formed project proposal with a total cost
estimate similar to that of the LHC, a moderate time scale for its
construction, and well-tested technologies for its accelerator and
detector designs.    The ILC is designed as a staged machine with its
first stage at 250~GeV.   Its design includes 
straightforward upgrade paths to extend this  initial configuration. The full ILC machine 
will be capable of running at center-of-mass energies from the $Z$
pole to 1~TeV, covering the production thresholds of:
 $Z$, $WW$, $Z$-Higgs, top-quark, top-Higgs and Higgs pair
 production.  The choice of energy can respond flexibly to new
 discoveries in particle physics, at the LHC, at the ILC, or at other
 facilities.

The   ILC detectors are designed to meet the challenges of high precision.
Taking advantage of the more benign environment of $\ee$ colliders,
they are designed for performance on charged-particle tracking,
heavy-flavor identification, and calorimetry that improve on
existing detectors by large factors.  These are essential capabilities
to confidently obtain the high-quality measurements that we seek. 

For first time in a collider for particle physics, the detectors will
 operate without any trigger system. Physics analyses and
 data-acquisition architecture will considerably benefit from this
 fact and consequently simplify with respect to present and past
 experiments. The overall computing costs should be an order of
 magnitude
smaller than those for the LHC.  The software and computing tools so far developed
 provide physics simulations and detector studies that give  solid
 predictions for the ILC performance toward its physics goals.  We
 have presented the results of those studies in this document.

The ILC will use polarised electron and positron beams. Beam polarisation brings both
 quantitative and qualitative advantages with respect to unpolarized
 $\ee$ colliders.   Polarisation enhances signal reactions and allows
 the measurement of
helicity-dependent observables, multiplying the physics
 output per unit of luminosity. It also allows suppression of
 backgrounds and accurate control of systematic errors, improving the
 robustness of high-precision measurements. 
 
The ILC machine is ready for construction.   We have described the
detailed design of the ILC, explaining how the performance of  each
component is supported by prototyping and, in most cases, by operational/industrial
experience.

The physics program begins with a stage at 250~GeV in the center of
mass.  At this stage, each Higgs boson is produced together with a $Z$ boson
at 110~GeV lab energy that serves to tag the event.  This allows
unambiguous,
model-independent measurements of the total cross section for Higgs
boson production and  the branching ratios for Higgs boson  decays.
It also gives a
tool for searches for exotic Higgs boson decays, including decays to
invisible or partially visible final states.

Measurements at the 250~GeV stage will improve current
measurements of $W$ boson couplings and SM quark and lepton couplings
by large factors beyond what is possible at the HL-LHC.

The simplicity of $\ee$ pair production allows the full set of
electroweak and Higgs processes at the ILC to be combined in a global
fit based on an Effective Field Theory description of modifications of
the Standard Model.   This framework is essentially model-independent
with respect to new physics.  Within this framework, the 250~GeV stage
of the ILC will measure the $Hbb$ couplings to 1\%, the
$HWW$ and $HZZ$ to 0.7\%, and all other important Higgs boson
couplings to levels close to 1\%.   These are the levels of precision
required to access new physics beyond the reach of direct searches at HL-LHC.

The first-stage ILC is intrinsically upgradable in energy and luminosity.
The accelerator and detectors are designed for operation up to a
center of mass energy of 1~TeV.   The technology, detector
performance, and physics for the 500~GeV stage has been described in
detail.  All of the measurements discussed in the previous paragraphs
benefit,
with the uncertainties in Higgs boson couplings decreasing by a factor
of 2.    The 500~GeV stage offers a program exploring 
the couplings of the top quark and thus a second, independent,
opportunity to probe for new physics through precision measurement.
It also offers the opportunity to search for pair-production of
elusive particles produced in electroweak interactions that are
challenging  to discover at the LHC. Running at the $Z$ pole
(ILC-GigaZ)
 is also possible. Essential observables implying
 lepton or quark left-right asymmetries can be 
measured with extremely high accuracy when 
combined with other measurements above the $Z$. 

The opportunities that the ILC gives to discover new physics are
robust, and the ILC measurements are improvable as the accelerator
moves from one energy stage to the next.

We have compared the projected ILC performance on Higgs boson
couplings to those put forward for other colliders.  The ILC will
provide a  significant---and necessary---step in precision beyond the
HL-LHC.    A number of other proposals for $\ee$ Higgs factories are
now under discussion.  We have shown that none expects a performance
significantly superior to the ILC even at its 250~GeV stage.   Also,
no other proposal has been designed and costed at the level of a
formal proposal.  Only ILC is on the table today.

 Finally, the ILC laboratory will provide a base for future proposals
 of $\ee$ and $\gamma\gamma$ colliders based on advanced high-gradient
 acceleration.   The ILC laboratory thus can expect a long lifetime,
 beyond our current horizon, in which it will continue to explore the
 frontier of fundamental physics.

Come join us!   It is time to make the International Linear Collider a
reality.

\bigskip 

\bigskip

\noindent {\bf Acknowledgements}

We are grateful to many funding agencies around the world for
supporting the research reported here and the preparation of this
document.  Among these are: the Deutsche Forschungsgemeinschaft
 (DFG, German Research 
Foundation) under Germany‘s Excellence Strategy
 – EXC 2121 "Quantum Universe" – 390833306, the Generalitat Valenciana under grant 
PROMETEO 2018/060 and the Spanish national program for particle 
physics under grant FPA2015-65652-C4-3-R, the Japan Society for the
 Promotion of Science (JSPS) under
Grants-in-Aid for Science Research 15H02083, 16H02173,  and 16H02176, and 
 the  US Department of Energy under   
 contracts DE--AC02--76SF00515, DE--AC05--06OR23177, and  DE--SC0017996.

\bibliography{ILC-supplement-refs,chapters/runscen-refs,chapters/machine-refs,chapters/software-refs,chapters/highenergy-refs,chapters/searches-refs}

\begin{thebibliography}{331}%
\makeatletter
\providecommand \@ifxundefined [1]{%
 \@ifx{#1\undefined}
}%
\providecommand \@ifnum [1]{%
 \ifnum #1\expandafter \@firstoftwo
 \else \expandafter \@secondoftwo
 \fi
}%
\providecommand \@ifx [1]{%
 \ifx #1\expandafter \@firstoftwo
 \else \expandafter \@secondoftwo
 \fi
}%
\providecommand \natexlab [1]{#1}%
\providecommand \enquote  [1]{``#1''}%
\providecommand \bibnamefont  [1]{#1}%
\providecommand \bibfnamefont [1]{#1}%
\providecommand \citenamefont [1]{#1}%
\providecommand \href@noop [0]{\@secondoftwo}%
\providecommand \href [0]{\begingroup \@sanitize@url \@href}%
\providecommand \@href[1]{\@@startlink{#1}\@@href}%
\providecommand \@@href[1]{\endgroup#1\@@endlink}%
\providecommand \@sanitize@url [0]{\catcode `\\12\catcode `\$12\catcode
  `\&12\catcode `\#12\catcode `\^12\catcode `\_12\catcode `\%12\relax}%
\providecommand \@@startlink[1]{}%
\providecommand \@@endlink[0]{}%
\providecommand \url  [0]{\begingroup\@sanitize@url \@url }%
\providecommand \@url [1]{\endgroup\@href {#1}{\urlprefix }}%
\providecommand \urlprefix  [0]{URL }%
\providecommand \Eprint [0]{\href }%
\providecommand \doibase [0]{http://dx.doi.org/}%
\providecommand \selectlanguage [0]{\@gobble}%
\providecommand \bibinfo  [0]{\@secondoftwo}%
\providecommand \bibfield  [0]{\@secondoftwo}%
\providecommand \translation [1]{[#1]}%
\providecommand \BibitemOpen [0]{}%
\providecommand \bibitemStop [0]{}%
\providecommand \bibitemNoStop [0]{.\EOS\space}%
\providecommand \EOS [0]{\spacefactor3000\relax}%
\providecommand \BibitemShut  [1]{\csname bibitem#1\endcsname}%
\let\auto@bib@innerbib\@empty
\bibitem [{\citenamefont {Behnke}\ \emph {et~al.}(2013)\citenamefont {Behnke},
  \citenamefont {Brau}, \citenamefont {Foster}, \citenamefont {Fuster},
  \citenamefont {Harrison}, \citenamefont {Paterson}, \citenamefont {Peskin},
  \citenamefont {Stanitzki}, \citenamefont {Walker},\ and\ \citenamefont
  {Yamamoto}}]{Behnke:2013xla}%
  \BibitemOpen
  \bibfield  {author} {\bibinfo {author} {\bibfnamefont {T.}~\bibnamefont
  {Behnke}}, \bibinfo {author} {\bibfnamefont {J.~E.}\ \bibnamefont {Brau}},
  \bibinfo {author} {\bibfnamefont {B.}~\bibnamefont {Foster}}, \bibinfo
  {author} {\bibfnamefont {J.}~\bibnamefont {Fuster}}, \bibinfo {author}
  {\bibfnamefont {M.}~\bibnamefont {Harrison}}, \bibinfo {author}
  {\bibfnamefont {J.~M.}\ \bibnamefont {Paterson}}, \bibinfo {author}
  {\bibfnamefont {M.}~\bibnamefont {Peskin}}, \bibinfo {author} {\bibfnamefont
  {M.}~\bibnamefont {Stanitzki}}, \bibinfo {author} {\bibfnamefont
  {N.}~\bibnamefont {Walker}}, \ and\ \bibinfo {author} {\bibfnamefont
  {H.}~\bibnamefont {Yamamoto}},\ }\href@noop {} {\  (\bibinfo {year}
  {2013})},\ \Eprint {http://arxiv.org/abs/1306.6327} {arXiv:1306.6327
  [physics.acc-ph]} \BibitemShut {NoStop}%
\bibitem [{\citenamefont {Evans}\ and\ \citenamefont
  {Michizono}(2017)}]{Evans:2017rvt}%
  \BibitemOpen
  \bibfield  {author} {\bibinfo {author} {\bibfnamefont {L.}~\bibnamefont
  {Evans}}\ and\ \bibinfo {author} {\bibfnamefont {S.}~\bibnamefont
  {Michizono}} (\bibinfo {collaboration} {Linear Collider Collaboration}),\
  }\href@noop {} {\  (\bibinfo {year} {2017})},\ \Eprint
  {http://arxiv.org/abs/1711.00568} {arXiv:1711.00568 [physics.acc-ph]}
  \BibitemShut {NoStop}%
\bibitem [{\citenamefont {Barklow}\ \emph
  {et~al.}(2018{\natexlab{a}})\citenamefont {Barklow}, \citenamefont {Fujii},
  \citenamefont {Jung}, \citenamefont {Karl}, \citenamefont {List},
  \citenamefont {Ogawa}, \citenamefont {Peskin},\ and\ \citenamefont
  {Tian}}]{Barklow:2017suo}%
  \BibitemOpen
  \bibfield  {author} {\bibinfo {author} {\bibfnamefont {T.}~\bibnamefont
  {Barklow}}, \bibinfo {author} {\bibfnamefont {K.}~\bibnamefont {Fujii}},
  \bibinfo {author} {\bibfnamefont {S.}~\bibnamefont {Jung}}, \bibinfo {author}
  {\bibfnamefont {R.}~\bibnamefont {Karl}}, \bibinfo {author} {\bibfnamefont
  {J.}~\bibnamefont {List}}, \bibinfo {author} {\bibfnamefont {T.}~\bibnamefont
  {Ogawa}}, \bibinfo {author} {\bibfnamefont {M.~E.}\ \bibnamefont {Peskin}}, \
  and\ \bibinfo {author} {\bibfnamefont {J.}~\bibnamefont {Tian}},\ }\href
  {\doibase 10.1103/PhysRevD.97.053003} {\bibfield  {journal} {\bibinfo
  {journal} {Phys. Rev.}\ }\textbf {\bibinfo {volume} {D97}},\ \bibinfo {pages}
  {053003} (\bibinfo {year} {2018}{\natexlab{a}})},\ \Eprint
  {http://arxiv.org/abs/1708.08912} {arXiv:1708.08912 [hep-ph]} \BibitemShut
  {NoStop}%
\bibitem [{\citenamefont {Fujii}\ \emph
  {et~al.}(2017{\natexlab{a}})\citenamefont {Fujii} \emph
  {et~al.}}]{Fujii:2017vwa}%
  \BibitemOpen
  \bibfield  {author} {\bibinfo {author} {\bibfnamefont {K.}~\bibnamefont
  {Fujii}} \emph {et~al.},\ }\href@noop {} {\  (\bibinfo {year}
  {2017}{\natexlab{a}})},\ \Eprint {http://arxiv.org/abs/1710.07621}
  {arXiv:1710.07621 [hep-ex]} \BibitemShut {NoStop}%
\bibitem [{\citenamefont {Abramowicz}\ \emph {et~al.}(2013)\citenamefont
  {Abramowicz} \emph {et~al.}}]{Behnke:2013lya}%
  \BibitemOpen
  \bibfield  {author} {\bibinfo {author} {\bibfnamefont {H.}~\bibnamefont
  {Abramowicz}} \emph {et~al.},\ }\bibfield  {booktitle} {\emph {\bibinfo
  {booktitle} {{The International Linear Collider Technical Design Report -
  Volume 4: Detectors}}},\ }\href@noop {} {\  (\bibinfo {year} {2013})},\
  \Eprint {http://arxiv.org/abs/1306.6329} {arXiv:1306.6329 [physics.ins-det]}
  \BibitemShut {NoStop}%
\bibitem [{\citenamefont {Asai}\ \emph {et~al.}(2017)\citenamefont {Asai},
  \citenamefont {Tanaka}, \citenamefont {Ushiroda}, \citenamefont {Nakao},
  \citenamefont {Tian}, \citenamefont {Kanemura}, \citenamefont {Matsumoto},
  \citenamefont {Shirai}, \citenamefont {Endo},\ and\ \citenamefont
  {Kakizaki}}]{Asai:2017pwp}%
  \BibitemOpen
  \bibfield  {author} {\bibinfo {author} {\bibfnamefont {S.}~\bibnamefont
  {Asai}}, \bibinfo {author} {\bibfnamefont {J.}~\bibnamefont {Tanaka}},
  \bibinfo {author} {\bibfnamefont {Y.}~\bibnamefont {Ushiroda}}, \bibinfo
  {author} {\bibfnamefont {M.}~\bibnamefont {Nakao}}, \bibinfo {author}
  {\bibfnamefont {J.}~\bibnamefont {Tian}}, \bibinfo {author} {\bibfnamefont
  {S.}~\bibnamefont {Kanemura}}, \bibinfo {author} {\bibfnamefont
  {S.}~\bibnamefont {Matsumoto}}, \bibinfo {author} {\bibfnamefont
  {S.}~\bibnamefont {Shirai}}, \bibinfo {author} {\bibfnamefont
  {M.}~\bibnamefont {Endo}}, \ and\ \bibinfo {author} {\bibfnamefont
  {M.}~\bibnamefont {Kakizaki}},\ }\href@noop {} {\  (\bibinfo {year}
  {2017})},\ \Eprint {http://arxiv.org/abs/1710.08639} {arXiv:1710.08639
  [hep-ex]} \BibitemShut {NoStop}%
\bibitem [{Adv(2018)}]{AdvPanel}%
  \BibitemOpen
  \href@noop {} {\enquote {\bibinfo {title} {{Summary of the ILC Advisory
  Panel's Discussions}},}\ }\bibinfo {howpublished}
  {\url{http://www.mext.go.jp/component/b\_menu/shingi/
  toushin/\__icsFiles/afieldfile/2018/09/20/1409220\_2\_1.pdf"}} (\bibinfo
  {year} {2018})\BibitemShut {NoStop}%
\bibitem [{\citenamefont {{{Science Council of Japan}}}(2012)}]{SJCreport}%
  \BibitemOpen
  \bibfield  {author} {\bibinfo {author} {\bibnamefont {{{Science Council of
  Japan}}}},\ }\href@noop {} {\enquote {\bibinfo {title} {Report on the
  international linear collider},}\ }\bibinfo {howpublished}
  {{\url{http://www.scj.go.jp/ja/info/kohyo/pdf/kohyo-24-k273.pdf}}; English
  translation of the executive summary at
  \url{http://newsline.linearcollider.org/2018/12/21/executive-summary-of-the-science-council-of-}
  {japans-report/}} (\bibinfo {year} {2012})\BibitemShut {NoStop}%
\bibitem [{\citenamefont {Aihara}\ \emph {et~al.}(2019)\citenamefont {Aihara}
  \emph {et~al.}}]{Aihara:2019gcq}%
  \BibitemOpen
  \bibfield  {author} {\bibinfo {author} {\bibfnamefont {H.}~\bibnamefont
  {Aihara}} \emph {et~al.} (\bibinfo {collaboration} {ILC}),\ }\href@noop {} {\
   (\bibinfo {year} {2019})},\ \Eprint {http://arxiv.org/abs/1901.09829}
  {arXiv:1901.09829 [hep-ex]} \BibitemShut {NoStop}%
\bibitem [{\citenamefont {Harrison}\ \emph
  {et~al.}(2013{\natexlab{a}})\citenamefont {Harrison}, \citenamefont {Ross},\
  and\ \citenamefont {Walker}}]{Harrison:2013nvaxxx}%
  \BibitemOpen
  \bibfield  {author} {\bibinfo {author} {\bibfnamefont {M.}~\bibnamefont
  {Harrison}}, \bibinfo {author} {\bibfnamefont {M.}~\bibnamefont {Ross}}, \
  and\ \bibinfo {author} {\bibfnamefont {N.}~\bibnamefont {Walker}},\
  }\bibfield  {booktitle} {\emph {\bibinfo {booktitle} {{Proceedings, 2013
  Community Summer Study on the Future of U.S. Particle Physics: Snowmass on
  the Mississippi (CSS2013): Minneapolis, MN, USA, Jul 29-Aug 6, 2013}}},\
  }\href@noop {} {\  (\bibinfo {year} {2013}{\natexlab{a}})},\ \Eprint
  {http://arxiv.org/abs/1308.3726} {arXiv:1308.3726 [physics.acc-ph]}
  \BibitemShut {NoStop}%
\bibitem [{\citenamefont {Asai}\ \emph {et~al.}(2012)\citenamefont {Asai} \emph
  {et~al.}}]{JAHEP:2012a}%
  \BibitemOpen
  \bibfield  {author} {\bibinfo {author} {\bibfnamefont {S.}~\bibnamefont
  {Asai}} \emph {et~al.} (\bibinfo {collaboration} {JAHEP Subcommittee on
  Future Projects of High Energy Physics}),\ }\href
  {http://www.jahep.org/office/doc/201202_hecsubc_report.pdf} {\enquote
  {\bibinfo {title} {The final report of the subcommittee on future projects of
  high energy physics},}\ }\bibinfo {howpublished}
  {\url{http://www.jahep.org/office/doc/201202_hecsubc_report.pdf}} (\bibinfo
  {year} {2012})\BibitemShut {NoStop}%
\bibitem [{\citenamefont {{Japan Association of High Energy Physicists
  (JAHEP)}}(2012)}]{JAHEP:2012b}%
  \BibitemOpen
  \bibfield  {author} {\bibinfo {author} {\bibnamefont {{Japan Association of
  High Energy Physicists (JAHEP)}}},\ }\href
  {http://www.jahep.org/office/doc/201210_ILC_staging_e.pdf} {\enquote
  {\bibinfo {title} {A proposal for a phased execution of the international
  linear collider project},}\ }\bibinfo {howpublished}
  {\url{http://www.jahep.org/office/doc/201210_ILC_staging_e.pdf}} (\bibinfo
  {year} {2012})\BibitemShut {NoStop}%
\bibitem [{\citenamefont {Sanuki}(2017)}]{bib:sanuki:desy2017}%
  \BibitemOpen
  \bibfield  {author} {\bibinfo {author} {\bibfnamefont {T.}~\bibnamefont
  {Sanuki}},\ }\href@noop {} {\enquote {\bibinfo {title} {Overview of the
  topography and geology of the {Kitakami} site for the {ILC}},}\ }\bibinfo
  {howpublished} {{Presentation at DESY, Feb. 14, 2017}} (\bibinfo {year}
  {2017})\BibitemShut {NoStop}%
\bibitem [{\citenamefont {Adolphsen}\ \emph
  {et~al.}(2013{\natexlab{a}})\citenamefont {Adolphsen} \emph
  {et~al.}}]{Adolphsen:2013jya}%
  \BibitemOpen
  \bibfield  {author} {\bibinfo {author} {\bibfnamefont {C.}~\bibnamefont
  {Adolphsen}} \emph {et~al.},\ }\href@noop {} {\  (\bibinfo {year}
  {2013}{\natexlab{a}})},\ \Eprint {http://arxiv.org/abs/1306.6353}
  {arXiv:1306.6353 [physics.acc-ph]} \BibitemShut {NoStop}%
\bibitem [{\citenamefont {Adolphsen}\ \emph
  {et~al.}(2013{\natexlab{b}})\citenamefont {Adolphsen} \emph
  {et~al.}}]{Adolphsen:2013kya}%
  \BibitemOpen
  \bibfield  {author} {\bibinfo {author} {\bibfnamefont {C.}~\bibnamefont
  {Adolphsen}} \emph {et~al.},\ }\href@noop {} {\  (\bibinfo {year}
  {2013}{\natexlab{b}})},\ \Eprint {http://arxiv.org/abs/1306.6328}
  {arXiv:1306.6328 [physics.acc-ph]} \BibitemShut {NoStop}%
\bibitem [{\citenamefont {Heuer}\ \emph {et~al.}(2006)\citenamefont {Heuer}
  \emph {et~al.}}]{Heuer:2006}%
  \BibitemOpen
  \bibfield  {author} {\bibinfo {author} {\bibfnamefont {R.~D.}\ \bibnamefont
  {Heuer}} \emph {et~al.},\ }\href
  {https://icfa.fnal.gov/wp-content/uploads/para-Nov20-final.pdf} {\enquote
  {\bibinfo {title} {Parameters for the linear collider},}\ }\bibinfo
  {howpublished}
  {\url{https://icfa.fnal.gov/wp-content/uploads/para-Nov20-final.pdf}}
  (\bibinfo {year} {2006})\BibitemShut {NoStop}%
\bibitem [{\citenamefont {Dugan}\ \emph {et~al.}(2014)\citenamefont {Dugan},
  \citenamefont {List},\ and\ \citenamefont {Walker}}]{Dugan:2014}%
  \BibitemOpen
  \bibfield  {author} {\bibinfo {author} {\bibfnamefont {M.}~\bibnamefont
  {Dugan}, \bibfnamefont {G.and~Harrison}}, \bibinfo {author} {\bibfnamefont
  {B.}~\bibnamefont {List}}, \ and\ \bibinfo {author} {\bibfnamefont
  {N.}~\bibnamefont {Walker}},\ }\href
  {http://edmsdirect.desy.de/item/D00000001046475} {\enquote {\bibinfo {title}
  {Implications of an energy-phased approach to the realization of the
  {ILC}},}\ }\bibinfo {howpublished}
  {\url{http://edmsdirect.desy.de/item/D00000001046475}} (\bibinfo {year}
  {2014})\BibitemShut {NoStop}%
\bibitem [{\citenamefont {White}(2014)}]{bib:cr-0002}%
  \BibitemOpen
  \bibfield  {author} {\bibinfo {author} {\bibfnamefont {K.}~\bibnamefont
  {Buesser}},\ }\href@noop {} {\enquote {\bibinfo {title} {{Change request
  ILC-CR-0002: Baseline optics to provide for a single $L*$}},}\ }\bibinfo
  {howpublished} {{\url{http://edmsdirect.desy.de/item/D00000001100205}}}
  (\bibinfo {year} {2014})\BibitemShut {NoStop}%
\bibitem [{\citenamefont {Buesser}(2014)}]{bib:cr-0003}%
  \BibitemOpen
  \bibfield  {author} {\bibinfo {author} {\bibfnamefont {K.}~\bibnamefont
  {Buesser}},\ }\href@noop {} {\enquote {\bibinfo {title} {{Change request
  ILC-CR-0003: Detector hall with vertical shaft access}},}\ }\bibinfo
  {howpublished} {{\url{http://edmsdirect.desy.de/item/D00000001084745}}}
  (\bibinfo {year} {2014})\BibitemShut {NoStop}%
\bibitem [{\citenamefont {Paterson}\ \emph {et~al.}(2015)\citenamefont
  {Paterson}, \citenamefont {Kuchler}, \citenamefont {Solyak},\ and\
  \citenamefont {Sanami}}]{bib:cr-0012}%
  \BibitemOpen
  \bibfield  {author} {\bibinfo {author} {\bibfnamefont {E.}~\bibnamefont
  {Paterson}}, \bibinfo {author} {\bibfnamefont {V.}~\bibnamefont {Kuchler}},
  \bibinfo {author} {\bibfnamefont {N.}~\bibnamefont {Solyak}}, \ and\ \bibinfo
  {author} {\bibfnamefont {T.}~\bibnamefont {Sanami}},\ }\href@noop {}
  {\enquote {\bibinfo {title} {{Change request ILC-CR-0012: Reduction of width
  of linac shield wall and tunnel cross--section}},}\ }\bibinfo {howpublished}
  {{\url{http://edmsdirect.desy.de/item/D00000001127835}}} (\bibinfo {year}
  {2015})\BibitemShut {NoStop}%
\bibitem [{\citenamefont {Yokoya}\ \emph {et~al.}(2016)\citenamefont {Yokoya},
  \citenamefont {List},\ and\ \citenamefont {Paterson}}]{bib:cr-0013}%
  \BibitemOpen
  \bibfield  {author} {\bibinfo {author} {\bibfnamefont {K.}~\bibnamefont
  {Yokoya}}, \bibinfo {author} {\bibfnamefont {B.}~\bibnamefont {List}}, \ and\
  \bibinfo {author} {\bibfnamefont {E.}~\bibnamefont {Paterson}},\ }\href@noop
  {} {\enquote {\bibinfo {title} {{Change request ILC-CR-0013: Update of the
  ILC beam dump specifications}},}\ }\bibinfo {howpublished}
  {{\url{http://edmsdirect.desy.de/item/D00000001145035}}} (\bibinfo {year}
  {2016})\BibitemShut {NoStop}%
\bibitem [{\citenamefont {Yokoya}(2017)}]{bib:cr-0016}%
  \BibitemOpen
  \bibfield  {author} {\bibinfo {author} {\bibfnamefont {K.}~\bibnamefont
  {Yokoya}},\ }\href@noop {} {\enquote {\bibinfo {title} {{Change request
  ILC-CR-0016: Luminosity improvement at 250 GeV}},}\ }\bibinfo {howpublished}
  {{\url{http://edmsdirect.desy.de/item/D00000001159725}}} (\bibinfo {year}
  {2017})\BibitemShut {NoStop}%
\bibitem [{\citenamefont {{Positron Working Group}}(2018)}]{PWG:2018a}%
  \BibitemOpen
  \bibfield  {author} {\bibinfo {author} {\bibnamefont {{Positron Working
  Group}}},\ }\href@noop {} {\enquote {\bibinfo {title} {{Report on the ILC
  positron source}},}\ }\bibinfo {howpublished}
  {\url{http://edmsdirect.desy.de/item/D00000001165115}} (\bibinfo {year}
  {2018})\BibitemShut {NoStop}%
\bibitem [{\citenamefont {{International Committee for Future Accelerators
  (ICFA)}}(2017)}]{ICFA:2017}%
  \BibitemOpen
  \bibfield  {author} {\bibinfo {author} {\bibnamefont {{International
  Committee for Future Accelerators (ICFA)}}},\ }\href
  {https://icfa.fnal.gov/wp-content/uploads/ICFA-Statement-Nov2017.pdf}
  {\enquote {\bibinfo {title} {{ICFA statement on the ILC operating at 250 GeV
  as a Higgs boson factory}},}\ }\bibinfo {howpublished}
  {\url{https://icfa.fnal.gov/wp-content/uploads/ICFA-Statement-Nov2017.pdf}}
  (\bibinfo {year} {2017})\BibitemShut {NoStop}%
\bibitem [{bib({\natexlab{a}})}]{bib:ttc}%
  \BibitemOpen
  \href@noop {} {\enquote {\bibinfo {title} {{TESLA Technology
  Collaboration}},}\ }\bibinfo {howpublished} {\url{http://tesla-new.desy.de/}}
  ({\natexlab{a}})\BibitemShut {NoStop}%
\bibitem [{\citenamefont {Schreiber}\ and\ \citenamefont
  {Faatz}(2015)}]{schreiber_faatz_2015}%
  \BibitemOpen
  \bibfield  {author} {\bibinfo {author} {\bibfnamefont {S.}~\bibnamefont
  {Schreiber}}\ and\ \bibinfo {author} {\bibfnamefont {B.}~\bibnamefont
  {Faatz}},\ }\href {\doibase 10.1017/hpl.2015.16} {\bibfield  {journal}
  {\bibinfo  {journal} {High Power Laser Science and Engineering}\ }\textbf
  {\bibinfo {volume} {3}},\ \bibinfo {pages} {E20} (\bibinfo {year}
  {2015})}\BibitemShut {NoStop}%
\bibitem [{\citenamefont {Vogt}\ \emph {et~al.}(2018)\citenamefont {Vogt} \emph
  {et~al.}}]{Vogt:2018wvy}%
  \BibitemOpen
  \bibfield  {author} {\bibinfo {author} {\bibfnamefont {M.}~\bibnamefont
  {Vogt}} \emph {et~al.},\ }in\ \href {\doibase
  10.18429/JACoW-IPAC2018-TUPMF090} {\emph {\bibinfo {booktitle} {{Proceedings,
  9th International Particle Accelerator Conference (IPAC 2018), Vancouver, BC,
  Apr 29--May 4, 2018}}}}\ (\bibinfo {year} {2018})\ pp.\ \bibinfo {pages}
  {1481--1484 (TUPMF090)},\ \bibinfo {note}
  {doi:10.18429/JACoW-IPAC2018-TUPMF090}\BibitemShut {NoStop}%
\bibitem [{bib({\natexlab{b}})}]{bib:xfel}%
  \BibitemOpen
  \href@noop {} {\enquote {\bibinfo {title} {{European XFEL}},}\ }\bibinfo
  {howpublished} {\url{https://www.xfel.eu/}} ({\natexlab{b}})\BibitemShut
  {NoStop}%
\bibitem [{bib({\natexlab{c}})}]{bib:lcls-ii}%
  \BibitemOpen
  \href@noop {} {\enquote {\bibinfo {title} {{LCLS-II}: A world-class discovery
  machine},}\ }\bibinfo {howpublished}
  {\url{https://lcls.slac.stanford.edu/lcls-ii}} ({\natexlab{c}})\BibitemShut
  {NoStop}%
\bibitem [{\citenamefont {Zhao}\ \emph {et~al.}(2018)\citenamefont {Zhao},
  \citenamefont {Wang}, \citenamefont {Yang},\ and\ \citenamefont
  {Yin}}]{Zhao:2018lcl}%
  \BibitemOpen
  \bibfield  {author} {\bibinfo {author} {\bibfnamefont {Z.}~\bibnamefont
  {Zhao}}, \bibinfo {author} {\bibfnamefont {D.}~\bibnamefont {Wang}}, \bibinfo
  {author} {\bibfnamefont {Z.-H.}\ \bibnamefont {Yang}}, \ and\ \bibinfo
  {author} {\bibfnamefont {L.}~\bibnamefont {Yin}},\ }in\ \href {\doibase
  10.18429/JACoW-FEL2017-MOP055} {\emph {\bibinfo {booktitle} {{Proceedings,
  38th International Free Electron Laser Conference, FEL2017}}}}\ (\bibinfo
  {year} {2018})\ pp.\ \bibinfo {pages} {182--184 (MOP055)},\ \bibinfo {note}
  {doi:10.18429/JACoW-FEL2017-MOP055}\BibitemShut {NoStop}%
\bibitem [{\citenamefont {Geng}\ \emph {et~al.}(2015)\citenamefont {Geng} \emph
  {et~al.}}]{Geng:2015glc}%
  \BibitemOpen
  \bibfield  {author} {\bibinfo {author} {\bibfnamefont {R.-L.}\ \bibnamefont
  {Geng}} \emph {et~al.},\ }in\ \href {\doibase
  10.18429/JACoW-IPAC2015-WEPWI013} {\emph {\bibinfo {booktitle} {{Proceedings,
  6th International Particle Accelerator Conference (IPAC 2015): Richmond, VA,
  May 3--8, 2015}}}}\ (\bibinfo {year} {2015})\ p.\ \bibinfo {pages}
  {WEPWI013},\ \bibinfo {note}
  {doi:10.18429/JACoW-IPAC2015-WEPWI013}\BibitemShut {NoStop}%
\bibitem [{\citenamefont {Adolphsen}(2011)}]{Adolphsen:2011a}%
  \BibitemOpen
  \bibfield  {author} {\bibinfo {author} {\bibfnamefont {C.}~\bibnamefont
  {Adolphsen}},\ }\href@noop {} {\enquote {\bibinfo {title} {{Optimum ML cavity
  performance: gradient, Q0, and other ML parameters}},}\ }\bibinfo
  {howpublished} {Presentation at 2011 Linear Collider Workshop of the Americas
  (ALCPG11), Eugene, OR, Mar 19--23, 2011,
  \url{https://agenda.linearcollider.org/event/4572/}} (\bibinfo {year}
  {2011})\BibitemShut {NoStop}%
\bibitem [{\citenamefont {Eremeev}\ \emph {et~al.}(2007)\citenamefont
  {Eremeev}, \citenamefont {Geng}, \citenamefont {Padamsee},\ and\
  \citenamefont {Shemelin}}]{Eremeev:2007zza}%
  \BibitemOpen
  \bibfield  {author} {\bibinfo {author} {\bibfnamefont {G.~V.}\ \bibnamefont
  {Eremeev}}, \bibinfo {author} {\bibfnamefont {R.~L.}\ \bibnamefont {Geng}},
  \bibinfo {author} {\bibfnamefont {H.}~\bibnamefont {Padamsee}}, \ and\
  \bibinfo {author} {\bibfnamefont {V.~D.}\ \bibnamefont {Shemelin}},\ }in\
  \href {\doibase 10.1109/PAC.2007.4441242} {\emph {\bibinfo {booktitle}
  {{Proceedings, 22nd Particle Accelerators Conference (PAC'07), Albuquerque,
  NM, Jun 25--29, 2007}}}}\ (\bibinfo {year} {2007})\ pp.\ \bibinfo {pages}
  {2337--2339 (WEPMS006)},\ \bibinfo {note}
  {doi:10.1109/PAC.2007.4441242}\BibitemShut {NoStop}%
\bibitem [{\citenamefont {Padamsee}\ \emph {et~al.}(2008)\citenamefont
  {Padamsee}, \citenamefont {Knobloch},\ and\ \citenamefont
  {Hays}}]{Padamsee:1998vf}%
  \BibitemOpen
  \bibfield  {author} {\bibinfo {author} {\bibfnamefont {H.}~\bibnamefont
  {Padamsee}}, \bibinfo {author} {\bibfnamefont {J.}~\bibnamefont {Knobloch}},
  \ and\ \bibinfo {author} {\bibfnamefont {T.}~\bibnamefont {Hays}},\
  }\href@noop {} {\emph {\bibinfo {title} {{RF superconductivity for
  accelerators}}}},\ \bibinfo {edition} {2nd}\ ed.\ (\bibinfo  {publisher}
  {Wiley},\ \bibinfo {year} {2008})\BibitemShut {NoStop}%
\bibitem [{\citenamefont {Grassellino}\ \emph {et~al.}(2018)\citenamefont
  {Grassellino} \emph {et~al.}}]{Grassellino:2018tqg}%
  \BibitemOpen
  \bibfield  {author} {\bibinfo {author} {\bibfnamefont {A.}~\bibnamefont
  {Grassellino}} \emph {et~al.},\ }\href@noop {} {\  (\bibinfo {year}
  {2018})},\ \Eprint {http://arxiv.org/abs/1806.09824} {arXiv:1806.09824
  [physics.acc-ph]} \BibitemShut {NoStop}%
\bibitem [{\citenamefont {Gurevich}(2017)}]{Gurevich:2017vnn}%
  \BibitemOpen
  \bibfield  {author} {\bibinfo {author} {\bibfnamefont {A.}~\bibnamefont
  {Gurevich}},\ }\href {\doibase 10.1088/1361-6668/30/3/034004} {\bibfield
  {journal} {\bibinfo  {journal} {Supercond. Sci. Technol.}\ }\textbf {\bibinfo
  {volume} {30}},\ \bibinfo {pages} {034004} (\bibinfo {year}
  {2017})}\BibitemShut {NoStop}%
\bibitem [{\citenamefont {Kubo}(2017)}]{Kubo:2017cww}%
  \BibitemOpen
  \bibfield  {author} {\bibinfo {author} {\bibfnamefont {T.}~\bibnamefont
  {Kubo}},\ }\bibfield  {booktitle} {\emph {\bibinfo {booktitle} {{Proceedings,
  International Workshop on Future Linear Colliders 2016 (LCWS2016), Morioka,
  Japan, Dec 5--9, 2016}}},\ }\href@noop {} {\  (\bibinfo {year} {2017})},\
  \Eprint {http://arxiv.org/abs/1703.05645} {arXiv:1703.05645 [physics.acc-ph]}
  \BibitemShut {NoStop}%
\bibitem [{\citenamefont {Singer}\ \emph {et~al.}(2016)\citenamefont {Singer}
  \emph {et~al.}}]{Singer:2016fbf}%
  \BibitemOpen
  \bibfield  {author} {\bibinfo {author} {\bibfnamefont {W.}~\bibnamefont
  {Singer}} \emph {et~al.},\ }\href {\doibase
  10.1103/PhysRevAccelBeams.19.092001} {\bibfield  {journal} {\bibinfo
  {journal} {Phys. Rev. Accel. Beams}\ }\textbf {\bibinfo {volume} {19}},\
  \bibinfo {pages} {092001} (\bibinfo {year} {2016})},\ \bibinfo {note}
  {doi:10.1103/PhysRevAccelBeams.19.092001}\BibitemShut {NoStop}%
\bibitem [{\citenamefont {Reschke}\ \emph {et~al.}(2017)\citenamefont {Reschke}
  \emph {et~al.}}]{Reschke:2017gjp}%
  \BibitemOpen
  \bibfield  {author} {\bibinfo {author} {\bibfnamefont {D.}~\bibnamefont
  {Reschke}} \emph {et~al.},\ }\href {\doibase
  10.1103/PhysRevAccelBeams.20.042004} {\bibfield  {journal} {\bibinfo
  {journal} {Phys. Rev. Accel. Beams}\ }\textbf {\bibinfo {volume} {20}},\
  \bibinfo {pages} {042004} (\bibinfo {year} {2017})},\ \bibinfo {note}
  {doi:10.1103/PhysRevAccelBeams.20.042004}\BibitemShut {NoStop}%
\bibitem [{\citenamefont {Walker}\ and\ \citenamefont
  {Kostin}(2017)}]{bib:Walker:2017.lcws}%
  \BibitemOpen
  \bibfield  {author} {\bibinfo {author} {\bibfnamefont {N.~J.}\ \bibnamefont
  {Walker}}\ and\ \bibinfo {author} {\bibfnamefont {D.}~\bibnamefont
  {Kostin}},\ }\href
  {https://agenda.linearcollider.org/event/7645/contributions/39682/} {\enquote
  {\bibinfo {title} {{The European XFEL -- experience and lessons learned}},}\
  }\bibinfo {howpublished} {{Presentation, International Workshop on Future
  Linear Colliders (LCWS2017), Strasbourg, France, Oct 22--27, 2017,
  \url{https://agenda.linearcollider.org/event/7645/}}} (\bibinfo {year}
  {2017})\BibitemShut {NoStop}%
\bibitem [{\citenamefont {Grassellino}\ \emph {et~al.}(2017)\citenamefont
  {Grassellino} \emph {et~al.}}]{Grassellino:2017bod}%
  \BibitemOpen
  \bibfield  {author} {\bibinfo {author} {\bibfnamefont {A.}~\bibnamefont
  {Grassellino}} \emph {et~al.},\ }\href {\doibase 10.1088/1361-6668/aa7afe}
  {\bibfield  {journal} {\bibinfo  {journal} {Supercond. Sci. Technol.}\
  }\textbf {\bibinfo {volume} {30}},\ \bibinfo {pages} {094004} (\bibinfo
  {year} {2017})},\ \bibinfo {note} {doi:10.1088/1361-6668/aa7afe},\ \Eprint
  {http://arxiv.org/abs/1701.06077} {arXiv:1701.06077 [physics.acc-ph]}
  \BibitemShut {NoStop}%
\bibitem [{\citenamefont {Grassellino}\ \emph {et~al.}(2013)\citenamefont
  {Grassellino} \emph {et~al.}}]{Grassellino:2013nza}%
  \BibitemOpen
  \bibfield  {author} {\bibinfo {author} {\bibfnamefont {A.}~\bibnamefont
  {Grassellino}} \emph {et~al.},\ }\href {\doibase
  10.1088/0953-2048/26/10/102001} {\bibfield  {journal} {\bibinfo  {journal}
  {Supercond. Sci. Technol.}\ }\textbf {\bibinfo {volume} {26}},\ \bibinfo
  {pages} {102001} (\bibinfo {year} {2013})},\ \Eprint
  {http://arxiv.org/abs/1306.0288} {arXiv:1306.0288 [physics.acc-ph]}
  \BibitemShut {NoStop}%
\bibitem [{\citenamefont {Wu}\ \emph {et~al.}(2018)\citenamefont {Wu} \emph
  {et~al.}}]{Wu:2018qyl}%
  \BibitemOpen
  \bibfield  {author} {\bibinfo {author} {\bibfnamefont {G.}~\bibnamefont {Wu}}
  \emph {et~al.},\ }\href@noop {} {\  (\bibinfo {year} {2018})},\ \Eprint
  {http://arxiv.org/abs/1812.09368} {arXiv:1812.09368 [physics.acc-ph]}
  \BibitemShut {NoStop}%
\bibitem [{\citenamefont {Dhakal}\ \emph {et~al.}(2018)\citenamefont {Dhakal},
  \citenamefont {Chetri}, \citenamefont {Balachandran}, \citenamefont {Lee},\
  and\ \citenamefont {Ciovati}}]{Dhakal:2017xxq}%
  \BibitemOpen
  \bibfield  {author} {\bibinfo {author} {\bibfnamefont {P.}~\bibnamefont
  {Dhakal}}, \bibinfo {author} {\bibfnamefont {S.}~\bibnamefont {Chetri}},
  \bibinfo {author} {\bibfnamefont {S.}~\bibnamefont {Balachandran}}, \bibinfo
  {author} {\bibfnamefont {P.~J.}\ \bibnamefont {Lee}}, \ and\ \bibinfo
  {author} {\bibfnamefont {G.}~\bibnamefont {Ciovati}},\ }\href {\doibase
  10.1103/PhysRevAccelBeams.21.032001} {\bibfield  {journal} {\bibinfo
  {journal} {Phys. Rev. Accel. Beams}\ }\textbf {\bibinfo {volume} {21}},\
  \bibinfo {pages} {032001} (\bibinfo {year} {2018})},\ \Eprint
  {http://arxiv.org/abs/1711.03991} {arXiv:1711.03991 [physics.acc-ph]}
  \BibitemShut {NoStop}%
\bibitem [{\citenamefont {Koufalis}\ \emph {et~al.}(2017)\citenamefont
  {Koufalis} \emph {et~al.}}]{Koufalis:2017blg}%
  \BibitemOpen
  \bibfield  {author} {\bibinfo {author} {\bibfnamefont {P.}~\bibnamefont
  {Koufalis}} \emph {et~al.},\ }in\ \href {\doibase
  10.18429/JACoW-LINAC2016-TUPRC025} {\emph {\bibinfo {booktitle}
  {{Proceedings, 28th International Linear Accelerator Conference (LINAC16):
  East Lansing, MI, Sep 25--30, 2016}}}}\ (\bibinfo {year} {2017})\ pp.\
  \bibinfo {pages} {472--474 (TUPRC025)}\BibitemShut {NoStop}%
\bibitem [{\citenamefont {Umemori}\ \emph {et~al.}(2018)\citenamefont {Umemori}
  \emph {et~al.}}]{bib:Umemori:2018.lcws}%
  \BibitemOpen
  \bibfield  {author} {\bibinfo {author} {\bibfnamefont {K.}~\bibnamefont
  {Umemori}} \emph {et~al.},\ }\href
  {https://indico-lcagenda-201605.s3.cern.ch/event/7889/contribution/42431/52008-51119-N-infusion_KEK_umemori.pdf}
  {\enquote {\bibinfo {title} {{Recent SRF activities toward high--Q/high--G at
  KEK}},}\ }\bibinfo {howpublished} {{Presentation, International Workshop on
  Future Linear Colliders (LCWS2018), Arlington, TX, Oct 22--26, 2018,
  \url{https://indico-lcagenda-201605.s3.cern.ch/event/7889/}}} (\bibinfo
  {year} {2018})\BibitemShut {NoStop}%
\bibitem [{\citenamefont {Wenskat}\ \emph {et~al.}(2018)\citenamefont {Wenskat}
  \emph {et~al.}}]{Wenskat:2018zco}%
  \BibitemOpen
  \bibfield  {author} {\bibinfo {author} {\bibfnamefont {M.}~\bibnamefont
  {Wenskat}} \emph {et~al.},\ }in\ \href {\doibase
  10.18429/JACoW-LINAC2018-TH2A01} {\emph {\bibinfo {booktitle} {{Proceedings,
  29th International Linear Accelerator Conference (LINAC18): Beijing, China,
  Sep 16--21, 2018}}}}\ (\bibinfo {year} {2018})\ pp.\ \bibinfo {pages}
  {652--657 (TH2A01)},\ \bibinfo {note}
  {doi:10.18429/JACoW-LINAC2018-TH2A01}\BibitemShut {NoStop}%
\bibitem [{\citenamefont {Romanenko}\ \emph {et~al.}(2014)\citenamefont
  {Romanenko}, \citenamefont {Grassellino}, \citenamefont {Barkov},
  \citenamefont {Suter}, \citenamefont {Salman},\ and\ \citenamefont
  {Prokscha}}]{Romanenko:2013saa}%
  \BibitemOpen
  \bibfield  {author} {\bibinfo {author} {\bibfnamefont {A.}~\bibnamefont
  {Romanenko}}, \bibinfo {author} {\bibfnamefont {A.}~\bibnamefont
  {Grassellino}}, \bibinfo {author} {\bibfnamefont {F.}~\bibnamefont {Barkov}},
  \bibinfo {author} {\bibfnamefont {A.}~\bibnamefont {Suter}}, \bibinfo
  {author} {\bibfnamefont {Z.}~\bibnamefont {Salman}}, \ and\ \bibinfo {author}
  {\bibfnamefont {T.}~\bibnamefont {Prokscha}},\ }\href {\doibase
  10.1063/1.4866013} {\bibfield  {journal} {\bibinfo  {journal} {Appl. Phys.
  Lett.}\ }\textbf {\bibinfo {volume} {104}},\ \bibinfo {pages} {072601}
  (\bibinfo {year} {2014})},\ \Eprint {http://arxiv.org/abs/1311.7153}
  {arXiv:1311.7153 [cond-mat.supr-con]} \BibitemShut {NoStop}%
\bibitem [{\citenamefont {Geng}(2006)}]{Geng:2006wf}%
  \BibitemOpen
  \bibfield  {author} {\bibinfo {author} {\bibfnamefont {R.}~\bibnamefont
  {Geng}},\ }\href {\doibase 10.1016/j.physc.2006.03.029} {\bibfield  {journal}
  {\bibinfo  {journal} {Physica}\ }\textbf {\bibinfo {volume} {C441}},\
  \bibinfo {pages} {145} (\bibinfo {year} {2006})}\BibitemShut {NoStop}%
\bibitem [{\citenamefont {Li}\ \emph {et~al.}(2008)\citenamefont {Li},
  \citenamefont {Adolphsen} \emph {et~al.}}]{Li:2008a}%
  \BibitemOpen
  \bibfield  {author} {\bibinfo {author} {\bibfnamefont {Z.}~\bibnamefont
  {Li}}, \bibinfo {author} {\bibfnamefont {C.}~\bibnamefont {Adolphsen}},
  \emph {et~al.},\ }in\ \href
  {http://accelconf.web.cern.ch/AccelConf/LINAC08/papers/thp038.pdf} {\emph
  {\bibinfo {booktitle} {{Proceedings, 24th Linear Accelerator Conference
  (LINAC08), Victoria, BC, Sep 29--Oct 3, 2008}}}}\ (\bibinfo {year} {2008})\
  pp.\ \bibinfo {pages} {867--869 (THP038)}\BibitemShut {NoStop}%
\bibitem [{\citenamefont {Geng}(2018)}]{bib:Geng:2018.lcws}%
  \BibitemOpen
  \bibfield  {author} {\bibinfo {author} {\bibfnamefont {R.-L.}\ \bibnamefont
  {Geng}},\ }\href
  {https://indico-lcagenda-201605.s3.cern.ch/event/7889/contribution/42585/52060-51171-LCWS2018_Geng.pdf}
  {\enquote {\bibinfo {title} {{Low-Surface-Field (LSF) shape cavity
  development: recent results and future plan}},}\ }\bibinfo {howpublished}
  {{Presentation, International Workshop on Future Linear Colliders (LCWS2018),
  Arlington, TX, Oct 22--26, 2018,
  \url{https://indico-lcagenda-201605.s3.cern.ch/event/7889/}}} (\bibinfo
  {year} {2018})\BibitemShut {NoStop}%
\bibitem [{\citenamefont {Kubo}\ and\ \citenamefont
  {Yamamoto}(2018)}]{bib:Kubo:2018.ttc}%
  \BibitemOpen
  \bibfield  {author} {\bibinfo {author} {\bibfnamefont {T.}~\bibnamefont
  {Kubo}}\ and\ \bibinfo {author} {\bibfnamefont {A.}~\bibnamefont
  {Yamamoto}},\ }\href
  {https://indico.desy.de/indico/event/20010/session/34/contribution/57/material/slides/0.pdf}
  {\enquote {\bibinfo {title} {{Comments on Niobium RRR and SRF cavity
  performance for ILC}},}\ }\bibinfo {howpublished} {{Presentation, TTC meeting
  2018, Saitama, Japan, Jun 26--29, 2018,
  \url{https://indico.desy.de/indico/event/20010/}}} (\bibinfo {year}
  {2018})\BibitemShut {NoStop}%
\bibitem [{\citenamefont {Kneisel}\ \emph {et~al.}(2015)\citenamefont {Kneisel}
  \emph {et~al.}}]{Kneisel:2014uqa}%
  \BibitemOpen
  \bibfield  {author} {\bibinfo {author} {\bibfnamefont {P.}~\bibnamefont
  {Kneisel}} \emph {et~al.},\ }\href {\doibase 10.1016/j.nima.2014.11.083}
  {\bibfield  {journal} {\bibinfo  {journal} {Nucl. Instrum. Meth.}\ }\textbf
  {\bibinfo {volume} {A774}},\ \bibinfo {pages} {133} (\bibinfo {year}
  {2015})}\BibitemShut {NoStop}%
\bibitem [{\citenamefont {Reschke}\ \emph {et~al.}(2011)\citenamefont {Reschke}
  \emph {et~al.}}]{Reschke:2011a}%
  \BibitemOpen
  \bibfield  {author} {\bibinfo {author} {\bibfnamefont {D.}~\bibnamefont
  {Reschke}} \emph {et~al.},\ }in\ \href
  {http://accelconf.web.cern.ch/AccelConf/SRF2011/papers/tupo046.pdf} {\emph
  {\bibinfo {booktitle} {{Proceedings, 15th International Conference on RF
  Superconductivity (SRF 2011), Chicago, IL, Jul 25-29, 2011}}}}\ (\bibinfo
  {year} {2011})\ pp.\ \bibinfo {pages} {490--494 (TUPO046)}\BibitemShut
  {NoStop}%
\bibitem [{\citenamefont {Dhakal}\ \emph {et~al.}(2015)\citenamefont {Dhakal},
  \citenamefont {Ciovati},\ and\ \citenamefont {Myneni}}]{Dhakal:2015xac}%
  \BibitemOpen
  \bibfield  {author} {\bibinfo {author} {\bibfnamefont {P.}~\bibnamefont
  {Dhakal}}, \bibinfo {author} {\bibfnamefont {G.}~\bibnamefont {Ciovati}}, \
  and\ \bibinfo {author} {\bibfnamefont {G.~R.}\ \bibnamefont {Myneni}},\
  }\bibfield  {booktitle} {\emph {\bibinfo {booktitle} {{Proceedings, JLab
  Ingot Niobium CRADA Workshop: Newport News, Virginia, December 4, 2015}}},\
  }\href {\doibase 10.1063/1.4935323} {\bibfield  {journal} {\bibinfo
  {journal} {AIP Conf. Proc.}\ }\textbf {\bibinfo {volume} {1687}},\ \bibinfo
  {pages} {030002} (\bibinfo {year} {2015})}\BibitemShut {NoStop}%
\bibitem [{\citenamefont {Kaabi}\ \emph {et~al.}(2013)\citenamefont {Kaabi}
  \emph {et~al.}}]{Kaabi:2013wna}%
  \BibitemOpen
  \bibfield  {author} {\bibinfo {author} {\bibfnamefont {W.}~\bibnamefont
  {Kaabi}} \emph {et~al.},\ }in\ \href
  {http://JACoW.org/IPAC2013/papers/wepwo001.pdf} {\emph {\bibinfo {booktitle}
  {{Proceedings, 4th International Particle Accelerator Conference (IPAC 2013),
  Shanghai, China, May 12-17, 2013}}}}\ (\bibinfo {year} {2013})\ pp.\ \bibinfo
  {pages} {2310--2312 (WEPWO001)}\BibitemShut {NoStop}%
\bibitem [{\citenamefont {Sierra}\ \emph {et~al.}(2017)\citenamefont {Sierra}
  \emph {et~al.}}]{Sierra:2017wyc}%
  \BibitemOpen
  \bibfield  {author} {\bibinfo {author} {\bibfnamefont {S.}~\bibnamefont
  {Sierra}} \emph {et~al.},\ }in\ \href {\doibase
  10.18429/JACoW-LINAC2016-TUPLR048} {\emph {\bibinfo {booktitle}
  {{Proceedings, 28th International Linear Accelerator Conference (LINAC16),
  East Lansing, MI, Sep 25--30, 2016}}}}\ (\bibinfo {year} {2017})\ pp.\
  \bibinfo {pages} {572--574 (TUPLR048)},\ \bibinfo {note}
  {doi:10.18429/JACoW-LINAC2016-TUPLR048}\BibitemShut {NoStop}%
\bibitem [{\citenamefont {Reschke}\ \emph {et~al.}(2018)\citenamefont
  {Reschke}, \citenamefont {Decking}, \citenamefont {Walker},\ and\
  \citenamefont {Weise}}]{Reschke:2018ywk}%
  \BibitemOpen
  \bibfield  {author} {\bibinfo {author} {\bibfnamefont {D.}~\bibnamefont
  {Reschke}}, \bibinfo {author} {\bibfnamefont {W.}~\bibnamefont {Decking}},
  \bibinfo {author} {\bibfnamefont {N.}~\bibnamefont {Walker}}, \ and\ \bibinfo
  {author} {\bibfnamefont {H.}~\bibnamefont {Weise}},\ }in\ \href {\doibase
  10.18429/JACoW-SRF2017-MOXA02} {\emph {\bibinfo {booktitle} {{Proceedings,
  18th International Conference on RF Superconductivity (SRF2017): Lanzhou,
  China, Jul 17--21, 2017}}}}\ (\bibinfo {year} {2018})\ pp.\ \bibinfo {pages}
  {1--5 (MOXA02)},\ \bibinfo {note}
  {doi:10.18429/JACoW-SRF2017-MOXA02}\BibitemShut {NoStop}%
\bibitem [{\citenamefont {Berry}\ and\ \citenamefont
  {Napoly}(2017)}]{Berry:2017gpt}%
  \BibitemOpen
  \bibfield  {author} {\bibinfo {author} {\bibfnamefont {S.}~\bibnamefont
  {Berry}}\ and\ \bibinfo {author} {\bibfnamefont {O.}~\bibnamefont {Napoly}},\
  }in\ \href {\doibase 10.18429/JACoW-LINAC2016-WE1A02} {\emph {\bibinfo
  {booktitle} {{Proceedings, 28th International Linear Accelerator Conference
  (LINAC16): East Lansing, Michigan, September 25-30, 2016}}}}\ (\bibinfo
  {year} {2017})\ pp.\ \bibinfo {pages} {646--650 (WE1A02)},\ \bibinfo {note}
  {doi:10.18429/JACoW-LINAC2016-WE1A02}\BibitemShut {NoStop}%
\bibitem [{\citenamefont {Peterson}\ \emph {et~al.}(2011)\citenamefont
  {Peterson} \emph {et~al.}}]{Peterson:2011zz}%
  \BibitemOpen
  \bibfield  {author} {\bibinfo {author} {\bibfnamefont {T.~J.}\ \bibnamefont
  {Peterson}} \emph {et~al.},\ }\bibfield  {booktitle} {\emph {\bibinfo
  {booktitle} {{Transactions of the Cryogenic Engineering Conference (CEC
  2011): Spokane, WA, Jun 13--17, 2011}}},\ }\href {\doibase 10.1063/1.4707088}
  {\bibfield  {journal} {\bibinfo  {journal} {AIP Conf. Proc.}\ }\textbf
  {\bibinfo {volume} {1434}},\ \bibinfo {pages} {1575} (\bibinfo {year}
  {2011})},\ \Eprint {http://arxiv.org/abs/1209.2405} {arXiv:1209.2405
  [physics.acc-ph]} \BibitemShut {NoStop}%
\bibitem [{\citenamefont {Weise}(2014)}]{Weise:2014zqa}%
  \BibitemOpen
  \bibfield  {author} {\bibinfo {author} {\bibfnamefont {H.}~\bibnamefont
  {Weise}},\ }in\ \href {\doibase 10.18429/JACoW-IPAC2014-WEIB03} {\emph
  {\bibinfo {booktitle} {{Proceedings, 5th International Particle Accelerator
  Conference (IPAC 2014): Dresden, Germany, Jun 15--20, 2014}}}}\ (\bibinfo
  {year} {2014})\ pp.\ \bibinfo {pages} {1923--1928 (WEIB03)},\ \bibinfo {note}
  {doi:10.18429/JACoW-IPAC2014-WEIB03}\BibitemShut {NoStop}%
\bibitem [{\citenamefont {Kostin}\ \emph {et~al.}(2009)\citenamefont {Kostin},
  \citenamefont {Moeller}, \citenamefont {Goessel},\ and\ \citenamefont
  {Jensch}}]{Kostin:2009a}%
  \BibitemOpen
  \bibfield  {author} {\bibinfo {author} {\bibfnamefont {D.}~\bibnamefont
  {Kostin}}, \bibinfo {author} {\bibfnamefont {W.-D.}\ \bibnamefont {Moeller}},
  \bibinfo {author} {\bibfnamefont {A.}~\bibnamefont {Goessel}}, \ and\
  \bibinfo {author} {\bibfnamefont {K.}~\bibnamefont {Jensch}},\ }in\ \href
  {http://accelconf.web.cern.ch/AccelConf/SRF2009/papers/TUPPO005.PDF} {\emph
  {\bibinfo {booktitle} {{Proceedings,14th International Conference on RF
  Superconductivity (SRF2009), Berlin, Sep 20--25, 2009}}}}\ (\bibinfo {year}
  {2009})\ pp.\ \bibinfo {pages} {180--184 (TUPPO005)}\BibitemShut {NoStop}%
\bibitem [{\citenamefont {Broemmelsiek}\ \emph {et~al.}(2018)\citenamefont
  {Broemmelsiek} \emph {et~al.}}]{Broemmelsiek:2018iqr}%
  \BibitemOpen
  \bibfield  {author} {\bibinfo {author} {\bibfnamefont {D.}~\bibnamefont
  {Broemmelsiek}} \emph {et~al.},\ }\href {\doibase 10.1088/1367-2630/aaec57}
  {\bibfield  {journal} {\bibinfo  {journal} {New J. Phys.}\ }\textbf {\bibinfo
  {volume} {20}},\ \bibinfo {pages} {113018} (\bibinfo {year} {2018})},\
  \Eprint {http://arxiv.org/abs/1808.03208} {arXiv:1808.03208 [physics.acc-ph]}
  \BibitemShut {NoStop}%
\bibitem [{\citenamefont {Yamamoto}\ \emph {et~al.}(2018)\citenamefont
  {Yamamoto} \emph {et~al.}}]{Yamamoto:2018kml}%
  \BibitemOpen
  \bibfield  {author} {\bibinfo {author} {\bibfnamefont {Y.}~\bibnamefont
  {Yamamoto}} \emph {et~al.},\ }in\ \href {\doibase
  10.18429/JACoW-SRF2017-THYA02} {\emph {\bibinfo {booktitle} {{Proceedings,
  18th International Conference on RF Superconductivity (SRF2017), Lanzhou,
  China, Jul 17--21, 2017}}}}\ (\bibinfo {year} {2018})\ pp.\ \bibinfo {pages}
  {722--728 (THYA02)},\ \bibinfo {note}
  {doi:10.18429/JACoW-SRF2017-THYA02}\BibitemShut {NoStop}%
\bibitem [{\citenamefont {Kasprzak}\ \emph {et~al.}(2018)\citenamefont
  {Kasprzak} \emph {et~al.}}]{Kasprzak:2018kkr}%
  \BibitemOpen
  \bibfield  {author} {\bibinfo {author} {\bibfnamefont {K.}~\bibnamefont
  {Kasprzak}} \emph {et~al.},\ }in\ \href {\doibase
  10.18429/JACoW-SRF2017-MOPB106} {\emph {\bibinfo {booktitle} {{Proceedings,
  18th International Conference on RF Superconductivity (SRF2017), Lanzhou,
  China, July 17-21, 2017}}}}\ (\bibinfo {year} {2018})\ pp.\ \bibinfo {pages}
  {312--316 ({MOPB106})},\ \bibinfo {note}
  {doi:10.18429/JACoW-SRF2017-MOPB106}\BibitemShut {NoStop}%
\bibitem [{\citenamefont {{S1Global collaboration}}(2012)}]{bib:s1g}%
  \BibitemOpen
  \bibfield  {author} {\bibinfo {author} {\bibnamefont {{S1Global
  collaboration}}},\ }\href@noop {} {\enquote {\bibinfo {title} {{S1Global
  report}},}\ }\bibinfo {howpublished}
  {\url{http://edmsdirect.desy.de/item/D00000001005135}} (\bibinfo {year}
  {2012})\BibitemShut {NoStop}%
\bibitem [{\citenamefont {Kemp}\ \emph {et~al.}(2011)\citenamefont {Kemp} \emph
  {et~al.}}]{Kemp:2011zz}%
  \BibitemOpen
  \bibfield  {author} {\bibinfo {author} {\bibfnamefont {M.~A.}\ \bibnamefont
  {Kemp}} \emph {et~al.},\ }in\ \href {\doibase 10.1109/PPC.2011.6191686}
  {\emph {\bibinfo {booktitle} {{Proceedings, 18th IEEE International Pulsed
  Power Conference (PPC11), Chicago, IL, Jun 19-23, 2011}}}}\ (\bibinfo {year}
  {2011})\ pp.\ \bibinfo {pages} {1582--1589},\ \bibinfo {note}
  {doi:10.1109/PPC.2011.6191686}\BibitemShut {NoStop}%
\bibitem [{\citenamefont {Gaudreau}\ \emph {et~al.}(2014)\citenamefont
  {Gaudreau}, \citenamefont {Silverman}, \citenamefont {Simpson},\ and\
  \citenamefont {Casey}}]{Gaudreau:2014pza}%
  \BibitemOpen
  \bibfield  {author} {\bibinfo {author} {\bibfnamefont {M.~P.~J.}\
  \bibnamefont {Gaudreau}}, \bibinfo {author} {\bibfnamefont {N.}~\bibnamefont
  {Silverman}}, \bibinfo {author} {\bibfnamefont {B.}~\bibnamefont {Simpson}},
  \ and\ \bibinfo {author} {\bibfnamefont {J.}~\bibnamefont {Casey}},\ }in\
  \href {http://jacow.org/IPAC2014/papers/mopme082.pdf} {\emph {\bibinfo
  {booktitle} {{Proceedings, 5th International Particle Accelerator Conference
  (IPAC 2014), Dresden, Germany, Jun 15-20, 2014}}}}\ (\bibinfo {year} {2014})\
  pp.\ \bibinfo {pages} {562--563 (MOPME082)}\BibitemShut {NoStop}%
\bibitem [{\citenamefont {Syratchev}(2015)}]{Syratchev:2015a}%
  \BibitemOpen
  \bibfield  {author} {\bibinfo {author} {\bibfnamefont {I.}~\bibnamefont
  {Syratchev}},\ }\href
  {https://indico.cern.ch/event/336335/contributions/789041/attachments/657768/904332/CLIC_WS_2015_01.pdf}
  {\enquote {\bibinfo {title} {{Introduction to the High Efficiency
  International Klystron Activity HEIKA}},}\ }\bibinfo {howpublished}
  {Presentation, CLIC workshop 2015, Geneva, Switzerland, Jan 26--30, 2015,
  \url{https://indico.cern.ch/event/336335/}} (\bibinfo {year}
  {2015})\BibitemShut {NoStop}%
\bibitem [{\citenamefont {Gerigk}(2018)}]{Gerigk:2018ebm}%
  \BibitemOpen
  \bibfield  {author} {\bibinfo {author} {\bibfnamefont {F.}~\bibnamefont
  {Gerigk}},\ }in\ \href {\doibase 10.18429/JACoW-IPAC2018-MOYGB1} {\emph
  {\bibinfo {booktitle} {{Proceedings, 9th International Particle Accelerator
  Conference (IPAC 2018), Vancouver, BC, Canada, Apr 29--May 4, 2018}}}}\
  (\bibinfo {year} {2018})\ pp.\ \bibinfo {pages} {12--17 (MOYGB1)},\ \bibinfo
  {note} {doi:10.18429/JACoW-IPAC2018-MOYGB1}\BibitemShut {NoStop}%
\bibitem [{\citenamefont {Guzilov}(2013)}]{Guzilov:2014a}%
  \BibitemOpen
  \bibfield  {author} {\bibinfo {author} {\bibfnamefont {I.~A.}\ \bibnamefont
  {Guzilov}},\ }in\ \href {\doibase 10.1109/IVESC.2014.6891996} {\emph
  {\bibinfo {booktitle} {Proceedings, 10th International Vacuum Electron
  Sources Conference (IVESC), Jun 30--Jul 4, 2014}}}\ (\bibinfo {year} {2013})\
  p.\ \bibinfo {pages} {6891996},\ \bibinfo {note}
  {doi:10.1109/IVESC.2014.6891996}\BibitemShut {NoStop}%
\bibitem [{\citenamefont {Constable}\ \emph {et~al.}(2017)\citenamefont
  {Constable} \emph {et~al.}}]{Constable:2017hha}%
  \BibitemOpen
  \bibfield  {author} {\bibinfo {author} {\bibfnamefont {D.}~\bibnamefont
  {Constable}} \emph {et~al.},\ }in\ \href {\doibase
  10.18429/JACoW-eeFACT2016-WET3AH2} {\emph {\bibinfo {booktitle}
  {{Proceedings, 58th ICFA Advanced Beam Dynamics Workshop on High Luminosity
  Circular $e^+ e^-$ Colliders (eeFACT2016), Daresbury, UK, Oct 24--27,
  2016}}}}\ (\bibinfo {year} {2017})\ pp.\ \bibinfo {pages} {185--187
  (WET3AH2)},\ \bibinfo {note}
  {doi:10.18429/JACoW-eeFACT2016-WET3AH2}\BibitemShut {NoStop}%
\bibitem [{\citenamefont {Baikov}\ \emph {et~al.}(2015)\citenamefont {Baikov},
  \citenamefont {Marrelli},\ and\ \citenamefont {Syratchev}}]{Baikov:2015bif}%
  \BibitemOpen
  \bibfield  {author} {\bibinfo {author} {\bibfnamefont {A.~Y.}\ \bibnamefont
  {Baikov}}, \bibinfo {author} {\bibfnamefont {C.}~\bibnamefont {Marrelli}}, \
  and\ \bibinfo {author} {\bibfnamefont {I.}~\bibnamefont {Syratchev}},\ }\href
  {\doibase 10.1109/TED.2015.2464096} {\bibfield  {journal} {\bibinfo
  {journal} {IEEE Trans. Electron. Dev.}\ }\textbf {\bibinfo {volume} {62}},\
  \bibinfo {pages} {3406} (\bibinfo {year} {2015})}\BibitemShut {NoStop}%
\bibitem [{\citenamefont {Jensen}(2016)}]{Jensen:2016a}%
  \BibitemOpen
  \bibfield  {author} {\bibinfo {author} {\bibfnamefont {E.}~\bibnamefont
  {Jensen}},\ }\href
  {https://indico.cern.ch/event/472685/contributions/2196873/attachments/1294428/1929442/Towards_Higher_Efficiency_RF_Systems.pdf}
  {\enquote {\bibinfo {title} {{Recent developments towards very high
  efficiency klystrons}},}\ }\bibinfo {howpublished} {Presentation, 9th CW and
  high average RF power workshop, Grenoble, France, Jun 20--24, 2016,
  \url{https://indico.cern.ch/event/472685/}} (\bibinfo {year}
  {2016})\BibitemShut {NoStop}%
\bibitem [{\citenamefont {Nakai}(2016)}]{bib:cr-0014}%
  \BibitemOpen
  \bibfield  {author} {\bibinfo {author} {\bibfnamefont {H.}~\bibnamefont
  {Nakai}},\ }\href@noop {} {\enquote {\bibinfo {title} {{Change request
  ILC-CR-0014: Cryogenic layout}},}\ }\bibinfo {howpublished}
  {{\url{http://edmsdirect.desy.de/item/D00000001146525}}} (\bibinfo {year}
  {2016})\BibitemShut {NoStop}%
\bibitem [{\citenamefont {Alley}\ \emph {et~al.}(1995)\citenamefont {Alley}
  \emph {et~al.}}]{Alley:1995ia}%
  \BibitemOpen
  \bibfield  {author} {\bibinfo {author} {\bibfnamefont {R.}~\bibnamefont
  {Alley}} \emph {et~al.},\ }\href {\doibase 10.1016/0168-9002(95)00450-5}
  {\bibfield  {journal} {\bibinfo  {journal} {Nucl. Instrum. Meth.}\ }\textbf
  {\bibinfo {volume} {A365}},\ \bibinfo {pages} {1} (\bibinfo {year}
  {1995})}\BibitemShut {NoStop}%
\bibitem [{\citenamefont {Alexander}\ \emph {et~al.}(2009)\citenamefont
  {Alexander} \emph {et~al.}}]{Alexander:2009nb}%
  \BibitemOpen
  \bibfield  {author} {\bibinfo {author} {\bibfnamefont {G.}~\bibnamefont
  {Alexander}} \emph {et~al.},\ }\href {\doibase 10.1016/j.nima.2009.07.091}
  {\bibfield  {journal} {\bibinfo  {journal} {Nucl. Instrum. Meth.}\ }\textbf
  {\bibinfo {volume} {A610}},\ \bibinfo {pages} {451} (\bibinfo {year}
  {2009})},\ \Eprint {http://arxiv.org/abs/0905.3066} {arXiv:0905.3066
  [physics.ins-det]} \BibitemShut {NoStop}%
\bibitem [{\citenamefont {Moffeit}\ \emph {et~al.}(2005)\citenamefont {Moffeit}
  \emph {et~al.}}]{Moffeit:2005pb}%
  \BibitemOpen
  \bibfield  {author} {\bibinfo {author} {\bibfnamefont {K.}~\bibnamefont
  {Moffeit}} \emph {et~al.},\ }\href@noop {} {\enquote {\bibinfo {title} {{Spin
  rotation schemes at the ILC for two interaction regions and positron
  polarization with both helicities}},}\ }\bibinfo {howpublished}
  {SLAC-TN-05-045} (\bibinfo {year} {2005})\BibitemShut {NoStop}%
\bibitem [{\citenamefont {Malysheva}\ \emph {et~al.}(2016)\citenamefont
  {Malysheva} \emph {et~al.}}]{Malysheva:2016jdr}%
  \BibitemOpen
  \bibfield  {author} {\bibinfo {author} {\bibfnamefont {L.~I.}\ \bibnamefont
  {Malysheva}} \emph {et~al.},\ }\bibfield  {booktitle} {\emph {\bibinfo
  {booktitle} {{Proceedings, International Workshop on Future Linear Colliders
  (LCWS15): Whistler, BC, Canada, Nov. 2-6, 2015}}},\ }\href@noop {} {\
  (\bibinfo {year} {2016})},\ \Eprint {http://arxiv.org/abs/1602.09050}
  {arXiv:1602.09050 [physics.acc-ph]} \BibitemShut {NoStop}%
\bibitem [{\citenamefont {Billing}\ \emph {et~al.}(2011)\citenamefont {Billing}
  \emph {et~al.}}]{Billing:2011zc}%
  \BibitemOpen
  \bibfield  {author} {\bibinfo {author} {\bibfnamefont {M.~G.}\ \bibnamefont
  {Billing}} \emph {et~al.},\ }in\ \href
  {http://accelconf.web.cern.ch/AccelConf/PAC2011/papers/WEP022.PDF} {\emph
  {\bibinfo {booktitle} {{Proceedings, 24th Particle accelerator Conference
  (PAC'11), New York, NY, Mar 28--Apr 1, 2011}}}}\ (\bibinfo {year} {2011})\
  pp.\ \bibinfo {pages} {1540--1542 (WEP022)}\BibitemShut {NoStop}%
\bibitem [{\citenamefont {Conway}\ \emph {et~al.}(2012)\citenamefont {Conway},
  \citenamefont {Li},\ and\ \citenamefont {Palmer}}]{Conway:2012zza}%
  \BibitemOpen
  \bibfield  {author} {\bibinfo {author} {\bibfnamefont {J.~V.}\ \bibnamefont
  {Conway}}, \bibinfo {author} {\bibfnamefont {Y.}~\bibnamefont {Li}}, \ and\
  \bibinfo {author} {\bibfnamefont {M.~A.}\ \bibnamefont {Palmer}},\ }in\ \href
  {http://accelconf.web.cern.ch/AccelConf/IPAC2012/papers/TUPPR062.PDF} {\emph
  {\bibinfo {booktitle} {{Proceedings, 3rd International Conference on Particle
  accelerator (IPAC 2012), New Orleans, LA, May 2-25, 2012}}}}\ (\bibinfo
  {year} {2012})\ pp.\ \bibinfo {pages} {1960--1962 (TUPPR062)}\BibitemShut
  {NoStop}%
\bibitem [{\citenamefont {Naito}\ \emph {et~al.}(2010)\citenamefont {Naito}
  \emph {et~al.}}]{Naito:2010zzb}%
  \BibitemOpen
  \bibfield  {author} {\bibinfo {author} {\bibfnamefont {T.}~\bibnamefont
  {Naito}} \emph {et~al.},\ }in\ \href
  {http://accelconf.web.cern.ch/AccelConf/IPAC10/papers/weobmh02.pdf} {\emph
  {\bibinfo {booktitle} {{Proceedings, 1st International Particle Accelerator
  Conference (IPAC'10), Kyoto, Japan, May 23--28, 2010}}}}\ (\bibinfo {year}
  {2010})\ pp.\ \bibinfo {pages} {2386--2388 (WEOBMH02)}\BibitemShut {NoStop}%
\bibitem [{\citenamefont {Emma}\ \emph {et~al.}(1996)\citenamefont {Emma},
  \citenamefont {Raubenheimer},\ and\ \citenamefont
  {Zimmermann}}]{Emma:1995kf}%
  \BibitemOpen
  \bibfield  {author} {\bibinfo {author} {\bibfnamefont {P.}~\bibnamefont
  {Emma}}, \bibinfo {author} {\bibfnamefont {T.}~\bibnamefont {Raubenheimer}},
  \ and\ \bibinfo {author} {\bibfnamefont {F.}~\bibnamefont {Zimmermann}},\
  }in\ \href
  {http://accelconf.web.cern.ch/AccelConf/p95/ARTICLES/RPC/RPC03.PDF} {\emph
  {\bibinfo {booktitle} {{Proceedings, 16th Particle Accelerator Conference and
  International Conference on High-Energy Accelerators, (HEACC 1995), Dallas,
  TX, May 1-5, 1995}}}}\ (\bibinfo {year} {1996})\ pp.\ \bibinfo {pages}
  {704--706 (RPC03)}\BibitemShut {NoStop}%
\bibitem [{\citenamefont {Walker}(2015)}]{bib:cr-0010}%
  \BibitemOpen
  \bibfield  {author} {\bibinfo {author} {\bibfnamefont {N.}~\bibnamefont
  {Walker}},\ }\href@noop {} {\enquote {\bibinfo {title} {{Change request
  ILC-CR-0010: Proposal to include bunch compressor sections into main linac
  accelerator system}},}\ }\bibinfo {howpublished}
  {{\url{http://edmsdirect.desy.de/item/D00000001119175}}} (\bibinfo {year}
  {2015})\BibitemShut {NoStop}%
\bibitem [{\citenamefont {Raimondi}\ and\ \citenamefont
  {Seryi}(2001)}]{Raimondi:2000cx}%
  \BibitemOpen
  \bibfield  {author} {\bibinfo {author} {\bibfnamefont {P.}~\bibnamefont
  {Raimondi}}\ and\ \bibinfo {author} {\bibfnamefont {A.}~\bibnamefont
  {Seryi}},\ }\href {\doibase 10.1103/PhysRevLett.86.3779} {\bibfield
  {journal} {\bibinfo  {journal} {Phys. Rev. Lett.}\ }\textbf {\bibinfo
  {volume} {86}},\ \bibinfo {pages} {3779} (\bibinfo {year}
  {2001})}\BibitemShut {NoStop}%
\bibitem [{\citenamefont {Grishanov}\ \emph {et~al.}(2005)\citenamefont
  {Grishanov} \emph {et~al.}}]{Grishanov:2005ek}%
  \BibitemOpen
  \bibfield  {author} {\bibinfo {author} {\bibfnamefont {B.~I.}\ \bibnamefont
  {Grishanov}} \emph {et~al.} (\bibinfo {collaboration} {ATF2}),\ }\href@noop
  {} {\enquote {\bibinfo {title} {{ATF2 Proposal}},}\ }\bibinfo {howpublished}
  {{SLAC-R-771, CERN-AB-2005-035, DESY-05-148, KEK-REPORT-2005-2}} (\bibinfo
  {year} {2005})\BibitemShut {NoStop}%
\bibitem [{\citenamefont {Grishanov}\ \emph {et~al.}(2006)\citenamefont
  {Grishanov} \emph {et~al.}}]{Grishanov:2006kx}%
  \BibitemOpen
  \bibfield  {author} {\bibinfo {author} {\bibfnamefont {B.~I.}\ \bibnamefont
  {Grishanov}} \emph {et~al.},\ }\href@noop {} {\  (\bibinfo {year} {2006})},\
  \Eprint {http://arxiv.org/abs/physics/0606194} {arXiv:physics/0606194
  [physics]} \BibitemShut {NoStop}%
\bibitem [{\citenamefont {Okugi}(2017)}]{Okugi:2017jji}%
  \BibitemOpen
  \bibfield  {author} {\bibinfo {author} {\bibfnamefont {T.}~\bibnamefont
  {Okugi}},\ }in\ \href {\doibase 10.18429/JACoW-LINAC2016-MO3A02} {\emph
  {\bibinfo {booktitle} {{Proceedings, 28th International Linear Accelerator
  Conference (LINAC16), East Lansing, MI, Sep 25-30, 2016}}}}\ (\bibinfo {year}
  {2017})\ pp.\ \bibinfo {pages} {27--31 (MO3A02)},\ \bibinfo {note}
  {doi:10.18429/JACoW-LINAC2016-MO3A02}\BibitemShut {NoStop}%
\bibitem [{\citenamefont {White}\ \emph {et~al.}(2014)\citenamefont {White}
  \emph {et~al.}}]{White:2014vwa}%
  \BibitemOpen
  \bibfield  {author} {\bibinfo {author} {\bibfnamefont {G.~R.}\ \bibnamefont
  {White}} \emph {et~al.} (\bibinfo {collaboration} {ATF2}),\ }\href {\doibase
  10.1103/PhysRevLett.112.034802} {\bibfield  {journal} {\bibinfo  {journal}
  {Phys. Rev. Lett.}\ }\textbf {\bibinfo {volume} {112}},\ \bibinfo {pages}
  {034802} (\bibinfo {year} {2014})}\BibitemShut {NoStop}%
\bibitem [{\citenamefont {Apsimon}\ \emph {et~al.}(2018)\citenamefont {Apsimon}
  \emph {et~al.}}]{Apsimon:2018bpq}%
  \BibitemOpen
  \bibfield  {author} {\bibinfo {author} {\bibfnamefont {R.~J.}\ \bibnamefont
  {Apsimon}} \emph {et~al.},\ }\href {\doibase
  10.1103/PhysRevAccelBeams.21.122802} {\bibfield  {journal} {\bibinfo
  {journal} {Phys. Rev. Accel. Beams}\ }\textbf {\bibinfo {volume} {21}},\
  \bibinfo {pages} {122802} (\bibinfo {year} {2018})},\ \Eprint
  {http://arxiv.org/abs/1812.08432} {arXiv:1812.08432 [physics.acc-ph]}
  \BibitemShut {NoStop}%
\bibitem [{\citenamefont {Ramjiawan}\ \emph {et~al.}(2018)\citenamefont
  {Ramjiawan} \emph {et~al.}}]{Ramjiawan:2018egu}%
  \BibitemOpen
  \bibfield  {author} {\bibinfo {author} {\bibfnamefont {R.}~\bibnamefont
  {Ramjiawan}} \emph {et~al.},\ }in\ \href {\doibase
  10.18429/JACoW-IPAC2018-WEPAL025} {\emph {\bibinfo {booktitle} {{Proceedings,
  9th International Particle Accelerator Conference (IPAC 2018): Vancouver, BC,
  Ap 29--May 4, 2018}}}}\ (\bibinfo {year} {2018})\ pp.\ \bibinfo {pages}
  {2212--2215 (WEPAL025)},\ \bibinfo {note}
  {doi:10.18429/JACoW-IPAC2018-WEPAL025}\BibitemShut {NoStop}%
\bibitem [{\citenamefont {Latina}\ and\ \citenamefont
  {Faus-Golfe}(2018)}]{bib:atf2esu}%
  \BibitemOpen
  \bibfield  {author} {\bibinfo {author} {\bibfnamefont {A.}~\bibnamefont
  {Latina}}\ and\ \bibinfo {author} {\bibfnamefont {A.}~\bibnamefont
  {Faus-Golfe}},\ }\href@noop {} {\enquote {\bibinfo {title} {{ATF2}},}\
  }\bibinfo {howpublished} {Input paper for European Strategy for Particle
  Physics Update} (\bibinfo {year} {2018})\BibitemShut {NoStop}%
\bibitem [{\citenamefont {Walz}\ \emph {et~al.}(1967)\citenamefont {Walz},
  \citenamefont {Lucas}, \citenamefont {Weidner}, \citenamefont {Vetterlein},\
  and\ \citenamefont {Seppi}}]{Walz:1967nz}%
  \BibitemOpen
  \bibfield  {author} {\bibinfo {author} {\bibfnamefont {D.~R.}\ \bibnamefont
  {Walz}}, \bibinfo {author} {\bibfnamefont {L.~R.}\ \bibnamefont {Lucas}},
  \bibinfo {author} {\bibfnamefont {H.~A.}\ \bibnamefont {Weidner}}, \bibinfo
  {author} {\bibfnamefont {R.~J.}\ \bibnamefont {Vetterlein}}, \ and\ \bibinfo
  {author} {\bibfnamefont {E.~J.}\ \bibnamefont {Seppi}},\ }\bibfield
  {booktitle} {\emph {\bibinfo {booktitle} {{2nd IEEE Particle Accelerator
  Conference Washington, D.C., March 1-3, 1967}}},\ }\href {\doibase
  10.1109/TNS.1967.4324681} {\bibfield  {journal} {\bibinfo  {journal} {IEEE
  Trans. Nucl. Sci.}\ }\textbf {\bibinfo {volume} {14}},\ \bibinfo {pages}
  {923} (\bibinfo {year} {1967})}\BibitemShut {NoStop}%
\bibitem [{\citenamefont {Boogert}\ \emph {et~al.}(2009)\citenamefont {Boogert}
  \emph {et~al.}}]{Boogert:2009ir}%
  \BibitemOpen
  \bibfield  {author} {\bibinfo {author} {\bibfnamefont {S.}~\bibnamefont
  {Boogert}} \emph {et~al.},\ }\href {\doibase 10.1088/1748-0221/4/10/P10015}
  {\bibfield  {journal} {\bibinfo  {journal} {JINST}\ }\textbf {\bibinfo
  {volume} {4}},\ \bibinfo {pages} {P10015} (\bibinfo {year} {2009})},\ \Eprint
  {http://arxiv.org/abs/0904.0122} {arXiv:0904.0122 [physics.ins-det]}
  \BibitemShut {NoStop}%
\bibitem [{\citenamefont {Vormwald}\ \emph {et~al.}(2016)\citenamefont
  {Vormwald}, \citenamefont {List},\ and\ \citenamefont
  {Vauth}}]{Vormwald:2015hla}%
  \BibitemOpen
  \bibfield  {author} {\bibinfo {author} {\bibfnamefont {B.}~\bibnamefont
  {Vormwald}}, \bibinfo {author} {\bibfnamefont {J.}~\bibnamefont {List}}, \
  and\ \bibinfo {author} {\bibfnamefont {A.}~\bibnamefont {Vauth}},\ }\href
  {\doibase 10.1088/1748-0221/11/01/P01014} {\bibfield  {journal} {\bibinfo
  {journal} {JINST}\ }\textbf {\bibinfo {volume} {11}},\ \bibinfo {pages}
  {P01014} (\bibinfo {year} {2016})},\ \Eprint
  {http://arxiv.org/abs/1509.03178} {arXiv:1509.03178 [physics.ins-det]}
  \BibitemShut {NoStop}%
\bibitem [{\citenamefont {Sanuki}(2018{\natexlab{a}})}]{bib:sanuki:lcws2018}%
  \BibitemOpen
  \bibfield  {author} {\bibinfo {author} {\bibfnamefont {T.}~\bibnamefont
  {Sanuki}},\ }\href
  {https://agenda.linearcollider.org/event/7826/contributions/41496/attachments/33213/50725/180531.pdf}
  {\enquote {\bibinfo {title} {Tunnel floor vibration issues},}\ }\bibinfo
  {howpublished} {{Presentation, Asian Linear Collider Workshop (ALCW2018),
  Fukuoka, Japan, May 28--Jun 2, 2018.
  \url{https://agenda.linearcollider.org/event/7826/}}} (\bibinfo {year}
  {2018}{\natexlab{a}})\BibitemShut {NoStop}%
\bibitem [{\citenamefont {Harrison}\ \emph
  {et~al.}(2013{\natexlab{b}})\citenamefont {Harrison}, \citenamefont {Ross},\
  and\ \citenamefont {Walker}}]{Harrison:2013nva}%
  \BibitemOpen
  \bibfield  {author} {\bibinfo {author} {\bibfnamefont {M.}~\bibnamefont
  {Harrison}}, \bibinfo {author} {\bibfnamefont {M.}~\bibnamefont {Ross}}, \
  and\ \bibinfo {author} {\bibfnamefont {N.}~\bibnamefont {Walker}},\ }in\
  \href@noop {} {\emph {\bibinfo {booktitle} {{Proceedings, 2013 Community
  Summer Study on the Future of U.S. Particle Physics: Snowmass on the
  Mississippi (CSS2013): Minneapolis, MN, USA, July 29-August 6, 2013}}}}\
  (\bibinfo {year} {2013})\ \Eprint {http://arxiv.org/abs/1308.3726}
  {arXiv:1308.3726 [physics.acc-ph]} \BibitemShut {NoStop}%
\bibitem [{\citenamefont {{ILC Strategy Council}}(2014)}]{ILCSC:2014a}%
  \BibitemOpen
  \bibfield  {author} {\bibinfo {author} {\bibnamefont {{ILC Strategy
  Council}}},\ }\href@noop {} {\enquote {\bibinfo {title} {{Announcement of the
  results of the ILC candidate site evaluation in Japan}},}\ }\bibinfo
  {howpublished} {Press release 28.8.2014,
  \url{http://ilc-str.jp/topics/2013/08281826/}} (\bibinfo {year}
  {2014})\BibitemShut {NoStop}%
\bibitem [{\citenamefont {Warmbein}(2014)}]{Warmbein:2014a}%
  \BibitemOpen
  \bibfield  {author} {\bibinfo {author} {\bibfnamefont {B.}~\bibnamefont
  {Warmbein}},\ }\href@noop {} {\enquote {\bibinfo {title} {The road to
  {Kitakami}},}\ }\bibinfo {howpublished} {ILC Newsline Feb. 20, 2014,
  \url{http://newsline.linearcollider.org/2014/02/20/the-road-to-kitakami/}}
  (\bibinfo {year} {2014})\BibitemShut {NoStop}%
\bibitem [{\citenamefont {Sanuki}(2015)}]{Sanuki:2015a}%
  \BibitemOpen
  \bibfield  {author} {\bibinfo {author} {\bibfnamefont {T.}~\bibnamefont
  {Sanuki}},\ }\href@noop {} {\enquote {\bibinfo {title} {New developments at
  the {Kitakami} site},}\ }\bibinfo {howpublished} {Presentation at Linear
  Collider Workshop 2015 (LCWS15), Whistler, BC, Canada, Nov 1--7, 2015,
  \url{https://agenda.linearcollider.org/event/6662/}} (\bibinfo {year}
  {2015})\BibitemShut {NoStop}%
\bibitem [{\citenamefont {Sanuki}\ and\ \citenamefont
  {Sekine}(2018)}]{Sanuki:2018b}%
  \BibitemOpen
  \bibfield  {author} {\bibinfo {author} {\bibfnamefont {T.}~\bibnamefont
  {Sanuki}}\ and\ \bibinfo {author} {\bibfnamefont {I.}~\bibnamefont
  {Sekine}},\ }\href@noop {} {\enquote {\bibinfo {title} {Seismic base
  isolation for detectors and accelerator},}\ }\bibinfo {howpublished}
  {Presentation at International Workshop on Future Linear Colliders, LCWS2018,
  Arlington, TX, Oct 22-26, 2018,
  \url{https://agenda.linearcollider.org/event/7889/}} (\bibinfo {year}
  {2018})\BibitemShut {NoStop}%
\bibitem [{\citenamefont {Sanuki}(2018{\natexlab{b}})}]{Sanuki:2018a}%
  \BibitemOpen
  \bibfield  {author} {\bibinfo {author} {\bibfnamefont {T.}~\bibnamefont
  {Sanuki}},\ }\href@noop {} {\enquote {\bibinfo {title} {Tunnel floor
  vibration issue},}\ }\bibinfo {howpublished} {Presentation at Asian Linear
  Collider Workshop (ALCW 2018), Fukuoka, Japan, May 28--Jun 2, 2018,
  \url{https://agenda.linearcollider.org/event/7826/}} (\bibinfo {year}
  {2018}{\natexlab{b}})\BibitemShut {NoStop}%
\bibitem [{\citenamefont {{OECD}}(2018)}]{OECD:2018}%
  \BibitemOpen
  \bibfield  {author} {\bibinfo {author} {\bibnamefont {{OECD}}},\ }\href@noop
  {} {\enquote {\bibinfo {title} {Prices and purchasing power parities
  ({PPP})},}\ }\bibinfo {howpublished}
  {\url{http://www.oecd.org/sdd/prices-ppp/}} (\bibinfo {year}
  {2018})\BibitemShut {NoStop}%
\bibitem [{\citenamefont {Eurostat}(2012)}]{Eurostat:2012}%
  \BibitemOpen
  \bibfield  {author} {\bibinfo {author} {\bibnamefont {Eurostat}},\
  }\href@noop {} {\emph {\bibinfo {title} {{Eurostat-OECD methodological manual
  on Purchasing Power Parities}}}},\ \bibinfo {edition} {2012th}\ ed.\
  (\bibinfo  {publisher} {Luxembourg: Publications Office of the European
  Union},\ \bibinfo {year} {2012})\BibitemShut {NoStop}%
\bibitem [{\citenamefont {{ILC Advisory Panel}}(2018)}]{ILCAP:2018}%
  \BibitemOpen
  \bibfield  {author} {\bibinfo {author} {\bibnamefont {{ILC Advisory
  Panel}}},\ }\href@noop {} {\enquote {\bibinfo {title} {Summary of the {ILC}
  advisory panel's discussions to date after revision},}\ }\bibinfo
  {howpublished} {Report, Jul 4, 2018,
  \url{http://www.mext.go.jp/component/b_menu/shingi/toushin/__icsFiles/afieldfile/2018/09/20/1409220_2_1.pdf}}
  (\bibinfo {year} {2018})\BibitemShut {NoStop}%
\bibitem [{\citenamefont {Habermehl}(2018)}]{Habermehl:417605}%
  \BibitemOpen
  \bibfield  {author} {\bibinfo {author} {\bibfnamefont {M.}~\bibnamefont
  {Habermehl}},\ }\emph {\bibinfo {title} {{D}ark {M}atter at the
  {I}nternational {L}inear {C}ollider}},\ \href {\doibase
  10.3204/PUBDB-2018-05723} {\bibinfo {type} {Dissertation}},\ \bibinfo
  {school} {Universität Hamburg}, \bibinfo {address} {Hamburg} (\bibinfo
  {year} {2018}),\ \bibinfo {note} {dissertation, Universität Hamburg,
  2018}\BibitemShut {NoStop}%
\bibitem [{\citenamefont {Barklow}\ \emph {et~al.}(2015)\citenamefont
  {Barklow}, \citenamefont {Brau}, \citenamefont {Fujii}, \citenamefont {Gao},
  \citenamefont {List}, \citenamefont {Walker},\ and\ \citenamefont
  {Yokoya}}]{Barklow:2015tja}%
  \BibitemOpen
  \bibfield  {author} {\bibinfo {author} {\bibfnamefont {T.}~\bibnamefont
  {Barklow}}, \bibinfo {author} {\bibfnamefont {J.}~\bibnamefont {Brau}},
  \bibinfo {author} {\bibfnamefont {K.}~\bibnamefont {Fujii}}, \bibinfo
  {author} {\bibfnamefont {J.}~\bibnamefont {Gao}}, \bibinfo {author}
  {\bibfnamefont {J.}~\bibnamefont {List}}, \bibinfo {author} {\bibfnamefont
  {N.}~\bibnamefont {Walker}}, \ and\ \bibinfo {author} {\bibfnamefont
  {K.}~\bibnamefont {Yokoya}},\ }\href@noop {} {\  (\bibinfo {year} {2015})},\
  \Eprint {http://arxiv.org/abs/1506.07830} {arXiv:1506.07830 [hep-ex]}
  \BibitemShut {NoStop}%
\bibitem [{\citenamefont {Asner}\ \emph {et~al.}(2013)\citenamefont {Asner}
  \emph {et~al.}}]{Asner:2013psa}%
  \BibitemOpen
  \bibfield  {author} {\bibinfo {author} {\bibfnamefont {D.~M.}\ \bibnamefont
  {Asner}} \emph {et~al.},\ }in\ \href
  {http://www.slac.stanford.edu/econf/C1307292/docs/submittedArxivFiles/1310.0763.pdf}
  {\emph {\bibinfo {booktitle} {{Proceedings, 2013 Community Summer Study on
  the Future of U.S. Particle Physics: Snowmass on the Mississippi (CSS2013):
  Minneapolis, MN, USA, July 29-August 6, 2013}}}}\ (\bibinfo {year} {2013})\
  \Eprint {http://arxiv.org/abs/1310.0763} {arXiv:1310.0763 [hep-ph]}
  \BibitemShut {NoStop}%
\bibitem [{\citenamefont {Karl}(2019)}]{bib:PhDRobert}%
  \BibitemOpen
  \bibfield  {author} {\bibinfo {author} {\bibfnamefont {R.}~\bibnamefont
  {Karl}},\ }\emph {\bibinfo {title} {From the Machine-Detector Interface to
  Electroweak Precision Measurements at the ILC --- Beam Gas Backgrounds, Beam
  Polarization and Triple Gauge Couplings}},\ \href@noop {} {\bibinfo {type}
  {Dissertation in preparation}},\ \bibinfo  {school} {Universität Hamburg},
  \bibinfo {address} {Hamburg} (\bibinfo {year} {2019}),\ \bibinfo {note}
  {dissertation, Universität Hamburg, 2019}\BibitemShut {NoStop}%
\bibitem [{\citenamefont {Fujii}\ \emph {et~al.}()\citenamefont {Fujii} \emph
  {et~al.}}]{Fujii:2018mli}%
  \BibitemOpen
  \bibfield  {author} {\bibinfo {author} {\bibfnamefont {K.}~\bibnamefont
  {Fujii}} \emph {et~al.},\ }\href@noop {} {\ }\Eprint
  {http://arxiv.org/abs/1801.02840} {arXiv:1801.02840 [hep-ph]} \BibitemShut
  {NoStop}%
\bibitem [{\citenamefont {Durieux}\ \emph {et~al.}(2017)\citenamefont
  {Durieux}, \citenamefont {Grojean}, \citenamefont {Gu},\ and\ \citenamefont
  {Wang}}]{Durieux:2017rsg}%
  \BibitemOpen
  \bibfield  {author} {\bibinfo {author} {\bibfnamefont {G.}~\bibnamefont
  {Durieux}}, \bibinfo {author} {\bibfnamefont {C.}~\bibnamefont {Grojean}},
  \bibinfo {author} {\bibfnamefont {J.}~\bibnamefont {Gu}}, \ and\ \bibinfo
  {author} {\bibfnamefont {K.}~\bibnamefont {Wang}},\ }\href {\doibase
  10.1007/JHEP09(2017)014} {\bibfield  {journal} {\bibinfo  {journal} {JHEP}\
  }\textbf {\bibinfo {volume} {09}},\ \bibinfo {pages} {014} (\bibinfo {year}
  {2017})},\ \Eprint {http://arxiv.org/abs/1704.02333} {arXiv:1704.02333
  [hep-ph]} \BibitemShut {NoStop}%
\bibitem [{\citenamefont {Aarons}\ \emph {et~al.}()\citenamefont {Aarons} \emph
  {et~al.}}]{Phinney:2007gp}%
  \BibitemOpen
  \bibfield  {author} {\bibinfo {author} {\bibfnamefont {G.}~\bibnamefont
  {Aarons}} \emph {et~al.},\ }\href@noop {} {\emph {\bibinfo {title} {ILC
  Reference Design Report Volume 3 - Accelerator}}},\ \bibinfo {type} {Tech.
  Rep.},\ \Eprint {http://arxiv.org/abs/0712.2361} {arXiv:0712.2361
  [physics.acc-ph]} \BibitemShut {NoStop}%
\bibitem [{\citenamefont {Jeans}(SLAC)}]{bib:jeans_awlc17}%
  \BibitemOpen
  \bibfield  {author} {\bibinfo {author} {\bibfnamefont {D.}~\bibnamefont
  {Jeans}},\ }\href@noop {} {\enquote {\bibinfo {title} {Ilc beam paramaters at
  250\,gev: Implications for physics and detectors},}\ }\bibinfo {howpublished}
  {{\url{https://agenda.linearcollider.org/event/7507/contributions/39291/}}}
  (\bibinfo {year} {presentation at AWLC 2017, SLAC})\BibitemShut {NoStop}%
\bibitem [{\citenamefont {Arbey}\ \emph {et~al.}(2015)\citenamefont {Arbey}
  \emph {et~al.}}]{Moortgat-Picka:2015yla}%
  \BibitemOpen
  \bibfield  {author} {\bibinfo {author} {\bibfnamefont {A.}~\bibnamefont
  {Arbey}} \emph {et~al.},\ }\href {\doibase 10.1140/epjc/s10052-015-3511-9}
  {\bibfield  {journal} {\bibinfo  {journal} {Eur. Phys. J.}\ }\textbf
  {\bibinfo {volume} {C75}},\ \bibinfo {pages} {371} (\bibinfo {year}
  {2015})},\ \Eprint {http://arxiv.org/abs/1504.01726} {arXiv:1504.01726
  [hep-ph]} \BibitemShut {NoStop}%
\bibitem [{\citenamefont {Aurand}\ \emph {et~al.}(2009)\citenamefont {Aurand}
  \emph {et~al.}}]{Aurand:2009kp}%
  \BibitemOpen
  \bibfield  {author} {\bibinfo {author} {\bibfnamefont {B.}~\bibnamefont
  {Aurand}} \emph {et~al.},\ }in\ \href
  {http://www-public.slac.stanford.edu/sciDoc/docMeta.aspx?slacPubNumber=SLAC-PUB-14776}
  {\emph {\bibinfo {booktitle} {{Linear colliders. Proceedings, International
  Linear Collider Workshop, LCWS08, and International Linear Collider Meeting,
  ILC08, Chicago, USA, Novermber 16-20, 2008}}}}\ (\bibinfo {year} {2009})\
  \Eprint {http://arxiv.org/abs/0903.2959} {arXiv:0903.2959 [physics.acc-ph]}
  \BibitemShut {NoStop}%
\bibitem [{\citenamefont {Moortgat-Pick}\ \emph {et~al.}(2008)\citenamefont
  {Moortgat-Pick} \emph {et~al.}}]{MoortgatPick:2005cw}%
  \BibitemOpen
  \bibfield  {author} {\bibinfo {author} {\bibfnamefont {G.}~\bibnamefont
  {Moortgat-Pick}} \emph {et~al.},\ }\href {\doibase
  10.1016/j.physrep.2007.12.003} {\bibfield  {journal} {\bibinfo  {journal}
  {Phys. Rept.}\ }\textbf {\bibinfo {volume} {460}},\ \bibinfo {pages} {131}
  (\bibinfo {year} {2008})},\ \Eprint {http://arxiv.org/abs/hep-ph/0507011}
  {arXiv:hep-ph/0507011 [hep-ph]} \BibitemShut {NoStop}%
\bibitem [{\citenamefont {Haber}(1994)}]{Haber:1994mt}%
  \BibitemOpen
  \bibfield  {author} {\bibinfo {author} {\bibfnamefont {H.~E.}\ \bibnamefont
  {Haber}},\ }in\ \href@noop {} {\emph {\bibinfo {booktitle} {{Joint
  U.S.-Polish Workshop on Physics from Planck Scale to Electro-Weak Scale (SUSY
  94) Warsaw, Poland, September 21-24, 1994}}}}\ (\bibinfo {year} {1994})\ pp.\
  \bibinfo {pages} {1--16},\ \bibinfo {note} {[,1(1994)]},\ \Eprint
  {http://arxiv.org/abs/hep-ph/9501320} {arXiv:hep-ph/9501320 [hep-ph]}
  \BibitemShut {NoStop}%
\bibitem [{\citenamefont {Lepage}\ \emph {et~al.}(2014)\citenamefont {Lepage},
  \citenamefont {Mackenzie},\ and\ \citenamefont {Peskin}}]{Lepage:2014fla}%
  \BibitemOpen
  \bibfield  {author} {\bibinfo {author} {\bibfnamefont {G.~P.}\ \bibnamefont
  {Lepage}}, \bibinfo {author} {\bibfnamefont {P.~B.}\ \bibnamefont
  {Mackenzie}}, \ and\ \bibinfo {author} {\bibfnamefont {M.~E.}\ \bibnamefont
  {Peskin}},\ }\href@noop {} {\  (\bibinfo {year} {2014})},\ \Eprint
  {http://arxiv.org/abs/1404.0319} {arXiv:1404.0319 [hep-ph]} \BibitemShut
  {NoStop}%
\bibitem [{\citenamefont {Wells}\ and\ \citenamefont
  {Zhang}(2018)}]{Wells:2017vla}%
  \BibitemOpen
  \bibfield  {author} {\bibinfo {author} {\bibfnamefont {J.~D.}\ \bibnamefont
  {Wells}}\ and\ \bibinfo {author} {\bibfnamefont {Z.}~\bibnamefont {Zhang}},\
  }\href {\doibase 10.1007/JHEP05(2018)182} {\bibfield  {journal} {\bibinfo
  {journal} {JHEP}\ }\textbf {\bibinfo {volume} {05}},\ \bibinfo {pages} {182}
  (\bibinfo {year} {2018})},\ \Eprint {http://arxiv.org/abs/1711.04774}
  {arXiv:1711.04774 [hep-ph]} \BibitemShut {NoStop}%
\bibitem [{\citenamefont {Cahill-Rowley}\ \emph {et~al.}(2014)\citenamefont
  {Cahill-Rowley}, \citenamefont {Hewett}, \citenamefont {Ismail},\ and\
  \citenamefont {Rizzo}}]{Cahill-Rowley:2014wba}%
  \BibitemOpen
  \bibfield  {author} {\bibinfo {author} {\bibfnamefont {M.}~\bibnamefont
  {Cahill-Rowley}}, \bibinfo {author} {\bibfnamefont {J.}~\bibnamefont
  {Hewett}}, \bibinfo {author} {\bibfnamefont {A.}~\bibnamefont {Ismail}}, \
  and\ \bibinfo {author} {\bibfnamefont {T.}~\bibnamefont {Rizzo}},\ }\href
  {\doibase 10.1103/PhysRevD.90.095017} {\bibfield  {journal} {\bibinfo
  {journal} {Phys. Rev.}\ }\textbf {\bibinfo {volume} {D90}},\ \bibinfo {pages}
  {095017} (\bibinfo {year} {2014})},\ \Eprint {http://arxiv.org/abs/1407.7021}
  {arXiv:1407.7021 [hep-ph]} \BibitemShut {NoStop}%
\bibitem [{\citenamefont {Kanemura}\ \emph {et~al.}(2015)\citenamefont
  {Kanemura}, \citenamefont {Kikuchi},\ and\ \citenamefont
  {Yagyu}}]{Kanemura:2015mxa}%
  \BibitemOpen
  \bibfield  {author} {\bibinfo {author} {\bibfnamefont {S.}~\bibnamefont
  {Kanemura}}, \bibinfo {author} {\bibfnamefont {M.}~\bibnamefont {Kikuchi}}, \
  and\ \bibinfo {author} {\bibfnamefont {K.}~\bibnamefont {Yagyu}},\ }\href
  {\doibase 10.1016/j.nuclphysb.2015.04.015} {\bibfield  {journal} {\bibinfo
  {journal} {Nucl. Phys.}\ }\textbf {\bibinfo {volume} {B896}},\ \bibinfo
  {pages} {80} (\bibinfo {year} {2015})},\ \Eprint
  {http://arxiv.org/abs/1502.07716} {arXiv:1502.07716 [hep-ph]} \BibitemShut
  {NoStop}%
\bibitem [{\citenamefont {Di~Vita}\ \emph {et~al.}(2017)\citenamefont
  {Di~Vita}, \citenamefont {Grojean}, \citenamefont {Panico}, \citenamefont
  {Riembau},\ and\ \citenamefont {Vantalon}}]{DiVita:2017eyz}%
  \BibitemOpen
  \bibfield  {author} {\bibinfo {author} {\bibfnamefont {S.}~\bibnamefont
  {Di~Vita}}, \bibinfo {author} {\bibfnamefont {C.}~\bibnamefont {Grojean}},
  \bibinfo {author} {\bibfnamefont {G.}~\bibnamefont {Panico}}, \bibinfo
  {author} {\bibfnamefont {M.}~\bibnamefont {Riembau}}, \ and\ \bibinfo
  {author} {\bibfnamefont {T.}~\bibnamefont {Vantalon}},\ }\href {\doibase
  10.1007/JHEP09(2017)069} {\bibfield  {journal} {\bibinfo  {journal} {JHEP}\
  }\textbf {\bibinfo {volume} {09}},\ \bibinfo {pages} {069} (\bibinfo {year}
  {2017})},\ \Eprint {http://arxiv.org/abs/1704.01953} {arXiv:1704.01953
  [hep-ph]} \BibitemShut {NoStop}%
\bibitem [{\citenamefont {Contino}\ \emph {et~al.}(2013)\citenamefont
  {Contino}, \citenamefont {Ghezzi}, \citenamefont {Grojean}, \citenamefont
  {Muhlleitner},\ and\ \citenamefont {Spira}}]{Contino:2013kra}%
  \BibitemOpen
  \bibfield  {author} {\bibinfo {author} {\bibfnamefont {R.}~\bibnamefont
  {Contino}}, \bibinfo {author} {\bibfnamefont {M.}~\bibnamefont {Ghezzi}},
  \bibinfo {author} {\bibfnamefont {C.}~\bibnamefont {Grojean}}, \bibinfo
  {author} {\bibfnamefont {M.}~\bibnamefont {Muhlleitner}}, \ and\ \bibinfo
  {author} {\bibfnamefont {M.}~\bibnamefont {Spira}},\ }\href {\doibase
  10.1007/JHEP07(2013)035} {\bibfield  {journal} {\bibinfo  {journal} {JHEP}\
  }\textbf {\bibinfo {volume} {07}},\ \bibinfo {pages} {035} (\bibinfo {year}
  {2013})},\ \Eprint {http://arxiv.org/abs/1303.3876} {arXiv:1303.3876
  [hep-ph]} \BibitemShut {NoStop}%
\bibitem [{\citenamefont {Han}\ \emph {et~al.}(2003)\citenamefont {Han},
  \citenamefont {Logan}, \citenamefont {McElrath},\ and\ \citenamefont
  {Wang}}]{Han:2003gf}%
  \BibitemOpen
  \bibfield  {author} {\bibinfo {author} {\bibfnamefont {T.}~\bibnamefont
  {Han}}, \bibinfo {author} {\bibfnamefont {H.~E.}\ \bibnamefont {Logan}},
  \bibinfo {author} {\bibfnamefont {B.}~\bibnamefont {McElrath}}, \ and\
  \bibinfo {author} {\bibfnamefont {L.-T.}\ \bibnamefont {Wang}},\ }\href
  {\doibase 10.1016/j.physletb.2004.10.021, 10.1016/S0370-2693(03)00657-9}
  {\bibfield  {journal} {\bibinfo  {journal} {Phys. Lett.}\ }\textbf {\bibinfo
  {volume} {B563}},\ \bibinfo {pages} {191} (\bibinfo {year} {2003})},\
  \bibinfo {note} {[Erratum: Phys. Lett.B603,257(2004)]},\ \Eprint
  {http://arxiv.org/abs/hep-ph/0302188} {arXiv:hep-ph/0302188 [hep-ph]}
  \BibitemShut {NoStop}%
\bibitem [{\citenamefont {Agashe}\ \emph {et~al.}(2005)\citenamefont {Agashe},
  \citenamefont {Contino},\ and\ \citenamefont {Pomarol}}]{Agashe:2004rs}%
  \BibitemOpen
  \bibfield  {author} {\bibinfo {author} {\bibfnamefont {K.}~\bibnamefont
  {Agashe}}, \bibinfo {author} {\bibfnamefont {R.}~\bibnamefont {Contino}}, \
  and\ \bibinfo {author} {\bibfnamefont {A.}~\bibnamefont {Pomarol}},\ }\href
  {\doibase 10.1016/j.nuclphysb.2005.04.035} {\bibfield  {journal} {\bibinfo
  {journal} {Nucl. Phys.}\ }\textbf {\bibinfo {volume} {B719}},\ \bibinfo
  {pages} {165} (\bibinfo {year} {2005})},\ \Eprint
  {http://arxiv.org/abs/hep-ph/0412089} {arXiv:hep-ph/0412089 [hep-ph]}
  \BibitemShut {NoStop}%
\bibitem [{\citenamefont {Cepeda}\ \emph {et~al.}(2019)\citenamefont {Cepeda}
  \emph {et~al.}}]{Cepeda:2019klc}%
  \BibitemOpen
  \bibfield  {author} {\bibinfo {author} {\bibfnamefont {M.}~\bibnamefont
  {Cepeda}} \emph {et~al.} (\bibinfo {collaboration} {Physics of the HL-LHC
  Working Group}),\ }\href@noop {} {\  (\bibinfo {year} {2019})},\ \Eprint
  {http://arxiv.org/abs/1902.00134} {arXiv:1902.00134 [hep-ph]} \BibitemShut
  {NoStop}%
\bibitem [{\citenamefont {Curtin}\ \emph {et~al.}(2014)\citenamefont {Curtin}
  \emph {et~al.}}]{Curtin:2013fra}%
  \BibitemOpen
  \bibfield  {author} {\bibinfo {author} {\bibfnamefont {D.}~\bibnamefont
  {Curtin}} \emph {et~al.},\ }\href {\doibase 10.1103/PhysRevD.90.075004}
  {\bibfield  {journal} {\bibinfo  {journal} {Phys. Rev.}\ }\textbf {\bibinfo
  {volume} {D90}},\ \bibinfo {pages} {075004} (\bibinfo {year} {2014})},\
  \Eprint {http://arxiv.org/abs/1312.4992} {arXiv:1312.4992 [hep-ph]}
  \BibitemShut {NoStop}%
\bibitem [{\citenamefont {Liu}\ \emph {et~al.}(2017)\citenamefont {Liu},
  \citenamefont {Wang},\ and\ \citenamefont {Zhang}}]{Liu:2016zki}%
  \BibitemOpen
  \bibfield  {author} {\bibinfo {author} {\bibfnamefont {Z.}~\bibnamefont
  {Liu}}, \bibinfo {author} {\bibfnamefont {L.-T.}\ \bibnamefont {Wang}}, \
  and\ \bibinfo {author} {\bibfnamefont {H.}~\bibnamefont {Zhang}},\ }\href
  {\doibase 10.1088/1674-1137/41/6/063102} {\bibfield  {journal} {\bibinfo
  {journal} {Chin. Phys.}\ }\textbf {\bibinfo {volume} {C41}},\ \bibinfo
  {pages} {063102} (\bibinfo {year} {2017})},\ \Eprint
  {http://arxiv.org/abs/1612.09284} {arXiv:1612.09284 [hep-ph]} \BibitemShut
  {NoStop}%
\bibitem [{\citenamefont {Fujii}\ \emph {et~al.}(2015)\citenamefont {Fujii}
  \emph {et~al.}}]{Fujii:2015jha}%
  \BibitemOpen
  \bibfield  {author} {\bibinfo {author} {\bibfnamefont {K.}~\bibnamefont
  {Fujii}} \emph {et~al.},\ }\href@noop {} {\  (\bibinfo {year} {2015})},\
  \Eprint {http://arxiv.org/abs/1506.05992} {arXiv:1506.05992 [hep-ex]}
  \BibitemShut {NoStop}%
\bibitem [{\citenamefont {Grzadkowski}\ \emph {et~al.}(2010)\citenamefont
  {Grzadkowski}, \citenamefont {Iskrzynski}, \citenamefont {Misiak},\ and\
  \citenamefont {Rosiek}}]{Grzadkowski:2010es}%
  \BibitemOpen
  \bibfield  {author} {\bibinfo {author} {\bibfnamefont {B.}~\bibnamefont
  {Grzadkowski}}, \bibinfo {author} {\bibfnamefont {M.}~\bibnamefont
  {Iskrzynski}}, \bibinfo {author} {\bibfnamefont {M.}~\bibnamefont {Misiak}},
  \ and\ \bibinfo {author} {\bibfnamefont {J.}~\bibnamefont {Rosiek}},\ }\href
  {\doibase 10.1007/JHEP10(2010)085} {\bibfield  {journal} {\bibinfo  {journal}
  {JHEP}\ }\textbf {\bibinfo {volume} {10}},\ \bibinfo {pages} {085} (\bibinfo
  {year} {2010})},\ \Eprint {http://arxiv.org/abs/1008.4884} {arXiv:1008.4884
  [hep-ph]} \BibitemShut {NoStop}%
\bibitem [{\citenamefont {Peskin}\ and\ \citenamefont
  {Takeuchi}(1990)}]{Peskin:1990zt}%
  \BibitemOpen
  \bibfield  {author} {\bibinfo {author} {\bibfnamefont {M.~E.}\ \bibnamefont
  {Peskin}}\ and\ \bibinfo {author} {\bibfnamefont {T.}~\bibnamefont
  {Takeuchi}},\ }\href {\doibase 10.1103/PhysRevLett.65.964} {\bibfield
  {journal} {\bibinfo  {journal} {Phys. Rev. Lett.}\ }\textbf {\bibinfo
  {volume} {65}},\ \bibinfo {pages} {964} (\bibinfo {year} {1990})}\BibitemShut
  {NoStop}%
\bibitem [{\citenamefont {Hagiwara}\ \emph {et~al.}(1987)\citenamefont
  {Hagiwara}, \citenamefont {Peccei}, \citenamefont {Zeppenfeld},\ and\
  \citenamefont {Hikasa}}]{Hagiwara:1986vm}%
  \BibitemOpen
  \bibfield  {author} {\bibinfo {author} {\bibfnamefont {K.}~\bibnamefont
  {Hagiwara}}, \bibinfo {author} {\bibfnamefont {R.~D.}\ \bibnamefont
  {Peccei}}, \bibinfo {author} {\bibfnamefont {D.}~\bibnamefont {Zeppenfeld}},
  \ and\ \bibinfo {author} {\bibfnamefont {K.}~\bibnamefont {Hikasa}},\ }\href
  {\doibase 10.1016/0550-3213(87)90685-7} {\bibfield  {journal} {\bibinfo
  {journal} {Nucl. Phys.}\ }\textbf {\bibinfo {volume} {B282}},\ \bibinfo
  {pages} {253} (\bibinfo {year} {1987})}\BibitemShut {NoStop}%
\bibitem [{\citenamefont {Falkowski}\ \emph {et~al.}(2017)\citenamefont
  {Falkowski}, \citenamefont {Gonzalez-Alonso}, \citenamefont {Greljo},
  \citenamefont {Marzocca},\ and\ \citenamefont {Son}}]{Falkowski:2016cxu}%
  \BibitemOpen
  \bibfield  {author} {\bibinfo {author} {\bibfnamefont {A.}~\bibnamefont
  {Falkowski}}, \bibinfo {author} {\bibfnamefont {M.}~\bibnamefont
  {Gonzalez-Alonso}}, \bibinfo {author} {\bibfnamefont {A.}~\bibnamefont
  {Greljo}}, \bibinfo {author} {\bibfnamefont {D.}~\bibnamefont {Marzocca}}, \
  and\ \bibinfo {author} {\bibfnamefont {M.}~\bibnamefont {Son}},\ }\href
  {\doibase 10.1007/JHEP02(2017)115} {\bibfield  {journal} {\bibinfo  {journal}
  {JHEP}\ }\textbf {\bibinfo {volume} {02}},\ \bibinfo {pages} {115} (\bibinfo
  {year} {2017})},\ \Eprint {http://arxiv.org/abs/1609.06312} {arXiv:1609.06312
  [hep-ph]} \BibitemShut {NoStop}%
\bibitem [{\citenamefont {Funatsu}\ \emph {et~al.}(2017)\citenamefont
  {Funatsu}, \citenamefont {Hatanaka}, \citenamefont {Hosotani},\ and\
  \citenamefont {Orikasa}}]{Funatsu:2017nfm}%
  \BibitemOpen
  \bibfield  {author} {\bibinfo {author} {\bibfnamefont {S.}~\bibnamefont
  {Funatsu}}, \bibinfo {author} {\bibfnamefont {H.}~\bibnamefont {Hatanaka}},
  \bibinfo {author} {\bibfnamefont {Y.}~\bibnamefont {Hosotani}}, \ and\
  \bibinfo {author} {\bibfnamefont {Y.}~\bibnamefont {Orikasa}},\ }\href
  {\doibase 10.1016/j.physletb.2017.10.068} {\bibfield  {journal} {\bibinfo
  {journal} {Phys. Lett.}\ }\textbf {\bibinfo {volume} {B775}},\ \bibinfo
  {pages} {297} (\bibinfo {year} {2017})},\ \Eprint
  {http://arxiv.org/abs/1705.05282} {arXiv:1705.05282 [hep-ph]} \BibitemShut
  {NoStop}%
\bibitem [{\citenamefont {Yoon}\ and\ \citenamefont
  {Peskin}(2018)}]{Yoon:2018xud}%
  \BibitemOpen
  \bibfield  {author} {\bibinfo {author} {\bibfnamefont {J.}~\bibnamefont
  {Yoon}}\ and\ \bibinfo {author} {\bibfnamefont {M.~E.}\ \bibnamefont
  {Peskin}},\ }\href@noop {} {\  (\bibinfo {year} {2018})},\ \Eprint
  {http://arxiv.org/abs/1811.07877} {arXiv:1811.07877 [hep-ph]} \BibitemShut
  {NoStop}%
\bibitem [{\citenamefont {Fujii}\ \emph
  {et~al.}(2017{\natexlab{b}})\citenamefont {Fujii} \emph
  {et~al.}}]{Fujii:2017ekh}%
  \BibitemOpen
  \bibfield  {author} {\bibinfo {author} {\bibfnamefont {K.}~\bibnamefont
  {Fujii}} \emph {et~al.},\ }\href@noop {} {\  (\bibinfo {year}
  {2017}{\natexlab{b}})},\ \Eprint {http://arxiv.org/abs/1702.05333}
  {arXiv:1702.05333 [hep-ph]} \BibitemShut {NoStop}%
\bibitem [{\citenamefont {Assmann}\ \emph {et~al.}(1999)\citenamefont {Assmann}
  \emph {et~al.}}]{Assmann:1998qb}%
  \BibitemOpen
  \bibfield  {author} {\bibinfo {author} {\bibfnamefont {R.}~\bibnamefont
  {Assmann}} \emph {et~al.},\ }\href {\doibase 10.1007/s100529801030}
  {\bibfield  {journal} {\bibinfo  {journal} {Eur. Phys. J.}\ }\textbf
  {\bibinfo {volume} {C6}},\ \bibinfo {pages} {187} (\bibinfo {year}
  {1999})}\BibitemShut {NoStop}%
\bibitem [{\citenamefont {Baer}\ \emph {et~al.}(2013)\citenamefont {Baer},
  \citenamefont {Barklow}, \citenamefont {Fujii}, \citenamefont {Gao},
  \citenamefont {Hoang}, \citenamefont {Kanemura}, \citenamefont {List},
  \citenamefont {Logan}, \citenamefont {Nomerotski}, \citenamefont {Perelstein}
  \emph {et~al.}}]{Baer:2013cma}%
  \BibitemOpen
  \bibfield  {author} {\bibinfo {author} {\bibfnamefont {H.}~\bibnamefont
  {Baer}}, \bibinfo {author} {\bibfnamefont {T.}~\bibnamefont {Barklow}},
  \bibinfo {author} {\bibfnamefont {K.}~\bibnamefont {Fujii}}, \bibinfo
  {author} {\bibfnamefont {Y.}~\bibnamefont {Gao}}, \bibinfo {author}
  {\bibfnamefont {A.}~\bibnamefont {Hoang}}, \bibinfo {author} {\bibfnamefont
  {S.}~\bibnamefont {Kanemura}}, \bibinfo {author} {\bibfnamefont
  {J.}~\bibnamefont {List}}, \bibinfo {author} {\bibfnamefont {H.~E.}\
  \bibnamefont {Logan}}, \bibinfo {author} {\bibfnamefont {A.}~\bibnamefont
  {Nomerotski}}, \bibinfo {author} {\bibfnamefont {M.}~\bibnamefont
  {Perelstein}},  \emph {et~al.},\ }\href@noop {} {\  (\bibinfo {year}
  {2013})},\ \Eprint {http://arxiv.org/abs/1306.6352} {arXiv:1306.6352
  [hep-ph]} \BibitemShut {NoStop}%
\bibitem [{\citenamefont {Linssen}\ \emph {et~al.}(2012)\citenamefont
  {Linssen}, \citenamefont {Miyamoto}, \citenamefont {Stanitzki},\ and\
  \citenamefont {Weerts}}]{Linssen:2012hp}%
  \BibitemOpen
  \bibfield  {author} {\bibinfo {author} {\bibfnamefont {L.}~\bibnamefont
  {Linssen}}, \bibinfo {author} {\bibfnamefont {A.}~\bibnamefont {Miyamoto}},
  \bibinfo {author} {\bibfnamefont {M.}~\bibnamefont {Stanitzki}}, \ and\
  \bibinfo {author} {\bibfnamefont {H.}~\bibnamefont {Weerts}},\ }\href
  {\doibase 10.5170/CERN-2012-003} {\  (\bibinfo {year} {2012}),\
  10.5170/CERN-2012-003},\ \Eprint {http://arxiv.org/abs/1202.5940}
  {arXiv:1202.5940 [physics.ins-det]} \BibitemShut {NoStop}%
\bibitem [{\citenamefont {de~Blas}\ \emph {et~al.}(2018)\citenamefont {de~Blas}
  \emph {et~al.}}]{deBlas:2018mhx}%
  \BibitemOpen
  \bibfield  {author} {\bibinfo {author} {\bibfnamefont {J.}~\bibnamefont
  {de~Blas}} \emph {et~al.},\ }\bibfield  {booktitle} {\emph {\bibinfo
  {booktitle} {{The CLIC Potential for New Physics}}},\ }\href@noop {} {\
  (\bibinfo {year} {2018})},\ \Eprint {http://arxiv.org/abs/1812.02093}
  {arXiv:1812.02093 [hep-ph]} \BibitemShut {NoStop}%
\bibitem [{\citenamefont {Roloff}\ \emph {et~al.}(2018)\citenamefont {Roloff},
  \citenamefont {Franceschini}, \citenamefont {Schnoor},\ and\ \citenamefont
  {Wulzer}}]{Roloff:2018dqu}%
  \BibitemOpen
  \bibfield  {author} {\bibinfo {author} {\bibfnamefont {P.}~\bibnamefont
  {Roloff}}, \bibinfo {author} {\bibfnamefont {R.}~\bibnamefont
  {Franceschini}}, \bibinfo {author} {\bibfnamefont {U.}~\bibnamefont
  {Schnoor}}, \ and\ \bibinfo {author} {\bibfnamefont {A.}~\bibnamefont
  {Wulzer}} (\bibinfo {collaboration} {CLIC, CLICdp}),\ }\href@noop {} {\
  (\bibinfo {year} {2018})},\ \Eprint {http://arxiv.org/abs/1812.07986}
  {arXiv:1812.07986 [hep-ex]} \BibitemShut {NoStop}%
\bibitem [{\citenamefont {Aicheler}\ \emph {et~al.}(2012)\citenamefont
  {Aicheler}, \citenamefont {Burrows}, \citenamefont {Draper}, \citenamefont
  {Garvey}, \citenamefont {Lebrun}, \citenamefont {Peach}, \citenamefont
  {Phinney}, \citenamefont {Schmickler}, \citenamefont {Schulte},\ and\
  \citenamefont {Toge}}]{Aicheler:2012bya}%
  \BibitemOpen
  \bibfield  {author} {\bibinfo {author} {\bibfnamefont {M.}~\bibnamefont
  {Aicheler}}, \bibinfo {author} {\bibfnamefont {P.}~\bibnamefont {Burrows}},
  \bibinfo {author} {\bibfnamefont {M.}~\bibnamefont {Draper}}, \bibinfo
  {author} {\bibfnamefont {T.}~\bibnamefont {Garvey}}, \bibinfo {author}
  {\bibfnamefont {P.}~\bibnamefont {Lebrun}}, \bibinfo {author} {\bibfnamefont
  {K.}~\bibnamefont {Peach}}, \bibinfo {author} {\bibfnamefont
  {N.}~\bibnamefont {Phinney}}, \bibinfo {author} {\bibfnamefont
  {H.}~\bibnamefont {Schmickler}}, \bibinfo {author} {\bibfnamefont
  {D.}~\bibnamefont {Schulte}}, \ and\ \bibinfo {author} {\bibfnamefont
  {N.}~\bibnamefont {Toge}},\ }\href {\doibase 10.5170/CERN-2012-007} {\
  (\bibinfo {year} {2012}),\ 10.5170/CERN-2012-007}\BibitemShut {NoStop}%
\bibitem [{\citenamefont {Cros}\ and\ \citenamefont
  {Muggli}(2017)}]{Cros:2017jxp}%
  \BibitemOpen
  \bibfield  {author} {\bibinfo {author} {\bibfnamefont {B.}~\bibnamefont
  {Cros}}\ and\ \bibinfo {author} {\bibfnamefont {P.}~\bibnamefont {Muggli}},\
  }\bibfield  {booktitle} {\emph {\bibinfo {booktitle} {{}Towards a Proposal
  for an Advanced Linear Collider: Report on the Advanced and Novel
  Accelerators for High Energy Physics Roadmap Workshop, CERN}},\ }\href@noop
  {} {\  (\bibinfo {year} {2017})}\BibitemShut {NoStop}%
\bibitem [{\citenamefont {Cros}\ and\ \citenamefont
  {Muggli}(2019)}]{Cros:2019tns}%
  \BibitemOpen
  \bibfield  {author} {\bibinfo {author} {\bibfnamefont {B.}~\bibnamefont
  {Cros}}\ and\ \bibinfo {author} {\bibfnamefont {P.}~\bibnamefont {Muggli}}
  (\bibinfo {collaboration} {ALEGRO}),\ }\bibfield  {booktitle} {\emph
  {\bibinfo {booktitle} {ALEGRO input for the 2020 update of the European
  Strategy}},\ }\href@noop {} {\  (\bibinfo {year} {2019})},\ \Eprint
  {http://arxiv.org/abs/1901.08436} {arXiv:1901.08436 [physics.acc-ph]}
  \BibitemShut {NoStop}%
\bibitem [{ALE(2019)}]{ALEGRO:2019alc}%
  \BibitemOpen
  \href@noop {} {\  (\bibinfo {year} {2019})},\ \Eprint
  {http://arxiv.org/abs/1901.10370} {arXiv:1901.10370 [physics.acc-ph]}
  \BibitemShut {NoStop}%
\bibitem [{\citenamefont {Sakharov}(1967)}]{Sakharov:1967dj}%
  \BibitemOpen
  \bibfield  {author} {\bibinfo {author} {\bibfnamefont {A.~D.}\ \bibnamefont
  {Sakharov}},\ }\href {\doibase 10.1070/PU1991v034n05ABEH002497} {\bibfield
  {journal} {\bibinfo  {journal} {Pisma Zh. Eksp. Teor. Fiz.}\ }\textbf
  {\bibinfo {volume} {5}},\ \bibinfo {pages} {32} (\bibinfo {year} {1967})},\
  \bibinfo {note} {[Usp. Fiz. Nauk161,no.5,61(1991)]}\BibitemShut {NoStop}%
\bibitem [{\citenamefont {Morrissey}\ and\ \citenamefont
  {Ramsey-Musolf}(2012)}]{Morrissey:2012db}%
  \BibitemOpen
  \bibfield  {author} {\bibinfo {author} {\bibfnamefont {D.~E.}\ \bibnamefont
  {Morrissey}}\ and\ \bibinfo {author} {\bibfnamefont {M.~J.}\ \bibnamefont
  {Ramsey-Musolf}},\ }\href {\doibase 10.1088/1367-2630/14/12/125003}
  {\bibfield  {journal} {\bibinfo  {journal} {New J. Phys.}\ }\textbf {\bibinfo
  {volume} {14}},\ \bibinfo {pages} {125003} (\bibinfo {year} {2012})},\
  \Eprint {http://arxiv.org/abs/1206.2942} {arXiv:1206.2942 [hep-ph]}
  \BibitemShut {NoStop}%
\bibitem [{\citenamefont {{Strube, Jan and Titov, Maxim}}(2018)}]{RDliaision}%
  \BibitemOpen
  \bibfield  {author} {\bibinfo {author} {\bibnamefont {{Strube, Jan and Titov,
  Maxim}}},\ }\href@noop {} {\enquote {\bibinfo {title} {{Linear Collider R\&D
  liaison report}},}\ }\bibinfo {howpublished}
  {\url{http://www.linearcollider.org/P-D/Working-groups/Detector-R-D-liaison}}
  (\bibinfo {year} {2018})\BibitemShut {NoStop}%
\bibitem [{\citenamefont {Sinev}\ \emph {et~al.}(2015)\citenamefont {Sinev},
  \citenamefont {Brau}, \citenamefont {Strom}, \citenamefont {Baltay},
  \citenamefont {Emmet},\ and\ \citenamefont {Rabinowitz}}]{Sinev:2015iwr}%
  \BibitemOpen
  \bibfield  {author} {\bibinfo {author} {\bibfnamefont {N.}~\bibnamefont
  {Sinev}}, \bibinfo {author} {\bibfnamefont {J.}~\bibnamefont {Brau}},
  \bibinfo {author} {\bibfnamefont {D.}~\bibnamefont {Strom}}, \bibinfo
  {author} {\bibfnamefont {C.}~\bibnamefont {Baltay}}, \bibinfo {author}
  {\bibfnamefont {W.}~\bibnamefont {Emmet}}, \ and\ \bibinfo {author}
  {\bibfnamefont {D.}~\bibnamefont {Rabinowitz}},\ }\bibfield  {booktitle}
  {\emph {\bibinfo {booktitle} {{Proceedings, 24th International Workshop on
  Vertex Detectors (Vertex 2015): Santa Fe, New Mexico, USA, June 1-5,
  2015}}},\ }\href {\doibase 10.22323/1.254.0038} {\bibfield  {journal}
  {\bibinfo  {journal} {PoS}\ }\textbf {\bibinfo {volume} {VERTEX2015}},\
  \bibinfo {pages} {038} (\bibinfo {year} {2015})}\BibitemShut {NoStop}%
\bibitem [{\citenamefont {{J. Barkeloo, J. Brau, M. Breidenbach, R. Frey, D.
  Freytag, C. Gallagher, R. Herbst, M. Oriunno, B. Reese, A. Steinhebel and D.
  Strom}}(2019)}]{calor:2018}%
  \BibitemOpen
  \bibfield  {author} {\bibinfo {author} {\bibnamefont {{J. Barkeloo, J. Brau,
  M. Breidenbach, R. Frey, D. Freytag, C. Gallagher, R. Herbst, M. Oriunno, B.
  Reese, A. Steinhebel and D. Strom}}},\ }\href@noop {} {\enquote {\bibinfo
  {title} {{A silicon-tungsten electromagnetic calorimeter with integrated
  electronics for the International Linear Collider}},}\ }\bibinfo
  {howpublished} {{Proceedings of the 18th International Conference on
  Calorimetry in Particle Physics - CALOR 2018, Journal of Physics: Conference
  Series}} (\bibinfo {year} {2019})\BibitemShut {NoStop}%
\bibitem [{\citenamefont {Brau}\ \emph {et~al.}(2012)\citenamefont {Brau} \emph
  {et~al.}}]{Brau:2013yb}%
  \BibitemOpen
  \bibfield  {author} {\bibinfo {author} {\bibfnamefont {J.}~\bibnamefont
  {Brau}} \emph {et~al.},\ }in\ \href {\doibase 10.1109/NSSMIC.2012.6551433}
  {\emph {\bibinfo {booktitle} {{Proceedings, 2012 IEEE Nuclear Science
  Symposium and Medical Imaging Conference (NSS/MIC 2012), SLAC-PUB-15285}}}}\
  (\bibinfo {year} {2012})\ pp.\ \bibinfo {pages} {1857--1860}\BibitemShut
  {NoStop}%
\bibitem [{\citenamefont {Steinhebel}\ and\ \citenamefont
  {Brau}(2017)}]{Steinhebel:2017qze}%
  \BibitemOpen
  \bibfield  {author} {\bibinfo {author} {\bibfnamefont {A.}~\bibnamefont
  {Steinhebel}}\ and\ \bibinfo {author} {\bibfnamefont {J.}~\bibnamefont
  {Brau}},\ }in\ \href@noop {} {\emph {\bibinfo {booktitle} {{Proceedings,
  International Workshop on Future Linear Colliders 2016 (LCWS2016): Morioka,
  Iwate, Japan, December 05-09, 2016}}}}\ (\bibinfo {year} {2017})\ \Eprint
  {http://arxiv.org/abs/1703.08605} {arXiv:1703.08605 [physics.ins-det]}
  \BibitemShut {NoStop}%
\bibitem [{\citenamefont {Parker}\ \emph {et~al.}(2010)\citenamefont {Parker},
  \citenamefont {Mikhailichenko}, \citenamefont {Buesser}, \citenamefont
  {Hauptman}, \citenamefont {Tauchi}, \citenamefont {Burrows}, \citenamefont
  {Markiewicz}, \citenamefont {Oriunno},\ and\ \citenamefont
  {Seryi}}]{Parker:2009zz}%
  \BibitemOpen
  \bibfield  {author} {\bibinfo {author} {\bibfnamefont {B.}~\bibnamefont
  {Parker}}, \bibinfo {author} {\bibfnamefont {A.}~\bibnamefont
  {Mikhailichenko}}, \bibinfo {author} {\bibfnamefont {K.}~\bibnamefont
  {Buesser}}, \bibinfo {author} {\bibfnamefont {J.}~\bibnamefont {Hauptman}},
  \bibinfo {author} {\bibfnamefont {T.}~\bibnamefont {Tauchi}}, \bibinfo
  {author} {\bibfnamefont {P.}~\bibnamefont {Burrows}}, \bibinfo {author}
  {\bibfnamefont {T.}~\bibnamefont {Markiewicz}}, \bibinfo {author}
  {\bibfnamefont {M.}~\bibnamefont {Oriunno}}, \ and\ \bibinfo {author}
  {\bibfnamefont {A.}~\bibnamefont {Seryi}},\ }in\ \href
  {http://www-public.slac.stanford.edu/sciDoc/docMeta.aspx?slacPubNumber=slac-pub-13657}
  {\emph {\bibinfo {booktitle} {{Particle accelerator. Proceedings, 23rd
  Conference, PAC'09, Vancouver, Canada, May 4-8, 2009}}}}\ (\bibinfo {year}
  {2010})\ p.\ \bibinfo {pages} {WE6PFP078}\BibitemShut {NoStop}%
\bibitem [{\citenamefont {Buesser}(2012)}]{Buesser:2012et}%
  \BibitemOpen
  \bibfield  {author} {\bibinfo {author} {\bibfnamefont {K.}~\bibnamefont
  {Buesser}},\ }in\ \href@noop {} {\emph {\bibinfo {booktitle} {{International
  Workshop on Future Linear Colliders (LCWS11) Granada, Spain, September 26-30,
  2011}}}}\ (\bibinfo {year} {2012})\ \Eprint {http://arxiv.org/abs/1201.5807}
  {arXiv:1201.5807 [physics.ins-det]} \BibitemShut {NoStop}%
\bibitem [{\citenamefont {Abe}\ \emph {et~al.}(2010)\citenamefont {Abe} \emph
  {et~al.}}]{Abe:2010aa}%
  \BibitemOpen
  \bibfield  {author} {\bibinfo {author} {\bibfnamefont {T.}~\bibnamefont
  {Abe}} \emph {et~al.} (\bibinfo {collaboration} {Linear Collider ILD Concept
  Group -}),\ }\href {\doibase 10.2172/975166} {\  (\bibinfo {year} {2010}),\
  10.2172/975166},\ \Eprint {http://arxiv.org/abs/1006.3396} {arXiv:1006.3396
  [hep-ex]} \BibitemShut {NoStop}%
\bibitem [{\citenamefont {{iLCSoft authors}}(2016)}]{bib:ilcsoft}%
  \BibitemOpen
  \bibfield  {author} {\bibinfo {author} {\bibnamefont {{iLCSoft authors}}},\
  }\href@noop {} {\enquote {\bibinfo {title} {{iLCSoft Project Page}},}\
  }\bibinfo {howpublished} {\url{https://github.com/iLCSoft}} (\bibinfo {year}
  {2016})\BibitemShut {NoStop}%
\bibitem [{\citenamefont {Gaede}\ \emph {et~al.}(2003)\citenamefont {Gaede},
  \citenamefont {Behnke}, \citenamefont {Graf},\ and\ \citenamefont
  {Johnson}}]{Gaede:2003ip}%
  \BibitemOpen
  \bibfield  {author} {\bibinfo {author} {\bibfnamefont {F.}~\bibnamefont
  {Gaede}}, \bibinfo {author} {\bibfnamefont {T.}~\bibnamefont {Behnke}},
  \bibinfo {author} {\bibfnamefont {N.}~\bibnamefont {Graf}}, \ and\ \bibinfo
  {author} {\bibfnamefont {T.}~\bibnamefont {Johnson}},\ }\bibfield
  {booktitle} {\emph {\bibinfo {booktitle} {{Proceedings, 13th International
  Conference on Computing in High-Enery and Nuclear Physics (CHEP 2003): La
  Jolla, California, March 24-28, 2003}}},\ }\href@noop {} {\bibfield
  {journal} {\bibinfo  {journal} {eConf}\ }\textbf {\bibinfo {volume}
  {C0303241}},\ \bibinfo {pages} {TUKT001} (\bibinfo {year} {2003})},\ \Eprint
  {http://arxiv.org/abs/physics/0306114} {arXiv:physics/0306114 [physics]}
  \BibitemShut {NoStop}%
\bibitem [{\citenamefont {Gaede}(2006)}]{Gaede:2006pj}%
  \BibitemOpen
  \bibfield  {author} {\bibinfo {author} {\bibfnamefont {F.}~\bibnamefont
  {Gaede}},\ }\bibfield  {booktitle} {\emph {\bibinfo {booktitle} {{Advanced
  computing and analysis techniques in physics research. Proceedings, 10th
  International Workshop, ACAT05, Zeuthen, Germany, May 22-27, 2005}}},\ }\href
  {\doibase 10.1016/j.nima.2005.11.138} {\bibfield  {journal} {\bibinfo
  {journal} {Nucl. Instrum. Meth.}\ }\textbf {\bibinfo {volume} {A559}},\
  \bibinfo {pages} {177} (\bibinfo {year} {2006})}\BibitemShut {NoStop}%
\bibitem [{\citenamefont {Frank}\ \emph {et~al.}(2014)\citenamefont {Frank},
  \citenamefont {Gaede}, \citenamefont {Grefe},\ and\ \citenamefont
  {Mato}}]{Frank:2014zya}%
  \BibitemOpen
  \bibfield  {author} {\bibinfo {author} {\bibfnamefont {M.}~\bibnamefont
  {Frank}}, \bibinfo {author} {\bibfnamefont {F.}~\bibnamefont {Gaede}},
  \bibinfo {author} {\bibfnamefont {C.}~\bibnamefont {Grefe}}, \ and\ \bibinfo
  {author} {\bibfnamefont {P.}~\bibnamefont {Mato}},\ }\bibfield  {booktitle}
  {\emph {\bibinfo {booktitle} {{Proceedings, 20th International Conference on
  Computing in High Energy and Nuclear Physics (CHEP 2013): Amsterdam, The
  Netherlands, October 14-18, 2013}}},\ }\href {\doibase
  10.1088/1742-6596/513/2/022010} {\bibfield  {journal} {\bibinfo  {journal}
  {J. Phys. Conf. Ser.}\ }\textbf {\bibinfo {volume} {513}},\ \bibinfo {pages}
  {022010} (\bibinfo {year} {2014})}\BibitemShut {NoStop}%
\bibitem [{\citenamefont {Frank}\ \emph {et~al.}(2015)\citenamefont {Frank},
  \citenamefont {Gaede}, \citenamefont {Nikiforou}, \citenamefont {Petric},\
  and\ \citenamefont {Sailer}}]{Frank:2015ivo}%
  \BibitemOpen
  \bibfield  {author} {\bibinfo {author} {\bibfnamefont {M.}~\bibnamefont
  {Frank}}, \bibinfo {author} {\bibfnamefont {F.}~\bibnamefont {Gaede}},
  \bibinfo {author} {\bibfnamefont {N.}~\bibnamefont {Nikiforou}}, \bibinfo
  {author} {\bibfnamefont {M.}~\bibnamefont {Petric}}, \ and\ \bibinfo {author}
  {\bibfnamefont {A.}~\bibnamefont {Sailer}},\ }\bibfield  {booktitle} {\emph
  {\bibinfo {booktitle} {{Proceedings, 21st International Conference on
  Computing in High Energy and Nuclear Physics (CHEP 2015): Okinawa, Japan,
  April 13-17, 2015}}},\ }\href {\doibase 10.1088/1742-6596/664/7/072017}
  {\bibfield  {journal} {\bibinfo  {journal} {J. Phys. Conf. Ser.}\ }\textbf
  {\bibinfo {volume} {664}},\ \bibinfo {pages} {072017} (\bibinfo {year}
  {2015})}\BibitemShut {NoStop}%
\bibitem [{\citenamefont {Agostinelli}\ \emph {et~al.}(2003)\citenamefont
  {Agostinelli} \emph {et~al.}}]{Agostinelli:2002hh}%
  \BibitemOpen
  \bibfield  {author} {\bibinfo {author} {\bibfnamefont {S.}~\bibnamefont
  {Agostinelli}} \emph {et~al.} (\bibinfo {collaboration} {GEANT4}),\ }\href
  {\doibase 10.1016/S0168-9002(03)01368-8} {\bibfield  {journal} {\bibinfo
  {journal} {Nucl. Instrum. Meth.}\ }\textbf {\bibinfo {volume} {A506}},\
  \bibinfo {pages} {250} (\bibinfo {year} {2003})}\BibitemShut {NoStop}%
\bibitem [{\citenamefont {Kilian}\ \emph {et~al.}(2011)\citenamefont {Kilian},
  \citenamefont {Ohl},\ and\ \citenamefont {Reuter}}]{Kilian:2007gr}%
  \BibitemOpen
  \bibfield  {author} {\bibinfo {author} {\bibfnamefont {W.}~\bibnamefont
  {Kilian}}, \bibinfo {author} {\bibfnamefont {T.}~\bibnamefont {Ohl}}, \ and\
  \bibinfo {author} {\bibfnamefont {J.}~\bibnamefont {Reuter}},\ }\href
  {\doibase 10.1140/epjc/s10052-011-1742-y} {\bibfield  {journal} {\bibinfo
  {journal} {Eur. Phys. J.}\ }\textbf {\bibinfo {volume} {C71}},\ \bibinfo
  {pages} {1742} (\bibinfo {year} {2011})},\ \Eprint
  {http://arxiv.org/abs/0708.4233} {arXiv:0708.4233 [hep-ph]} \BibitemShut
  {NoStop}%
\bibitem [{\citenamefont {Sjostrand}\ \emph {et~al.}(2006)\citenamefont
  {Sjostrand}, \citenamefont {Mrenna},\ and\ \citenamefont
  {Skands}}]{Sjostrand:2006za}%
  \BibitemOpen
  \bibfield  {author} {\bibinfo {author} {\bibfnamefont {T.}~\bibnamefont
  {Sjostrand}}, \bibinfo {author} {\bibfnamefont {S.}~\bibnamefont {Mrenna}}, \
  and\ \bibinfo {author} {\bibfnamefont {P.~Z.}\ \bibnamefont {Skands}},\
  }\href {\doibase 10.1088/1126-6708/2006/05/026} {\bibfield  {journal}
  {\bibinfo  {journal} {JHEP}\ }\textbf {\bibinfo {volume} {05}},\ \bibinfo
  {pages} {026} (\bibinfo {year} {2006})},\ \Eprint
  {http://arxiv.org/abs/hep-ph/0603175} {arXiv:hep-ph/0603175 [hep-ph]}
  \BibitemShut {NoStop}%
\bibitem [{\citenamefont {Schulte}(1998)}]{Schulte:1998au}%
  \BibitemOpen
  \bibfield  {author} {\bibinfo {author} {\bibfnamefont {D.}~\bibnamefont
  {Schulte}},\ }\bibfield  {booktitle} {\emph {\bibinfo {booktitle}
  {{Proceedings, 1998 International Computational Accelerator Physics
  Conference}}},\ }\href@noop {} {\bibfield  {journal} {\bibinfo  {journal}
  {eConf}\ }\textbf {\bibinfo {volume} {C980914}},\ \bibinfo {pages} {127}
  (\bibinfo {year} {1998})},\ \bibinfo {note} {[,127(1998)]}\BibitemShut
  {NoStop}%
\bibitem [{\citenamefont {Sch{\"u}tz}(2017)}]{Schutz:2017ihd}%
  \BibitemOpen
  \bibfield  {author} {\bibinfo {author} {\bibfnamefont {A.}~\bibnamefont
  {Sch{\"u}tz}},\ }in\ \href@noop {} {\emph {\bibinfo {booktitle}
  {{Proceedings, International Workshop on Future Linear Colliders 2016
  (LCWS2016): Morioka, Iwate, Japan, December 05-09, 2016}}}}\ (\bibinfo {year}
  {2017})\ \Eprint {http://arxiv.org/abs/1703.05737} {arXiv:1703.05737
  [physics.ins-det]} \BibitemShut {NoStop}%
\bibitem [{\citenamefont {Chen}\ \emph {et~al.}(1994)\citenamefont {Chen},
  \citenamefont {Barklow},\ and\ \citenamefont {Peskin}}]{Chen:1993dba}%
  \BibitemOpen
  \bibfield  {author} {\bibinfo {author} {\bibfnamefont {P.}~\bibnamefont
  {Chen}}, \bibinfo {author} {\bibfnamefont {T.~L.}\ \bibnamefont {Barklow}}, \
  and\ \bibinfo {author} {\bibfnamefont {M.~E.}\ \bibnamefont {Peskin}},\
  }\href {\doibase 10.1103/PhysRevD.49.3209} {\bibfield  {journal} {\bibinfo
  {journal} {Phys. Rev.}\ }\textbf {\bibinfo {volume} {D49}},\ \bibinfo {pages}
  {3209} (\bibinfo {year} {1994})},\ \Eprint
  {http://arxiv.org/abs/hep-ph/9305247} {arXiv:hep-ph/9305247 [hep-ph]}
  \BibitemShut {NoStop}%
\bibitem [{\citenamefont {{lcgeo authors}}(2016)}]{bib:lcgeo}%
  \BibitemOpen
  \bibfield  {author} {\bibinfo {author} {\bibnamefont {{lcgeo authors}}},\
  }\href@noop {} {\enquote {\bibinfo {title} {{lcgeo Project Page}},}\
  }\bibinfo {howpublished} {\url{https://github.com/iLCSoft/lcgeo}} (\bibinfo
  {year} {2016})\BibitemShut {NoStop}%
\bibitem [{\citenamefont {Mora~de Freitas}\ and\ \citenamefont
  {Videau}(2002)}]{MoradeFreitas:2002kj}%
  \BibitemOpen
  \bibfield  {author} {\bibinfo {author} {\bibfnamefont {P.}~\bibnamefont
  {Mora~de Freitas}}\ and\ \bibinfo {author} {\bibfnamefont {H.}~\bibnamefont
  {Videau}},\ }in\ \href
  {http://www-library.desy.de/cgi-bin/showprep.pl?lc-tool03-010} {\emph
  {\bibinfo {booktitle} {{Linear colliders. Proceedings, International Workshop
  on physics and experiments with future electron-positron linear colliders,
  LCWS 2002, Seogwipo, Jeju Island, Korea, August 26-30, 2002}}}}\ (\bibinfo
  {year} {2002})\ pp.\ \bibinfo {pages} {623--627}\BibitemShut {NoStop}%
\bibitem [{\citenamefont {{SLIC authors}}()}]{bib:slic}%
  \BibitemOpen
  \bibfield  {author} {\bibinfo {author} {\bibnamefont {{SLIC authors}}},\
  }\href@noop {} {\enquote {\bibinfo {title} {{SLIC Project Page}},}\ }\bibinfo
  {howpublished}
  {\url{https://confluence.slac.stanford.edu/display/ilc/SLIC}}\BibitemShut
  {NoStop}%
\bibitem [{\citenamefont {Li}\ \emph {et~al.}(2014)\citenamefont {Li},
  \citenamefont {Fujii},\ and\ \citenamefont {Gao}}]{Li:2013cxa}%
  \BibitemOpen
  \bibfield  {author} {\bibinfo {author} {\bibfnamefont {B.}~\bibnamefont
  {Li}}, \bibinfo {author} {\bibfnamefont {K.}~\bibnamefont {Fujii}}, \ and\
  \bibinfo {author} {\bibfnamefont {Y.}~\bibnamefont {Gao}},\ }\href {\doibase
  10.1016/j.cpc.2013.11.003} {\bibfield  {journal} {\bibinfo  {journal}
  {Comput. Phys. Commun.}\ }\textbf {\bibinfo {volume} {185}},\ \bibinfo
  {pages} {754} (\bibinfo {year} {2014})},\ \Eprint
  {http://arxiv.org/abs/1305.7300} {arXiv:1305.7300 [physics.ins-det]}
  \BibitemShut {NoStop}%
\bibitem [{\citenamefont {Gaede}\ \emph {et~al.}(2014)\citenamefont {Gaede},
  \citenamefont {Aplin}, \citenamefont {Glattauer}, \citenamefont {Rosemann},\
  and\ \citenamefont {Voutsinas}}]{Gaede:2014aza}%
  \BibitemOpen
  \bibfield  {author} {\bibinfo {author} {\bibfnamefont {F.}~\bibnamefont
  {Gaede}}, \bibinfo {author} {\bibfnamefont {S.}~\bibnamefont {Aplin}},
  \bibinfo {author} {\bibfnamefont {R.}~\bibnamefont {Glattauer}}, \bibinfo
  {author} {\bibfnamefont {C.}~\bibnamefont {Rosemann}}, \ and\ \bibinfo
  {author} {\bibfnamefont {G.}~\bibnamefont {Voutsinas}},\ }\bibfield
  {booktitle} {\emph {\bibinfo {booktitle} {{Proceedings, 20th International
  Conference on Computing in High Energy and Nuclear Physics (CHEP 2013):
  Amsterdam, The Netherlands, October 14-18, 2013}}},\ }\href {\doibase
  10.1088/1742-6596/513/2/022011} {\bibfield  {journal} {\bibinfo  {journal}
  {J. Phys. Conf. Ser.}\ }\textbf {\bibinfo {volume} {513}},\ \bibinfo {pages}
  {022011} (\bibinfo {year} {2014})}\BibitemShut {NoStop}%
\bibitem [{\citenamefont {{LCSim authors}}()}]{bib:LCSim}%
  \BibitemOpen
  \bibfield  {author} {\bibinfo {author} {\bibnamefont {{LCSim authors}}},\
  }\href@noop {} {\enquote {\bibinfo {title} {{LCSim Project Page}},}\
  }\bibinfo {howpublished} {\url{http://www.lcsim.org/sites/lcsim}}\BibitemShut
  {NoStop}%
\bibitem [{\citenamefont {Marshall}\ and\ \citenamefont
  {Thomson}(2015)}]{Marshall:2015rfa}%
  \BibitemOpen
  \bibfield  {author} {\bibinfo {author} {\bibfnamefont {J.~S.}\ \bibnamefont
  {Marshall}}\ and\ \bibinfo {author} {\bibfnamefont {M.~A.}\ \bibnamefont
  {Thomson}},\ }\href {\doibase 10.1140/epjc/s10052-015-3659-3} {\bibfield
  {journal} {\bibinfo  {journal} {Eur. Phys. J.}\ }\textbf {\bibinfo {volume}
  {C75}},\ \bibinfo {pages} {439} (\bibinfo {year} {2015})},\ \Eprint
  {http://arxiv.org/abs/1506.05348} {arXiv:1506.05348 [physics.data-an]}
  \BibitemShut {NoStop}%
\bibitem [{\citenamefont {Suehara}\ and\ \citenamefont
  {Tanabe}(2016)}]{Suehara:2015ura}%
  \BibitemOpen
  \bibfield  {author} {\bibinfo {author} {\bibfnamefont {T.}~\bibnamefont
  {Suehara}}\ and\ \bibinfo {author} {\bibfnamefont {T.}~\bibnamefont
  {Tanabe}},\ }\href {\doibase 10.1016/j.nima.2015.11.054} {\bibfield
  {journal} {\bibinfo  {journal} {Nucl. Instrum. Meth.}\ }\textbf {\bibinfo
  {volume} {A808}},\ \bibinfo {pages} {109} (\bibinfo {year} {2016})},\ \Eprint
  {http://arxiv.org/abs/1506.08371} {arXiv:1506.08371 [physics.ins-det]}
  \BibitemShut {NoStop}%
\bibitem [{\citenamefont {Cacciari}(2006)}]{Cacciari:2006sm}%
  \BibitemOpen
  \bibfield  {author} {\bibinfo {author} {\bibfnamefont {M.}~\bibnamefont
  {Cacciari}},\ }in\ \href@noop {} {\emph {\bibinfo {booktitle} {{Deep
  inelastic scattering. Proceedings, 14th International Workshop, DIS 2006,
  Tsukuba, Japan, April 20-24, 2006}}}}\ (\bibinfo {year} {2006})\ pp.\
  \bibinfo {pages} {487--490},\ \bibinfo {note} {[,125(2006)]},\ \Eprint
  {http://arxiv.org/abs/hep-ph/0607071} {arXiv:hep-ph/0607071 [hep-ph]}
  \BibitemShut {NoStop}%
\bibitem [{\citenamefont {Berggren}(2012)}]{Berggren:2012ar}%
  \BibitemOpen
  \bibfield  {author} {\bibinfo {author} {\bibfnamefont {M.}~\bibnamefont
  {Berggren}},\ }in\ \href@noop {} {\emph {\bibinfo {booktitle} {{International
  Workshop on Future Linear Colliders (LCWS11) Granada, Spain, September 26-30,
  2011}}}}\ (\bibinfo {year} {2012})\ \Eprint {http://arxiv.org/abs/1203.0217}
  {arXiv:1203.0217 [physics.ins-det]} \BibitemShut {NoStop}%
\bibitem [{\citenamefont {Abdallah}\ \emph {et~al.}(2003)\citenamefont
  {Abdallah} \emph {et~al.}}]{Abdallah:2003xe}%
  \BibitemOpen
  \bibfield  {author} {\bibinfo {author} {\bibfnamefont {J.}~\bibnamefont
  {Abdallah}} \emph {et~al.} (\bibinfo {collaboration} {DELPHI}),\ }\href
  {\doibase 10.1140/epjc/s2003-01355-5} {\bibfield  {journal} {\bibinfo
  {journal} {Eur. Phys. J.}\ }\textbf {\bibinfo {volume} {C31}},\ \bibinfo
  {pages} {421} (\bibinfo {year} {2003})},\ \Eprint
  {http://arxiv.org/abs/hep-ex/0311019} {arXiv:hep-ex/0311019 [hep-ex]}
  \BibitemShut {NoStop}%
\bibitem [{\citenamefont {Aaboud}\ \emph
  {et~al.}(2018{\natexlab{a}})\citenamefont {Aaboud} \emph
  {et~al.}}]{Aaboud:2018zhk}%
  \BibitemOpen
  \bibfield  {author} {\bibinfo {author} {\bibfnamefont {M.}~\bibnamefont
  {Aaboud}} \emph {et~al.} (\bibinfo {collaboration} {ATLAS}),\ }\href
  {\doibase 10.1016/j.physletb.2018.09.013} {\bibfield  {journal} {\bibinfo
  {journal} {Phys. Lett.}\ }\textbf {\bibinfo {volume} {B786}},\ \bibinfo
  {pages} {59} (\bibinfo {year} {2018}{\natexlab{a}})},\ \Eprint
  {http://arxiv.org/abs/1808.08238} {arXiv:1808.08238 [hep-ex]} \BibitemShut
  {NoStop}%
\bibitem [{\citenamefont {Sirunyan}\ \emph
  {et~al.}(2018{\natexlab{a}})\citenamefont {Sirunyan} \emph
  {et~al.}}]{Sirunyan:2018kst}%
  \BibitemOpen
  \bibfield  {author} {\bibinfo {author} {\bibfnamefont {A.~M.}\ \bibnamefont
  {Sirunyan}} \emph {et~al.} (\bibinfo {collaboration} {CMS}),\ }\href
  {\doibase 10.1103/PhysRevLett.121.121801} {\bibfield  {journal} {\bibinfo
  {journal} {Phys. Rev. Lett.}\ }\textbf {\bibinfo {volume} {121}},\ \bibinfo
  {pages} {121801} (\bibinfo {year} {2018}{\natexlab{a}})},\ \Eprint
  {http://arxiv.org/abs/1808.08242} {arXiv:1808.08242 [hep-ex]} \BibitemShut
  {NoStop}%
\bibitem [{\citenamefont {Ogawa}(2018)}]{Ogawa:2018}%
  \BibitemOpen
  \bibfield  {author} {\bibinfo {author} {\bibfnamefont {T.}~\bibnamefont
  {Ogawa}},\ }\href@noop {} {\enquote {\bibinfo {title} {Sensitivity to
  anomalous vvh couplings induced by dimension-6 operators at the ilc},}\
  }\bibinfo {howpublished} {{PhD Thesis, Sokendai / KEK}} (\bibinfo {year}
  {2018})\BibitemShut {NoStop}%
\bibitem [{\citenamefont {Catani}\ \emph {et~al.}(1993)\citenamefont {Catani},
  \citenamefont {Dokshitzer}, \citenamefont {Seymour},\ and\ \citenamefont
  {Webber}}]{Catani:1993hr}%
  \BibitemOpen
  \bibfield  {author} {\bibinfo {author} {\bibfnamefont {S.}~\bibnamefont
  {Catani}}, \bibinfo {author} {\bibfnamefont {Y.~L.}\ \bibnamefont
  {Dokshitzer}}, \bibinfo {author} {\bibfnamefont {M.~H.}\ \bibnamefont
  {Seymour}}, \ and\ \bibinfo {author} {\bibfnamefont {B.~R.}\ \bibnamefont
  {Webber}},\ }\href {\doibase 10.1016/0550-3213(93)90166-M} {\bibfield
  {journal} {\bibinfo  {journal} {Nucl. Phys.}\ }\textbf {\bibinfo {volume}
  {B406}},\ \bibinfo {pages} {187} (\bibinfo {year} {1993})}\BibitemShut
  {NoStop}%
\bibitem [{\citenamefont {Cacciari}\ \emph {et~al.}(2008)\citenamefont
  {Cacciari}, \citenamefont {Salam},\ and\ \citenamefont
  {Soyez}}]{Cacciari:2008gp}%
  \BibitemOpen
  \bibfield  {author} {\bibinfo {author} {\bibfnamefont {M.}~\bibnamefont
  {Cacciari}}, \bibinfo {author} {\bibfnamefont {G.~P.}\ \bibnamefont {Salam}},
  \ and\ \bibinfo {author} {\bibfnamefont {G.}~\bibnamefont {Soyez}},\ }\href
  {\doibase 10.1088/1126-6708/2008/04/063} {\bibfield  {journal} {\bibinfo
  {journal} {JHEP}\ }\textbf {\bibinfo {volume} {04}},\ \bibinfo {pages} {063}
  (\bibinfo {year} {2008})},\ \Eprint {http://arxiv.org/abs/0802.1189}
  {arXiv:0802.1189 [hep-ph]} \BibitemShut {NoStop}%
\bibitem [{\citenamefont {Boronat}\ \emph {et~al.}(2015)\citenamefont
  {Boronat}, \citenamefont {Fuster}, \citenamefont {Garcia}, \citenamefont
  {Ros},\ and\ \citenamefont {Vos}}]{Boronat:2014hva}%
  \BibitemOpen
  \bibfield  {author} {\bibinfo {author} {\bibfnamefont {M.}~\bibnamefont
  {Boronat}}, \bibinfo {author} {\bibfnamefont {J.}~\bibnamefont {Fuster}},
  \bibinfo {author} {\bibfnamefont {I.}~\bibnamefont {Garcia}}, \bibinfo
  {author} {\bibfnamefont {E.}~\bibnamefont {Ros}}, \ and\ \bibinfo {author}
  {\bibfnamefont {M.}~\bibnamefont {Vos}},\ }\href {\doibase
  10.1016/j.physletb.2015.08.055} {\bibfield  {journal} {\bibinfo  {journal}
  {Phys. Lett.}\ }\textbf {\bibinfo {volume} {B750}},\ \bibinfo {pages} {95}
  (\bibinfo {year} {2015})},\ \Eprint {http://arxiv.org/abs/1404.4294}
  {arXiv:1404.4294 [hep-ex]} \BibitemShut {NoStop}%
\bibitem [{\citenamefont {Catani}\ \emph {et~al.}(1991)\citenamefont {Catani},
  \citenamefont {Dokshitzer}, \citenamefont {Olsson}, \citenamefont {Turnock},\
  and\ \citenamefont {Webber}}]{Catani:1991hj}%
  \BibitemOpen
  \bibfield  {author} {\bibinfo {author} {\bibfnamefont {S.}~\bibnamefont
  {Catani}}, \bibinfo {author} {\bibfnamefont {Y.~L.}\ \bibnamefont
  {Dokshitzer}}, \bibinfo {author} {\bibfnamefont {M.}~\bibnamefont {Olsson}},
  \bibinfo {author} {\bibfnamefont {G.}~\bibnamefont {Turnock}}, \ and\
  \bibinfo {author} {\bibfnamefont {B.~R.}\ \bibnamefont {Webber}},\ }\href
  {\doibase 10.1016/0370-2693(91)90196-W} {\bibfield  {journal} {\bibinfo
  {journal} {Phys. Lett.}\ }\textbf {\bibinfo {volume} {B269}},\ \bibinfo
  {pages} {432} (\bibinfo {year} {1991})}\BibitemShut {NoStop}%
\bibitem [{\citenamefont {Yan}\ \emph {et~al.}(2016)\citenamefont {Yan},
  \citenamefont {Watanuki}, \citenamefont {Fujii}, \citenamefont {Ishikawa},
  \citenamefont {Jeans}, \citenamefont {Strube}, \citenamefont {Tian},\ and\
  \citenamefont {Yamamoto}}]{Yan:2016xyx}%
  \BibitemOpen
  \bibfield  {author} {\bibinfo {author} {\bibfnamefont {J.}~\bibnamefont
  {Yan}}, \bibinfo {author} {\bibfnamefont {S.}~\bibnamefont {Watanuki}},
  \bibinfo {author} {\bibfnamefont {K.}~\bibnamefont {Fujii}}, \bibinfo
  {author} {\bibfnamefont {A.}~\bibnamefont {Ishikawa}}, \bibinfo {author}
  {\bibfnamefont {D.}~\bibnamefont {Jeans}}, \bibinfo {author} {\bibfnamefont
  {J.}~\bibnamefont {Strube}}, \bibinfo {author} {\bibfnamefont
  {J.}~\bibnamefont {Tian}}, \ and\ \bibinfo {author} {\bibfnamefont
  {H.}~\bibnamefont {Yamamoto}},\ }\href {\doibase 10.1103/PhysRevD.94.113002}
  {\bibfield  {journal} {\bibinfo  {journal} {Phys. Rev.}\ }\textbf {\bibinfo
  {volume} {D94}},\ \bibinfo {pages} {113002} (\bibinfo {year} {2016})},\
  \Eprint {http://arxiv.org/abs/1604.07524} {arXiv:1604.07524 [hep-ex]}
  \BibitemShut {NoStop}%
\bibitem [{\citenamefont {Tomita}(2015)}]{Tomita:2015}%
  \BibitemOpen
  \bibfield  {author} {\bibinfo {author} {\bibfnamefont {T.}~\bibnamefont
  {Tomita}},\ }\href@noop {} {\enquote {\bibinfo {title} {Hadronic higgs recoil
  mass study with ild at 250gev},}\ }\bibinfo {howpublished} {{presentation at
  Asian Linear Collider Workshop, KEK, Tsukuba, Japan, April 19-24, 2015,
  \url{https://agenda.linearcollider.org/event/6557/contributions/31831/}}}
  (\bibinfo {year} {2015})\BibitemShut {NoStop}%
\bibitem [{\citenamefont {Thomson}(2016)}]{Thomson:2015jda}%
  \BibitemOpen
  \bibfield  {author} {\bibinfo {author} {\bibfnamefont {M.}~\bibnamefont
  {Thomson}},\ }\href {\doibase 10.1140/epjc/s10052-016-3911-5} {\bibfield
  {journal} {\bibinfo  {journal} {Eur. Phys. J.}\ }\textbf {\bibinfo {volume}
  {C76}},\ \bibinfo {pages} {72} (\bibinfo {year} {2016})},\ \Eprint
  {http://arxiv.org/abs/1509.02853} {arXiv:1509.02853 [hep-ex]} \BibitemShut
  {NoStop}%
\bibitem [{\citenamefont {Miyamoto}(2013)}]{Miyamoto:2013zva}%
  \BibitemOpen
  \bibfield  {author} {\bibinfo {author} {\bibfnamefont {A.}~\bibnamefont
  {Miyamoto}},\ }\href@noop {} {\  (\bibinfo {year} {2013})},\ \Eprint
  {http://arxiv.org/abs/1311.2248} {arXiv:1311.2248 [hep-ex]} \BibitemShut
  {NoStop}%
\bibitem [{\citenamefont {Duerig}\ \emph {et~al.}(2014)\citenamefont {Duerig},
  \citenamefont {Fujii}, \citenamefont {List},\ and\ \citenamefont
  {Tian}}]{Durig:2014lfa}%
  \BibitemOpen
  \bibfield  {author} {\bibinfo {author} {\bibfnamefont {C.}~\bibnamefont
  {Duerig}}, \bibinfo {author} {\bibfnamefont {K.}~\bibnamefont {Fujii}},
  \bibinfo {author} {\bibfnamefont {J.}~\bibnamefont {List}}, \ and\ \bibinfo
  {author} {\bibfnamefont {J.}~\bibnamefont {Tian}},\ }in\ \href@noop {} {\emph
  {\bibinfo {booktitle} {{International Workshop on Future Linear Colliders
  (LCWS13) Tokyo, Japan, November 11-15, 2013}}}}\ (\bibinfo {year} {2014})\
  \Eprint {http://arxiv.org/abs/1403.7734} {arXiv:1403.7734 [hep-ex]}
  \BibitemShut {NoStop}%
\bibitem [{\citenamefont {Tian}(2017)}]{Tian:2017}%
  \BibitemOpen
  \bibfield  {author} {\bibinfo {author} {\bibfnamefont {J.}~\bibnamefont
  {Tian}},\ }\href@noop {} {\enquote {\bibinfo {title} {Update of
  $e^+e^−\to\nu\bar{\nu}h$ analysis},}\ }\bibinfo {howpublished}
  {{presentation at ILD Analysis and Software Meeting on July 19, 2017,
  \url{https://agenda.linearcollider.org/event/7703}}} (\bibinfo {year}
  {2017})\BibitemShut {NoStop}%
\bibitem [{\citenamefont {Ono}\ and\ \citenamefont
  {Miyamoto}(2013)}]{Ono:2013sea}%
  \BibitemOpen
  \bibfield  {author} {\bibinfo {author} {\bibfnamefont {H.}~\bibnamefont
  {Ono}}\ and\ \bibinfo {author} {\bibfnamefont {A.}~\bibnamefont {Miyamoto}},\
  }\href {\doibase 10.1140/epjc/s10052-013-2343-8} {\bibfield  {journal}
  {\bibinfo  {journal} {Eur. Phys. J.}\ }\textbf {\bibinfo {volume} {C73}},\
  \bibinfo {pages} {2343} (\bibinfo {year} {2013})},\ \Eprint
  {http://arxiv.org/abs/1207.0300} {arXiv:1207.0300 [hep-ex]} \BibitemShut
  {NoStop}%
\bibitem [{\citenamefont {Ono}(2012)}]{Ono:2012}%
  \BibitemOpen
  \bibfield  {author} {\bibinfo {author} {\bibfnamefont {H.}~\bibnamefont
  {Ono}},\ }\href@noop {} {\enquote {\bibinfo {title} {Higgs branching fraction
  study},}\ }\bibinfo {howpublished} {{presentation at the KILC2012 workshop,
  April 2012,
  \url{https://agenda.linearcollider.org/event/5414/contributions/23402}}}
  (\bibinfo {year} {2012})\BibitemShut {NoStop}%
\bibitem [{\citenamefont {Barklow}(2017)}]{Barklow:2017}%
  \BibitemOpen
  \bibfield  {author} {\bibinfo {author} {\bibfnamefont {T.}~\bibnamefont
  {Barklow}},\ }\href@noop {} {\enquote {\bibinfo {title} {New ild/sid results
  on higgs boson physics},}\ }\bibinfo {howpublished} {{presentation at The
  Americas Workshop on Linear Colliders 2017 (AWLC17), June 26-30, 2017,
  \url{https://agenda.linearcollider.org/event/7507/contributions/39121/}}}
  (\bibinfo {year} {2017})\BibitemShut {NoStop}%
\bibitem [{\citenamefont {Liao}(2017)}]{Liao:2017}%
  \BibitemOpen
  \bibfield  {author} {\bibinfo {author} {\bibfnamefont {L.}~\bibnamefont
  {Liao}},\ }\href@noop {} {\enquote {\bibinfo {title} {Study of br($h\to
  ww^*$) at cepc},}\ }\bibinfo {howpublished} {{presentation at The 51st
  General Meeting of ILC Physics Subgroup, April 15, 2017,
  \url{https://agenda.linearcollider.org/event/7617}}} (\bibinfo {year}
  {2017})\BibitemShut {NoStop}%
\bibitem [{\citenamefont {Kawada}\ \emph {et~al.}(2015)\citenamefont {Kawada},
  \citenamefont {Fujii}, \citenamefont {Suehara}, \citenamefont {Takahashi},\
  and\ \citenamefont {Tanabe}}]{Kawada:2015wea}%
  \BibitemOpen
  \bibfield  {author} {\bibinfo {author} {\bibfnamefont {S.-i.}\ \bibnamefont
  {Kawada}}, \bibinfo {author} {\bibfnamefont {K.}~\bibnamefont {Fujii}},
  \bibinfo {author} {\bibfnamefont {T.}~\bibnamefont {Suehara}}, \bibinfo
  {author} {\bibfnamefont {T.}~\bibnamefont {Takahashi}}, \ and\ \bibinfo
  {author} {\bibfnamefont {T.}~\bibnamefont {Tanabe}},\ }\href {\doibase
  10.1140/epjc/s10052-015-3854-2} {\bibfield  {journal} {\bibinfo  {journal}
  {Eur. Phys. J.}\ }\textbf {\bibinfo {volume} {C75}},\ \bibinfo {pages} {617}
  (\bibinfo {year} {2015})},\ \Eprint {http://arxiv.org/abs/1509.01885}
  {arXiv:1509.01885 [hep-ex]} \BibitemShut {NoStop}%
\bibitem [{\citenamefont {Ishikawa}(2014)}]{Ishikawa:2014}%
  \BibitemOpen
  \bibfield  {author} {\bibinfo {author} {\bibfnamefont {A.}~\bibnamefont
  {Ishikawa}},\ }\href@noop {} {\enquote {\bibinfo {title} {Search for
  invisible higgs decays at the ilc},}\ }\bibinfo {howpublished} {{presentation
  at Linear Collider Workshop, Belgrade, Serbia, October 5-10, 2014,
  \url{http://agenda.linearcollider.org/event/6389/session/0/contribution/140}}}
  (\bibinfo {year} {2014})\BibitemShut {NoStop}%
\bibitem [{\citenamefont {Tian}(2015{\natexlab{a}})}]{Tian:2015}%
  \BibitemOpen
  \bibfield  {author} {\bibinfo {author} {\bibfnamefont {J.}~\bibnamefont
  {Tian}},\ }\href@noop {} {\enquote {\bibinfo {title} {Higgs projections},}\
  }\bibinfo {howpublished} {{presentation at Asian Linear Collider Workshop,
  KEK, Tsukuba, Japan, April 19-24, 2015,
  \url{https://agenda.linearcollider.org/event/
  6557/session/12/contribution/129}}} (\bibinfo {year}
  {2015}{\natexlab{a}})\BibitemShut {NoStop}%
\bibitem [{\citenamefont {Kato}(2016)}]{Kato:2016}%
  \BibitemOpen
  \bibfield  {author} {\bibinfo {author} {\bibfnamefont {Y.}~\bibnamefont
  {Kato}},\ }\href@noop {} {\enquote {\bibinfo {title} {Search for invisible
  higgs decays at the ilc},}\ }\bibinfo {howpublished} {{presentation at Linear
  Collider Workshop 2016, Morioka, December 4-9, 2016,
  \url{https://agenda.linearcollider.org/event/7371/contributions/37892/}}}
  (\bibinfo {year} {2016})\BibitemShut {NoStop}%
\bibitem [{\citenamefont {Kawada}\ \emph {et~al.}(2018)\citenamefont {Kawada},
  \citenamefont {List},\ and\ \citenamefont {Berggren}}]{Kawada:2018wyz}%
  \BibitemOpen
  \bibfield  {author} {\bibinfo {author} {\bibfnamefont {S.-I.}\ \bibnamefont
  {Kawada}}, \bibinfo {author} {\bibfnamefont {J.}~\bibnamefont {List}}, \ and\
  \bibinfo {author} {\bibfnamefont {M.}~\bibnamefont {Berggren}},\ }in\
  \href@noop {} {\emph {\bibinfo {booktitle} {{International Workshop on Future
  Linear Collider (LCWS2017) Strasbourg, France, October 23-27, 2017}}}}\
  (\bibinfo {year} {2018})\ \Eprint {http://arxiv.org/abs/1801.07966}
  {arXiv:1801.07966 [hep-ex]} \BibitemShut {NoStop}%
\bibitem [{\citenamefont {Calancha}(2013)}]{Calancha:2013}%
  \BibitemOpen
  \bibfield  {author} {\bibinfo {author} {\bibfnamefont {C.}~\bibnamefont
  {Calancha}},\ }\href@noop {} {\enquote {\bibinfo {title} {Full simulation
  study on h to gamma gamma with the ild detector},}\ }\bibinfo {howpublished}
  {{presentation at Linear Collider Workshop 2013, Tokyo,
  \url{http://agenda.linearcollider.org/event/6000/session/31/contribution/180}}}
  (\bibinfo {year} {2013})\BibitemShut {NoStop}%
\bibitem [{\citenamefont {Fujii}\ and\ \citenamefont
  {Tian}(2018)}]{Fujii:2018}%
  \BibitemOpen
  \bibfield  {author} {\bibinfo {author} {\bibfnamefont {K.}~\bibnamefont
  {Fujii}}\ and\ \bibinfo {author} {\bibfnamefont {J.}~\bibnamefont {Tian}},\
  }\href@noop {} {\enquote {\bibinfo {title} {Study of h to z gamma branching
  ratio at the ilc 250 gev},}\ }\bibinfo {howpublished} {{presentation at
  Linear Collider Workshop 2018, Arlington,
  \url{https://agenda.linearcollider.org/event/7889/timetable/}}} (\bibinfo
  {year} {2018})\BibitemShut {NoStop}%
\bibitem [{\citenamefont {Jeans}(2016)}]{Jeans:2015vaa}%
  \BibitemOpen
  \bibfield  {author} {\bibinfo {author} {\bibfnamefont {D.}~\bibnamefont
  {Jeans}},\ }\href {\doibase 10.1016/j.nima.2015.11.030} {\bibfield  {journal}
  {\bibinfo  {journal} {Nucl. Instrum. Meth.}\ }\textbf {\bibinfo {volume}
  {A810}},\ \bibinfo {pages} {51} (\bibinfo {year} {2016})},\ \Eprint
  {http://arxiv.org/abs/1507.01700} {arXiv:1507.01700 [hep-ex]} \BibitemShut
  {NoStop}%
\bibitem [{\citenamefont {Jeans}\ and\ \citenamefont
  {Wilson}(2018)}]{Jeans:2018anq}%
  \BibitemOpen
  \bibfield  {author} {\bibinfo {author} {\bibfnamefont {D.}~\bibnamefont
  {Jeans}}\ and\ \bibinfo {author} {\bibfnamefont {G.~W.}\ \bibnamefont
  {Wilson}},\ }\href {\doibase 10.1103/PhysRevD.98.013007} {\bibfield
  {journal} {\bibinfo  {journal} {Phys. Rev.}\ }\textbf {\bibinfo {volume}
  {D98}},\ \bibinfo {pages} {013007} (\bibinfo {year} {2018})},\ \Eprint
  {http://arxiv.org/abs/1804.01241} {arXiv:1804.01241 [hep-ex]} \BibitemShut
  {NoStop}%
\bibitem [{\citenamefont {Ogawa}\ \emph {et~al.}(2017)\citenamefont {Ogawa},
  \citenamefont {Fujii},\ and\ \citenamefont {Tian}}]{Ogawa:2017bmg}%
  \BibitemOpen
  \bibfield  {author} {\bibinfo {author} {\bibfnamefont {T.}~\bibnamefont
  {Ogawa}}, \bibinfo {author} {\bibfnamefont {K.}~\bibnamefont {Fujii}}, \ and\
  \bibinfo {author} {\bibfnamefont {J.}~\bibnamefont {Tian}},\ }\bibfield
  {booktitle} {\emph {\bibinfo {booktitle} {{Proceedings, 2017 European
  Physical Society Conference on High Energy Physics (EPS-HEP 2017): Venice,
  Italy, July 5-12, 2017}}},\ }\href {\doibase 10.22323/1.314.0322} {\bibfield
  {journal} {\bibinfo  {journal} {PoS}\ }\textbf {\bibinfo {volume}
  {EPS-HEP2017}},\ \bibinfo {pages} {322} (\bibinfo {year} {2017})},\ \Eprint
  {http://arxiv.org/abs/1712.09772} {arXiv:1712.09772 [hep-ex]} \BibitemShut
  {NoStop}%
\bibitem [{\citenamefont {Abramowicz}\ \emph {et~al.}(2010)\citenamefont
  {Abramowicz} \emph {et~al.}}]{Abramowicz:2010bg}%
  \BibitemOpen
  \bibfield  {author} {\bibinfo {author} {\bibfnamefont {H.}~\bibnamefont
  {Abramowicz}} \emph {et~al.},\ }\href {\doibase
  10.1088/1748-0221/5/12/P12002} {\bibfield  {journal} {\bibinfo  {journal}
  {JINST}\ }\textbf {\bibinfo {volume} {5}},\ \bibinfo {pages} {P12002}
  (\bibinfo {year} {2010})},\ \Eprint {http://arxiv.org/abs/1009.2433}
  {arXiv:1009.2433 [physics.ins-det]} \BibitemShut {NoStop}%
\bibitem [{\citenamefont {Bozovic-Jelisavcic}\ \emph
  {et~al.}(2014)\citenamefont {Bozovic-Jelisavcic}, \citenamefont {Lukic},
  \citenamefont {Pandurovic},\ and\ \citenamefont
  {Smiljanic}}]{Bozovic-Jelisavcic:2014aza}%
  \BibitemOpen
  \bibfield  {author} {\bibinfo {author} {\bibfnamefont {I.}~\bibnamefont
  {Bozovic-Jelisavcic}}, \bibinfo {author} {\bibfnamefont {S.}~\bibnamefont
  {Lukic}}, \bibinfo {author} {\bibfnamefont {M.}~\bibnamefont {Pandurovic}}, \
  and\ \bibinfo {author} {\bibfnamefont {I.}~\bibnamefont {Smiljanic}},\ }in\
  \href@noop {} {\emph {\bibinfo {booktitle} {{International Workshop on Future
  Linear Colliders (LCWS13) Tokyo, Japan, November 11-15, 2013}}}}\ (\bibinfo
  {year} {2014})\ \Eprint {http://arxiv.org/abs/1403.7348} {arXiv:1403.7348
  [physics.acc-ph]} \BibitemShut {NoStop}%
\bibitem [{\citenamefont {Lukic}\ and\ \citenamefont
  {Smiljanic}(2013)}]{Lukic:2013fw}%
  \BibitemOpen
  \bibfield  {author} {\bibinfo {author} {\bibfnamefont {B.-J. I. P.~M.}\
  \bibnamefont {Lukic}, \bibfnamefont {S.}}\ and\ \bibinfo {author}
  {\bibfnamefont {I.}~\bibnamefont {Smiljanic}},\ }\href {\doibase
  10.1088/1748-0221/8/05/P05008} {\bibfield  {journal} {\bibinfo  {journal}
  {JINST}\ }\textbf {\bibinfo {volume} {8}},\ \bibinfo {pages} {P05008}
  (\bibinfo {year} {2013})},\ \Eprint {http://arxiv.org/abs/1301.1449}
  {arXiv:1301.1449 [physics.acc-ph]} \BibitemShut {NoStop}%
\bibitem [{\citenamefont {List}\ \emph {et~al.}(2015)\citenamefont {List},
  \citenamefont {Vauth},\ and\ \citenamefont {Vormwald}}]{List:2015lsa}%
  \BibitemOpen
  \bibfield  {author} {\bibinfo {author} {\bibfnamefont {J.}~\bibnamefont
  {List}}, \bibinfo {author} {\bibfnamefont {A.}~\bibnamefont {Vauth}}, \ and\
  \bibinfo {author} {\bibfnamefont {B.}~\bibnamefont {Vormwald}},\ }\href
  {\doibase 10.1088/1748-0221/10/05/P05014} {\bibfield  {journal} {\bibinfo
  {journal} {JINST}\ }\textbf {\bibinfo {volume} {10}},\ \bibinfo {pages}
  {P05014} (\bibinfo {year} {2015})},\ \Eprint
  {http://arxiv.org/abs/1502.06955} {arXiv:1502.06955 [physics.ins-det]}
  \BibitemShut {NoStop}%
\bibitem [{\citenamefont {Beckmann}\ \emph {et~al.}(2014)\citenamefont
  {Beckmann}, \citenamefont {List}, \citenamefont {Vauth},\ and\ \citenamefont
  {Vormwald}}]{Beckmann:2014mka}%
  \BibitemOpen
  \bibfield  {author} {\bibinfo {author} {\bibfnamefont {M.}~\bibnamefont
  {Beckmann}}, \bibinfo {author} {\bibfnamefont {J.}~\bibnamefont {List}},
  \bibinfo {author} {\bibfnamefont {A.}~\bibnamefont {Vauth}}, \ and\ \bibinfo
  {author} {\bibfnamefont {B.}~\bibnamefont {Vormwald}},\ }\href {\doibase
  10.1088/1748-0221/9/07/P07003} {\bibfield  {journal} {\bibinfo  {journal}
  {JINST}\ }\textbf {\bibinfo {volume} {9}},\ \bibinfo {pages} {P07003}
  (\bibinfo {year} {2014})},\ \Eprint {http://arxiv.org/abs/1405.2156}
  {arXiv:1405.2156 [physics.acc-ph]} \BibitemShut {NoStop}%
\bibitem [{\citenamefont {Blondel}\ \emph {et~al.}(2019)\citenamefont
  {Blondel}, \citenamefont {Freitas}, \citenamefont {Gluza}, \citenamefont
  {Riemann}, \citenamefont {Heinemeyer}, \citenamefont {Jadach},\ and\
  \citenamefont {Janot}}]{Blondel:2019qlh}%
  \BibitemOpen
  \bibfield  {author} {\bibinfo {author} {\bibfnamefont {A.}~\bibnamefont
  {Blondel}}, \bibinfo {author} {\bibfnamefont {A.}~\bibnamefont {Freitas}},
  \bibinfo {author} {\bibfnamefont {J.}~\bibnamefont {Gluza}}, \bibinfo
  {author} {\bibfnamefont {T.}~\bibnamefont {Riemann}}, \bibinfo {author}
  {\bibfnamefont {S.}~\bibnamefont {Heinemeyer}}, \bibinfo {author}
  {\bibfnamefont {S.}~\bibnamefont {Jadach}}, \ and\ \bibinfo {author}
  {\bibfnamefont {P.}~\bibnamefont {Janot}},\ }\href@noop {} {\  (\bibinfo
  {year} {2019})},\ \Eprint {http://arxiv.org/abs/1901.02648} {arXiv:1901.02648
  [hep-ph]} \BibitemShut {NoStop}%
\bibitem [{\citenamefont {Abe}\ \emph {et~al.}(1997)\citenamefont {Abe} \emph
  {et~al.}}]{Abe:1996nj}%
  \BibitemOpen
  \bibfield  {author} {\bibinfo {author} {\bibfnamefont {K.}~\bibnamefont
  {Abe}} \emph {et~al.} (\bibinfo {collaboration} {SLD}),\ }\href {\doibase
  10.1103/PhysRevLett.78.2075} {\bibfield  {journal} {\bibinfo  {journal}
  {Phys. Rev. Lett.}\ }\textbf {\bibinfo {volume} {78}},\ \bibinfo {pages}
  {2075} (\bibinfo {year} {1997})},\ \Eprint
  {http://arxiv.org/abs/hep-ex/9611011} {arXiv:hep-ex/9611011 [hep-ex]}
  \BibitemShut {NoStop}%
\bibitem [{\citenamefont {Dürig}(2016)}]{Duerig:2016dvi}%
  \BibitemOpen
  \bibfield  {author} {\bibinfo {author} {\bibfnamefont {C.~F.}\ \bibnamefont
  {Dürig}},\ }\emph {\bibinfo {title} {{Measuring the Higgs Self-coupling at
  the International Linear Collider}}},\ \href
  {http://bib-pubdb1.desy.de/search?cc=Publication+Database&of=hd&p=reportnumber:DESY-THESIS-2016-027}
  {Ph.D. thesis},\ \bibinfo  {school} {Hamburg U.}, \bibinfo {address}
  {Hamburg} (\bibinfo {year} {2016})\BibitemShut {NoStop}%
\bibitem [{\citenamefont {Tian}(2013)}]{Tian:2013qmi}%
  \BibitemOpen
  \bibfield  {author} {\bibinfo {author} {\bibfnamefont {J.}~\bibnamefont
  {Tian}},\ }in\ \href@noop {} {\emph {\bibinfo {booktitle} {{Helmholtz
  Alliance Linear Collider Forum: Proceedings of the Workshops Hamburg, Munich,
  Hamburg 2010-2012, Germany}}}},\ \bibinfo {organization} {DESY}\ (\bibinfo
  {publisher} {DESY},\ \bibinfo {address} {Hamburg},\ \bibinfo {year} {2013})\
  pp.\ \bibinfo {pages} {224--247}\BibitemShut {NoStop}%
\bibitem [{\citenamefont {Kurata}(2013)}]{KurataHHH}%
  \BibitemOpen
  \bibfield  {author} {\bibinfo {author} {\bibfnamefont {M.}~\bibnamefont
  {Kurata}},\ }\bibfield  {booktitle} {\emph {\bibinfo {booktitle}
  {www-flc.desy.de/lcnotes/notes/LC-REP-2013-025.pdf}},\ }\href@noop {} {\
  (\bibinfo {year} {2013})}\BibitemShut {NoStop}%
\bibitem [{\citenamefont {Roloff}\ \emph {et~al.}(2019)\citenamefont {Roloff},
  \citenamefont {Schnoor}, \citenamefont {Simoniello},\ and\ \citenamefont
  {Xu}}]{Roloff:2019crr}%
  \BibitemOpen
  \bibfield  {author} {\bibinfo {author} {\bibfnamefont {P.}~\bibnamefont
  {Roloff}}, \bibinfo {author} {\bibfnamefont {U.}~\bibnamefont {Schnoor}},
  \bibinfo {author} {\bibfnamefont {R.}~\bibnamefont {Simoniello}}, \ and\
  \bibinfo {author} {\bibfnamefont {B.}~\bibnamefont {Xu}} (\bibinfo
  {collaboration} {CLICdp}),\ }\href@noop {} {\  (\bibinfo {year} {2019})},\
  \Eprint {http://arxiv.org/abs/1901.05897} {arXiv:1901.05897 [hep-ex]}
  \BibitemShut {NoStop}%
\bibitem [{\citenamefont {Tian}(2015{\natexlab{b}})}]{TianHHH:2015}%
  \BibitemOpen
  \bibfield  {author} {\bibinfo {author} {\bibfnamefont {J.}~\bibnamefont
  {Tian}},\ }\href@noop {} {\enquote {\bibinfo {title} {Zhh cross section
  analysis},}\ }\bibinfo {howpublished} {{presentation at Linear Collider
  Workshop 2015, Whistler, Canada, November 1-7, 2015,
  \url{https://agenda.linearcollider.org/event/6662/contributions/32620/}}}
  (\bibinfo {year} {2015}{\natexlab{b}})\BibitemShut {NoStop}%
\bibitem [{\citenamefont {McCullough}(2014)}]{McCullough:2013rea}%
  \BibitemOpen
  \bibfield  {author} {\bibinfo {author} {\bibfnamefont {M.}~\bibnamefont
  {McCullough}},\ }\href {\doibase 10.1103/PhysRevD.90.015001,
  10.1103/PhysRevD.92.039903} {\bibfield  {journal} {\bibinfo  {journal} {Phys.
  Rev.}\ }\textbf {\bibinfo {volume} {D90}},\ \bibinfo {pages} {015001}
  (\bibinfo {year} {2014})},\ \bibinfo {note} {[Erratum: Phys.
  Rev.D92,no.3,039903(2015)]},\ \Eprint {http://arxiv.org/abs/1312.3322}
  {arXiv:1312.3322 [hep-ph]} \BibitemShut {NoStop}%
\bibitem [{\citenamefont {Barklow}\ \emph
  {et~al.}(2018{\natexlab{b}})\citenamefont {Barklow}, \citenamefont {Fujii},
  \citenamefont {Jung}, \citenamefont {Peskin},\ and\ \citenamefont
  {Tian}}]{Barklow:2017awn}%
  \BibitemOpen
  \bibfield  {author} {\bibinfo {author} {\bibfnamefont {T.}~\bibnamefont
  {Barklow}}, \bibinfo {author} {\bibfnamefont {K.}~\bibnamefont {Fujii}},
  \bibinfo {author} {\bibfnamefont {S.}~\bibnamefont {Jung}}, \bibinfo {author}
  {\bibfnamefont {M.~E.}\ \bibnamefont {Peskin}}, \ and\ \bibinfo {author}
  {\bibfnamefont {J.}~\bibnamefont {Tian}},\ }\href {\doibase
  10.1103/PhysRevD.97.053004} {\bibfield  {journal} {\bibinfo  {journal} {Phys.
  Rev.}\ }\textbf {\bibinfo {volume} {D97}},\ \bibinfo {pages} {053004}
  (\bibinfo {year} {2018}{\natexlab{b}})},\ \Eprint
  {http://arxiv.org/abs/1708.09079} {arXiv:1708.09079 [hep-ph]} \BibitemShut
  {NoStop}%
\bibitem [{\citenamefont {Di~Vita}\ \emph {et~al.}(2018)\citenamefont
  {Di~Vita}, \citenamefont {Durieux}, \citenamefont {Grojean}, \citenamefont
  {Gu}, \citenamefont {Liu}, \citenamefont {Panico}, \citenamefont {Riembau},\
  and\ \citenamefont {Vantalon}}]{DiVita:2017vrr}%
  \BibitemOpen
  \bibfield  {author} {\bibinfo {author} {\bibfnamefont {S.}~\bibnamefont
  {Di~Vita}}, \bibinfo {author} {\bibfnamefont {G.}~\bibnamefont {Durieux}},
  \bibinfo {author} {\bibfnamefont {C.}~\bibnamefont {Grojean}}, \bibinfo
  {author} {\bibfnamefont {J.}~\bibnamefont {Gu}}, \bibinfo {author}
  {\bibfnamefont {Z.}~\bibnamefont {Liu}}, \bibinfo {author} {\bibfnamefont
  {G.}~\bibnamefont {Panico}}, \bibinfo {author} {\bibfnamefont
  {M.}~\bibnamefont {Riembau}}, \ and\ \bibinfo {author} {\bibfnamefont
  {T.}~\bibnamefont {Vantalon}},\ }\href {\doibase 10.1007/JHEP02(2018)178}
  {\bibfield  {journal} {\bibinfo  {journal} {JHEP}\ }\textbf {\bibinfo
  {volume} {02}},\ \bibinfo {pages} {178} (\bibinfo {year} {2018})},\ \Eprint
  {http://arxiv.org/abs/1711.03978} {arXiv:1711.03978 [hep-ph]} \BibitemShut
  {NoStop}%
\bibitem [{\citenamefont {Marchesini}(2011)}]{Marchesini:94888}%
  \BibitemOpen
  \bibfield  {author} {\bibinfo {author} {\bibfnamefont {I.}~\bibnamefont
  {Marchesini}},\ }\emph {\bibinfo {title} {{T}riple {G}auge {C}ouplings and
  {P}olarization at the ${ILC}\\and$ {L}eakage in a {H}ighly {G}ranular
  {C}alorimeter}},\ \href {http://bib-pubdb1.desy.de/record/94888} {\bibinfo
  {type} {Dr.}},\ \bibinfo  {school} {Hamburg University}, \bibinfo {address}
  {Hamburg} (\bibinfo {year} {2011}),\ \bibinfo {note} {hamburg University,
  Diss., 2011}\BibitemShut {NoStop}%
\bibitem [{\citenamefont {Rosca}(2016)}]{Rosca:2016hcq}%
  \BibitemOpen
  \bibfield  {author} {\bibinfo {author} {\bibfnamefont {A.}~\bibnamefont
  {Rosca}},\ }in\ \href {\doibase 10.1016/j.nuclphysbps.2015.09.362} {\emph
  {\bibinfo {booktitle} {37th International Conference on High Energy
  Physics}}},\ Vol.\ \bibinfo {volume} {273-275}\ (\bibinfo {address}
  {Valencia, Spain},\ \bibinfo {year} {2016})\ pp.\ \bibinfo {pages}
  {2226--2231}\BibitemShut {NoStop}%
\bibitem [{\citenamefont {Schael}\ \emph {et~al.}(2005)\citenamefont {Schael}
  \emph {et~al.}}]{Schael:2004tq}%
  \BibitemOpen
  \bibfield  {author} {\bibinfo {author} {\bibfnamefont {S.}~\bibnamefont
  {Schael}} \emph {et~al.} (\bibinfo {collaboration} {ALEPH}),\ }\href
  {\doibase 10.1016/j.physletb.2005.03.058} {\bibfield  {journal} {\bibinfo
  {journal} {Phys.Lett.B}\ }\textbf {\bibinfo {volume} {614}},\ \bibinfo
  {pages} {7} (\bibinfo {year} {2005})}\BibitemShut {NoStop}%
\bibitem [{\citenamefont {Karl}()}]{Karl:2017let}%
  \BibitemOpen
  \bibfield  {author} {\bibinfo {author} {\bibfnamefont {R.}~\bibnamefont
  {Karl}},\ }in\ \href {\doibase 10.22323/1.314.0763} {\emph {\bibinfo
  {booktitle} {2017 European Physical Society Conference on High Energy
  Physics}}},\ Vol.\ \bibinfo {volume} {EPS-HEP2017}\ (\bibinfo  {publisher}
  {SISSA})\ p.\ \bibinfo {pages} {763}\BibitemShut {NoStop}%
\bibitem [{\citenamefont {Barklow}(1995)}]{Barklow:1995sk}%
  \BibitemOpen
  \bibfield  {author} {\bibinfo {author} {\bibfnamefont {T.}~\bibnamefont
  {Barklow}},\ }in\ \href {\doibase 10.1063/1.49314} {\emph {\bibinfo
  {booktitle} {International Symposium on Vector Boson Self-interactions
  (WGZ95) (To be followed by Topics in Particle Physics with CMS, Los Angeles,
  CA, 6-8 Feb 1995)}}},\ Vol.\ \bibinfo {volume} {350}\ (\bibinfo {address}
  {Los Angeles, United States},\ \bibinfo {year} {1995})\ pp.\ \bibinfo {pages}
  {307--322}\BibitemShut {NoStop}%
\bibitem [{\citenamefont {Diehl}\ and\ \citenamefont
  {Nachtmann}(1998)}]{Diehl:1997ft}%
  \BibitemOpen
  \bibfield  {author} {\bibinfo {author} {\bibfnamefont {M.}~\bibnamefont
  {Diehl}}\ and\ \bibinfo {author} {\bibfnamefont {O.}~\bibnamefont
  {Nachtmann}},\ }\href {\doibase 10.1007/BF01245807} {\bibfield  {journal}
  {\bibinfo  {journal} {Eur.Phys.J.C}\ }\textbf {\bibinfo {volume} {1}},\
  \bibinfo {pages} {177} (\bibinfo {year} {1998})},\ \Eprint
  {http://arxiv.org/abs/hep-ph/9702208} {arXiv:hep-ph/9702208} \BibitemShut
  {NoStop}%
\bibitem [{\citenamefont {Diehl}\ \emph {et~al.}(2003)\citenamefont {Diehl},
  \citenamefont {Nachtmann},\ and\ \citenamefont {Nagel}}]{Diehl:2002nj}%
  \BibitemOpen
  \bibfield  {author} {\bibinfo {author} {\bibfnamefont {M.}~\bibnamefont
  {Diehl}}, \bibinfo {author} {\bibfnamefont {O.}~\bibnamefont {Nachtmann}}, \
  and\ \bibinfo {author} {\bibfnamefont {F.}~\bibnamefont {Nagel}},\ }\href
  {\doibase 10.1140/epjc/s2002-01096-y} {\bibfield  {journal} {\bibinfo
  {journal} {Eur.Phys.J.C}\ }\textbf {\bibinfo {volume} {27}},\ \bibinfo
  {pages} {375} (\bibinfo {year} {2003})},\ \Eprint
  {http://arxiv.org/abs/hep-ph/0209229} {arXiv:hep-ph/0209229} \BibitemShut
  {NoStop}%
\bibitem [{\citenamefont {Aad}\ \emph {et~al.}(2015)\citenamefont {Aad} \emph
  {et~al.}}]{Aad:2014mda}%
  \BibitemOpen
  \bibfield  {author} {\bibinfo {author} {\bibfnamefont {G.}~\bibnamefont
  {Aad}} \emph {et~al.} (\bibinfo {collaboration} {ATLAS}),\ }\href {\doibase
  10.1007/JHEP01(2015)049} {\bibfield  {journal} {\bibinfo  {journal} {JHEP}\
  }\textbf {\bibinfo {volume} {01}},\ \bibinfo {pages} {049} (\bibinfo {year}
  {2015})},\ \Eprint {http://arxiv.org/abs/1410.7238} {arXiv:1410.7238
  [hep-ex]} \BibitemShut {NoStop}%
\bibitem [{\citenamefont {Moenig}(CERN)}]{bib:HLLHCMoenig}%
  \BibitemOpen
  \bibfield  {author} {\bibinfo {author} {\bibfnamefont {K.}~\bibnamefont
  {Moenig}},\ }\href@noop {} {\enquote {\bibinfo {title} {Atlas \& cms physics
  prospects for the high-luminosity lhc},}\ }\bibinfo {howpublished}
  {{\url{https://cds.cern.ch/record/1510150/files/ATL-PHYS-SLIDE-2013-042.pdf}}}
  (\bibinfo {year} {presentation at CLIC Workshop 2013, CERN})\BibitemShut
  {NoStop}%
\bibitem [{\citenamefont {Karl}(2017)}]{bib:TGC_EPS17}%
  \BibitemOpen
  \bibfield  {author} {\bibinfo {author} {\bibfnamefont {R.}~\bibnamefont
  {Karl}},\ }\href {{https://pos.sissa.it/314/763/}} {\  (\bibinfo {year}
  {2017})}\BibitemShut {NoStop}%
\bibitem [{\citenamefont {{ALEPH, DELPHI, L3, OPAL, and the LEP TGC Working
  Group}}(2005)}]{bib:LEPTGC}%
  \BibitemOpen
  \bibfield  {author} {\bibinfo {author} {\bibnamefont {{ALEPH, DELPHI, L3,
  OPAL, and the LEP TGC Working Group}}},\ }\href@noop {} {\enquote {\bibinfo
  {title} {{A combination of results on charged triple gauge boson couplings
  measured by the LEP experiments}},}\ } (\bibinfo {year} {2005})\BibitemShut
  {NoStop}%
\bibitem [{\citenamefont {{LHC collaborations}}(2018)}]{bib:LHCTGC}%
  \BibitemOpen
  \bibfield  {author} {\bibinfo {author} {\bibnamefont {{LHC
  collaborations}}},\ }\href
  {{https://twiki.cern.ch/twiki/bin/view/CMSPublic/PhysicsResultsSMPaTGC}}
  {\enquote {\bibinfo {title} {{LHC limits on anomalous triple and quartic
  gauge couplings in comparison of the running experiments}},}\ } (\bibinfo
  {year} {2018})\BibitemShut {NoStop}%
\bibitem [{\citenamefont {Freitas}\ \emph {et~al.}()\citenamefont {Freitas},
  \citenamefont {Hagiwara}, \citenamefont {Heinemeyer}, \citenamefont
  {Langacker}, \citenamefont {Moenig}, \citenamefont {Tanabashi},\ and\
  \citenamefont {Wilson}}]{Freitas:2013xga}%
  \BibitemOpen
  \bibfield  {author} {\bibinfo {author} {\bibfnamefont {A.}~\bibnamefont
  {Freitas}}, \bibinfo {author} {\bibfnamefont {K.}~\bibnamefont {Hagiwara}},
  \bibinfo {author} {\bibfnamefont {S.}~\bibnamefont {Heinemeyer}}, \bibinfo
  {author} {\bibfnamefont {P.}~\bibnamefont {Langacker}}, \bibinfo {author}
  {\bibfnamefont {K.}~\bibnamefont {Moenig}}, \bibinfo {author} {\bibfnamefont
  {M.}~\bibnamefont {Tanabashi}}, \ and\ \bibinfo {author} {\bibfnamefont
  {G.}~\bibnamefont {Wilson}},\ }in\ \href
  {http://www.slac.stanford.edu/econf/C1307292/docs/submittedArxivFiles/1307.3962.pdf}
  {\emph {\bibinfo {booktitle} {Community Summer Study 2013}}}\ (\bibinfo
  {address} {Minneapolis, United States})\ \Eprint
  {http://arxiv.org/abs/1307.3962} {arXiv:1307.3962 [hep-ph]} \BibitemShut
  {NoStop}%
\bibitem [{\citenamefont {Wilson}(2016{\natexlab{a}})}]{Wilson:2016tto}%
  \BibitemOpen
  \bibfield  {author} {\bibinfo {author} {\bibfnamefont {G.~W.}\ \bibnamefont
  {Wilson}},\ }in\ \href {\doibase 10.22323/1.282.0688} {\emph {\bibinfo
  {booktitle} {38th International Conference on High Energy Physics}}},\ Vol.\
  \bibinfo {volume} {ICHEP2016}\ (\bibinfo  {publisher} {SISSA},\ \bibinfo
  {address} {Chicago, United States},\ \bibinfo {year} {2016})\ p.\ \bibinfo
  {pages} {688}\BibitemShut {NoStop}%
\bibitem [{\citenamefont {Wilson}(2016{\natexlab{b}})}]{Wilson:2016hne}%
  \BibitemOpen
  \bibfield  {author} {\bibinfo {author} {\bibfnamefont {G.~W.}\ \bibnamefont
  {Wilson}},\ }in\ \href@noop {} {\emph {\bibinfo {booktitle} {{Proceedings,
  International Workshop on Future Linear Colliders (LCWS15): Whistler, B.C.,
  Canada, November 02-06, 2015}}}}\ (\bibinfo {year} {2016})\ \Eprint
  {http://arxiv.org/abs/1603.06016} {arXiv:1603.06016 [hep-ex]} \BibitemShut
  {NoStop}%
\bibitem [{\citenamefont {Bilokin}\ \emph {et~al.}()\citenamefont {Bilokin},
  \citenamefont {Pöschl},\ and\ \citenamefont {Richard}}]{Bilokin:2017lco}%
  \BibitemOpen
  \bibfield  {author} {\bibinfo {author} {\bibfnamefont {S.}~\bibnamefont
  {Bilokin}}, \bibinfo {author} {\bibfnamefont {R.}~\bibnamefont {Pöschl}}, \
  and\ \bibinfo {author} {\bibfnamefont {F.}~\bibnamefont {Richard}},\
  }\href@noop {} {\ }\Eprint {http://arxiv.org/abs/1709.04289}
  {arXiv:1709.04289 [hep-ex]} \BibitemShut {NoStop}%
\bibitem [{\citenamefont {Deguchi}( USA)}]{bib:deguchi_lcws18}%
  \BibitemOpen
  \bibfield  {author} {\bibinfo {author} {\bibfnamefont {Y.}~\bibnamefont
  {Deguchi}},\ }\href@noop {} {\enquote {\bibinfo {title} {Study of fermion
  pair production events at the ilc with center of mass energy of 250 gev},}\
  }\bibinfo {howpublished}
  {{\url{https://agenda.linearcollider.org/event/7889/contributions/42502/}}}
  (\bibinfo {year} {presentation at LCWS 2018, University of Texas, Arlington,
  USA})\BibitemShut {NoStop}%
\bibitem [{\citenamefont {Yamashiro}\ \emph {et~al.}(2018)\citenamefont
  {Yamashiro}, \citenamefont {Kawagoe}, \citenamefont {Suehara}, \citenamefont
  {Yoshioka}, \citenamefont {Fujii},\ and\ \citenamefont
  {Miyamoto}}]{Yamashiro:2018ant}%
  \BibitemOpen
  \bibfield  {author} {\bibinfo {author} {\bibfnamefont {H.}~\bibnamefont
  {Yamashiro}}, \bibinfo {author} {\bibfnamefont {K.}~\bibnamefont {Kawagoe}},
  \bibinfo {author} {\bibfnamefont {T.}~\bibnamefont {Suehara}}, \bibinfo
  {author} {\bibfnamefont {T.}~\bibnamefont {Yoshioka}}, \bibinfo {author}
  {\bibfnamefont {K.}~\bibnamefont {Fujii}}, \ and\ \bibinfo {author}
  {\bibfnamefont {A.}~\bibnamefont {Miyamoto}},\ }in\ \href@noop {} {\emph
  {\bibinfo {booktitle} {{International Workshop on Future Linear Collider
  (LCWS2017) Strasbourg, France, October 23-27, 2017}}}}\ (\bibinfo {year}
  {2018})\ \Eprint {http://arxiv.org/abs/1801.04671} {arXiv:1801.04671
  [hep-ex]} \BibitemShut {NoStop}%
\bibitem [{\citenamefont {Ueno}( USA)}]{bib:ueno_lcws18}%
  \BibitemOpen
  \bibfield  {author} {\bibinfo {author} {\bibfnamefont {T.}~\bibnamefont
  {Ueno}},\ }\href@noop {} {\enquote {\bibinfo {title} {Measurement of the
  left-right asymmetry in $e^+e^- \to \gamma z$ at the 250 gev ilc},}\
  }\bibinfo {howpublished}
  {{\url{https://agenda.linearcollider.org/event/7889/contributions/42503/}}}
  (\bibinfo {year} {presentation at LCWS 2018, University of Texas, Arlington,
  USA})\BibitemShut {NoStop}%
\bibitem [{\citenamefont {Blondel}(1988)}]{Blondel:1987wr}%
  \BibitemOpen
  \bibfield  {author} {\bibinfo {author} {\bibfnamefont {A.}~\bibnamefont
  {Blondel}},\ }\href {\doibase 10.1016/0370-2693(88)90869-6} {\bibfield
  {journal} {\bibinfo  {journal} {Phys. Lett.}\ }\textbf {\bibinfo {volume}
  {B202}},\ \bibinfo {pages} {145} (\bibinfo {year} {1988})},\ \bibinfo {note}
  {[Erratum: Phys. Lett.208,531(1988)]}\BibitemShut {NoStop}%
\bibitem [{\citenamefont {Monig}(2001)}]{Monig:2001db}%
  \BibitemOpen
  \bibfield  {author} {\bibinfo {author} {\bibfnamefont {K.}~\bibnamefont
  {Monig}},\ }\bibfield  {booktitle} {\emph {\bibinfo {booktitle}
  {{Proceedings, APS / DPF / DPB Summer Study on the Future of Particle Physics
  (Snowmass 2001), Snowmass, Colorado, 30 Jun - 21 Jul 2001}}},\ }\href@noop {}
  {\bibfield  {journal} {\bibinfo  {journal} {eConf}\ }\textbf {\bibinfo
  {volume} {C010630}},\ \bibinfo {pages} {E3009} (\bibinfo {year}
  {2001})}\BibitemShut {NoStop}%
\bibitem [{\citenamefont {Suehara}(2009)}]{Suehara:2009nj}%
  \BibitemOpen
  \bibfield  {author} {\bibinfo {author} {\bibfnamefont {T.}~\bibnamefont
  {Suehara}},\ }in\ \href@noop {} {\emph {\bibinfo {booktitle} {{8th General
  Meeting of the ILC Physics Subgroup Tsukuba, Japan, January 21, 2009}}}}\
  (\bibinfo {year} {2009})\ \Eprint {http://arxiv.org/abs/0909.2398}
  {arXiv:0909.2398 [hep-ex]} \BibitemShut {NoStop}%
\bibitem [{\citenamefont {Tran}\ \emph {et~al.}(2016)\citenamefont {Tran},
  \citenamefont {Balagura}, \citenamefont {Boudry}, \citenamefont {Brient},\
  and\ \citenamefont {Videau}}]{Tran:2015nxa}%
  \BibitemOpen
  \bibfield  {author} {\bibinfo {author} {\bibfnamefont {T.~H.}\ \bibnamefont
  {Tran}}, \bibinfo {author} {\bibfnamefont {V.}~\bibnamefont {Balagura}},
  \bibinfo {author} {\bibfnamefont {V.}~\bibnamefont {Boudry}}, \bibinfo
  {author} {\bibfnamefont {J.-C.}\ \bibnamefont {Brient}}, \ and\ \bibinfo
  {author} {\bibfnamefont {H.}~\bibnamefont {Videau}},\ }\href {\doibase
  10.1140/epjc/s10052-016-4315-2} {\bibfield  {journal} {\bibinfo  {journal}
  {Eur. Phys. J.}\ }\textbf {\bibinfo {volume} {C76}},\ \bibinfo {pages} {468}
  (\bibinfo {year} {2016})},\ \Eprint {http://arxiv.org/abs/1510.05224}
  {arXiv:1510.05224 [physics.ins-det]} \BibitemShut {NoStop}%
\bibitem [{\citenamefont {Bilokin}(2017)}]{bilokin:tel-01946099}%
  \BibitemOpen
  \bibfield  {author} {\bibinfo {author} {\bibfnamefont {S.}~\bibnamefont
  {Bilokin}},\ }\emph {\bibinfo {title} {{Hadronic showers in a highly granular
  silicon-tungsten calorimeter and production of bottom and top quarks at the
  ILC}}},\ \href {https://tel.archives-ouvertes.fr/tel-01946099} {\bibinfo
  {type} {Theses}},\ \bibinfo  {school} {{Universit{\'e} Paris-Saclay}}
  (\bibinfo {year} {2017})\BibitemShut {NoStop}%
\bibitem [{\citenamefont {Agashe}\ \emph {et~al.}(2013)\citenamefont {Agashe}
  \emph {et~al.}}]{Agashe:2013hma}%
  \BibitemOpen
  \bibfield  {author} {\bibinfo {author} {\bibfnamefont {K.}~\bibnamefont
  {Agashe}} \emph {et~al.} (\bibinfo {collaboration} {Top Quark Working
  Group}),\ }in\ \href
  {http://www.slac.stanford.edu/econf/C1307292/docs/Top-21.pdf} {\emph
  {\bibinfo {booktitle} {{Proceedings, 2013 Community Summer Study on the
  Future of U.S. Particle Physics: Snowmass on the Mississippi (CSS2013):
  Minneapolis, MN, USA, July 29-August 6, 2013}}}}\ (\bibinfo {year} {2013})\
  \Eprint {http://arxiv.org/abs/1311.2028} {arXiv:1311.2028 [hep-ph]}
  \BibitemShut {NoStop}%
\bibitem [{\citenamefont {Vos}\ \emph {et~al.}(2016)\citenamefont {Vos} \emph
  {et~al.}}]{Vos:2016til}%
  \BibitemOpen
  \bibfield  {author} {\bibinfo {author} {\bibfnamefont {M.}~\bibnamefont
  {Vos}} \emph {et~al.},\ }\href@noop {} {\  (\bibinfo {year} {2016})},\
  \Eprint {http://arxiv.org/abs/1604.08122} {arXiv:1604.08122 [hep-ex]}
  \BibitemShut {NoStop}%
\bibitem [{\citenamefont {Abramowicz}\ \emph {et~al.}(2018)\citenamefont
  {Abramowicz} \emph {et~al.}}]{Abramowicz:2018rjq}%
  \BibitemOpen
  \bibfield  {author} {\bibinfo {author} {\bibfnamefont {H.}~\bibnamefont
  {Abramowicz}} \emph {et~al.} (\bibinfo {collaboration} {CLICdp}),\
  }\href@noop {} {\  (\bibinfo {year} {2018})},\ \Eprint
  {http://arxiv.org/abs/1807.02441} {arXiv:1807.02441 [hep-ex]} \BibitemShut
  {NoStop}%
\bibitem [{\citenamefont {Amjad}\ \emph {et~al.}(2015)\citenamefont {Amjad}
  \emph {et~al.}}]{Amjad:2015mma}%
  \BibitemOpen
  \bibfield  {author} {\bibinfo {author} {\bibfnamefont {M.~S.}\ \bibnamefont
  {Amjad}} \emph {et~al.},\ }\href {\doibase 10.1140/epjc/s10052-015-3746-5}
  {\bibfield  {journal} {\bibinfo  {journal} {Eur. Phys. J.}\ }\textbf
  {\bibinfo {volume} {C75}},\ \bibinfo {pages} {512} (\bibinfo {year}
  {2015})},\ \Eprint {http://arxiv.org/abs/1505.06020} {arXiv:1505.06020
  [hep-ex]} \BibitemShut {NoStop}%
\bibitem [{\citenamefont {Bernreuther}\ \emph {et~al.}(2018)\citenamefont
  {Bernreuther}, \citenamefont {Chen}, \citenamefont {García}, \citenamefont
  {Perelló}, \citenamefont {Poeschl}, \citenamefont {Richard}, \citenamefont
  {Ros},\ and\ \citenamefont {Vos}}]{Bernreuther:2017cyi}%
  \BibitemOpen
  \bibfield  {author} {\bibinfo {author} {\bibfnamefont {W.}~\bibnamefont
  {Bernreuther}}, \bibinfo {author} {\bibfnamefont {L.}~\bibnamefont {Chen}},
  \bibinfo {author} {\bibfnamefont {I.}~\bibnamefont {García}}, \bibinfo
  {author} {\bibfnamefont {M.}~\bibnamefont {Perelló}}, \bibinfo {author}
  {\bibfnamefont {R.}~\bibnamefont {Poeschl}}, \bibinfo {author} {\bibfnamefont
  {F.}~\bibnamefont {Richard}}, \bibinfo {author} {\bibfnamefont
  {E.}~\bibnamefont {Ros}}, \ and\ \bibinfo {author} {\bibfnamefont
  {M.}~\bibnamefont {Vos}},\ }\href {\doibase 10.1140/epjc/s10052-018-5625-3}
  {\bibfield  {journal} {\bibinfo  {journal} {Eur. Phys. J.}\ }\textbf
  {\bibinfo {volume} {C78}},\ \bibinfo {pages} {155} (\bibinfo {year}
  {2018})},\ \Eprint {http://arxiv.org/abs/1710.06737} {arXiv:1710.06737
  [hep-ex]} \BibitemShut {NoStop}%
\bibitem [{\citenamefont {Devetak}\ \emph {et~al.}(2011)\citenamefont
  {Devetak}, \citenamefont {Nomerotski},\ and\ \citenamefont
  {Peskin}}]{Devetak:2010na}%
  \BibitemOpen
  \bibfield  {author} {\bibinfo {author} {\bibfnamefont {E.}~\bibnamefont
  {Devetak}}, \bibinfo {author} {\bibfnamefont {A.}~\bibnamefont {Nomerotski}},
  \ and\ \bibinfo {author} {\bibfnamefont {M.}~\bibnamefont {Peskin}},\ }\href
  {\doibase 10.1103/PhysRevD.84.034029} {\bibfield  {journal} {\bibinfo
  {journal} {Phys. Rev.}\ }\textbf {\bibinfo {volume} {D84}},\ \bibinfo {pages}
  {034029} (\bibinfo {year} {2011})},\ \Eprint {http://arxiv.org/abs/1005.1756}
  {arXiv:1005.1756 [hep-ex]} \BibitemShut {NoStop}%
\bibitem [{\citenamefont {Durieux}\ \emph
  {et~al.}(2018{\natexlab{a}})\citenamefont {Durieux}, \citenamefont
  {Perelló}, \citenamefont {Vos},\ and\ \citenamefont
  {Zhang}}]{Durieux:2018tev}%
  \BibitemOpen
  \bibfield  {author} {\bibinfo {author} {\bibfnamefont {G.}~\bibnamefont
  {Durieux}}, \bibinfo {author} {\bibfnamefont {M.}~\bibnamefont {Perelló}},
  \bibinfo {author} {\bibfnamefont {M.}~\bibnamefont {Vos}}, \ and\ \bibinfo
  {author} {\bibfnamefont {C.}~\bibnamefont {Zhang}},\ }\href {\doibase
  10.1007/JHEP10(2018)168} {\bibfield  {journal} {\bibinfo  {journal} {JHEP}\
  }\textbf {\bibinfo {volume} {10}},\ \bibinfo {pages} {168} (\bibinfo {year}
  {2018}{\natexlab{a}})},\ \Eprint {http://arxiv.org/abs/1807.02121}
  {arXiv:1807.02121 [hep-ph]} \BibitemShut {NoStop}%
\bibitem [{\citenamefont {Baak}\ \emph {et~al.}(2014)\citenamefont {Baak},
  \citenamefont {Cúth}, \citenamefont {Haller}, \citenamefont {Hoecker},
  \citenamefont {Kogler}, \citenamefont {Mönig}, \citenamefont {Schott},\ and\
  \citenamefont {Stelzer}}]{Baak:2014ora}%
  \BibitemOpen
  \bibfield  {author} {\bibinfo {author} {\bibfnamefont {M.}~\bibnamefont
  {Baak}}, \bibinfo {author} {\bibfnamefont {J.}~\bibnamefont {Cúth}},
  \bibinfo {author} {\bibfnamefont {J.}~\bibnamefont {Haller}}, \bibinfo
  {author} {\bibfnamefont {A.}~\bibnamefont {Hoecker}}, \bibinfo {author}
  {\bibfnamefont {R.}~\bibnamefont {Kogler}}, \bibinfo {author} {\bibfnamefont
  {K.}~\bibnamefont {Mönig}}, \bibinfo {author} {\bibfnamefont
  {M.}~\bibnamefont {Schott}}, \ and\ \bibinfo {author} {\bibfnamefont
  {J.}~\bibnamefont {Stelzer}} (\bibinfo {collaboration} {Gfitter Group}),\
  }\href {\doibase 10.1140/epjc/s10052-014-3046-5} {\bibfield  {journal}
  {\bibinfo  {journal} {Eur. Phys. J.}\ }\textbf {\bibinfo {volume} {C74}},\
  \bibinfo {pages} {3046} (\bibinfo {year} {2014})},\ \Eprint
  {http://arxiv.org/abs/1407.3792} {arXiv:1407.3792 [hep-ph]} \BibitemShut
  {NoStop}%
\bibitem [{\citenamefont {Degrassi}\ \emph {et~al.}(2012)\citenamefont
  {Degrassi}, \citenamefont {Di~Vita}, \citenamefont {Elias-Miro},
  \citenamefont {Espinosa}, \citenamefont {Giudice}, \citenamefont {Isidori},\
  and\ \citenamefont {Strumia}}]{Degrassi:2012ry}%
  \BibitemOpen
  \bibfield  {author} {\bibinfo {author} {\bibfnamefont {G.}~\bibnamefont
  {Degrassi}}, \bibinfo {author} {\bibfnamefont {S.}~\bibnamefont {Di~Vita}},
  \bibinfo {author} {\bibfnamefont {J.}~\bibnamefont {Elias-Miro}}, \bibinfo
  {author} {\bibfnamefont {J.~R.}\ \bibnamefont {Espinosa}}, \bibinfo {author}
  {\bibfnamefont {G.~F.}\ \bibnamefont {Giudice}}, \bibinfo {author}
  {\bibfnamefont {G.}~\bibnamefont {Isidori}}, \ and\ \bibinfo {author}
  {\bibfnamefont {A.}~\bibnamefont {Strumia}},\ }\href {\doibase
  10.1007/JHEP08(2012)098} {\bibfield  {journal} {\bibinfo  {journal} {JHEP}\
  }\textbf {\bibinfo {volume} {08}},\ \bibinfo {pages} {098} (\bibinfo {year}
  {2012})},\ \Eprint {http://arxiv.org/abs/1205.6497} {arXiv:1205.6497
  [hep-ph]} \BibitemShut {NoStop}%
\bibitem [{\citenamefont {Buras}(2009)}]{Buras:2009if}%
  \BibitemOpen
  \bibfield  {author} {\bibinfo {author} {\bibfnamefont {A.~J.}\ \bibnamefont
  {Buras}},\ }\bibfield  {booktitle} {\emph {\bibinfo {booktitle}
  {{Proceedings, Europhysics Conference on High energy physics (EPS-HEP 2009):
  Cracow, Poland, July 16-22, 2009}}},\ }\href {\doibase 10.22323/1.084.0024}
  {\bibfield  {journal} {\bibinfo  {journal} {PoS}\ }\textbf {\bibinfo {volume}
  {EPS-HEP2009}},\ \bibinfo {pages} {024} (\bibinfo {year} {2009})},\ \Eprint
  {http://arxiv.org/abs/0910.1032} {arXiv:0910.1032 [hep-ph]} \BibitemShut
  {NoStop}%
\bibitem [{\citenamefont {Langacker}(1994)}]{Langacker:1994vf}%
  \BibitemOpen
  \bibfield  {author} {\bibinfo {author} {\bibfnamefont {P.}~\bibnamefont
  {Langacker}},\ }in\ \href@noop {} {\emph {\bibinfo {booktitle} {{Proceedings,
  Tennessee International Symposium on Radiative Corrections: Status and
  Outlook, Gatlinburg, Tennessee June 27 - July 1, 1994}}}}\ (\bibinfo {year}
  {1994})\ pp.\ \bibinfo {pages} {415--437},\ \Eprint
  {http://arxiv.org/abs/hep-ph/9411247} {arXiv:hep-ph/9411247 [hep-ph]}
  \BibitemShut {NoStop}%
\bibitem [{\citenamefont {Gusken}\ \emph {et~al.}(1985)\citenamefont {Gusken},
  \citenamefont {Kuhn},\ and\ \citenamefont {Zerwas}}]{Gusken:1985nf}%
  \BibitemOpen
  \bibfield  {author} {\bibinfo {author} {\bibfnamefont {S.}~\bibnamefont
  {Gusken}}, \bibinfo {author} {\bibfnamefont {J.~H.}\ \bibnamefont {Kuhn}}, \
  and\ \bibinfo {author} {\bibfnamefont {P.~M.}\ \bibnamefont {Zerwas}},\
  }\href {\doibase 10.1016/0370-2693(85)90983-9} {\bibfield  {journal}
  {\bibinfo  {journal} {Phys. Lett.}\ }\textbf {\bibinfo {volume} {155B}},\
  \bibinfo {pages} {185} (\bibinfo {year} {1985})}\BibitemShut {NoStop}%
\bibitem [{\citenamefont {Strassler}\ and\ \citenamefont
  {Peskin}(1991)}]{Strassler:1990nw}%
  \BibitemOpen
  \bibfield  {author} {\bibinfo {author} {\bibfnamefont {M.~J.}\ \bibnamefont
  {Strassler}}\ and\ \bibinfo {author} {\bibfnamefont {M.~E.}\ \bibnamefont
  {Peskin}},\ }\href {\doibase 10.1103/PhysRevD.43.1500} {\bibfield  {journal}
  {\bibinfo  {journal} {Phys. Rev.}\ }\textbf {\bibinfo {volume} {D43}},\
  \bibinfo {pages} {1500} (\bibinfo {year} {1991})}\BibitemShut {NoStop}%
\bibitem [{\citenamefont {Beneke}\ \emph {et~al.}(2015)\citenamefont {Beneke},
  \citenamefont {Kiyo}, \citenamefont {Marquard}, \citenamefont {Penin},
  \citenamefont {Piclum},\ and\ \citenamefont {Steinhauser}}]{Beneke:2015kwa}%
  \BibitemOpen
  \bibfield  {author} {\bibinfo {author} {\bibfnamefont {M.}~\bibnamefont
  {Beneke}}, \bibinfo {author} {\bibfnamefont {Y.}~\bibnamefont {Kiyo}},
  \bibinfo {author} {\bibfnamefont {P.}~\bibnamefont {Marquard}}, \bibinfo
  {author} {\bibfnamefont {A.}~\bibnamefont {Penin}}, \bibinfo {author}
  {\bibfnamefont {J.}~\bibnamefont {Piclum}}, \ and\ \bibinfo {author}
  {\bibfnamefont {M.}~\bibnamefont {Steinhauser}},\ }\href {\doibase
  10.1103/PhysRevLett.115.192001} {\bibfield  {journal} {\bibinfo  {journal}
  {Phys. Rev. Lett.}\ }\textbf {\bibinfo {volume} {115}},\ \bibinfo {pages}
  {192001} (\bibinfo {year} {2015})},\ \Eprint
  {http://arxiv.org/abs/1506.06864} {arXiv:1506.06864 [hep-ph]} \BibitemShut
  {NoStop}%
\bibitem [{\citenamefont {Hoang}\ and\ \citenamefont
  {Stahlhofen}(2014)}]{Hoang:2013uda}%
  \BibitemOpen
  \bibfield  {author} {\bibinfo {author} {\bibfnamefont {A.~H.}\ \bibnamefont
  {Hoang}}\ and\ \bibinfo {author} {\bibfnamefont {M.}~\bibnamefont
  {Stahlhofen}},\ }\href {\doibase 10.1007/JHEP05(2014)121} {\bibfield
  {journal} {\bibinfo  {journal} {JHEP}\ }\textbf {\bibinfo {volume} {05}},\
  \bibinfo {pages} {121} (\bibinfo {year} {2014})},\ \Eprint
  {http://arxiv.org/abs/1309.6323} {arXiv:1309.6323 [hep-ph]} \BibitemShut
  {NoStop}%
\bibitem [{\citenamefont {Bach}\ \emph {et~al.}(2018)\citenamefont {Bach},
  \citenamefont {Nejad}, \citenamefont {Hoang}, \citenamefont {Kilian},
  \citenamefont {Reuter}, \citenamefont {Stahlhofen}, \citenamefont {Teubner},\
  and\ \citenamefont {Weiss}}]{Bach:2017ggt}%
  \BibitemOpen
  \bibfield  {author} {\bibinfo {author} {\bibfnamefont {F.}~\bibnamefont
  {Bach}}, \bibinfo {author} {\bibfnamefont {B.~C.}\ \bibnamefont {Nejad}},
  \bibinfo {author} {\bibfnamefont {A.}~\bibnamefont {Hoang}}, \bibinfo
  {author} {\bibfnamefont {W.}~\bibnamefont {Kilian}}, \bibinfo {author}
  {\bibfnamefont {J.}~\bibnamefont {Reuter}}, \bibinfo {author} {\bibfnamefont
  {M.}~\bibnamefont {Stahlhofen}}, \bibinfo {author} {\bibfnamefont
  {T.}~\bibnamefont {Teubner}}, \ and\ \bibinfo {author} {\bibfnamefont
  {C.}~\bibnamefont {Weiss}},\ }\href {\doibase 10.1007/JHEP03(2018)184}
  {\bibfield  {journal} {\bibinfo  {journal} {JHEP}\ }\textbf {\bibinfo
  {volume} {03}},\ \bibinfo {pages} {184} (\bibinfo {year} {2018})},\ \Eprint
  {http://arxiv.org/abs/1712.02220} {arXiv:1712.02220 [hep-ph]} \BibitemShut
  {NoStop}%
\bibitem [{\citenamefont {Marquard}\ \emph {et~al.}(2015)\citenamefont
  {Marquard}, \citenamefont {Smirnov}, \citenamefont {Smirnov},\ and\
  \citenamefont {Steinhauser}}]{Marquard:2015qpa}%
  \BibitemOpen
  \bibfield  {author} {\bibinfo {author} {\bibfnamefont {P.}~\bibnamefont
  {Marquard}}, \bibinfo {author} {\bibfnamefont {A.~V.}\ \bibnamefont
  {Smirnov}}, \bibinfo {author} {\bibfnamefont {V.~A.}\ \bibnamefont
  {Smirnov}}, \ and\ \bibinfo {author} {\bibfnamefont {M.}~\bibnamefont
  {Steinhauser}},\ }\href {\doibase 10.1103/PhysRevLett.114.142002} {\bibfield
  {journal} {\bibinfo  {journal} {Phys. Rev. Lett.}\ }\textbf {\bibinfo
  {volume} {114}},\ \bibinfo {pages} {142002} (\bibinfo {year} {2015})},\
  \Eprint {http://arxiv.org/abs/1502.01030} {arXiv:1502.01030 [hep-ph]}
  \BibitemShut {NoStop}%
\bibitem [{\citenamefont {Martinez}\ and\ \citenamefont
  {Miquel}(2003)}]{Martinez:2002st}%
  \BibitemOpen
  \bibfield  {author} {\bibinfo {author} {\bibfnamefont {M.}~\bibnamefont
  {Martinez}}\ and\ \bibinfo {author} {\bibfnamefont {R.}~\bibnamefont
  {Miquel}},\ }\href {\doibase 10.1140/epjc/s2002-01094-1} {\bibfield
  {journal} {\bibinfo  {journal} {Eur. Phys. J.}\ }\textbf {\bibinfo {volume}
  {C27}},\ \bibinfo {pages} {49} (\bibinfo {year} {2003})},\ \Eprint
  {http://arxiv.org/abs/hep-ph/0207315} {arXiv:hep-ph/0207315 [hep-ph]}
  \BibitemShut {NoStop}%
\bibitem [{\citenamefont {Horiguchi}\ \emph {et~al.}(2013)\citenamefont
  {Horiguchi}, \citenamefont {Ishikawa}, \citenamefont {Suehara}, \citenamefont
  {Fujii}, \citenamefont {Sumino}, \citenamefont {Kiyo},\ and\ \citenamefont
  {Yamamoto}}]{Horiguchi:2013wra}%
  \BibitemOpen
  \bibfield  {author} {\bibinfo {author} {\bibfnamefont {T.}~\bibnamefont
  {Horiguchi}}, \bibinfo {author} {\bibfnamefont {A.}~\bibnamefont {Ishikawa}},
  \bibinfo {author} {\bibfnamefont {T.}~\bibnamefont {Suehara}}, \bibinfo
  {author} {\bibfnamefont {K.}~\bibnamefont {Fujii}}, \bibinfo {author}
  {\bibfnamefont {Y.}~\bibnamefont {Sumino}}, \bibinfo {author} {\bibfnamefont
  {Y.}~\bibnamefont {Kiyo}}, \ and\ \bibinfo {author} {\bibfnamefont
  {H.}~\bibnamefont {Yamamoto}},\ }\href@noop {} {\  (\bibinfo {year}
  {2013})},\ \Eprint {http://arxiv.org/abs/1310.0563} {arXiv:1310.0563
  [hep-ex]} \BibitemShut {NoStop}%
\bibitem [{\citenamefont {Seidel}\ \emph {et~al.}(2013)\citenamefont {Seidel},
  \citenamefont {Simon}, \citenamefont {Tesar},\ and\ \citenamefont
  {Poss}}]{Seidel:2013sqa}%
  \BibitemOpen
  \bibfield  {author} {\bibinfo {author} {\bibfnamefont {K.}~\bibnamefont
  {Seidel}}, \bibinfo {author} {\bibfnamefont {F.}~\bibnamefont {Simon}},
  \bibinfo {author} {\bibfnamefont {M.}~\bibnamefont {Tesar}}, \ and\ \bibinfo
  {author} {\bibfnamefont {S.}~\bibnamefont {Poss}},\ }\href {\doibase
  10.1140/epjc/s10052-013-2530-7} {\bibfield  {journal} {\bibinfo  {journal}
  {Eur. Phys. J.}\ }\textbf {\bibinfo {volume} {C73}},\ \bibinfo {pages} {2530}
  (\bibinfo {year} {2013})},\ \Eprint {http://arxiv.org/abs/1303.3758}
  {arXiv:1303.3758 [hep-ex]} \BibitemShut {NoStop}%
\bibitem [{\citenamefont {Simon}(2017)}]{Simon:2016pwp}%
  \BibitemOpen
  \bibfield  {author} {\bibinfo {author} {\bibfnamefont {F.}~\bibnamefont
  {Simon}},\ }\bibfield  {booktitle} {\emph {\bibinfo {booktitle}
  {{Proceedings, 38th International Conference on High Energy Physics (ICHEP
  2016): Chicago, IL, USA, August 3-10, 2016}}},\ }\href {\doibase
  10.22323/1.282.0872} {\bibfield  {journal} {\bibinfo  {journal} {PoS}\
  }\textbf {\bibinfo {volume} {ICHEP2016}},\ \bibinfo {pages} {872} (\bibinfo
  {year} {2017})},\ \Eprint {http://arxiv.org/abs/1611.03399} {arXiv:1611.03399
  [hep-ex]} \BibitemShut {NoStop}%
\bibitem [{\citenamefont {Hesari}\ \emph {et~al.}(2014)\citenamefont {Hesari},
  \citenamefont {Khanpour}, \citenamefont {Khatiri~Yanehsari},\ and\
  \citenamefont {Mohammadi~Najafabadi}}]{Hesari:2014eua}%
  \BibitemOpen
  \bibfield  {author} {\bibinfo {author} {\bibfnamefont {H.}~\bibnamefont
  {Hesari}}, \bibinfo {author} {\bibfnamefont {H.}~\bibnamefont {Khanpour}},
  \bibinfo {author} {\bibfnamefont {M.}~\bibnamefont {Khatiri~Yanehsari}}, \
  and\ \bibinfo {author} {\bibfnamefont {M.}~\bibnamefont
  {Mohammadi~Najafabadi}},\ }\href {\doibase 10.1155/2014/476490} {\bibfield
  {journal} {\bibinfo  {journal} {Adv. High Energy Phys.}\ }\textbf {\bibinfo
  {volume} {2014}},\ \bibinfo {pages} {476490} (\bibinfo {year} {2014})},\
  \Eprint {http://arxiv.org/abs/1412.8572} {arXiv:1412.8572 [hep-ex]}
  \BibitemShut {NoStop}%
\bibitem [{\citenamefont {Zarnecki}\ and\ \citenamefont {van~der
  Kolk}(2018)}]{Zarnecki:2018wsw}%
  \BibitemOpen
  \bibfield  {author} {\bibinfo {author} {\bibfnamefont {A.~F.}\ \bibnamefont
  {Zarnecki}}\ and\ \bibinfo {author} {\bibfnamefont {N.}~\bibnamefont {van~der
  Kolk}} (\bibinfo {collaboration} {CLICdp}),\ }in\ \href@noop {} {\emph
  {\bibinfo {booktitle} {{International Workshop on Future Linear Collider
  (LCWS2017) Strasbourg, France, October 23-27, 2017}}}}\ (\bibinfo {year}
  {2018})\ \Eprint {http://arxiv.org/abs/1801.04585} {arXiv:1801.04585
  [hep-ex]} \BibitemShut {NoStop}%
\bibitem [{\citenamefont {Richard}(2014)}]{Richard:2014upa}%
  \BibitemOpen
  \bibfield  {author} {\bibinfo {author} {\bibfnamefont {F.}~\bibnamefont
  {Richard}},\ }\href@noop {} {\  (\bibinfo {year} {2014})},\ \Eprint
  {http://arxiv.org/abs/1403.2893} {arXiv:1403.2893 [hep-ph]} \BibitemShut
  {NoStop}%
\bibitem [{\citenamefont {Barducci}\ \emph {et~al.}(2015)\citenamefont
  {Barducci}, \citenamefont {De~Curtis}, \citenamefont {Moretti},\ and\
  \citenamefont {Pruna}}]{Barducci:2015aoa}%
  \BibitemOpen
  \bibfield  {author} {\bibinfo {author} {\bibfnamefont {D.}~\bibnamefont
  {Barducci}}, \bibinfo {author} {\bibfnamefont {S.}~\bibnamefont {De~Curtis}},
  \bibinfo {author} {\bibfnamefont {S.}~\bibnamefont {Moretti}}, \ and\
  \bibinfo {author} {\bibfnamefont {G.~M.}\ \bibnamefont {Pruna}},\ }\href
  {\doibase 10.1007/JHEP08(2015)127} {\bibfield  {journal} {\bibinfo  {journal}
  {JHEP}\ }\textbf {\bibinfo {volume} {08}},\ \bibinfo {pages} {127} (\bibinfo
  {year} {2015})},\ \Eprint {http://arxiv.org/abs/1504.05407} {arXiv:1504.05407
  [hep-ph]} \BibitemShut {NoStop}%
\bibitem [{\citenamefont {Durieux}\ and\ \citenamefont
  {Matsedonskyi}(2018)}]{Durieux:2018ekg}%
  \BibitemOpen
  \bibfield  {author} {\bibinfo {author} {\bibfnamefont {G.}~\bibnamefont
  {Durieux}}\ and\ \bibinfo {author} {\bibfnamefont {O.}~\bibnamefont
  {Matsedonskyi}},\ }\href@noop {} {\  (\bibinfo {year} {2018})},\ \Eprint
  {http://arxiv.org/abs/1807.10273} {arXiv:1807.10273 [hep-ph]} \BibitemShut
  {NoStop}%
\bibitem [{\citenamefont {Baur}\ \emph {et~al.}(2005)\citenamefont {Baur},
  \citenamefont {Juste}, \citenamefont {Orr},\ and\ \citenamefont
  {Rainwater}}]{Baur:2004uw}%
  \BibitemOpen
  \bibfield  {author} {\bibinfo {author} {\bibfnamefont {U.}~\bibnamefont
  {Baur}}, \bibinfo {author} {\bibfnamefont {A.}~\bibnamefont {Juste}},
  \bibinfo {author} {\bibfnamefont {L.~H.}\ \bibnamefont {Orr}}, \ and\
  \bibinfo {author} {\bibfnamefont {D.}~\bibnamefont {Rainwater}},\ }\href
  {\doibase 10.1103/PhysRevD.71.054013} {\bibfield  {journal} {\bibinfo
  {journal} {Phys. Rev.}\ }\textbf {\bibinfo {volume} {D71}},\ \bibinfo {pages}
  {054013} (\bibinfo {year} {2005})},\ \Eprint
  {http://arxiv.org/abs/hep-ph/0412021} {arXiv:hep-ph/0412021 [hep-ph]}
  \BibitemShut {NoStop}%
\bibitem [{\citenamefont {Baur}\ \emph {et~al.}(2006)\citenamefont {Baur},
  \citenamefont {Juste}, \citenamefont {Rainwater},\ and\ \citenamefont
  {Orr}}]{Baur:2005wi}%
  \BibitemOpen
  \bibfield  {author} {\bibinfo {author} {\bibfnamefont {U.}~\bibnamefont
  {Baur}}, \bibinfo {author} {\bibfnamefont {A.}~\bibnamefont {Juste}},
  \bibinfo {author} {\bibfnamefont {D.}~\bibnamefont {Rainwater}}, \ and\
  \bibinfo {author} {\bibfnamefont {L.~H.}\ \bibnamefont {Orr}},\ }\href
  {\doibase 10.1103/PhysRevD.73.034016} {\bibfield  {journal} {\bibinfo
  {journal} {Phys. Rev.}\ }\textbf {\bibinfo {volume} {D73}},\ \bibinfo {pages}
  {034016} (\bibinfo {year} {2006})},\ \Eprint
  {http://arxiv.org/abs/hep-ph/0512262} {arXiv:hep-ph/0512262 [hep-ph]}
  \BibitemShut {NoStop}%
\bibitem [{\citenamefont {Abramowicz}\ \emph {et~al.}(2017)\citenamefont
  {Abramowicz} \emph {et~al.}}]{Abramowicz:2016zbo}%
  \BibitemOpen
  \bibfield  {author} {\bibinfo {author} {\bibfnamefont {H.}~\bibnamefont
  {Abramowicz}} \emph {et~al.},\ }\href {\doibase
  10.1140/epjc/s10052-017-4968-5} {\bibfield  {journal} {\bibinfo  {journal}
  {Eur. Phys. J.}\ }\textbf {\bibinfo {volume} {C77}},\ \bibinfo {pages} {475}
  (\bibinfo {year} {2017})},\ \Eprint {http://arxiv.org/abs/1608.07538}
  {arXiv:1608.07538 [hep-ex]} \BibitemShut {NoStop}%
\bibitem [{\citenamefont {Azzi}\ \emph {et~al.}(2019)\citenamefont {Azzi} \emph
  {et~al.}}]{Azzi:2019yne}%
  \BibitemOpen
  \bibfield  {author} {\bibinfo {author} {\bibfnamefont {P.}~\bibnamefont
  {Azzi}} \emph {et~al.} (\bibinfo {collaboration} {HL-LHC, HE-LHC Working
  Group}),\ }\href@noop {} {\  (\bibinfo {year} {2019})},\ \Eprint
  {http://arxiv.org/abs/1902.04070} {arXiv:1902.04070 [hep-ph]} \BibitemShut
  {NoStop}%
\bibitem [{\citenamefont {Boselli}\ \emph {et~al.}(2018)\citenamefont
  {Boselli}, \citenamefont {Hunter},\ and\ \citenamefont
  {Mitov}}]{Boselli:2018zxr}%
  \BibitemOpen
  \bibfield  {author} {\bibinfo {author} {\bibfnamefont {S.}~\bibnamefont
  {Boselli}}, \bibinfo {author} {\bibfnamefont {R.}~\bibnamefont {Hunter}}, \
  and\ \bibinfo {author} {\bibfnamefont {A.}~\bibnamefont {Mitov}},\
  }\href@noop {} {\  (\bibinfo {year} {2018})},\ \Eprint
  {http://arxiv.org/abs/1805.12027} {arXiv:1805.12027 [hep-ph]} \BibitemShut
  {NoStop}%
\bibitem [{\citenamefont {Sirunyan}\ \emph
  {et~al.}(2018{\natexlab{b}})\citenamefont {Sirunyan} \emph
  {et~al.}}]{Sirunyan:2018hoz}%
  \BibitemOpen
  \bibfield  {author} {\bibinfo {author} {\bibfnamefont {A.~M.}\ \bibnamefont
  {Sirunyan}} \emph {et~al.} (\bibinfo {collaboration} {CMS}),\ }\href
  {\doibase 10.1103/PhysRevLett.120.231801} {\bibfield  {journal} {\bibinfo
  {journal} {Phys. Rev. Lett.}\ }\textbf {\bibinfo {volume} {120}},\ \bibinfo
  {pages} {231801} (\bibinfo {year} {2018}{\natexlab{b}})},\ \Eprint
  {http://arxiv.org/abs/1804.02610} {arXiv:1804.02610 [hep-ex]} \BibitemShut
  {NoStop}%
\bibitem [{\citenamefont {Aaboud}\ \emph
  {et~al.}(2018{\natexlab{b}})\citenamefont {Aaboud} \emph
  {et~al.}}]{Aaboud:2018urx}%
  \BibitemOpen
  \bibfield  {author} {\bibinfo {author} {\bibfnamefont {M.}~\bibnamefont
  {Aaboud}} \emph {et~al.} (\bibinfo {collaboration} {ATLAS}),\ }\href
  {\doibase 10.1016/j.physletb.2018.07.035} {\bibfield  {journal} {\bibinfo
  {journal} {Phys. Lett.}\ }\textbf {\bibinfo {volume} {B784}},\ \bibinfo
  {pages} {173} (\bibinfo {year} {2018}{\natexlab{b}})},\ \Eprint
  {http://arxiv.org/abs/1806.00425} {arXiv:1806.00425 [hep-ex]} \BibitemShut
  {NoStop}%
\bibitem [{\citenamefont {Yonamine}\ \emph {et~al.}(2011)\citenamefont
  {Yonamine}, \citenamefont {Ikematsu}, \citenamefont {Tanabe}, \citenamefont
  {Fujii}, \citenamefont {Kiyo}, \citenamefont {Sumino},\ and\ \citenamefont
  {Yokoya}}]{Yonamine:2011jg}%
  \BibitemOpen
  \bibfield  {author} {\bibinfo {author} {\bibfnamefont {R.}~\bibnamefont
  {Yonamine}}, \bibinfo {author} {\bibfnamefont {K.}~\bibnamefont {Ikematsu}},
  \bibinfo {author} {\bibfnamefont {T.}~\bibnamefont {Tanabe}}, \bibinfo
  {author} {\bibfnamefont {K.}~\bibnamefont {Fujii}}, \bibinfo {author}
  {\bibfnamefont {Y.}~\bibnamefont {Kiyo}}, \bibinfo {author} {\bibfnamefont
  {Y.}~\bibnamefont {Sumino}}, \ and\ \bibinfo {author} {\bibfnamefont
  {H.}~\bibnamefont {Yokoya}},\ }\href {\doibase 10.1103/PhysRevD.84.014033}
  {\bibfield  {journal} {\bibinfo  {journal} {Phys. Rev.}\ }\textbf {\bibinfo
  {volume} {D84}},\ \bibinfo {pages} {014033} (\bibinfo {year} {2011})},\
  \Eprint {http://arxiv.org/abs/1104.5132} {arXiv:1104.5132 [hep-ph]}
  \BibitemShut {NoStop}%
\bibitem [{\citenamefont {Price}\ \emph {et~al.}(2015)\citenamefont {Price},
  \citenamefont {Roloff}, \citenamefont {Strube},\ and\ \citenamefont
  {Tanabe}}]{Price:2014oca}%
  \BibitemOpen
  \bibfield  {author} {\bibinfo {author} {\bibfnamefont {T.}~\bibnamefont
  {Price}}, \bibinfo {author} {\bibfnamefont {P.}~\bibnamefont {Roloff}},
  \bibinfo {author} {\bibfnamefont {J.}~\bibnamefont {Strube}}, \ and\ \bibinfo
  {author} {\bibfnamefont {T.}~\bibnamefont {Tanabe}},\ }\href {\doibase
  10.1140/epjc/s10052-015-3532-4} {\bibfield  {journal} {\bibinfo  {journal}
  {Eur. Phys. J.}\ }\textbf {\bibinfo {volume} {C75}},\ \bibinfo {pages} {309}
  (\bibinfo {year} {2015})},\ \Eprint {http://arxiv.org/abs/1409.7157}
  {arXiv:1409.7157 [hep-ex]} \BibitemShut {NoStop}%
\bibitem [{\citenamefont {Han}\ \emph {et~al.}(2000)\citenamefont {Han},
  \citenamefont {Huang}, \citenamefont {Lin}, \citenamefont {Wang},\ and\
  \citenamefont {Zhang}}]{Han:1999xd}%
  \BibitemOpen
  \bibfield  {author} {\bibinfo {author} {\bibfnamefont {T.}~\bibnamefont
  {Han}}, \bibinfo {author} {\bibfnamefont {T.}~\bibnamefont {Huang}}, \bibinfo
  {author} {\bibfnamefont {Z.~H.}\ \bibnamefont {Lin}}, \bibinfo {author}
  {\bibfnamefont {J.~X.}\ \bibnamefont {Wang}}, \ and\ \bibinfo {author}
  {\bibfnamefont {X.}~\bibnamefont {Zhang}},\ }\href {\doibase
  10.1103/PhysRevD.61.015006} {\bibfield  {journal} {\bibinfo  {journal} {Phys.
  Rev.}\ }\textbf {\bibinfo {volume} {D61}},\ \bibinfo {pages} {015006}
  (\bibinfo {year} {2000})},\ \Eprint {http://arxiv.org/abs/hep-ph/9908236}
  {arXiv:hep-ph/9908236 [hep-ph]} \BibitemShut {NoStop}%
\bibitem [{\citenamefont {Azatov}\ \emph {et~al.}(2016)\citenamefont {Azatov},
  \citenamefont {Grojean}, \citenamefont {Paul},\ and\ \citenamefont
  {Salvioni}}]{Azatov:2016xik}%
  \BibitemOpen
  \bibfield  {author} {\bibinfo {author} {\bibfnamefont {A.}~\bibnamefont
  {Azatov}}, \bibinfo {author} {\bibfnamefont {C.}~\bibnamefont {Grojean}},
  \bibinfo {author} {\bibfnamefont {A.}~\bibnamefont {Paul}}, \ and\ \bibinfo
  {author} {\bibfnamefont {E.}~\bibnamefont {Salvioni}},\ }\href {\doibase
  10.1007/JHEP09(2016)123} {\bibfield  {journal} {\bibinfo  {journal} {JHEP}\
  }\textbf {\bibinfo {volume} {09}},\ \bibinfo {pages} {123} (\bibinfo {year}
  {2016})},\ \Eprint {http://arxiv.org/abs/1608.00977} {arXiv:1608.00977
  [hep-ph]} \BibitemShut {NoStop}%
\bibitem [{\citenamefont {Durieux}\ \emph
  {et~al.}(2018{\natexlab{b}})\citenamefont {Durieux}, \citenamefont {Gu},
  \citenamefont {Vryonidou},\ and\ \citenamefont {Zhang}}]{Durieux:2018ggn}%
  \BibitemOpen
  \bibfield  {author} {\bibinfo {author} {\bibfnamefont {G.}~\bibnamefont
  {Durieux}}, \bibinfo {author} {\bibfnamefont {J.}~\bibnamefont {Gu}},
  \bibinfo {author} {\bibfnamefont {E.}~\bibnamefont {Vryonidou}}, \ and\
  \bibinfo {author} {\bibfnamefont {C.}~\bibnamefont {Zhang}},\ }\href
  {\doibase 10.1088/1674-1137/42/12/123107} {\bibfield  {journal} {\bibinfo
  {journal} {Chin. Phys.}\ }\textbf {\bibinfo {volume} {C42}},\ \bibinfo
  {pages} {123107} (\bibinfo {year} {2018}{\natexlab{b}})},\ \Eprint
  {http://arxiv.org/abs/1809.03520} {arXiv:1809.03520 [hep-ph]} \BibitemShut
  {NoStop}%
\bibitem [{\citenamefont {Hartland}\ \emph {et~al.}(2019)\citenamefont
  {Hartland}, \citenamefont {Maltoni}, \citenamefont {Nocera}, \citenamefont
  {Rojo}, \citenamefont {Slade}, \citenamefont {Vryonidou},\ and\ \citenamefont
  {Zhang}}]{Hartland:2019bjb}%
  \BibitemOpen
  \bibfield  {author} {\bibinfo {author} {\bibfnamefont {N.~P.}\ \bibnamefont
  {Hartland}}, \bibinfo {author} {\bibfnamefont {F.}~\bibnamefont {Maltoni}},
  \bibinfo {author} {\bibfnamefont {E.~R.}\ \bibnamefont {Nocera}}, \bibinfo
  {author} {\bibfnamefont {J.}~\bibnamefont {Rojo}}, \bibinfo {author}
  {\bibfnamefont {E.}~\bibnamefont {Slade}}, \bibinfo {author} {\bibfnamefont
  {E.}~\bibnamefont {Vryonidou}}, \ and\ \bibinfo {author} {\bibfnamefont
  {C.}~\bibnamefont {Zhang}},\ }\href@noop {} {\  (\bibinfo {year} {2019})},\
  \Eprint {http://arxiv.org/abs/1901.05965} {arXiv:1901.05965 [hep-ph]}
  \BibitemShut {NoStop}%
\bibitem [{CEP(2018)}]{CEPCStudyGroup:2018ghi}%
  \BibitemOpen
  \href@noop {} {\  (\bibinfo {year} {2018})},\ \Eprint
  {http://arxiv.org/abs/1811.10545} {arXiv:1811.10545 [hep-ex]} \BibitemShut
  {NoStop}%
\bibitem [{\citenamefont {Benedikt}\ \emph {et~al.}(2019)\citenamefont
  {Benedikt} \emph {et~al.}}]{Benedikt:2018qee}%
  \BibitemOpen
  \bibfield  {author} {\bibinfo {author} {\bibfnamefont {M.}~\bibnamefont
  {Benedikt}} \emph {et~al.},\ }\bibfield  {booktitle} {\emph {\bibinfo
  {booktitle} {{Future Circular Collider. Vol. 2: the Lepton Collider}}},\
  }\href@noop {} {\  (\bibinfo {year} {2019})}\BibitemShut {NoStop}%
\bibitem [{\citenamefont {Charles}\ \emph {et~al.}(2018)\citenamefont {Charles}
  \emph {et~al.}}]{Charles:2018vfv}%
  \BibitemOpen
  \bibfield  {author} {\bibinfo {author} {\bibfnamefont {T.~K.}\ \bibnamefont
  {Charles}} \emph {et~al.} (\bibinfo {collaboration} {CLICdp, CLIC}),\ }\href
  {\doibase 10.23731/CYRM-2018-002} {\bibfield  {journal} {\bibinfo  {journal}
  {CERN Yellow Rep. Monogr.}\ }\textbf {\bibinfo {volume} {1802}},\ \bibinfo
  {pages} {1} (\bibinfo {year} {2018})},\ \Eprint
  {http://arxiv.org/abs/1812.06018} {arXiv:1812.06018 [physics.acc-ph]}
  \BibitemShut {NoStop}%
\bibitem [{\citenamefont {Sirunyan}\ \emph
  {et~al.}(2018{\natexlab{c}})\citenamefont {Sirunyan} \emph
  {et~al.}}]{Sirunyan:2018koj}%
  \BibitemOpen
  \bibfield  {author} {\bibinfo {author} {\bibfnamefont {A.~M.}\ \bibnamefont
  {Sirunyan}} \emph {et~al.} (\bibinfo {collaboration} {CMS}),\ }\href@noop {}
  {\bibfield  {journal} {\bibinfo  {journal} {Submitted to: Eur. Phys. J.}\ }
  (\bibinfo {year} {2018}{\natexlab{c}})},\ \Eprint
  {http://arxiv.org/abs/1809.10733} {arXiv:1809.10733 [hep-ex]} \BibitemShut
  {NoStop}%
\bibitem [{\citenamefont {Berggren}\ \emph {et~al.}(2016)\citenamefont
  {Berggren}, \citenamefont {Cakir}, \citenamefont {Krücker}, \citenamefont
  {List}, \citenamefont {Melzer-Pellmann}, \citenamefont {Safarzadeh~Samani},
  \citenamefont {Seitz},\ and\ \citenamefont {Wayand}}]{Berggren:2015qua}%
  \BibitemOpen
  \bibfield  {author} {\bibinfo {author} {\bibfnamefont {M.}~\bibnamefont
  {Berggren}}, \bibinfo {author} {\bibfnamefont {A.}~\bibnamefont {Cakir}},
  \bibinfo {author} {\bibfnamefont {D.}~\bibnamefont {Krücker}}, \bibinfo
  {author} {\bibfnamefont {J.}~\bibnamefont {List}}, \bibinfo {author}
  {\bibfnamefont {I.~A.}\ \bibnamefont {Melzer-Pellmann}}, \bibinfo {author}
  {\bibfnamefont {B.}~\bibnamefont {Safarzadeh~Samani}}, \bibinfo {author}
  {\bibfnamefont {C.}~\bibnamefont {Seitz}}, \ and\ \bibinfo {author}
  {\bibfnamefont {S.}~\bibnamefont {Wayand}},\ }\href {\doibase
  10.1140/epjc/s10052-016-3914-2} {\bibfield  {journal} {\bibinfo  {journal}
  {Eur. Phys. J.}\ }\textbf {\bibinfo {volume} {C76}},\ \bibinfo {pages} {183}
  (\bibinfo {year} {2016})},\ \Eprint {http://arxiv.org/abs/1508.04383}
  {arXiv:1508.04383 [hep-ph]} \BibitemShut {NoStop}%
\bibitem [{\citenamefont {Berggren}(2013)}]{Berggren:2013vna}%
  \BibitemOpen
  \bibfield  {author} {\bibinfo {author} {\bibfnamefont {M.}~\bibnamefont
  {Berggren}},\ }in\ \href@noop {} {\emph {\bibinfo {booktitle} {{Proceedings,
  2013 Community Summer Study on the Future of U.S. Particle Physics: Snowmass
  on the Mississippi (CSS2013): Minneapolis, MN, USA, July 29-August 6,
  2013}}}}\ (\bibinfo {year} {2013})\ \Eprint {http://arxiv.org/abs/1308.1461}
  {arXiv:1308.1461 [hep-ph]} \BibitemShut {NoStop}%
\bibitem [{\citenamefont {Aaboud}\ \emph
  {et~al.}(2018{\natexlab{c}})\citenamefont {Aaboud} \emph
  {et~al.}}]{Aaboud:2017leg}%
  \BibitemOpen
  \bibfield  {author} {\bibinfo {author} {\bibfnamefont {M.}~\bibnamefont
  {Aaboud}} \emph {et~al.} (\bibinfo {collaboration} {ATLAS}),\ }\href
  {\doibase 10.1103/PhysRevD.97.052010} {\bibfield  {journal} {\bibinfo
  {journal} {Phys. Rev.}\ }\textbf {\bibinfo {volume} {D97}},\ \bibinfo {pages}
  {052010} (\bibinfo {year} {2018}{\natexlab{c}})},\ \Eprint
  {http://arxiv.org/abs/1712.08119} {arXiv:1712.08119 [hep-ex]} \BibitemShut
  {NoStop}%
\bibitem [{\citenamefont {Aad}\ \emph {et~al.}(2014)\citenamefont {Aad} \emph
  {et~al.}}]{Aad:2014vma}%
  \BibitemOpen
  \bibfield  {author} {\bibinfo {author} {\bibfnamefont {G.}~\bibnamefont
  {Aad}} \emph {et~al.} (\bibinfo {collaboration} {ATLAS}),\ }\href {\doibase
  10.1007/JHEP05(2014)071} {\bibfield  {journal} {\bibinfo  {journal} {JHEP}\
  }\textbf {\bibinfo {volume} {05}},\ \bibinfo {pages} {071} (\bibinfo {year}
  {2014})},\ \Eprint {http://arxiv.org/abs/1403.5294} {arXiv:1403.5294
  [hep-ex]} \BibitemShut {NoStop}%
\bibitem [{\citenamefont {Sirunyan}\ \emph
  {et~al.}(2018{\natexlab{d}})\citenamefont {Sirunyan} \emph
  {et~al.}}]{Sirunyan:2018nwe}%
  \BibitemOpen
  \bibfield  {author} {\bibinfo {author} {\bibfnamefont {A.~M.}\ \bibnamefont
  {Sirunyan}} \emph {et~al.} (\bibinfo {collaboration} {CMS}),\ }\href@noop {}
  {\bibfield  {journal} {\bibinfo  {journal} {Submitted to: Phys. Lett.}\ }
  (\bibinfo {year} {2018}{\natexlab{d}})},\ \Eprint
  {http://arxiv.org/abs/1806.05264} {arXiv:1806.05264 [hep-ex]} \BibitemShut
  {NoStop}%
\bibitem [{\citenamefont {Aaboud}\ \emph
  {et~al.}(2018{\natexlab{d}})\citenamefont {Aaboud} \emph
  {et~al.}}]{Aaboud:2017vwy}%
  \BibitemOpen
  \bibfield  {author} {\bibinfo {author} {\bibfnamefont {M.}~\bibnamefont
  {Aaboud}} \emph {et~al.} (\bibinfo {collaboration} {ATLAS}),\ }\href
  {\doibase 10.1103/PhysRevD.97.112001} {\bibfield  {journal} {\bibinfo
  {journal} {Phys. Rev.}\ }\textbf {\bibinfo {volume} {D97}},\ \bibinfo {pages}
  {112001} (\bibinfo {year} {2018}{\natexlab{d}})},\ \Eprint
  {http://arxiv.org/abs/1712.02332} {arXiv:1712.02332 [hep-ex]} \BibitemShut
  {NoStop}%
\bibitem [{\citenamefont {Sirunyan}\ \emph
  {et~al.}(2018{\natexlab{e}})\citenamefont {Sirunyan} \emph
  {et~al.}}]{Sirunyan:2018vjp}%
  \BibitemOpen
  \bibfield  {author} {\bibinfo {author} {\bibfnamefont {A.~M.}\ \bibnamefont
  {Sirunyan}} \emph {et~al.} (\bibinfo {collaboration} {CMS}),\ }\href
  {\doibase 10.1007/JHEP05(2018)025} {\bibfield  {journal} {\bibinfo  {journal}
  {JHEP}\ }\textbf {\bibinfo {volume} {05}},\ \bibinfo {pages} {025} (\bibinfo
  {year} {2018}{\natexlab{e}})},\ \Eprint {http://arxiv.org/abs/1802.02110}
  {arXiv:1802.02110 [hep-ex]} \BibitemShut {NoStop}%
\bibitem [{\citenamefont {Aaboud}\ \emph {et~al.}(2017)\citenamefont {Aaboud}
  \emph {et~al.}}]{Aaboud:2017ayj}%
  \BibitemOpen
  \bibfield  {author} {\bibinfo {author} {\bibfnamefont {M.}~\bibnamefont
  {Aaboud}} \emph {et~al.} (\bibinfo {collaboration} {ATLAS}),\ }\href
  {\doibase 10.1007/JHEP12(2017)085} {\bibfield  {journal} {\bibinfo  {journal}
  {JHEP}\ }\textbf {\bibinfo {volume} {12}},\ \bibinfo {pages} {085} (\bibinfo
  {year} {2017})},\ \Eprint {http://arxiv.org/abs/1709.04183} {arXiv:1709.04183
  [hep-ex]} \BibitemShut {NoStop}%
\bibitem [{\citenamefont {Heister}\ \emph
  {et~al.}(2002{\natexlab{a}})\citenamefont {Heister} \emph
  {et~al.}}]{Heister:2001nk}%
  \BibitemOpen
  \bibfield  {author} {\bibinfo {author} {\bibfnamefont {A.}~\bibnamefont
  {Heister}} \emph {et~al.} (\bibinfo {collaboration} {ALEPH}),\ }\href
  {\doibase 10.1016/S0370-2693(01)01494-0} {\bibfield  {journal} {\bibinfo
  {journal} {Phys. Lett.}\ }\textbf {\bibinfo {volume} {B526}},\ \bibinfo
  {pages} {206} (\bibinfo {year} {2002}{\natexlab{a}})},\ \Eprint
  {http://arxiv.org/abs/hep-ex/0112011} {arXiv:hep-ex/0112011 [hep-ex]}
  \BibitemShut {NoStop}%
\bibitem [{\citenamefont {Achard}\ \emph {et~al.}(2004)\citenamefont {Achard}
  \emph {et~al.}}]{Achard:2003ge}%
  \BibitemOpen
  \bibfield  {author} {\bibinfo {author} {\bibfnamefont {P.}~\bibnamefont
  {Achard}} \emph {et~al.} (\bibinfo {collaboration} {L3}),\ }\href {\doibase
  10.1016/j.physletb.2003.10.010} {\bibfield  {journal} {\bibinfo  {journal}
  {Phys. Lett.}\ }\textbf {\bibinfo {volume} {B580}},\ \bibinfo {pages} {37}
  (\bibinfo {year} {2004})},\ \Eprint {http://arxiv.org/abs/hep-ex/0310007}
  {arXiv:hep-ex/0310007 [hep-ex]} \BibitemShut {NoStop}%
\bibitem [{\citenamefont {Abbiendi}\ \emph
  {et~al.}(2004{\natexlab{a}})\citenamefont {Abbiendi} \emph
  {et~al.}}]{Abbiendi:2003ji}%
  \BibitemOpen
  \bibfield  {author} {\bibinfo {author} {\bibfnamefont {G.}~\bibnamefont
  {Abbiendi}} \emph {et~al.} (\bibinfo {collaboration} {OPAL}),\ }\href
  {\doibase 10.1140/epjc/s2003-01466-y} {\bibfield  {journal} {\bibinfo
  {journal} {Eur. Phys. J.}\ }\textbf {\bibinfo {volume} {C32}},\ \bibinfo
  {pages} {453} (\bibinfo {year} {2004}{\natexlab{a}})},\ \Eprint
  {http://arxiv.org/abs/hep-ex/0309014} {arXiv:hep-ex/0309014 [hep-ex]}
  \BibitemShut {NoStop}%
\bibitem [{\citenamefont {Heister}\ \emph
  {et~al.}(2002{\natexlab{b}})\citenamefont {Heister} \emph
  {et~al.}}]{Heister:2002hp}%
  \BibitemOpen
  \bibfield  {author} {\bibinfo {author} {\bibfnamefont {A.}~\bibnamefont
  {Heister}} \emph {et~al.} (\bibinfo {collaboration} {ALEPH}),\ }\href
  {\doibase 10.1016/S0370-2693(02)01827-0} {\bibfield  {journal} {\bibinfo
  {journal} {Phys. Lett.}\ }\textbf {\bibinfo {volume} {B537}},\ \bibinfo
  {pages} {5} (\bibinfo {year} {2002}{\natexlab{b}})},\ \Eprint
  {http://arxiv.org/abs/hep-ex/0204036} {arXiv:hep-ex/0204036 [hep-ex]}
  \BibitemShut {NoStop}%
\bibitem [{\citenamefont {Abbiendi}\ \emph {et~al.}(2002)\citenamefont
  {Abbiendi} \emph {et~al.}}]{Abbiendi:2002mp}%
  \BibitemOpen
  \bibfield  {author} {\bibinfo {author} {\bibfnamefont {G.}~\bibnamefont
  {Abbiendi}} \emph {et~al.} (\bibinfo {collaboration} {OPAL}),\ }\href
  {\doibase 10.1016/S0370-2693(02)02808-3, 10.1016/S0370-2693(02)02593-5}
  {\bibfield  {journal} {\bibinfo  {journal} {Phys. Lett.}\ }\textbf {\bibinfo
  {volume} {B545}},\ \bibinfo {pages} {272} (\bibinfo {year} {2002})},\
  \bibinfo {note} {[Erratum: Phys. Lett.B548,258(2002)]},\ \Eprint
  {http://arxiv.org/abs/hep-ex/0209026} {arXiv:hep-ex/0209026 [hep-ex]}
  \BibitemShut {NoStop}%
\bibitem [{\citenamefont {Abbiendi}\ \emph
  {et~al.}(2004{\natexlab{b}})\citenamefont {Abbiendi} \emph
  {et~al.}}]{Abbiendi:2003sc}%
  \BibitemOpen
  \bibfield  {author} {\bibinfo {author} {\bibfnamefont {G.}~\bibnamefont
  {Abbiendi}} \emph {et~al.} (\bibinfo {collaboration} {OPAL}),\ }\href
  {\doibase 10.1140/epjc/s2004-01758-8} {\bibfield  {journal} {\bibinfo
  {journal} {Eur. Phys. J.}\ }\textbf {\bibinfo {volume} {C35}},\ \bibinfo
  {pages} {1} (\bibinfo {year} {2004}{\natexlab{b}})},\ \Eprint
  {http://arxiv.org/abs/hep-ex/0401026} {arXiv:hep-ex/0401026 [hep-ex]}
  \BibitemShut {NoStop}%
\bibitem [{\citenamefont {Heister}\ \emph
  {et~al.}(2002{\natexlab{c}})\citenamefont {Heister} \emph
  {et~al.}}]{Heister:2002mn}%
  \BibitemOpen
  \bibfield  {author} {\bibinfo {author} {\bibfnamefont {A.}~\bibnamefont
  {Heister}} \emph {et~al.} (\bibinfo {collaboration} {ALEPH}),\ }\href
  {\doibase 10.1016/S0370-2693(02)01584-8} {\bibfield  {journal} {\bibinfo
  {journal} {Phys. Lett.}\ }\textbf {\bibinfo {volume} {B533}},\ \bibinfo
  {pages} {223} (\bibinfo {year} {2002}{\natexlab{c}})},\ \Eprint
  {http://arxiv.org/abs/hep-ex/0203020} {arXiv:hep-ex/0203020 [hep-ex]}
  \BibitemShut {NoStop}%
\bibitem [{\citenamefont {Acciarri}\ \emph {et~al.}(2000)\citenamefont
  {Acciarri} \emph {et~al.}}]{Acciarri:1999km}%
  \BibitemOpen
  \bibfield  {author} {\bibinfo {author} {\bibfnamefont {M.}~\bibnamefont
  {Acciarri}} \emph {et~al.} (\bibinfo {collaboration} {L3}),\ }\href {\doibase
  10.1016/S0370-2693(99)01388-X} {\bibfield  {journal} {\bibinfo  {journal}
  {Phys. Lett.}\ }\textbf {\bibinfo {volume} {B472}},\ \bibinfo {pages} {420}
  (\bibinfo {year} {2000})},\ \Eprint {http://arxiv.org/abs/hep-ex/9910007}
  {arXiv:hep-ex/9910007 [hep-ex]} \BibitemShut {NoStop}%
\bibitem [{LEP({\natexlab{a}})}]{LEPSUSYWG/04-01.1}%
  \BibitemOpen
  \href
  {http://lepsusy.web.cern.ch/lepsusy/www/sleptons_summer04/slep_final.html}
  {\emph {\bibinfo {title} {Combined LEP Selectron/Smuon/Stau Results,
  $183$-$208$ GeV}}},\ \bibinfo {type} {Tech. Rep.}\ \bibinfo {number}
  {LEPSUSYWG/04-01.1}\BibitemShut {NoStop}%
\bibitem [{LEP({\natexlab{b}})}]{LEPSUSYWG/04-02.1}%
  \BibitemOpen
  \href
  {http://lepsusy.web.cern.ch/lepsusy/www/squarks_summer04/stop_combi_208_final.html}
  {\emph {\bibinfo {title} {Combined LEP stop and sbottom Results $183$-$208$
  GeV}}},\ \bibinfo {type} {Tech. Rep.}\ \bibinfo {number}
  {LEPSUSYWG/04-02.1}\BibitemShut {NoStop}%
\bibitem [{LEP({\natexlab{c}})}]{LEPSUSYWG/02-04.1}%
  \BibitemOpen
  \href
  {http://lepsusy.web.cern.ch/lepsusy/www/inoslowdmsummer02/charginolowdm_pub.html}
  {\emph {\bibinfo {title} {Combined LEP Chargino Results, up to $208$ GeV for
  low DM}}},\ \bibinfo {type} {Tech. Rep.}\ \bibinfo {number}
  {LEPSUSYWG/02-04.1}\BibitemShut {NoStop}%
\bibitem [{LEP({\natexlab{d}})}]{LEPSUSYWG/01-03.1}%
  \BibitemOpen
  \href
  {http://lepsusy.web.cern.ch/lepsusy/www/inos_moriond01/charginos_pub.html}
  {\emph {\bibinfo {title} {Combined LEP Chargino Results, up to $208$ GeV for
  large $m_0$}}},\ \bibinfo {type} {Tech. Rep.}\ \bibinfo {number}
  {LEPSUSYWG/01-03.1}\BibitemShut {NoStop}%
\bibitem [{\citenamefont {Bechtle}\ \emph {et~al.}(2010)\citenamefont
  {Bechtle}, \citenamefont {Berggren}, \citenamefont {List}, \citenamefont
  {Schade},\ and\ \citenamefont {Stempel}}]{Bechtle:2009em}%
  \BibitemOpen
  \bibfield  {author} {\bibinfo {author} {\bibfnamefont {P.}~\bibnamefont
  {Bechtle}}, \bibinfo {author} {\bibfnamefont {M.}~\bibnamefont {Berggren}},
  \bibinfo {author} {\bibfnamefont {J.}~\bibnamefont {List}}, \bibinfo {author}
  {\bibfnamefont {P.}~\bibnamefont {Schade}}, \ and\ \bibinfo {author}
  {\bibfnamefont {O.}~\bibnamefont {Stempel}},\ }\href {\doibase
  10.1103/PhysRevD.82.055016} {\bibfield  {journal} {\bibinfo  {journal} {Phys.
  Rev.}\ }\textbf {\bibinfo {volume} {D82}},\ \bibinfo {pages} {055016}
  (\bibinfo {year} {2010})},\ \Eprint {http://arxiv.org/abs/0908.0876}
  {arXiv:0908.0876 [hep-ex]} \BibitemShut {NoStop}%
\bibitem [{\citenamefont {Caiazza}(2018)}]{Caiazza:416980}%
  \BibitemOpen
  \bibfield  {author} {\bibinfo {author} {\bibfnamefont {S.}~\bibnamefont
  {Caiazza}},\ }\emph {\bibinfo {title} {{T}he {G}rid{GEM} module for the {ILD}
  {TPC} $\&$ {A} new algorithm for kinematic edge determination}},\ \href
  {\doibase 10.3204/PUBDB-2018-05289} {\bibinfo {type} {Dissertation}},\
  \bibinfo  {school} {Universität Hamburg}, \bibinfo {address} {Hamburg}
  (\bibinfo {year} {2018}),\ \bibinfo {note} {dissertation, Universität
  Hamburg, 2018}\BibitemShut {NoStop}%
\bibitem [{\citenamefont {Berggren}\ \emph {et~al.}(2013)\citenamefont
  {Berggren}, \citenamefont {Brümmer}, \citenamefont {List}, \citenamefont
  {Moortgat-Pick}, \citenamefont {Robens}, \citenamefont {Rolbiecki},\ and\
  \citenamefont {Sert}}]{Berggren:2013vfa}%
  \BibitemOpen
  \bibfield  {author} {\bibinfo {author} {\bibfnamefont {M.}~\bibnamefont
  {Berggren}}, \bibinfo {author} {\bibfnamefont {F.}~\bibnamefont {Brümmer}},
  \bibinfo {author} {\bibfnamefont {J.}~\bibnamefont {List}}, \bibinfo {author}
  {\bibfnamefont {G.}~\bibnamefont {Moortgat-Pick}}, \bibinfo {author}
  {\bibfnamefont {T.}~\bibnamefont {Robens}}, \bibinfo {author} {\bibfnamefont
  {K.}~\bibnamefont {Rolbiecki}}, \ and\ \bibinfo {author} {\bibfnamefont
  {H.}~\bibnamefont {Sert}},\ }\href {\doibase 10.1140/epjc/s10052-013-2660-y}
  {\bibfield  {journal} {\bibinfo  {journal} {Eur. Phys. J.}\ }\textbf
  {\bibinfo {volume} {C73}},\ \bibinfo {pages} {2660} (\bibinfo {year}
  {2013})},\ \Eprint {http://arxiv.org/abs/1307.3566} {arXiv:1307.3566
  [hep-ph]} \BibitemShut {NoStop}%
\bibitem [{\citenamefont {Baer}\ \emph
  {et~al.}(2016{\natexlab{a}})\citenamefont {Baer}, \citenamefont {Berggren},
  \citenamefont {Fujii}, \citenamefont {Lehtinen}, \citenamefont {List},
  \citenamefont {Tanabe},\ and\ \citenamefont {Yan}}]{Baer:2016new}%
  \BibitemOpen
  \bibfield  {author} {\bibinfo {author} {\bibfnamefont {H.}~\bibnamefont
  {Baer}}, \bibinfo {author} {\bibfnamefont {M.}~\bibnamefont {Berggren}},
  \bibinfo {author} {\bibfnamefont {K.}~\bibnamefont {Fujii}}, \bibinfo
  {author} {\bibfnamefont {S.-L.}\ \bibnamefont {Lehtinen}}, \bibinfo {author}
  {\bibfnamefont {J.}~\bibnamefont {List}}, \bibinfo {author} {\bibfnamefont
  {T.}~\bibnamefont {Tanabe}}, \ and\ \bibinfo {author} {\bibfnamefont
  {J.}~\bibnamefont {Yan}},\ }\bibfield  {booktitle} {\emph {\bibinfo
  {booktitle} {{Proceedings, 38th International Conference on High Energy
  Physics (ICHEP 2016): Chicago, IL, USA, August 3-10, 2016}}},\ }\href
  {\doibase 10.22323/1.282.0156} {\bibfield  {journal} {\bibinfo  {journal}
  {PoS}\ }\textbf {\bibinfo {volume} {ICHEP2016}},\ \bibinfo {pages} {156}
  (\bibinfo {year} {2016}{\natexlab{a}})},\ \Eprint
  {http://arxiv.org/abs/1611.02846} {arXiv:1611.02846 [hep-ph]} \BibitemShut
  {NoStop}%
\bibitem [{\citenamefont {Chera}(2018)}]{Chera:402736}%
  \BibitemOpen
  \bibfield  {author} {\bibinfo {author} {\bibfnamefont {M.}~\bibnamefont
  {Chera}},\ }\emph {\bibinfo {title} {{P}article {F}low: {F}rom {F}irst
  {P}rinciples to {G}augino {P}roperty {D}etermination at the {ILC}}},\ \href
  {\doibase 10.3204/PUBDB-2018-01897} {\bibinfo {type} {Dissertation}},\
  \bibinfo  {school} {Universität Hamburg}, \bibinfo {address} {Hamburg}
  (\bibinfo {year} {2018}),\ \bibinfo {note} {dissertation, Universität
  Hamburg, 2018}\BibitemShut {NoStop}%
\bibitem [{\citenamefont {Aaboud}\ \emph
  {et~al.}(2018{\natexlab{e}})\citenamefont {Aaboud} \emph
  {et~al.}}]{Aaboud:2018jiw}%
  \BibitemOpen
  \bibfield  {author} {\bibinfo {author} {\bibfnamefont {M.}~\bibnamefont
  {Aaboud}} \emph {et~al.} (\bibinfo {collaboration} {ATLAS}),\ }\href
  {\doibase 10.1140/epjc/s10052-018-6423-7} {\bibfield  {journal} {\bibinfo
  {journal} {Eur. Phys. J.}\ }\textbf {\bibinfo {volume} {C78}},\ \bibinfo
  {pages} {995} (\bibinfo {year} {2018}{\natexlab{e}})},\ \Eprint
  {http://arxiv.org/abs/1803.02762} {arXiv:1803.02762 [hep-ex]} \BibitemShut
  {NoStop}%
\bibitem [{\citenamefont {Sirunyan}\ \emph
  {et~al.}(2018{\natexlab{f}})\citenamefont {Sirunyan} \emph
  {et~al.}}]{Sirunyan:2018ubx}%
  \BibitemOpen
  \bibfield  {author} {\bibinfo {author} {\bibfnamefont {A.~M.}\ \bibnamefont
  {Sirunyan}} \emph {et~al.} (\bibinfo {collaboration} {CMS}),\ }\href
  {\doibase 10.1007/JHEP03(2018)160} {\bibfield  {journal} {\bibinfo  {journal}
  {JHEP}\ }\textbf {\bibinfo {volume} {03}},\ \bibinfo {pages} {160} (\bibinfo
  {year} {2018}{\natexlab{f}})},\ \Eprint {http://arxiv.org/abs/1801.03957}
  {arXiv:1801.03957 [hep-ex]} \BibitemShut {NoStop}%
\bibitem [{ATL(2018)}]{ATL-PHYS-PUB-2018-048}%
  \BibitemOpen
  \href {http://cds.cern.ch/record/2651927} {\emph {\bibinfo {title}
  {{Prospects for searches for staus, charginos and neutralinos at the high
  luminosity LHC with the ATLAS Detector}}}},\ \bibinfo {type} {Tech. Rep.}\
  \bibinfo {number} {ATL-PHYS-PUB-2018-048}\ (\bibinfo  {institution} {CERN},\
  \bibinfo {address} {Geneva},\ \bibinfo {year} {2018})\BibitemShut {NoStop}%
\bibitem [{\citenamefont {Abbiendi}\ \emph
  {et~al.}(2003{\natexlab{a}})\citenamefont {Abbiendi} \emph
  {et~al.}}]{Abbiendi:2002vz}%
  \BibitemOpen
  \bibfield  {author} {\bibinfo {author} {\bibfnamefont {G.}~\bibnamefont
  {Abbiendi}} \emph {et~al.} (\bibinfo {collaboration} {OPAL}),\ }\href
  {\doibase 10.1140/epjc/s2003-01237-x} {\bibfield  {journal} {\bibinfo
  {journal} {Eur. Phys. J.}\ }\textbf {\bibinfo {volume} {C29}},\ \bibinfo
  {pages} {479} (\bibinfo {year} {2003}{\natexlab{a}})},\ \Eprint
  {http://arxiv.org/abs/hep-ex/0210043} {arXiv:hep-ex/0210043 [hep-ex]}
  \BibitemShut {NoStop}%
\bibitem [{\citenamefont {de~Vries}\ \emph {et~al.}(2015)\citenamefont
  {de~Vries} \emph {et~al.}}]{deVries:2015hva}%
  \BibitemOpen
  \bibfield  {author} {\bibinfo {author} {\bibfnamefont {K.~J.}\ \bibnamefont
  {de~Vries}} \emph {et~al.},\ }\href {\doibase 10.1140/epjc/s10052-015-3599-y}
  {\bibfield  {journal} {\bibinfo  {journal} {Eur. Phys. J.}\ }\textbf
  {\bibinfo {volume} {C75}},\ \bibinfo {pages} {422} (\bibinfo {year}
  {2015})},\ \Eprint {http://arxiv.org/abs/1504.03260} {arXiv:1504.03260
  [hep-ph]} \BibitemShut {NoStop}%
\bibitem [{\citenamefont {Lehtinen}(2018)}]{Lehtinen:415433}%
  \BibitemOpen
  \bibfield  {author} {\bibinfo {author} {\bibfnamefont {S.-L.}\ \bibnamefont
  {Lehtinen}},\ }\emph {\bibinfo {title} {{S}upersymmetry parameter
  determination at the {I}nternational {L}inear {C}ollider}},\ \href {\doibase
  10.3204/PUBDB-2018-04509} {\bibinfo {type} {Dissertation}},\ \bibinfo
  {school} {Universität Hamburg}, \bibinfo {address} {Hamburg} (\bibinfo
  {year} {2018}),\ \bibinfo {note} {dissertation, Universität Hamburg,
  2018}\BibitemShut {NoStop}%
\bibitem [{\citenamefont {Ade}\ \emph {et~al.}(2016)\citenamefont {Ade} \emph
  {et~al.}}]{Ade:2015xua}%
  \BibitemOpen
  \bibfield  {author} {\bibinfo {author} {\bibfnamefont {P.~A.~R.}\
  \bibnamefont {Ade}} \emph {et~al.} (\bibinfo {collaboration} {Planck}),\
  }\href {\doibase 10.1051/0004-6361/201525830} {\bibfield  {journal} {\bibinfo
   {journal} {Astron. Astrophys.}\ }\textbf {\bibinfo {volume} {594}},\
  \bibinfo {pages} {A13} (\bibinfo {year} {2016})},\ \Eprint
  {http://arxiv.org/abs/1502.01589} {arXiv:1502.01589 [astro-ph.CO]}
  \BibitemShut {NoStop}%
\bibitem [{\citenamefont {Choi}\ \emph {et~al.}(2001)\citenamefont {Choi},
  \citenamefont {Kalinowski}, \citenamefont {Moortgat-Pick},\ and\
  \citenamefont {Zerwas}}]{Choi:2001ww}%
  \BibitemOpen
  \bibfield  {author} {\bibinfo {author} {\bibfnamefont {S.~Y.}\ \bibnamefont
  {Choi}}, \bibinfo {author} {\bibfnamefont {J.}~\bibnamefont {Kalinowski}},
  \bibinfo {author} {\bibfnamefont {G.~A.}\ \bibnamefont {Moortgat-Pick}}, \
  and\ \bibinfo {author} {\bibfnamefont {P.~M.}\ \bibnamefont {Zerwas}},\
  }\href {\doibase 10.1007/s100520100808} {\bibfield  {journal} {\bibinfo
  {journal} {Eur. Phys. J.}\ }\textbf {\bibinfo {volume} {C22}},\ \bibinfo
  {pages} {563} (\bibinfo {year} {2001})},\ \bibinfo {note} {[Addendum: Eur.
  Phys. J.C23,769(2002)]},\ \Eprint {http://arxiv.org/abs/hep-ph/0108117}
  {arXiv:hep-ph/0108117 [hep-ph]} \BibitemShut {NoStop}%
\bibitem [{\citenamefont {Bae}\ \emph {et~al.}(2015)\citenamefont {Bae},
  \citenamefont {Baer},\ and\ \citenamefont {Serce}}]{Bae:2014yta}%
  \BibitemOpen
  \bibfield  {author} {\bibinfo {author} {\bibfnamefont {K.~J.}\ \bibnamefont
  {Bae}}, \bibinfo {author} {\bibfnamefont {H.}~\bibnamefont {Baer}}, \ and\
  \bibinfo {author} {\bibfnamefont {H.}~\bibnamefont {Serce}},\ }\href
  {\doibase 10.1103/PhysRevD.91.015003} {\bibfield  {journal} {\bibinfo
  {journal} {Phys. Rev.}\ }\textbf {\bibinfo {volume} {D91}},\ \bibinfo {pages}
  {015003} (\bibinfo {year} {2015})},\ \Eprint {http://arxiv.org/abs/1410.7500}
  {arXiv:1410.7500 [hep-ph]} \BibitemShut {NoStop}%
\bibitem [{\citenamefont {Baer}\ \emph
  {et~al.}(2016{\natexlab{b}})\citenamefont {Baer}, \citenamefont {Barger},
  \citenamefont {Savoy},\ and\ \citenamefont {Tata}}]{Baer:2016usl}%
  \BibitemOpen
  \bibfield  {author} {\bibinfo {author} {\bibfnamefont {H.}~\bibnamefont
  {Baer}}, \bibinfo {author} {\bibfnamefont {V.}~\bibnamefont {Barger}},
  \bibinfo {author} {\bibfnamefont {M.}~\bibnamefont {Savoy}}, \ and\ \bibinfo
  {author} {\bibfnamefont {X.}~\bibnamefont {Tata}},\ }\href {\doibase
  10.1103/PhysRevD.94.035025} {\bibfield  {journal} {\bibinfo  {journal} {Phys.
  Rev.}\ }\textbf {\bibinfo {volume} {D94}},\ \bibinfo {pages} {035025}
  (\bibinfo {year} {2016}{\natexlab{b}})},\ \Eprint
  {http://arxiv.org/abs/1604.07438} {arXiv:1604.07438 [hep-ph]} \BibitemShut
  {NoStop}%
\bibitem [{\citenamefont {Baer}\ \emph {et~al.}(2005)\citenamefont {Baer},
  \citenamefont {Mustafayev}, \citenamefont {Profumo}, \citenamefont
  {Belyaev},\ and\ \citenamefont {Tata}}]{Baer:2005bu}%
  \BibitemOpen
  \bibfield  {author} {\bibinfo {author} {\bibfnamefont {H.}~\bibnamefont
  {Baer}}, \bibinfo {author} {\bibfnamefont {A.}~\bibnamefont {Mustafayev}},
  \bibinfo {author} {\bibfnamefont {S.}~\bibnamefont {Profumo}}, \bibinfo
  {author} {\bibfnamefont {A.}~\bibnamefont {Belyaev}}, \ and\ \bibinfo
  {author} {\bibfnamefont {X.}~\bibnamefont {Tata}},\ }\href {\doibase
  10.1088/1126-6708/2005/07/065} {\bibfield  {journal} {\bibinfo  {journal}
  {JHEP}\ }\textbf {\bibinfo {volume} {07}},\ \bibinfo {pages} {065} (\bibinfo
  {year} {2005})},\ \Eprint {http://arxiv.org/abs/hep-ph/0504001}
  {arXiv:hep-ph/0504001 [hep-ph]} \BibitemShut {NoStop}%
\bibitem [{\citenamefont {Baer}\ \emph
  {et~al.}(2016{\natexlab{c}})\citenamefont {Baer}, \citenamefont {Barger},\
  and\ \citenamefont {Savoy}}]{Baer:2015rja}%
  \BibitemOpen
  \bibfield  {author} {\bibinfo {author} {\bibfnamefont {H.}~\bibnamefont
  {Baer}}, \bibinfo {author} {\bibfnamefont {V.}~\bibnamefont {Barger}}, \ and\
  \bibinfo {author} {\bibfnamefont {M.}~\bibnamefont {Savoy}},\ }\href
  {\doibase 10.1103/PhysRevD.93.035016} {\bibfield  {journal} {\bibinfo
  {journal} {Phys. Rev.}\ }\textbf {\bibinfo {volume} {D93}},\ \bibinfo {pages}
  {035016} (\bibinfo {year} {2016}{\natexlab{c}})},\ \Eprint
  {http://arxiv.org/abs/1509.02929} {arXiv:1509.02929 [hep-ph]} \BibitemShut
  {NoStop}%
\bibitem [{\citenamefont {Abdallah}\ \emph {et~al.}(2005)\citenamefont
  {Abdallah} \emph {et~al.}}]{Abdallah:2003np}%
  \BibitemOpen
  \bibfield  {author} {\bibinfo {author} {\bibfnamefont {J.}~\bibnamefont
  {Abdallah}} \emph {et~al.} (\bibinfo {collaboration} {DELPHI}),\ }\href
  {\doibase 10.1140/epjc/s2004-02051-8} {\bibfield  {journal} {\bibinfo
  {journal} {Eur. Phys. J.}\ }\textbf {\bibinfo {volume} {C38}},\ \bibinfo
  {pages} {395} (\bibinfo {year} {2005})},\ \Eprint
  {http://arxiv.org/abs/hep-ex/0406019} {arXiv:hep-ex/0406019 [hep-ex]}
  \BibitemShut {NoStop}%
\bibitem [{\citenamefont {Abdallah}\ \emph {et~al.}(2009)\citenamefont
  {Abdallah} \emph {et~al.}}]{Abdallah:2008aa}%
  \BibitemOpen
  \bibfield  {author} {\bibinfo {author} {\bibfnamefont {J.}~\bibnamefont
  {Abdallah}} \emph {et~al.} (\bibinfo {collaboration} {DELPHI}),\ }\href
  {\doibase 10.1140/epjc/s10052-009-0874-9} {\bibfield  {journal} {\bibinfo
  {journal} {Eur. Phys. J.}\ }\textbf {\bibinfo {volume} {C60}},\ \bibinfo
  {pages} {17} (\bibinfo {year} {2009})},\ \Eprint
  {http://arxiv.org/abs/0901.4486} {arXiv:0901.4486 [hep-ex]} \BibitemShut
  {NoStop}%
\bibitem [{\citenamefont {Fox}\ \emph {et~al.}(2011)\citenamefont {Fox},
  \citenamefont {Harnik}, \citenamefont {Kopp},\ and\ \citenamefont
  {Tsai}}]{Fox:2011fx}%
  \BibitemOpen
  \bibfield  {author} {\bibinfo {author} {\bibfnamefont {P.~J.}\ \bibnamefont
  {Fox}}, \bibinfo {author} {\bibfnamefont {R.}~\bibnamefont {Harnik}},
  \bibinfo {author} {\bibfnamefont {J.}~\bibnamefont {Kopp}}, \ and\ \bibinfo
  {author} {\bibfnamefont {Y.}~\bibnamefont {Tsai}},\ }\href {\doibase
  10.1103/PhysRevD.84.014028} {\bibfield  {journal} {\bibinfo  {journal} {Phys.
  Rev.}\ }\textbf {\bibinfo {volume} {D84}},\ \bibinfo {pages} {014028}
  (\bibinfo {year} {2011})},\ \Eprint {http://arxiv.org/abs/1103.0240}
  {arXiv:1103.0240 [hep-ph]} \BibitemShut {NoStop}%
\bibitem [{\citenamefont {Bartels}\ \emph {et~al.}(2012)\citenamefont
  {Bartels}, \citenamefont {Berggren},\ and\ \citenamefont
  {List}}]{Bartels:2012ex}%
  \BibitemOpen
  \bibfield  {author} {\bibinfo {author} {\bibfnamefont {C.}~\bibnamefont
  {Bartels}}, \bibinfo {author} {\bibfnamefont {M.}~\bibnamefont {Berggren}}, \
  and\ \bibinfo {author} {\bibfnamefont {J.}~\bibnamefont {List}},\ }\href
  {\doibase 10.1140/epjc/s10052-012-2213-9} {\bibfield  {journal} {\bibinfo
  {journal} {Eur. Phys. J.}\ }\textbf {\bibinfo {volume} {C72}},\ \bibinfo
  {pages} {2213} (\bibinfo {year} {2012})},\ \Eprint
  {http://arxiv.org/abs/1206.6639} {arXiv:1206.6639 [hep-ex]} \BibitemShut
  {NoStop}%
\bibitem [{\citenamefont {Baltz}\ \emph {et~al.}(2006)\citenamefont {Baltz},
  \citenamefont {Battaglia}, \citenamefont {Peskin},\ and\ \citenamefont
  {Wizansky}}]{Baltz:2006fm}%
  \BibitemOpen
  \bibfield  {author} {\bibinfo {author} {\bibfnamefont {E.~A.}\ \bibnamefont
  {Baltz}}, \bibinfo {author} {\bibfnamefont {M.}~\bibnamefont {Battaglia}},
  \bibinfo {author} {\bibfnamefont {M.~E.}\ \bibnamefont {Peskin}}, \ and\
  \bibinfo {author} {\bibfnamefont {T.}~\bibnamefont {Wizansky}},\ }\href
  {\doibase 10.1103/PhysRevD.74.103521} {\bibfield  {journal} {\bibinfo
  {journal} {Phys. Rev.}\ }\textbf {\bibinfo {volume} {D74}},\ \bibinfo {pages}
  {103521} (\bibinfo {year} {2006})},\ \Eprint
  {http://arxiv.org/abs/hep-ph/0602187} {arXiv:hep-ph/0602187 [hep-ph]}
  \BibitemShut {NoStop}%
\bibitem [{\citenamefont {Abbiendi}\ \emph
  {et~al.}(2003{\natexlab{b}})\citenamefont {Abbiendi} \emph
  {et~al.}}]{Abbiendi:2002in}%
  \BibitemOpen
  \bibfield  {author} {\bibinfo {author} {\bibfnamefont {G.}~\bibnamefont
  {Abbiendi}} \emph {et~al.} (\bibinfo {collaboration} {OPAL}),\ }\href
  {\doibase 10.1140/epjc/s2003-01139-y} {\bibfield  {journal} {\bibinfo
  {journal} {Eur. Phys. J.}\ }\textbf {\bibinfo {volume} {C27}},\ \bibinfo
  {pages} {483} (\bibinfo {year} {2003}{\natexlab{b}})},\ \Eprint
  {http://arxiv.org/abs/hep-ex/0209068} {arXiv:hep-ex/0209068 [hep-ex]}
  \BibitemShut {NoStop}%
\bibitem [{\citenamefont {Wang}(2018)}]{yanichep}%
  \BibitemOpen
  \bibfield  {author} {\bibinfo {author} {\bibfnamefont {Y.}~\bibnamefont
  {Wang}},\ }in\ \href@noop {} {\emph {\bibinfo {booktitle} {{39th
  International Conference on High Energy Physics (ICHEP) }}}}\ (\bibinfo
  {year} {2018})\ p.\ \bibinfo {pages} {PoS(ICHEP2018)630}\BibitemShut
  {NoStop}%
\bibitem [{\citenamefont {Wang}(2019)}]{yanlcws18}%
  \BibitemOpen
  \bibfield  {author} {\bibinfo {author} {\bibfnamefont {Y.}~\bibnamefont
  {Wang}},\ }in\ \href@noop {} {\emph {\bibinfo {booktitle} {{Linear Collider
  Work Shop (LCWS2018) }}}}\ (\bibinfo {year} {2019})\BibitemShut {NoStop}%
\end{thebibliography}%

\end{document}
%